# Isospin-Asymmetry Dependence

of the

# Thermodynamic

# Nuclear Equation of State

in

# Many-Body Perturbation Theory

by

Corbinian Wellenhofer

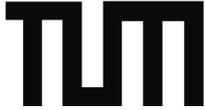
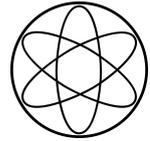

Technische Universität München
Physik Department
Institut für Theoretische Physik T39

# Isospin-Asymmetry Dependence

of the

# Thermodynamic

# Nuclear Equation of State

in

# Many-Body Perturbation Theory

Corbinian Wellenhofer

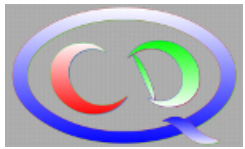 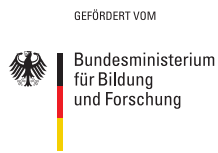 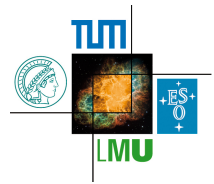 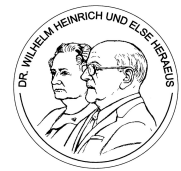

This work has been supported in part by the DFG and the NSFC through funds provided to the Sino-German CRC 110 "Symmetries and the Emergence of Structure in QCD", the BMBF, the Excellence Cluster "Origin and Structure of the Universe", and the Wilhelm und Else Heraeus-Stiftung.

# Abstract

The computation of the thermodynamic properties of nuclear matter is a central task of theoretical nuclear physics. The nuclear equation of state is an essential quantity in nuclear astrophysics and governs the properties of neutron stars and core-collapse supernovæ. The framework of chiral effective field theory provides the basis for the description of nuclear interactions in terms of a systematic low-energy expansion. In this thesis, we apply chiral two- and three-nucleon interactions in perturbative many-body calculations of the thermodynamic equation of state of infinite homogeneous nuclear matter. The conceptual issues that arise concerning the consistent generalization of the standard zero-temperature form of many-body perturbation theory to finite temperatures are investigated in detail. The structure of many-body perturbation theory at higher orders is examined, in particular concerning the role of the so-called anomalous contributions. The first-order nuclear liquid-gas phase transition is analyzed with respect to its dependence on temperature and the neutron-to-proton ratio. Furthermore, the convergence behavior of the expansion of the equation of state in terms of the isospin asymmetry is examined. It is shown that the expansion coefficients beyond the quadratic order diverge in the zero-temperature limit, implying a nonanalytic form of the isospin-asymmetry dependence at low temperatures. This behavior is associated with logarithmic terms in the isospin-asymmetry dependence at zero temperature.

# Zusammenfassung

Die Berechnung der thermodynamischen Eigenschaften der Kernmaterie ist ein grundlegendes Problem der theoretischen Kernphysik. Die nukleare Zustandsgleichung ist eine essentielle Größe der nuklearen Astrophysik und bestimmt die Eigenschaften von Neutronensternen und Kernkollaps-Supernovae. Die chirale effektive Feldtheorie stellt die Basis für die Beschreibung der Kernkräfte in der Form einer systematischen Entwicklung für niedrige Energieskalen dar. Ausgehend von chiralen Wechselwirkungen zwischen zwei und drei Nukleonen wird in der vorliegenden Arbeit die thermodynamische Zustandsgleichung von unendlicher gleichförmiger Kernmaterie mittels der Vielteilchenstörungstheorie berechnet. Die konzeptuellen Aspekte hinsichtlich der konsistenten thermodynamischen Verallgemeinerung der gewöhnlichen Form der Vielteilchenstörungstheorie bei verschwindender Temperatur werden ausführlich behandelt. Die Struktur der Vielteilchenstörungstheorie bei höhere Ordnung wird untersucht, insbesondere hinsichtlich der Rolle der sogenannten anomalen Beiträge. Der nukleare Phasenübergang erster Ordnung von einer Kernflüssigkeit zu einem wechselwirkenden Nukleonengas wird hinsichtlich seiner Abhängigkeit von der Temperatur und dem Verhältnis der Neutronen- und Protonendichten analysiert. Des Weiteren wird das Konvergenzverhalten der Entwicklung der Zustandsgleichung in der Isospin-Asymmetrie untersucht. Es wird gezeigt, dass die Entwicklungskoeffizienten höherer Ordnung im Grenzfall verschwindender Temperatur divergieren, was eine nichtanalytische Abhängigkeit von der Isospin-Asymmetrie impliziert. Dieses Verhalten geht mit logarithmischen Termen in der Abhängigkeit von der Isospin-Asymmetrie bei verschwindender Temperatur einher.

*"There is a fascination in dealing with nuclear processes, with nuclear matter with its tremendous density; a matter, however, that is inert on earth but is not inert at all in most other large accumulations of matter in the universe. The dynamics of nuclear matter is probably much more essential to the life of the universe than our terrestrial atomic and molecular physics. After all, what is that physics? It deals with the electron shells around nuclei that are only formed at very low temperatures on a few outlying planets where the conditions are just right–where the temperature is not too high, low enough to form those electron shells but high enough to have them react with each other. These conditions are possible only because of the nearness of a nuclear fire. Under the influence of that nuclear fire, self-reproducing units were formed here on earth. And after billions of years of benign radiation from the solar furnace, thinking beings evolved who investigate the processes that may be nearer to the heart of the universe than the daily world in which we live."*
– Victor F. Weisskopf

# Publications

Many of the results presented in this thesis have been obtained in collaboration with the coauthors of the articles listed below. The description of the results has been slightly altered and expanded in many cases, but often the previously published text is maintained to a certain degree.

- C. Wellenhofer, J. W. Holt, and N. Kaiser, *Divergence of the isospin-asymmetry expansion of the nuclear equation of state in many-body perturbation theory*, Phys. Rev. C, 93 (2016), p. 055802.

- C. Wellenhofer, J. W. Holt, and N. Kaiser, *Thermodynamics of isospin-asymmetric nuclear matter from chiral effective field theory*, Phys. Rev. C, 92 (2015), p. 015801.

- C. Wellenhofer, J. W. Holt, N. Kaiser, and W. Weise, *Nuclear thermodynamics from chiral low-momentum interactions*, Phys. Rev. C, 89 (2014), p. 064009.

# Contents

# Introduction

In the final stage of the life of a massive star, when all its internal nuclear fuel is exhausted, thermonuclear fusion in the stellar core comes to an end. The star can then no longer support itself against its own gravitational pressure, which leads to a sudden collapse of the core accompanied by the violent repulsion of the star's outer layers. The final product of this event—a core-collapse supernova—is a compact stellar remnant referred to as a neutron star.[1] With a mass that typically lies in the range of 1-2 solar masses ($M_\odot$) and a radius of only about 10-14 km, neutron stars are composed of the densest material known to exist in the universe: homogeneous nuclear matter,—a dense fluid of strongly-interacting nucleons (neutrons and protons),—the highly compressed and macroscopically extended form of the matter inside atomic nuclei. The present thesis is concerned with the study of the thermodynamic properties of this matter, i.e., with the computation and investigation of the nuclear equation of state (EoS).[2]

In neutron-star matter the electrostatic repulsion of protons is blocked by a charge-neutralizing background of electrons (and muons). This leads to the notion of (infinite) nuclear matter as the theoretical idealization of the matter inside atomic nuclei where finite-size effects are neglected and only the strong interaction is taken into account.[3] In this respect, (including the Coulomb energy) atomic nuclei can, as a first approximation, be modelled as self-bound liquid drops of nuclear matter [400]. In turn, some properties of nuclear matter can be inferred from extrapolating properties of atomic nuclei (bulk limit). The most well-established nuclear bulk property is the nuclear saturation point: isospin-symmetric nuclear matter at zero temperature should be self-bound at a nucleon density of $\rho_{\rm sat} \simeq 0.17\,{\rm fm}^{-3}$ and an energy per nucleon of $\bar E_{0,\rm sat} \simeq -16$ MeV (cf. e.g., Ref. [42]).

As for any fluid that is self-bound at low temperatures, (infinite) nuclear matter is subject to a liquid-gas phase transition. The instability of expanding (thermally excited) nuclear matter with respect to a phase separation has been linked to the underlying mechanism of multifragmentation reactions observed in intermediate-energy heavy-ion collision experiments [268, 23, 83, 299, 298, 426, 121, 93, 327, 194, 406, 280, 269].

Presumably the most prominent application of the nuclear EoS lies in the domain of astrophysics, i.e., in the modelling of neutron stars and in simulations of core-collapse supernovæ and binary neutron-star mergers. Since they are rotating and strongly magnetized objects, neutron stars emit beams of electromagnetic radiation that make them detectable as pulsars. The EoS of neutron-rich nuclear matter determines—via the Tolman-Oppenheimer-Volkoff equations [319, 393]—the mass-radius relation of neutron stars. In this respect, the recent observation and precise mass-measurement of two-solar-mass neutron stars—PSR J1614-2230 with a mass of $(1.97 \pm 0.04)\,M_\odot$ [100] and PSR J0348+0432 with $(2.01 \pm 0.04)\,M_\odot$ [11]—places strong constraints on the EoS of dense neutron-rich matter (at zero temperature).

---

[1] To be precise, neutron stars represent only one of the final stages in stellar evolution, i.e., for progenitor stars with masses (roughly) in the range $10 M_\odot \lesssim M \lesssim 25 M_\odot$. Stars with masses below about $10 M_\odot$ form white dwarfs at the end of their life cycle, and for stellar masses above about $25 M_\odot$ core collapse leads to a black hole or leaves no remnant (pair-instability supernova) [193].

The problem of calculating the nuclear EoS can (to a certain extent) be separated into two aspects: the description of the strong nuclear interactions, and the computation of the EoS from these interactions. Nowadays, the fundamental theory of the strong interaction is unambiguously considered to be the theory of quarks and gluons called quantum chromodynamics (QCD), and the view of the strong *nuclear* interaction is that of a large-distance (low-energy) residual interaction emerging (in an intricate and not yet completely understood way) from that fundamental theory. Related to this, QCD is a strongly-coupled theory at nuclear energy scales, and thus an approach towards the nuclear many-body problem that uses QCD *directly* is unfeasible (with the exception of numerical lattice simulations). Instead, an effective description of the (residual) strong nuclear interaction in terms of an appropriate large-distance approximation is needed.

A systematic approach based on general principles towards such an effective description of the nuclear interaction is provided by chiral effective field theory ($\chi$EFT), the effective field theory of low-energy QCD. In $\chi$EFT, the interactions of nucleons are organized in a hierarchical expansion that naturally includes multi-nucleon interactions. The low-energy constants parametrizing the short-range part of these interactions are generally fixed by fits to nucleon-nucleon scattering observables and properties of light nuclei. In that sense, (and with some further qualifications, cf. Secs. 1.4 and 1.5), employing $\chi$EFT interactions in many-body calculations amounts to a prediction of the nuclear EoS from an underlying (effective) microscopic theory.

Traditionally, the nuclear many-body problem has been complicated by the nonperturbative features (related to the presence of high-momentum components) of phenomenological (i.e., not based on general EFT principles) high-precision models of the nuclear interaction. In contrast, in an EFT the ultraviolet momentum cutoff $\Lambda$ is a variable parameter, and employing $\chi$EFT interactions with suitably low cutoffs enables the use of many-body perturbation theory. From a given large-cutoff model of the nuclear interaction (chiral or phenomenological), low-momentum interactions can also be derived by means of a renormalization-group (RG) evolution. These novel developments ($\chi$EFT, RG) have opened the way towards a systematic investigation of the nuclear many-body problem.

By now, chiral low-momentum interactions have been used in perturbative nuclear matter calculations by numerous authors [251, 106, 185, 191, 386, 91, 90, 107, 109]. In particular, in Ref. [251] it was found that many of the equations of state commonly used in nuclear astrophysics applications are inconsistent with perturbative $\chi$EFT-based calculations of the neutron-matter EoS. All of these calculations—except for Refs. [107, 109] where the zero-temperature EoS of isospin-asymmetric matter (ANM) was computed, and Ref. [394] where neutron matter at finite temperature was examined—have been restricted to zero temperature and either isospin-symmetric nuclear matter (SNM) or pure neutron matter (PNM). For astrophysical simulations of core-collapse supernovæ and proto-neutron star dynamics, however, a global thermodynamic nuclear EoS is required, i.e., an EoS that covers a wide range of temperatures, densities, and isospin-asymmetries (roughly, $0 \leq T \lesssim 100$ MeV, $0 \leq \rho \lesssim (4-6)\rho_{\text{sat}}$, and $0 \leq \delta \leq 1$; cf. Ref. [313]). The dependence of the EoS (given in terms of the free energy per particle) of homogeneous nuclear matter on the isospin asymmetry $\delta = (\rho_n - \rho_p)/\rho$ (with $\rho_{n/p}$ the neutron/proton density, and $\rho = \rho_n + \rho_p$ the total nucleon density; $T$ is the temperature) can, as a first approximation, be assumed to have a parabolic form:

$$\bar{F}(T, \rho, \delta) \simeq \bar{F}(T, \rho, \delta = 0) + \bar{F}_{\text{sym}}(T, \rho)\,\delta^2.$$

---

[3] In this thesis, we use the expression *equation of state* to denote the set of relations between thermodynamic variables that specify the complete thermodynamic information about the nuclear many-body system in equilibrium; for the most part of this thesis, this corresponds to the free energy per particle of infinite homogeneous nuclear matter as a function of its natural variables: $\bar{F}(T, \rho_\text{n}, \rho_\text{p})$, or equivalently, $\bar{F}(T, \rho, \delta)$.

Within this approximation, the isospin-asymmetry dependence of the EoS is associated with the difference of the free energy per particle of SNM ($\delta = 0$) and PNM ($\delta = 1$), i.e., with $\bar{F}_{\text{sym}}(T, \rho) := \bar{F}(T, \rho, \delta = 1) - \bar{F}(T, \rho, \delta = 0)$. The quantity $\bar{F}_{\text{sym}}$ is called the *symmetry free energy*. Results from fits to nuclear binding energies and various nuclear many-body calculations indicate an indeed approximately quadratic dependence on $\delta$ of the free energy per particle (at zero temperature), and a quadratic dependence is routinely assumed in studies of (e.g.,) neutron-star properties and neutron-rich atomic nuclei [192, 386, 189, 190, 371, 78]. Even so, it has been shown that higher-order corrections with respect to the isospin-asymmetry dependence can still have a significant influence on various properties of (e.g.,) neutron stars [359, 368, 68]. A particular objective of the present thesis is therefore to examine in more detail the accuracy of the parabolic isospin-asymmetry approximation for different temperatures and densities, and to investigate the question how improved parametrizations of the isospin-asymmetry dependence can be constructed.

Nevertheless, based on its observed reasonable accuracy at zero temperature, the parabolic approximation sets the strategy for the investigation of the thermodynamic nuclear EoS. After having set up the computational framework, the first step is to study the nuclear EoS for the limiting cases SNM ($\delta = 0$) and PNM ($\delta = 1$). Then, one extracts the symmetry free energy $\bar{F}_{\text{sym}}(T, \rho)$ and examines its density and temperature dependence. This sets the basis for a detailed study of the isospin-asymmetry dependence of the EoS. In detail, the present thesis is structured as follows.

- In **Chapter 1** we summarize the main properties of QCD relevant for nuclear energy scales. This then leads us to introduce the general framework of $\chi$EFT, which provides the basis for our examination of various methods to construct effective low-momentum nuclear potentials. The application of these potentials in nuclear many-body calculations is discussed, and finally, we take a look at recent research results in the nuclear many-body problem.

- **Chapter 2** starts with a short overview of the standard formulation of many-body perturbation theory (MBPT) at zero temperature ($T = 0$). The consistent generalization of MBPT to finite temperatures is (in fact) nontrivial, and is investigated in the remainder of the chapter. The structure of finite-temperature MBPT at higher orders is studied, in particular, we investigate the cancellation of the so-called anomalous contributions in terms of the self-consistent renormalization of the single-particle basis.

- In **Chapter 3** we begin with the actual numerical nuclear many-body calculations. We compute the thermodynamic EoS of isospin-symmetric nuclear matter (SNM) using various sets of chiral low-momentum (two- and three-nucleon) potentials in second-order

---
[3] Since the nuclear interaction is finite-ranged, the thermodynamic limit exists (only) in case where the electrostatic repulsion of protons is "switched off", cf. Refs. [179, 306, 307, 271, 402] and also [112, 84].

MBPT. We examine the model dependence of the results, and benchmark against available empirical constraints. The sets of potentials which lead to consistent results are then used to compute the EoS of pure neutron matter (PNM). Finally, the symmetry free energy is extracted, and its density and temperature dependence is studied.

- In **Chapter 4** we first review the general principles involved in the thermodynamic analysis of the nuclear liquid-gas phase transition, and then examine the dependence of the associated instability region on temperature $T$ and the isospin asymmetry $\delta = (\rho_n - \rho_p)/\rho$. In particular, we determine the trajectories of the critical temperature $T_c(\delta)$ and the fragmentation temperature $T_F(\delta)$.

- Finally, in **Chapter 5** we investigate in detail the dependence of the nuclear EoS (as obtained in MBPT) on the isospin asymmetry $\delta$ for different temperatures $T$ and nucleon densities $\rho$. For the most part, we focus on the investigation of the Maclaurin expansion of the free energy per particle $\bar{F}(T, \rho, \delta)$ in terms of $\delta$. It is shown that this expansion constitutes a divergent asymptotic series at low temperatures, implying a non-polynomial form of the isospin-asymmetry dependence in this regime. Moreover, it is shown that at zero temperature this nonanalytic behavior is associated with logarithmic terms of the form $\delta^{2n \geq 4} \ln|\delta|$. The coefficient of the leading logarithmic term $\delta^4 \ln|\delta|$ is extracted, and it is shown that the inclusion of this terms overall leads to a considerably improved description of the isospin-asymmetry dependence of the zero-temperature EoS. Furthermore, we identify additional nonanalyticities of the $\delta$ dependence at vanishing proton fraction.

The thesis is concluded thereafter. Three appendices are attached; in particular, in the first appendix we provide a detailed discussion of the noninteracting nucleon gas, including the application of MBPT for the derivation of relativistic correction terms.

# 1. Nuclear Interactions and Many-Body Problem

The modern theory of nuclear interactions is based on chiral effective field theory ($\chi$EFT), the large-distance realization of the (fundamental) theory of the strong interactions of quarks and gluons ("color force") called quantum chromodynamics (QCD). A short overview of QCD is given in **Section 1.1** of this chapter. At energy scales $Q \lesssim 1$ GeV, QCD is strongly-coupled and features highly nonperturbative phenomena such as color confinement and spontaneous chiral symmetry breaking. Color confinement implies that quarks and gluons cannot appear as isolated particles. Instead, the effective degrees of freedom at large length scales are hadrons: color-neutral bound-states of quarks and gluons, such as the neutron, the proton, and in particular also the pions (the Nambu-Goldstone bosons of the spontaneously broken chiral symmetry).

Since it is an emergent strong-coupling phenomenon, an *explicit* description in terms of QCD of the nuclear interaction appears unfeasible. However, at sufficiently low energy scales the quark-gluon substructure of nucleons is not resolved. This means that they can be described as point particles, and a Lagrangian that governs their interactions can be set up in terms of a low-energy expansion. The general principles involved in the construction of this Lagrangian (the Lagrangian of $\chi$EFT) are discussed briefly in **Sec. 1.2**. In **Sec. 1.3** we then examine the hierarchy of effective two- and multi-nucleon interactions that emerges from $\chi$EFT.

In an EFT, the ultraviolet momentum cutoff $\Lambda$ is a variable parameter. Employing low cutoff scales $\Lambda \lesssim 450$ MeV has the benefit that the nonperturbative short-distance features of the nuclear interactions (which are otherwise necessary for a realistic description) are evaded. Different methods to construct such low-momentum interactions are outlined in **Sec. 1.4**, where we also present the sets of chiral two- and three-nucleon interactions used in this thesis. The application of these interactions in nuclear many-body calculations is discussed in **Sec. 1.5**. In **Sec. 1.5**, we also look at results from recent research activities that serve as a motivation for and a guide towards the subsequent chapters.

## 1.1. Quantum Chromodynamics

QCD is a quantum gauge field theory with gauge group SU(3)$_{\text{color}}$ ("color group"). It describes the dynamics of massive spin-1/2 fermions called quarks and massless spin-1 bosons called gluons, the SU(3)$_{\text{color}}$ gauge bosons. Historically, QCD emerged from the *quark Model* [160, 436, 437] in which hadrons are classified as bound-states of *three quarks* (baryons) or *quark-antiquark pairs* (mesons) according to the representations of the "flavor group" SU(3)$_F$.[1] In particular, the nucleon forms an isospin doublet with respect to the SU(2) subgroup of SU(3)$_F$. In its modern understanding, the dynamics of hadrons is understood as an emergent phenomenon of strongly-coupled QCD, and their classification scheme according to SU(3)$_F$ is traced back to the underlying flavor symmetry of the QCD Lagrangian. In the following, we briefly describe how this comes about.

*1. Nuclear Interactions and Many-Body Problem*

## 1.1.1. Asymptotic Freedom and Color Confinement

Related to the picture of hadrons as quark-gluon bound-states are two key properties of QCD: *asymptotic freedom* and *color confinement*. Asymptotic freedom is fundamental for the application of perturbative QCD in high-energy physics, and the property of color confinement is the quintessence of the emergence of nuclear and hadronic physics from QCD. In the following, after having introduced the QCD Lagrangian, we briefly discuss these two features, and their relation to the renormalization-group flow of the strong coupling constant *g*.

***QCD Lagrangian***. The QCD Lagrangian $\mathscr{L}_{\text{QCD}}$ is uniquely determined by the requirement of local gauge invariance under the nonabelian group $\text{SU}(3)_{\text{color}}$ together with the general principles of relativistic quantum field theory and the criterion of renormalizability. It is given by[2]

$$\mathscr{L}_{\text{QCD}} = \sum_{f=u,d,s,c,b,t} \bar{\psi}_f \left( i\slashed{D} - m_f \right) \psi_f - \frac{1}{4} \mathcal{G}^a_{\mu\nu} \mathcal{G}^{\mu\nu}_a, \quad (1.1)$$

where where $\slashed{D} = \gamma^\mu D_\mu$, with $\gamma^\mu$ the usual Dirac matrices. For each flavor $f \in \{u, d, s, c, b, t\}$ the quark field $\psi_f$ has three color components (corresponding to the fundamental triplet representation of $\text{SU}(3)_{\text{color}}$), i.e., $\psi_f = (\psi_f^{\text{red}}, \psi_f^{\text{blue}}, \psi_f^{\text{green}})$. The gauge-covariant derivative $D_\mu$ is given by

$$D_\mu = \partial_\mu - igA_\mu, \qquad A_\mu = t_a A^a_\mu, \quad (1.2)$$

where $A^a_\mu$, $a \in \{1, \dots, 8\}$, are the eight gluon fields (corresponding to the adjoint representation of $\text{SU}(3)_{\text{color}}$), and $t_a = \lambda_a/2$ are the generators of $\text{SU}(3)_{\text{color}}$ (with $\lambda_a$ the usual Gell-Mann matrices). The parameter *g* is the strong coupling constant. The gluon field strength tensor $\mathcal{G}^a_{\mu\nu}$ is given by

$$\mathcal{G}^a_{\mu\nu} = \partial_\mu A^a_\nu - \partial_\nu A^a_\mu + g f^{abc} A_{\mu,b} A_{\nu,c}. \quad (1.3)$$

where $f^{abc}$ are the structure constants of the Lie algebra $\mathfrak{su}(3)$. Note that the last term in Eq. (1.3) would be absent in an abelian gauge theory. It gives rise to gluonic self-interactions, i.e., in perturbation theory there are vertices involving three or four gluons, as shown in Fig. 1.1.

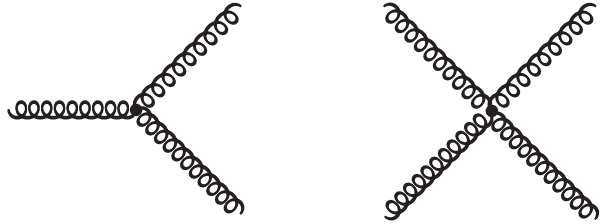

**Figure 1.1.:** Three-gluon vertex (order *g*) and four-gluon vertex (order $g^2$)

---

[1] Note that the *quarks* ("constituent quarks") associated with this classification scheme have to be distinguished from the "elementary" quarks of high-energy QCD. The *quarks* associated with the *quark model* constitute only an approximative (but very useful) concept. Instead, hadrons themselves must be regarded as the large-distance realization of QCD; they are bound-states of strongly-interacting quarks and gluons, and their structure is more involved than the *quark* model assumes. In particular, "exotic" hadrons which are not allowed in the *quark* model have been observed [316].

[2] The so-called $\Theta$ term associated with the strong CP problem [21, 337, 413] is omitted here (it is not relevant for the present discussion). We note also that if the criterion of renormalizability is given up, additional terms are possible, but these terms would become relevant only at very high energies [414, 418, 413, 412].

***Asymptotic Freedom***. In addition to invariance under (local) gauge transformations as well as Poincaré and CPT transformations, for $m_f = 0$ the QCD action $\int d^4 x \mathscr{L}_{\text{QCD}}$ is invariant under the rescaling

$$x^\mu \to \lambda x^\mu, \quad A \to \lambda^{-1} A, \quad \psi \to \lambda^{-3/2} \psi. \tag{1.4}$$

This scale invariance is however present only at the classical level but not in the full quantum theory. The QCD vacuum can be pictured as a medium populated by "virtual" particle-antiparticle fluctuations, and the properties of particles (excitations of the vacuum) are influenced by this medium, leading to the general feature of a quantum field theory that the values of the coupling constants (and any other parameters of the theory) depend on the energy scale $Q$ at which they are measured.

In quantum electrodynamics (i.e., in an *abelian* U(1) gauge theory) the strength of the fields produced by a charged particle is also modified by such "virtual" effecs (*vaccum polarization*), and the effective charge is screened at large distances, i.e., its value decreases with distance. In QCD, however, the coupling $g$ does not increase with decreasing distance but instead tends to zero at very short length scales (this feature is also known as *antiscreening*): the theory is *asymptotically free*. Asymptotic freedom is a distinct feature of relativistic *nonabelian* gauge theories [87].

The scale dependence (renormalization-group flow) of $g$ is described by the beta function (cf. e.g, [305, 89])

$$\beta(g) = \frac{\partial g(Q)}{\partial \ln Q}. \tag{1.5}$$

Eq. (1.5) is called the *renormalization-group equation* for the strong coupling constant. In the regime where the coupling $g$ is weak, the beta function can be calculated in perturbation theory. The leading term in the expansion $\beta(g) = \beta_0 g^3 + \beta_1 g^5 + \ldots$ is given by [175, 176, 177, 333]

$$\beta_0 = -\frac{1}{16\pi^2}\left(11 - \frac{2}{3}N_f\right), \tag{1.6}$$

where $N_f \in [3, 6]$ is the number of flavors that have to be taken into account at a given energy scale. Setting $\beta(g) = \beta_0 g^3$, Eq. (1.5) can be integrated, yielding the leading-order result for the running of the QCD fine-structure "constant" $\alpha_s = g^2/4\pi$:

$$\boxed{\alpha_s(Q) = \frac{\alpha_s(Q_0)}{1 - 4\pi\beta_0 \alpha_s(Q_0) \ln\left(\frac{Q^2}{Q_0^2}\right)}} \tag{1.7}$$

This equation defines the value of $\alpha_s$ at a given energy scale $Q$, based on a normalization point $Q_0$ for which its value $\alpha_s(Q_0)$ needs to be inferred from experimental measurements. For $N_f < 33/2$ the beta function is negative, thus, for QCD asymptotic freedom is realized: the QCD fine-structure constant $\alpha_s(Q)$ decreases logarithmically with increasing $Q$. The scale dependence of $\alpha_s(Q)$ is depicted in Fig. 1.2.

*1. Nuclear Interactions and Many-Body Problem*

***Color Confinement***. As seen above, the strong coupling constant increases with decreasing energy scales. The consequence of this is that the dynamics of quarks and gluons becomes very complicated (and cannot be treated in perturbation theory) at low energies $Q \lesssim 1\,\text{GeV}$ or, equivalently, large length scales $\ell \gtrsim 0.2\,\text{fm}$. Associated with this *nonperturbative* regime of QCD is the **confinement of quarks and gluons into color-neutral bound-states (the hadrons).** In particular, quarks and gluons do not exist as isolated particles, and thus can be observed only indirectly.[3]

The confining feature of QCD can be visualized by the following picture: as one increases the distance between two quarks, narrow color flux lines are formed between the two quarks, which tend to hold them together, and at a certain point break, leading to the formation of quark-antiquark pairs, which together with the original two quarks then form hadrons. This process (hadronization) is observed in high-energy collision experiments in the form of "jets": narrow cones of hadrons emitted from the collision center. Further evidence for confinement comes from the large-$N_c$ approximation of QCD [379, 422, 86], from the study of QCD-type theories in lower space-time dimensions [380], and from numerical studies (in terms of lattice QCD simulations [316]) of the heavy quark-antiquark ($Q\bar{Q}$) system where the $Q\bar{Q}$ potential was observed to be Coulombic at short distances and linearly increasing at long distances [17]:

$$V_{Q\bar{Q}} \sim -\frac{e}{r} + \sigma\, r. \quad (1.8)$$

The linear term proportional to $\sigma$ (the "string tension") renders the separation of the $Q\bar{Q}$ pair energetically impossible.

Although various qualitative ideas have been conceived,[4] the precise mechanism for color confinement is still not fully understood at a theoretical level. In particular, no mathematical proof (associated with the "mass-gap problem" [140]) that QCD is confining exists.

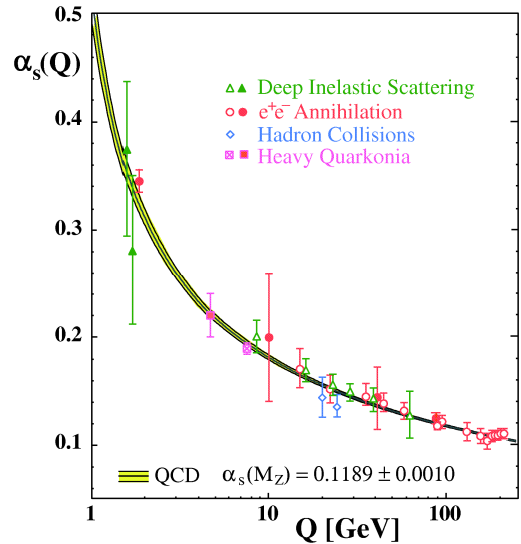

**Figure 1.2.:** (From [43]) Scale dependence of the QCD fine-structure constant $\alpha_s(Q)$ established by various types of measurements, compared to the QCD prediction based on $\alpha_s(M_Z) = 0.1189 \pm 0.00010$, with $M_Z \simeq 91.2\,\text{GeV}$ the Z-boson mass.

---

[3] See e.g., Refs. [316, 305] for discussions of indirect evidence (from the analysis of data from high-energy collision experiments) for quarks and gluons.

[4] Another noteworthy one is the formation of chromoelectric flux tubes between color charges (dual Meissner effect) [303, 284, 382]. For further ideas about confinement, cf. [334, 420] and the reviews [10, 285, 363].

## 1.1.2. Spontaneous Chiral Symmetry Breaking

In addition to confinement, the nonperturbative low-energy regime of QCD features another property that is essential for nuclear and hadronic physics: the spontaneous breaking of (approximate) chiral symmetry (S$\chi$SB).[5] Here, we briefly describe this feature and its relation to the properties of the hadron spectrum, in particular nucleons and pions.

*Chiral Limit.* The QCD Lagrangian has an approximate global U(2)$_L$ × U(2)$_R$ symmetry in flavor space, where "approximate" means that the symmetry is an exact symmetry of $\mathscr{L}_{\text{QCD}}$ in the *chiral limit* where the masses of the two lightest quarks ($m_u = 2.3^{+0.7}_{-0.5}$ MeV and $m_d = 4.8^{+0.5}_{-0.3}$ MeV at a scale $Q \simeq 2\text{GeV}$ [316]) are set to zero, i.e., $m_f = 0$ for $f = u, d$.[6] Writing $\psi = (\psi_u, \psi_d)$, the chiral limit of the QCD Lagrangian (restricted to $f = u, d$) reads

$$\mathscr{L}_{\text{QCD}} \xrightarrow{\text{chiral limit}} \bar{\psi}_L i\slashed{D} \psi_L + \bar{\psi}_R i\slashed{D} \psi_R - \frac{1}{4}\mathcal{G}^a_{\mu\nu}\mathcal{G}^{\mu\nu}_a, \tag{1.9}$$

where the two (left- and right-handed) chiral components of the quark fields $\psi_{L/R} = \frac{1}{2}(1 \mp \gamma_5)\psi$ have been separated. The Lagrangian given by Eq. (1.9) is invariant under the global transformations

$$\begin{pmatrix}\psi_L \\ \psi_R\end{pmatrix} \xrightarrow{\text{U(2)}_L} \begin{pmatrix}L\psi_L \\ \psi_R\end{pmatrix}, \qquad \begin{pmatrix}\psi_L \\ \psi_R\end{pmatrix} \xrightarrow{\text{U(2)}_R} \begin{pmatrix}\psi_L \\ R\psi_R\end{pmatrix}, \tag{1.10}$$

where $L, R \in $ U(2). This U(2)$_L$ × U(2)$_R$ symmetry can be decomposed as U(1)$_L$ × U(1)$_R$ × SU(2)$_L$ × SU(2)$_R$, where the transformations under the subgroups are given by Eq. (1.9) but with $L, R \in $ U(1) and $L, R \in $ SU(2), respectively. Furthermore, the abelian part U(1)$_L$ × U(1)$_R$ is isomorphic to U(1)$_V$ × U(1)$_A$. Here, the subscript "*V*" denotes a *vector symmetry* (a symmetry that does not distinguish between the left- and right-handed components of the quark field), and the subscript "*A*" labels an *axial symmetry* (a symmetry which treats fields with different chirality in an opposite way). The transformations under U(1)$_V$ and U(1)$_A$ are given by

$$\psi_{L,R} \xrightarrow{\text{U(1)}_V} \exp(i\Theta)\psi_{L,R}, \qquad \psi_{L,R} \xrightarrow{\text{U(1)}_A} \exp(i\Theta\gamma_5)\psi_{L,R}. \tag{1.11}$$

The group **SU(2)$_L$ × SU(2)$_R$** is generally called the **chiral symmetry group**. A transformation under its diagonal subgroup SU(2)$_V$ is given by

$$\begin{pmatrix}\psi_L \\ \psi_R\end{pmatrix} \xrightarrow{\text{SU(2)}_V} \begin{pmatrix}V\psi_L \\ V\psi_R\end{pmatrix}, \tag{1.12}$$

where $V \in $ SU(2). Note that U(1)$_V$ (corresponding to baryon number conservation[7]) is a symmetry of the Lagrangian also for physical quark masses, $m_u \neq m_d \neq 0$. U(1)$_A$, SU(2)$_L$ and SU(2)$_R$ are exact symmetries of $\mathscr{L}_{\text{QCD}}$ only in the chiral limit, but SU(2)$_V$ (the isospin symmetry) is exact also in the **isospin limit** of equal quark masses $m_u = m_d \neq 0$.

---

[5] Color confinement and spontaneous chiral symmetry breaking (S$\chi$SB) are thought to be connected [22, 75] (cf. also [362]), but their exact relation is not yet fully understood.

[6] It should be noted that this analysis can be extended to include the strange quark; i.e., setting $m_f = 0$ for $f = u, d, s$, a global U(3)$_L$ × U(3)$_R$ symmetry emerges, the SU(3)$_V$ subgroup of which is the flavor group of the *quark* model. Since in this thesis we are concerned with *nucleonic* matter (only), we restrict the discussion to the two-flavor case. The interactions of strange hadrons in nuclear matter are of interest in astrophysics [180, 169], and their description in terms of SU(3) $\chi$EFT is pursued actively [329, 330].

***Symmetry Breaking Pattern***. Regarding the (approximate) global U(2)$_L$ × U(2)$_R$ symmetry, QCD has the following symmetry breaking pattern

$$\begin{aligned}
\text{U}(2)_L \times \text{U}(2)_R &= \text{U}(1)_V \times \text{U}(1)_A \times \overbrace{\text{SU}(2)_L \times \text{SU}(2)_R}^{\text{chiral symmetry}} \\
&\xrightarrow{\text{anomaly}} \text{U}(1)_V \times \text{SU}(2)_L \times \text{SU}(2)_R \\
&\xrightarrow{\text{spontaneous}} \text{U}(1)_V \times \text{SU}(2)_V
\end{aligned} \quad (1.13)$$

The group $U(1)_A$, although it is a symmetry of the QCD Lagrangian (i.e., it is a symmetry at the classical level), it is not a symmetry of QCD (i.e., the fully quantum theory). Its breaking is due to the Adler-Bell-Jackiw anomaly of the axial current [7, 36] and instantons [381, 413].

At low energies, **chiral symmetry is "spontaneously broken" to SU(2)$_V$**: SU(2)$_L$×SU(2)$_R$ is a symmetry of QCD (i.e., the full quantum theory), but only its diagonal subgroup SU(2)$_V$ is manifest as a symmetry of the physical states present at low energies (i.e., the low-energy vacuum $|0\rangle$ and the hadron spectrum). In particular, the low-energy vacuum can be considered to be populated by quark-antiquark fluctuations, and the quark-antiquark correlator $\bar{\psi}\psi = \bar{\psi}_u\psi_u + \bar{\psi}_d\psi_d$ has a nonvanishing vacuum expectation value $\langle 0|\bar{\psi}\psi|0\rangle$, the so-called chiral condensate, which serves as an order parameter for S$\chi$SB [since $\bar{\psi}\psi = \bar{\psi}_L\psi_R + \bar{\psi}_R\psi_L$ is invariant under SU(2)$_V$ but not under SU(2)$_L$ × SU(2)$_R$]. See (e.g.,) Refs. [86, 413, 390] and the appendix A.2 for more details.

***Pions as Nambu-Goldstone Bosons***. A quantum field theory with a spontaneously broken symmetry has to obey the Nambu-Goldstone theorem [172, 173, 302, 413]:

> If there is a continuous symmetry group $G$ (with generators $Q_i$, $i = 1, ..., N_G$) under which the Lagrangian is invariant, but the vacuum is invariant only under a subgroup $H \subset G$ (with generators $Q_i$, $i = 1, ..., N_H$), then there must exist $N_G - N_H$ massless states with zero spin.

These Nambu-Goldstone bosons then carry the quantum numbers of the "broken" charges $Q_i$, $i = N_H + 1, ..., N_G$. This theorem has important implications for the hadron spectrum. To see this, we consider the Noether currents $j^a_{L,\mu}$ and $j^a_{R,\mu}$ associated with the components $SU(2)_L$ and $SU(2)_R$ of the chiral symmetry group:

$$j^a_{L,\mu} = \bar{\psi}_L(x)\gamma_\mu \frac{\tau^a}{2}\psi_L(x), \qquad j^a_{R,\mu} = \bar{\psi}_R(x)\gamma_\mu \frac{\tau^a}{2}\psi_R(x), \quad (1.14)$$

with $\boldsymbol{\tau} = (\tau^1, \tau^2, \tau^3)$ the isospin Pauli matrices. The vector and axial-vector currents are given by

$$j^a_{V,\mu} = j^a_{R,\mu} + j^a_{L,\mu} = \bar{\psi}(x)\gamma_\mu \frac{\tau^a}{2}\psi(x), \qquad j^a_{A,\mu} = j^a_{R,\mu} - j^a_{L,\mu} = \bar{\psi}(x)\gamma_\mu\gamma_5 \frac{\tau^a}{2}\psi(x), \quad (1.15)$$

and the associated charges are given by

$$Q^a_V = \int d^3x\, \psi^\dagger(x)\frac{\tau^a}{2}\psi(x), \qquad Q^a_A = \int d^3x\, \psi^\dagger(x)\gamma_5 \frac{\tau^a}{2}\psi(x). \quad (1.16)$$

---

[7] We note that in the Standard Model (QCD plus electroweak theory) the U(1)$_V$ symmetry is anomalously broken by nonperturbative effects in the electroweak sector ("sphalerons"), cf. [252, 240, 413]

In the case of S$\chi$SB the "broken" generators are the axial charges $Q_A^a$. Hence, the associated Nambu-Goldstone bosons are pseudoscalar mesons, identified with the pions $\pi^+$, $\pi^-$, and $\pi^0$. Because chiral symmetry is also *explicitly broken* by the nonvanishing masses of up and down quarks, pions have a small (compared to the other hadrons) but nonzero mass. In this sense they are *pseudo Nambu-Goldstone bosons*. At small momenta they are weakly interacting particles. This can be seen by considering the action of the Hamiltonian $\mathcal{H}_{\text{QCD}}$ on a state $|\pi^n\rangle$ with $n$ pions:

$$\mathcal{H}_{\text{QCD}} |\pi^n\rangle \sim \mathcal{H}_{\text{QCD}} \left(Q_A^{a_1} \cdots Q_A^{a_n}\right) |0\rangle = \left(Q_A^{a_1} \cdots Q_A^{a_n}\right) \mathcal{H}_{\text{QCD}} |0\rangle = 0. \qquad (1.17)$$

Thus, in the case of vanishing momenta, the $n$-pion state is energetically degenerate with the vacuum $|0\rangle$.

The special nature of the pions as pseudo Nambu-Goldstone bosons is manifest in the Gell-Mann–Oakes–Renner relation [162], which connects the chiral condensate, the quark masses, the pion mass $m_\pi \simeq 138$ MeV, and the pion decay constant $f_\pi \simeq 92.4$ MeV:

$$m_\pi^2 f_\pi^2 = -\frac{1}{2}(m_u + m_d) \langle \bar{\psi}\psi \rangle + O(m_{u,d}^2). \qquad (1.18)$$

Eq. (1.18) implies that, just like the chiral condensate, the pion decay constant $f_\pi$ is a measure for the strength of S$\chi$SB. The associated characteristic scale is $\Lambda_\chi \sim 4\pi f_\pi \sim 1$ GeV (cf. e.g., Refs. [413, 356] for more details).

## 1.2. Chiral Effective Field Theory

In the following we briefly describe the general principles involved in the construction of the Lagrangian of $\chi$EFT, and give its leading terms. More detailed reviews on $\chi$EFT and the general principles of EFTs can be found in Refs. [130, 356, 40, 119, 283, 157, 267, 413, 305, 332]. Additional details are also given in the appendix A.2.

*General Principles.* The notion *effective field theory* (EFT) corresponds to a general quantum-field theoretical framework that yields, for a given system, an effective (i.e., not fundamental) description in terms of the degrees of freedom which are "active" at low energies. In most cases, constructing an EFT corresponds to "integrating out" the details of the high-energy physics: in the low-energy regime, short-distance effects are not resolved and can be included through contact terms. By definition, an EFT is useful only for energies below a characteristic breakdown scale $\Lambda_{\text{EFT}}$ above which the high-energy physics becomes relevant.

The general principles governing the construction of an EFT can be formulated as follows [283, 356]:

- Identify the relevant energy scales and appropriate degrees of freedom.
- Identify the relevant symmetries and investigate if and how they are broken.
- Construct the most general Lagrangian consistent with these symmetries and symmetry breaking patterns.
- Design an organizational scheme that can distinguish between more and less important terms.
- Evaluate Feynman diagrams for the process under consideration up to the desired accuracy.

# 1. Nuclear Interactions and Many-Body Problem

In general, an EFT is not renormalizable in the sense that the divergences occuring in perturbation theory (via loop diagrams) can be removed by a redefinition of a finite number of parameters. Instead, an EFT has to be renormalized order by order, and additional counterterms (contact terms) are needed at each order. The unknown constants parametrizing the contact terms are called **low-energy constants (LECs)**, and the values of these LECs have to be fixed either by a fit to experimental data or by matching their values to an underlying high-energy theory. See e.g., Refs. [412, 413, 40] for more details.

*$\chi$EFT Lagrangian.* Since the interactions between pions vanish at zero momentum transfer (in the chiral limit) and the pion mass signifies the explicit breaking of chiral symmetry (cf. Sec. 1.1.2), the Lagrangian of $\chi$EFT is arranged in powers of derivatives of pion fields and powers of pion masses. Formally, it can be written as

$$\mathscr{L}_{\chi\text{EFT}} = \mathscr{L}_{\pi\pi} + \mathscr{L}_{\pi N} + \mathscr{L}_{NN} + \ldots, \tag{1.19}$$

where $\mathscr{L}_{\pi\pi}$ describes the dynamics of pions ($\pi$) without nucleons (N), $\mathscr{L}_{\pi N}$ collects $\pi$N interactions, $\mathscr{L}_{NN}$ the NN contact terms, etc.; the ellipsis represents terms with three or more nucleon fields. The relativistic treatment of nucleons leads to certain problems (the time-derivative of the relativistic nucleon field generates the large factor $E \simeq M$), which can be avoided by treating nucleons as heavy static sources [222, 39] ("extreme nonrelativistic limit", cf. the appendix A.2 for details). For the organization of nuclear interactions it is useful to organize the various terms of the effective Lagrangian according to the so-called "index of interaction" $\Delta$, which is given by

$$\Delta = d + \frac{n}{2} - 2, \tag{1.20}$$

where the "chiral dimension" $d$ counts the number of derivatives or pion mass insertions, and $n$ is the number of nucleon fields. The leading-order part of $\mathscr{L}_{\text{EFT}}$ in this organization scheme reads[8]

$$\begin{aligned}\mathscr{L}_{\chi\text{EFT}}^{\Delta=0} =& \frac{1}{2}\partial_\mu\boldsymbol{\pi}\cdot\partial^\mu\boldsymbol{\pi} - \frac{1}{2}m_\pi^2\boldsymbol{\pi}^2 + \frac{1-4\alpha}{2f_\pi^2}\left(\boldsymbol{\pi}\cdot\partial_\mu\boldsymbol{\pi}\right)\left(\boldsymbol{\pi}\cdot\partial^\mu\boldsymbol{\pi}\right) - \frac{\alpha}{f_\pi^2}\boldsymbol{\pi}^2\partial_\mu\boldsymbol{\pi}\cdot\partial^\mu\boldsymbol{\pi} + \frac{8\alpha-1}{8f_\pi^2}m_\pi^2\boldsymbol{\pi}^4 \\ &+ N^\dagger\left(i\partial_0 - \frac{g_A}{2f_\pi}\boldsymbol{\tau}\cdot(\vec{\sigma}\cdot\vec{\nabla})\boldsymbol{\pi} - \frac{1}{4f_\pi^2}\boldsymbol{\tau}\cdot(\boldsymbol{\pi}\times\partial_0\boldsymbol{\pi})\right)N \\ &- N^\dagger\left(\frac{g_A(4\alpha-1)}{4f_\pi^3}(\boldsymbol{\tau}\cdot\boldsymbol{\pi})\bigl[\boldsymbol{\pi}\cdot(\vec{\sigma}\cdot\vec{\nabla})\boldsymbol{\pi}\bigr] + \frac{g_A\alpha}{2f_\pi^3}\boldsymbol{\pi}^2\bigl[\boldsymbol{\tau}\cdot(\vec{\sigma}\cdot\vec{\nabla})\boldsymbol{\pi}\bigr]\right)N \\ &- \frac{1}{2}C_S(N^\dagger N)(N^\dagger N) - \frac{1}{2}C_T(N^\dagger\vec{\sigma}N)\cdot(N^\dagger\vec{\sigma}N) + \ldots,\end{aligned} \tag{1.21}$$

where $M \simeq 938.9$ MeV is the average nucleon mass, $N = (p,n)$ is the (heavy) nucleon field vector, and the vector $\boldsymbol{\pi} = (\frac{1}{\sqrt{2}}(\pi^+ + i\pi^-), \frac{i}{\sqrt{2}}(\pi^+ - i\pi^-), \pi^0)$ collects the pion fields. The spin Pauli-matrix vector $\vec{\sigma} = (\sigma_1, \sigma_2, \sigma_3)$ is always multiplied with another vector quantity, and the Pauli matrices in the vector $\vec{\sigma}$ act on the spinor part of $N$. The product of an isospin Pauli-matrix vector $\boldsymbol{\tau} = (\tau^1, \tau^2, \tau^3)$ with a pion field vector $\boldsymbol{\pi}$ is given by

$$\boldsymbol{\tau}\cdot\boldsymbol{\pi} = \tau_1\frac{1}{\sqrt{2}}(\pi^+ + i\pi^-) + \tau_2\frac{i}{\sqrt{2}}(\pi^+ - i\pi^-) + \tau_3\pi^0 = \sqrt{2}(\tau_-\pi^+ + \tau_+\pi^-) + \tau_3\pi^0. \tag{1.22}$$

---

[8] In Eq. (1.21), the coefficient $\alpha$ is arbitrary, where different values of $\alpha$ correspond to different parametrizations of the pion-field matrix $U$, cf. Eq. (A.39). The $\alpha$ dependence of $\mathscr{L}_{\chi\text{EFT}}$ affects only (unobservable) off-shell amplitudes, whereas physical on-shell amplitudes are invariant under a change of parametrization [130, 283].

The operators $\tau_\pm = \frac{1}{2}(\tau_1 \pm i\tau_2)$ and $\tau_3$ act on nucleon states in the following way:

$$\begin{aligned} \tau_- |p\rangle &= |n\rangle, & \tau_- |n\rangle &= 0, \\ \tau_+ |p\rangle &= 0, & \tau_+ |n\rangle &= |p\rangle, \\ \tau_3 |p\rangle &= |p\rangle, & \tau_3 |n\rangle &= -|n\rangle. \end{aligned} \quad (1.23)$$

In Eq. (1.21), the parameter $g_A \simeq 1.27$ is the nucleon axial-vector coupling constant [316]; it can be measured in neutron beta decay $n \to p + e^- + \bar{\nu}_e$ [170, 413, 40]. The low-energy constants $C_S$ and $C_T$ can be fixed by fits to nucleon-nucleon scattering observables. The subleading terms are given by (with additional LECs $c_{1,2,3,4}$ and $D, E$)[9]

$$\begin{aligned} \mathcal{L}_{\chi\text{EFT}}^{A=1} =& N^\dagger \left( \frac{\vec{\nabla}^2}{2M} - \frac{ig_A}{4Mf_\pi} \vec{\tau} \cdot \left[ \vec{\sigma} \cdot (\overleftarrow{\vec{\nabla}} \partial_0 \boldsymbol{\pi} - \partial_0 \boldsymbol{\pi} \overrightarrow{\vec{\nabla}}) \right] - \frac{i}{8Mf_\pi^2} \vec{\tau} \cdot \left[ \overleftarrow{\vec{\nabla}} \cdot (\boldsymbol{\pi} \times \vec{\nabla} \boldsymbol{\pi}) - (\boldsymbol{\pi} \times \vec{\nabla} \boldsymbol{\pi}) \cdot \overrightarrow{\vec{\nabla}} \right] \right) N \\ & + N^\dagger \left( 4c_1 m_\pi^2 - \frac{2c_1}{f_\pi^2} m_\pi^2 \boldsymbol{\pi}^2 + \left( c_2 - \frac{g_A^2}{8M} \right) \frac{1}{f_\pi^2} (\partial_0 \boldsymbol{\pi} \cdot \partial_0 \boldsymbol{\pi}) \right. \\ & \left. + \frac{c_3}{f_\pi^2} (\partial_\mu \boldsymbol{\pi} \cdot \partial^\mu \boldsymbol{\pi}) - \left( c_4 + \frac{1}{4M} \right) \frac{1}{2f_\pi^2} \epsilon^{ijk} \epsilon^{abc} \sigma^i \tau^a (\partial^i \pi^b)(\partial^k \pi^c) \right) N \\ & - \frac{D}{4f_\pi} (N^\dagger N) \left( N^\dagger \left[ \vec{\tau} \cdot (\vec{\sigma} \cdot \vec{\nabla}) \boldsymbol{\pi} \right] N \right) - \frac{1}{2} E (N^\dagger N)(N^\dagger \vec{\tau} N) \cdot (N^\dagger \vec{\tau} N) + \dots \end{aligned} \quad (1.24)$$

In Eqs. (1.21) and (1.24), the ellipses represent terms that are not relevant for the construction of nuclear potentials up to fourth order (N3LO, cf. Sec. 1.3).

As for any EFT, the Lagrangian $\mathcal{L}_{\chi\text{EFT}}$ has infinitely many terms, and thus gives rise to an unlimited number of (increasingly complicated) interactions between pions and nucleons. However, since these interactions are hierarchically ordered, this must not be regarded as a flaw. Concerning the description of the low-energy dynamics of nucleons and pions up to a certain degree of accuracy, $\chi$EFT is a well-controlled theory. Its limited range of applicability and limited precision[10] are what makes it an effective theory.

## 1.3. Hierarchy of Chiral Nuclear Interactions

For a given system, the usefulness ("effectiveness") of a description in terms of an EFT relies on the availability of two distinct energy scales that separate large- and short-distance effects. In the case of low-energy nuclear (or hadronic) physics, these scales are given by the symmetry breaking scale $\Lambda_\chi \sim 4\pi f_\pi \sim 1\,\text{GeV}$—the breakdown scale ("hard scale") of $\chi$EFT—, and a "soft scale" $Q$ associated with the pion mass $m_\pi$ or a (small) nucleon momentum.[11] This separation of scales implies the hierachical ordering of chiral nuclear interactions in terms of powers of the expansion parameter $Q/\Lambda_\chi$. For a given diagram contributing to the nuclear interaction, it follows from the rules of covariant perturbation theory that (see Ref. [413])

- an intermediate nucleon line counts as $Q^{-1}$,
- an intermediate pion line counts as $Q^{-2}$,
- each derivative in any vertex counts as $Q$,
- each four-momentum integration counts as $Q^4$.

Using this dimensional analysis and some topological identities, one finds the following formula for the chiral power $\nu$ of a diagram involving a given number of nucleons [409, 410, 411, 283]:

$$\nu = -2 + \sum_i \kappa_i, \qquad \kappa_i = d_i - \frac{3}{2} n_i + \pi_i - 4, \tag{1.25}$$

where $d_i$ is the number of derivatives or pion mass insertions, $n_i$ is the number of nucleon fields, and $\pi_i$ the number of pion fields at the vertex $i$. The hierarchy of nuclear interactions arising from this power counting is depicted in Fig. 1.3 (contributions at order $\nu = 1$ vanish due to parity and time-reversal invariance [283, 409]).

|  | NN Interactions | 3N Interactions | 4N Interactions |
|---|---|---|---|
| **LO** $(Q/\Lambda_\chi)^0$ | diagrams | — | — |
| **NLO** $(Q/\Lambda_\chi)^2$ | diagrams | — | — |
| **N2LO** $(Q/\Lambda_\chi)^3$ | diagrams $+\ldots$ | diagrams | — |
| **N3LO** $(Q/\Lambda_\chi)^4$ | diagrams $+\ldots$ | diagrams $+\ldots$ | diagrams $+\ldots$ |

**Figure 1.3.:** Hierarchy of chiral nuclear interactions [130]. Solid lines represent nucleons, and dashed lines pions. Tiny dots, large solid dots, large solid squares and large crossed squares denote vertices with $\Delta = 0, 1, 2$ and $4$, respectively.

---

[9] In Eq. (1.24), gradients with large (right- and left-handed) vector arrows [$\vec{\nabla}$ and $\overleftarrow{\nabla}$] act on the nucleon fields to the right and left, respectively.

[10] To be precise, the property of limited precision is not peculiar to effective field theories. Since perturbation theory (the expansion in terms of powers of the coupling constant $g$) in general corresponds to an asymptotic expansion with zero radius of convergence [146, 388, 118, 377, 136], also renormalizable theories can be evaluated only to limited precision (even when the coupling is weak).

[11] The separation of scales associated with S$\chi$SB is manifest in the large gap between the (average) pion mass ($m_\pi \simeq 138$ MeV) and the masses of the lightest vector mesons, i.e., the rho ($m_\rho \simeq 775$ MeV) and the omega meson ($m_\omega \simeq 782$ MeV). Note also that the large mass of the $\eta'$ meson (which is close to the $\eta_0$, the "would-be" Nambu-Goldstone boson of spontaneous U(1)$_A$ breaking), $m_{\eta'} \simeq 960$ MeV, is explained in terms of the axial anomaly, see Sec. 1.1.2 and [413, 316].

*1. Nuclear Interactions and Many-Body Problem*

## 1.3.1. Two-Nucleon Interaction

For the various chiral NN interactions (evaluated in the heavy-baryon formalism) up to a given order in the power counting, one can construct an effective NN potential (and similar for 3N, 4N , etc.). The potential matrix elements are defined as the sum of the diagrammatic amplitudes of the irreducible components of the chiral interactions [410]. The reducible components corresponding to purely nucleonic intermediate states are then reproduced by iterating the NN potential in the Lippmann-Schwinger equation. The contributions to the chiral NN potential can be organized as follows:

$$V_{\text{NN}} = V_{\text{c.t.}} + V_{1\pi} + V_{2\pi} + \ldots \tag{1.26}$$

The contact ("c.t."), one-pion exchange ("$1\pi$") and two-pion exchange ("$2\pi$") contributions up to next-to-next-to-next-to-leading order (N3LO, $\nu = 4$) are ordered in the following way [283]:

$$V_{\text{c.t.}} = V_{\text{c.t.}}^{(0)} + V_{\text{c.t.}}^{(2)} + V_{\text{c.t.}}^{(4)}, \tag{1.27}$$

$$V_{1\pi} = V_{1\pi}^{(0)} + V_{1\pi}^{(2)} + V_{1\pi}^{(3)} + V_{2\pi}^{(4)}, \tag{1.28}$$

$$V_{2\pi} = V_{2\pi}^{(2)} + V_{2\pi}^{(3)} + V_{2\pi}^{(4)}, \tag{1.29}$$

where the superscript denotes the order $\nu$ with respect to the chiral power counting. (At N3LO there are also the first three-pion exchange interactions, which are however negligible [283]). The reducible part of two-pion exchange arises from the "planar box diagram" at next-to-leading order (NLO). The corresponding time-ordered graphs are shown in Fig. 1.4. Notably, as a consequence of the nonrelativistic treatment of nucleons ($1/M$ expansion) the irreducible component of the planar box diagram receives a contribution also from the four graphs with reducible time-orderings, see Ref. [236] for details.

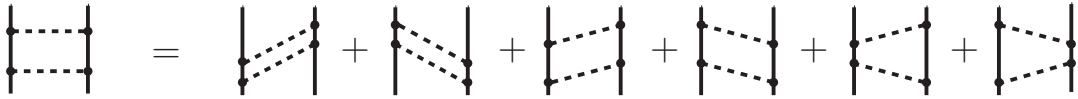

**Figure 1.4.:** Planar box diagram: covariant Feynman diagram (left-hand side of the equal sign) and corresponding time-ordered graphs (right-hand side). As noted in the text, the irreducible component of the diagram receives, in addition to the contribution from the first two graphs with irreducible time-orderings, also a contribution the four graphs with reducible time-orderings.

The nucleon-nucleon potential $V_{\text{NN}}$ is an operator acting on (physical) two-nucleon states $|\varphi\rangle$, i.e.,

$$|\varphi\rangle = \int d^3x_1 d^3x_2 \, |\vec{x}_1 \vec{x}_2\rangle \, \varphi(\vec{x}_1, \vec{x}_2) = \int d^3k_1 d^3k_2 \, |\vec{k}_1 \vec{k}_2\rangle \, \tilde{\varphi}(\vec{k}_1, \vec{k}_2), \tag{1.30}$$

where the coordinate-space [$\varphi(\vec{x}_1, \vec{x}_2)$] and momentum-space [$\tilde{\varphi}(\vec{k}_1, \vec{k}_2)$] wave functions are spin-isospin spinors. The potential can thus be represented as an integral operator acting on momentum eigenstates $|\vec{k}_1 \vec{k}_2\rangle$, i.e.,

$$V_{\text{NN}} = \int d^3k_1' d^3k_2' d^3k_1 d^3k_2 \, \langle \vec{k}_1' \vec{k}_2' | V_{\text{NN}} | \vec{k}_1 \vec{k}_2 \rangle \, |\vec{k}_1' \vec{k}_2'\rangle \langle \vec{k}_1 \vec{k}_2|, \tag{1.31}$$

*1. Nuclear Interactions and Many-Body Problem*

where the matrix elements $\langle \vec{k}'_1 \vec{k}'_2 | V_{NN} | \vec{k}_1 \vec{k}_2 \rangle$ are operators in spin and isospin space. By translational and Galilei invariance, the matrix elements are diagonal with respect to the total momenta $\vec{K} = \vec{k}_1 + \vec{k}_2$ and $\vec{K}' = \vec{k}'_1 + \vec{k}'_2$ and depend only on the relative momenta $\vec{p} = (\vec{k}_1 - \vec{k}_2)/2$ and $\vec{p}' = (\vec{k}'_1 - \vec{k}'_2)/2$. Therefore, we can write the matrix elements as

$$\langle \vec{k}'_1 \vec{k}'_2 | V_{NN} | \vec{k}_1 \vec{k}_2 \rangle = \delta(\vec{k}'_1 + \vec{k}'_2 - \vec{k}_1 - \vec{k}_2) \underbrace{\langle \vec{p}' | V_{NN} | \vec{p} \rangle}_{\equiv V_{NN}(\vec{p}', \vec{p})}. \tag{1.32}$$

Instead of using relative momenta as independent variables, one can use the momentum transfer $\vec{q} = \vec{k}_1 - \vec{k}'_1 = \vec{p} - \vec{p}'$ and the average relative momentum $\vec{k} = (\vec{p} + \vec{p}')/2 = (\vec{k}_1 - \vec{k}_2)/2$. The contribution to the potential from the leading-order (LO) contact interaction is momentum independent, and is given by

$$V^{(0)}_{c.t.}(\vec{p}', \vec{p}) = C_S + C_T \, \vec{\sigma}_1 \cdot \vec{\sigma}_2. \tag{1.33}$$

For the subleading contact interactions, cf. (e.g.,) Refs. [283, 384]. The leading-order one-pion exchange potential, which represents the contribution to the NN interaction of longest range, is given by

$$V^{(0)}_{1\pi}(\vec{p}', \vec{p}) = -\frac{g_A^2}{4f_\pi^2} \tau_1 \cdot \tau_2 \frac{(\vec{\sigma}_1 \cdot \vec{q})(\vec{\sigma}_2 \cdot \vec{q})}{q^2 + m_\pi^2}. \tag{1.34}$$

At NLO the one-pion exchange potential gets renormalized by several one-loop graphs and counter-term insertions; the relevant diagrams are shown in Fig. 1.5. The first two diagrams at the top of Fig. 1.5 renormalize the nucleon propagator and the next two the pion propagator. The four diagrams at the bottom renormalize the pion-nucleon coupling constant. In the one-loop diagrams, all vertices are from the leading order Lagrangian $\mathscr{L}^{\Delta=0}$ given in Eq. (1.21), whereas the counter-term insertions come from $\mathscr{L}^{\Delta=2}$. At N2LO and N3LO, further corrections arise; they renormalize various LECs, the pion-mass, and the pion-nucleon coupling,[12] but do not generate a pion-nucleon form-factor [283]. Therefore the one-pion exchange contributions maintain the structure given by Eq. (1.34) up to (at least) N3LO. Two-pion exchange starts at NLO and involves additional loop diagrams which have to be regularized. For more details regarding the different NLO, N2LO and N3LO contribution to the NN potential, see Refs. [129, 283, 236, 227, 226].[13] At N3LO, the chiral NN potential becomes a "high-precision" potential: the description of NN scattering phase shifts reaches a precision comparable to "conventional" phenomenological high-precision NN potentials such as the CD-Bonn potential [282] or the AV18 potential [421].[14]

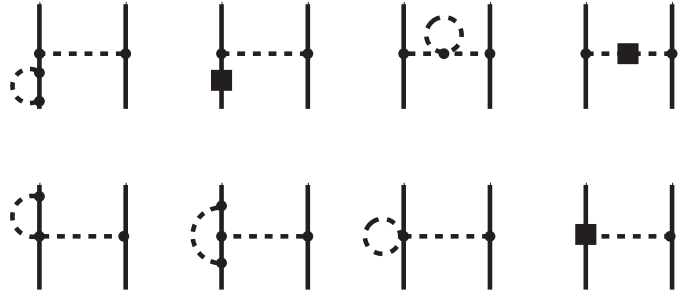

**Figure 1.5.:** One-pion exchange diagrams at order $(Q/\Lambda_\chi)^2$

---

[12] At NLO this correction involves the so-called Goldberger-Treimann discrepancy [133, 283]; it leads to $g_A \to$ 1.29.

[13] Recently, also the N4LO contributions (as well as the dominant N5LO terms) have been worked out [123, 124].

[14] See Ref. [281] for details and a comparison of chiral and conventional high-precision potentials.

*1. Nuclear Interactions and Many-Body Problem*

The general form of the NN potential consistent with rotational invariance, exchange symmetry, parity and time-reversal invariance, Hermiticity, and invariance under rotations in isospin space (isospin symmetry) is given by [174]

$$\begin{aligned}V_{\text{NN}}(\vec{p}',\vec{p}) =& V_C + \boldsymbol{\tau}_1 \cdot \boldsymbol{\tau}_2 W_C \\ &+ (V_\sigma + \boldsymbol{\tau}_1 \cdot \boldsymbol{\tau}_2 W_S)\, \vec{\sigma}_1 \cdot \vec{\sigma}_2 \\ &+ (V_{\sigma q} + \boldsymbol{\tau}_1 \cdot \boldsymbol{\tau}_2 W_T)(\vec{\sigma}_1 \cdot \vec{q})(\vec{\sigma}_2 \cdot \vec{q}) \\ &+ (V_{LS} + \boldsymbol{\tau}_1 \cdot \boldsymbol{\tau}_2 W_{LS}) i(\vec{\sigma}_1 + \vec{\sigma}_2)\cdot(\vec{q}\times\vec{k}) \\ &+ (V_{\sigma L} + \boldsymbol{\tau}_1 \cdot \boldsymbol{\tau}_2 W_{\sigma L})\, \vec{\sigma}_1\cdot(\vec{q}\times\vec{k})\, \vec{\sigma}_2\cdot(\vec{q}\times\vec{k}) \\ &+ (V_{\sigma k} + \boldsymbol{\tau}_1 \cdot \boldsymbol{\tau}_2 W_{T2})(\vec{\sigma}_1\cdot\vec{k})(\vec{\sigma}_2\cdot\vec{k}), \end{aligned} \quad (1.35)$$

where the coefficients $V_\alpha$ and $W_\alpha$, $\alpha \in \{C, S, \sigma q, LS, \sigma L, \sigma k\}$, are functions of $q = |\vec{q}|$, $k = |\vec{k}|$ and $|\vec{q}\times\vec{k}|$. The *on-shell* potential matrix elements can be expressed solely in terms of the central (C), spin ($\sigma$), tensor[15] ($\sigma q$), spin-orbit (LS) and quadratic spin-orbit ($\sigma L$) parts of the potential, with coefficients that depend only on $q = |\vec{q}|$ and $k = |\vec{k}|$. This follows from the identity [174]

$$\begin{aligned}(\vec{\sigma}_1\cdot\vec{k})(\vec{\sigma}_2\cdot\vec{k})q^2 =& -(\vec{\sigma}_1\cdot\vec{q})(\vec{\sigma}_2\cdot\vec{q})k^2 - (\vec{\sigma}_1\cdot\vec{q}\times\vec{k})(\vec{\sigma}_2\cdot\vec{q}\times\vec{k}) + (\vec{\sigma}_1\cdot\vec{\sigma}_2)(\vec{q}\times\vec{k})^2 \\ &+ (\vec{q}\cdot\vec{k})[(\vec{\sigma}_1\times\vec{q})(\vec{\sigma}_2\cdot\vec{k}) + (\vec{\sigma}_1\times\vec{k})(\vec{\sigma}_2\cdot\vec{q})],\end{aligned} \quad (1.36)$$

and the fact that for on-shell scattering it is $(\vec{q}\times\vec{k})^2 = q^2 k^2$ and $(\vec{q}\cdot\vec{k})^2 = 0$:

$$(\vec{q}\times\vec{k})\cdot(\vec{q}\times\vec{k}) = q^2 k^2 - (\vec{q}\cdot\vec{k})^2 = q^2 k^2 - \underbrace{[p^2 - (p')^2]^2}_{=0\text{ (on-shell)}}. \quad (1.37)$$

The isospin structure of the NN potential given by Eq. (1.35) is such that the nn and pp interactions and the np interactions in the triplet channels with total isospin $T = 1$ are identical.[16] This situation (isospin symmetry) is also referred to as "charge independence". If isospin-symmetry breaking effects are taken into account than additional isospin structures appear as compared to Eq. (1.35), i.e., $\tau_1^3 \tau_2^3$ (charge-independence breaking), $\tau_1^3 + \tau_2^3$ (charge-symmetry breaking without isospin mixing), and $\tau_1^3 - \tau_2^3$ as well as $[\boldsymbol{\tau}_1 \times \boldsymbol{\tau}_2]^3$ (charge-symmetry breaking and isospin mixing) [403, 131, 295]. Here, "charge symmetry" refers to the situation where the nn and pp interactions are the same, but not necessarily equal to the np interactions in isospin triplet channels, and "isospin mixing" refers to the mixing between $T = 1$ and $T = 0$ states in the np system.

The parts of $V_{\text{NN}}(\vec{p}',\vec{p})$ that depend only on $\vec{q} = \vec{k}_1 - \vec{k}'_1$ lead to *local* coordinate-space potential matrix elements: using $\langle \vec{x}|\vec{k}\rangle = e^{-i\vec{k}\cdot\vec{x}}/(2\pi)^{3/2}$ we obtain

$$\langle \vec{x}'_1 \vec{x}'_2 | V_{\text{NN}} | \vec{x}_1 \vec{x}_2 \rangle = \frac{1}{(2\pi)^6} \int d^3k_1 d^3k'_2 d^3k'_1 d^3k_2 \, \langle \vec{k}'_1 \vec{k}'_2 | V_{\text{NN}} | \vec{k}_1 \vec{k}_2\rangle \, e^{-i(\vec{k}'_1\cdot\vec{x}'_1 + \vec{k}'_2\cdot\vec{x}'_2)} e^{i(\vec{k}_1\cdot\vec{x}_1 + \vec{k}_2\cdot\vec{x}_2)}, \quad (1.38)$$

which for $\langle \vec{k}'_1 \vec{k}'_2 | V_{\text{NN}} | \vec{k}_1 \vec{k}_2 \rangle = \delta(\vec{k}'_1 + \vec{k}'_2 - \vec{k}_1 - \vec{k}_2) V_{\text{NN}}(\vec{q})$ leads to a local potential, i.e.,

$$\langle \vec{x}'_1 \vec{x}'_2 | V_{\text{NN}} | \vec{x}_1 \vec{x}_2 \rangle = \frac{1}{(2\pi)^6} \int d^3k'_1 d^3k'_2 d^3k_1 \, V_{\text{NN}}(\vec{q})\, e^{-i(\vec{k}'_1\cdot\vec{x}'_1 + \vec{k}'_2\cdot\vec{x}'_2)} e^{i[\vec{k}_1\cdot\vec{x}_1 + (\vec{k}'_1 + \vec{k}'_2 - \vec{k}_1)\cdot\vec{x}_2]}$$

---

[15] Note that the "tensor" part of the potential is often associated with the tensor operator $S_{12} = 3(\vec{\sigma}_1\cdot\vec{q})(\vec{\sigma}_2\cdot\vec{q}) - \vec{\sigma}_1\cdot\vec{\sigma}_2$ instead of the $\sigma q$ part of Eq. (1.35), cf. (e.g.,) Ref. [138].

[16] Note however that since $\boldsymbol{\tau}_1\cdot\boldsymbol{\tau}_2 = -3$ for $T = 0$ but $\boldsymbol{\tau}_1\cdot\boldsymbol{\tau}_2 = 1$ for $T = 1$, the np interactions in the isospin singlet singlet ($T = 0$) and isospin triplet ($T = 1$) channels are different.

$$= \frac{1}{(2\pi)^6} \int d^3k_1 d^3q \, d^3k V_{\rm NN}(\vec{q}) \, e^{-i[(\vec{k}_1-\vec{q})\cdot\vec{x}_1' + (\vec{k}_1-\vec{k})\cdot\vec{x}_2']} \, e^{i[\vec{k}_1\cdot\vec{x}_1 + (\vec{k}_1-\vec{q}-\vec{k})\cdot\vec{x}_2]}$$

$$= \delta(\vec{x}_1' - \vec{x}_1)\delta(\vec{x}_2' - \vec{x}_2) \underbrace{\int d^3q V_{\rm NN}(\vec{q}) \, e^{i\vec{q}\cdot(\vec{x}_1-\vec{x}_2)}}_{V_{\rm NN}(\vec{x}_1-\vec{x}_2)}. \tag{1.39}$$

Terms that depend also on $\vec{k}$ lead to a *nonlocal* potential. For such a potential, the coordinate-space Schrödinger equation has the form

$$T_{\rm kin}\varphi(\vec{x}_1,\vec{x}_2) + \int d^3x_1' d^3x_2' \, \langle \vec{x}_1, \vec{x}_2 | V_{\rm NN} | \vec{x}_1' \vec{x}_2' \rangle \, \varphi(\vec{x}_1'\vec{x}_2') = i\partial_t \, \varphi(\vec{x}_1, \vec{x}_2), \tag{1.40}$$

where $T_{\rm kin} = \vec{k}_1^2/(2M) + \vec{k}_2^2/(2M)$, with $\vec{k}_{1,2} = -i\vec{\nabla}_{1,2}$. It will be instructive to rewrite this in terms relative $\vec{r}^{(\prime)} = \vec{x}_1^{(\prime)} - \vec{x}_2^{(\prime)}$ and center-of-mass coordinates $\vec{R}^{(\prime)} = (\vec{x}_1^{(\prime)} + \vec{x}_2^{(\prime)})/2$. In terms of these coordinates, the two-nucleon state vector $|\varphi\rangle$ is given by

$$|\varphi\rangle = \int d^3r d^3R \, |\vec{r}\vec{R}\rangle \, \varphi(\vec{r}, \vec{R}). \tag{1.41}$$

For the potential matrix elements in terms of $|\vec{r}\vec{R}\rangle$ states one obtains

$$\langle \vec{r}'\vec{R}' | V_{\rm NN} | \vec{r}\vec{R} \rangle = \frac{1}{(2\pi)^6} \int d^3p' d^3K' d^3p \, d^3K \, \delta(\vec{K}' - \vec{K}) V_{\rm NN}(\vec{p}', \vec{p}) \, e^{-i(\vec{p}'\cdot\vec{r}' + \vec{K}'\cdot\vec{R}')} \, e^{i(\vec{p}\cdot\vec{r} + \vec{K}\cdot\vec{R})}$$

$$= \delta(\vec{R}' - \vec{R}) \underbrace{\frac{1}{(2\pi)^3} \int d^3p' d^3p \, V_{\rm NN}(\vec{p}', \vec{p}) \, e^{-i\vec{p}'\cdot\vec{r}'} \, e^{i\vec{p}\cdot\vec{r}}}_{\langle \vec{r}' | V_{\rm NN} | \vec{r} \rangle \equiv V_{\rm NN}(\vec{r}',\vec{r})}. \tag{1.42}$$

Using the ansatz $\varphi(\vec{r}, \vec{R}) = \phi(\vec{r})\xi(\vec{R})$, the relative and center-of-mass parts of the Schrödinger equation decouple, leading to

$$T_{\rm rel} \, \phi(\vec{r}) + \int d^3r' \, \langle \vec{r}' | V_{\rm NN} | \vec{r} \rangle \, \phi(\vec{r}') = i\partial_t \, \phi(\vec{r}), \tag{1.43}$$

$$T_R \, \xi(\vec{R}) = i\partial_t \, \xi(\vec{R}), \tag{1.44}$$

where $T_{\rm rel} = \vec{p}^2/(M)$ and $T_R = \vec{K}^2/(4M)$, with $\vec{p} = -i\vec{\nabla}_r$ and $\vec{K} = -i\vec{\nabla}_R$.

For a general NN potential, its nonlocal part can be associated with a *velocity dependence*. This can be seen by expanding the wave-function $\phi(\vec{r}')$ as [344]

$$\phi(\vec{r}') = \phi(\vec{r}) + (\vec{r}' - \vec{r}) \cdot \vec{\nabla}_r \phi(\vec{r}) + \ldots = \mathcal{N}[e^{-i(\vec{r}'-\vec{r})\cdot\vec{p}}] \, \phi(\vec{r}), \tag{1.45}$$

where "$\mathcal{N}[\,]$" (normal ordering) means that the momentum operators should not act on the coordinates in the expansion of the exponent. This shows that the potential can be represented as an operator $V_{\rm NN}(\vec{p}, \vec{r})$ given by

$$V_{\rm NN}(\vec{p}, \vec{r}) = \int d^3r' V_{\rm NN}(\vec{r}', \vec{r}) \mathcal{N}[e^{-i(\vec{r}'-\vec{r})\cdot\vec{p}}], \tag{1.46}$$

and the coordinate-space Schrödinger equation can be written as

$$T_{\rm rel} \, \phi(\vec{r}) + V_{\rm NN}(\vec{p}, \vec{r}) \, \phi(\vec{r}) = i\partial_t \, \phi(\vec{r}). \tag{1.47}$$

For the general form of $V_{\rm NN}(\vec{p}, \vec{r})$ consistent with translational and Galilei invariance, rotational invariance, exchange symmetry, parity and time-reversal invariance, Hermiticity, and isospin symmetry, see Ref. [315].

## 1.3.2. Multi-Nucleon Interactions

The leading-order contribution to the three-nucleon (3N) interactions appears at N2LO (in the case of $\chi$EFT without explicit $\Delta(1232)$-isobars) [127, 411, 134], and consists of the three diagrams shown in Fig. (1.3): a short-range 3N contact interaction, an intermediate-range one-pion echange interaction, and a long-range two-pion exchange interaction. The contact interaction arises from the last term of $\mathscr{L}_{\chi\text{EFT}}^{\Delta=1}$ [Eq. (1.24)] proportional to the low-energy constant $E$, and is given by

$$V_{3N,\text{c.t.}}^{(3)} = \frac{E}{2} \sum_{j\neq k} (\boldsymbol{\tau}_j \cdot \boldsymbol{\tau}_k). \tag{1.48}$$

The one-pion exchange diagram proportional to the low-energy constant $D$ arises from the second-last term of $\mathscr{L}_{\chi\text{EFT}}^{\Delta=1}$, is given by

$$V_{3N,1\pi}^{(3)} = -D \frac{g_A}{8 f_\pi^2} \sum_{i\neq j\neq k} \frac{\vec{\sigma}_j \cdot \vec{q}_j}{\vec{q}_j^2 + m_\pi^2} (\vec{\sigma}_i \cdot \vec{q}_j)(\boldsymbol{\tau}_i \cdot \boldsymbol{\tau}_j), \tag{1.49}$$

where $\vec{q}_i = \vec{k}_i' - \vec{k}_i$, with $i = 1, 2, 3$. The low-energy constants $E$ and $D$ are usually rewritten in terms of dimensionless constants $c_E$ and $c_D$ as [53, 209]

$$E = \frac{c_E}{f_\pi^4 \tilde{\Lambda}_\chi}, \qquad D = \frac{c_D}{f_\pi^2 \tilde{\Lambda}_\chi}, \tag{1.50}$$

where we set $\tilde{\Lambda}_\chi = 700$ MeV. Finally, the two-pion exchange contribution composes itself from the terms proportional to $c_{1,3,4}$ in $\mathscr{L}_{\chi\text{EFT}}^{\Delta=1}$, and has the form

$$V_{3N,2\pi}^{(3)} = \frac{1}{2}\left(\frac{g_A}{2f_\pi}\right)^2 \sum_{i\neq j\neq k} \frac{(\vec{\sigma}_i \cdot \vec{q}_i)(\vec{\sigma}_j \cdot \vec{q}_j)}{(\vec{q}_i^2 + m_\pi^2)(\vec{q}_i^2 + m_\pi^2)} F_{ijk}^{ab} \tau_i^a \tau_j^b, \tag{1.51}$$

where $F_{ijk}^{ab}$ is given by (with $\epsilon^{abc}$ the Levi-Civita tensor)

$$F_{ijk}^{ab} = \delta^{ab}\left(-\frac{4c_1 m_\pi^2}{f_\pi^2} + \frac{2c_3}{f_\pi^2}(\vec{q}_i \cdot \vec{q}_j)\right) + \frac{c_4}{f_\pi^2}\sum_c \epsilon^{abc}\tau_k^c \vec{\sigma}_k \cdot (\vec{q}_i \times \vec{q}_j), \tag{1.52}$$

resulting in $c_1$ and $c_3$ terms proportional to $\boldsymbol{\tau}_i \cdot \boldsymbol{\tau}_j$ and a $c_4$ term proportional to $(\boldsymbol{\tau}_i \times \boldsymbol{\tau}_j) \cdot \boldsymbol{\tau}_k$. At N3LO, there are additional 3N diagrams, and also the first 4N interactions. The N3LO multi-nucleon diagrams do not involve new LECs, i.e., they are completely predicted from parameters appearing in the LO two-nucleon sector [128, 384]. The N3LO three-nucleon interactions can be organized as

$$V_{3N}^{(4)} = V_{2\pi}^{(4)} + V_{2\pi-1\pi}^{(4)} + V_{\text{ring}}^{(4)} + V_{1\pi-\text{c.t.}}^{(4)} + V_{2\pi-\text{c.t.}}^{(4)} + V_{1/m}^{(4)}, \tag{1.53}$$

where the different terms correspond to different types of diagrams (cf. [219, 38, 37] and also [384] for details and explicit expressions). The various N3LO four-nucleon interactions have been examined in Ref. [128].

At present, most nuclear many-body calculations with chiral interactions have employed the N2LO three-nucleon potential but no higher-order multi-nucleon interactions. Progress towards the implementation of higher-order multi-nucleon interactions has however been achieved in recent works [386, 251, 106, 188].

## 1.4. Low-Momentum Chiral Nuclear Potentials

Since $\chi$EFT is an effective theory valid only at energy scales below the breakdown scale $\Lambda_\chi \sim 1$ GeV, chiral nuclear potentials should to be regularized at a scale $\Lambda < \Lambda_\chi$ [283, 132]. Consequently, the values of the low-energy constants (LECs) parametrizing the interactions depend on the employed cutoff scale $\Lambda$. The LECs are generally fixed by fits to nucleon-nucleon (NN) phase shifts and properties of light nuclei, and different values for the LECs can emerge depending on the fitting scheme.[17] Various sets of LECs can be found in the literature (cf. e.g., Refs. [67, 135, 283, 343]), all of which lead (by construction) to similar results in the few-body sector. As a consequence, there are many different chiral potentials (different LECs, and different cutoff scales) that can give an accurate description of few-nucleon observables. However, this model independence cannot be expected to be maintained in the nuclear many-body sector, concerning both cutoff variations and fitting uncertainties, in particular since many-body calculations are necessarily approximative. In order to obtain conclusive results, one should therefore use a variety of many-body methods in combination with a large number of different potentials (corresponding to different fitting schemes, different regularization methods, different cutoff scales, and different orders in the chiral power counting).[18]

For large cutoffs (or equivalently, high resolution scales), however, in general a very strong NN potential results at short distances (some details are discussed below), which impedes convergence in various many-body frameworks [53]. It is therefore expedient to use potentials constructed at low cutoff scales $\Lambda \lesssim (450 - 500)$ MeV. In particular, by employing such low-momentum nuclear potentials, many-body perturbation theory becomes a valid approach to the nuclear many-body problem [55]. Regarding the construction of low-momentum potentials, in the literature much emphasis has been put on the application of renormalization-group (RG) methods that by construction leave few-nucleon low-energy observables invariant [53, 154, 56, 153, 223]. In our many-body calculations we use potentials regularized "by hand" via (re)fitting the LECs (the "n3lo" ones in Table 1.1) as well as potentials based on RG methods (the "VLK" ones).

The properties of the five different sets of NN and 3N potentials used in this thesis are summarized in Table 1.1. The LECs parametrizing the potentials have all been fixed to few-nucleon observables. In that sense, with the above qualifications considered, the results obtained from these potentials in many-body calculations can be regarded as genuine predictions without additional fine-tuning. The five potential sets of Table 1.1 provide an adequate basis for exploring a significant variety of aspects (concerning many-body uncertainties) associated with the choice of resolution scale and LECs. Of particular interest will be the comparison of nuclear potentials defined at the same resolution scale but constructed via RG methods or by refitting LECs.

---

[17] The uncertainties of the LECs associated with the fitting procedure must be strictly distinguished from their (natural) scale dependence. Recently, a more systematic approach (based on Baysian statistics) regarding the estimation of the LECs from few-nucleon data has been illustrated in Ref. [415] (cf. also [155]). Moreover, it may eventually be possible to additionally constrain the values of the LECs by matching to lattice QCD simulations [127, 145, 196]. These future developments can be expected to help constrain the (current) uncertainties in the values of the LECs related to their determination via different fitting schemes.

[18] In this thesis, we do not consider potentials constructed at different chiral orders. More specifically, we use the N2LO three-nucleon interactions in combination with (various) N3LO two-nucleon potentials. Initial studies concerning the order-by-order (in terms of the chiral power counting) convergence of nuclear many-body calculations have been performed in [354, 355]. The consistent implementation of the N3LO chiral multi-nucleon interactions remains a challenge in contemporary nuclear many-body theory, but progress toward this end is

|         | $\Lambda$ (fm$^{-1}$) | $n$      | $c_E$   | $c_D$   | $c_1$ (GeV$^{-1}$) | $c_3$ (GeV$^{-1}$) | $c_4$ (GeV$^{-1}$) |
|---------|-----------------------|----------|---------|---------|---------------------|---------------------|---------------------|
| n3lo500 | 2.5                   | 2        | -0.205  | -0.20   | -0.81               | -3.2                | 5.4                 |
| n3lo450 | 2.3                   | 3        | -0.106  | -0.24   | -0.81               | -3.4                | 3.4                 |
| n3lo414 | 2.1                   | 10       | -0.072  | -0.4    | -0.81               | -3.0                | 3.4                 |
| VLK23   | 2.3                   | $\infty$ | -0.822  | -2.785  | -0.76               | -4.78               | 3.96                |
| VLK21   | 2.1                   | $\infty$ | -0.625  | -2.062  | -0.76               | -4.78               | 3.96                |

**Table 1.1.:** Parameters of the different sets of chiral low-momentum NN (N3LO) and 3N (N2LO) potentials used in this thesis, see text for details.

**"n3lo" Potentials (n3lo414, n3lo450, n3lo500).** Regarding the NN potential, a common way to enforce a restriction to low momentum scales is to employ a smooth regulator function of the form

$$f(p, p') = \exp\left[-(p/\Lambda)^{2n} - (p'/\Lambda)^{2n}\right]. \tag{1.54}$$

The widely used N3LO two-nucleon potential constructed by Entem and Machleidt [125] (the NN part of the NN+3N potential set denoted by "n3lo500") uses the values $\Lambda = 500$ MeV and $n = 2$. In addition, in this thesis we employ also chiral two-nucleon potentials constructed using regulators with $\Lambda = 450$ MeV and $n = 3$ as well as $\Lambda = 414$ MeV and $n = 10$. In each case, the NN potential is combined with the next-to-next-to-leading order (N2LO) chiral three-nucleon interaction. We hereafter denote these three sets of chiral two- and three-nucleon potentials by "n3lo500", "n3lo450" and "n3lo414". For each of these potential sets, the two-nucleon LECs have been fixed by fits to NN phase shifts, and the 3N contact LECs, $c_E$ and $c_D$, have been adjusted to reproduce the triton ($^3$H) binding energy and its (beta-decay) lifetime (see Refs. [91, 90] for more details). With the cutoff scale and the regulator width taken from the respective NN regulator [and using the 3N regulator given by Eq. (3.34)], the different 3N potentials are completely determined by the values of $c_E$, $c_D$ and $c_{1,3,4}$. Because of the different regulating functions used in the respective potentials, different values of these LECs emerge from fits to few-nucleon low-energy observables. The resulting values for the LECs that parametrize the various N2LO three-nucleon diagrams are given in the first three rows of Table 1.1.

**"VLK" Potentials (VLK21, VLK23).** As discussed above, an alternative scheme for obtaining low-momentum nuclear interactions is to employ renormalization-group (RG) techniques that by construction leave low-energy observables invariant. In the case of an evolution of the NN potential based on half-on-shell $K$-matrix equivalence the resulting potential is usually denoted by $V_{\text{low-}k}(\Lambda)$, with $\Lambda$ being now a sharp cutoff in momentum space [i.e., $n = \infty$ in terms of Eq. (1.54)]. The basic mechanism of the evolution is described below. The $K$-matrix method has the advantage of producing low-momentum NN potentials directly through the evolution of partial-wave matrix elements, however, the inclusion of "induced" multi-nucleon interactions is crucial (see below for discussion). In view of this, Nogga *et al.* [312] have used the chiral N2LO 3N interactions with the values of the $c_{1,3,4}$ LECs equal to the ones extracted by the Nijmegen group in an analysis of NN scattering data [343, 342][19] and determined $c_E$ and $c_D$ by fitting to

---

being achieved [106, 251, 386, 188].

[19] We note here that the methods used by the Nijmegen group to determine $c_{3,4}$ have been criticized in Ref. [126]. Nevertheless, using the Nijmegen values for $c_{3,4}$ (in the case of the VLK three-nucleon potentials) in nuclear many-body calculations can still be instructive—and was, in fact, found to be so in this thesis, cf. Secs. 3.2 and 3.3—concerning the impact of LEC uncertainties on nuclear matter properties.

the binding energies of $^3$H, $^3$He and $^4$He. The resulting LECs for two different $V_{\text{low-}k}$ potentials (both constructed by evolving the n3lo500 NN potential) can be found in the last two rows of Table 1.1. The two sets of $V_{\text{low-}k}$ two-nucleon and chiral N2LO three-nucleon potentials are denoted by "VLK21" ($\Lambda = 2.1 \text{ fm}^{-1}$) and "VLK21" ($\Lambda = 2.3 \text{ fm}^{-1}$), respectively.

*Renormalization-Group (RG) Evolution.* For large cutoff scales, two-nucleon potentials that reproduce NN scattering phase shifts well typically have a strong short-range repulsive component (the "hard core"), and strong short-range tensor interactions [53].[20] These features give rise to a significant "coupling" of high- and low-momentum modes, i.e., the presence of large off-diagonal matrix elements. The "nonperturbativeness" of this coupling is manifest in the intermediate-state summations in perturbation theory. The perturbation series for the $K$-matrix[21] for NN scattering ("Born series") in a given partial wave is given by

$$K_{\text{NN}}(p', p) = V_{\text{NN}}(p', p) + \frac{M}{2\pi^2} \fint_0^\infty dq\, q^2 \frac{V_{\text{NN}}(p', q) V_{\text{NN}}(q, p)}{p^2 - q^2} + \ldots, \quad (1.55)$$

where the dashed integral denotes the Cauchy principal value. The second order term in Eq. (1.55) involves an integral over off-diagonal matrix elements and thus includes (off-diagonal) high-momentum modes. As evident from the RG analysis of the $K$-matrix conducted below, the high-momentum modes are not relevant for a reliable description of low-energy NN scattering, and can be integrated out (i.e., decoupled from the low-momentum modes, and thus, completely decoupled from low-energy NN observables) in terms of an evolution to low resolution scales.

*K-Matrix Equivalence.* We now briefly describe the evolution based on half-on-shell $K$-matrix equivalence[22] of a given input NN potential (e.g., the n3lo500 two-nucleon potential) to lower cutoff scales. We start with the integral equation for the $K$-matrix:

$$K_{\text{NN}}(p', p; p^2) = V_{\text{NN}}(p', p) + \frac{M}{2\pi^2} \fint_0^\infty dq\, q^2 \frac{V_{\text{NN}}(p', q) K_{\text{NN}}(q, p; p^2)}{p^2 - q^2}. \quad (1.56)$$

We write $(p', p; p^2)$ to indicate that the in-going nucleons with relative momentum $p = |\vec{p}|$ are on-shell, while the out-going nucleons with relative momentum $p' = |\vec{p}'|$ are not. The

---

[20] An additional source of nonperturbative behavior is the presence of few-nucleon bound (or nearly-bound) states. In the case of the NN channel, the convergence behavior of the Born series (for free space, or for NN scattering in a nuclear medium) can be studied in terms of the eigenvalue of $V_{\text{NN}} G_0$ (cf. [408]), where the (free) nucleon propagator is given by $G_0(E) = 1/(E - H)$ [free-space] and $G_0(E) = Q_F/(E - H)$ [in-medium], respectively (with $Q_F$ the Pauli-blocking operator). This analysis has been performed in Ref. [55] (see also [52, 53]): it was found that for sufficiently low cutoffs, the in-medium Born series becomes strongly convergent if the medium is dense enough to suppress to deuteron bound-state (via Pauli blocking). The in-medium properties of few-nucleon bound-states in nuclear matter have been investigated also in Ref. [398] (and also in [212, 348, 349, 350]) using two different methods (a quantum statistical approach, and RMF); it was found that light nuclei (i.e., $^3$H, $^3$He, etc.) are strongly suppressed (Pauli-blocked) for densities $\rho \gtrsim 0.1 \text{ fm}^{-3}$. Crucially, the presence of light nuclei in homogeneous nuclear matter (often referred to as *light clusters* in this context) should be distinguished from the clustering of neutron-star matter at subsaturation densities (i.e., in the crust) associated with the nuclear liquid-gas instability and Coulomb forces. The properties of neutron-star matter at subsaturation densities are discussed further at the beginning of Chap. 4.

[21] The $K$-matrix (also known as "reactance matrix", "reaction matrix", and "Heitler's matrix") corresponds (basically) to the real version of the $T$-matrix, and is defined as $K = (T^{-1} + i)^{-1} = \tan \delta$, with $\delta$ the phase shift for the partial wave under consideration (see Refs. [309, 383, 282] for more details).

[22] We follow [55] in the description. See also Refs. [56, 53] for a description of alternative methods (not based on $K$-matrix equivalence) that lead to the same $V_{\text{low-}k}$.

# 1. Nuclear Interactions and Many-Body Problem

low-momentum version of Eq. (1.56) is obtained by imposing a sharp cutoff $\Lambda$:

$$K_{\text{low-}k}(p', p; p^2) = V_{\text{low-}k}(p', p) + \frac{M}{2\pi^2} \int_0^\Lambda dq\, q^2 \frac{V_{\text{low-}k}(p', q) K_{\text{low-}k}(q, p; p^2)}{p^2 - q^2}. \quad (1.57)$$

We demand the matching condition

$$K_{\text{low-}k}(p', p; p^2) = K_{\text{NN}}(p', p; p^2) \qquad \forall\ p', p < \Lambda, \quad (1.58)$$

in order to ensure that $V_{\text{low-}k}$ gives the same results for low-energy (two-nucleon) observables as the input potential $V_{\text{NN}}$. This implies that $V_{\text{low-}k}$ depends on $\Lambda$, and differentiating Eq. (1.57) with respect to $\Lambda$ leads to the following flow equation for $V_{\text{low-}k}(\Lambda)$:

$$\frac{d}{d\Lambda} V_{\text{low-}k}(p', p) = \frac{M}{2\pi^2} \frac{V_{\text{low-}k}(p', \Lambda) K_{\text{low-}k}(\Lambda, p; \Lambda^2)}{1 - (p/\Lambda)^2}. \quad (1.59)$$

Note that the K-matrix on the right-hand side of Eq. (1.59) is left side half-on-shell, as opposed to the K-matrices in Eqs. (1.56) and (1.57), and therefore not RG-invariant. Because Eq. (1.59) is asymmetric with respect to $p$ and $p'$, it generates a non-Hermitian $V_{\text{low-}k}$. This deficiency can be cured by working with a symmetrized version of Eq. (1.59):

$$\frac{d}{d\Lambda} V_{\text{low-}k}(p', p) = \frac{M}{4\pi^2} \left( \frac{V_{\text{low-}k}(p', \Lambda) K_{\text{low-}k}(\Lambda, p; \Lambda^2)}{1 - (p/\Lambda)^2} + \frac{K_{\text{low-}k}(p', \Lambda; \Lambda^2) V_{\text{low-}k}(\Lambda, p)}{1 - (p'/\Lambda)^2} \right), \quad (1.60)$$

which preserves the on-shell $K$-matrix [54]. The low-momentum potential $V_{\text{low-}k}(\Lambda)$ can now be obtained by integrating the flow equation numerically, using the input potential as the initial condition. The resulting RG flow of the triplet $S$-wave ($^3S_1$) matrix elements is depicted in Fig. 1.6. Notably, for cutoffs in the range $\Lambda \lesssim 2\,\text{fm}^{-1}$, the RG evolved potential has been found to be approximately universal [53, 54], i.e., independent of the input potential.

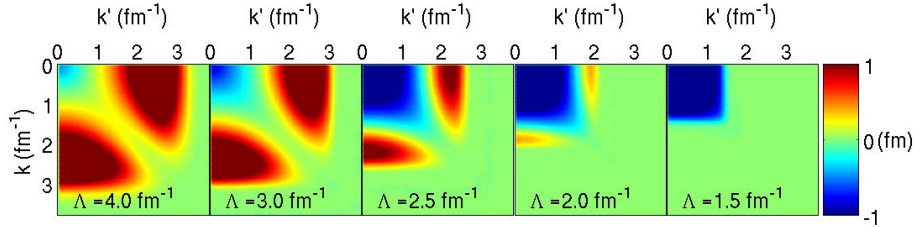

**Figure 1.6.:** (From [53]) $V_{\text{low-}k}$ evolution of momentum-space matrix elements ($^3S_1$-channel) of the chiral N3LO two-nucleon potential from Ref. [129].

*"Induced" Multi-Nucleon Interactions.* For applications outside of the NN sector, an evolution of the NN potential based on $K$-matrix equivalence implies certain problems concerning the consistent implementation of multi-nucleon interactions. From a general viewpoint, multi-nucleon interactions are required in an effective low-energy description of the nuclear interaction to parametrize unresolved short-distance effects in multi-nucleon channels. For instance, in 3N scattering diagrams generated by iterated NN interactions, resonances can appear in the intermediate states ("virtual excitations", cf. Fig. 1 in [182]). A consistent evolution of the NN interaction to lower cutoffs has the same effect: high-momentum modes are not taken into account explicitly, but included via 3N contact terms, i.e., $c_E$ and $c_D$ would have to be refit to be consistent with an NN potential evolved to lower cutoffs. In addition, of course, the evolved NN potential corresponds to new values of the NN LECs (as evident in the different "n3lo" values for $c_{1,3,4}$ in Table 1.1)—but these new values are not known if an RG evolution of partial-wave matrix elements is employed. In Ref. [312] (see also [55]) the (somewhat provisional) approach was taken to use for each cutoff $\Lambda$ the (same) Nijmegen values of $c_{1,3,4}$ and adjust $c_E(\Lambda)$ and $c_D(\Lambda)$ to the $^3$H, $^3$He and $^4$He binding energies. A better method (SRG) is discussed below.

To summarize, additional 3N interactions (in terms of new values of the relevant LECs, $c_{E,D,1,2,3}$ in the N2LO case) are "induced" when intermediate-state excitations (such as the $\Delta$-isobar) and high momentum modes are integrated out.[23]

*Similarity Renormalization Group.* A much celebrated RG method is the "similarity renormalization group" (SRG) [55, 223, 184, 153, 154, 168], which is based on a continuous sequence of unitary transformations (with flow parameter $\lambda$) of the nuclear Hamiltonian, i.e.,

$$H_\lambda = U_\lambda H U_\lambda \equiv T_{\rm rel} + V_\lambda, \qquad (1.61)$$

where $H = T_{\rm rel} + V$ is the input Hamiltonian with relative kinetic energy operator $T_{\rm rel}$ and nuclear potential $V$. The SRG method is more flexible (compared to the $K$-matrix method), since there are many possible choices for the unitary operator $U_\lambda$. Concerning the application of the SRG method to the NN interaction, in most cases a choice was made that drives $V_{\rm NN}$ towards a band-diagonal form [53, 153], cf. Fig. 1.7. The SRG method can evolve also multi-nucleon potentials to lower resolution scales, i.e., $V = V_{\rm NN} + V_{\rm 3N} + \ldots$ in Eq. (1.61); thus, "induced" multi-nucleon interactions can be taken into account explicitly.[24]

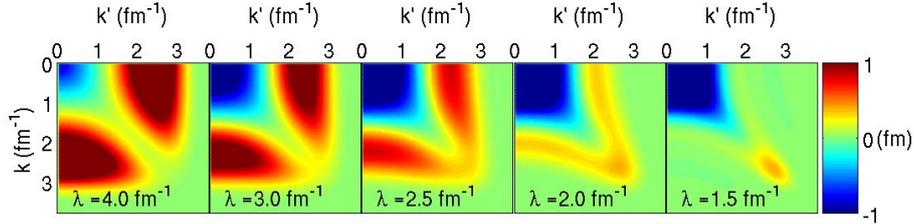

**Figure 1.7.:** (From [53]) SRG evolution of momentum-space matrix elements ($^3S_1$-channel) of the chiral N3LO two-nucleon potential from Ref. [129].

## 1.5. Nuclear Many-Body Problem and Astrophysics

We have completed our discussion of the construction of chiral low-momentum two- and three-nucleon potentials at this point. The question now is how to calculate from these potentials the thermodynamic properties of the nuclear many-body system? There exist various (complementary) methods that allow to compute nuclear many-body properties from a given nuclear potential. The most basic one is *many-body perturbation theory* (MBPT), which treats the nuclear potentials as a perturbation with respect to a (solvable) reference state (e.g., the noninteracting Fermi gas). Regarding "conventional" high-precision NN potentials (e.g., CD-Bonn, AV18) the application of MBPT is impeded by the nonperturbative properties of these potentials, but, as discussed in Sec. 1.4, with chiral potentials constructed at low resolution scales MBPT can be expected to be a valid and systematic approach. In this thesis we use MBPT to compute the thermodynamic properties of infinite nuclear matter. The general framework of MBPT (for both zero temperature, $T = 0$, and finite temperature, $T \neq 0$) is investigated in detail in the next chapter. Three other methods that have been used for nuclear matter calculations with (chiral) nuclear potentials are:

- the *Brueckner-Hartree-Fock* (BHF) method: The first-order approximation in a general zero-temperature scheme—often referred to as the "Brueckner-Goldstone theory"—that was designed to deal with the nonperturbative short-distance features of conventional NN potentials by (formally) resumming to all orders the contributions associated with in-medium scattering of excited states above the Fermi sphere (i.e., the *particle-particle* ladder diagrams of MBPT, cf. Sec. 2.3.2), see Refs. [95, 42, 338, 101, 171] for details. The resummation is carried out by constructing an effective *in-medium* interaction called the $G$-matrix (or "Brueckner reaction matrix"), which is defined in terms of the operator equation (the "Bethe-Goldstone equation")

$$G = V_{\text{NN}} + V_{\text{NN}} \frac{Q_F}{\mathcal{E}_0 - \mathcal{T}} G, \quad (1.62)$$

where $\mathcal{E}_0$ is the ground-state energy of a (solvable) reference system with (one-body) Hamiltonian $\mathcal{T}$, and $Q_F$ is the Pauli-blocking operator. The original perturbation series of MBPT in terms of $V_{\text{NN}}$ is then replaced by a modified series (involving $G$ instead of $V_{\text{NN}}$) to avoid the double-counting of diagrams (called the "hole-line expansion" or "Brueckner-Bethe-Goldstone expansion").[25] Nuclear many-body calculations within this scheme have been restricted mostly to the BHF level.[26] BHF calculations with chiral potentials were performed, e.g., in Refs. [275, 274]. The Brueckner theory has been generalized to finite temperatures in different ways [50, 48, 15, 266, 345], and was applied in thermodynamic nuclear many-body calculations (e.g.,) in Refs. [16, 266, 345].

---

[23] These two effects are (of course) not really separable; this point is emphasized (e.g.,) in Ref. [53].
[24] The consistent SRG evolution of both NN and 3N potentials has been achieved only relatively recently, cf. [223, 184, 154]. In this thesis we do not use SRG evolved potentials. We reiterate that nevertheless, the potential sets of Table 1.1 provide an adequate basis for the investigation of the model dependence of results in nuclear many-body calculations.

- the *Self-Consistent Green's Functions* (SCGF) method: The SCGF method is based on the self-consistent computation of the in-medium propagators or Green's functions in Fourier (Matsubara) space, corresponding to the resummation to all orders of (a given class of) perturbative contributions to the propagators (cf. e.g., [225, 367, 346, 71]). The thermodynamic information is then extracted from the propagators in terms of various identities [143, 102]. SCGF calculations with chiral nuclear potentials have been performed, e.g., in Refs. [73, 106], in the ladder approximation; this generalizes (in a sense) the (finite-temperature) BHF method such that also *hole-hole* ladders are (formally) resummed and several identities connecting thermodynamic quantities and single-particle properties are preserved ("conserving approximation") [346, 30].

---

[25] However, it should be noted here that, if based on a large-cutoff (free-space) NN potential, the $G$-matrix still contains a significant coupling between low- and high-momentum modes, see Ref. [53] for details.

[26] Higher order contributions have been included, e.g., in [366]. It should be noted that Brueckner-theory calculations based on a high-precision NN potential only have failed to reproduce the empirical saturation point of isospin-symmetric nuclear matter. The obtained results form a band (the "Coester band" [85]) that misses the empirical saturation point. This deficiency can be improved by including 3N interactions, or relativistic effects ("Dirac-Brueckner-Hartree-Fock"), cf., e.g., Ref. [62, 353]. We note also that in Ref. [274] the empirical saturation point has been reproduced in BHF by using chiral NN and 3N potentials with explicit $\Delta$-isobars and tuning only $c_E$ and $c_D$. Note that the $\Delta$-N mass splitting ($m_\Delta - m_N \simeq 293$ MeV) is not a large quantity in $\chi$EFT. This point is discussed further below, and in Sec. 3.3.3.

- *Quantum Monte Carlo* (QMC) methods: QMC encompasses a large family of methods based on solving the (many-body) Schrödinger equation (for a large but finite number of particles) computationally in terms of a repeated random sampling (*Monte Carlo method*). As such, QMC methods are truly nonperturbative, but in general the computational resources required are rather large (as compared to MBPT). Different QMC methods have been used in (zero-temperature) nuclear many-body calculations with chiral potentials, cf. e.g., Refs. [352, 347, 165, 424, 385].

In addition to MBPT and the above three methods, further many-body methods have been developed and applied in nuclear matter calculations; e.g., the coupled-cluster theory [181, 187, 361], variational approaches [8, 300], and nuclear lattice simulations [130]. Compared to MBPT, the great majority of these methods are (however) computationally much more expensive, in particular if high numerical precision is required.

At this point one should note that the use of *free-space* nuclear potentials—i.e., potentials derived from diagrams describing (NN, 3N, etc.) scattering processes that take place in vacuum— to compute nuclear many-body properties raises some concerns. The structure and effective interactions of nucleons embedded in a dense nuclear medium deviate (to a certain extent) from those in free space, and additional scales are involved in the nuclear many-body problem as compared to the case of few nucleons in free space.[27] Such nontrivial medium effects are (if at all) only partially included in the various many-body frameworks; [e.g., in MBPT particular classes of diagrams correspond to corrections to the nucleon mass in the presence of the medium (cf. Sec. 2.5.5), and in the Brueckner-Goldstone framework the $G$-matrix is an effective in-medium potential].

The question that arises is as to whether a more manifest inclusion of nucleon structure effects is possible (and expedient for a proper treatment of the nuclear many-body problem). At the EFT level, changes in the structure of nucleons can be accounted for by including (in addition to nucleons and pions) also nucleon excited states (e.g., the $\Delta$-resonance [316]) as degrees of freedom. Nuclear many-body calculations using nuclear $\chi$EFT-based potentials with explicit $\Delta$-isobars included have been performed recently in Ref. [274]. Since most of the available chiral nuclear potentials are based on pion-nucleon $\chi$EFT, we do not consider $\Delta$-improved potentials in this thesis.

As a related medium effect, we note that in (perturbative) nuclear matter calculations the impact of 3N interactions is rather large (cf. Sec. 3.3 and (e.g.,) [90, 91]) compared to what would be expected from the chiral hierarchy. This may be seen as an indication that, regarding the chiral power counting, a more consistent approach should address the new scales associated with the nuclear medium, i.e., the Fermi momentum and excitations of nucleons. Such an *in-medium chiral perturbation theory* based on an expansion around the zero-temperature ground-state of nuclear matter has been developed and studied in [318, 317, 254, 293, 245, 253], cf. also [130, 156]. A different approach was taken in [279, 237, 151] where the perturbative many-body diagrams have been ordered in terms of pion masses and momentum and the Fermi momentum (as well as the nucleon-$\Delta$ mass splitting, in the $\Delta$-full version [151]); the framework has been extended to finite temperatures in Ref. [150] (see also the review [209]).

---

[27] These effects become particularly evident by adopting the fundamental point of view that hadrons are bound-states of strongly-interacting quarks and gluons. Clearly, for densities considerably above nuclear saturation density a description of dense matter in terms of hadronic point particles becomes less meaningful or even genuinely wrong (for the case of a transition to deconfined QCD matter at these densities); cf. (e.g.) Refs. [149, 169] and [31, 195] for general aspects and Refs. [210, 389] for different perspectives on this point.

# 1. Nuclear Interactions and Many-Body Problem

Altogether, these concerns should not be thought of as invalidating an approach based on free-space potentials (constructed from chiral pion-nucleon dynamics). But it should be clear that this method must not be considered as the prior route to a realistic description of nuclear thermodynamics, and should be complemented by other approaches. Furthermore, the above discussion makes evident the importance of empirical constraints that can be used as benchmarks for nuclear many-body calculations. In the present thesis, we will benchmark our nuclear many-body calculations against various empirical constraints, in particular (of course) the empirical saturation point of isospin-symmetric nuclear matter. An important recent contribution regarding constraints on the EoS of dense neutron-rich matter comes from the observation and precise measurement of two-solar-mass neutron stars [100, 11]. Any EoS that leads to a neutron-star mass-radius relation $M(R)$ which does not support a neutron star with $M \simeq 2M_\odot$ is ruled out, which at first glance appears to favor neutron-star models with primarily nucleonic degrees of freedom [100, 197, 256, 190, 192], and challenges models that include substantial contributions from "exotic" (non-nucleonic) hadronic matter, i.e., hyperons, pion or kaon condensates, or deconfined quark matter ("hybrid star").[28] "Exotic" hadrons or deconfined quark matter may however still be present in the inner core of the star [288, 289, 370, 248]. As an example, we show in Fig. 1.8 the $M(R)$ plot of Ref. [100] where the $M(R)$ results for various equations of state based on purely nucleonic degrees of freedom as well as nucleons plus "exotic" hadrons are compared. Included are also two equations of state for compact stars composed entirely of self-bound quark matter ("strange stars" [51, 423, 9, 180], a so far only theoretical concept).

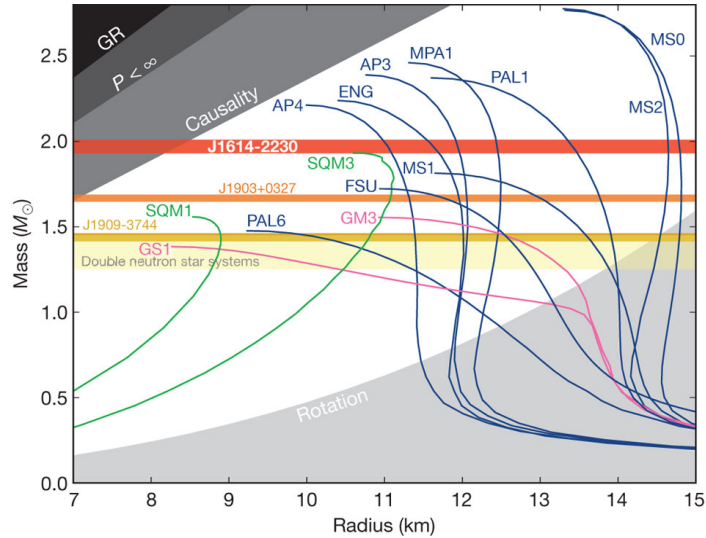

**Figure 1.8.:** (From [100]) Neutron-star mass-radius relation $M(R)$ for several equations of state (blue lines: nucleonic, pink lines: nucleonic plus "exotic", green lines: strange stars; see Refs. [100, 259] for details). The horizontal bands show constraints from neutron-star measurements, including the two-solar-mass PSR J1614-2230. The regions colored in different shades of gray correspond to general constraints from neutron-star dynamics and general relativity (including the "causality border" where the speed of sound in the star begins to exceed the speed of light, cf. [122] for discussion). Any EoS which leads to a $M(R)$ line that does not intersect the PSR J1614-2230 band is ruled out.

---

[28] In particular, in order to support neutron stars with $M \simeq 2M_\odot$, the EoS has to be sufficiently *stiff* (i.e., involve a high incompressibility) at high densities, which is complicated by introducing additional degrees of freedom.

## 1. Nuclear Interactions and Many-Body Problem

In Ref. [352], the mass-radius relation corresponding to the EoS of pure neutron matter (protons and electrons in the star have been neglected) computed within this thesis has been constructed. The results are shown in Fig. 1.9. Since it is based on (chiral) low-momentum interactions, this EoS is by construction restricted to densities below about twice nuclear saturation density, $\rho \lesssim 2\,\rho_{\text{sat}}$, which lies below the central densities reached in massive neutron stars $\rho \sim (4-6)\,\rho_{\text{sat}}$ [260, 189, 197, 417]. In view of this issue, several phenomenological mean-field models—relativistic mean-field (RMF) models based on phenomenological nucleon-meson Lagrangians [360, 341, 169] as well as energy-density functionals based on phenomenological Skyrme interactions [374]—have been benchmarked to the thermodynamic nuclear EoS of isospin-symmetric nuclear matter and pure neutron matter derived within this thesis [352]. The neutron-star matter EoS and mass-radius relation obtained from the constrained models (reaching to higher densities) are also shown in Fig. 1.9. One sees that none of the models is consistent with the two-solar-mass constraint. However, since for the high densities required for two-solar-mass neutron stars there are no substantial constraints (in addition to the condition that $M(R)$ reaches $2M_\odot$ for $R \simeq (10-14)$ km, cf. Refs. [396, 263, 261]) on the properties of dense baryonic matter, this failure should not be taken too seriously. But it suggests to consider (in addition to modifying the high-density behavior of the mean-field models used in [352][29]) additional constraints for mean-field models in terms of the isospin-asymmetry dependent nuclear EoS.

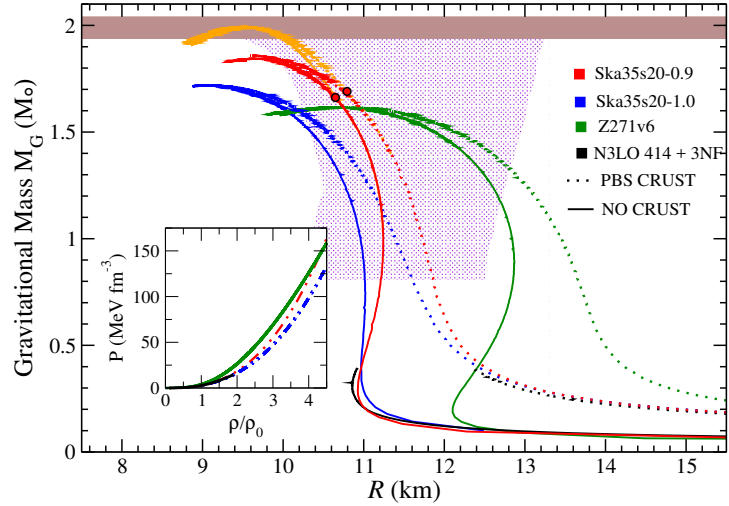

**Figure 1.9.:** (From [352]) Neutron-star mass-radius relation and pressure as a function of nucleon density (inset) obtained from the nuclear EoS ("N3LO414 + 3NF") computed within this thesis as well as results from different mean-field models (red, blue and green lines) benchmarked to the thermodynamic "N3LO414 + 3NF" results, see text and Ref. [352] for more details. Since they are based on low-momentum interactions, the "N3LO414 + 3NF" results are restricted to densities $0 \le \rho \lesssim 2\rho_{\text{sat}}$ and (as a consequence) to masses $M \lesssim 0.5\,M_\odot$. The solid lines show the results where the neutron-star crust has been modelled as a homogeneous fluid, and the dashed lines the results where a more realistic model of the crust is included. The yellow lines show the results based on the "Ska35s20-09" EoS where for $\rho \ge 4.5\,\rho_{\text{sat}}$ the EoS is replaced by that of a liquid with constant speed of sound equal to the speed of light.

---

[29] See also [189, 190, 251] where an alternative approach towards the extension to high-densities of the EoS from chiral nuclear potentials in MBPT was considered, i.e., a parametrization using piecewise polytropes.

## 1. Nuclear Interactions and Many-Body Problem

After their creation in core-collapse supernovæ, (proto-)neutron stars cool rapidly to temperatures below the MeV scale (cf. [322, 335, 427, 180]) where the free energy density of nuclear matter can be identified with the ground-state energy density to very good accuracy. The observation of two-solar-mass neutron stars thus sets constraints on the nuclear EoS at zero temperature. Concerning the finite-temperature EoS, a major application lies in the domain of (computationally very demanding) simulations of core-collapse supernovæ and binary neutron-star mergers (cf., e.g., Refs. [294, 28, 27, 224, 25]), where (basically) the nuclear EoS serves as an input for solving (numerically) the gravitational and hydrodynamics problems involved. Such simulations are important for understanding the details of the collapse mechanism, general aspects of heavy-element nucleosynthesis [387, 294, 224], and for the interpretation of expected observations of gravitational wave forms linked to binary neutron star (or neutron star—black hole) mergers [26, 372, 340].[30] A noteworthy study regarding the nuclear-physics side of these issues was performed in Ref. [251], where the results for the ground-state energy density of pure neutron matter obtained from various chiral nuclear potentials (including generous uncertainty estimates for the contribution from 3N interactions) in MBPT were compared to several phenomenological equations of state commonly used in nuclear astrophysics. The results of this study are shown in Fig. 1.10. One sees that most of the phenomenological equations of state are inconsistent with the results obtained from the chiral nuclear potentials. The analysis of Ref. [251] as well as the vast majority of nuclear many-body calculations with $\chi$EFT-based potentials in general, however, were restricted to the zero-temperature case and either PNM or SNM. Extending these studies to finite temperature and variable neutron-to-proton ratio is a major motivation, and thus, the central objective of this thesis is the investigation of **the thermodynamic equation of state of isospin-asymmetric nuclear matter from low-momentum chiral nuclear potentials in many-body perturbation theory**.

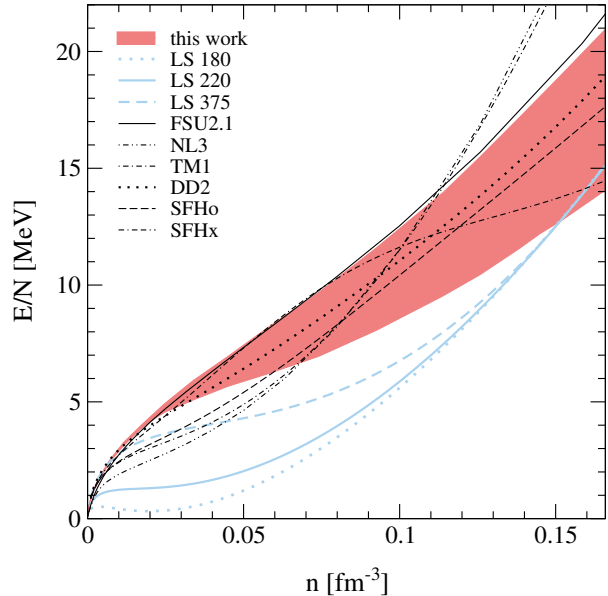

**Figure 1.10.:** (From [251]) Comparison of the zero-temperature EoS of pure neutron matter (ground-state energy per particle "E/N" as function of neutron density "n") from perturbative calculations using chiral nuclear potentials (red band) with phenomenological equations of state commonly used in core-collapse supernova simulations (blue and black lines), cf. Ref. [251] for details.

---

[30] For the case of black hole—black hole mergers, gravitational waves have been observed recently [1, 2].

# 2. Many-Body Perturbation Theory

The approach to the thermodynamic nuclear many-body problem (i.e., the problem of computing the EoS of infinite homogeneous nuclear matter) used in this thesis is many-body perturbation theory (MBPT). This approach is motivated by the use of (chiral) nuclear interactions constructed at low resolution scales, for which a perturbative treatment is expected to be valid. MBPT is based on a separation of the Hamiltonian $\mathcal{H} = \mathcal{T} + \mathcal{V}$ into a reference Hamiltonian $\mathcal{T}$ and a perturbation $\mathcal{V}$. The many-body problem is then set up as an expansion in terms of $\mathcal{V}$, where all contributions are evaluated with respect to eigenstates of $\mathcal{T}$. In this chapter, we investigate in detail this framework (MBPT), both at zero temperature and at finite temperature.

To set up the notation, we define $|\Psi_p\rangle$ as the energy eigenstates of a many-fermion system with Hamiltonian $\mathcal{H} = \mathcal{T} + \mathcal{V}$, i.e.,

$$\mathcal{H}|\Psi_p\rangle = (\mathcal{T} + \mathcal{V})|\Psi_p\rangle = E_p|\Psi_p\rangle. \tag{2.1}$$

Furthermore, we define $|\Phi_p\rangle$ as the energy eigenstates of the unperturbed system with Hamiltonian $\mathcal{T}$, i.e.,

$$\mathcal{T}|\Phi_p\rangle = \mathcal{E}_p|\Phi_p\rangle. \tag{2.2}$$

The unperturbed Hamiltonian is diagonalized by a set of orthonormal single-particle states $|\varphi_i\rangle$, i.e.,[1]

$$\mathcal{T} = \sum_i \varepsilon_i\, a_i^\dagger a_i, \tag{2.3}$$

with single-particle energies $\varepsilon_i$. The creation and destruction operators $a_i^\dagger$ and $a_i$ are defined by creating or destroying single-particle states $|\varphi_i\rangle$ in the Fock states $|\Phi_p\rangle = |\eta_1 \eta_2 \ldots \eta_\infty; N_p\rangle$, with occupation numbers $\eta_i \in \{0, 1\}$, where $N_p = \sum_i \eta_i$ is the number of particles in the state $|\Phi_p\rangle$.

It will be useful to start by taking $\mathcal{T}$ as the noninteracting Hamiltonian, so $\mathcal{V}$ is the interaction Hamiltonian. For clarity and without loss of generality, (for the most part of this chapter) we consider two-body (2B) interactions only, i.e., the interaction Hamiltonian is

$$\mathcal{V} = \frac{1}{2!}\sum_{ijkl} \langle \varphi_i \varphi_j | V_{2\mathrm{B}} | \varphi_k \varphi_l \rangle\, a_i^\dagger a_j^\dagger a_l a_k \equiv \frac{1}{2!}\sum_{ijkl} V_{2\mathrm{B}}^{ij,kl}\, a_i^\dagger a_j^\dagger a_l a_k. \tag{2.4}$$

Using this setup, in **Sec. 2.1** we then construct the zero-temperature perturbation series for the ground-state energy[2]

$$E_0(\varepsilon_F, \Omega) = \mathcal{E}_0(\varepsilon_F, \Omega) + \lambda E_{0;1}(\varepsilon_F, \Omega) + \lambda^2 E_{0;2}(\varepsilon_F, \Omega) + \lambda^3 E_{0;3}(\varepsilon_F, \Omega) + \mathcal{O}(\lambda^4), \tag{2.5}$$

---

[1] The states $|\varphi_i\rangle$ are solutions to the single-particle Schrödinger equation under the boundary conditions imposed by the confining volume $\Omega$.

[2] The artifical parameter $\lambda$ (the "interaction strength") is introduced for bookkeeping reasons only.

where $\mathcal{E}_0(\varepsilon_F, \Omega)$ is the energy of the unperturbed ground state $|\Phi_0\rangle$. The Fermi energy $\varepsilon_F$ is in one-to-one correspondence with the number of particles in the system via $N = \sum_i \Theta(\varepsilon_F - \varepsilon_i)$, and $\Omega$ is the confining volume. In **Sec. 2.2** we then derive the perturbation series for the grand-canonical potential

$$A(T, \mu, \Omega) = \mathcal{A}(T, \mu, \Omega) + \lambda A_1(T, \mu, \Omega) + \lambda^2 A_2(T, \mu, \Omega) + \lambda^3 A_3(T, \mu, \Omega) + O(\lambda^4), \qquad (2.6)$$

where $T$ is the temperature and $\mu$ is the chemical potential. To describe an infinite homogeneous system one can fix the single-particle states $|\varphi_i\rangle$ by using periodic boundary conditions in a cubic box, and then (for the actual numerical many-body calculations) take the thermodynamic limit.[3] Crucial concerning the thermodynamic limit is the appearance of energy denominators in the expressions for (certain) higher-order perturbative contributions to $E_0(\varepsilon_F, \Omega)$ and $A(T, \mu, \Omega)$, respectively. In the ground-state formalism, the energy denominators vanish (only) at boundary of the respective integration regions,[4] but in the grand-canonical case they can be zero in the interior of the integrals; the corresponding energy-denominator poles are however ficticious, since they are not present if all perturbative contributions whose diagrammatic representations transform into each other under **cyclic vertex permutations** are evaluated together. The different behavior of the energy denominators at zero and at finite temperature will play a major role in Chap. 5.

The various contributions to $E_0(\varepsilon_F, \Omega)$ and $A(T, \mu, \Omega)$ are analyzed in terms of diagrams in **Sec. 2.3**. We will distinguish between "skeleton" and "non-skeleton" diagrams, where non-skeletons are diagrams that involve several skeleton parts connected via "articulation lines". All diagrams that appear in the ground-state formalism are also present in the grand-canonical case, however, in the grand-canonical formalism there are additional types of non-skeletons, called "anomalous diagrams". The skeletons and non-skeletons that appear also in the ground-state formalism are correspondingly called "normal". In both cases, additional non-skeletons arise if an additional one-body operator is added to the perturbation Hamiltonian $\mathcal{V}$. Classes of contributions that have the same structure as (a subset of) the additional ones arising in that way are referred to as "insertions". In the ground-state formalism, there is exactly one insertion type, referred to as "one-loop" insertions: contributions from (normal) non-skeletons with first-order parts. In the grand-canonical (and the canonical) formalism there are additional insertion structures, and a major part of this chapter deals with the problem of identifying and interpreting this feature.

In the zero-temperature formalism the expression associated with an individual diagram is unique, but in the grand-canonical formalism this is not the case: there are several ways individual diagrams can be evaluated. We will consider three different prescriptions to evaluate diagrams, referred to as the **"cyclic formula"**, the **"direct formula"**, and the **"reduced formula"**.[5] Concerning the expressions that emerge from these formulas, the ones obtained for (certain) non-skeletons require special attention, i.e.,

- non-skeletons that involve an even number of identical energy denominators: the expressions for these diagrams obtained from the direct or the reduced formula diverge in the thermodynamic limit. This is due to contributions from the vicinity of symmetric energy-denominator poles and therefore.

---

[3] In this setup, the various contributions in Eqs. (2.5) and (2.6) all scale linearly with $\Omega$, leading to perturbation series for the ground-state energy density and the grand-canonical potential density, respectively, cf. the beginning of Sec. 2.3.

[4] This holds for the perturbative contributions to the ground-state energy but not for the "self-energies" $X_{[n]}$. This is why the regularization symbol "$R$" in Eq. (2.8) appears also in the zero-temperature case.

[5] For the sum of all contributions that emerge from each other under cyclic permutations of interaction operators in the ensemble average, the three formulas lead to equivalent (but not necessarily identical) expressions.

- non-skeletons with more than two skeleton parts: for each formula, these diagrams lead to expressions that diverge in the zero-temperature limit.[6]

These two features are artifacts: if the cyclic formula (where each diagram is evaluated together with its cyclic permutations) is used, no energy-denominator poles appear, and for the sum of all (normal and anomalous) non-skeletons at each order the $T \to 0$ limit is nonsingular. Thus, there are no serious obstacles if one treats consistently all contributions at a given perturbative order. However, the above two features do become relevant if the single-particle basis is "renormalized" in terms of a self-consistent single-particle potential beyond Hartree-Fock. More specifically, we will consider the self-consistent "renormalization" of $\mathcal{T}$ and $\mathcal{V}$ according to

$$\mathcal{T} \to \mathcal{T} + \mathcal{X}_{[n]}^{(\aleph)}, \qquad \text{and} \qquad \mathcal{V} \to \mathcal{T} - \mathcal{X}_{[n]}^{(\aleph)}, \tag{2.7}$$

with one-body operators $\mathcal{X}_{[n]}^{(\aleph)} = \sum_i X_{[n];i}^{(\aleph)} a_i^\dagger a_i = \sum_{\nu=1}^n \sum_i X_{\nu;i}^{(\aleph)} a_i^\dagger a_i$, where $X_{\nu;i}^{(\aleph)}$ is defined via

$$X_{\nu;i}(\varepsilon_F) = \frac{\delta E_{0;\nu,\text{non-insertion}}^R[\Theta_i^-; \{\varepsilon_{X_{[n]};i}\}]}{\delta \Theta_i^-}, \qquad \text{and} \qquad X_{\nu;i}^{\aleph}(T,\mu) = \frac{\delta A_{\nu,\text{non-insertion}}^{R,\aleph}[f_i^-; \{\varepsilon_{X_{[n]}^\aleph;i}\}]}{\delta f_i^-}, \tag{2.8}$$

respectively. Here, "$\nu$" refers to the perturbative order, "$\aleph$" refers to the particular way the non-insertion contributions are evaluated in the grand-canonical case ("direct" or "reduced"), "$R$" refers to a regularization of terms with symmetric energy-denominator poles required to make the thermodynamic limit well-defined, and $\varepsilon_{X_{[n]}^{(\aleph)};i}$ are the single-particle energies in the $\mathcal{T} + \mathcal{X}_{[n]}^{(\aleph)}$ reference system.[7]

In the zero-temperature case, identifying the effect of Eq. (2.7) is straightforward: single-particle energies in the energy-denominators are renormalized (in terms of $\varepsilon_{X_{[n]};i}$), and there are additional (normal) non-skeletons with $-X_2, \ldots, -X_n$ vertices, which have the structure of "one-loop" insertions. The $-X_1$ vertices are cancelled by the previous "one-loop" insertions, and $-X_2, \ldots, -X_n$ vertices can be cancelled by expanding the energy denominators about the Hartree-Fock ones.[8] In the grand-canonical (and the canonical) case, the situation is more involved; in particular, additional cancellations appear (whose precise form depends on the choice of $\aleph$), but to identify these cancellations a reorganization of the grand-canonical (canonical) perturbation series is required. More specifically, in **Sec. 2.4**, we will reorganize the grand-canonical perturbation series by evaluating the various perturbative contribution in terms of **cumulants** (truncated correlation functions). In the cumulant formalism (i) the expressions for normal non-skeletons are such that all Fermi-Dirac distributions correspond to distinct single-particle states, and (ii) the contributions from anomalous diagrams are zero; instead, there now additional contributions from unlinked diagrams that are "connected" via higher cumulants: these contributions correspond to the additional insertions present at finite temperature.

---

[6] In the case of the reduced formula, only anomalous non-skeletons with more than two skeleton parts contain terms that diverge as $T \to 0$.

[7] For a discrete spectrum, the regularization "$R$" is not required, but the perturbation series corresponding to the $\{\mathcal{T} + \mathcal{X}_{[n]}^{(\aleph)}, \mathcal{V} - \mathcal{X}_{[n]}^{(\aleph)}\}$ setups with $n \geq 2$ are then not well-behaved for large systems. Notably, however, in the finite-temperature case a different "regularization" would be needed for a discrete spectrum, i.e., one that explicitly excludes the contributions from "accidentally" vanishing energy denominators; otherwise, there are terms that diverge in the zero-temperature limit, see Sec. 2.3.4 for details.

[8] In other terms, in the zero-temperature case the $\{\mathcal{T} + \mathcal{X}_1, \mathcal{V} - \mathcal{X}_1\}$ setup is equivalent to the $\{\mathcal{T}, \mathcal{V}\}$ one with

Furthermore, in **Sec. 2.4** we show that the cumulant formalism can be generalized to construct (starting from the canonical ensemble) a perturbation series for the free energy

$$F(T, \tilde{\mu}, \Omega) = \mathcal{F}(T, \tilde{\mu}, \Omega) + \lambda F_1(T, \tilde{\mu}, \Omega) + \lambda^2 F_2(T, \tilde{\mu}, \Omega) + \lambda^3 F_3(T, \tilde{\mu}, \Omega) + O(\lambda^4), \qquad (2.9)$$

where the auxiliary parameter $\tilde{\mu}$ is fixed by the condition $\sum_i \tilde{f}_i^- = N$ and therefore satisfies $\tilde{\mu} \xrightarrow{T \to 0} \varepsilon_F$ (here, $\tilde{f}_i^-$ denotes the Fermi-Dirac distribution with $\tilde{\mu}$ as the chemical potential). Compared to the grand-canonical perturbation series, the perturbation series for the free energy involves additional higher-cumulant contributions, called **"correlation bonds"**. The origin of these terms is the "substitution" of $\tilde{\mu}$ for the particle number $N$ in Eq. (2.9) in terms of a Legendre transformation of cumulants. It can be seen that if the reduced formula is used the sum of the two types of higher-cumulant contributions vanishes at each order in the zero-temperature limit (in the "isotropic case"). Since in the "reduced" case the expressions for normal diagrams have the same form as in the ground-state formalism (in the cumulant formalism), it follows that Eq. (2.9) provides a consistent thermodynamic generalization of the ground-state formalism. *This is not the case for the grand-canonical perturbation series*, which in general does not lead to consistent results, as discussed in **Sec. 2.5**. Moreover, in **Sec. 2.5** we also examine in more detail the effect of the self-consistent "renormalization" of $\mathcal{T}$ and $\mathcal{V}$, in both the grand-canonical and the canonical formalism. In particular, we examine the case where (formally) $n = \infty$ in Eq. (2.7), referred to as the "fully-renormalized" case. We will find that (only) the "fully-renormalized" grand-canonical and canonical perturbation series are equivalent, and for $\aleph$ = "reduced" one obtains "statistical quasiparticle" [96, 32, 74, 48, 47] relations for the particle number and the entropy.

## 2.1. Zero-Temperature Formalism

For the derivation of the perturbation series for the ground-state energy $E_0(\varepsilon_F, \Omega)$ one considers the adiabatic "switching on and off" of the interaction; i.e., the Hamiltonian is modified according to $\mathcal{H} \to \mathcal{H}_\epsilon(t) = \mathcal{T} + e^{-\epsilon|t|}\mathcal{V}$, where $\epsilon$ is infinitesimal. Under the condition that the ground state $|\Psi_0\rangle$ of the many-fermion system **evolves adiabatically**[9] from the ground state of the unperturbed system $|\Phi_0\rangle$, the Gell-Mann–Low theorem [161] (cf. also Ref. [143] pp.61-64) gives the following formula for the energy shift of the ground state, $\Delta E_0 = E_0 - \mathcal{E}_0 = \langle\Psi_0|\mathcal{H}|\Psi_0\rangle / \langle\Psi_0|\Psi_0\rangle - \langle\Phi_0|\mathcal{T}|\Phi_0\rangle / \langle\Phi_0|\Phi_0\rangle$:

$$\Delta E_0 = \lim_{\epsilon \to 0^+} \frac{\langle\Phi_0|\mathcal{U}_\epsilon(\infty)\mathcal{V}|\Phi_0\rangle}{\langle\Phi_0|\mathcal{U}_\epsilon(\infty)|\Phi_0\rangle}. \qquad (2.10)$$

The Dyson operator $\mathcal{U}_\epsilon(t) = e^{i\mathcal{T}t} e^{-i\mathcal{H}_\epsilon t}$ is formally given by the following series:

$$\mathcal{U}_\epsilon(t) = \sum_{n=0}^{\infty} \frac{(-i)^n}{n!} \int_0^t dt_n \cdots dt_1 \; e^{-\epsilon(t_n+\ldots+t_1)} \mathcal{P}[\mathcal{V}_I(t_n)\cdots\mathcal{V}_I(t_1)], \qquad (2.11)$$

with $\mathcal{P}[\ldots]$ the time-ordered product, and $\mathcal{V}_I(t)$ the interaction-picture representation of $\mathcal{V}$:

$$\mathcal{V}_I(t) = e^{i\mathcal{T}t} \mathcal{V} e^{-i\mathcal{T}t} = \frac{1}{2!} \sum_{ijkl} V_{2B}^{ij,kl} \; a_{I,i}^\dagger(t) a_{I,j}^\dagger(t) a_{I,l}(t) a_{I,k}(t), \qquad (2.12)$$

---

resummed "one-loop" insertions, and the $\{\mathcal{T} + \mathcal{X}_{[n]}, \mathcal{V} - \mathcal{X}_{[n]}\}$ setups with $n \geq 2$ are somewhat contrived in that case, since they lead to additional contributions but do not cancel any previous ones.

with $a_{I,i}^{(\dagger)}(t) = e^{i\mathcal{T}t} a_i^{(\dagger)} e^{-i\mathcal{T}t}$. For clarity, we restrict the discussion to the case of a (fermionic) system with only one species (i.e., a pure substance). The general case is readily obtained by introducing for each particle species $\xi$ separate particle numbers $N_\xi$ and Fermi energies $\varepsilon_{F;\xi}$.

## 2.1.1. Linked-Cluster Theorem

From Eqs. 2.10 and 2.11, the expression for the energy shift $\Delta E_0$ is given by

$$\Delta E_0 = \lim_{\epsilon \to 0^+} \frac{\sum_{n=0}^{\infty} \frac{(-i)^n}{n!} \int_0^\infty dt_n \cdots dt_1 \, e^{-\epsilon(t_n+\ldots+t_1)} \langle \Phi_0 | \mathcal{P}[\mathcal{V}_I(t_n) \cdots \mathcal{V}_I(t_1)] \mathcal{V}_I(0) | \Phi_0 \rangle}{\sum_{n=0}^{\infty} \frac{(-i)^n}{n!} \int_0^\infty dt_n \cdots dt_1 \, e^{-\epsilon(t_n+\ldots+t_1)} \langle \Phi_0 | \mathcal{P}[\mathcal{V}_I(t_n) \cdots \mathcal{V}_I(t_1)] | \Phi_0 \rangle}. \tag{2.13}$$

To evaluate (the numerator and denominator of) Eq. (2.13) one needs to evaluate the ground-state expectation values of a time-ordered product of creation and destruction operators, i.e.,

$$\langle \Phi_0 | \, \mathcal{P}[\, a_{I,i_n}^\dagger(t_n) a_{I,j_n}^\dagger(t_n) a_{I,k_n}(t_n) a_{I,l_n}(t_n) \, \cdots \, a_{I,i_1}^\dagger(t_1) a_{I,j_1}^\dagger(t_1) a_{I,k_1}(t_1) a_{I,l_1}(t_1) \,] \, | \Phi_0 \rangle. \tag{2.14}$$

By Wick's theorem, Eq. (2.14) is equal to all fully-contracted versions of the time-ordered string of creation and destruction operators, keeping track of the signs coming from the fermionic anticommutation relations while rearranging the operators to ensure the contracted terms are adjacent in the string. To evaluate the contractions, we note that

$$\frac{\partial}{\partial t} a_{I,i}^\dagger(t) = e^{i\mathcal{T}t} [i\mathcal{T}, a_i^\dagger]_- e^{-i\mathcal{T}t} = i\varepsilon_i a_i \quad \Rightarrow \quad a_{I,i}^\dagger(t) = a_i^\dagger \, e^{i\varepsilon_i t}, \tag{2.15}$$

$$\frac{\partial}{\partial t} a_{I,k}(t) = e^{i\mathcal{T}t} [i\mathcal{T}, a_k]_- e^{-i\mathcal{T}t} = -i\varepsilon_k a_k \quad \Rightarrow \quad a_{I,k}(t) = a_k \, e^{-i\varepsilon_k t}. \tag{2.16}$$

Using the fact that the unperturbed ground state $|\Phi_0\rangle$ (the filled Fermi sea) corresponds to the occupation of all single-particle states with energies $\varepsilon_i$ below the Fermi energy $\varepsilon_F$, one finds that the contractions of time-independent creation and destruction operators are given by (cf. e.g., [178] pp.205-217)

$$\underline{a_i^\dagger a_k} = \delta_{ik} \Theta(\varepsilon_F - \varepsilon_i) =: \delta_{ik} \Theta_i^-, \qquad \underline{a_k a_i^\dagger} = \delta_{ik} (1 - \Theta(\varepsilon_F - \varepsilon_i)) =: \delta_{ik} \Theta_i^+. \tag{2.17}$$

The different ways of contracting the creation and destruction operators in Eq. (2.14) can be represented by so-called **Hugenholtz diagrams**: for each quadruple of creation and destruction operators $a_{i_\nu}^\dagger a_{j_\nu}^\dagger a_{k_\nu} a_{l_\nu}$ at time $t_\nu$ one draws a black dot (vertex), a green line for every $\underline{a^\dagger a}$ type contraction, and a black line for every $\underline{a a^\dagger}$ type contraction. For example, the four diagrams in Fig. 2.1 represent possible contributions to the third summand in the numerator and the fourth summand in the denominator in Eq. (2.13). Diagram (a) in Fig. 2.1 is a *linked cluster*, i.e., there are no unconnected parts. In contrast, diagrams (b,c,d) have each one vertex that is not linked to any other vertex and thus are composed of two separate linked clusters. The green lines in Fig. 2.1 are called *hole* lines, since a contraction $\underline{a_{i_\nu}^\dagger a_{i_{\nu'}}}$ can be associated with the "destruction" of a single-particle state $|\phi_{i_\nu}\rangle = |\phi_{i_{\nu'}}\rangle$ in the filled Fermi sea at time $t_\nu$ and its recreation at (a later or equal) time $t_{\nu'}$. Similarly, the black lines are called *particle* lines: a contraction $\underline{a_{i_\nu} a_{i_{\nu'}}^\dagger}$ is

---

[9] This assumption is justified only for the "isotropic case", i.e., for a infinite homogeneous system with rotationally invariant interactions (and no external potential), see Secs. 2.4.6 and 2.5 and Refs. [246, 218]. Here, we assume that the adiabatically evolved state coincides with the true ground state of the interacting system.

## 2. Many-Body Perturbation Theory

associated with the propagation of a fermion in an unoccupied state $|\phi_{i_\nu}\rangle = |\phi_{i_{\nu'}}\rangle$ above the filled Fermi sea.

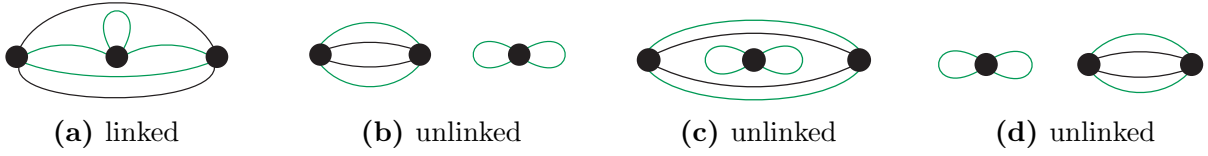

**(a)** linked     **(b)** unlinked     **(c)** unlinked     **(d)** unlinked

**Figure 2.1.:** Hugenholtz diagrams representing possible (linked or unlinked) contractions of three quadruples of creation and destruction operators. Green lines represent *holes*, black lines *particles*.

We now denote by $\mathit{\Gamma}_n$ a linked-cluster contribution associated with the expectation value of $\mathcal{P}[\mathcal{V}_I(t_n)\cdots\mathcal{V}_I(t_1)]$, and by $\mathit{\Gamma}_{n_1^{k_1},\ldots,n_\nu^{k_\nu}}$ the contribution given by the sum of all diagrams consisting of $k_1 n_1 + \ldots + k_\nu n_\nu$ separate linked clusters (with $n_1^{k_1} + \ldots + n_\nu^{k_\nu} = n$), where $k_i$ is the number of subclusters of order $n_i$. The number of ways the $n$ interaction operators can be partitioned into the subgroups specified by $\mathit{\Gamma}_{n_1^{\alpha_1},\ldots,n_\nu^{\alpha_\nu}}$ is given by [6, 178, 143]

$$\frac{1}{\alpha_1!\cdots\alpha_\nu!}\frac{n!}{(n_1!)^{\alpha_1}\cdots(n_\nu!)^{\alpha_\nu}}. \tag{2.18}$$

For instance, $\mathit{\Gamma}_{1^1,2^1}$ is given by the three diagrams (b,c,d) of Fig. 2.1, which all give equivalent contributions[10] since different time orderings are summed over in Eq. (2.11). Eqs. (2.11) and (2.18) show that the expression for $\mathit{\Gamma}_{n_1^{\alpha_1},\ldots,n_\nu^{\alpha_\nu}}$ is given by

$$\boxed{\mathit{\Gamma}_{n_1^{\alpha_1},\ldots,n_\nu^{\alpha_\nu}} = \frac{1}{\alpha_1!\cdots\alpha_\nu!}(\mathit{\Gamma}_{n_1})^{\alpha_1}\cdots(\mathit{\Gamma}_{n_\nu})^{\alpha_\nu}} \tag{2.19}$$

Eq. (2.19) is called the **factorization theorem**. Denoting by $\mathit{\Gamma}_n^{(0)}$ a linked-cluster contribution corresponding to $\langle\mathcal{P}[\mathcal{V}_I(t_n)\cdots\mathcal{V}_I(t_1)]\mathcal{V}_I(0)\rangle$, Eq. (2.13) is given by

$$\Delta E_0 = \frac{\left(\mathit{\Gamma}_1^{(0)} + \mathit{\Gamma}_2^{(0)} + \mathit{\Gamma}_3^{(0)} + \ldots\right)\left(\sum\frac{1}{\alpha_1!\alpha_2!\cdots}(\mathit{\Gamma}_1)^{\alpha_1}(\mathit{\Gamma}_2)^{\alpha_2}\cdots\right)}{\sum\frac{1}{\alpha_1!\alpha_2!\cdots}(\mathit{\Gamma}_1)^{\alpha_1}(\mathit{\Gamma}_2)^{\alpha_2}\cdots} = \sum_{n=0}^\infty \mathit{\Gamma}_n^{(0)}. \tag{2.20}$$

This is known as the **linked-cluster theorem** for the perturbation series for the ground-state energy shift $\Delta E_0$. The expression for $\Delta E_0$ is therefore given by

$$\boxed{\Delta E_0 = \lim_{\epsilon\to 0^+}\sum_{n=0}^\infty\frac{(-i)^n}{n!}\int_0^\infty dt_n\cdots dt_1\ e^{-\epsilon(t_n+\ldots+t_1)}\langle\Phi_0|\mathcal{P}[\mathcal{V}_I(t_n)\cdots\mathcal{V}_I(t_1)]\mathcal{V}_I(0)|\Phi_0\rangle_{\text{linked}}}$$

(2.21)

---

[10] Note that if the time-ordering had been fixed [as in Eq. (2.22)], this would not be the case. The time-ordered expressions for diagrams (b,d) differ from the one for diagram (c), see Eqs. (2.197) and (2.198).

## 2.1.2. Time-Independent Representation

To carry out the time integrals, we consider the equivalent form of Eq. (2.21) with fixed time-ordering:

$$\Delta E_0 = \lim_{\epsilon \to 0^+} \sum_{n=0}^{\infty} (-i)^n \int_{\infty > t_n > \ldots > t_1 > 0} dt_n \cdots dt_1 \; e^{-\epsilon(t_n + \ldots + t_1)} \langle \Phi_0 | \mathcal{V}_I(t_n) \cdots \mathcal{V}_I(t_1) \mathcal{V}_I(0) | \Phi_0 \rangle_{\text{linked}} .$$
(2.22)

The ground-state expectation value in Eq. (2.22) can be written as

$$\left\langle \Phi_0 \big| \mathcal{V}_n \, e^{-i\Delta\mathcal{E}_n t_n} \cdots \mathcal{V}_1 \, e^{-i\Delta\mathcal{E}_1 t_1} \mathcal{V}_0 \big| \Phi_0 \right\rangle_{\text{linked}} , \tag{2.23}$$

where $\Delta\mathcal{E}_\nu = \varepsilon_{k_\nu} + \varepsilon_{l_\nu} - \varepsilon_{i_\nu} - \varepsilon_{j_\nu}$ is the excitation energy associated with the interaction operator $\mathcal{V}_\nu$.[11] In terms of intermediate states $|\Phi_\nu\rangle$ (i.e, the states of the system between times $t_\nu$ and $t_{\nu-1}$) and corresponding energy levels $\mathcal{E}_\nu$, Eq. (2.23) reads

$$\langle \Phi_0 | \mathcal{V}_n \, e^{-i(\mathcal{E}_n - \mathcal{E}_0) t_n} | \Phi_n \rangle \cdots \langle \Phi_2 | \mathcal{V}_1 \, e^{-i(\mathcal{E}_1 - \mathcal{E}_2) t_1} | \Phi_1 \rangle \langle \Phi_1 | \mathcal{V}_0 | \Phi_0 \rangle , \tag{2.24}$$

i.e., $\Delta\mathcal{E}_n = \mathcal{E}_n - \mathcal{E}_0$ and $\Delta\mathcal{E}_{\nu \neq n} = \mathcal{E}_\nu - \mathcal{E}_{\nu-1}$. For example, for the diagram shown in Fig. 2.2 one has[12]

$$\mathcal{E}_3 = \Delta\mathcal{E}_3 + \mathcal{E}_0 = \varepsilon_k + \varepsilon_l - \varepsilon_i - \varepsilon_j + \mathcal{E}_0,$$
$$\mathcal{E}_2 = \Delta\mathcal{E}_2 + \mathcal{E}_3 = \Delta\mathcal{E}_3 + \Delta\mathcal{E}_2 + \mathcal{E}_0 = \varepsilon_p + \varepsilon_q + \varepsilon_k + \varepsilon_l - \varepsilon_a - \varepsilon_b - \varepsilon_i - \varepsilon_j + \mathcal{E}_0,$$
$$\mathcal{E}_1 = \Delta\mathcal{E}_1 + \mathcal{E}_2 = \underbrace{\Delta\mathcal{E}_3 + \Delta\mathcal{E}_2 + \Delta\mathcal{E}_1}_{-\Delta\mathcal{E}_0} + \mathcal{E}_0 = \underbrace{\varepsilon_p + \varepsilon_q - \varepsilon_i - \varepsilon_j}_{-\Delta\mathcal{E}_0} + \mathcal{E}_0.$$

Inserting $n-1$ factors $1 = e^{i\mathcal{E}_0(t_\nu - t_\nu)}$, where $\nu = 1, \ldots, n-1$, Eq. (2.23) becomes

$$\left\langle \Phi_0 \big| \mathcal{V}_n \, e^{-i(\mathcal{E}_n - \mathcal{E}_0)(t_n - t_{n-1})} \cdots \mathcal{V}_2 \, e^{-i(\mathcal{E}_2 - \mathcal{E}_0)(t_2 - t_1)} \mathcal{V}_1 \, e^{-i(\mathcal{E}_1 - \mathcal{E}_0) t_1} \mathcal{V}_0 \big| \Phi_0 \right\rangle_{\text{linked}} . \tag{2.25}$$

This suggests to introduce the $n$ new variables $\zeta_1, \ldots, \zeta_n$ defined as $\zeta_\nu = t_\nu - t_{\nu-1}$ (where $t_0 = 0$), i.e.,

$$\left\langle \Phi_0 \big| \mathcal{V}_n \, e^{-i\tilde{\mathcal{E}}_n \zeta_n} \cdots \mathcal{V}_1 \, e^{-i\tilde{\mathcal{E}}_1 \zeta_1} \mathcal{V}_0 \big| \Phi_0 \right\rangle_{\text{linked}} , \tag{2.26}$$

where $\tilde{\mathcal{E}}_\nu = \mathcal{E}_\nu - \mathcal{E}_0$ are the energies of the intermediate states relative to the (unperturbed) ground-state energy. The variables $\zeta_\nu$ can be represented by red dashed lines ("time-cuts") that intersect the various contraction lines, as shown in Fig. 2.2. For each $\zeta_\nu$ the corresponding energy $\tilde{\mathcal{E}}_\nu$ is the sum of the energies of the intersected *particle* lines minus the sum of the energies of the intersected *hole* lines.

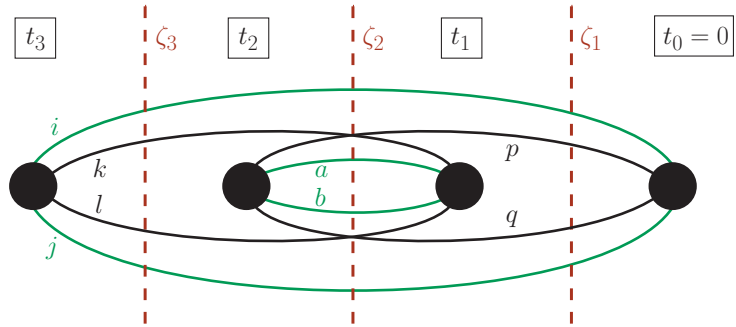

**Figure 2.2.:** "Time-cuts" (red dashed lines) in a linked cluster of fourth order.

---

[11] The subscript "$\nu$" in $\mathcal{V}_\nu$ refers only to the labeling of single-particle states.
[12] Note that the subscript "0" in $\mathcal{E}_0$ refers to both the ground state and the time $t_0 = 0$.

Substituting the "time-cut" variables $\zeta_\nu$, Eq. (2.21) is given by

$$\Delta E_0 = \lim_{\epsilon \to 0^+} \sum_{n=0}^\infty (-i)^n \int_0^\infty d\zeta_n \cdots d\zeta_1 \, e^{-\epsilon(n\zeta_n + \ldots + \zeta_1)} \left\langle \Phi_0 \left| \mathcal{V}_n e^{-i\mathcal{E}_n \zeta_n} \cdots \mathcal{V}_1 e^{-i\mathcal{E}_1 \zeta_1} \mathcal{V}_0 \right| \Phi_0 \right\rangle_{\text{linked}}. \quad (2.27)$$

The time integrations can now be carried out (the integrals converge because of the adiabatic exponential); letting $\epsilon \to 0$ afterwards, one obtains the following final expression for $\Delta E_0$:

$$\boxed{\Delta E_0 = \sum_{n=0}^\infty (-1)^n \left\langle \Phi_0 \left| \mathcal{V}_n \frac{1}{\mathcal{E}_n} \cdots \mathcal{V}_1 \frac{1}{\mathcal{E}_1} \mathcal{V}_0 \right| \Phi_0 \right\rangle_{\text{linked}}} \quad (2.28)$$

Eq. (2.28) was first derived by Goldstone [171], and is often called the **Goldstone formula**.[13]

## 2.2. Grand-Canonical Perturbation Theory

Following closely the methods established in the previous section, we now derive the perturbation series for the grand-canonical potential $A(T, \mu, \Omega) = \mathcal{A}(T, \mu, \Omega) + \Delta A(T, \mu, \Omega)$, where the unperturbed contribution is given by (cf. e.g., Ref. [15] pp.67-77)

$$\boxed{\mathcal{A}(T, \mu, \Omega) = -\frac{1}{\beta} \sum_i \ln\left(1 + e^{-\beta(\varepsilon_i - \mu)}\right)} \quad (2.29)$$

Here, $\beta = 1/T$ is the inverse temperature. In principle, the derivation of the perturbation series for $A(T, \mu, \Omega)$ can be completely taken over to the canonical ensemble, yielding a perturbation series for the free energy $F(T, N, \Omega)$, cf., e.g., Refs. [137, 167, 324]. By definition, the states $|\Phi_p\rangle = |\eta_1 \eta_2 \ldots \eta_\infty; N_p\rangle$ involved in the evaluation of the perturbative contributions to $F(T, N, \Omega)$ are subject to the constraint $N_p = N$. The constraint $N_p = N$ renders this naive canonical perturbation theory practically useless, since it implies that single-particle states cannot be summed independently. This is why the standard form of finite-temperature MBPT is (for all practical purposes) restricted to the grand-canonical ensemble.

### 2.2.1. Linked-Cluster Theorem

The grand-canonical partition function $Y(T, \mu, \Omega)$ is given by

$$Y = \sum_p \left\langle \Psi_p \left| e^{-\beta(\mathcal{H} - \mu N_p)} \right| \Psi_p \right\rangle = \sum_p \left\langle \Psi_p \left| e^{-\beta(\mathcal{T} - \mu N_p)} \mathcal{U}(\beta) \right| \Psi_p \right\rangle, \quad (2.30)$$

---

[13] See also Refs. [60, 218, 391, 101] for discussions regarding the relation of Eq. (2.28) with the "standard" (i.e., not based on QFT methods) Rayleigh-Schrödinger (RS) and self-consistent Brillouin-Wigner (BW) perturbation series: the Goldstone formula can be considered more convenient as the (equivalent) RS for technical reasons (linked-cluster theorem), and the BW has deficiencies concerning its convergence behavior in the many-body sector. Moreover, we note that in Refs. [218, 217] a derivation ("resolvent method") of the Goldstone formula that does not rely on the adiabatic "switching on and off" of the interaction is discussed. The resolvent method involves the expansion of the interacting ground-state $|\Psi_0\rangle$ in terms of the unperturbed states $|\Phi_p\rangle$; as discussed in [218, 217], the coefficients in this expansion vanish in the thermodynamic limit, i.e., for an infinite system the eigenstates of $\mathcal{T}$ and $\mathcal{H}$ do not belong to the same Hilbert space, and in particular, $\langle \Psi_0 | \Phi_0 \rangle = 0$ in that case (note that in the "standard' RS method, $\langle \Psi_0 | \Phi_0 \rangle \neq 0$ is postulated). Note also that in the adiabatic formalism one finds that $\langle \Psi_0 | \Phi_0 \rangle \propto e^{\alpha \epsilon^{-1}}$ in general, cf. Ref. [143] pp.61-64 (see also Ref. [179] p.55) and Ref. [115] pp.359-376.

where the finite-temperature Dyson operator $\mathscr{U}(\beta) = e^{-\beta\mathcal{T}} e^{\beta\mathcal{H}}$ is given by Wick rotating Eq. (2.11) to imaginary time ($t \rightarrow -i\tau$) and setting $\epsilon = 0$, i.e.,

$$\mathscr{U}(\beta) = \sum_{n=0}^{\infty} \frac{(-1)^n}{n!} \int_0^\beta d\tau_n \cdots d\tau_1 \, \mathcal{P}[\mathcal{V}_I(\tau_n) \cdots \mathcal{V}_I(\tau_1)]. \tag{2.31}$$

Changing the basis used to evaluate the partition function we obtain

$$Y = \sum_p \left\langle \Phi_p \left| e^{-\beta(\mathcal{T}-\mu\mathcal{N})} \mathscr{U}(\beta) \right| \Phi_p \right\rangle \equiv \mathrm{Tr}\left[ e^{-\beta(\mathcal{T}-\mu\mathcal{N})} \mathscr{U}(\beta) \right], \tag{2.32}$$

where $\mathcal{N} = \sum_i a_i^\dagger a_i$ is the number operator. Introducing the statistical operator corresponding to the unperturbed system $\varrho = \mathcal{Y}^{-1} e^{-\beta(\mathcal{T}-\mu\mathcal{N})}$, with $\mathcal{Y} = \sum_p \left\langle \Phi_p \left| e^{-\beta(\mathcal{T}-\mu\mathcal{N})} \right| \Phi_p \right\rangle$ the unperturbed partition function, one obtains the following expression for the difference of grand-canonical potentials $\Delta A = A - \mathcal{A}$:

$$\Delta A = -\frac{1}{\beta} \ln \overbrace{\langle \mathscr{U}(\beta) \rangle}^{\mathrm{Tr}[\varrho \mathscr{U}(\beta)]} = -\frac{1}{\beta} \ln \left[ \sum_{n=1}^{\infty} \frac{(-1)^n}{n!} \int_0^\beta d\tau_n \cdots d\tau_1 \, \langle \mathcal{P}[\mathcal{V}_I(\tau_n) \cdots \mathcal{V}_I(\tau_1)] \rangle \right], \tag{2.33}$$

where $\langle \ldots \rangle$ denotes the ensemble average in the unperturbed system. The finite-temperature interaction-picture creation and destruction operators are given by [cf. Eqs. (2.15) and (2.16)]

$$a_{I,i}^\dagger(\tau) = a_i^\dagger \, e^{\tau\varepsilon_i}, \qquad a_{I,k}(\tau) = a_k \, e^{-\tau\varepsilon_k}. \tag{2.34}$$

The finite-temperature version of Wick's theorem states that the ensemble average of a string of creation and destruction operators is given by the sum of all fully-contracted versions of that string. From the fermionic commutation relations and the cyclic property of the trace one finds that the nonvanishing contractions are given by (cf. [143] pp.238-240)

$$\underline{a_i^\dagger a_k} = \delta_{ik} f_i^-, \qquad \underline{a_k a_i^\dagger} = \delta_{ik} f_i^+. \tag{2.35}$$

where $f_i^- = 1/[1 + \exp(\beta(\varepsilon_i - \mu))]$ is the Fermi-Dirac distribution function, and $f_i^+ = 1 - f_i^- = 1/[1 + \exp(-\beta(\varepsilon_i - \mu))]$. Denoting again the contribution with $n$ linked vertices by $\Gamma_n$, one obtains the finite-temperature version of the linked-cluster theorem:

$$\Delta A = -\frac{1}{\beta} \ln \left[ \sum \frac{1}{\alpha_1! \alpha_2! \cdots} (\Gamma_1)^{\alpha_1} (\Gamma_2)^{\alpha_2} \cdots \right] = -\frac{1}{\beta} \ln \left[ e^{\Gamma_1 + \Gamma_2 + \cdots} \right] = -\frac{1}{\beta} \sum_{n=1}^{\infty} \Gamma_n. \tag{2.36}$$

Hence, the counterpart of Eq. (2.21) for the grand-canonical potential is

$$\boxed{\Delta A = -\frac{1}{\beta} \sum_{n=1}^{\infty} \frac{(-1)^n}{n!} \int_0^\beta d\tau_n \cdots d\tau_1 \, \langle \mathcal{P}[\mathcal{V}_I(\tau_n) \cdots \mathcal{V}_I(\tau_1)] \rangle_{\mathrm{linked}}} \tag{2.37}$$

Note that the derivation of the grand-canonical perturbation series involves no adiabatic hypothesis (no "switching on and off" of the interaction).

## 2. Many-Body Perturbation Theory

### 2.2.2. Time-Independent Representations

The time-ordered version of Eq. (2.37) is

$$\Delta A = \frac{1}{\beta} \sum_{n=0}^{\infty} (-1)^n \int_{\beta > \tau_n > \ldots > \tau_0 > 0} d\tau_n \cdots d\tau_0 \; \langle \mathcal{V}_I(\tau_n) \cdots \mathcal{V}_I(\tau_0) \rangle_{\text{linked}}, \tag{2.38}$$

where we have relabeled dummy indices for convenience, i.e., to match those used in Fig. 2.2. For reasons that will become evident in the end of this section we now cast Eq. (2.38) in a different form. From the cyclic property of the trace we obtain

$$\int_{\beta > \tau_n > \ldots > \tau_0 > 0} d\tau_n \cdots d\tau_0 \, \text{Tr} \left[ \varrho \mathcal{V}_n \, e^{-\Delta \mathcal{E}_n \tau_n} \, \mathcal{V}_{n-1} \, e^{-\Delta \mathcal{E}_{n-1} \tau_{n-1}} \cdots \mathcal{V}_0 \, e^{-\Delta \mathcal{E}_0 \tau_0} \right]$$

$$= \int_{\beta > \tau_n > \ldots > \tau_0 > 0} d\tau_n \cdots d\tau_0 \, \text{Tr} \left[ \varrho \mathcal{V}_{n-1} \, e^{-\Delta \mathcal{E}_{n-1} \tau_{n-1}} \cdots \mathcal{V}_0 \, e^{-\Delta \mathcal{E}_0 \tau_0} \, \mathcal{V}_n \, e^{-\Delta \mathcal{E}_n \tau_n} \right] e^{\Delta \mathcal{E}_n \beta}$$

$$= \int_{\beta > \tau_0 > \tau_n > \ldots > \tau_2 > 0} d\tau_n \cdots d\tau_0 \, \text{Tr} \left[ \varrho \mathcal{V}_n \, e^{-\Delta \mathcal{E}_n \tau_n} \, \mathcal{V}_{n-1} \, e^{-\Delta \mathcal{E}_{n-1} \tau_{n-1}} \cdots \mathcal{V}_0 \, e^{-\Delta \mathcal{E}_0 \tau_0} \right] e^{\Delta \mathcal{E}_0 \beta}, \tag{2.39}$$

where in the first step we have used the relations $\varrho a_i = a_i \varrho \, e^{(\varepsilon_i - \mu)\beta}$ and $\varrho a_i^\dagger = a_i^\dagger \varrho \, e^{-(\varepsilon_i - \mu)\beta}$ (cf. [143] pp.233-239), and in the second step we have relabeled indices. This procedure can be continued further until after in total $n+1$ cyclic exchanges the original expression is reproduced. The expressions obtained by cyclic exchanges are all equivalent, so we can write Eq. (2.38) as

$$\Delta A = \frac{1}{\beta} \sum_{n=0}^{\infty} \frac{(-1)^n}{n+1} \left( \sum_{\nu=0}^{n} \int_{\mathfrak{D}_\nu} d\tau_n \cdots d\tau_0 \; e^{-(\Delta \mathcal{E}_{\nu+1} + \ldots \Delta \mathcal{E}_n)(\beta - \tau_0)} \, \langle \mathcal{V}_I(\tau_n) \cdots \mathcal{V}_I(\tau_0) \rangle_{\text{linked}} \right), \tag{2.40}$$

where we have used that $\Delta \mathcal{E}_0 + \ldots + \Delta \mathcal{E}_n = 0$ to rewrite the additional energy exponential obtained from the cyclic exchanges. The different integration domains $\mathfrak{D}_0, \ldots, \mathfrak{D}_n$ are given by

$$\mathfrak{D}_\nu : \; \beta > \tau_\nu > \tau_1 > \tau_0 > \tau_n > \ldots > \tau_{\nu+1} > 0. \tag{2.41}$$

The integral to be evaluated for the contribution with domain $\mathfrak{D}_\nu$ is

$$\int_{\mathfrak{D}_\nu} d\tau_n \cdots d\tau_0 \; e^{-\Delta \mathcal{E}_n x_n - \ldots - \Delta \mathcal{E}_1 x_1}, \tag{2.42}$$

where we have introduced for every integration domain specific variables $x_1, \ldots, x_n$ defined as $x_\alpha = \tau_\alpha - \tau_0$ for $\alpha = 1, \ldots, \nu$ and $x_\alpha = \tau_\alpha - \tau_0 + \beta$ for $\alpha = \nu + 1, \ldots, n$. With these variables all integration domains have the same integrand. In terms of the variables $x_\nu$, the integration domains are given by

$$\mathfrak{D}_\nu : \; \beta - \tau_0 > x_\nu > \ldots > x_1 > 0 > x_n - \beta > \ldots > x_{\nu+1} - \beta > -\tau_0. \tag{2.43}$$

This is equivalent to

$$\mathfrak{D}_\nu : \; \beta > \tau_0 > 0 \quad \& \quad \beta > x_n > \ldots > x_{\nu+1} > \beta - \tau_0 > x_\nu > \ldots > x_1 > 0. \tag{2.44}$$

Hence, the sum of the $n$ integration domains is given by

$$\sum_{\nu=0}^{n} \mathfrak{D}_\nu : \; \beta > \tau_0 > 0 \quad \& \quad \beta > x_n > \ldots > x_1 > 0. \tag{2.45}$$

## 2. Many-Body Perturbation Theory

The $\tau_0$ integral can now be carried out, giving a factor $\beta$. Renaming the variables $x_\nu \to \tau_\nu$ we obtain the following formula for the perturbative contribution to the grand-canonical potential:

$$\Delta A = \sum_{n=0}^{\infty} \frac{(-1)^n}{n+1} \int_{\beta > \tau_n > \ldots > \tau_1 > 0} d\tau_n \cdots d\tau_1 \, \langle \mathcal{V}_I(\tau_n) \cdots \mathcal{V}_I(\tau_1) \mathcal{V}_I(0) \rangle_{\text{linked}} \qquad (2.46)$$

To evaluate the (imaginary-)time integrals, we proceed analogous to the zero-temperature case. In terms of intermediate states, the ensemble average in Eq. (2.46) reads

$$\sum_p \langle \Phi_p | \varrho \mathcal{V}_n \, e^{-(\mathcal{E}_n - \mathcal{E}_p)\tau_n} | \Phi_n \rangle \cdots \langle \Phi_2 | \mathcal{V}_1 \, e^{-(\mathcal{E}_1 - \mathcal{E}_2)\tau_1} | \Phi_1 \rangle \langle \Phi_1 | \mathcal{V}_0 | \Phi_p \rangle. \qquad (2.47)$$

Introducing again $n$ "time-cut" variables $\zeta_1, \ldots, \zeta_n$ defined as $\zeta_\nu = \tau_\nu - \tau_{\nu-1}$ (with $\tau_0 = 0$) we obtain [cf. Eqs. (2.25) and (2.26]

$$\sum_p \langle \Phi_p | \varrho \mathcal{V}_n \, e^{-\mathscr{E}_n \zeta_n} \cdots \mathcal{V}_1 \, e^{-\mathscr{E}_1 \zeta_1} \mathcal{V}_0 | \Phi_p \rangle_{\text{linked}}, \qquad (2.48)$$

where $\mathscr{E}_\nu = \mathcal{E}_\nu - \mathcal{E}_p$ are the intermediate-state energies relative to the state $|\Phi_p\rangle$. With these substitutions the expression for $\Delta A$ becomes

$$\Delta A = \sum_{n=0}^{\infty} \frac{(-1)^n}{n+1} \int_{\beta > \zeta_1 + \ldots + \zeta_n} d\zeta_n \cdots d\zeta_1 \, \langle \mathcal{V}_n \, e^{-\zeta_n \mathscr{E}_n} \cdots \mathcal{V}_1 \, e^{-\zeta_1 \mathscr{E}_1} \mathcal{V}_0 \rangle_{\text{linked}}. \qquad (2.49)$$

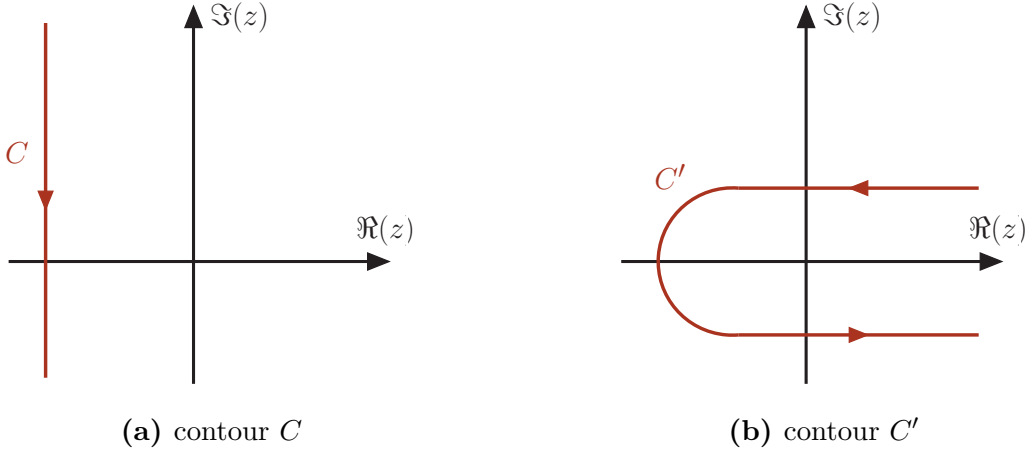

**(a)** contour $C$      **(b)** contour $C'$

**Figure 2.3.:** Contours $C$ and $C'$ in the complex plane.

The restriction on the integration region in Eq. (2.49) can be removed by inserting the integral

$$\frac{1}{2\pi i} \oint_C \frac{dz}{z} \, e^{z(\zeta_1 + \ldots + \zeta_n - \beta)} = \Theta(\beta - \zeta_1 - \ldots - \zeta_n), \qquad (2.50)$$

where the contour $C$ is running parallel to the imaginary axis from $i\infty - a$ to $-i\infty - a$, as shown in plot (a) in Fig 2.3. The expression for $\Delta A$ is then given by

$$\Delta A = \sum_{n=0}^{\infty} \frac{(-1)^n}{n+1} \frac{1}{2\pi i} \int_0^\infty d\zeta_n \cdots d\zeta_1 \oint_C \frac{dz}{z} \, e^{z(\zeta_1 + \ldots + \zeta_n - \beta)} \langle \mathcal{V}_n \, e^{-\zeta_n \mathscr{E}_n} \cdots \mathcal{V}_1 \, e^{-\zeta_1 \mathscr{E}_1} \mathcal{V}_0 \rangle_{\text{linked}}. \qquad (2.51)$$

## 2. Many-Body Perturbation Theory

The contour $C$ can be continuously deformed into the contour $C'$ given by the infinite semicircle displayed in plot (b) in Fig 2.3. If $a$ is chosen such that $a > \mathcal{E}_\nu$, $\forall \nu$, then all $\zeta_\nu$ integrals are convergent, and the integration order can be freely interchanged. Carrying out the $\zeta_\nu$ integrals leads to

$$\Delta A = \sum_{n=0}^{\infty} \frac{(-1)^n}{n+1} \frac{1}{2\pi i} \oint_{C'} dz \frac{e^{-\beta z}}{z} \left\langle \mathcal{V}_n \frac{1}{\mathcal{E}_n - z} \cdots \mathcal{V}_1 \frac{1}{\mathcal{E}_1 - z} \mathcal{V}_0 \right\rangle_{\text{linked}} \tag{2.52}$$

This formula [and Eqs. (2.53) and (2.55)] was first derived by Bloch and de Dominicis [49]. The following expression results from Eq. (2.38) by carrying out the integrals in the same way:

$$\Delta A = -\frac{1}{\beta} \sum_{n=0}^{\infty} \frac{(-1)^n}{2\pi i} \oint_{C'} dz \frac{e^{-\beta z}}{z^2} \left\langle \mathcal{V}_n \frac{1}{\mathcal{E}_n - z} \cdots \mathcal{V}_1 \frac{1}{\mathcal{E}_1 - z} \mathcal{V}_0 \right\rangle_{\text{linked}} \tag{2.53}$$

We refer to Eq. (2.52) as the **"cyclic formula"**, and to Eq. (2.53) as the **"direct formula"**. The cyclic formula is symmetric with respect to intermediate states: following Ref. [49], we substitute $\eta = z + \mathcal{E}_p$ in the cyclic formula and find

$$\Delta A = \sum_{n=0}^{\infty} \frac{(-1)^n}{n+1} \frac{1}{2\pi i} \oint_{C'} d\eta\, e^{-\beta(\eta - \mathcal{E}_0)} \left\langle \mathcal{V}_n \frac{1}{\mathcal{E}_n - \eta} \cdots \mathcal{V}_1 \frac{1}{\mathcal{E}_1 - \eta} \mathcal{V}_0 \frac{1}{\mathcal{E}_0 - \eta} \right\rangle_{\text{linked}}. \tag{2.54}$$

Note that here, $\mathcal{E}_0$ is not the unperturbed ground-state energy but the energy of the state associated with $\mathcal{V}_0$. This is the energy of to the "external" state $|\Phi_p\rangle$, i.e., $\mathcal{E}_0 \equiv \mathcal{E}_p$, and $\eta = \mathcal{E}_0$ corresponds to $z = 0$ in the cyclic formula. Eq. (2.54) shows that if all intermediate-state energies $\mathcal{E}_\nu$ are distinct, the contributions from the $n+1$ poles in the cyclic formula are all equivalent, and equal to $n + 1$ times the contribution from $z = 0$. When $\mathcal{O}$ intermediate-state energies are identical there is a pole of order $\mathcal{O}$, which appears $\mathcal{O}$ times as $z = 0$ if "cyclically related" diagrams are summed; this leads to

$$\Delta A = \sum_{n=0}^{\infty} (-1)^n \frac{1}{\mathcal{O}} \operatorname*{Res}_{z=0} \frac{e^{-\beta z}}{z} \left\langle \mathcal{V}_n \frac{1}{\mathcal{E}_n - z} \cdots \mathcal{V}_1 \frac{1}{\mathcal{E}_1 - z} \mathcal{V}_0 \right\rangle_{\text{linked}} \tag{2.55}$$

where $\mathcal{O}$ is the order of the pole at $z = 0$. We refer to Eq. (2.55) as the **"reduced formula"**.[14]

One can evaluate diagrams using either the direct, the cyclic, or the reduced formula, but *consistency requires that all "cyclically related" diagrams are evaluated in the same way*. For individual diagrams the expressions obtained from the three formulas are equivalent (but not identical) only if the diagrams are invariant under cyclic vertex permutations. For the sum of all cyclic permutations of a given (non-invariant) diagram they give equivalent (but not identical) results. The expressions obtained from the cyclic formula for "cyclically related" diagrams are (by construction) identical, and *for the sum of all "cyclically related" diagrams* the $n + 1$ poles in the cyclic formula all give equivalent (but not identical) expressions. For diagrams with an even number of identical energy denominators, the direct and the reduced formula lead to expressions that diverge in the thermodynamic limit (due to contributions from the vicinity of symmetric energy-denominator poles).[15] These singularities are however ficticious, since they do not appear if the cyclic formula is used: Eq. (2.54) is regular in the limit where a certain number of energies $\mathcal{E}_0, \ldots, \mathcal{E}_n$ become equal, as can be seen by (iteratively) applying l'Hôpital's rule. More specifically, it is

$$\lim_{\mathcal{E}_\nu \to \mathcal{E}_{\nu'}} \oint_{C'} d\eta \frac{e^{-\beta(\eta - \mathcal{E}_0)}}{(\mathcal{E}_n - \eta) \cdots (\mathcal{E}_0 - \eta)} = \oint_{C'} d\eta \frac{e^{-\beta(\eta - \mathcal{E}_0)}}{(\mathcal{E}_n - \eta) \cdots (\mathcal{E}_0 - \eta)}\bigg|_{\mathcal{E}_\nu = \mathcal{E}_{\nu'}}, \tag{2.56}$$

$$\lim_{\mathcal{E}_\nu \to \mathcal{E}_{\nu''}} \oint_{C'} d\eta \frac{e^{-\beta(\eta - \mathcal{E}_0)}}{(\mathcal{E}_n - \eta) \cdots (\mathcal{E}_0 - \eta)}\bigg|_{\mathcal{E}_\nu = \mathcal{E}_{\nu'}} = \oint_{C'} d\eta \frac{e^{-\beta(\eta - \mathcal{E}_0)}}{(\mathcal{E}_n - \eta) \cdots (\mathcal{E}_0 - \eta)}\bigg|_{\mathcal{E}_\nu = \mathcal{E}_{\nu'} = \mathcal{E}_{\nu''}}, \tag{2.57}$$

etc. In the case of the direct and the reduced formula, the left sides of these equations are singular.

---

[14] Note that the reduced formula leads to expressions that are similar to the ones obtained in the ground-state formalism. Compared to the Goldstone formula, for a given (normal) diagram Eq. (2.55 gives additional contributions from "accidentally" vanishing energy denominators, but these contributions are irrelevant in the thermodynamic limit [see Sec. 2.3.4]. Contributions with $\mathcal{O} > 1$ appear only for anomalous diagrams whose defining feature is the presence of "identically" vanishing energy denominators.

## 2.3. Diagrammatic Analysis

In this section we examine in more detail the various diagrammatic contributions (and corresponding analytic expressions) to the perturbation part of the ground-state energy $E_0(\varepsilon_F, \Omega)$ and the grand-canonical potential $A(T, \mu, \Omega)$, respectively. For the most part, we consider two-body (2B) interactions only. The generalization to $\mathcal{N}$-body interactions (with $\mathcal{N} > 2$) does not involve any additional conceptual complications. The three-body (3B) case $\mathcal{N} = 3$ is examined briefly in Sec. 2.3.7).

At this point, we specify the single-particle basis. For a "sufficiently large" homogeneous system where boundary effects are negligible, one can adopt periodic boundary conditions on a cube of side $\Omega^{1/3}$ ("Born–van Karman conditions").[16] The noninteracting single-particle states are then given by plane-wave states, i.e.,

$$|\varphi_i\rangle \sim |\varphi_{\vec{k}}\rangle, \quad \text{with} \quad \langle \vec{x} | \varphi_{\vec{k}} \rangle = \varphi_{\vec{k}}(\vec{x}) = \frac{1}{\sqrt{\Omega}} e^{i\vec{k} \cdot \vec{x}}, \tag{2.58}$$

with momentum quantum numbers $\vec{k} = \frac{2\pi}{\Omega^{1/3}}(n_1, n_2, n_3)$, $n_i \in \mathbb{Z}$. The corresponding single-particle energies are

$$\varepsilon_i \sim \varepsilon(|\vec{k}|), \quad \text{with} \quad \varepsilon(|\vec{k}|) = \langle \varphi_{\vec{k}} | T_{\text{kin}} | \varphi_{\vec{k}} \rangle = \frac{|\vec{k}|^2}{2M}. \tag{2.59}$$

The potential matrix elements are given by [cf. Eq. (1.32)]

$$V_{2B}^{ij,kl} \equiv V_{2B}^{ij,kl} \delta_{k+l,i+j} \sim \langle \varphi_{\vec{k}_1'} \varphi_{\vec{k}_2'} | V_{2B} | \varphi_{\vec{k}_1} \varphi_{\vec{k}_2} \rangle = \frac{1}{\Omega} \delta_{\vec{k}_1' + \vec{k}_2', \vec{k}_1 + \vec{k}_2} V_{2B}(\vec{p}', \vec{p}). \tag{2.60}$$

In the thermodynamic limit where the spectrum becomes continuous, the Kronecker delta $\delta_{\vec{k}_1' + \vec{k}_2' - \vec{k}_1 - \vec{k}_2}$ in Eq. (2.61) is replaced by a delta function according to

$$\delta_{\vec{k}_1' + \vec{k}_2', \vec{k}_1 + \vec{k}_2} \xrightarrow{\{N, \Omega\} \to \infty} \frac{(2\pi)^3}{\Omega} \delta(\vec{k}_1' + \vec{k}_2' - \vec{k}_1 - \vec{k}_2), \tag{2.61}$$

---

[15] This situation occurs for non-skeleton diagrams that impose an even number identical of energy denominators. Note that the energy-denominator poles themselves are excluded since the contour integration (residue sum) in Eqs. (2.52), (2.53) and (2.55) is performed outside the ensemble average.

[16] For a discussion of this choice of the single-particle basis, see (e.g.,) Ref. [82].

and sums over states are replaced by integrals, i.e.,

$$\sum_i \xrightarrow{\{N,\Omega\}\to\infty} \frac{\Omega}{(2\pi)^3} \int d^3k_i. \tag{2.62}$$

For convenience, we will often stick to the condensed "index notation". To point out explicitly that a certain relation implies a continuous spectrum, we will sometimes use the following shorthand notation

$$\sum_i \xrightarrow{\{N,\Omega\}\to\infty} \int_i. \tag{2.63}$$

A diagram with $N$ two-body vertices involves $2N$ contractions, thus there are $4N - 2N = 2N$ index summations, which by Eq. (2.62) give a scale factor $\Omega^{2N}$. The $N$ different potential matrix elements bring a scale factor $\Omega^{-N}$, and between each two vertices there is momentum conservation, thus for a linked cluster there are $N - 1$ delta functions, contributing a scale factor $\Omega^{-(N-1)}$. Altogether, we find the required linear scaling of *linked* clusters for a homogeneous system in the thermodynamic limit, i.e., $\Omega^{2N}\Omega^{-N}\Omega^{-(N-1)} = \Omega$. Similarly, one finds that the "self-energies" $X_{[n]}$ (i.e., functional derivatives of linked clusters) scale as $\Omega^0 = 1$, as required.

### 2.3.1. Classification of Hugenholtz Diagrams

Denoting by $R_h$ and $R_p$ the number of outgoing *hole* and *particle* lines connected to the right of a given vertex, and by $L_h$ and $R_p$ the ones to the left, all vertices are subject to the condition

$$R_p - R_h + L_h - L_p = 0. \tag{2.64}$$

The first-order ("Hartree-Fock") diagram and the two second-order diagrams are shown in Fig. 2.4 below. At third order there in total fourteen different diagrams, which are displayed in Figs. 2.5, 2.7, and 2.8. One can distinguish between the following classes of diagrams:

- **skeletons:** diagrams that cannot be disconnected by cutting only two lines.

- **normal non-skeletons:** diagrams that can be disconnected by cutting either two *hole* or two *particle* lines, but not by cutting one *hole* and one *particle* line.

- **anomalous non-skeletons:** diagrams that can be disconnected by cutting one *hole* and one *particle* line.

Skeletons and normal non-skeletons are referred to as **normal diagrams**, and anomalous non-skeletons are also called **anomalous diagrams**. The severable lines of non-skeleton diagrams are called "articulation lines". Cutting all articulation lines (and suitably reconnecting the cutted lines), a given non-skeleton can be associated with a number of skeleton parts. Each skeleton part can be associated with either an "anomalous" connection or with "normal" connection. Normal non-skeletons involve always at least one higher-order (i.e., beyond first order) skeleton part. A given articulation line can be involved in various subdiagram connections ("cycles", cf. Fig. 2.7), in particular, in both "anomalous" connection and "normal" connections.

By Eq. (2.60), the states corresponding to the *hole* and *particle* articulation line associated with an anomalous diagram must coincide. Since at zero temperature the momentum distributions for *holes* and *particles* have zero overlap, $\Theta_i^- \Theta_i^+ = \Theta_i^-(1 - \Theta_i^-) = 0$, the contributions from

## 2. Many-Body Perturbation Theory

anomalous diagrams are zero in the ground-state formalism. Anomalous diagrams however do contribute in the grand-canonical formalism, giving rise to terms of the form [cf. e.g., Eq. (2.113)]

$$\beta f_i^- f_i^+ = \frac{\partial f_i^-}{\partial \mu} \xrightarrow{T \to 0} \frac{\partial \Theta(\mu - \varepsilon_i)}{\partial \mu} = \delta(\mu - \varepsilon_i); \qquad (2.65)$$

i.e., there is a nonzero contribution to $A(T = 0, \mu, \Omega)$. Such terms, i.e., terms of the form $\sim \beta^n \partial^n f_i^- / \partial \mu^n$, are called **anomalous terms**. In addition, we will find that in finite-temperature MBPT beyond second order, there are terms $\sim \beta^K \partial^n f_i^- / \partial \mu^n$ with $K < n$. Such terms, which vanish in the zero-temperature limit, are referred to as **"pseudo-anomalous" terms**. Anomalous and "pseudo-anomalous" terms are collectively referred as **anomalous contributions**.[17]

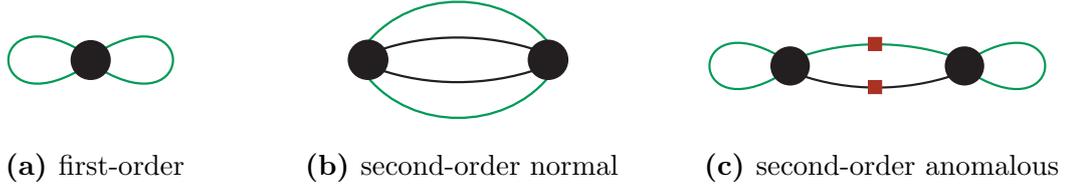

(a) first-order        (b) second-order normal        (c) second-order anomalous

**Figure 2.4.:** Hugenholtz diagrams at first and second order. In the case of the second-order anomalous diagram the red squares mark the two articulation lines.

*Higher-Order Skeletons.* Skeletons beyond second order can be classified according to their intermediate vertices and their behavior under cyclic vertex permutations. The three third-order skeletons are shown in Fig. 2.5; they belong to the following three classes of (cyclically invariant) diagrams:

- **$hh$-ladders:** intermediate vertices have only *hole* lines, with $(L_h, L_p, R_h, R_p) = (2, 0, 2, 0)$.

- **$pp$-ladders:** intermediate vertices have only *particle* lines, with $(L_h, L_p, R_h, R_p) = (0, 2, 0, 2)$.

- **ring diagrams:** intermediate vertices have each two *hole* and two *particle* lines, with $(L_h, L_p, R_h, R_p) = (1, 1, 1, 1)$.

At higher orders, however, many additional skeletons that do not fall into this classification scheme appear. At fourth order there are already 24 such additional skeletons, which are shown in Fig. 2.6.[18] The 24 "extra" skeletons at fourth order can be divided into eight different types (A1, A2, A3, B1, B2, C1, C2, C3) according to their transformation behavior under cyclic vertex permutations; note that they contain already all additional possibilities for intermediate vertices without instantaneous contractions $(L_h, L_p, R_h, R_p) = (0, 0, 2, 2), (2, 2, 0, 0), (1, 0, 2, 1), (0, 1, 1, 2), (2, 1, 1, 0), (1, 2, 0, 1)$.

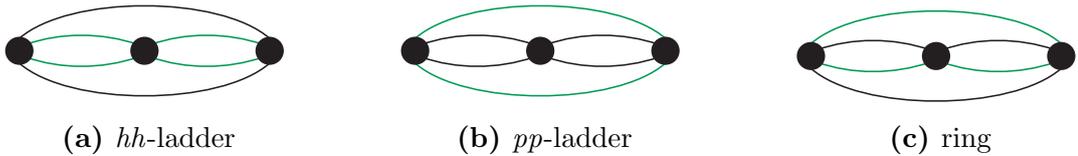

(a) $hh$-ladder        (b) $pp$-ladder        (c) ring

**Figure 2.5.:** Ladder and ring diagrams at third order.

---

[17] Note that the $T \to 0$ limit of anomalous terms is discontinuous for a discrete spectrum, cf. also Ref. [246]. This feature can be associated with the fact that the chemical potential at zero temperature is quantized for a discrete spectrum.

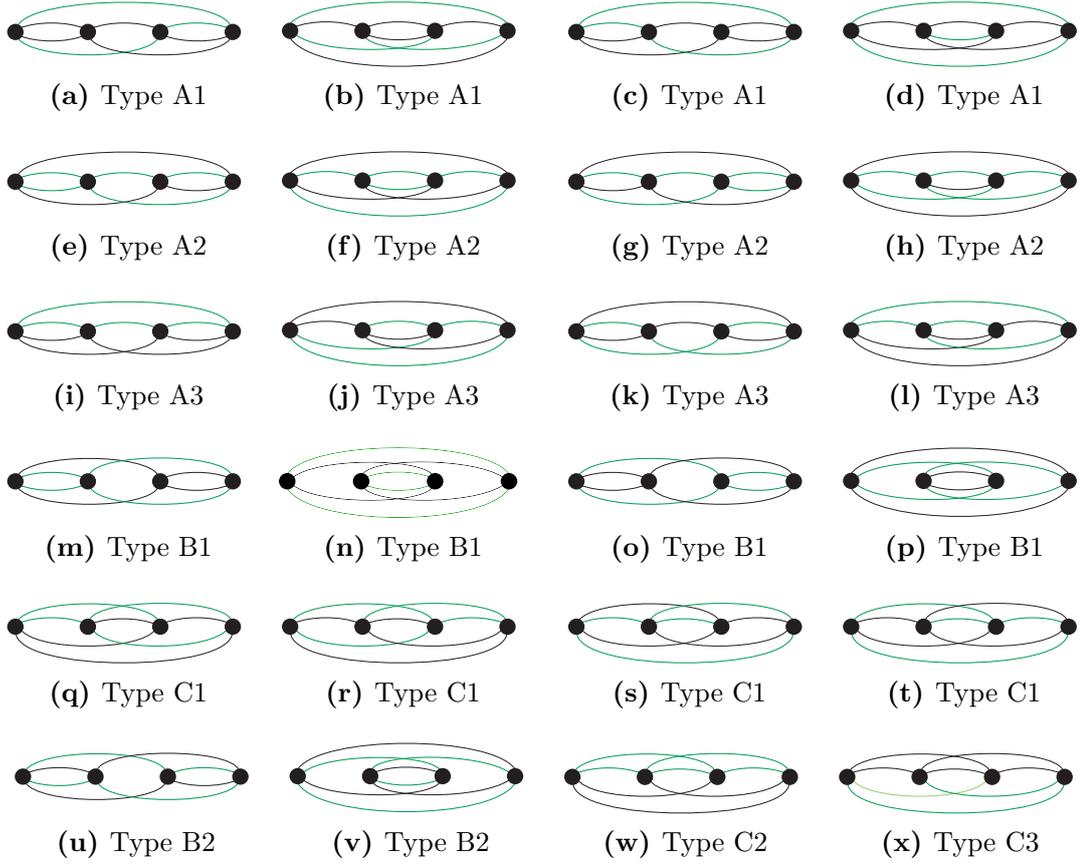

**Figure 2.6.:** "Extra" skeletons at fourth order. The diagrams labeled "A1", "A2", "A3", "B1" and "C1" correspond to quadruples of diagrams connected via cyclic vertex permutations. The two "B2" diagrams are each invariant under two consecutive permutations, and transform into each other under one cyclic vertex permutation; the diagrams "C2" and "C3" are invariant under cyclic vertex permutations.

*Self-Consistent Renormalization.* The essential difference between skeletons and non-skeletons is that skeletons can (with the exception of the first-order diagram) involve only vertices corresponding to $\mathcal{N}$-body operators with $\mathcal{N} \geq 2$, but non-skeletons also vertices with $\mathcal{N} = 1$. Consider the case where the reference and perturbation Hamiltonians, $\mathcal{T}$ and $\mathcal{V}$, are redefined in terms of an additional one-body Hamiltonian $\mathcal{X} = \sum_\alpha X_\alpha a^\dagger_\alpha a_\alpha$ according to[19]

$$\mathcal{T} \to \mathcal{T} + \mathcal{X}, \quad \text{and} \quad \mathcal{V} \to \mathcal{V} - \mathcal{X}. \tag{2.66}$$

The new perturbation series about $\mathcal{T} + \mathcal{X}$ involves exactly the same skeletons as the one about $\mathcal{T}$, but there are additional non-skeletons with $-X_\alpha$ vertices. Classes of contributions that have the same structure as (a subset of) the additional ones arising due to the change from the $\{\mathcal{T}, \mathcal{V}\}$ to the $\{\mathcal{T} + \mathcal{X}, \mathcal{V} - \mathcal{X}\}$ setup are referred to as **"insertions"**. In the ground-state formalism there is only insertion type, referred to as (normal) "one-loop" insertions: (normal) non-skeletons with first-order skeleton parts ("one-loop" vertices). In particular, *normal non-skeletons without*

---

[18] See also Fig. 3 in Ref. [373] or Fig. 6.1. in Ref. [378], which include the fourth-order normal and anomalous non-skeletons appearing in Hartree-Fock perturbation theory (HFPT), i.e., the perturbation expansion about the self-consistent Hartree-Fock Hamiltonian $\mathcal{T}_{X_{[1]}} = \mathcal{T} + \mathcal{X}_1^{[1]}$. The total number of HFPT diagrams at order $n = 2, 3, 4, 5, 6$ is $N_{\text{HF}}(n) = 1, 3, 39, 840, 27300$, cf. [Online Encyclopedia of Integer Sequences (OEIS)].

*first-order parts do not contain insertion structures.* This crucial feature is discussed in Sec. 2.3.5.

In the grand-canonical formalism, however, there are additional insertion structures, but to identify these structures a reorganization of the contributions associated with normal and anomalous non-skeletons is required. Following Ref. [18], we refer to this reorganization as **"disentanglement"**. The "disentanglement" to all orders is achieved by reevaluating the ensemble averages in the respective formula in terms of cumulants (truncated correlation functions). There are two essentially different ways the "disentanglement" can be performed, corresponding to the use of either the direct of the reduced formula, and in each case the additional insertion structures are associated with anomalous contributions (in the "reduced" case, with anomalous terms only).

The point is now that insertions can be removed through a suitable choice of the one-body Hamiltonian $\mathcal{X} = \sum_\alpha X_\alpha a^\dagger_\alpha a_\alpha$, i.e., by identifying $X_\alpha$ with functional derivatives of non-insertion contributions.

*Ground-State Formalism.* Consider in the ground-state formalism the case where $X_i$ is given by the expression for the self-consistent Hartree-Fock self-energy

$$X_{1;\alpha}(e_F) = \frac{\delta E_{0;1}[\Theta^-_\alpha]}{\delta \Theta^-_\alpha} = \sum_\beta \bar{V}_{2B}^{\alpha\beta,\alpha\beta} \, \Theta^-_\beta, \tag{2.67}$$

where $e_F$ is the Fermi energy corresponding to the ground-state associated with $\mathcal{T} + \mathcal{X}_1$. It is easy to see that the perturbation series about $\mathcal{T} + \mathcal{X}_1$ has the same structure as the one about $\mathcal{T}$, except that there are no (normal) non-skeletons with "one-loop" vertices: these are cancelled by the additional $-X_{1;\alpha}$ insertions.

In the "isotropic case" the unperturbed single-particle states in the $\{\mathcal{T} + \mathcal{X}_1, \mathcal{V} - \mathcal{X}_1\}$ setup are again plane waves, i.e.,

$$\langle \vec{x} | \varphi_\alpha \rangle \xrightarrow{\{N,\Omega\}\to\infty} \varphi_{\vec{k}}(\vec{x}) = \frac{1}{\sqrt{\Omega}} e^{i\vec{k}\cdot\vec{x}}. \tag{2.68}$$

This can be seen as follows. Given Eq. (2.68) holds, the Fermi energy $e_F$ is fixed by

$$N = \sum_\alpha \Theta(e_F - \varepsilon_\alpha - X_{1;\alpha}) \xrightarrow{\{N,\Omega\}\to\infty} \int \frac{\Omega}{(2\pi)^3} d^3k \, \Theta(e_F - \varepsilon(|\vec{k}|) - X_1(e_F;|\vec{k}|)), \tag{2.69}$$

where we have used that (by rotational invariance) $X_1$ depends only on the magnitude $|\vec{k}|$ of the plane-wave momentum $\vec{k}$. Eq. (2.69) then defined a Fermi momentum $k_F$ up to which states are "summed". The Fermi energy $\varepsilon_F$ in the $\{\mathcal{T}, \mathcal{V}\}$ setup is fixed by

$$N = \sum_i \Theta(\varepsilon_F - \varepsilon_i) \xrightarrow{\{N,\Omega\}\to\infty} \int \frac{\Omega}{(2\pi)^3} d^3k \, \Theta(\varepsilon_F - \varepsilon(|\vec{k}|)), \tag{2.70}$$

which (since $N$ is fixed) defines the same Fermi momentum $k_F$ (for a different Fermi energy). In each case, the distribution functions are then given by $\Theta(k_F - k)$. This implies (for a given value of $N$) that $X_{1;\alpha}(e_F) = X_{1;i}(\varepsilon_F)$, i.e.,

$$X_{1;\alpha}(e_F) \xrightarrow{\{N,\Omega\}\to\infty} X_{1;i}(\varepsilon_F) = \frac{\delta E_{0;1}[\Theta^-_i]}{\delta \Theta^-_i} = \sum_a \bar{V}_{2B}^{ia,ia} \, \Theta^-_a. \tag{2.71}$$

---

[19] The bold-faced greek indices "$\boldsymbol{\alpha}$" refer to the single-particle states associated with the new reference Hamiltonian $\mathcal{T} + \mathcal{X}$.

It is then easy to see that Eq. (2.68) gives the solutions to the unperturbed Schrödinger equation in the $\{\mathcal{T} + \mathcal{X}_1, \mathcal{V} - \mathcal{X}_1\}$ setup, with energy eigenvalues

$$e_i = \varepsilon_i + X_{1;i}(\varepsilon_F). \tag{2.72}$$

In the "isotropic case", for a given diagram without first-order parts, the change from $\{\mathcal{T}, \mathcal{V}\}$ to $\{\mathcal{T} + \mathcal{X}_1, \mathcal{V} - \mathcal{X}_1\}$ is then equivalent to the resummation (in the $\{\mathcal{T}, \mathcal{V}\}$ setup) of all its higher-order "cousins" with additional "one-loop" vertices, which renormalizes the single-particle energies in the energy denominators according to

$$\mathcal{E}(\{\varepsilon_i\}) \to \mathcal{E}(\{\varepsilon_i + X_{1;i}(\varepsilon_F)\}) \equiv \mathcal{E}(\{\varepsilon_{X_{1;i}}(\varepsilon_F)\}). \tag{2.73}$$

This resummation is possible only in the case of "one-loop" insertions: normal non-skeletons without first-order parts are not of the simple insertion type, and contain no resummable structures. In other terms, the change from $\{\mathcal{T} + \mathcal{X}_1, \mathcal{V} - \mathcal{X}_1\}$ to $\{\mathcal{T} + \mathcal{X}_{[n]}, \mathcal{V} - \mathcal{X}_{[n]}\}$, where

$$\mathcal{X}_{[n]} = \sum_{\nu=1}^{n} \mathcal{X}_\nu = \sum_{\nu=1}^{n} \sum_i X_{\nu;i} a_i^\dagger a_i, \tag{2.74}$$

leads to new non-skeletons with $-X_{\nu \geq 2}$ insertions but does not remove any previous diagrams. Note that the energy eigenvalue equation for the $\{\mathcal{T} + \mathcal{X}_{[n]}, \mathcal{V} - \mathcal{X}_{[n]}\}$ setup becomes a self-consistent one for with $n \geq 2$. This is due to the presence of the energy denominators in the expression for $X_{\nu \geq 2}$, i.e, in the $n = 2$ case we have

$$e_i = \varepsilon_i + X_{1;i}(\varepsilon_F) + X_{2;i}(\varepsilon_F; \{e_a\}), \tag{2.75}$$

where

$$X_{[2];a}(\varepsilon_F; \{e_a\}) = \frac{\delta E_{0;2,\text{normal}}[\Theta_a^-; \{e_a\}]}{\delta \Theta_a^-} = \int_{jkl} \frac{H_{ajkl}}{e_k + e_l - e_a - e_j} + \int_{ijk} \frac{P_{ijka}}{e_k + e_a - e_i - e_j}, \tag{2.76}$$

with

$$H_{ajkl} = -\frac{1}{2} \bar{V}_{2B}^{aj,kl} \bar{V}_{2B}^{kl,aj} \Theta_j^- \Theta_k^+ \Theta_l^+, \qquad P_{ijka} = \frac{1}{2} \bar{V}_{2B}^{ij,ka} \bar{V}_{2B}^{ka,ij} \Theta_i^- \Theta_j^- \Theta_k^+. \tag{2.77}$$

Denoting the solution of Eq. (2.75) by $\varepsilon_{X_{[2];i}}(\varepsilon_F)$, the energy denominators in the $\{\mathcal{T} + \mathcal{X}_{[2]}, \mathcal{V} - \mathcal{X}_{[2]}\}$ setup are given by

$$\mathcal{E}(\{\varepsilon_{X_{[2];i}}(\varepsilon_F)\}). \tag{2.78}$$

In the zero-temperature case, the $\{\mathcal{T} + \mathcal{X}_{[n]}, \mathcal{V} - \mathcal{X}_{[n]}\}$ setups with $n \geq 2$ are however somewhat contrived: no previous diagrams are cancelled in that case, and the resummation of the additional $-X_2, \ldots, -X_n$ vertices in terms of a geometric series leads back to Hartree-Fock energy denominators, i.e., to the $\{\mathcal{T} + \mathcal{X}_1, \mathcal{V} - \mathcal{X}_1\}$ setup.

*Grand-Canonical Formalism.* The generalization of Eq. (2.67) to the grand-canonical case is given by

$$X_{1;\alpha}(T, \mu) = \frac{\delta A_1[f_\alpha^-]}{\delta f_\alpha^-} = \sum_\beta \bar{V}_{2B}^{\alpha\beta,\alpha\beta} f_\alpha^-, \tag{2.79}$$

and the change from $\{\mathcal{T},\mathcal{V}\}$ to $\{\mathcal{T} + \mathcal{X}_1, \mathcal{V} - \mathcal{X}_1\}$ removes all (normal and anomalous) non-skeletons with "one-loop" vertices.

In the "isotropic case", the solutions to the unperturbed Schrödinger equation in the $\{\mathcal{T} + \mathcal{X}_1, \mathcal{V} - \mathcal{X}_1\}$ setup are still plane waves, but the single-particle energies are now given by the self-consistent equation

$$e_i = \varepsilon_i + X_{1;i}[f^-_{e_a}(T,\mu)], \tag{2.80}$$

where $f^-_{e_i}(T,\mu)$ denotes the Fermi-Dirac distribution with $e_i$ as the single-particle energies. We will denote the solution to Eq. (2.80) as $\varepsilon_{X_1;i}(T,\mu)$, and the corresponding Fermi-Dirac distribution as $f^-_{X_1,i}(T,\mu)$. The "isotropic case" of Eq. (2.79) is then given by

$$X_{1;\alpha}(T,\mu) \xrightarrow{\{N,\Omega\}\to\infty} X_{1;i}(T,\mu) = \frac{\delta A_1[f^-_i;\{\varepsilon_{X_1;i}\}]}{\delta f^-_i} = \sum_i \bar{V}^{ia,ia}_{2\mathrm{B}} f^-_{X_1,i}. \tag{2.81}$$

Thus, contrary to the zero-temperature case where the distribution functions are invariant under the self-consistent renormalization of the single-particle basis (in the "isotropic case"), in the grand-canonical formalism the change from $\{\mathcal{T},\mathcal{V}\}$ to $\{\mathcal{T} + \mathcal{X}_1, \mathcal{V} - \mathcal{X}_1\}$ renormalizes not only the energy denominators but also the Fermi-Dirac distributions. Furthermore, in the grand-canonical formalism the change from $\{\mathcal{T} + \mathcal{X}_1, \mathcal{V} - \mathcal{X}_1\}$ to $\{\mathcal{T} + \mathcal{X}^\aleph_{[n]}, \mathcal{V} - \mathcal{X}^\aleph_{[n]}\}$ does remove certain (anomalous) contributions, but this is obscured in the standard "contraction" formalism. After the reorganization of the grand-canonical series in terms of cumulants, however, identifying the effect of the additional anomalous $-X_{\nu\geq 2}$ insertions is straightforward: in the cumulant formalism, the anomalous contributions are represented by unlinked normal diagrams connected via "insertion-lines" ("higher-cumulant connections"), i.e., symbolically for the case of three diagrams $\varGamma_{n_1}$, $\varGamma_{n_2}$, and $\varGamma_{n_3}$:

$$[\varGamma_{n_1}\text{—}\varGamma_{n_2}\text{—}\varGamma_{n_3}]^\aleph. \tag{2.82}$$

In the standard "contraction" formalism, consistency requires that all contributions related via cyclic vertex permutations are evaluated in the same way. By construction, the cumulant formalism "messes up" the relations between "cyclically related" diagrams (cf. Sec. 2.4.3); one then needs to evaluate all higher-cumulant contributions and normal (non-skeleton) diagrams "$\mathscr{D}$" at a given order in the same way, i.e., symbolically

$$\boxed{[\varGamma_{n_1}\text{—}\varGamma_{n_2}\text{—}\varGamma_{n_3}]^\aleph + [\mathscr{D}_{n_1+n_2+n_3}]^\aleph} \tag{2.83}$$

For $\aleph =$ "direct" it is straightforward to show (using time-dependent methods, cf. Sec. 2.5.1)) that the following factorization property holds (where all relative time-orderings of the subdiagrams are summed):

$$\boxed{[\varGamma_{n_1}\text{—}\varGamma_{n_2}\text{—}\varGamma_{n_3}]^{\mathrm{direct}} = -\beta[\varGamma^{\mathrm{direct}}_{n_1}\text{—}\varGamma^{\mathrm{direct}}_{n_2}\text{—}\varGamma^{\mathrm{direct}}_{n_3}]} \tag{2.84}$$

This contribution is then cancelled by the corresponding contribution with a $-X^{\mathrm{direct}}_{n_1}$ vertex, i.e.,

$$(-X^{\mathrm{direct}}_{n_1})\text{—}\varGamma^{\mathrm{direct}}_{n_2}\text{—}\varGamma^{\mathrm{direct}}_{n_3}. \tag{2.85}$$

If $-X_{n_3}^{\text{direct}}$ vertices are present, then there are also the contributions

$$\Gamma_{n_1}^{\text{direct}}\!\!-\!\!\Gamma_{n_2}^{\text{direct}}\!\!-\!\!(-X_{n_3}^{\text{direct}}), \tag{2.86}$$

$$(-X_{n_1}^{\text{direct}})\!\!-\!\!\Gamma_{n_2}^{\text{direct}}\!\!-\!\!(-X_{n_1}^{\text{direct}}). \tag{2.87}$$

which also cancel each other. If the insertion-lines are placed on different lines of $\Gamma_{n_2}^{\text{direct}}$, then these are the only cases that appear, since $\Gamma_{n_2}^{\text{direct}}$ cannot be replaced by a one-body vertex; in particular, in that case the contribution given by Eq. (2.82) cannot be cancelled by including $X_{n_2}^{\text{direct}}$ vertices. Only if the two insertion-lines are connected to the same line, there are also contributions where $\Gamma_{n_2}^{\text{direct}}$ is replaced by $-X_{n_2}^{\text{direct}}$.

Notably, in the $\{\mathcal{T} + \mathcal{X}_{[n]}^{\text{direct}}, \mathcal{V} - \mathcal{X}_{[n]}^{\text{direct}}\}$ setup with $n \geq 6$ the zero-temperature limit of the (formal) expressions for the "self-energies" and the various perturbative contributions *does not exist*. This is due to the feature that normal non-skeletons without first-order parts are not insertions, and the fact that the "direct" expressions for normal non-skeletons with $K$ skeleton parts involve terms that scale as $\beta^{K-2}$ (cf. the last paragraph of Sec. 2.3.4).

For $\aleph$ = "reduced" it is straightforward to show that the following factorization property holds (cf. Sec. 2.4.3)

$$\boxed{[\Gamma_{n_1}\!\!-\!\!\Gamma_{n_2}\!\!-\!\!\Gamma_{n_3}]^{\text{reduced}} = -\beta[\Gamma_{n_1}^{\text{reduced}}\!\!-\!\!\Gamma_{n_2}^{\text{reduced}}\!\!-\!\!\Gamma_{n_3}^{\text{reduced}} + \mathcal{R}]} \tag{2.88}$$

where $\mathcal{R} \xrightarrow{T \to 0} 0$ (i.e. $\mathcal{R}$ is "pseudo-anomalous"). For the factorized part of the higher-cumulant terms (which is anomalous) the same cancellation combinatorics as in the "direct" case applies, but with "reduced" self-energies. In contrast to the $\aleph$ = "direct" case, if $\aleph$ = "reduced" the $\{\mathcal{T}+\mathcal{X}_{[n]}^{\text{reduced}}, \mathcal{V}-\mathcal{X}_{[n]}^{\text{reduced}}\}$ setups have a well-behaved zero-temperature limit (for all values of $n$), and the $T \to 0$ limit of the "fully-renormalized" perturbation series reproduces the Goldstone formula in the "isotropic case", cf. Sec. 2.5.2.

In Refs. [96, 20] it is claimed that the "pseudo-anomalous" term in Eq. (2.88) is in fact zero, i.e.,

$$\boxed{\mathcal{R} = 0} \tag{2.89}$$

We will show that $\mathcal{R} = 0$ only for selected contributions (cf. Sec. 2.4.3), but leave a full proof to future research.[20] If $\mathcal{R} = 0$ to all orders, then the expression for the entropy that results from the "fully-renormalized" perturbation series (for $\aleph$ = "reduced") matches the corresponding one for a free Fermi gas. This corresponds to the notion of "statistical quasiparticles".

For both the $\aleph$ = "direct" and the $\aleph$ = "reduced" case (as well as the zero-temperature case), the existence of the thermodynamic limit in the $\{\mathcal{T} + \mathcal{X}_{[n]}^{\aleph}, \mathcal{V} - \mathcal{X}_{[n]}^{\aleph}\}$ setups with $n \geq 4$ requires a regularization procedure ("R"), since in that case certain anomalous contributions are absent, which would however be required to cancel contributions (from non-skeletons) from the vicinity of symmetric energy-denominator poles;[21] this is discussed in Sec. 2.3.6.

---

[20] Note added in corrected version: the property $\mathcal{R} = 0$ is (of course) a straightforward consequence of the linked-cluster theorem (in other terms, the factorization property Eq. (2.84) holds for any $\aleph$).

[21] The expressions for $X_{\nu \geq 4}^{(\aleph)}$ also need to be regularized, but this is included in the regularization procedure. To be precuse, we note here also that the regularization is also required for terms *anti*symmetric poles, i.e., to remove the ambiguity associated with the integration order.

***"Self-Energies".*** As evident from the above discussion, in the "isotropic case" the energy eigenvalue equation for the $\{\mathcal{T} + \mathcal{X}^{(\aleph)}_{[n]}, \mathcal{V} - \mathcal{X}^{(\aleph)}_{[n]}\}$ setup is given by

$$e_i = \varepsilon_i + X_{1;i}(\varepsilon_F) + \sum_{\nu=2}^{n} X_{\nu;i}(\varepsilon_F; \{e_a\}), \qquad \text{and} \qquad e_i = \varepsilon_i + X_{1;i}[f^-_{e_i}(T,\mu)] + \sum_{\nu=2}^{n} X^\aleph_{\nu;i}[f^-_{e_i}(T,\mu)], \tag{2.90}$$

in the zero-temperature and grand-canonical case, respectively.[22] We denote the solutions to these equations as $e_i = \varepsilon_{X^{(\aleph)}_{[n]};i}$ for each case. The self-consistent "self-energy" of order $n$ is then given by

$$X^{(\aleph)}_{[n]} = \sum_{\nu=1}^{n} X^{(\aleph)}_\nu, \tag{2.91}$$

where the expressions for $X^{(\aleph)}_{\nu;i}$ are

$$X_{\nu;i}(\varepsilon_F) = \frac{\delta E^R_{0;\nu,\text{non-insertion}}[\Theta^-_i; \{\varepsilon_{X_{[n]};i}\}]}{\delta \Theta^-_i}, \qquad \text{and} \qquad X^\aleph_{\nu;i}(T,\mu) = \frac{\delta A^{R,\aleph}_{\nu,\text{non-insertion}}[f^-_i; \{\varepsilon_{X^\aleph_{[n]};i}\}]}{\delta f^-_i}. \tag{2.92}$$

Here, "$R$" refers to the regularization procedure specified in Sec. 2.3.6, and "non-insertion" refers to contributions that are not of the insertion type, i.e., contributions (evaluated in terms of cumulants) from skeletons and non-skeletons without first-order parts, i.e., **normal non-skeletons with** $-X_2, \ldots, -X_{n-1}$ **vertices are not counted as insertions** and are therefore included in Eq. (2.92).[23]

For the following discussion it will be useful to define also the *perturbative* equivalent of these self-energies, i.e., we define

$$S_{\nu;i}(\varepsilon_F) = \frac{\delta E^R_{0;\nu,\text{non-insertion}}[\Theta^-_i; \{\varepsilon_i\}]}{\delta \Theta^-_i}, \qquad \text{and} \qquad S^\aleph_{\nu;i}(T,\mu) = \frac{\delta A^{R,\aleph}_{\nu,\text{non-insertion}}[f^-_i; \{\varepsilon_i\}]}{\delta f^-_i}. \tag{2.93}$$

The main reason for the introduction of the perturbative self-energies is that it allows to write down more concise expressions for non-skeleton contributions. Note that $X_{1,i}(\varepsilon_F) = S_{1,i}(\varepsilon_F)$, which is a special feature of the Hartree-Fock case at zero-temperature (in the "isotropic case").

### 2.3.2. Evaluation of Hugenholtz Diagrams

The contractions specified by a given Hugenholtz diagram can be carried out in different ways, e.g., in the case of the first-order (or Hartree-Fock) contribution one has the (direct and exchange) contributions

$$\sum_{ijkl} V^{ij,kl}_{2B} \left( a^\dagger_i a^\dagger_j a_l a_k + a^\dagger_i a^\dagger_j a_l a_k \right) = \sum_{ijkl} \left[ V^{ij,kl}_{2B} \delta_{ik}\delta_{jl} - V^{ij,kl}_{2B} \delta_{il}\delta_{jk} \right] \times \begin{cases} \Theta^-_i \Theta^-_j \\ f^-_i f^-_j \end{cases}. \tag{2.94}$$

---

[22] The distinction between different $\aleph$'s in the grand-canonical case is relevant only for non-insertion contributions from (normal) non-skeletons, which appear first at order $\nu = 4$.

[23] Note that this implies that $\{\mathcal{T} + \mathcal{X}_{[\infty]}, \mathcal{V} - \mathcal{X}_{[\infty]}\} \equiv \{\mathcal{T} + \mathcal{X}_1, \mathcal{V} - \mathcal{X}_1\}$ in the zero-temperature case.

The potential matrix elements satisfy the exchange symmetry $V_{2B}^{ij,kl} = \mathscr{P}_L \mathscr{P}_R V_{2B}^{ij,kl}$, with exchange operators $\mathscr{P}_{L/R} = -P_{12}$,[24] and Hermiticity implies that $V_{2B}^{ij,kl} = [V_{2B}^{kl,ij}]^*$. The two contributions given by Eq. (2.94) can be written in compact form by introducing the **antisymmetrized two-body potential** $\bar{V}_{2B} = \mathscr{A}_L^{2B} V_{2B}$, with antisymmetrization operator $\mathscr{A}_{(L)}^{2B} = (1 - P_{12})_{(L)}$. The expressions for the first-order contributions to the ground-state energy and the grand-canonical potential are then given by[25]

$$E_{0;1} = \frac{1}{2} \sum_{ij} \bar{V}_{2B}^{ij,ij} \Theta_i^- \Theta_j^-, \qquad A_1 = \frac{1}{2} \sum_{ij} \bar{V}_{2B}^{ij,ij} f_i^- f_j^-. \tag{2.95}$$

In general, for a given diagram the overall sign factor is given by $(-1)^{\ell+h}$, where $\ell$ is the number of loops formed by different *hole* and *particle* lines in the diagram (loops formed by lines of one type only are not counted) plus the number of instantaneous contractions, and $h$ is the number of *hole* lines in the diagram (cf. e.g., Ref. [178] pp.227-231). The number of all possible contractions can be identified by applying to the matrix element associated with each vertex the permutation operators $\mathscr{P}_{L/R}$, modulo permutations that lead to expressions which are equivalent under the mutual relabeling of indices. For instance, the contractions for the second-order anomalous diagram are given by[26]

$$\sum_{\text{contractions}} = (-1)^{3+3} \left[ \mathscr{P}_2 V_{2B}^{ij,jl} \right] \left[ \mathscr{P}_1 V_{2B}^{la,ja} \right] \tag{2.96}$$

with $\mathscr{P}_{1/2} = 1 + \mathscr{P}_L + \mathscr{P}_R + \mathscr{P}_L \mathscr{P}_R = \mathscr{A}_L^{2B}(1 + \mathscr{P}_R)$, corresponding to in total 16 different contractions. By contrast, the second-order normal diagram involves only 4 different contractions:

$$\sum_{\text{contractions}} = (-1)^{2+2} \left[ \mathscr{P}_2 V_{2B}^{ij,kl} \right] \left[ \mathscr{P}_1 V_{2B}^{kl,ij} \right] \tag{2.97}$$

where $\mathscr{P}_{1/2} = 1 + \mathscr{P}_L = \mathscr{A}_L^{2B}$. For a given diagram the antisymmetrizer $\mathscr{A}_L^{2B}$ always leads to a nonequivalent contribution; the additional factor left after extracting the antisymmetrizers $\mathscr{A}_L^{2B}$ is the "*multiplicity*" $\mathcal{M}$ of the diagram. In the cases above, $\mathcal{M}_{2,\text{normal}} = 1$ and $\mathcal{M}_{2,\text{anomalous}} = 4$. For a given non-skeleton only the multiplicity of the corresponding skeletons has to be determined (which can be done by counting pairs of equivalent lines, cf. [378] pp.356-362); the multiplicity of the whole diagram is equal to the product of the multiplicities of the various skeleton-parts multiplied with the number of possible ways the articulation lines can be placed:

$$\mathcal{M}_{\text{non-skeleton}} = \mathcal{M}_{\text{articulation}} \times \prod_{\text{parts}} \mathcal{M}_{\text{parts}}. \tag{2.98}$$

For instance, in the case of the second-order anomalous diagram the two subskeletons (first-order diagrams) have $\mathcal{M} = 1$, and there are four different but equivalent ways of placing the articulation lines, thus $\mathcal{M}_{2,\text{anomalous}} = 4 \times 1 = 4$.

---

[24] The subscripts $L$ and $R$ indicate that the Pauli exchange operator $P_{12}$ acts on the index tuple corresponding to the creation (L) and destruction operators (R), respectively: $\mathscr{P}_L V_{2B}^{ij,kl} = -V_{2B}^{ji,kl}$ and $\mathscr{P}_R V_{2B}^{ij,kl} = -V_{2B}^{ij,lk}$. The distinction between $L$ and $R$ permutations is introduced for convenience, i.e., to identify in a simple way the number of equivalent contractions for a given diagram.

[25] The different ways the contractions can be realized for a given Hugenholtz diagrams can be visualized by so-called "Goldstone diagrams" where each Hugenholtz vertex $a_i^\dagger a_j^\dagger a_k a_l$ is split up in two Goldstone vertices $a_i^\dagger a_k$ and $a_j^\dagger a_l$ connected by an interaction line. For the various Goldstone diagrams for the first- and second-order contributions, cf. (e.g.,) Fig. 20.15 in Ref. [178].

[26] The subscripts in $\mathscr{P}_{1/2}$ refer to the labeling of Hugenholtz vertices (cf. Fig. 2.2). For the diagrams considered here it is $\mathscr{P}_1 = \mathscr{P}_2$, but this is not the case in general; e.g., in the case of diagram (a) of Fig. 2.8 two of the permutation operators are given by $\mathscr{P}_{3,1} = \mathscr{A}_L^{2B}$ and one by $\mathscr{P}_2 = \mathscr{A}_L^{2B}(1 + \mathscr{P}_R)$. In most cases, the particular form of the $\mathscr{P}$ operators is (of course) not unique.

### 2.3.3. Ground-State Energy up to Third Order

Here, we give the expressions for the contributions up to third order in the zero-temperature formalism (in the $\{\mathcal{T}, \mathcal{V}\}$ setup, referred to as the "bare" case from now on). In addition, we examine the resummation of "one-loop" insertions in terms of a geometric series.

***Skeletons.*** The expression for the second-order normal diagram ($\mathcal{M} = 1$, $\ell = 2$, $h = 2$) is

$$E_{0;2,\text{normal}} = -\frac{1}{4} \sum_{ijkl} \bar{V}_{2\text{B}}^{ij,kl} \bar{V}_{2\text{B}}^{kl,ij} \frac{\Theta_i^- \Theta_j^- \Theta_k^+ \Theta_l^+}{\varepsilon_k + \varepsilon_l - \varepsilon_i - \varepsilon_j}. \tag{2.99}$$

The expressions for the *hh*-ladder ($\mathcal{M} = 1$, $\ell = 2$, $h = 4$), the *pp*-ladder ($\mathcal{M} = 1$, $\ell = 2$, $h = 2$), and the ring diagram ($\mathcal{M} = 8$, $\ell = 3$, $h = 3$) are given by

$$E_{0;3,hh\text{-ladder}} = \frac{1}{8} \sum_{ijabkl} \bar{V}_{2\text{B}}^{ij,kl} \bar{V}_{2\text{B}}^{ab,ij} \bar{V}_{2\text{B}}^{kl,ab} \frac{\Theta_i^- \Theta_j^- \Theta_a^- \Theta_b^- \Theta_k^+ \Theta_l^+}{(\varepsilon_k + \varepsilon_l - \varepsilon_i - \varepsilon_j)(\varepsilon_k + \varepsilon_l - \varepsilon_a - \varepsilon_b)}, \tag{2.100}$$

$$E_{0;3,pp\text{-ladder}} = \frac{1}{8} \sum_{ijklpq} \bar{V}_{2\text{B}}^{ij,kl} \bar{V}_{2\text{B}}^{kl,pq} \bar{V}_{2\text{B}}^{pq,ij} \frac{\Theta_i^- \Theta_j^- \Theta_k^+ \Theta_l^+ \Theta_p^+ \Theta_q^+}{(\varepsilon_k + \varepsilon_l - \varepsilon_i - \varepsilon_j)(\varepsilon_p + \varepsilon_q - \varepsilon_i - \varepsilon_j)}, \tag{2.101}$$

$$E_{0;3,\text{ring}} = \sum_{ijaklp} \bar{V}_{2\text{B}}^{ij,kl} \bar{V}_{2\text{B}}^{ka,ip} \bar{V}_{2\text{B}}^{pl,aj} \frac{\Theta_i^- \Theta_j^- \Theta_a^- \Theta_k^+ \Theta_l^+ \Theta_p^+}{(\varepsilon_k + \varepsilon_l - \varepsilon_i - \varepsilon_j)(\varepsilon_p + \varepsilon_l - \varepsilon_a - \varepsilon_j)}. \tag{2.102}$$

These expressions are readily generalized for higher-order ladder and ring diagrams. In the thermodynamic limit the state sums become integrals, and the integral kernels of Eqs. (2.99)-(2.102) diverge at the boundary of the respective integration regions.

***Non-Skeletons with First-Order Parts.*** The remaining two normal third-order diagrams are the two normal "one-loop" diagrams shown in Figs. 2.8 (a) and (d). The sum of these two contributions is given by ($\mathcal{M}_{\text{articulation}} = 2$, $\ell = 3$, $h = 4$ and $\mathcal{M}_{\text{articulation}} = 2$, $\ell = 3$, $h = 3$)

$$E_{0;3,\text{one-loop}} = -\frac{1}{4} \sum_{ijkl} \bar{V}_{2\text{B}}^{ij,kl} \bar{V}_{2\text{B}}^{kl,ij} \frac{\Theta_i^- \Theta_j^- \Theta_k^+ \Theta_l^+}{(\varepsilon_k + \varepsilon_l - \varepsilon_i - \varepsilon_j)^2} \Big( \Theta_i^- X_{1;i} + \Theta_j^- X_{1;j} - \Theta_k^+ X_{1;k} - \Theta_l^+ X_{1;l} \Big), \tag{2.103}$$

where $X_{1;a}(\varepsilon_F, \Omega) = \sum_a \bar{V}_{2\text{B}}^{ab,ab} \Theta_b^-$ is the first-order self-energy. Using $\Theta_i^- \Theta_i^- = \Theta_i^-$ and $\Theta_k^+ \Theta_k^+ = \Theta_k^+$, the sum of all second-order normal diagrams with different numbers of "one-loop" insertions is given by multiplying the integrand in Eq. (2.99) with a factor

$$\sum_{n=0}^{\infty} \left( \frac{X_{1;i} + X_{1;j} - X_{1;k} - X_{1;l}}{\varepsilon_k + \varepsilon_l - \varepsilon_i - \varepsilon_j} \right)^n = \left( 1 - \frac{X_{1;i} + X_{1;j} - X_{1;k} - X_{1;l}}{\varepsilon_k + \varepsilon_l - \varepsilon_i - \varepsilon_j} \right)^{-1}. \tag{2.104}$$

Hence, the resummation of "one-loop" insertions leads to the renormalization of the single-particle energies in the energy denominator of Eq. (2.99) according to

$$\varepsilon_i \to \varepsilon_{X_{1;i}} = \varepsilon_i + X_{1;i}. \tag{2.105}$$

This straightforward resummation of (normal) "one-loop" insertions in terms of a geometric series is possible only in the ground-state formalism. The grand-canonical case is examined in Secs. 2.3.4 and 2.4.3.

### 2.3.4. Grand-Canonical Potential up to Third Order

Here, we examine the (formal) expressions for the various contributions up to third order in the grand-canonical formalism, in the "bare" case, and evaluated in the standard way ("contraction formalism"). In particular, we discuss certain intricacies associated with energy denominators as well as the appearance of terms that diverge in the zero-temperature limit.

***Skeletons.*** The second- and third-order skeletons are invariant under cyclic permutations, thus the different formulas give equivalent (but not identical) expressions for these diagrams. For the case where the energy denominator is nonzero, the expression obtained from the direct formula for the second-order normal diagram is given by

$$A_{2,\text{normal}}^{\text{direct},\mathscr{E}\neq 0} = -\frac{(-1)}{\beta}\langle\mathcal{V}_{\text{2B}}\mathcal{V}_{\text{2B}}\rangle_{2,\text{normal}}\left[\underset{z=0}{\text{Res}}\frac{e^{-\beta z}}{z^2}\frac{1}{\mathscr{E}-z} + \underset{z=\mathscr{E}}{\text{Res}}\frac{e^{-\beta z}}{z^2}\frac{1}{\mathscr{E}-z}\right]$$
$$= -\frac{1}{4}\sum_{ijkl}^{\mathscr{E}\neq 0}\bar{V}_{\text{2B}}^{ij,kl}\bar{V}_{\text{2B}}^{kl,ij}\left[\frac{f_i^- f_j^- f_k^+ f_l^+}{\varepsilon_k+\varepsilon_l-\varepsilon_i-\varepsilon_j} - \frac{1}{\beta}\frac{f_i^- f_j^- f_k^+ f_l^+ - f_i^+ f_j^+ f_k^- f_l^-}{(\varepsilon_k+\varepsilon_l-\varepsilon_i-\varepsilon_j)^2}\right], \quad (2.106)$$

where we have used $f_i^- f_j^- f_k^+ f_l^+ e^{-\beta\mathscr{E}} = f_i^+ f_j^+ f_k^- f_l^-$. The part of the integrand proportional to $\beta^{-1}$ gives a vanishing contribution, as can be seen by relabeling indices $i\to k$, $j\to l$. The remaining part coincides with the expression obtained from the reduced formula, and the index sum becomes a principal value integral in the thermodynamic limit.[27] From the cyclic formula one obtains for the second-order normal diagram the expression

$$A_{2,\text{normal}}^{\text{cyclic},\mathscr{E}\neq 0} = \frac{(-1)}{2}\langle\mathcal{V}_{\text{2B}}\mathcal{V}_{\text{2B}}\rangle_{2,\text{normal}}\left[\underset{z=0}{\text{Res}}\frac{e^{-\beta z}}{z}\frac{1}{\mathscr{E}-z} + \underset{z=\mathscr{E}}{\text{Res}}\frac{e^{-\beta z}}{z}\frac{1}{\mathscr{E}-z}\right]$$
$$= -\frac{1}{8}\sum_{ijkl}^{\mathscr{E}\neq 0}\bar{V}_{\text{2B}}^{ij,kl}\bar{V}_{\text{2B}}^{kl,ij}\frac{f_i^- f_j^- f_k^+ f_l^+ - f_i^+ f_j^+ f_k^- f_l^-}{\varepsilon_k+\varepsilon_l-\varepsilon_i-\varepsilon_j}. \quad (2.107)$$

The contributions from the residues at $z=0$ and $z=\mathscr{E}$ are equivalent, as can be seen by relabeling indices $i\to k$, $j\to l$. However, keeping both contributions has the advantage that in that case the integrand is regular at $\mathscr{E}=0$, i.e.,[28]

$$(f_i^- f_j^- f_k^+ f_l^+ - f_i^+ f_j^+ f_k^- f_l^-)\frac{1}{\mathscr{E}} = f_i^- f_j^- f_k^+ f_l^+\frac{1-e^{-\beta\mathscr{E}}}{\mathscr{E}} \xrightarrow{\mathscr{E}\to 0} \beta f_i^- f_j^- f_k^+ f_l^+. \quad (2.108)$$

This reproduces the expression obtained (from each formula) by evaluating the contour integral for $\mathscr{E}=0$:

$$A_{2,\text{normal}}^{\mathscr{E}=0} = -\frac{\beta}{8}\sum_{ijkl}^{\mathscr{E}=0}\bar{V}_{\text{2B}}^{ij,kl}\bar{V}_{\text{2B}}^{kl,ij}\, f_i^- f_j^- f_k^+ f_l^+. \quad (2.109)$$

This contribution can therefore be included by letting $\mathscr{E}\to 0$ in Eq. (2.107).[29] Such contributions from "accidentally" vanishing denominators diverge in the zero-temperature limit $\beta\to\infty$.

---

[27] The vanishing part propertional to $\beta^{-1}$ however does play a certain role in the resummation of "one-loop" insertions if the direct formula is used, cf. Sec. 2.4.3.

[28] Note that the zero-temperature limit of terms $\sim e^{-\beta\mathscr{E}}$ must always be evaluated together with the respective product of Fermi-Dirac distributions, i.e., $f_i^+ f_j^+ f_k^- f_l^-\, e^{-\beta\mathscr{E}} \xrightarrow{T\to 0} \Theta_i^- \Theta_j^- \Theta_k^+ \Theta_l^+$.

[29] This is general, cf. Eqs. (2.56) and (2.57).

For a continuous spectrum these contributions have measure zero, but in the discrete situation they would have to be excluded "by hand" to obtain well-behaved results at low temperatures.[30] From now on we consider only the expressions obtained for nonzero energy denominators. At higher orders, "accidentally" vanishing denominators arise not only for vanishing $\mathscr{E}$'s but also when different $\mathscr{E}$'s coincide. For instance, the "cyclic" expressions for the third-order ladder and ring diagrams are given by

$$A_{3,hh\text{-ladder}}^{\text{cyclic}} = \frac{1}{24} \sum_{ijabkl} \bar{V}_{2B}^{ij,kl} \bar{V}_{2B}^{ab,ij} \bar{V}_{2B}^{kl,ab} \, f_i^- f_j^- f_a^- f_b^- f_k^+ f_l^+$$
$$\times \left[ \frac{1}{\mathscr{E}_1 \mathscr{E}_2} - \frac{e^{-\beta \mathscr{E}_1}}{\mathscr{E}_1(\mathscr{E}_2 - \mathscr{E}_1)} - \frac{e^{-\beta \mathscr{E}_2}}{\mathscr{E}_2(\mathscr{E}_1 - \mathscr{E}_2)} \right]_{\substack{\mathscr{E}_1=\varepsilon_k+\varepsilon_l-\varepsilon_i-\varepsilon_j \\ \mathscr{E}_2=\varepsilon_k+\varepsilon_l-\varepsilon_a-\varepsilon_b}} \quad (2.110)$$

$$A_{3,pp\text{-ladder}}^{\text{cyclic}} = \frac{1}{24} \sum_{ijklpq} \bar{V}_{2B}^{ij,kl} \bar{V}_{2B}^{kl,pq} \bar{V}_{2B}^{pq,ij} \, f_i^- f_j^- f_k^+ f_l^+ f_p^+ f_q^+$$
$$\times \left[ \frac{1}{\mathscr{E}_1 \mathscr{E}_2} - \frac{e^{-\beta \mathscr{E}_1}}{\mathscr{E}_1(\mathscr{E}_2 - \mathscr{E}_1)} - \frac{e^{-\beta \mathscr{E}_2}}{\mathscr{E}_2(\mathscr{E}_1 - \mathscr{E}_2)} \right]_{\substack{\mathscr{E}_1=\varepsilon_k+\varepsilon_l-\varepsilon_i-\varepsilon_j \\ \mathscr{E}_2=\varepsilon_p+\varepsilon_q-\varepsilon_i-\varepsilon_j}}, \quad (2.111)$$

$$A_{3,\text{ring}}^{\text{cyclic}} = \frac{1}{3} \sum_{ijaklp} \bar{V}_{2B}^{ij,kl} \bar{V}_{2B}^{ka,ip} \bar{V}_{2B}^{pl,aj} \, f_i^- f_j^- f_a^- f_k^+ f_l^+ f_p^+$$
$$\times \left[ \frac{1}{\mathscr{E}_1 \mathscr{E}_2} - \frac{e^{-\beta \mathscr{E}_1}}{\mathscr{E}_1(\mathscr{E}_2 - \mathscr{E}_1)} - \frac{e^{-\beta \mathscr{E}_2}}{\mathscr{E}_2(\mathscr{E}_1 - \mathscr{E}_2)} \right]_{\substack{\mathscr{E}_1=\varepsilon_k+\varepsilon_l-\varepsilon_i-\varepsilon_j \\ \mathscr{E}_2=\varepsilon_p+\varepsilon_l-\varepsilon_a-\varepsilon_j}}, \quad (2.112)$$

where the expressions in the square brackets [...] are regular (as expected) in the (single or combined) limits $\mathscr{E}_1 \to 0$, $\mathscr{E}_2 \to 0$, and $\mathscr{E}_1 \to \mathscr{E}_2$, e.g., $[\ldots] \xrightarrow{\mathscr{E}_1 \to 0, \mathscr{E}_2 \to 0} \beta^2/2$. In each case, *for a discrete spectrum* (see Sec. 2.3.5) the three terms in [...] are equivalent.[31]

***Non-Skeletons with First-Order Parts Only.*** There is one such diagram at second order [diagram (c) of Fig. 2.4], and five such diagrams at third order (shown in Fig. 2.7), which we tentatively label according to the number of instantaneous contractions ("loops"). Diagrams of this kind either transform into each other under cyclic permutations of the interaction vertices (e.g., the two-loop diagrams), or are invariant under cyclic permutations (e.g., the second-order anomalous and three-loop diagrams), so the cyclic, the direct, and the reduced formula are equivalent in that case.[32] The expressions for the second-order anomalous diagram, the sum of the three two-loop diagrams, as well as the sum of the two three-loop diagrams are given by

$$A_{2,\text{anomalous}} = -\frac{\beta}{2} \sum_{ijb} \bar{V}_{2B}^{ij,ij} \bar{V}_{2B}^{jb,jb} \, f_i^- f_j^- f_j^+ f_b^-, \quad (2.113)$$

$$A_{3,\text{two-loop}} = \frac{\beta^2}{2} \sum_{ijab} \bar{V}_{2B}^{ij,ij} \bar{V}_{2B}^{ja,ja} \bar{V}_{2B}^{ab,ab} \, f_i^- f_j^- f_j^+ f_a^- f_a^+ f_b^-, \quad (2.114)$$

$$A_{3,\text{three-loop}} = \frac{\beta^2}{6} \sum_{ijab} \bar{V}_{2B}^{ij,ij} \bar{V}_{2B}^{ja,ja} \bar{V}_{2B}^{jb,jb} \, f_i^- f_j^- f_j^+ f_a^- f_b^- (f_j^+ - f_j^-), \quad (2.115)$$

where we have used that $\mathcal{M}_{\text{articulation}} = 8$, $\ell = 4$, $h = 4$ for diagrams (a,b,c), $\mathcal{M}_{\text{articulation}} = 8$, $\ell = 4$, $h = 5$ for diagram (d) and $\mathcal{M}_{\text{articulation}} = 8$, $\ell = 4$, $h = 4$ for diagram (e) of Fig. 2.7.

---

[30] Note that the contributions from "accidentally" vanishing denominators do not appear if Wick's theorem is applied not to the ensemble average but to the individual terms in the partition sum, cf. Ref. [277].

[31] For instance, in the ring contribution the second term in [...] transforms into the first one under $i \to l$, $j \to k$.

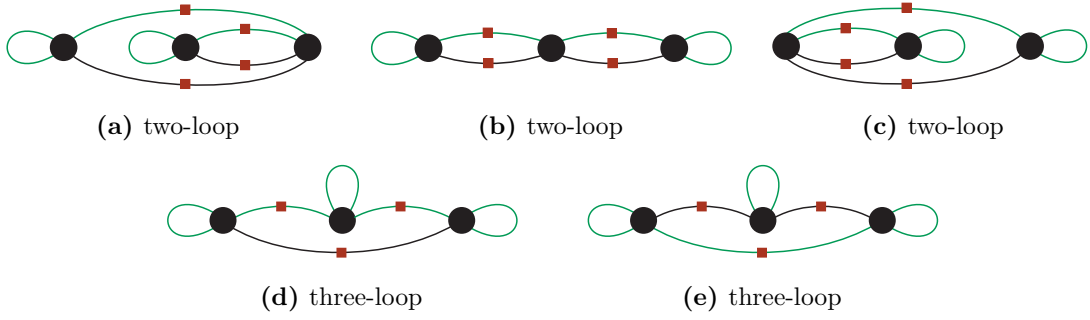

(a) two-loop    (b) two-loop    (c) two-loop

(d) three-loop    (e) three-loop

**Figure 2.7.:** Third-order anomalous diagrams involving only first-order "parts". Articulation lines are marked by red squares. Note that in the case of the "three-loop" diagrams, the articulation lines form a "cycle".

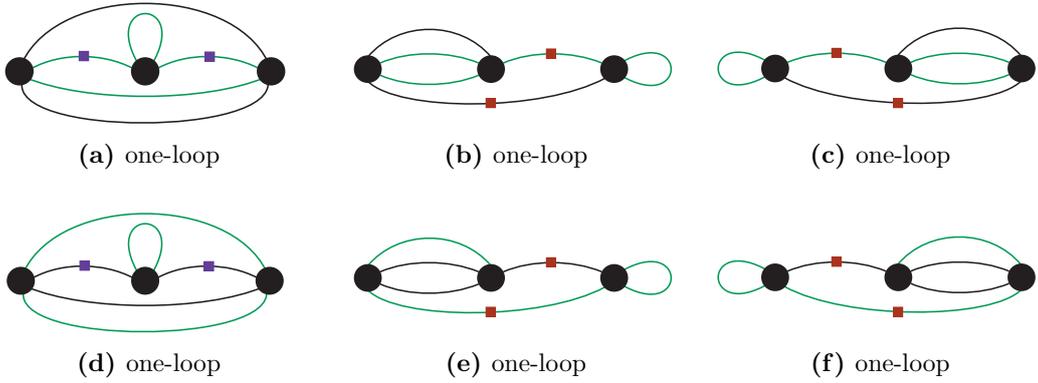

(a) one-loop    (b) one-loop    (c) one-loop

(d) one-loop    (e) one-loop    (f) one-loop

**Figure 2.8.:** Third-order diagrams corresponding to normal (a,d) and anomalous (b,c,e,f) "one-loop" insertions on the second-order diagram. Articulation lines are marked by red (anomalous) and black (self-energy) squares. The upper and lower three diagrams transform into each other under cyclic permutations.

*Non-Skeletons with First- and Higher-Order Parts, "Disentanglement".* The remaining six third-order diagrams are shown in Fig. 2.8;[33] they correspond to normal and anomalous "one-loop" insertions onto the second-order normal diagram. The diagrams (a,b,c) transform into each other via cyclic vertex permutations, and similar for (d,e,f). In the following, we examine (assuming a discrete spectrum) the formal expressions obtained for these diagrams. Using the direct formula, and suitably labeling indices, we obtain for the sum of the four anomalous "one-loop" diagrams (b,c,e,f) the expression

$$
\begin{aligned}
A^{\text{direct}}_{3,\text{one-loop(b+c+e+f)}} &= 2 \times \left\{ -\frac{(-1)^2}{\beta} \langle \mathcal{V}_{2B} \mathcal{V}_{2B} \mathcal{V}_{2B} \rangle_{(b)+(e)} \left[ \operatorname*{Res}_{z=0} \frac{e^{-\beta z}}{z^2(-z)(\mathcal{E}-z)} + \operatorname*{Res}_{z=\mathcal{E}} \frac{e^{-\beta z}}{z^2(-z)(\mathcal{E}-z)} \right] \right\} \\
&= \frac{1}{4} \sum_{ijkl} \bar{V}^{ij,kl}_{2B} \bar{V}^{kl,ij}_{2B} f_i^- f_j^- f_k^+ f_l^+ \left[ \overbrace{-\frac{2}{\beta}\left(\frac{1-e^{-\beta\mathcal{E}}}{\mathcal{E}^3}\right) + \frac{2}{\mathcal{E}^2}}^{\text{pseudo-anomalous}} \overbrace{-\frac{\beta}{\mathcal{E}}}^{\text{anomalous}} \right] \\
&\quad \times \left( f_i^+ S_{1;i} + f_j^+ S_{1;j} - f_k^- S_{1;k} - f_l^- S_{1;l} \right).
\end{aligned}
\tag{2.116}
$$

---

[32] This can be seen explicitly from Eqs. (2.52) and (2.53), since for vanishing excitation energies the only effect of the additional factor $(-\beta^{-1}z^{-1})$ in the direct formula is to reproduce the factor $(n+1)^{-1}$ in the cyclic formula.

[33] Each diagram could also be counted as two diagrams (with half multiplicity); e.g., in the case of diagram (a) there is one with a "one-loop" insertion onto the upper line and one with an insertion onto the lower line (corresponding to insertions on the line with index $i$ and $j$, respectively).

2. Many-Body Perturbation Theory

where $\mathscr{E} = \varepsilon_k + \varepsilon_l - \varepsilon_i - \varepsilon_j$. Only the last term in the square brackets is an anomalous term of the form $\partial f_i^- / \partial \mu$; the other terms involve $f_i^- f_i^+$ factors, but vanish in the zero-temperature limit; we call terms of this kind "*pseudo-anomalous terms*". For the sum of the two normal "one-loop" diagrams from Fig. 2.8, the direct formula gives the expression

$$A^{\text{direct}}_{3,\text{one-loop(a+d)}} = -\frac{(-1)^2}{\beta} \langle \mathcal{V}_{2B} \mathcal{V}_{2B} \mathcal{V}_{2B} \rangle_{(a)+(d)} \left[ \operatorname*{Res}_{z=0} \frac{e^{-\beta z}}{z^2} \frac{1}{(\mathscr{E}-z)^2} + \operatorname*{Res}_{z=\mathscr{E}} \frac{e^{-\beta z}}{z^2} \frac{1}{(\mathscr{E}-z)^2} \right]$$

$$= \frac{1}{4} \sum_{ijkl} \bar{V}^{ij,kl}_{2B} \bar{V}^{kl,ij}_{2B} f_i^- f_j^- f_k^+ f_l^+ \left[ \overbrace{\frac{2}{\beta}\left(\frac{1-e^{-\beta\mathscr{E}}}{\mathscr{E}^3}\right) - \frac{e^{-\beta\mathscr{E}}}{\mathscr{E}^2}}^{\text{pseudo-anomalous}} \overbrace{-\frac{1}{\mathscr{E}^2}}^{\text{double-normal}} \right]$$

$$\times \left( f_i^- S_{1;i} + f_j^- S_{1;j} - f_k^+ S_{1;k} - f_l^+ S_{1;l} \right). \quad (2.117)$$

Here, "*double-normal*" refers to the presence of a squared ("double") Fermi-Dirac distribution, i.e., $f_i^- f_i^-$ in the first term. The sum of all third-order self-energy and anomalous "one-loop" diagrams is equivalent to three times the expression obtained from the cyclic formula for the sum of (for instance) diagrams (a) and (b), which leads to[34]

$$A^{(\text{cyclic})}_{3,\text{one-loop(a+b+c+d+e+f)}} = -\frac{1}{4} \sum_{ijkl} \bar{V}^{ij,kl}_{2B} \bar{V}^{kl,ij}_{2B} \frac{f_i^- f_j^- f_k^+ f_l^+}{\mathscr{E}^2} \left( \overbrace{1}^{\text{double-n.}} \overbrace{-e^{-\beta\mathscr{E}}}^{\text{pseudo-anom.}} \overbrace{-\beta\mathscr{E} \, e^{-\beta\mathscr{E}}}^{\text{anom.}} \right)$$

$$\times \left( f_i^- S_{1;i} + f_j^- S_{1;j} - f_k^+ S_{1;k} - f_l^+ S_{1;l} \right), \quad (2.118)$$

which (as expected) is well-defined in the thermodynamic limit, in contrast to Eqs. (2.116), (2.118) and (2.119), which involve symmetric poles. The remaining "pseudo-anomalous" term in Eq. (2.118) can be removed by means of $f_i^- f_i^- = f_i^- - f_i^- f_i^+$ and $f_k^+ f_k^+ = f_k^+ - f_k^- f_k^+$. This leads to

$$A^{\text{"disentangled"}}_{3,\text{one-loop(a+b+c+d+e+f)}} = -\frac{1}{4} \sum_{ijkl} \bar{V}^{ij,kl}_{2B} \bar{V}^{kl,ij}_{2B} \frac{f_i^- f_j^- f_k^+ f_l^+}{\mathscr{E}^2} \left( \overbrace{S_{1;i} + S_{1;j} - S_{1;k} - S_{1;l}}^{\text{normal}} \right.$$

$$\left. \overbrace{- \beta\mathscr{E} \left( f_i^+ S_{1;i} + f_j^+ S_{1;j} - f_k^- S_{1;k} - f_l^- S_{1;l} \right)}^{\text{anomalous}} \right). \quad (2.119)$$

This represents the "disentangled" form (for $\aleph$ = "reduced") of the contribution from the third-order "one-loop" diagrams. In particular, the "normal" part of Eq. (2.119) has the same form as the expression for the normal "one-loop" diagrams in the ground-state formalism, but without "double-normal" terms.

Finally, we note that the contributions from non-skeletons with more than two parts involve terms that diverge in the zero-temperature limit (for each formula). More specifically, for such contributions there are higher-order derivatives of the exponential $e^{-\beta z}$, leading to ("non-anomalous") terms $\sim \beta^{K \geq 1}$. For instance, for the case of the second-order normal diagram with two normal "one-loop" insertions the contour integral in the "direct" case is given by[35]

$$-\frac{1}{\beta} \operatorname*{Res}_{z=0} \frac{e^{-\beta z}}{z^2} \frac{1}{(\mathscr{E}-z)^3} = \frac{\beta}{2!} \frac{e^{-\beta\mathscr{E}}}{\mathscr{E}^2} + O(\beta^0). \quad (2.120)$$

---

[34] To obtain Eq. (2.118) explicitly as the sum of Eqs. (2.116) and (2.117) one needs to identify various cancellations that become apparent only after suitably relabeling the indices of various terms.

[35] In the $T \to 0$ limit the exponential term needs to be evaluated together with the Fermi-Dirac distributions.

Such ("non-anomalous") $\beta^{K\geq 1}$ terms are artifacts. In the present case, they are not present if the reduced formula is used, and thus cancel in the cyclic sum of diagrams. In general, however, terms that diverge in the zero-temperature limit arise also if the reduced formula is used (i.e., for anomalous diagrams with more than two parts), and the "cancellation" of these terms involves the sum of various (not cyclically related) normal and anomalous non-skeletons at a given order. In the non-skeleton sum, the $\beta^{K\geq 1}$ *terms then correspond to anomalous terms of order K*, i.e., terms that involve derivatives of order $K$ of Fermi-Dirac distributions. The "cancellation" of the ("non-anomalous") $\beta^{K\geq 1}$ terms to all orders follows from Eq. (2.88) and the fact that for normal non-skeletons the reduced formula leads to terms that are nonsingular at zero temperature. The "cancellation" of the $\beta^{K\geq 1}$ terms is however "dissolved" in the $\{\mathcal{T} + \mathcal{X}_{[n]}^{\text{direct}}, \mathcal{V} - \mathcal{X}_{[n]}^{\text{direct}}\}$ setups with $n \geq 6$. This is due to the fact that normal non-skeletons without first-order parts are not insertions, in contrast to higher-cumulant contributions without first-order parts. At sixth order in the perturbation series, the first normal non-skeletons with more than two higher-order parts appear, i.e., normal non-skeletons with three second order parts (one or two of which can be $-X_2$ vertices). If the direct formula is used, such diagrams lead to ("non-anomalous") $\beta^{K\geq 1}$ terms in the (formal) expressions for the self-energies. This feature corresponds to the notion that for the $\{\mathcal{T} + \mathcal{X}_{[n]}^{\text{direct}}, \mathcal{V} - \mathcal{X}_{[n]}^{\text{direct}}\}$ setups with $n \geq 6$ the zero-temperature limit does not exist.[36]

### 2.3.5. Normal Non-Skeletons without First-Order Parts

Here we show that normal non-skeletons without first-order parts are not insertions. Restricting the discussion to the ground-state formalism (without loss of generality), we examine the case of the fourth-order non-skeletons with two second-order parts, denoted as "$\Gamma_{2,2}$". The expressions for these diagrams can be expressed in terms of the second-order *perturbative* self-energy, i.e.,

$$S_{2;a}(\varepsilon_F) = \frac{\delta E_{0;2,\text{normal}}[\Theta_a^-;\{\varepsilon_a\}]]}{\delta \Theta_a^-} = \overbrace{\sum_{jkl} \frac{H_{ajkl}}{\varepsilon_k + \varepsilon_l - \varepsilon_a - \varepsilon_j}}^{S_{2;a}^H} + \overbrace{\sum_{ijk} \frac{P_{ijka}}{\varepsilon_k + \varepsilon_a - \varepsilon_i - \varepsilon_j}}^{S_{2;a}^P}. \quad (2.121)$$

where the functions $H_{ajkl}$ and $P_{ijka}$ are given by Eq. (2.77). The six $\Gamma_{2,2}$ diagrams are shown in Fig. 2.9.[37] The expression for the sum of the first two diagrams (a,b) is then given by

$$E_{0;4,\Gamma_{2,2}(a+b)} = -\frac{1}{4} \sum_{abmn} \bar{V}_{2B}^{ab,mn} \bar{V}_{2B}^{mn,ab} \Theta_a^- \Theta_b^- \Theta_m^+ \Theta_n^+ \underbrace{\sum_{jkl} H_{ajkl} \left[\frac{1}{\mathcal{E}_1(\mathcal{E}_2 + \mathcal{E}_1)\mathcal{E}_1} + \overbrace{\frac{1}{\mathcal{E}_1(\mathcal{E}_2 + \mathcal{E}_1)\mathcal{E}_2}}^{1/[(\mathcal{E}_1)^2 \mathcal{E}_2]}\right]}_{\frac{1}{(\mathcal{E}_1)^2} S_{2;a}^H},$$

(2.122)

where $\mathcal{E}_1 = \varepsilon_m + \varepsilon_n - \varepsilon_a - \varepsilon_b$ is corresponds to the "large" part (large vertices in Fig. 2.9), and $\mathcal{E}_2 = \varepsilon_k + \varepsilon_l - \varepsilon_a - \varepsilon_j$ to the "small" part (small vertices). The expression for the sum of the diagrams (c,d) is similar, but involves $S_{2;a}^P$ instead of $S_{2;a}^H$. Both expressions have a "factorized" form similar to the one of the normal "one-loop" diagrams [cf. Eq. (2.103)]; i.e., they have the form of the first term in a geometric series in terms of $\frac{1}{\mathcal{E}_1}S^H_{2;a}$ and $\frac{1}{\mathcal{E}_1}S^P_{2;a}$, respectively. However, this series would renormalize *hole* and *particle* energies differently, i.e., according to $S^H_{2;a}$ and $S^P_{2;a}$, respectively. This is in contrast to the effect of additional normal non-skeletons from $-X[n]$ vertices, which are always of the "one-loop" insertion form, corresponding to a geometric series in $\frac{1}{\mathcal{E}_1}X_{[n]}$. Furthermore, inspecting the corresponding diagrams with multiple second-order normal subdiagrams one finds that they do not "factorize". In fact, already in the "$\Gamma_{2,2}$" case there are additional "nonfactorizable" contributions where the energy-denominator sum does not simplify: diagrams (e) and (f) of Fig. 2.9.

---

[36] In the $\{\mathcal{T} + \mathcal{X}_{[n]}^{\text{direct}}, \mathcal{V} - \mathcal{X}_{[n]}^{\text{direct}}\}$ setups with $2 \leq n < 6$, there are also $\beta^{K\geq 1}$ terms, i.e., the ones from normal non-skeletons with (multiple) $-X_2, \ldots, X_n$ vertices, but the $-X_2, \ldots, X_n$ vertices can be cancelled by expanding the energy denominators about the Hartree-Fock ones.

[37] Note that diagram (a) is a cyclic permutation of diagram (b), and similar for diagrams (c,d). Diagrams (e) and (f) are each one cyclic permutation of themselves. The other six cyclic vertex permutations are anomalous diagrams.

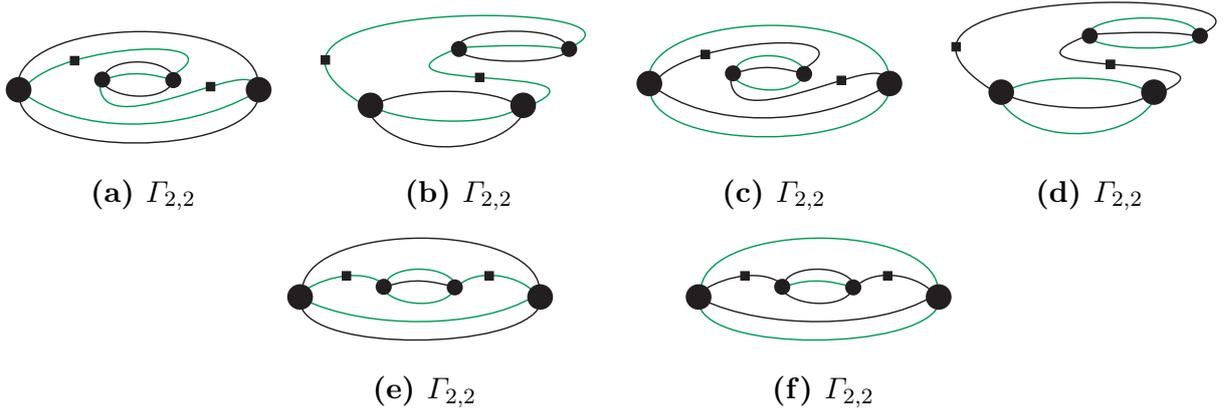

(a) $\Gamma_{2,2}$   (b) $\Gamma_{2,2}$   (c) $\Gamma_{2,2}$   (d) $\Gamma_{2,2}$

(e) $\Gamma_{2,2}$   (f) $\Gamma_{2,2}$

**Figure 2.9.:** The six "$\Gamma_{2,2}$" diagrams. For clarity, in each case one of the two skeleton parts is drawn with smaller vertex dots.

### 2.3.6. Regularization of Energy Denominators and Thermodynamic Limit

For a continuous spectrum, to obtain nonsingular expressions certain diagrams need to be evaluated "together with their cyclic permutations" (i.e., the cyclic formula has to be used). This requirement however severly restricts the manipulations allowed on the expressions for individual contributions. In particular, the cumulant formalism ("disentanglement") and the "renormalization" of $\mathcal{T}$ and $\mathcal{V}$ in terms of higher-order ($\nu \geq 4$) "self-energies" are well-defined for a discrete spectrum, but these manipulations involve terms (with symmetric energy-denominator poles) that diverge in the thermodynamic limit. For a continuous spectrum, these manipulations therefore require a regularization procedure. This regularization must be such that it *leaves the cyclic formula invariant*. We adopt the following procedure: for a given diagram, we add in the energy denominators to each independent $\mathcal{E}_\nu$ an infinitesimal imaginary part $i\xi_\nu$, and then take the average with respect to $\text{sgn}(\xi_\nu)$, i.e., for the case where there are no linear combinations of $\mathcal{E}$'s in the energy denominators[38] one has (where $\xi_1 \neq \xi_2 \neq \ldots \neq \xi_n$)

$$\frac{1}{(\mathcal{E}_1)^q_1 \cdots (\mathcal{E}_n)^q_n} \xrightarrow{\text{regularization (R)}} \frac{1}{(\mathcal{E}^R_1)^q_1 \cdots (\mathcal{E}^R_n)^q_n} := \frac{1}{2^n} \sum_{\text{sgn}(\xi_1),\ldots,\text{sgn}(\xi_n)} \frac{1}{(\mathcal{E}_1 + i\xi_1)^{q_1} \cdots (\mathcal{E}_n + i\xi_n)^{q_n}}$$
(2.123)

---

[38] For the case with linear combination of energy denominators one fixes the magnitude of the infinitesimal imaginary parts such that $|\xi_\nu| \neq \xi_{\nu'} \; \forall \{\nu, \nu'\}$. Averaging with respect to $\text{sgn}(\xi_\nu)$ then leads again to finite-part integrals.

In the thermodynamic limit, this is equivalent to

$$\frac{1}{(\mathcal{E}_1)_1^q \cdots (\mathcal{E}_n)_n^q} \xrightarrow{\text{regularization (R)}} \frac{\mathscr{P}}{(\mathcal{E}_1)^{q_1}} \cdots \frac{\mathscr{P}}{\mathcal{E}_n)^{q_n}}, \tag{2.124}$$

where the *finite part* $\mathscr{P}$ is defined via (where $a < x < b$)

$$\mathscr{P} \int_a^b dt \frac{f(t)}{(t-x)^q} = \frac{1}{(q-1)!} \frac{d^q}{dx^q} \int_a^b dt \frac{f(t)}{t-x} = \frac{1}{(q-1)!} \frac{d^q}{dx^q} \left[ \int_a^{x-\varepsilon} dt \frac{f(t)}{t-x} + \int_{x+\varepsilon}^b dt \frac{f(t)}{t-x} \right]. \tag{2.125}$$

For instance, for $q = 1$, $\mathscr{P}$ is the Cauchy principal value; for $q = 2$, $\mathscr{P}$ is Hadamard's finite part. The equivalence of Eqs. (2.123) and (2.124) follows from the generalized Plemelj identity [19]

$$\frac{1}{(\mathcal{E} + i\xi)^q} = \frac{\mathscr{P}}{(\mathcal{E})_1^q} + \frac{i\pi(-1)^q}{(q-1)!} \text{sgn}(\xi) \delta^{(q-1)}(\mathcal{E}). \tag{2.126}$$

The expressions obtained from the cyclic formula are invariant under the regularization procedure "R" defined by Eq. (2.123) and (2.124), respectively, as required. The *proper* "direct" and "reduced" expressions *for a continuous spectrum* are then constructed by (suitably) separating the various terms in the regularized "cyclic" expressions. This procudure involves a quite tricky point [249, 429, 242] that leads to correction terms to the expressions obtained from the "direct" and "reduced" formula, Eqs. (2.53) and (2.55), respectively: *the decomposition of the regularized cyclic expressions and the thermodynamic limit do not commute*. In particular, performing the decomposition *before* taking the thermodynamic limit in general leads to wrong results. This is because the value of an integral that involves several principal parts depends on the integration order. The cyclic formula does not depend on the integration order, but the thermodynamic limit of Eqs. (2.53) and (2.55) does, and depending on the integration order, different correction terms are needed.

To illustrate this point, we consider the regularized cyclic expression for the third-order *hh*-ladder diagram:

$$A_{3,hh\text{-ladder}}^R \sim \frac{1}{3} \sum_{ijabkl} [\ldots] \left\{ \frac{\mathscr{P}}{\mathcal{E}_1} \frac{\mathscr{P}}{\mathcal{E}_2} - e^{-\beta \mathcal{E}_1} \frac{\mathscr{P}}{\mathcal{E}_1} \frac{\mathscr{P}}{\mathcal{E}_2 - \mathcal{E}_1} - e^{-\beta \mathcal{E}_2} \frac{\mathscr{P}}{\mathcal{E}_2} \frac{\mathscr{P}}{\mathcal{E}_1 - \mathcal{E}_2} \right\}. \tag{2.127}$$

Separating the three summands and relabelling indices leads to

$$A_{3,hh\text{-ladder}}^R \sim \frac{1}{3} \left\{ \sum_{ijabkl} + \sum_{klijab} + \sum_{abklij} \right\} [\ldots] \frac{\mathscr{P}}{\mathcal{E}_1} \frac{\mathscr{P}}{\mathcal{E}_2}. \tag{2.128}$$

For a discrete spectrum the three terms in the above formula are identical, leading to the "reduced" formula results for $A_{3,hh\text{-ladder}}^R$. Performing then the thermodynamic limit, this leads to the (incorrect) result

$$A_{3,hh\text{-ladder}}^{R,\text{wrong}} \sim \int_{ijabkl} [\ldots] \frac{\mathscr{P}}{\mathcal{E}_1} \frac{\mathscr{P}}{\mathcal{E}_2}. \tag{2.129}$$

Instead, performing the thermodynamic limit already in Eq. (2.127) and then separating the summands and relabelling indices leads to

$$A_{3,hh\text{-ladder}}^R \sim \frac{1}{3} \left\{ \int_{ijabkl} + \int_{klijab} + \int_{abklij} \right\} [\ldots] \frac{\mathscr{P}}{\mathcal{E}_1} \frac{\mathscr{P}}{\mathcal{E}_2}. \tag{2.130}$$

## 2. Many-Body Perturbation Theory

By the Poincaré-Bertrand transformation formula [242], this is equivalent to

$$A^R_{3,hh\text{-ladder}} \sim \int_{ijabkl} [\ldots] \left\{ \frac{\mathscr{P}}{\mathscr{E}_1} \frac{\mathscr{P}}{\mathscr{E}_2} - \frac{\pi^2}{3} \delta(\mathscr{E}_1)\delta(\mathscr{E}_2) \right\} \xrightarrow{T \to 0} \int_{ijabkl} [\ldots] \frac{1}{\mathscr{E}_1 \mathscr{E}_2}, \quad (2.131)$$

which deviates from Eq. (2.129) in terms of the additional delta function term. This term vanishes in the $T \to 0$ limit, since in that limit the energy-denominator poles lie on the boundary of the integration region.

The third-order $hh$-ladder diagram is invariant under cyclic permutations, and Eqs. (2.127), (2.130), and (2.131) are equivalent. The distinction between "cyclic", "direct" and "reduced" expressions comes with the disentanglement of higher-order non-skeleton diagrams where not all parts of the respective cyclic expressions are considered. The procedure to construct the proper disentangled "direct" and "reduced" expressions for a continuous spectrum can be formulated as follows. We start with the sum $\mathcal{L}$ of all diagrams $\mathcal{L}_m$ in a given cyclic group, all evaluated with the cyclic formula ($\aleph$ = "cyclic"), and fix the integration order $ij\ldots$ for all diagrams. Averaging over all corresponding "cyclic" permutations of the integration order $C[ij\ldots]$, the restriction $\aleph$ = "cyclic" can be omitted,[39] i.e., symbolically,

$$\mathcal{L} = \int_{ij\ldots} \sum_m \mathcal{L}_m^{\aleph=\text{"cyclic"}} = \frac{1}{|C|} \int_{C[ij\ldots]} \sum_m \mathcal{L}_m^{\aleph}. \quad (2.132)$$

We now apply the disentanglement, which leads to

$$\mathcal{L} = \frac{1}{|C|} \int_{C[ij\ldots]} \left\{ \sum_m \mathscr{D}_m^{\aleph} + \{\boldsymbol{\Gamma}^{\aleph}\!\!-\!\!\boldsymbol{\Gamma}^{\aleph}\} \right\}, \quad (2.133)$$

where $\mathscr{D}_m$ denotes the (disentangled) normal diagrams, and $\{\boldsymbol{\Gamma}^{\aleph}\!\!-\!\!\boldsymbol{\Gamma}^{\aleph}\}$ the higher-cumulant terms. Applying the regularization, these two terms can be separated, i.e.,

$$\boxed{\mathcal{L} = \frac{1}{|C|} \int_{C[ij\ldots]} \sum_m \mathscr{D}_m^{R,\aleph} + \frac{1}{|C|} \int_{C[ij\ldots]} \{\boldsymbol{\Gamma}^{R,\aleph}\!\!-\!\!\boldsymbol{\Gamma}^{R,\aleph}\}} \quad (2.134)$$

This formula specifies the rules to construct the ("direct" and "reduced") higher-order renormalizations of the single-particle basis (i.e., in terms of functional derivatives of the first term in Eq. (2.134)) for a continuous spectrum.

### 2.3.7. Contributions from Three-Body Interactions

With three-body interactions included the interaction Hamiltonian is $\mathcal{V} = \mathcal{V}_{2B} + \mathcal{V}_{3B}$, with

$$\mathcal{V}_{3B} = \frac{1}{3!} \sum_{ijaklm} V_{3B}^{ija,klp} \, a_i^\dagger a_j^\dagger a_a^\dagger a_p a_l a_k. \quad (2.135)$$

The matrix elements $V_{3B}^{ija,klm} \equiv V_{3B}^{ija,klp} \, \delta_{k+l+p,i+j+a}$ satisfy the exchange symmetries $V_{3B}^{ija,klm} = \mathcal{Q}_L \mathcal{Q}_R V_{3B}^{ija,klm}$, with exchange operators $\mathcal{Q}_{L/R} \in \{-P_{12}, -P_{13}, -P_{23}, P_{12}P_{13}, P_{12}P_{23}\}$, and by time-reversal invariance they obey $V_{3B}^{ija,klm} = V_{3B}^{klm,ija}$. The rules for the evaluation of diagrams are similar to the two-body case. In particular, we introduce the antisymmetrized three-body potential $\bar{V}_{3B} = \mathscr{A}^{3B}_{(L)} V_{3B}$, with antisymmetrization operator $\mathscr{A}^{3B}_{(L)} = (1-P_{12})_{(L)}(1-P_{13}-P_{23})_{(L)}$.

---

[39] To be precise, it would be sufficient to average the integration order with respect to the subset of cyclic permutations that leave the diagrams invariant.

*Reducible Vertices.* There is one important simplification for three-body vertices that have at least one instantaneous contraction (which we call "reducible vertices"). For instance, the diagram (cf. Fig. 2.10) for the first-order three-body contribution, $E^{3B}_{0;1} = \langle \Phi_0 | \mathcal{V}_{3B} | \Phi_0 \rangle$ and $A^{3B}_1 = \langle \mathcal{V}_{3B} \rangle$, respectively, is given by a three-body vertex with three instantaneous contractions; we denote this as $\mathcal{N}_{3B} = 3$. The expressions for $E^{3B}_{0;1}$ and $A^{3B}_1$ are given by (using $\mathcal{M} = 1$, $\ell = 3$, $h = 3$)

$$E^{3B}_{0;1} = \frac{1}{6} \sum_{ija} \bar{V}^{ija,ija}_{3B} \, \Theta^-_i \Theta^-_j \Theta^-_a, \qquad A^{3B}_1 = \frac{1}{6} \sum_{ija} \bar{V}^{ija,ija}_{3B} \, f^-_i f^-_j f^-_a. \qquad (2.136)$$

These expressions can be written in terms of an effective in-medium two-body potential $\bar{V}_{DD2B}$ generated from the genuine three-body potential, i.e.,

$$\boxed{\bar{V}^{ij,kl}_{DD2B} = \sum_a \bar{V}^{aij,akl}_{3B} \times \begin{cases} \Theta^-_a \\ f^-_a \end{cases}} \qquad (2.137)$$

Note that the effective two-body potential $\bar{V}^{ij,kl}_{DD2B}$ is a function of the macroscopic parameters ("DD" stands for "density dependent"). In terms of $\bar{V}^{ij,kl}_{DD2B}$ the first-order contributions read

$$E^{3B}_{0;1} = \frac{1}{2} \sum_{ij} \frac{\bar{V}^{ij,ij}_{DD2B}}{3} \Theta^-_i \Theta^-_j, \qquad A^{3B}_1 = \frac{1}{2} \sum_{ij} \frac{\bar{V}^{ij,ij}_{DD2B}}{3} f^-_i f^-_j. \qquad (2.138)$$

A diagram that involves three-body vertices of the reducible kind only has the same structure as the corresponding diagram with only two-body vertices. With three-body interactions included, the contribution from all diagrams with two-body and/or reducible three-body vertices is then given by substituting in the expressions for the "pure" two-body diagrams for the matrix elements $\bar{V}^{ij,kl}_{2B}$ the quantity $\bar{V}^{ij,kl}_{2B} + \alpha \bar{V}^{ij,kl}_{DD2B}$, where $\alpha = 1/\mathcal{N}_{3B}$.[40] For instance, for each of the two second-order two-body (2B-2B) diagram there are then in total four diagrams (3B-3B, 2B-3B, 3B-2B, 2B-2B); the 3B-3B normal and the 2B-3B anomalous diagram are shown in Fig. 2.10. Except for the first-order diagram where $\mathcal{N}_{3B} = 3$, all reducible three-body vertices have either one or two instantaneous contractions, $\mathcal{N}_{3B} \in \{1, 2\}$. For skeletons only vertices with $\mathcal{N}_{3B} = 1$ are possible, since a vertex with $\mathcal{N}_{3B} = 2$ can be disconnected by cutting the two other lines.

To summarize, for the case of reducible diagrams, i.e., diagrams that involve three-body vertices of the reducible kind only, the results obtained in an analysis of "pure" two-body contributions can be immediately taken over to the case where three-body interactions are included.

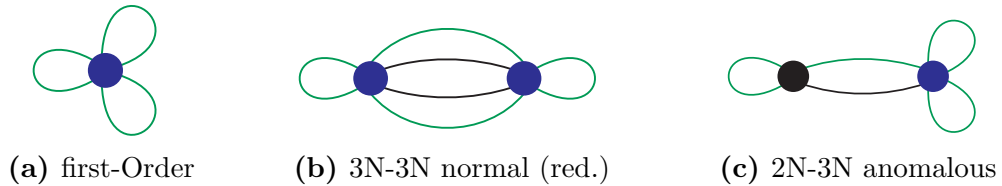

(a) first-Order    (b) 3N-3N normal (red.)    (c) 2N-3N anomalous

**Figure 2.10.:** Selected Hugenholtz diagrams with reducible three-body vertices (blue dots).

---

[40] To proof the relation $\alpha = 1/\mathcal{N}_{3B}$, consider an arbitrary two-body skeleton, and replace one vertex by an $\mathcal{N}_{3B} = 1$ vertex; with $\alpha = 1$ the multiplicity of the diagram is unchanged, which establishes $\alpha = 1/\mathcal{N}_{3B}$ for skeletons. The general relation $\alpha = 1/\mathcal{N}_{3B}$ then follows by considering the various possibilities for the linking of the contraction lines of $\mathcal{N}_{3B} = 1, 2, 3$ vertices.

***Irreducible Vertices.*** There are of course also three-body vertices without instantaneous contractions ($\mathcal{N}_{3B} = 0$). Here, no such simplification is possible; these vertices are accordingly called "irreducible". The first diagram with irreducible vertices is the irreducible 3B-3B diagram shown in Fig. 2.11. The sum of contractions for this diagram is given by

$$\sum_{\text{contractions}} = (-1)^{3+3} \mathscr{A}_L^{3B} V_{3B}^{ija,klm} \mathscr{A}_L^{3B} V_{3B}^{klm,ija}, \qquad (2.139)$$

i.e., $\mathcal{M} = 1$, $\ell = 3$ and $h = 3$. The evaluation of the expressions for the irreducible 3B-3B diagram is otherwise similar to the one for the second-order normal (2B-2B) diagram, and we obtain (using the cyclic formula)

$$A_{2,\text{normal(irr.)}}^{3B-3B,\text{cyclic}} = -\frac{1}{72} \sum_{ijaklm} \bar{V}_{3B}^{ija,klm} \bar{V}_{3B}^{klm,ija} \frac{f_i^- f_j^- f_a^- f_k^+ f_l^+ f_m^+ - f_k^- f_l^- f_m^- f_i^+ f_j^+ f_a^+}{\varepsilon_k + \varepsilon_l + \varepsilon_m - \varepsilon_i - \varepsilon_j - \varepsilon_a}. \qquad (2.140)$$

At higher orders there are also diagrams (skeleton and non-skeleton) that involve both two-body (or reducible three-body) vertices and irreducible three-body vertices.

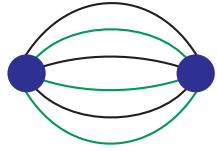

(a) 3B-3B normal (irr.)

**Figure 2.11.:** Irreducible contribution from three-body interactions at second-order.

## 2.4. Cumulants and Canonical Perturbation Theory

We now introduce a different (but equivalent) way of evaluating the grand-canonical perturbation series: the cumulant ("semi-invariant") formalism of Brout and Englert [65].[41] The cumulant formalism leads to the "disentanglement" of normal and anomalous non-skeleton contributions, in the sense that no "double-normal" terms appear and for anomalous contributions the "time-ordering" constraint associated with the standard representation in terms of articulation lines is removed.[42] In addition, the cumulant formalism provides a way to construct a "workable" perturbation series for the canonical ensemble, which reproduces the ground-state perturbation series in the zero-temperature limit (in the "isotropic case").

### 2.4.1. Cumulant Formalism

The formal expression for the perturbative contribution to the grand-canonical potential [cf. Eq. (2.33)] is given by $\Delta A = -\frac{1}{\beta} \ln[1 - \beta \sum_{n=1}^{\infty} \lambda^n A_n]$. Expanding the logarithm we find

$$\Delta A = \sum_{n=1}^{\infty} \lambda^n \sum_{k} \sum_{\{a_i\},\{b_i\}} \beta^{b_1+\ldots+b_k-1} \binom{b_1+\ldots+b_k}{b_1,\ldots,b_k} \frac{(A_{a_1})^{b_1} \cdots (A_{a_k})^{b_k}}{b_1+\ldots+b_k}\bigg|_{a_1 b_1+\ldots+a_k b_k = n}, \qquad (2.141)$$

---

[41] The paper by Brout and Englert [65] (see also Refs. [63, 65, 64, 214]) actually discusses exclusively the generalized case (correlation-bond formalism) to be discussed in Sec. 2.4.1. The "disentanglement" ("désenchevêtrement") of the grand-canonical perturbation series was first discussed by Balian, Bloch and de Dominicis [18].

## 2. Many-Body Perturbation Theory

where $A_\nu = -\frac{1}{\beta}\frac{(-1)^\nu}{\nu!}\int_0^\beta d\tau_\nu \cdots d\tau_1 \ \langle \mathcal{P}[\mathcal{V}_I(\tau_\nu)\cdots\mathcal{V}_I(\tau_1)]\rangle$. In Eq. (2.141), the quantity in large brackets is a multinomial coefficient. Note that the ensemble average contains no "linked" prescription, i.e., at this point we consider both linked and unlinked contributions to each of the $A_\nu$. In the standard approach ("contraction formalism") the unlinked terms are cancelled by the respective "product terms" (as required by the linked-cluster theorem), but we will find that certain unlinked contributions remain in cumulant formalism.

To set up the formalism, we define $\mathcal{G}_{i_1\ldots i_n}$ as the ensemble average of a fully-contracted (indicted by paired indices) sequence of creation and destruction operators, i.e.,

$$\mathcal{G}_{i_1\ldots i_n} = \langle a^\dagger_{i_1} a_{i_1} \cdots a^\dagger_{i_n} a_{i_n}\rangle, \tag{2.142}$$

where some of the index tuples may be identical. In Eq. (2.142), all contractions are of the *hole* type. For the case where there are also *particles* we introduce the notation

$$\mathcal{G}^{k_1\cdots k_m}_{i_1\cdots i_n} = \langle a^\dagger_{i_1} a_{i_1} \cdots a^\dagger_{i_n} a_{i_n} a_{k_1} a^\dagger_{k_1} \cdots a_{k_m} a^\dagger_{k_m}\rangle. \tag{2.143}$$

$\mathcal{G}^{k_1\cdots k_m}_{i_1\cdots i_n}$ can be expressed in terms of derivatives of the unperturbed partition function $\mathcal{Y}$:

$$\mathcal{G}^{k_1\cdots k_m}_{i_1\cdots i_n} = \frac{1}{\mathcal{Y}}\frac{\partial}{\partial[-\beta\varepsilon_{i_1}]}\cdots\frac{\partial}{\partial[-\beta\varepsilon_{i_n}]}\left(1-\frac{\partial}{\partial[-\beta\varepsilon_{k_1}]}\right)\cdots\left(1-\frac{\partial}{\partial[-\beta\varepsilon_{k_m}]}\right)\mathcal{Y}. \tag{2.144}$$

This shows that the upper indices can be "lowered" iteratively, i.e., $\mathcal{G}^{k_1\cdots k_m}_{i_1\cdots i_n} = \mathcal{G}^{k_1\cdots k_{m-1}}_{i_1\cdots i_n} - \mathcal{G}^{k_1\cdots k_{m-1}}_{i_1\cdots i_n k_m}$. The following general relation is readily verified:

$$\boxed{\mathcal{G}^{k_1\cdots k_m}_{i_1\cdots i_n} = \sum_{P\subset\{1,\ldots,m\}}(-1)^{|P|}\mathcal{G}_{i_1\cdots i_n\{k_\nu\}_{\nu\in P}}} \tag{2.145}$$

It is now sufficient to consider only *hole*-type $\mathcal{G}$'s. The cumulants $\mathcal{K}_{i_1\ldots i_n}$ are defined by

$$\boxed{\mathcal{K}_{i_1\ldots i_n} = \frac{\partial^n \ln\mathcal{Y}}{\partial[-\beta\varepsilon_{i_1}]\cdots\partial[-\beta\varepsilon_{i_n}]}} \tag{2.146}$$

Evaluating Eq. (2.146) iteratively determines the $\mathcal{K}$'s in terms of the $\mathcal{G}$'s:

$$\mathcal{K}_{i_1} = \mathcal{G}_{i_1}, \tag{2.147}$$
$$\mathcal{K}_{i_1 i_2} = \mathcal{G}_{i_1 i_2} - \mathcal{G}_{i_1}\mathcal{G}_{i_2}, \tag{2.148}$$
$$\mathcal{K}_{i_1 i_2 i_3} = \mathcal{G}_{i_1 i_2 i_3} - \mathcal{G}_{[i_1 i_2}\mathcal{G}_{i_3]} + 2\mathcal{G}_{i_1}\mathcal{G}_{i_2}\mathcal{G}_{i_3}, \tag{2.149}$$
$$\mathcal{K}_{i_1 i_2 i_3 i_4} = \mathcal{G}_{i_1 i_2 i_3 i_4} - \mathcal{G}_{[i_1 i_2 i_3}\mathcal{G}_{i_4]} - \mathcal{G}_{[i_1 i_2}\mathcal{G}_{i_3 i_4]} + 2\mathcal{G}_{[i_1 i_2}\mathcal{G}_{i_3}\mathcal{G}_{i_4]} - 3\mathcal{G}_{i_1}\mathcal{G}_{i_2}\mathcal{G}_{i_3}\mathcal{G}_{i_4}; \tag{2.150}$$

where the squared brackets imply all permutations among indices (modulo permutations that lead to equivalent expressions), e.g., $\mathcal{G}_{[i_1 i_2}\mathcal{G}_{i_3 i_4]} = \mathcal{G}_{i_1 i_2}\mathcal{G}_{i_3 i_4} + \mathcal{G}_{i_1 i_3}\mathcal{G}_{i_2 i_4} + \mathcal{G}_{i_1 i_4}\mathcal{G}_{i_2 i_3}$. The relations

---

[42] Both normal and anomalous non-skeletons impose a particular diagrammatic structure, i.e., a particular "time-ordering" of vertices (such that the articulation lines are either "normal" or "anomalous"). Note however that in the standard "contraction" formalism, anomalous contributions in general arise not only anomalous diagrams but also from normal non-skeletons, cf. Eqs. (2.117) and (2.118). In contrast, in the cumulant formalism, anomalous contributions arise only from higher cumulants.

between the $\mathcal{K}$'s and the $\mathcal{G}$'s can be inverted iteratively; the first few relations are

$$\mathcal{G}_{i_1} = \mathcal{K}_{i_1}, \tag{2.151}$$
$$\mathcal{G}_{i_1 i_2} = \mathcal{K}_{i_1} \mathcal{K}_{i_2} + \mathcal{K}_{i_1 i_2}, \tag{2.152}$$
$$\mathcal{G}_{i_1 i_2 i_3} = \mathcal{K}_{i_1} \mathcal{K}_{i_2} \mathcal{K}_{i_3} + \mathcal{K}_{[i_1 i_2} \mathcal{K}_{i_3]} + \mathcal{K}_{i_1 i_2 i_3}, \tag{2.153}$$
$$\mathcal{G}_{i_1 i_2 i_3 i_4} = \mathcal{K}_{i_1} \mathcal{K}_{i_2} \mathcal{K}_{i_3} \mathcal{K}_{i_4} + \mathcal{K}_{[i_1 i_2} \mathcal{K}_{i_3} \mathcal{K}_{i_4]} + \mathcal{K}_{[i_1 i_2} \mathcal{K}_{i_3 i_4]} + \mathcal{K}_{[i_1 i_2 i_3} \mathcal{K}_{i_4]} + \mathcal{K}_{i_1 i_2 i_3 i_4}; \tag{2.154}$$

and in general [142]

$$\boxed{\mathcal{G}_{i_1 \cdots i_n} = \sum_{\substack{P \in \text{ partitions} \\ \text{of } \{1,\ldots,n\}}} \prod_{I \in P} \mathcal{K}_{\{i_\nu\}_{\nu \in I}}} \tag{2.155}$$

Eqs. (2.145), (2.146) and (2.155) provide the basis for the evaluation of (grand-canonical) ensemble averages of creation and destruction operators in terms of cumulants.

## 2.4.2. Cumulant Representation of Linked Clusters

In the case of the grand-canonical ensemble the higher cumulants vanish for the case where not all indices are equal, i.e., $\mathcal{G}_{i_1 \cdots i_n} = \prod_{\nu=1}^n \mathcal{K}_{i_\nu}$ for $i_1 \neq i_2 \neq \ldots \neq i_n$. For skeletons (where all lines carry different indices) this implies that, concerning the evaluation in terms of cumulants, the obtained expression are the same as in the standard approach. This is however not the case for diagrams with repeated indices, i.e., normal and anomalous non-skeletons.

*Normal Non-Skeletons.* Such diagrams correspond to $\mathcal{G}$'s where repeated indices appear in the subscripts or in the superscripts. For repeated indices also higher cumulants contribute. Without loss of generality we consider the case where only one *hole* line carries insertions, i.e., we consider $\mathcal{G}^{k_1 \cdots k_m}_{i_1 \cdots i_n a \cdots a}$. Using the relations $\partial f_i^- / \partial(-\beta \varepsilon_i) = f_i^- f_i^+$ and $f_i^+ = 1 - f_i^-$ we find

$$\mathcal{G}^{k_1 \cdots k_m}_{i_1 \cdots i_n aa} \sim \mathcal{K}_a \mathcal{K}_a + \mathcal{K}_{aa} = f_a^- f_a^- + f_a^- f_a^+ = f_a^-, \tag{2.156}$$

$$\mathcal{G}^{k_1 \cdots k_m}_{i_1 \cdots i_n aaa} \sim \mathcal{K}_a \mathcal{K}_a \mathcal{K}_a + \binom{3}{1} \mathcal{K}_{aa} \mathcal{K}_a + \mathcal{K}_{aaa} = f_a^- f_a^- f_a^- + 3 f_a^- f_a^+ f_a^- + f_a^- f_a^+ (f_a^+ - f_a^-) = f_a^-. \tag{2.157}$$

One sees that, [in contrast to the situation in the contraction formalism, cf. Eq. (2.117)], the articulation lines carry only a single Fermi-Dirac distribution, i.e.,

$$\boxed{\mathcal{G}^{k_1 \cdots k_m}_{i_1 \cdots i_n a \cdots a} = \mathcal{G}^{k_1 \cdots k_m}_{i_1 \cdots i_n a}} \tag{2.158}$$

From Eq. (2.145) it then follows that also $\mathcal{G}^{k_1 \cdots k_m a \cdots a}_{i_1 \cdots i_n} = \mathcal{G}^{k_1 \cdots k_m a}_{i_1 \cdots i_n}$. Note that this implies that the resummation of normal "one-loop" insertions in terms of a geometric series has the same form as in the ground-state formalism *if the reduced formula is used* (but not if the direct formula is used, see below).

We will now rewrite Eq. (2.158) in a more explicit combinatorial form. For this, we introduce the notation $f \equiv f_a^-$ as well as $\mathcal{G}_n \equiv \mathcal{G}_{i_1 \cdots i_n}$ and $\mathcal{K}_n \equiv \mathcal{K}_{i_1 \cdots i_n}$, where $i_\nu = a \, \forall \nu$. It is easy to see that $\mathcal{K}_n$ can be written as

$$\mathcal{K}_n = \sum_{k=1}^n x_{k,n} f^k, \tag{2.159}$$

where the numbers $x_{k,n}$ satisfy the recursion relation

$$x_{k,n} = k\, x_{k,n-1} - (k-1)\, x_{k-1,n-1}, \tag{2.160}$$

with starting value $x_{1,1} = 1$. Note that $x_{1,n} = 1$ and $x_{k,n} = (-1)^{k-1} x_{k,n}$. It is also easy to see that $\mathcal{G}_n$ can be written as

$$\mathcal{G}_n = \sum_{\substack{\{m_\ell\} \\ \sum_{\ell=1}^L m_\ell = n \\ m_1 \geq m_2 \geq \ldots \geq m_L}} N_{\{m_\ell\}} \prod_{\ell=1}^L \mathcal{K}_{m_\ell}, \tag{2.161}$$

where the numbers $N_{\{m_\ell\}}$ are given by

$$N_{\{m_\ell\}} = \frac{\binom{n}{m_1}\binom{n-m_1}{m_2}\cdots\binom{n-m_1-\ldots-m_{L-2}}{m_{L-1}}}{\Theta_1! \cdots \Theta_M!}, \tag{2.162}$$

where $\Theta_1, \ldots, \Theta_M$ are the numbers of identical $m_\ell$'s in the set $\{m_\ell\}$.[43] Eq. (2.158) is then equivalent to

$$\boxed{\forall n \in \mathbb{N}: \qquad \mathcal{N}_{k,n} := \sum_{\substack{\{m_\ell\} \\ \sum_{\ell=1}^L m_\ell = n \\ m_1 \geq m_2 \geq \ldots \geq m_L}} N_{\{m_\ell\}} \prod_{\ell=1}^L x_{k_\ell, m_\ell} = 0, \quad \forall\; k = \prod_{\ell=1}^L k_\ell \in [2, n]} \tag{2.163}$$

together with $\mathcal{N}_{1,n} = 1$, which is however trivial since the only contribution to $\mathcal{N}_{1,n}$ is from $\mathcal{K}_n$, which has the term $x_{1,n} f = f$. For instance, for $n = 4$ it is[44]

- $\mathcal{N}_{2,4} = x_{2,4} + N_{\{3,1\}} x_{1,3} + N_{\{2,2\}} x_{1,2} = -7 + 4 + 3 = 0$.

- $\mathcal{N}_{3,4} = x_{3,4} + N_{\{3,1\}} x_{2,3} + N_{\{2,2\}} (x_{2,2} + x_{2,2}) + N_{\{2,1,1\}} = 12 - 12 - 6 + 6 = 0$.

- $\mathcal{N}_{4,4} = x_{4,4} + N_{\{3,1\}} x_{3,3} + N_{\{2,2\}} x_{2,2} x_{2,2} + N_{\{2,1,1\}} x_{2,2} + 1 = -6 + 8 + 3 - 6 + 1 = 0$.

For $n = 5$ it is

- $\mathcal{N}_{2,5} = x_{2,5} + N_{\{4,1\}} + N_{\{3,2\}} = -15 + 5 + 10 = 0$.

- $\mathcal{N}_{3,5} = x_{3,5} + N_{\{4,1\}} x_{2,4} + N_{\{3,2\}} (x_{2,3} + x_{2,2}) + N_{\{3,1,1\}} + N_{\{2,2,2\}} = 50 - 35 - 40 + 15 + 10 = 0$.

- $\mathcal{N}_{4,5} = x_{4,5} + N_{\{4,1\}} x_{3,4} + N_{\{3,2\}} (x_{3,3} + x_{2,3} x_{2,2}) + N_{\{3,1,1\}} x_{2,3} + N_{\{2,2,1\}} (x_{2,2} + x_{2,2}) = -60 + 60 + 50 - 30 - 30 + 10 = 0$.

- $\mathcal{N}_{5,5} = x_{5,5} + N_{\{4,1\}} x_{5,4} + N_{\{3,2\}} (x_{3,3} x_{2,2}) + N_{\{3,1,1\}} x_{3,3} + N_{\{2,2,1\}} x_{2,2} x_{2,2} + 1 = 24 - 30 - 20 + 15 + 20 - 10 + 1 = 0$.

---

[43] For instance, $\mathcal{G}_{14}$ has a contribution $N_{\{4,4,2,2,2\}} \mathcal{K}_4 \mathcal{K}_4 \mathcal{K}_2 \mathcal{K}_2 \mathcal{K}_2$, where $N_{\{4,4,2,2,2\}} = \frac{\binom{14}{4}\binom{10}{4}\binom{6}{2}\binom{4}{2}}{2!\,3!}$.

[44] We omit the trivial factors $N_{\{n\}} = 1$, $N_{\{1,\ldots,1\}} = 1$, and $x_{1,p} = 1$; e.g., with the trivial factors included, the expression for $\mathcal{N}_{3,4}$ reads $\mathcal{N}_{3,4} = N_{\{4\}} x_{3,4} + N_{\{3,1\}} x_{2,3} x_{1,1} + N_{\{2,2\}} (x_{2,2} x_{1,2} + x_{1,2} x_{2,2}) + N_{\{2,1,1\}} x_{1,2} x_{1,1} x_{1,1}$.

This will be taken as sufficient to "proof" Eq. (2.158) and Eq. (2.163), respectively.[45]

***Anomalous Non-Skeletons.*** The expressions that arise from anomalous diagrams correspond to $\mathcal{G}$'s with a certain number of identical upper and lower indices, i.e., $\mathcal{G}^{k_1\cdots k_m a\cdots a}_{i_1\cdots i_n a\cdots a}$. It is sufficient to consider the term $\mathcal{G}^{a\cdots a}_{i_1\cdots i_n a\cdots a}$, which by Eq. (2.145), using the previous relation $\mathcal{G}_{i_1\cdots i_n a\cdots a} = \mathcal{G}_{i_1\cdots i_n a}$ and denoting by $l$ is the number of $a$'s that appear as upper indices, is given by

$$\boxed{\mathcal{G}^{a\cdots a}_{i_1\cdots i_n a\cdots a} = \sum_{P\subset\{1,\ldots,l\}} (-1)^{|P|} \mathcal{G}_{i_1\cdots i_n a\cdots a\,\{a_\nu\}_{\nu\in P}} = \sum_{P\subset\{1,\ldots,l\}} (-1)^{|P|} \mathcal{G}_{i_1\cdots i_n a} = 0} \quad (2.164)$$

i.e., if evaluated in terms of cumulants, the contributions associated with anomalous *diagrams* are zero. In the next section we will find that the anomalous contributions arise instead from unlinked diagrams connected via higher cumulants.

## 2.4.3. Cumulant Representation of Unlinked Clusters

We now evaluate the remaining contributions in Eq. (2.141), i.e., the contributions from unlinked clusters and products of (linked or unlinked) clusters. In the standard approach these contributions cancel each other (by virtue of the linked-cluster theorem), but we will find that certain unlinked terms survive in the cumulant formalism, and that these terms generate the ("disentangled") anomalous contributions which are missing so far.

***Simply-Connected Diagrams.*** Each of the unlinked or product contributions contains terms that diverge superlinearly ($\propto \Omega^\nu$, $\nu \geq 2$) in the thermodynamic limit. These unphysical terms must still cancel each other in the cumulant formalism. To identify the cancellation of the superlinearly divergent terms, we write the expression for a general unlinked contribution to $\mathcal{G}^{klm\cdots}_{ija,\cdots}$ as

$$A_{n,\text{unlinked}} = \left[\frac{1}{\beta}\frac{1}{\alpha_1!\cdots\alpha_\nu!}(\Gamma_{n_1})^{\alpha_1}\cdots(\Gamma_{n_\nu})^{\alpha_\nu}\right]_{n_1^{\alpha_1}+\ldots+n_\nu^{\alpha_\nu}=n}. \quad (2.165)$$

The corresponding product terms are given by

$$\beta A_{n_1} A_{n-n_1,\text{unlinked}} = \left[\frac{1}{\beta}\frac{1}{(\alpha_1-1)!\cdots\alpha_\nu!}(\Gamma_{n_1})^{\alpha_1}\cdots(\Gamma_{n_\nu})^{\alpha_\nu}\right]_{n_1^{\alpha_1}+\ldots+n_\nu^{\alpha_m}=n},$$

$$\vdots$$

$$\beta^n (A_{n_1})^{\alpha_1}\cdots(A_{n_m})^{\alpha_\nu} = \left[\frac{1}{\beta}(\Gamma_{n_1})^{\alpha_1}\cdots(\Gamma_{n_\nu})^{\alpha_\nu}\right]_{n_1^{\alpha_1}+\ldots+n_\nu^{\alpha_\nu}=n}. \quad (2.166)$$

The linked-cluster theorem implies the cancellation of the unlinked and product terms for contributions to $\mathcal{G}^{kl\cdots}_{ij,\cdots}$ from first ("single-index") cumulants, i.e., the contributions from

$$\mathcal{G}^{klm\cdots}_{ija,\cdots} \sim \mathcal{K}_i \mathcal{K}_j \mathcal{K}_a \cdots \bar{\mathcal{K}}_k \bar{\mathcal{K}}_l \bar{\mathcal{K}}_m \cdots, \quad (2.167)$$

---

[45] The correctness of Eq. (2.158) and Eq. (2.163), respectively, to all orders can be inferred from Luttinger's alternative derivation [277] of the mean-occupation number formalism (the fully-renormalized $\aleph$ = "direct" case, cf. Sec. 2.5.1).

where we have used the notation $\bar{\mathcal{K}}_k = 1 - \mathcal{K}_k$. The next contributions are the ones with one double-index cumulant, i.e.,

$$\mathcal{G}^{klm...}_{ija,...} \sim \delta_{ij}\mathcal{K}_{ii}\mathcal{K}_a \cdots \bar{\mathcal{K}}_k\bar{\mathcal{K}}_l\bar{\mathcal{K}}_m \cdots + \{\text{similar terms}\}. \quad (2.168)$$

Note that whereas for a selection of $\alpha$ unlinked diagrams the terms with first cumulants diverge as $\Omega^\alpha$, because of the delta function in Eq. (2.168) the terms with double-index cumulants diverge as $\Omega^{\alpha-1}$, and similarly for other higher cumulants, e.g.,

$$\mathcal{G}^{klm...}_{ija,...} \sim \delta_{ij}\delta_{ia}\mathcal{K}_{iii} \cdots \bar{\mathcal{K}}_k\bar{\mathcal{K}}_l\bar{\mathcal{K}}_m \cdots + \{\text{similar terms}\}. \quad (2.169)$$

If (for a given higher cumulant) the paired indices belong to the same subdiagram, the linked-cluster theorem again applies. If the paired indices belong to different subdiagrams, then only those product terms contribute where the linked subdiagrams do not belong to different factors but to the same factor $A^{\alpha_\nu}_{n_\nu,\text{unlinked}}$. However, by the linked-cluster theorem all of these contributions cancel each other, since the subdiagrams with paired indices take the role of a linked contribution, and the same combinatorics applies. Hence, the first nonvanishing contribution is the one of order $\Omega$ where each subdiagram is connected to another one by a single pairing of indices, because in that case only the term $A_{n,\text{unlinked}}$ remains, and therefore no cancellation takes place. The remaining contributions are then all possible unlinked diagrams where the diagrams are simply-connected in terms of index pairings. The index pairings can be represented by "insertion-lines" ("higher-cumulant connections"); the diagrams contributing at third order are shown in Figs. 2.12 and 2.13, where the insertion-lines are colored in dark red.

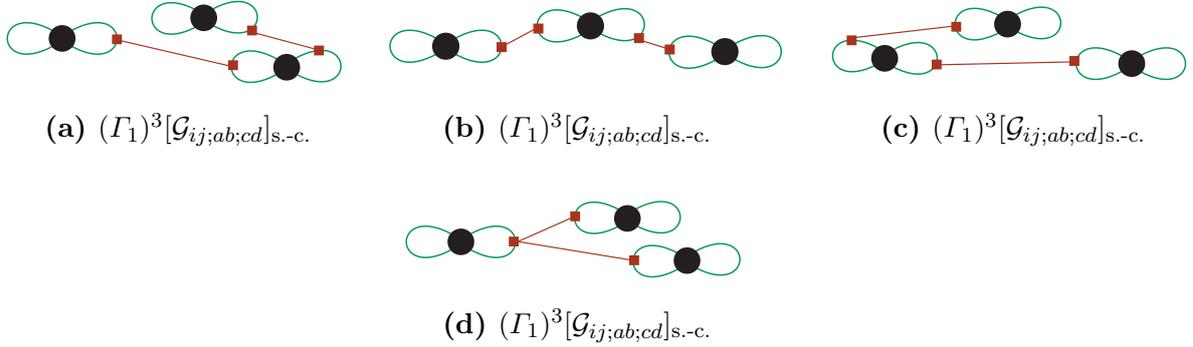

**(a)** $(\Gamma_1)^3[\mathcal{G}_{ij;ab;cd}]_{\text{s.-c.}}$   **(b)** $(\Gamma_1)^3[\mathcal{G}_{ij;ab;cd}]_{\text{s.-c.}}$   **(c)** $(\Gamma_1)^3[\mathcal{G}_{ij;ab;cd}]_{\text{s.-c.}}$

**(d)** $(\Gamma_1)^3[\mathcal{G}_{ij;ab;cd}]_{\text{s.-c.}}$

**Figure 2.12.:** Simply-connected diagrams involving three first-order diagrams, $(\Gamma_1)^3$. Diagrams (a,b,c) correspond to the three terms with two double-index cumulants, diagram (d) to the triple-index cumulant; each diagram represents $2^3$ different (equivalent) index pairings. The four diagrams are in one-to-one correspondence with the two-loop (a,b,c) and three-loop (d) diagrams of Fig. 2.7.

For a given simply-connected contribution with diagrams $(\Gamma_{n_1})^{\{k_{n_1}\}}_{\{i_{n_1}\}} \cdots (\Gamma_{n_\nu})^{\{k_{n_\nu}\}}_{\{i_{n_\nu}\}}$ (where $\{i_n\}$ and $\{k_n\}$ are the indices of the *hole* and *particle* lines for each diagram) we introduce the notation

$$\left[\mathcal{G}^{\{k_{n_1}\};...;\{k_{n_\nu}\}}_{\{i_{n_1}\};...;\{i_{n_\nu}\}}\right]_{\text{s.-c.}}, \quad (2.170)$$

where "s.-c." stands for "simply-connected". The simply-connected diagrams up to third order are given by

$$(\Gamma_1)_{ij}(\Gamma_1)_{ab}: \quad \left[\mathcal{G}_{ij;ab}\right]_{\text{s.c.}} = 2^2 \times \delta_{ia}\mathcal{K}_{ii}\mathcal{K}_j\mathcal{K}_b, \quad (2.171)$$

## 2. Many-Body Perturbation Theory

$$(\Gamma_{2,\text{normal}})_{ij}^{kl}(\Gamma_1)_{ab}: \quad \left[\mathcal{G}_{ij;ab}^{kl}\right]_{\text{s.c.}} = 2^2 \times \delta_{ia}\mathcal{K}_{ii}\mathcal{K}_j\mathcal{K}_b\bar{\mathcal{K}}_l\bar{\mathcal{K}}_k - 2^2 \times \delta_{ka}\mathcal{K}_{kk}\mathcal{K}_i\mathcal{K}_j\mathcal{K}_b\bar{\mathcal{K}}_l, \quad (2.172)$$

$$(\Gamma_1)_{ij}(\Gamma_1)_{ab}(\Gamma_1)_{cd}: \quad \left[\mathcal{G}_{ij;ab;cd}\right]_{\text{s.c.}} = 3 \times 2^3 \times \delta_{ia}\mathcal{K}_{ii}\mathcal{K}_{jc}\mathcal{K}_b\mathcal{K}_c + 2^3 \times \delta_{ia}\delta_{ic}\mathcal{K}_{iii}\mathcal{K}_j\mathcal{K}_b\mathcal{K}_d, \quad (2.173)$$

where the prefactors correspond to the number of equivalent pairings of indices for each contribution. The simply-connected diagrams for $(\Gamma_1)_{ij}(\Gamma_1)_{ab}(\Gamma_1)_{cd}$ are shown in Fig. 2.12. It is readily verified that the contributions from $(\Gamma_1)_{ij}(\Gamma_1)_{ab}$ and $(\Gamma_1)_{ij}(\Gamma_1)_{ab}(\Gamma_1)_{cd}$ reproduce the expressions for the second-order anomalous and the third-order two- and three-loop anomalous diagrams (cf. Sec. 2.3.4), respectively. The three diagrams for $(\Gamma_{2,\text{normal}})_{ij}^{kl}(\Gamma_1)_{ab}$ are shown in Fig. 2.13.

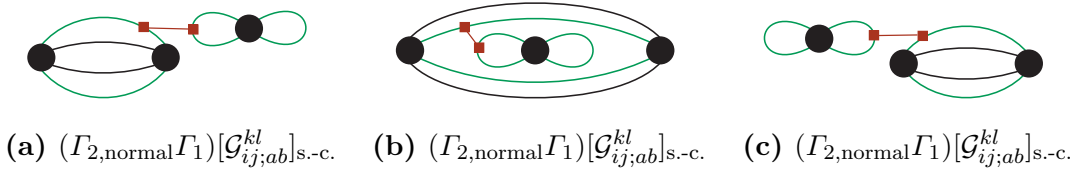

**(a)** $(\Gamma_{2,\text{normal}}\Gamma_1)[\mathcal{G}_{ij;ab}^{kl}]_{\text{s.-c.}}$    **(b)** $(\Gamma_{2,\text{normal}}\Gamma_1)[\mathcal{G}_{ij;ab}^{kl}]_{\text{s.-c.}}$    **(c)** $(\Gamma_{2,\text{normal}}\Gamma_1)[\mathcal{G}_{ij;ab}^{kl}]_{\text{s.-c.}}$

**Figure 2.13.:** Simply-connected diagrams for $\Gamma_{2,\text{normal}}\Gamma_1$ (only the cases where the insertion-line is on the upper *hole* line of $\Gamma_{2,\text{normal}}$ are shown). Diagrams (a) and (c) can be seen as the analogs of the third-order "one-loop" diagrams (e) and (f) of Fig. 2.8. Diagram (b) has no counterpart in the contraction formalism.

***"Direct" and "Reduced" Factorization.*** In the standard representation of "anomalous contributions" (cf. footnote[42]) in terms of non-skeletons with "anomalous" articulation lines, the particular "time-ordering" of diagram (b) of Fig. (2.13) does not occur. This lack of any "time-ordering" constraint is the crucial feature that leads to the "direct" factorization property [Eq. (2.84)], see Sec. 2.5.1 for the proof. The "reduced" factorization property [Eq. (2.88)] follows from the fact that for a simply-connected diagram with $N$ subclusters, there is a product of derivatives of Fermi-Dirac distributions

$$\prod_\nu \frac{\partial^{n_\nu} f_{i_\nu}^-}{\partial \mu^{n_\nu}}, \quad (2.174)$$

where $\sum_\nu n_\nu = N - 1$. To generate an anomalous term of order $N - 1$, this requires a factor $\beta^K$ with $K = N - 1$, while for $K < N - 1$ a "pseudo-anomalous" term results. If the subclusters are not overlapping in time, there are $N - 1$ "identically" vanishing energy denominators. The $z = 0$ residue in the "reduced" case is then of order $N - 1$, and the contribution where the $N - 1$ derivatives from the residue act on the exponential $e^{-\beta z}$ gives a factor $\beta^{N-1}$ multiplied with the "reduced" expressions for the subclusters, i.e., the factorized part of Eq. (2.88). The other terms from the $z = 0$ residue as well as the contributions with partially overlapping subclusters are all "pseudo-anomalous", and comprise the "$\mathcal{R}$" term in Eq. (2.88). We now show that $\mathcal{R} = 0$ for selected contributions. For the "one-loop" case (which is discussed below) $\mathcal{R} = 0$ follows from the resummability of normal and anomalous "one-loop" insertions in the "direct" case. In the "reduced" case, the normal "one-loop" insertions are resummable in terms of a geometric series, and the factorized "reduced" term is also resummable. In the case where there are several higher-order diagrams the cancellation is more involved. For instance, for $\Gamma_{2,\text{normal}}\Gamma_{2,\text{normal}}$ there are six different time-orderings. Denoting by $\mathcal{E}_{1,2}$ the energy denominators associated with the two diagrams, the residue sum for the two non-overlapping time-orderings (where $\mathscr{O} = 2$) is given by

$$2 \times \frac{1}{\mathscr{O}} \operatorname*{Res}_{z=0} \frac{e^{-\beta z}}{z(-z)} \frac{1}{(\mathscr{E}_1 - z)(\mathscr{E}_2 - z)} = \frac{\beta}{\mathscr{E}_1 \mathscr{E}_2} - \frac{1}{\mathscr{E}_1^2 \mathscr{E}_2} - \frac{1}{\mathscr{E}_1 \mathscr{E}_2^2}, \tag{2.175}$$

The $\mathcal{R}$ term is comprised of the second and third term in the above equation together with the contributions from the four overlapping time-orderings, i.e.,

$$\mathcal{R}[\boldsymbol{\Gamma}_{2,\text{normal}} \boldsymbol{\Gamma}_{2,\text{normal}}] \sim -\frac{1}{\mathscr{E}_1^2 \mathscr{E}_2} - \frac{1}{\mathscr{E}_1 \mathscr{E}_2^2} + \frac{2}{\mathscr{E}_1 (\mathscr{E}_1 + \mathscr{E}_2) \mathscr{E}_2} + \frac{1}{\mathscr{E}_1^2 (\mathscr{E}_1 + \mathscr{E}_2)} + \frac{1}{(\mathscr{E}_1 + \mathscr{E}_2) \mathscr{E}_2^2} = 0. \tag{2.176}$$

For the case of a third- and a second-order diagram, $\boldsymbol{\Gamma}_3 \boldsymbol{\Gamma}_{2,\text{normal}}$, there are three energy denominators, denoted by $\mathscr{E}_{1,2}$ (corresponding to $\boldsymbol{\Gamma}_3$, which can be either a ladder, a ring, or a normal "one-loop" diagram) and $\mathscr{E}_3$ (corresponding to $\boldsymbol{\Gamma}_{2,\text{normal}}$), and ten different time-orderings. In that case, the $\mathcal{R}$ term is given by

$$\begin{aligned}\mathcal{R}[\boldsymbol{\Gamma}_3 \boldsymbol{\Gamma}_{2,\text{normal}}] \sim &- \frac{1}{\mathscr{E}_1^2 \mathscr{E}_2 \mathscr{E}_3} - \frac{1}{\mathscr{E}_1 \mathscr{E}_2^2 \mathscr{E}_3} - \frac{1}{\mathscr{E}_1 \mathscr{E}_2 \mathscr{E}_3^2} \\ &+ \frac{1}{\mathscr{E}_1 \mathscr{E}_2 (\mathscr{E}_2 + \mathscr{E}_3) \mathscr{E}_3} + \frac{1}{\mathscr{E}_1 \mathscr{E}_2^2 (\mathscr{E}_2 + \mathscr{E}_3)} + \frac{1}{\mathscr{E}_1 \mathscr{E}_2 (\mathscr{E}_1 + \mathscr{E}_3) \mathscr{E}_3} + \frac{1}{\mathscr{E}_1^2 \mathscr{E}_2 (\mathscr{E}_1 + \mathscr{E}_3)} \\ &+ \frac{1}{(\mathscr{E}_1 + \mathscr{E}_3)(\mathscr{E}_2 + \mathscr{E}_3)} \left[ \frac{1}{\mathscr{E}_1 \mathscr{E}_2} + \frac{1}{\mathscr{E}_1 \mathscr{E}_3} + \frac{1}{\mathscr{E}_2 \mathscr{E}_3} + \frac{1}{\mathscr{E}_3^2} \right] \\ &= 0. \end{aligned} \tag{2.177}$$

Based on these findings we assume that $\mathcal{R} = 0$ to all orders. However, a general proof that $\mathcal{R} = 0$ should be attempted in future research (see footnote [20]).

***"One-Loop" Insertions.*** For higher-cumulant diagrams without higher-order parts such as those of Fig. 2.12 it does not matter which formula ("direct", "cyclic", "reduced") is used; the corresponding anomalous Hugenholtz diagrams transform into each other under cyclic vertex permutations, so the three formulas are equivalent for the time-ordered sum of the various higher-cumulant diagrams. This is *not* the case for (equal-index) higher-cumulant diagrams with higher-order parts, e.g., the $\boldsymbol{\Gamma}_{2,\text{normal}} \boldsymbol{\Gamma}_1$ diagrams of Fig. 2.12. The time-ordered sum of all $\boldsymbol{\Gamma}_{2,\text{normal}} \boldsymbol{\Gamma}_1$ diagrams does contain all their cyclic permutations, but this argument does not work here: in the contraction formalism, the contributions from these diagrams correspond to the ones from anomalous "one-loop" diagrams, which transform into normal "one-loop" diagrams under cyclic vertex permutations. In other terms, the correspondence between cyclic vertex permutations and cyclic permutations of interaction operators in the ensemble average is *not valid for equal-index higher-cumulant terms*. The absence of this correspondence is of course due to the evaluation of ensemble averages via cumulants.

It will now be instructive to evaluate the $\boldsymbol{\Gamma}_{2,\text{normal}} \boldsymbol{\Gamma}_1$ contribution as well as the normal "one-loop" contribution $\mathscr{D}_{3,\text{one-loop}}$ explicitly using both the reduced and the cyclic formula. In the "reduced" case, the sum of the residues for the non-overlapping (a,c) and overlapping (b) time-orderings are given by

$$2 \times \frac{1}{\mathscr{O}} \operatorname*{Res}_{z=0} \frac{e^{-\beta z}}{z} \frac{1}{(\mathscr{E} - z)(-z)} + \operatorname*{Res}_{z=0} \frac{e^{-\beta z}}{z} \frac{1}{(\mathscr{E} - z)^2} = \frac{\beta}{\mathscr{E}} - \frac{1}{\mathscr{E}^2} + \frac{1}{\mathscr{E}^2} = \frac{\beta}{\mathscr{E}}, \tag{2.178}$$

where we have used $\mathscr{O} = 2$ for the non-overlapping case. Eq. (2.178) reproduces the anomalous part of Eq. (2.119). The normal part is (of course) reproduced by $\mathscr{D}_{3,\text{one-loop}}^{\text{reduced}}$.

If the direct formula is used, the residue sum is given by

$$-\frac{1}{\beta}\left[2\times\underset{z=0}{\text{Res}}\frac{e^{-\beta z}}{z^2}\frac{1}{(\mathscr{E}-z)(-z)} + 2\times\underset{z=\mathscr{E}}{\text{Res}}\frac{e^{-\beta z}}{z^2}\frac{1}{(\mathscr{E}-z)(-z)} + \underset{z=0}{\text{Res}}\frac{e^{-\beta z}}{z^2}\frac{1}{(\mathscr{E}-z)^2} + \underset{z=\mathscr{E}}{\text{Res}}\frac{e^{-\beta z}}{z^2}\frac{1}{(\mathscr{E}-z)^2}\right]$$
$$= \frac{\beta}{\mathscr{E}} - \frac{1}{\mathscr{E}^2} + \frac{e^{-\beta\mathscr{E}}}{\mathscr{E}^2} + \ldots, \tag{2.179}$$

where the ellipses denote terms that vanish by relabeling indices. The first-term $\sim \beta$ leads to the anomalous part and the second term to the normal part of Eq. (2.119), as follows from

$$\sum_{ijkl}\left[-\frac{1}{\mathscr{E}^2} + \frac{e^{-\beta\mathscr{E}}}{\mathscr{E}^2}\right]f_i^- f_j^- f_k^+ f_l^+ \left(S_{1;i}f_i^+ + S_{1;j}f_j^+ - S_{1;k}f_k^- - S_{1;l}f_l^-\right)$$
$$= \sum_{ijkl}\frac{f_i^- f_j^- f_k^+ f_l^+}{\mathscr{E}^2}\left(S_{1;i} + S_{1;j} - S_{1;k} - S_{1;l}\right). \tag{2.180}$$

This implies that $\mathscr{D}_{3,\text{one-loop}}^{\text{direct}} = 0$, which can easily be seen explicity. The "direct" expression for the sum of all normal "one-loop" insertions can in fact be derived explicitly by noting that the time-dependent (time-ordered) "direct" expression for (e.g.,) the second-order normal diagram with $K = 0, \ldots, \infty$ normal "one-loop" insertions on a given line has the form

$$\sum_{K=0}^{\infty}\int_{\beta>\tau_{K+1}>\ldots>\tau_0>0}d\tau_{K+1}\cdots d\tau_0\,e^{-\mathscr{E}(\tau_{K+1}-\tau_0)}(-S)^K = \int_{\beta>\tau_{K+1}>\tau_0>0}d\tau_{K+1}d\tau_0\,e^{-(\mathscr{E}+S)(\tau_{K+1}-\tau_0)}, \tag{2.181}$$

where $S$ is shorthand for the first-order perturbative self-energy correction to the *particle* and *hole* energies. The remaining time integrals in the above equation are equivalent to the "direct" residue sum for the second-order normal diagram, leading to

$$\frac{\beta}{\mathscr{E}_S} - \frac{1}{\mathscr{E}_S^2} + \frac{e^{-\beta\mathscr{E}_S}}{\mathscr{E}_S^2} = \frac{\beta}{\mathscr{E}}\left(1 - \frac{S}{\mathscr{E}} + \frac{S^2}{\mathscr{E}^2} + \frac{S^3}{\mathscr{E}^3} + \ldots\right) - \frac{1}{\mathscr{E}^2}\left(1 - 2\frac{S}{\mathscr{E}} + 3\frac{S^2}{\mathscr{E}^2} + 4\frac{S^3}{\mathscr{E}^3} + \ldots\right)$$
$$+ \frac{e^{-\beta\mathscr{E}}}{\mathscr{E}^2}\left(1 - 2\frac{S}{\mathscr{E}} + 3\frac{S^2}{\mathscr{E}^2} + 4\frac{S^3}{\mathscr{E}^3} + \ldots\right)\left(1 - \beta S + \frac{\beta^2 S^2}{2!} - \frac{\beta^3 S^3}{3!} + \ldots\right). \tag{2.182}$$

Adding a factor $f_i^- f_j^- f_k^+ f_l^+$ and summing over indices, in the expanded case (right side of the above equation) the third-order terms $\sim S$ cancel each other. The terms that diverge as $T \to 0$ would be cancelled by the corresponding terms from diagrams with both normal and anomalous "one-loop" insertions. Correspondingly, the second and third term in the resummed version (left side) cancel if the $f_i^- f_j^- f_k^+ f_l^+$ factor is renormalized by resumming all anomalous "one-loop" insertions (in that case the self-energy becomes the self-consistent one, $S \to X$). However, terms that diverge as $T \to 0$ remain if higher-cumulant contributions associated with (certain) non-skeleton self-energies beyond sixth order are "resummed" and $\aleph =$ "direct".

### 2.4.4. Correlation-Bond Formalism

The cumulant formalism can be adapted to construct a "workable" perturbation series for the canonical ensemble.[46] We start with the standard finite-temperature perturbation series for the

## 2. Many-Body Perturbation Theory

interaction contribution to the free energy $\Delta F = F - \mathcal{F}$ (cf. Sec. 2.2), which can be written as

$$\Delta F = \sum_{n=1}^{\infty} \lambda^n \sum_k \sum_{\{a_i\},\{b_i\}} \beta^{b_1+\ldots+b_k-1} \binom{b_1+\ldots+b_k}{b_1,\ldots,b_k} \frac{(F_{a_1})^{b_1}\cdots(F_{a_k})^{b_k}}{b_1+\ldots+b_k}\bigg|_{a_1 b_1+\ldots+a_k b_k=n}, \quad (2.183)$$

where $F_\nu = -\frac{1}{\beta}\frac{(-1)^\nu}{\nu!}\int_0^\beta d\tau_\nu\cdots d\tau_1\ \langle\mathcal{P}[\mathcal{V}_I(\tau_\nu)\cdots\mathcal{V}_I(\tau_1)]\rangle_N$, with $\langle\ldots\rangle_N$ denoting the (unperturbed) canonical ensemble average involving states with $\langle\Phi_p|\mathcal{N}|\Phi_p\rangle = N$ (where $N$ is fixed). The crucial idea leading to the "proper" canonical perturbation series $F(T,\tilde{\mu},\Omega)$ is to evaluate Eq. (2.183) not directly, but in terms of a Legendre transform with respect to the (logarithm of) the unperturbed grand-canonical partition function, i.e., all ensemble averages are taken with respect to an unperturbed grand-canonical ensemble. Crudely speaking, this method effectively "shifts" the constraint $\langle\Phi_p|\mathcal{N}|\Phi_p\rangle = N$ to the level of diagrams, resulting in new contributions from (simply-connected) unlinked diagrams; we refer to these contributions as "correlation bonds" (slightly modifying the use of this term compared to Ref. [65]).

The canonical-ensemble average of a fully-contracted [*hole*-type, cf. Eq. (2.145)] string of creation and annihilation operators is given by

$$\mathcal{G}_{i_1\ldots i_n} = \langle a_{i_1}^\dagger a_{i_1}\cdots a_{i_n}^\dagger a_{i_n}\rangle_N = \frac{1}{\mathcal{Z}}\frac{\partial^n \mathcal{Z}}{\partial[-\beta\varepsilon_{i_1}]\cdots\partial[-\beta\varepsilon_{i_n}]}, \quad (2.184)$$

where $\mathcal{Z}(T,N,\Omega)$ is the partition function of the unperturbed canonical ensemble. The cumulants are now given by

$$\mathcal{K}_{i_1\ldots i_n} = \frac{\partial^n \ln \mathcal{Z}}{\partial[-\beta\varepsilon_{i_1}]\cdots\partial[-\beta\varepsilon_{i_n}]}. \quad (2.185)$$

The decisive step is now to evaluate the cumulants not directly but using the Legendre transformation

$$\boxed{\ln\mathcal{Z}(T,N,\Omega) = \ln\mathcal{Y}(T,\tilde{\mu},\Omega) - \tilde{\mu}\frac{\partial\ln\mathcal{Y}(T,\tilde{\mu},\Omega)}{\partial\tilde{\mu}}} \quad (2.186)$$

Here, $\tilde{\mu}$ is the chemical potential of an unperturbed grand-canonical system with the same mean particle number as the interacting canonical system, i.e., $N = -\frac{1}{\beta}\partial\ln\mathcal{Y}/\partial\tilde{\mu} = \sum_a \tilde{f}_a^-$. Here, $\tilde{f}_a^-$ denotes the Fermi-Dirac distribution with $\tilde{\mu}$ as the chemical potential, i.e., $\tilde{f}_a^- := f_a^-(T,\tilde{\mu},\Omega)$. With $N$ being fixed, $N = \sum_a \tilde{f}_a^-$ determines $\tilde{\mu}$ as a functional of the spectrum $\{\varepsilon_a\}$, which we bring into effect by means of an implicit equation:

$$\boxed{\mathcal{J}(\tilde{\mu},\{\varepsilon_a\}) := \sum_a \tilde{f}_a^- - N = 0} \quad (2.187)$$

---

[46] A different (and more well-known) method to derive $F(T,\tilde{\mu},\Omega)$ was introduced earlier by Kohn and Luttinger [246], see Sec. 2.5.3 for details. The Brout-Englert method (correlation-bond formalism) is conceptually clearer (since it starts from the canonical ensemble), and has the great benefit that it manifests the insertion nature of the additional (correlation-bond) contributions that arise due to the change from $A(T,\mu,\Omega)$ to $F(T,\tilde{\mu},\Omega)$. In particular, the zero-temperature limit of $F(T,\tilde{\mu},\Omega)$ involves the cancellation of "anomalous insertions" (higher cumulants with equal indices) and correlation bonds [see Sec. 2.4.6, and also Ref. [215]]; identifying these cancellations at higher orders is much more complicated in the Kohn-Luttinger formalism. We also note that in Ref. [214] it has been argued that the canonical-ensemble perturbation series $F(T,\tilde{\mu},\Omega)$ can also be derived from an asymptotic expansion of the Laplace transform of the grand partition function via contour integrals, cf. [214] for details.

Eq. (2.187) entails that (in contrast to the grand-canonical perturbation series) also higher cumulants with distinct indices contribute; these are the correlation-bond contributions.

To see how the correlation bonds arise we now derive the expressions for the first few cumulants. From Eqs. (2.185) and (2.186) the first ("single-index") cumulants are given by

$$\mathcal{K}_i = \frac{\partial \ln \mathcal{Y}}{\partial[-\beta \varepsilon_i]} + \frac{\partial \ln \mathcal{Y}}{\partial[\beta \tilde{\mu}]} \left( \frac{\partial[\beta \tilde{\mu}]}{\partial[-\beta \varepsilon_i]} \right)_{\mathcal{J}} + N \left( \frac{\partial[\beta \tilde{\mu}]}{\partial[-\beta \varepsilon_i]} \right)_{\mathcal{J}} = \tilde{f}_i^-. \tag{2.188}$$

The expression for $\mathcal{K}_{i_1 i_2}$ (with $i_1 \neq i_2$) is then given by

$$\mathcal{K}_{i_1 i_2} = \left( \frac{\partial \mathcal{K}_{i_1}}{\partial[-\beta \varepsilon_{i_2}]} \right)_{\mathcal{J}} = \frac{\partial \mathcal{K}_{i_1}}{\partial[\beta \tilde{\mu}]} \left( \frac{\partial[\beta \tilde{\mu}]}{\partial[-\beta \varepsilon_{i_2}]} \right)_{\mathcal{J}}. \tag{2.189}$$

The term $(\partial[\beta \tilde{\mu}]/\partial[-\beta \varepsilon_{i_2}])_{\mathcal{J}}$ is given by

$$\left( \frac{\partial[\beta \tilde{\mu}]}{\partial[-\beta \varepsilon_i]} \right)_{\mathcal{J}} = -\left[ \frac{\partial \mathcal{J}(\tilde{\mu}, \{\varepsilon_\alpha\})}{\partial[\beta \tilde{\mu}]} \right]^{-1} \frac{\partial \mathcal{J}(\tilde{\mu}, \{\varepsilon_\alpha\})}{\partial[-\beta \varepsilon_i]} = -\frac{\tilde{f}_i^- \tilde{f}_i^+}{\sum_\alpha \tilde{f}_\alpha^- \tilde{f}_\alpha^+} = O(1/N). \tag{2.190}$$

Hence, $\mathcal{K}_{i_1 i_2}$ is a function of $\varepsilon_{i_1}$, $\varepsilon_{i_2}$, $\tilde{\mu}(\{\varepsilon_\alpha\})$, and also explicitly of $\{\varepsilon_\alpha\}$ through the term $\partial \mathcal{J}/\partial[\beta \tilde{\mu}] = [\sum_\alpha \tilde{f}_\alpha^- \tilde{f}_\alpha^+]^{-1}$. The expression for $\mathcal{K}_{i_1 i_2 i_3}$ with $i_1 \neq i_2 \neq i_3$ is then given by

$$\mathcal{K}_{i_1 i_2 i_3} = \left( \frac{\partial \mathcal{K}_{i_1 i_2}}{\partial[-\beta \varepsilon_{i_3}]} \right)_{\mathcal{J}} = \frac{\partial \mathcal{K}_{i_1 i_2}}{\partial[\beta \tilde{\mu}]} \left( \frac{\partial[\beta \tilde{\mu}]}{\partial[-\beta \varepsilon_{i_3}]} \right)_{\mathcal{J}} - \frac{\partial \mathcal{K}_{i_1}}{\partial[\beta \tilde{\mu}]} \frac{\partial \mathcal{J}(\tilde{\mu}, \{\varepsilon_\alpha\})}{\partial[-\beta \varepsilon_{i_2}]} \frac{\partial}{\partial[-\beta \varepsilon_{i_3}]} \left[ \frac{\partial \mathcal{J}(\tilde{\mu}, \{\varepsilon_\alpha\})}{\partial[\beta \tilde{\mu}]} \right]^{-1}, \tag{2.191}$$

The higher $\mathcal{K}$'s then follow by iteration.[47] Both terms in Eq. (2.191) are of order $O(1/N^3)$; it is then straightforward to show that $[\mathcal{K}_{i_1 \ldots i_n}]_{i_a \neq i_b \ \forall a,b \in [1,n]} = O(1/N^{n-1})$. Thus, altogether, the relevant contributions from higher cumulants are again given by simply-connected diagrams: multiply-connected diagrams give a vanishing contribution to the free energy density in the thermodynamic limit.[48]

## 2.4.5. Free Energy up to Third Order

The perturbation series for $\Delta F$ involves the same diagrams as the one for $\Delta A$, but for each higher cumulant there are, in addition to the equal-index terms (anomalous contributions), several terms with a number of distinct indices, e.g., $\mathcal{K}_{ia} \sim \mathcal{K}_{ia} + \delta_{ia} \mathcal{K}_{ii}$, and $\mathcal{K}_{iac} \sim \mathcal{K}_{iac} + \delta_{ia} \mathcal{K}_{iic} + \delta_{ic} \mathcal{K}_{iai} + \delta_{ac} \mathcal{K}_{iaa} + \delta_{ia} \delta_{ic} \mathcal{K}_{iii}$. The expression for $F(T, \tilde{\mu}, \Omega)$ is therefore given by

$$F(T, \tilde{\mu}, \Omega) = \mathcal{F}(T, \tilde{\mu}, \Omega) + \Delta A(T, \tilde{\mu}, \Omega) + F_{\text{corr.-bond}}(T, \tilde{\mu}, \Omega), \tag{2.192}$$

where $F_{\text{corr.-bond}}(T, \tilde{\mu}, \Omega)$ denotes the additional contributions from higher cumulants with distinct indices. For orientation we give here the explicit expression for the first few cumulants with only distinct indices (in shorthand notation):

$$\mathcal{K}_{i_1 i_2} = \frac{1}{(-\omega)} \mathcal{K}_{i_1}^{(1)} \mathcal{K}_{i_2}^{(1)}, \tag{2.193}$$

---

[47] Note that the recursion formula for $\mathcal{K}_{i_1 \ldots i_n}$ given by Eq. (B.12) in [65] is not valid; e.g., for $\mathcal{K}_{i_1 i_2 i_3}$ it misses the second term in Eq. (2.191).

[48] In the grand-canonical case, the linked-cluster theorem implies that the multiply-connected diagrams in fact cancel, but the same cancellation is not evident here (i.e., for the canonical case).

$$\mathcal{K}_{i_1 i_2 i_3} = \frac{1}{(-\omega)^2} \mathcal{K}^{(2)}_{[i_1} \mathcal{K}^{(1)}_{i_2} \mathcal{K}^{(1)}_{i_3]} + \frac{\omega^{(1)}}{(-\omega)^3} \mathcal{K}^{(1)}_{i_1} \mathcal{K}^{(1)}_{i_2} \mathcal{K}^{(1)}_{i_3}, \tag{2.194}$$

$$\mathcal{K}_{i_1 i_2 i_3 i_4} = \frac{1}{(-\omega)^3} \mathcal{K}^{(3)}_{[i_1} \mathcal{K}^{(1)}_{i_2} \mathcal{K}^{(1)}_{i_3} \mathcal{K}^{(1)}_{i_4]} + \frac{2}{(-\omega)^3} \mathcal{K}^{(2)}_{[i_1} \mathcal{K}^{(2)}_{i_2} \mathcal{K}^{(1)}_{i_3} \mathcal{K}^{(1)}_{i_4]} + \left[\frac{1}{(-\omega)^3}\right]^{(1)} \mathcal{K}^{(2)}_{[i_1} \mathcal{K}^{(1)}_{i_2} \mathcal{K}^{(1)}_{i_3} \mathcal{K}^{(1)}_{i_4]}$$
$$+ \frac{1}{(-\omega)} \left[\frac{\omega^{[1]}}{(-\omega)^3}\right]^{(1)} \mathcal{K}^{(1)}_{i_1} \mathcal{K}^{(1)}_{i_2} \mathcal{K}^{(1)}_{i_3} \mathcal{K}^{(1)}_{i_4}, \tag{2.195}$$

where $\omega := \sum_\alpha \tilde{f}^-_\alpha \tilde{f}^+_\alpha$, and the superscripts "$(n)$" denote the $n^{\text{th}}$ derivative with respect to $\tilde{\mu}$. The derivative terms are given by $\omega^{(n)} = \frac{\partial^n \omega}{\partial [\beta\tilde{\mu}]^n} = \sum_\alpha \frac{\partial^{n+1} \tilde{f}^-_\alpha}{\partial [\beta\tilde{\mu}]^{n+1}}$ and $\mathcal{K}_i^{(n)} = \frac{\partial^n \mathcal{K}_i}{\partial [\beta\tilde{\mu}]^n} = \frac{\partial^n \tilde{f}^-_i}{\partial [\beta\tilde{\mu}]^n}$. The expressions for cumulants with a subset of equal indices are obtained from the distinct-index cumulants via $\mathcal{K}_{i_1 \cdots i_n aa} = \frac{\partial}{\partial[-\beta\varepsilon_a]} \mathcal{K}_{i_1 \cdots i_n a}$, e.g., $\mathcal{K}_{iaa} = \mathcal{K}_i^{(1)} \mathcal{K}_a^{(2)}/(-\omega)$.

The correlation-bond contributions are represented by the same diagrams as the anomalous contributions; e.g., the third-order contributions are given by the diagrams given in Figs. 2.12 and 2.13. None of these contributions involve equal-index cumulants connected to higher-order parts, so the direct, cyclic, and reduced formula are equivalent. For the contribution from two first-order diagrams one finds (from the cyclic formula):

$$F^{[(\Gamma_1)^2]}_{\text{corr.-bond}} = \underbrace{\frac{(-1)}{2} \operatorname*{Res}_{z=0} \frac{e^{-\beta z}}{z(-z)} \frac{1}{(2!)^2}}_{-\beta/8} \sum_{ijab} \bar{V}^{ij,ij}_{2\text{B}} \bar{V}^{ab,ab}_{2\text{B}} \underbrace{[\mathcal{G}_{ij;ab}]_{\text{s.-c.}}}_{2^2 \times \frac{\mathcal{K}^{(1)}_i \mathcal{K}^{(1)}_a \mathcal{K}_j \mathcal{K}_b}{(-\omega)}} = -\frac{(A^{[1]}_1)^2}{2\mathcal{A}^{[2]}}. \tag{2.196}$$

where we have introduced the notation $A_n^{[0]}(T,\tilde{\mu},\Omega) = A_n(T,\tilde{\mu},\Omega)$ and $A_n^{[m\neq 0]}(T,\tilde{\mu},\Omega) = [(m-1)!]^{-1} \partial^m A_n(T,\tilde{\mu},\Omega)/\partial \tilde{\mu}^m$, with $A_0 = \mathcal{A}$. For the sum of diagrams (a) and (c) as well as for diagram (b) one obtains (from the cyclic formula) the expressions

$$F^{[\Gamma_{2,\text{normal}} \Gamma_1]}_{\text{corr.-bond}\,\text{a+c}} = \frac{1}{12} \sum_{ijklab} \bar{V}^{ij,kl}_{2\text{B}} \bar{V}^{kl,ij}_{2\text{B}} \bar{V}^{ab,ab}_{2\text{B}} \frac{-1 + e^{-\beta\mathcal{E}} + \beta\mathcal{E}}{\mathcal{E}^2} [\mathcal{G}^{kl}_{ij;ab}]_{\text{s.-c.}}, \tag{2.197}$$

$$F^{[\Gamma_{2,\text{normal}} \Gamma_1]}_{\text{corr.-bond}\,\text{b}} = \frac{1}{24} \sum_{ijklab} \bar{V}^{ij,kl}_{2\text{B}} \bar{V}^{ab,ab}_{2\text{B}} \bar{V}^{kl,ij}_{2\text{B}} \frac{1 - e^{-\beta\mathcal{E}} - \beta\mathcal{E} e^{-\beta\mathcal{E}}}{\mathcal{E}^2} [\mathcal{G}^{kl}_{ij;ab}]_{\text{s.-c.}}, \tag{2.198}$$

The expression for $[\mathcal{G}^{kl}_{ij;ab}]_{\text{s.-c.}}$ [cf. Eq. (2.172)] is invariant under the multiplication with a factor $\exp(-\beta\mathcal{E})$ and the simultaneous relabeling of indices $ij \leftrightarrow kl$; the energy denominator is symmetric under $ij \leftrightarrow kl$ (and the potential matrix elements are invariant). Hence, the "pseudo-anomalous" terms in Eqs. (2.197) and (2.198) cancel. The sum of the remaining terms leads to

$$F^{[\Gamma_{2,\text{normal}} \Gamma_1]}_{\text{corr.-bond}} = -\frac{A^{[1]}_1 A^{[1]}_{2,\text{normal}}}{\mathcal{A}^{[2]}}. \tag{2.199}$$

The expression for the contributions from $(\Gamma_1)^3$ is given by (cyclic formula)

$$F^{[(\Gamma_1)^3]}_{\text{corr.-bond}} = \underbrace{\frac{(-1)^2}{3} \operatorname*{Res}_{z=0} \frac{e^{-\beta z}}{z(-z)^2} \frac{1}{(2!)^3}}_{\beta^2/48} \sum_{ijabcd} \frac{\bar{V}^{ij,ij}_{2\text{B}}}{2!} \frac{\bar{V}^{ab,ab}_{2\text{B}}}{2!} \frac{\bar{V}^{cd,cd}_{2\text{B}}}{2!} [\mathcal{G}_{ij;ab;cd}]_{\text{s.-c.}}. \tag{2.200}$$

The possible distinct-index cumulants for $[\mathcal{G}_{ij;ab;cd}]_{\text{s.-c.}}$ are given by $24 \times \mathcal{K}_{ia}\mathcal{K}_{jc}$ and $8 \times \mathcal{K}_{iac}$. The contribution from $24 \times \mathcal{K}_{ia}\mathcal{K}_{jc}$ is given by $\frac{(A_1^{[1]})^2}{2(\mathcal{A}^{[2]})^2} \beta^2 \sum_{ij} \bar{V}_{\text{2B}}^{ij,ij} \tilde{f}_i^- \tilde{f}_i^+ \tilde{f}_j^- \tilde{f}_j^+$. The two terms in the expression for the triple-index cumulant [cf. Eq. (2.194)] give the following contributions:

$$8 \times \frac{1}{(-\omega)^2} \mathcal{K}_{[i}^{(2)} \mathcal{K}_a^{(1)} \mathcal{K}_{c]}^{(1)} \sim \frac{(A_1^{[1]})^2}{2(\mathcal{A}^{[2]})^2} \beta^2 \sum_{ij} \bar{V}_{\text{2B}}^{ij,ij} \tilde{f}_i^- \tilde{f}_i^+ \tilde{f}_j^- (\tilde{f}_i^+ - \tilde{f}_i^-), \qquad (2.201)$$

$$8 \times \frac{\omega^{[1]}}{(-\omega)^3} \mathcal{K}_i^{(1)} \mathcal{K}_a^{[1]} \mathcal{K}_c^{[1]} \sim -\frac{(A_1^{[1]})^3 \mathcal{A}^{[3]}}{3(\mathcal{A}^{[2]})^3}. \qquad (2.202)$$

The contributions from $24 \times \mathcal{K}_{ia}\mathcal{K}_{jc}$ and the one from the first term of $8 \times \mathcal{K}_{iac}$ (note the additional factor 3 from the permutations) sum up to $(A_1^{[1]})^2 A_1^{[2]}/[2(\mathcal{A}^{[2]})^2]$. The remaining contributions from mixed equal-index–distinct-index cumulants are given by

$$24 \times \left( \delta_{ia} \mathcal{K}_{ii} \mathcal{K}_{jc} + \mathcal{K}_{ia} \delta_{jc} \mathcal{K}_{jj} \right) \sim \frac{A_1^{[1]}}{2(\mathcal{A}^{[2]})^2} \beta \sum_{ijc} \bar{V}_{\text{2B}}^{ij,ij} \bar{V}_{\text{2B}}^{jb,jb} \tilde{f}_i^- \tilde{f}_j^- \tilde{f}_j^+ \tilde{f}_b^- (2\tilde{f}_i^+), \qquad (2.203)$$

$$8 \times \left( \delta_{ia} \mathcal{K}_{iic} + \delta_{ic} \mathcal{K}_{iai} + \delta_{ac} \mathcal{K}_{iaa} \right) \sim \frac{A_1^{[1]}}{2(\mathcal{A}^{[2]})^2} \beta \sum_{ijc} \bar{V}_{\text{2B}}^{ij,ij} \bar{V}_{\text{2B}}^{jb,jb} \tilde{f}_i^- \tilde{f}_i^+ \tilde{f}_j^- \tilde{f}_b^- (\tilde{f}_i^+ - \tilde{f}_i^-). \qquad (2.204)$$

These two contributions add up to $-A_1^{[1]} A_{2,\text{anomalous}}^{[1]} / \mathcal{A}^{[2]}$. Overall, the contribution from $(\Gamma_1)^3$ is then given by

$$F_{\text{corr.-bond}}^{[(\Gamma_1)^3]} = -\frac{A_1^{[1]} A_{2,\text{anomalous}}^{[1]}}{\mathcal{A}^{[2]}} + \frac{(A_1^{[1]})^2 A_1^{[2]}}{2(\mathcal{A}^{[2]})^2} - \frac{(A_1^{[1]})^3 \mathcal{A}^{[3]}}{3(\mathcal{A}^{[2]})^3}. \qquad (2.205)$$

Note that terms $\propto \mathcal{A}^{[n\geq 3]}$ [or $\propto \omega^{(\nu \geq 1)}$ in Eqs. (2.194) and (2.195)] are "pseudo-anomalous", i.e.,

$$\mathcal{A}^{[n\geq 3]} \xrightarrow{T\to 0} -\sum_i \frac{\partial^{n-1} \Theta(\tilde{\mu} - \varepsilon_i)}{\partial \tilde{\mu}^{n-1}} \bigg|_{n\geq 3} = 0. \qquad (2.206)$$

### 2.4.6. Zero-Temperature Limit

If the reduced formula is used, in the cumulant formalism the expressions for normal diagrams match (in the zero-temperature limit) the corresponding ones in ground-state formalism. The zero-temperature limit of the perturbation series for the free energy $F(T, \tilde{\mu}, \Omega)$ is then given by

$$F(T, \tilde{\mu}, \Omega) \xrightarrow{T\to 0} E_0(\varepsilon_F, \Omega) + \delta E_0(\varepsilon_F, \Omega), \qquad (2.207)$$

where by definition $\tilde{\mu} \xrightarrow{T\to 0} \varepsilon_F$. The term $\delta E_0(\varepsilon_F, \Omega)$ is comprised entirely of anomalous terms, i.e., it corresponds to the sum of the factorized "reduced" parts of the higher-cumulant contributions up to the considered perturbative order. If the adiabatic zero-temperature formalism of Sec. 2.1 is valid it should be $\delta E_0(\varepsilon_F, \Omega) = 0$. In the following, we analyze selected contributions to $\delta E_0$ and show that they cancel (only) in the case of an infinite homogeneous system with rotationally invariant interactions ("isotropic case"). Note that in the following, if we refer to the contribution from a given higher-cumulant term we always mean the factorized part of Eq. (2.88).

## 2. Many-Body Perturbation Theory

For any quantity $g(\vec{k}_1, \ldots, \vec{k}_n)$ that depends on a set of single-particle momenta $\{\vec{k}_1, \ldots, \vec{k}_n\}$, we define its average with respect the (unperturbed) Fermi surface as

$$\left\langle g(\vec{k}_1, \ldots, \vec{k}_n) \right\rangle_F := \frac{\sum_{\vec{k}_1 \cdots \vec{k}_n} g(\vec{k}_1, \ldots, \vec{k}_n) \prod_{i=1}^n \delta(k_F - k_i)}{\sum_{\vec{k}_1 \cdots \vec{k}_n} \prod_{i=1}^n \delta(k_F - k_i)}, \qquad (2.208)$$

where $k_F$ is the Fermi momentum defined by $\varepsilon_F = k_F^2/(2M)$, and $k_i = |\vec{k}_i|$. Then, in the "isotropic case", the following factorization property holds

$$\boxed{\left\langle g_1(\vec{k}_1, \vec{q})\, g_2(\vec{k}_2, \vec{q}) \right\rangle_F = \left\langle g_1(\vec{k}_1, \vec{q}) \right\rangle_F \left\langle g_2(\vec{k}_2, \vec{q}) \right\rangle_F} \qquad (2.209)$$

This is because by rotational invariance, $g_1$ and $g_2$ can be written as functions of $k_1, q, \theta_1$ and $k_2, q, \theta_2$, respectively, where $\theta_{1,2} = \sphericalangle(\vec{k}_{1,2}, \vec{q})$. In the Fermi-surface average the two independent angles $\theta_1$ and $\theta_2$ are integrated over; carrying out the angular integrals in the Fermi-surface average of $g_1(\vec{k}_1, \vec{q})g_2(\vec{k}_2, \vec{q})$ then leads to an integrand $G_1(p, k)G_2(q, k)$ that has no angular dependence, i.e., in shorthand notation

$$\int_{\vec{k},\vec{p},\vec{q}} g_1(p,k,\theta)g_2(p,k,\theta')\, \delta_{k_1}\delta_{k_2}\delta_q = \int_{k,p,q} G_1(p,k)G_2(q,k)\, \delta_{k_1}\delta_{k_2}\delta_q = G_1(k_F, k_F)G_2(k_F, k_F), \qquad (2.210)$$

which proofs Eq. (2.209).

***Second Order.*** We now show that the second-order contribution to $\delta E_0$ vanishes in the isotropic case. The second-order contributions to $\delta E_0$ can be written as

$$F_{2,\text{anomalous}} \xrightarrow{T \to 0} -\frac{1}{2}\sum_i \frac{\delta A_1}{\delta \tilde{f}_i}\frac{\delta A_1}{\delta \tilde{f}_i}\delta_i, \qquad F_{2,\text{corr.-bond}} \xrightarrow{T \to 0} -\frac{1}{2}\sum_{ij}\frac{\delta A_1}{\delta \tilde{f}_i}\frac{\delta A_1}{\delta \tilde{f}_j}\frac{\delta_i \delta_j}{(-\omega_F)}, \qquad (2.211)$$

where $\omega_F := \sum_i \delta_i$ denotes the number of states on the Fermi surface, and $\delta_i := \delta(\varepsilon_F - \varepsilon_i)$. With $X_{1;i} := \delta A_1/\delta \tilde{f}_i$, Eq. (2.211) reads

$$F_{2,\text{anomalous}} \xrightarrow{T \to 0} -\frac{1}{2}\left\langle S_1^2 \right\rangle_F \omega_F, \qquad F_{2,\text{corr.-bond}} \xrightarrow{T \to 0} \frac{1}{2}\left\langle S_1 \right\rangle_F^2 \omega_F, \qquad (2.212)$$

where $\langle f \rangle_F := \sum_i \delta_i X_{1;i}/\sum_i \delta_i$. The vanishing of $\delta E_{0;2}$ in the "isotropic case" follows then as the trivial case of Eq. (2.209) where the two quantities in the Fermi-surface average are independent. Note also that in general $\delta E_{0;2} \leq 0$, i.e.,

$$\delta E_{0;2} = -\frac{\omega_F}{2}\left\langle (S_1 - \langle S_1 \rangle_F)^2 \right\rangle_F \leq 0. \qquad (2.213)$$

This shows that, at second order in MBPT, in the "anisotropic case" the energy of the state that adiabatically evolves from the unperturbed ground state (as calculated in the adiabatic zero-temperature formalism) is larger than the energy of the "true" ground-state (as obtained from the zero-temperature limit of the canonical perturbation series).

The generalization for the case of a system with multiple species (where for each species $\xi$ there is an associated Fermi energy $\varepsilon_{F,\xi}$) is straightforward. Because the higher cumulants can connect only equal-species lines, the contributions to $\delta E_{0;2}$ are now given by

$$F_{2,\text{anomalous}} \xrightarrow{T \to 0} -\frac{1}{2}\sum_{i,\xi} \frac{\delta A_1}{\delta \tilde{f}_{i,\xi}}\frac{\delta A_1}{\delta \tilde{f}_{i,\xi}}\delta_{i,\xi}, \qquad F_{2,\text{corr.-bond}} \xrightarrow{T \to 0} -\frac{1}{2}\sum_{ij,\xi}\frac{\delta A_1}{\delta \tilde{f}_{i,\xi}}\frac{\delta A_1}{\delta \tilde{f}_{j,\xi}}\frac{\delta_{i,\xi}\delta_{j,\xi}}{(-\omega_{F,\xi})}, \qquad (2.214)$$

where $\delta_{i,\xi} := \delta(\varepsilon_{F,\xi} - \varepsilon_i)$ and $\omega_{F,\xi} := \sum_\alpha \delta_{\alpha,\xi}$. With $S_{1;i,\xi} := \delta A_1/\delta \tilde{f}_{i,\xi}$ and $\langle f \rangle_{F,\xi} := \sum_i \delta_{i,\xi} f_i / \sum_i \delta_{i,\xi}$, Eq. (2.214) reads

$$F_{2,\text{anomalous}} \xrightarrow{T \to 0} -\frac{1}{2} \sum_\xi \left\langle (S_{1;\xi})^2 \right\rangle_{F,\xi} \omega_{F,\xi}, \qquad F_{2,\text{corr.-bond}} \xrightarrow{T \to 0} \frac{1}{2} \sum_\xi \left\langle S_{1;\xi} \right\rangle_{F,\xi}^2 \omega_{F,\xi}, \quad (2.215)$$

which again shows that $\delta E_{0;2}$ vanishes in the "isotropic case". In the following we restrict the discussion to the one-species case.

***Double-Index Cumulants.*** The above analysis is readily generalized for any contribution $\sim \Gamma_{\nu_1} \cdots \Gamma_{\nu_{N+1}}$, where all diagrams $\Gamma_{n_k}$ are connected via double-index cumulants $\mathcal{K}_{ij}$ only. Since there are $N + 1$ diagrams, one has $N$ double-index cumulants, which can be either equal-index or distinct-index ones. The contribution to $\delta E_{0;\sum \nu_i}$ with $N$ equal-index cumulants $\mathcal{K}_{i_1 i_1} \cdots \mathcal{K}_{i_N i_N}$ has the form

$$\mathcal{K}_{i_1 i_1} \cdots \mathcal{K}_{i_N i_N} \sim \sum_{i_1 \cdots i_N} g_{i_1 i_1 \cdots i_N}^{i_1 \cdots i_N} \delta_{i_1} \ldots \delta_{i_N} = \langle g \rangle_F \, \omega_F^N, \quad (2.216)$$

where $g_{i_1 i_1 \cdots i_N}^{i_1 \cdots i_N} = g_{1;i_1 i_1} \cdots g_{N;i_N i_N}$. The contribution where the cumulant with index $i_1$ is a distinct-index one has the form

$$\mathcal{K}_{i_1 a_1} \mathcal{K}_{i_2 i_2} \cdots \mathcal{K}_{i_N i_N} \sim \sum_{i_1 a_1 i_2 \cdots i_N} g_{i_1 i_2 \cdots i_N}^{a_1 i_2 \cdots i_N} \frac{\delta_{a_1}}{(-\omega_F)} \delta_{i_1} \ldots \delta_{i_N} = -\langle g \rangle_F \, \omega_F^N, \quad (2.217)$$

the one with two leading distinct-index cumulants is given by

$$\mathcal{K}_{i_1 a_1} \mathcal{K}_{i_2 a_2} \mathcal{K}_{i_3 i_3} \cdots \mathcal{K}_{i_N i_N} \sim \sum_{i_1 a_1 i_2 a_2 i_3 \cdots i_N} g_{i_1 i_2 i_3 \cdots i_N}^{a_1 a_2 i_3 \cdots i_N} \frac{\delta_{a_1}}{(-\omega_F)} \frac{\delta_{a_2}}{(-\omega_F)} \delta_{i_1} \ldots \delta_{i_N} = \langle g \rangle_F \, \omega_F^N, \quad (2.218)$$

etc. In Eqs. (2.216), (2.217) and (2.218), the last equality (which holds for the "isotropic case" only) corresponds to the generalization of Eq. (2.209) for multiple $g$'s.

There are $\binom{N}{1}$ contributions with one distinct-index cumulant, $\binom{N}{2}$ with two distinct-index cumulants, etc. The vanishing of any contribution to $\delta E_0$ that involves double-index cumulants only then follows from the identity

$$\sum_{k=0}^N (-1)^k \binom{N}{k} = 0. \quad (2.219)$$

In particular, this shows the cancellation of the third-order contributions from $\Gamma_{2,\text{normal}} \Gamma_1$ as well as the double-index ones from $(\Gamma_1)^3$. To proof that $\delta E_{0,3} = 0$ in the "isotropic case", it remains to show that also the triple-index contributions from $(\Gamma_1)^3$ cancel each other.

***Multi-Index Cumulants.*** The cancellation of contributions to $\delta E_0$ that involve cumulants with more than two indices is less straightforward. For instance, the contributions to $\delta E_{0,n+p+q}$ from three diagrams [$\sim \Gamma_n \Gamma_p \Gamma_q$] connected via triple-index cumulants are given by

$$\mathcal{K}_{iii} \sim \sum_i S_{n;i} S_{p;i} S_{q;i} \delta_i', \quad (2.220)$$

$$\mathcal{K}_{iia} \sim \sum_{ia} S_{n;i} S_{p;i} S_{q;a} \delta_i' \frac{\delta_a}{(-\omega_F)}, \quad (2.221)$$

$$\mathcal{K}_{iai} \sim \sum_{ia} S_{n;i} S_{p;a} S_{q;i} \delta'_i \frac{\delta_a}{(-\omega_F)}, \tag{2.222}$$

$$\mathcal{K}_{aii} \sim \sum_{ia} S_{n;a} S_{p;i} S_{q;i} \delta'_i \frac{\delta_a}{(-\omega_F)}, \tag{2.223}$$

$$\mathcal{K}_{iab} \sim \sum_{iab} S_{n;i} S_{p;a} S_{q;b} \frac{\delta_a \delta_b}{(-\omega_F)^2} \delta'_i + \sum_{iab} S_{n;i} S_{p;a} S_{q;b} \frac{\delta_i \delta_b}{(-\omega_F)^2} \delta'_a + \sum_{iab} S_{n;i} S_{p;a} S_{q;b} \frac{\delta_i \delta_a}{(-\omega_F)^2} \delta'_b, \tag{2.224}$$

where $\delta'_i := \partial \delta(\varepsilon_F - \varepsilon_i)/\partial \varepsilon_F = -\partial \delta(\varepsilon_F - \varepsilon_i)/\partial \varepsilon_i$. Note that we have omitted the "pseudo-anomalous" term from $\mathcal{K}_{iab}$ (which vanishes by itself in the zero-temperature limit), cf. Eq. (2.194). In the "isotropic case", the contributions with correlation-bonds simplify, i.e.,

$$\mathcal{K}_{iia} \sim -\langle S_q \rangle_F \int_i S_{n;i} S_{p;i} \delta'_i, \tag{2.225}$$

$$\mathcal{K}_{iai} \sim -\langle S_p \rangle_F \int_i S_{n;i} S_{q;i} \delta'_i, \tag{2.226}$$

$$\mathcal{K}_{aii} \sim -\langle S_n \rangle_F \int_i S_{p;i} S_{q;i} \delta'_i, \tag{2.227}$$

$$\mathcal{K}_{iab} \sim \langle S_p \rangle_F \langle S_q \rangle_F \int_i S_{n;i} \delta'_i + \langle S_n \rangle_F \langle S_q \rangle \int_i S_{p;i} \delta'_i + \langle S_n \rangle_F \langle S_p \rangle \int_i S_{q;i} \delta'_i. \tag{2.228}$$

By partial integration one finds that the contributions from $\mathcal{K}_{iii}$ is identical to the one from $\mathcal{K}_{iab}$, i.e.,

$$\mathcal{K}_{iii} \sim \int_i \delta_i \frac{\partial}{\partial \varepsilon_i} (S_{n;i} S_{p;i} S_{q;i}) = \langle S_p \rangle_F \langle S_q \rangle_F \int_i S_{n;i} \delta'_i + \langle S_n \rangle_F \langle S_q \rangle_F \int_i S_{p;i} \delta'_i + \langle S_n \rangle_F \langle S_p \rangle_F \int_i S_{q;i} \delta'_i. \tag{2.229}$$

Similarly, one finds that the contributions from $\mathcal{K}_{iia}$, $\mathcal{K}_{iai}$, and $\mathcal{K}_{aii}$ add up to twice the negative contribution from $\mathcal{K}_{iab}$, i.e.,

$$\mathcal{K}_{iia} \sim -\langle S_p \rangle_F \langle S_q \rangle_F \int_i S_{n;i} \delta'_i - \langle S_n \rangle_F \langle S_q \rangle_F \int_i S_{p;i} \delta'_i, \tag{2.230}$$

$$\mathcal{K}_{iai} \sim -\langle S_n \rangle_F \langle S_p \rangle_F \int_i S_{q;i} \delta'_i - \langle S_p \rangle_F \langle S_q \rangle_F \int_i S_{n;i} \delta'_i, \tag{2.231}$$

$$\mathcal{K}_{aii} \sim -\langle S_n \rangle_F \langle S_p \rangle_F \int_i S_{q;i} \delta'_i - \langle S_n \rangle_F \langle S_q \rangle_F \int_i S_{p;i} \delta'_i. \tag{2.232}$$

This shows the vanishing of any contribution to $\delta E_0$ from three diagrams connected with triple-index cumulants, in particular the third-order one from three first-order diagrams [$\sim (\Gamma_1)^3$].

The cancellations that lead to the vanishing of $\delta E_0$ become more involved for increasingly complicated multi-index cumulant contributions, i.e., for increasing numbers of indices and for contributions with multiple multi-index cumulants. No direct proof of the vanishing of $\delta E_0$ to all orders seems to be available. However, an indirect argument that $\delta E_0 = 0$ to all orders (in the "isotropic case") has been given by Luttinger and Ward [278] (see also Ref. [65]); this is discussed in Sec. 2.5.2.

## 2.5. Role of Higher-Cumulant Contributions

In rigorous statistical mechanics different thermostatistical ensembles are equivalent in the thermodynamic limit [402, 216, 84, 179, 92, 271, 202, 306, 307]. However, this equivalence need not apply for an approximative treatment via perturbation theory. The question arises how the perturbation series for the grand-canonical potential $A(T, \mu, \Omega)$ and the canonical one for the free energy $F(T, \tilde{\mu}, \Omega)$ are related to each other. In particular, we have seen in Sec. 2.4.6 that the zero-temperature limit of $F(T, \tilde{\mu}, \Omega)$ reproduces the adiabatic ground-state perturbation series (in the "isotropic case"), $F(T, \tilde{\mu}, \Omega) \xrightarrow{T \to 0} E_0(\varepsilon_F, \Omega)$, where the particle number is fixed, $N = \sum_i \tilde{f}_i^- = \sum_i \Theta_i^-$, and thus $\tilde{\mu} \xrightarrow{T \to 0} \varepsilon_F$. The expression for the free energy obtained from $A(T, \mu, \Omega)$ is given by

$$\check{F}(T, \mu, \Omega) = A(T, \mu, \Omega) + \mu \check{N}(T, \mu, \Omega), \qquad \text{with} \qquad \check{N}(T, \mu, \Omega) = -\frac{\partial A(T, \mu, \Omega)}{\partial \mu}. \qquad (2.233)$$

There is no reason to expect that $\check{F}(T = 0, \mu, \Omega) = E_0(\varepsilon_F, \Omega)$ for fixed $N = \check{N}(T = 0, \mu, \Omega) = \sum_i \Theta_i^-$, or more generally, that $\check{F}(T, \mu, \Omega) = F(T, \tilde{\mu}, \Omega)$ for fixed $N = \check{N}(T, \mu, \Omega) = \sum_i \tilde{f}_i^-$.

We are therefore lead to the task to determine which formalism should be used, the grand-canonical one, or the canonical formalism. Clearly, the investigation of this issue involves the analysis of the higher-cumulant contributions, which appear in a different form in the grand-canonical (anomalous insertions, i.e., higher cumulants with equal indices) and the canonical (anomalous insertions and correlation bonds) perturbation series, and in particular the effects of "removing" these contributions via the self-consistent "renormalization" of $\mathcal{T}$ and $\mathcal{V}$ according to $\mathcal{T} \to \mathcal{T} + \mathcal{X}_{[n]}^{\aleph}$ and $\mathcal{V} \to \mathcal{T} - \mathcal{X}_{[n]}^{\aleph}$, where $\mathcal{X}_{[n];i}^{\aleph} = \sum_i X_{[n];i}^{\aleph} = \sum_{\nu=1}^{n} \sum_i X_{\nu;i}^{\aleph}$, and[49]

$$X_{\nu;i}^{\aleph}(T, \mu, \Omega) = \frac{\delta A_{\nu,\text{non-insertion}}^{R,\aleph}[f_i^-; \{\varepsilon_{X_{[n];i}^{\aleph}}\}]}{\delta f_i^-}, \qquad X_{\nu;i}^{\aleph}(T, \tilde{\mu}_{X_{[n]}^{\aleph}}, \Omega) = \frac{\delta F_{\nu,\text{non-insertion}}^{R,\aleph}[\tilde{f}_i^-; \{\varepsilon_{X_{[n];i}^{\aleph}}\}]}{\delta \tilde{f}_i^-}, \qquad (2.234)$$

in the grand-canonical and the canonical case, respectively (see Sec. 2.3.1 for more details). Note that in the canonical case also the auxiliary chemical potential is "renormalized", i.e., $\tilde{\mu} \to \tilde{\mu}_{X_{[n]}^{\aleph}}$, where (by construction) $\sum_i \tilde{f}_i^- = \sum_i \tilde{f}_{X_{[n];i}^{\aleph}}^-$.

The change from the $\{\mathcal{T}, \mathcal{V}\}$ to the $\{\mathcal{T} + \mathcal{X}_{[n]}^{\aleph}, \mathcal{V} - \mathcal{X}_{[n]}^{\aleph}\}$ setup removes all higher-cumulant insertions (i.e. the factorized parts of the higher-cumulant terms) with "insertion diagrams" $\Gamma_1, \ldots \Gamma_n$ (in addition, normal "one-loop" insertions are removed), as depicted in Fig. 2.15. Insertion diagrams are identified as follows: consider three diagrams $\Gamma_{n_1}, \Gamma_{n_2},$ and $\Gamma_{n_3}$ connected via higher cumulants: $\Gamma_{n_1}$—$\Gamma_{n_2}$—$\Gamma_{n_3}$, where the insertion-lines are placed on *different* lines of $\Gamma_{n_2}$. Then $\Gamma_{n_1}$ and $\Gamma_{n_3}$ are insertion diagrams; this situation is depicted in Fig. 2.14 for the case of first-order diagrams. If the insertion diagrams are placed on the *same* line of $\Gamma_{n_2}$, then also $\Gamma_{n_2}$ is an insertion diagram. In other terms, for a given higher-cumulant chain, an insertion diagram is a diagram that can be replaced by a one-body vertex. The perturbation series in the $\{\mathcal{T} + \mathcal{X}_{[n]}^{\aleph}, \mathcal{V} - \mathcal{X}_{[n]}^{\aleph}\}$ setup is then given by all skeletons plus all normal non-skeletons without first-order ("one-loop") parts, plus higher-cumulant contributions that involve insertion diagrams of orders greater than $n$, plus additional normal non-skeletons with $-X_2^{\aleph}, \ldots, X_n^{\aleph}$ vertices.

---

[49] Here, "non-insertion" refers to contributions not of the insertion type, i.e., skeletons and normal-non-skeletons without first-oder parts, and normal non-skeletons with $-X_2, \ldots, -X_{n-1}$ vertices.

## 2. Many-Body Perturbation Theory

In a sense, the change from $\{\mathcal{T}+\mathcal{X}^{\aleph}_{[n-1]},\mathcal{V}-\mathcal{X}^{\aleph}_{[n-1]}\}$ to $\{\mathcal{T}+\mathcal{X}^{\aleph}_{[n]},\mathcal{V}-\mathcal{X}^{\aleph}_{[n]}\}$ corresponds to the "resummation" of higher-cumulant insertions with $\Gamma_n$ insertion diagrams. However, (except for $n=1$) this "resummation" involves not only the renormalization of the distribution functions (as in Sec. 2.5.1) but also the renormalization of energy denominators (in particular regarding the energy eigenvalue equations that determine the renormalized single-particle energies, cf. Sec. 2.3.1), as well as the emergence of new normal non-skeleton contributions from $-X^{\aleph}_n$ vertices.

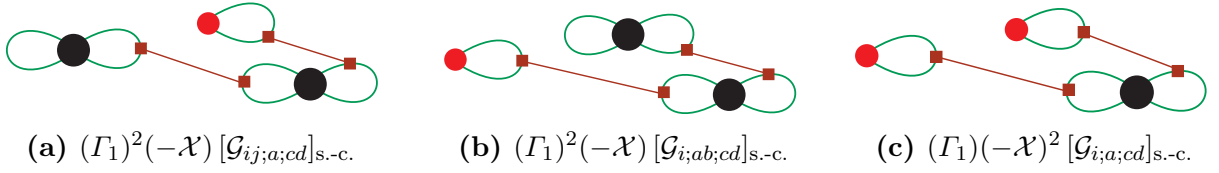

(a) $(\Gamma_1)^2(-\mathcal{X})\,[\mathcal{G}_{ij;a;cd}]_{\text{s.-c.}}$    (b) $(\Gamma_1)^2(-\mathcal{X})\,[\mathcal{G}_{i;ab;cd}]_{\text{s.-c.}}$    (c) $(\Gamma_1)(-\mathcal{X})^2\,[\mathcal{G}_{i;a;cd}]_{\text{s.-c.}}$

**Figure 2.14.:** Additional higher-cumulant contributions from $-X_1$ vertices (small red dots) for diagram (a) of Fig. 2.12.

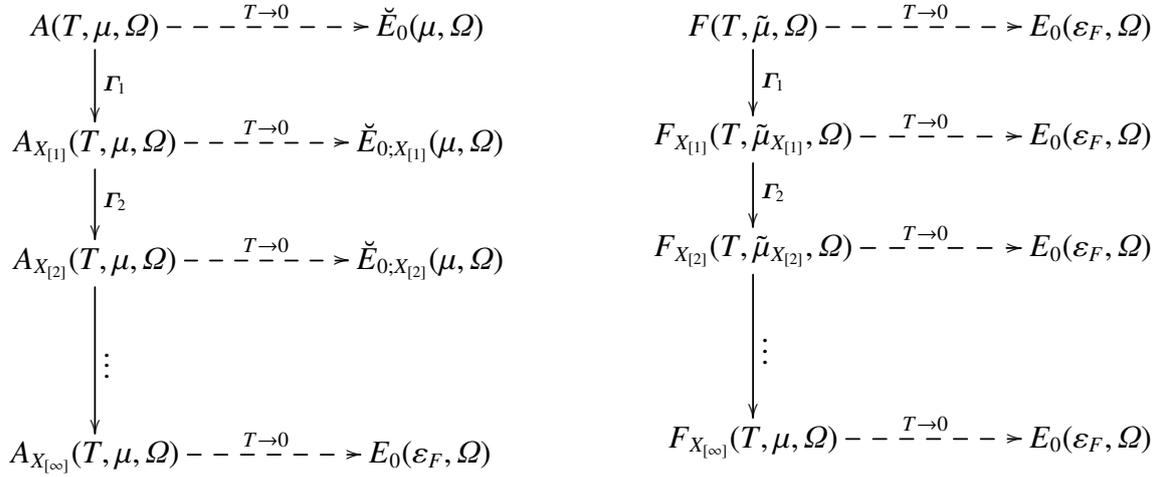

**Figure 2.15.:** Formal hierarchy of renormalized perturbation series in the grand-canonical and the canonical formalism. The downwards arrows denote the change from the $\{\mathcal{T}+\mathcal{X}_{[n-1]},\mathcal{V}]-\mathcal{X}_{[n-1]}\}$ setup to the $\{\mathcal{T}+\mathcal{X}_{[n]},\mathcal{V}]-\mathcal{X}_{[n]}\}$ setup, corresponding to the hierarchical "resummation" of higher-cumulant insertions with first order ($\Gamma_1$), second-order ($\Gamma_2$), etc. insertion diagrams. The subscripts "$X_{[n]}$" refers to the corresponding self-consistent renormalization of the single-particle energies and (in the canonical case) the auxiliary chemical potential. The dashed arrows denote the zero-temperature limit of the free energy at each level (in the "isotropic case", for the case where the reduced formula is used).

In the "fully-renormalized" ($n = \infty$) perturbation series no higher-cumulant contributions remain, and ensemble equivalence is recovered, i.e., the "fully-renormalized" canonical and grand-canonical perturbation series are equivalent. The zero-temperature limit of both series however does not exist if $\aleph = $ "direct" (in general, for $n \geq 6$, see the last paragraph of Sec. 2.3.4).[50] In contrast, if $\aleph = $ "reduced", the zero-temperature limit exists, and the "bare" and the various "renormalized" canonical perturbation series all have the same zero-temperature limit at each perturbative order (in the "isotropic case"): the Goldstone formula $E_0(\varepsilon_F, \Omega)$.[51] This is not the case for (the free energy obtained from) the various grand-canonical perturbation series, where the zero-temperature limit does not lead to the Goldstone formula, *with the exception of "fully-renormalized" series* (in the "isotropic case", for ℵ = "reduced"). This strongly suggests that the "bare" and the "only partially resummed" grand-canonical series are deficient. In the "bare" case, this is particularly evident for a system with a first-order liquid-gas phase transition (cf. Sec. 2.5.4). The deficiency of the (not "fully-renormalized") grand-canonical perturbation series can also be seen by analyzing in more detail the renormalization of the Fermi-Dirac distributions (cf. Sec. 2.5.5).

In the following, we examine these features (in particular, the properties of the "fully-renormalized" perturbation series) in more detail. Note that for notational convenience we leave out the regularization symbol "$R$" as well as the formula symbol "ℵ". Moreover, we will distinguish explicitly between ℵ = "direct" and ℵ = "reduced" only when necessary.

### 2.5.1. Resummation of Higher-Cumulant Contributions

Here, we show the resummation of higher-cumulant insertions renormalizes the distribution functions. Note that we use ℵ = "direct" throughout this section.

***Anomalous Insertions.*** The ("direct") time-dependent expression for a diagram of order $n$ (the "main" diagram) with a single diagram of order $p$ (the "sub" diagram) attached on the line with index $a$ (which can be a *hole* or a *particle* line) via $\mathcal{K}_{aa} = f_a^+ f_a^-$ is given by [cf. Eq. (2.37)]

$$\Gamma_n(\Gamma_p) \sim -\frac{1}{\beta}(-1)^n \int_{\beta>\tau_n>\ldots\tau_1>0} d\tau_n \cdots d\tau_1 \, (-1)^p \int_{\beta>\tau_{n+p}>\ldots\tau_{n+1}>0} d\tau_{n+p} \cdots d\tau_{n+1} \, \langle\ldots\rangle_{\text{main}(n)+\text{sub}(p)}\,, \quad (2.235)$$

where we have summed over possible time orderings of the insertion diagram relative to the normal diagram (the cumulant formalism is essential for this). Note that this proofs Eq. (2.84). Without loss of generality, in the following we consider the grand-canonical case, and denote by $A_{n,\text{main}} = \sum_{aijkl\ldots}(\ldots)_{\text{main}(n)}$ the expression for the "main" diagram. Carrying out the time integrals in Eq. (2.235) leads to[52]

$$\boxed{\Gamma_n(\Gamma_p) \sim (-\beta) \sum_{aijkl\ldots} \frac{\partial(\ldots)_{\text{main}(n)}}{\partial[-\beta\varepsilon_a]} \frac{\delta A_{p,\text{sub}}}{\delta f_a^-}} \quad (2.236)$$

This equation shows the insertion nature [as defined in Sec. 2.3.1] of the anomalous contributions, i.e., the contributions from higher cumulants with equal indices: the contribution from $\Gamma_n(\Gamma_p)$ has the same form as the corresponding one where the "sub" diagram $\Gamma_p$ is replaced by the one-body vertex associated with the operator $\mathcal{S}_p = \sum_i S_{p;i} a_i^\dagger a_i$, where

$$S_{p;a} = \frac{\delta A_{p,\text{sub}}}{\delta f_a^-}. \quad (2.237)$$

---

[50] Note that the nonexistence of the zero-temperature limit $n \geq 6$ is entirely due to the "resummation" of higher cumulants with equal indices. The "resummation" of correlation bonds does not lead to complications regarding the $T \to 0$ limit (this feature is also evident from the Kohn-Luttinger inversion method).

[51] The canonical "renormalized" perturbation series lead (for ℵ = "reduced") to the ground-state perturbation series with renormalized energy denominators and additional normal non-skeletons with $-X_2, \ldots, X_n$ vertices, but this is (formally) equivalent to the Goldstone formula (i.e., the "bare" ground-state perturbation series).

[52] Insertions on *particles* carry an additional minus sign, cf. Eq. (2.172), but this sign is cancelled by the additional minus sign needed to correct for the sign from $\partial f_k^+/\partial[-\beta\varepsilon_k] = -f_k^+ f_k^-$.

## 2. Many-Body Perturbation Theory

We can now show explicitly that the resummation of the chain of anomalous insertions renormalizes the single-particle energies in the Fermi-Dirac distributions. The derivative operator $\partial/\partial[-\beta\varepsilon_a]$ in Eq. (2.236) transforms the $f_a^-$ in $(\ldots)_{\mathrm{main}(n)}$ into $f_a^+ f_a^-$, and the functional derivative removes $f_a^-$ and "$\sum_a$" in the expression for the insertion.[53] The expression for a diagram of order $n$ with $m = \sum_{\nu=1}^L m_\nu$ insertions of order $k$ attached to the $1,\ldots,L$ lines of the diagram has the form

$$\Gamma_n(\Gamma_p)^m \sim \frac{(-\beta)^m}{m!} \sum_{a_1\ldots a_L} \sum_{ijkl\ldots} \Upsilon_p^m (\ldots)_{\mathrm{main}(n)}. \tag{2.238}$$

Summing over all possible $\{m_\nu\}$, the insertion operator $\Upsilon_p^m$ is given by

$$\Upsilon_p^m = \sum_{\{m_\nu\}} \binom{m}{m_1,\ldots,m_L} \prod_{\nu=1}^L \left[ S_{p;a_\nu} \frac{\partial}{\partial[-\beta\varepsilon_{a_\nu}]} \right]^{m_\nu} = \left[ \sum_{\nu=1}^L S_{p;a_\nu} \frac{\partial}{\partial[-\beta\varepsilon_{a_\nu}]} \right]^m, \tag{2.239}$$

where

$$S_{p;a} = \frac{\delta A_{\mathrm{sub}(p)}}{\delta f_a^-}. \tag{2.240}$$

From Eqs. (2.238) and (2.239) it follows that the sum over all numbers of insertions $m$ leads to the renormalization of the distribution functions (in the expression for the "main" diagram):

$$\sum_{m=0}^\infty \frac{1}{m!} \Upsilon_p^m \left\{ f_{a_\nu}^\mp \right\} = \exp\left[ \sum_{\nu=1}^L S_{p;a_\nu} \frac{\partial}{\partial \varepsilon_{a_\nu}} \right] \left\{ f_{a_\nu}^\mp \right\} = \left\{ \frac{1}{1 + \exp(\pm\beta(\varepsilon_{a_\nu} + S_{p;a_\nu} - \mu))} \right\} \equiv \left\{ f_{S_p;a_\nu}^\mp \right\}. \tag{2.241}$$

So far, we have considered only the case of "first-degree" insertions, i.e., the case where the sub diagrams do not carry insertions on themselves. The effect of "higher-degree" insertions can be identified by applying the "first-degree" analysis at each level. Since at each level the resummation of the insertions renormalized the Fermi-Dirac distribution, resumming the infinite chain of "higher-degree" insertions leads to the renormalization of the Fermi-Dirac distributions in terms of self-consistent self-energies.

***Self-Consistent Hartree-Fock.*** The partition of a given simply-connected diagram into a "main" diagram and a number of insertion diagrams is (of course) ambiguous. In particular, in the case where the "main" diagram and the insertion diagrams are of the same order, the factorization theorem [Eq. (2.19)] gives the "wrong" factor, e.g., for first-order diagrams $\Gamma_{1^{m+1}} = (\Gamma_1)^{m+1}/(m+1)! \neq \Gamma_1(\Gamma_1)^m/m!$, so the above analysis does not apply in this case. The precise effect (i.e., the structure of the resulting perturbation series) of the "resummation" of higher-cumulant insertions can only be indentified in terms of the "renormalization" of $\mathcal{T}$ and $\mathcal{V}$.

Nevertheless, it will be instructive to examine the resummation of higher-cumulant insertions explicitly for the case of anomalous insertions that involve only first-order diagrams. (In that case there is no energy-denominator renormalization and no additional non-skeletons with one-body vertices). To start, consider the contributions from a first-order diagram with $n$ "first-degree" first-order insertion diagrams on one of the two contraction lines [e.g., diagram (d) of Fig. 2.13]; in term of standard Hugenholtz diagrams, this corresponds to (the sum of) the $n$-loop diagrams of order $n$. The expression for contributions of this kind can be written as

$$A_{n,n\text{-loop}} = -\frac{1}{\beta} \frac{(-\beta)^n}{n!} \sum_a \left[ S_{1;a} \frac{\partial}{\partial[-\beta\varepsilon_a]} \right]^n \ln \mathcal{Y}. \tag{2.242}$$

---

[53] Note that the functional derivative $\delta/\delta f_a^-$ automatically includes the correct multiplicity factor for the placement of insertion-lines.

## 2. Many-Body Perturbation Theory

There is no "1-loop contribution"; the expression for first-order diagram (which has in fact two loops) has the different form $A_1 = \frac{1}{2} \sum_a \left[ S_{1;a} \frac{\partial}{\partial[-\beta \varepsilon_a]} \right] \ln \mathcal{Y} = \frac{1}{2} A\text{"1,1-loop"}$. The "0-loop contribution" however has the "correct" form $\mathcal{A} = -\beta^{-1} \ln \mathcal{Y} = A\text{"0,0-loop"}$. The sum of the noninteracting term, the first-order diagram, and the $n$-loop diagrams of order $n = 2, \ldots, \infty$ is then given by

$$\mathcal{A} + A_1 + \sum_{n=2}^{\infty} A_{n,n\text{-loop}} = -A_1 + \sum_{n=0}^{\infty} A_{n,n\text{-loop}} = -A_1 - \frac{1}{\beta} \sum_a \exp\left(-\beta S_{1;a} \frac{\partial}{\partial[-\beta \varepsilon_a]}\right) \ln \mathcal{Y}. \tag{2.243}$$

To evaluate the second term $\sum_{n=0}^{\infty} A_{n,n\text{-loop}}$ we consider the Taylor series of $\ln \mathcal{Y}$, i.e.,

$$\ln \mathcal{Y} = \sum_i \ln\left(1 + \exp\left(-\beta(\varepsilon_i - \mu)\right)\right) = \sum_i \sum_{\nu=1}^{\infty} \frac{1}{\nu!} \exp\left(\nu \beta(\varepsilon_i - \mu)\right). \tag{2.244}$$

Expanding the exponential in Eq. (2.243) then leads to

$$\mathcal{A} + A_1 + \sum_{n=2}^{\infty} A_{n,n\text{-loop}} = -A_1 + \mathcal{A}_{S_1}, \qquad \mathcal{A}_{S_1} = \frac{1}{\beta} \sum_i \ln\left[1 + \exp\left(-\beta(\varepsilon_i + S_{1;i} - \mu)\right)\right]; \tag{2.245}$$

i.e., the resummation of all $n$-loop diagrams of order $n = 2, \ldots, \infty$ leads to the renormalization of the noninteracting term, and changes the sign of the first-order contribution.

The expression for the self-consistently renormalized noninteracting term is given by

$$\mathcal{A}_{X_1} = \frac{1}{\beta} \sum_i \ln\left[1 + \exp\left(-\beta(\varepsilon_{X_1;i} - \mu)\right)\right] = \frac{1}{\beta} \sum_a \sum_\nu \frac{1}{\nu!} \left[-\beta X_{1;a} \frac{\partial}{\partial[-\beta \varepsilon_a]}\right]^\nu \ln \mathcal{Y}. \tag{2.246}$$

Inserting the diagrammatic expansion of $X_{1;a}$ one obtains all the diagrams of the type "nonskeleton diagrams with first-order parts only" ($A_{n,m\text{-loop}}, 2 \leq m \leq n$), but in general with different prefactors as those that appear in the original perturbation series. Given that the "remainder" $\mathcal{A}_{X_1} - \sum_{2 \leq m \leq n} A_{n,m\text{-loop}}$ has the correct combinatorial factors for the self-consistent renormalization of the negative first-order contribution, we have

$$\mathcal{A} + A_1 + \sum_{2 \leq m \leq n} A_{n,m\text{-loop}} = \mathcal{A}_{X_1} - A_{1;X_1}, \tag{2.247}$$

which corresponds to the self-consistent Hartree-Fock approximation for the grand-canonical potential. A direct combinatorial proof of Eq. (2.247) is unnecessary, since Eq. (2.247) follows trivially if one considers the change from the $\{\mathcal{T}, \mathcal{V}\}$ to the $\{\mathcal{T} + \mathcal{X}_1, \mathcal{V} - \mathcal{X}_1\}$ setup.

***Correlation Bonds.*** In the canonical case the "'resummation" of higher-cumulants includes also correlation bonds, i.e., the indices connected by cumulants need not be equal. Thus, instead of Eq. (2.241) we have

$$\sum_{m=0}^{\infty} \frac{1}{m!} \Upsilon_p^m \{f_{a_\nu}^{\mp}\} = \exp\left[\sum_{\nu=1}^{L} S_{p;b_\nu} \frac{\partial}{\partial \varepsilon_{b_\nu}}\right] \{f_{a_\nu}^{\mp}\} \equiv \{f_{S_p;a_\nu}^{\mp}\}. \tag{2.248}$$

where the action of a derivative operator with subindex "$\nu$" is restricted to quantities with the same "$\nu$". For a given line $\nu \in [1, L]$, with index denoted by "$i$", we expand the expontial

$$e^{-\beta \sum_a S_{p;a} \frac{\partial}{\partial[-\beta \varepsilon_a]}} \tilde{f}_i^{\mp} = \tilde{f}_i^{\mp} - \beta X_{p;i} \frac{\partial}{\partial[-\beta \varepsilon_i]} \tilde{f}_i^{\mp} - \beta \sum_a X_{p;a} \frac{\partial[\beta \tilde{\mu}]}{\partial[-\beta \varepsilon_a]} \frac{\partial}{\partial[\beta \tilde{\mu}]} \tilde{f}_i^{\mp} + \ldots \tag{2.249}$$

The next term in the exponential series involves also a contribution where the derivative acts on the term $\partial[\beta\tilde{\mu}]/\partial[-\beta\varepsilon_a]$, etc., inhibiting the explicit evaluation of Eq. (2.249). The effect of the "resummation" can only be identified indirectly in terms of a redefintion of $\mathcal{T}$ and $\mathcal{V}$. Since by construction, in the $\{\mathcal{T} + \mathcal{X}_{[n]}, \mathcal{V} - \mathcal{X}_{[n]}\}$ setup it is

$$N = \sum_i \tilde{f}^-_{X_{[p]};i}, \quad (2.250)$$

if follows that that resumming the additional correlation bonds corresponds to renormalizing the auxililary chemical potential, $\tilde{\mu} \to \tilde{\mu}_{X_{[p]}}$, such that the "particle number is conserved", i.e., such that $\tilde{\mu}_{X_{[p]}}$ is the same functional of the renormalized spectrum $\{\varepsilon_{X_{[p]};a}\}$ as $\tilde{\mu}$ is of the "bare" spectrum $\{\varepsilon_a\}$.

## 2.5.2. Statistical Quasiparticles and Kohn-Luttinger-Ward Theorem

Here, we examine the properties of the "fully-renormalized" grand-canonical and canonical perturbation series, respectively, for both $\aleph$ = "direct" and $\aleph$ = "reduced".

A discussed at the beginning of this section, the "fully-renormalized" grand-canonical perturbation series does not have any (factorited parts of) higher-cumulant contributions. However, it involves the additional contribution

$$-\langle \mathcal{X}_{[\infty]}\rangle = -\sum_i f^-_{X_{[\infty]};i} X_{[\infty];i}. \quad (2.251)$$

With $X_i \equiv X_{[\infty];i}$ for notational convenience, the expression for the "fully-renormalized" perturbation series is then given by[54]

$$\boxed{A_X(T,\mu,\Omega) = \underbrace{\frac{1}{\beta}\sum_a \ln(1 - f^-_{X;a}) - \sum_a f^-_{X;a} X_a}_{\mathcal{A}_X(T,\mu,\Omega)} + \mathcal{D}[f^-;\{\varepsilon_{X;i}\}]} \quad (2.252)$$

Since we are in the "fully-renormalized" situation, $\mathcal{D}$ is given by the sum of all normal diagrams that are not of the insertion type, plus additional normal non-skeletons with $-X_2, \ldots, X_\infty$ vertices. The self-consistent self-energy of order "$\infty$" satisfies the relation

$$X_i = \frac{\delta \mathcal{D}}{\delta f^-_{X;i}}. \quad (2.253)$$

The discussion of the canonical case is entirely similar, and the expression for the "fully-renormalized" canonical perturbation series is given by

$$\boxed{F_X(T,\tilde{\mu}_X,\Omega) = \mathcal{F}_X(T,\tilde{\mu}_X,\Omega) - \sum_a \tilde{f}^-_{X;a} X_a + \mathcal{D}[\tilde{f}^-;\{\varepsilon_{X;i}\}]} \quad (2.254)$$

By construction, it is $\sum_i \tilde{f}^-_{X;i} = N$. Below, we will show that also $\sum_i f^-_{X;i} = N$, for both $\aleph$ = "direct" and $\aleph$ = "reduced". Thus, $\tilde{\mu}_X = \mu$, and therefore[55]

$$F_X(T,\tilde{\mu}_X,\Omega) = \check{F}_X(T,\mu,\Omega), \quad (2.255)$$

---

[54] An equation of this form was first derived (using combinatorial arguments only, cf. Eq. (35) in Ref. [18]) by Balian, Bloch and de Dominicis [18], cf. also Refs. [48, 47, 45].

[55] For $\aleph$ = "direct" and $n \geq 6$ it should be $T \neq 0$ in Eq. (2.255).

*2. Many-Body Perturbation Theory*

where $\check{F}_X(T,\mu,\Omega)$ denotes the free energy obtained from Eq. (2.252). Thus, ensemble equivalence is recovered in the "fully-renormalized" situation, and the "bare" grand-canonical and canonical perturbation series can be seen to emerge from expanding the Fermi-Dirac distributions in the "fully-renormalized" series about the "bare" single-particle energies and about both the "bare" single-particle energies and the "bare" auxiliary chemical potential, respectively. The implications of this will be examined further in Sec. 2.5.5.

$\aleph$ = *"direct"*. Following Balian, Bloch and de Dominicis [18] we now derive expressions for various derived thermodynamic quantities in terms of the "fully-renormalized" distribution functions $f_X^-$ for $\aleph$ = "direct". We start by rewriting the expression for the "fully-renormalized" grand-canonical potential [Eq. (2.252)] as

$$A_X(T,\mu,\Omega) = \frac{1}{\beta}\sum_a \left[f_{X;a}^- \ln(f_{X;a}^-) + f_{X;a}^+ \ln(f_{X;a}^+)\right] + \sum_a (\varepsilon_a - \mu)f_{X;a}^- + \mathscr{D}[f^-;\{\varepsilon_{X;i}\}]. \quad (2.256)$$

Using $\ln(f_{X;a}^+/f_{X;a}^-) = -\beta\,(\varepsilon_a + X_a[f_{X;a}^-] - \mu)$ one finds that the functional derivative of $A_X$ with respect to $f_{X;a}^-$ vanishes[56]

$$\frac{\delta A_X}{\delta f_{X;a}^-} = \frac{1}{\beta}\ln(f_{X;a}^+/f_{X;a}^-) + \varepsilon_a - \mu + X_a[f_X^-] = 0. \quad (2.257)$$

From this, one finds that the grand-canonical expression for the particle number is given by

$$\check{N}_X(T,\mu,\Omega) = -\frac{\partial A_X}{\partial \mu} = -\frac{\bar{\partial} A_X}{\bar{\partial}\mu} + \sum_a \underbrace{\frac{\delta A_X}{\delta f_{X;a}^-}}_{=0} \frac{\partial f_{X;a}^-}{\partial \mu} = \sum_a f_{X;a}^-, \quad (2.258)$$

where $\bar{\partial}/\bar{\partial}\mu$ means that the derivative does not act on the distribution functions. Eq. (2.258) matches the corresponding relations for the free Fermi gas. In particular, Eq. (2.258) implies that the "fully-renormalized" distribution functions $f_{X;a}^-$ coincide with the exact mean occupation numbers of the unperturbed single-particle states [277] i.e.,

$$f_{X;i}^- = \text{Tr}[Y^{-1}\,e^{-\beta(\mathcal{H}-\mu\mathcal{N})}\,a_i^\dagger a_i]. \quad (2.259)$$

If $\aleph$ = "direct", the "fully-renormalized" entropy however does not have a quasiparticle form. This is due to the fact that for $\aleph$ = "direct", $\mathscr{D}$ has an explicit dependence on $T$, leading to

$$\check{S}_X(T,\mu,\Omega) = -\frac{\partial A_X}{\partial T} = -\frac{\bar{\partial} A_X}{\bar{\partial} T} + \sum_a \underbrace{\frac{\delta A_X}{\delta f_{X;a}^-}}_{=0} \frac{\partial f_{X;a}^-}{\partial T}$$

$$= -\sum_a \left[f_{X;a}^- \ln(f_{X;a}^-) + f_{X;a}^+ \ln(f_{X;a}^+)\right] - \frac{\bar{\partial}\mathscr{D}}{\bar{\partial} T}, \quad (2.260)$$

which deviates from the free Fermi gas expression in terms of the last term $\bar{\partial}\mathscr{D}/\bar{\partial} T$. Finally, from Eq. (2.258) one obtains for the "fully-renormalized" ("grand-canonical") free energy $\check{F}_X = \mu\check{N}_X + A_X$ the expression

$$\check{F}_X(T,\mu,\Omega) = \overbrace{\mu\sum_a f_{X;a}^- + \frac{1}{\beta}\sum_a \ln(1 - f_{X;a}^-)}^{\mathcal{F}_X(T,\mu,\Omega)} - \sum_a f_{X;a}^- X_a + \mathscr{D}[f^-;\{\varepsilon_{X;i}\}], \quad (2.261)$$

---

[56] A stationarity property holds also for $\check{F}_X(T,\mu,\Omega)$, i.e., $\check{F}_X(T,\mu,\Omega)$ is stationary with respect to variations of the distribution functions under the constraint $N = \sum_i f_{X;i}^-$.

which leads to the following expression for the internal energy $\check{E}_X = \check{F}_X + T\check{S}_X$:

$$\check{E}_X(T,\mu,\Omega) = \sum_a \varepsilon_a f^-_{X_a} + \mathscr{D}[f^-;\{\varepsilon_{X;i}\}] - T\frac{\bar{\partial}\mathscr{D}}{\bar{\partial}T}. \tag{2.262}$$

Note that, as a consequence of the $\bar{\partial}\mathscr{D}/\bar{\partial}T$ term, the functional derivative of $\check{E}_X(T,\mu,\Omega)$ does *not* reproduce the "fully-renormalized" single-particle energies.

$\aleph = $ *"reduced"*. If $\aleph = $ "reduced", then $\mathscr{D}$ has no explicit $T$ dependence. This leads to "statistical quasiparticle" relations, i.e., (in contrast to the "direct" case) the expression for both the particle number and the entropy match those for a free Fermi gas, and the functional derivative of $\check{E}_X$ reproduces the "fully-renormalized" single-particle energies

$$\frac{\delta \check{E}_X}{\delta f^-_{X_a}} = \varepsilon_a + X_a. \tag{2.263}$$

It should be noted that Eq. (2.263) and the existence of the "statistical quasiparticle" relation for the entropy rely on the assumption that the "pseudo-anomalous" remainder $\mathcal{R}$ vanishes to all orders [cf. Eq. (2.89)]. If $\mathcal{R} = 0$ does not hold, then the "statistical quasiparticle" relations are valid only asympotically as $T \to 0$.[57]

In the "isotropic case", $X_i(T,\mu,\Omega) \equiv X(T,\mu,|\vec{k}|)$, i.e., $X_i$ is a function of the magnitude $|\vec{k}|$ of the plane-wave three-momentum $\vec{k}$ only (at fixed $T$ and $\mu$). For fixed $T$ and $\mu$ the step function $\Theta(\mu - \varepsilon_i - X_i)$ then defines a certain momentum $k_F$ up to which states are summed, i.e., $\Theta(\mu - \varepsilon_i - X_i) \equiv \Theta(k_F - |\vec{k}|)$. The zero-temperature limit of the expression for the particle number is then given by

$$\check{N}(T=0,\mu,\Omega) = \frac{\Omega}{(2\pi)^3} \int d^3k \, \Theta(k_F - |\vec{k}|), \tag{2.264}$$

which matches the relation $N = \sum_i \Theta(\varepsilon_F - \varepsilon_i)$ in the adiabatic zero-temperature formalism, with $\varepsilon_F = k_F^2/(2M)$. Thus, we can write $\Theta(\mu - \varepsilon_i - X_i) = \Theta(\varepsilon_F - \varepsilon_i)$ and the zero-temperature limits of the "fully-renormalized" expressions for the internal energy and free energy are given by

$$\boxed{\check{E}_X(T,\mu,\Omega) \xrightarrow{T \to 0} \underbrace{\sum_i \varepsilon_i \Theta(\varepsilon_F - \varepsilon_i)}_{\mathcal{E}_0(\varepsilon_F,\Omega)} + \underbrace{\mathscr{D}[\Theta(\varepsilon_F - \varepsilon_i)]}_{\Delta E_0(\varepsilon_F,\Omega)} = E_0(\varepsilon_F,\Omega)} \tag{2.265}$$

This result was first derived by Luttinger and Ward [278]. We emphasize that Eq. (2.265) does not imply that the "bare" (i.e., "unrenormalized") grand-canonical perturbation series (or, for that matter, an only "partially renormalized" version) has the "correct" zero-temperature limit. The "fully-renormalized" version is only formally equivalent to the "bare" one, but in general leads to quite different results. This is in particular evident for a system with a first-order liquid-gas phase transition (cf. Sec. 2.5.4).

The indirect argument due to Luttinger and Ward [278] (sometimes referred to as the "Kohn-Luttinger-Ward theorem" [143, 394, 150, 144]) for the cancellation of the anomalous contributions (higher-cumulant contributions) in the "isotropic case" is now basically as follows: because $F(T,\tilde{\mu},\Omega) \xrightarrow{T \to 0} E_0(\varepsilon_F,\Omega) + \delta E_0(\varepsilon_F,\Omega)$ and $F_X(T,\tilde{\mu}_X,\Omega) \xrightarrow{T \to 0} E_0(\varepsilon_F,\Omega)$, and since $F_X(T,\tilde{\mu}_X,\Omega)$ follows from $F(T,\tilde{\mu},\Omega)$ in terms of a partial resummation, it should be $\delta E_0(\varepsilon_F,\Omega) = 0$ at each order.

---

[57] See footnote [20].

### 2.5.3. Kohn-Luttinger Inversion Method[58]

Eq. (2.255) suggests an alternative derivation of the "bare" canonical perturbation series $F(T,\tilde{\mu},\Omega)$ in terms of an expansion of the chemical potential $\mu$ in the (unrenormalized) "grand-canonical" free energy $\check{F}(T,\mu,\Omega)$ about $\tilde{\mu}$. To investigate this we start by considering a (formal) expansion of $\mu$ in terms of the interaction strength: $\mu = \mu_0 + \sum_{n=1}^{\infty} \lambda^n \mu_n$, where the unperturbed value $\mu_0$ will be qualified later. The expansions of $A(T,\mu,\Omega)$ and $\check{N}(T,\mu,\Omega) = -\partial A(T,\mu,\Omega)/\partial\mu$ about $\mu_0$ are then given by

$$A(T,\mu,\Omega) = \sum_{n=0}^{\infty} \lambda^n \sum_{\substack{k \\ a_1<...<a_k \\ \{b_i\},c,d}} \binom{b_1+\ldots+b_k}{b_1,\ldots,b_k} (\mu_{a_1})^{b_1}\cdots(\mu_{a_k})^{b_k} \frac{A_c^{[d]}}{d} \Bigg|_{b_1+\ldots+b_k=d}^{a_1+\ldots+a_k+c=n}, \quad (2.266)$$

$$\frac{\partial A(T,\mu,\Omega)}{\partial \mu} = \sum_{n=0}^{\infty} \lambda^n \sum_{\substack{k \\ a_1<...<a_k \\ \{b_i\},c,d}} \binom{b_1+\ldots+b_k}{b_1,\ldots,b_k} (\mu_{a_1})^{b_1}\cdots(\mu_{a_k})^{b_k} A_c^{[d+1]} \Bigg|_{b_1+\ldots+b_k=d}^{a_1+\ldots+a_k+c=n}, \quad (2.267)$$

where $A_c^{[0]}(T,\mu_0,\Omega) = A_c(T,\mu_0,\Omega)$ and $A_c^{[d\neq 0]}(T,\mu_0,\Omega) = [(d-1)!]^{-1}\partial^d A_c(T,\mu_0,\Omega)/\partial\tilde{\mu}^d$, with $A_0(T,\mu_0,\Omega) = \mathcal{A}(T,\mu_0,\Omega)$. The "chemical potential perturbations" $\mu_n$, $n \geq 1$, are now fixed by the requirement that all terms in Eq. (2.267) beyond the leading term $\mathcal{O}(\lambda^0)$ vanish, i.e.,

$$\check{N}(T,\mu,\Omega) = -\mathcal{A}^{[1]}(T,\mu_0,\Omega). \quad (2.268)$$

The expression for $\mu_n(T,\mu_0,\Omega)$ determined by Eqs. (2.267) and (2.268) is

$$\mu_n(T,\mu_0,\Omega) = -\frac{1}{\mathcal{A}^{[2]}} \sum_{n=0}^{\infty} \lambda^n \sum_{\substack{k \\ a_1<...<a_k<n \\ \{b_i\},c,d}} \binom{b_1+\ldots+b_k}{b_1,\ldots,b_k} \mu_{a_1}^{b_1}\cdots\mu_{a_k}^{b_k} A_c^{[d+1]} \Bigg|_{b_1+\ldots+b_k=d}^{a_1+\ldots+a_k+c=n}. \quad (2.269)$$

To obtain the explicit expression for $\mu_n(T,\mu_0,\Omega)$, all previous terms $\mu_\nu(T,\mu_0,\Omega)$, $1 \leq \nu < n$, are required.[59] From Eqs. (2.266), (2.268) and (2.269) the expansion of the grand-canonical free energy about $\mu_0$ is given by:[60]

$$\check{F}(T,\mu,\Omega) = \underbrace{\mathcal{A} - \mu_0 \mathcal{A}^{[1]}}_{\mathcal{F}(T,\mu_0,\Omega)} + \sum_{n=1}^{\infty} \lambda^n A_n + \underbrace{\sum_{n=2}^{\infty} \lambda^n \sum_{\substack{k \\ a_1<...<a_k<n \\ \{b_i\},c<n,d}} \binom{b_1+\ldots+b_k}{b_1,\ldots,b_k} \mu_{a_1}^{b_1}\cdots\mu_{a_k}^{b_k} \frac{A_c^{[d]}}{d} \Bigg|_{b_1+\ldots+b_k=d}^{a_1+\ldots+a_k+c=n}}_{F_{n,\text{counter}}(T,\mu_0,\Omega)}, \quad (2.270)$$

where we have separated the term with $c = n$ in the expansion of $A(T,\mu,\Omega)$. The additional derivative terms (starting at order $n = 2$) arising from the expansion will be called "counterterms". The explicit expressions for the first two counterterms are given by

$$F_{2,\text{counter}} = \frac{\mu_1^2 \mathcal{A}^{[2]}}{2} + \mu_1 A_1^{[1]} = -\frac{(A_1^{[1]})^2}{2\mathcal{A}^{[2]}}, \quad (2.271)$$

---

[58] This method was introduced by Kohn and Luttinger [246] as a means to (re)obtain the Goldstone formula from the zero-temperature limit of the ("bare") grand-canonical perturbation series. The method has been used in perturbative nuclear many-body calculations to second order in Refs. [150, 394, 144]. The designation "Kohn-Luttinger inversion method" has been adapted from R. J. Furnstahl [152].

[59] The first three expressions are given by $\mu_1(T,\mu_0,\Omega) = -\frac{A_1^{[1]}}{\mathcal{A}^{[2]}}$, $\mu_2(T,\mu_0,\Omega) = -\mu_1^2 \frac{\mathcal{A}^{[3]}}{\mathcal{A}^{[2]}} - \mu_1 \frac{A_1^{[2]}}{\mathcal{A}^{[2]}} - \frac{A_2^{[1]}}{\mathcal{A}^{[2]}}$, and $\mu_3(T,\mu_0,\Omega) = -2\mu_2\mu_1 \frac{\mathcal{A}^{[3]}}{\mathcal{A}^{[2]}} - \mu_1^3 \frac{\mathcal{A}^{[4]}}{\mathcal{A}^{[2]}} - \mu_2 \frac{A_1^{[2]}}{\mathcal{A}^{[2]}} - \mu_1^2 \frac{A_1^{[3]}}{\mathcal{A}^{[2]}} - \mu_1 \frac{A_2^{[2]}}{\mathcal{A}^{[2]}} - \frac{A_3^{[1]}}{\mathcal{A}^{[2]}}$.

[60] Note that in the expression for the free energy the term $-\mathcal{A}^{[1]} \sum_{n=1}^{\infty} \mu_n$ is cancelled by corresponding terms in the expansion of $A(T,\mu,\Omega)$, leading to the restrictions $n \geq 2$ and $\{a_i < n\}$ in the last term in Eq. (2.270).

$$F_{3,\text{counter}} = \mu_2 \underbrace{(\mu_1 \mathcal{A}^{[2]} + A_1^{[1]})}_{=0} + \frac{\mu_1^3 \mathcal{A}^{[3]}}{3} + \frac{\mu_1^2 A_1^{[2]}}{2} + \mu_1 A_2^{[1]} = -\frac{(A_1^{[1]})^3 \mathcal{A}^{[3]}}{3(\mathcal{A}^{[2]})^3} + \frac{(A_1^{[1]})^2 A_1^{[2]}}{2(\mathcal{A}^{[2]})^2} - \frac{A_1^{[1]} A_2^{[1]}}{\mathcal{A}^{[2]}},$$
(2.272)

where we have inserted the explicit expression for $\mu_1(T, \mu_0, \Omega)$.[61] One sees that at each order the sum of the counterterms coincides with the sum of the correlation-bond contributions of Sec. 2.4.5.

It should be emphasized here that the equivalence of $F(T, \tilde{\mu}, \Omega)$ and $A(T, \mu, \Omega)$ in terms of the above expansion is purely formal. Given an arbitrary truncation of the respective perturbation series, to obtain $F(T, \tilde{\mu}, \Omega)$ from $\check{F}(T, \mu, \Omega)$ requires the truncation of the Taylor expansions of $A(T, \mu, \Omega)$ and $\partial A(T, \mu, \Omega)/\partial \mu$ about $\mu_0$ at the same order $n$ the grand-canonical perturbation series is truncated at. The EoS associated with $F(T, \tilde{\mu}, \Omega)$ then deviates from the one corresponding to $A(T, \mu, \Omega)$ in terms of the neglected terms in the Taylor expansions. This truncation leads to the "inversion" of the grand-canonical perturbation series into the canonical $F(T, \tilde{\mu}, \Omega)$. In particular, the truncation amounts to Eq. (2.268) being interpreted not as an implicit equation for $\mu_0(T, \mu, \Omega)$ but for $\mu_0(T, N, \Omega)$, leading to the identification of $\mu_0$ with the auxiliary chemical potential $\tilde{\mu}(T, N, \Omega)$ [fixed by Eq. (2.187)] of the correlation-bond formalism.

## 2.5.4. Single-Phase Constraint

The use of many-body perturbation theory, i.e., the use of a noninteracting single-particle basis, corresponds to the assumption that the average distribution of particles is the same as in the absence of interactions.[62] In the case of a system with a liquid-gas phase transition, this corresponds to the system being homogeneous also in the liquid-gas coexistence region: the system is under the *single-phase constraint*. The single-phase constraint implies the existence of a region in the interior of the coexistence region, the so-called spinodal region, where the free energy $F(T, N, \Omega)$ is a concave function of $\Omega$ (at fixed $N$) and of $N$ (at fixed $\Omega$), i.e., $\frac{\partial^2 F(T,N,\Omega)}{\partial \Omega^2} = -\frac{\partial P(T,N,\Omega)}{\partial \Omega} = \frac{1}{\Omega \kappa_T} > 0$ and $\frac{\partial^2 F(T,N,\Omega)}{\partial N^2} = \frac{\partial \mu(T,N,\Omega)}{\partial N} = \frac{\Omega}{N^2 \kappa_T} > 0$, where $\kappa_T$ is the isothermal compressibility).[63] Thus, in the spinodal region the stability requirement $\kappa_T^{-1} < 0$ is violated, which entails that the system is unstable with respect to infinitesimal density fluctuations (cf. Refs. [4, 99, 44] for more details). In general, a Legendre transformation is invertible only for convex functions (cf. Ref. [433]). Legendre-transforming the single-phase constrained (canonical) free energy $F(T, N, \Omega)$ leads to a grand-canonical potential $A(T, \mu, \Omega)$ that is multivalued with respect to $\mu$, as depicted in Fig. 2.16.[64]

Since for the single-phase constrained system the liquid-gas coexistence region corresponds to a multivalued grand-canonical potential, the "bare" grand-canonical perturbation series $A(T, \mu, \Omega)$ (a single-valued function) cannot lead to an EoS with a phase transition [cf. also the discussion below Eq. (4.31)]. In particular, given that $E_0(\varepsilon_F, \Omega)$ has a spinodal region, this means that the equation of state corresponding to $A(T, \mu, \Omega)$ cannot have the "correct" zero-temperature limit at any order in many-body perturbation theory. This deficiency need not pertain for the renormalized versions of the grand-canonical perturbation series [given by $A_{X_{[1]}}$, $A_{X_{[2]}}$, etc., in Fig. 2.15]: a self-consistent procedure may very well have multiple solutions. In fact, the "fully-renormalized" version, $A_X(T, \mu, \Omega)$, does reproduce the Goldstone formula in the zero-temperature limit (in the "isotropic case", for $\aleph$ = "reduced"). This may seem paradoxical, since $A_X(T, \mu, \Omega)$ results (in a sense) from a partial resummation of the "bare" series $A(T, \mu, \Omega)$.

---

[61] The cancellation in Eq. (2.272) is general; using Eq. (2.270) it is straightforward to show that the expression for a counterterm of order $n$ involves only factors $\mu_a^b$ with $a \leq n/2$.

However, there is no paradox: in general, (many-body) perturbation theory yields only divergent asymptotic series [14, 392, 377, 136], so there is no reason why a partially resummed version may not deviate qualitatively from the original series.

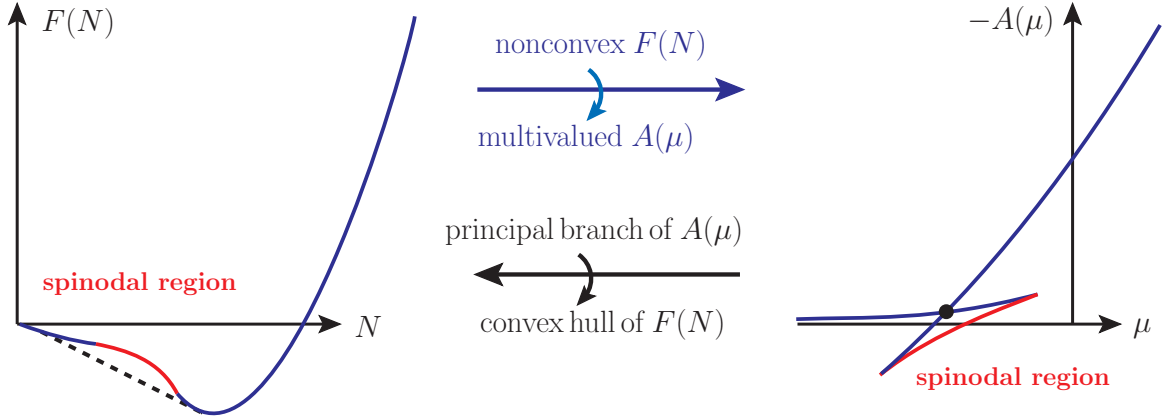

**Figure 2.16.:** Sketch of the Legendre transformations between the free energy $F(T, N, \Omega)$ and the grand-canonical potential $A(T, \mu, \Omega)$ for a one-component system with a liquid-gas phase transition ($T$ and $\Omega$ are fixed). The blue lines (red in the spinodal region) correspond to the single-phase constrained system, and the black dashed lines to the stable equilibrium configuration. Between the *spinodal* (the boundary of the spinodal region) and the *binodal* (the coexistence boundary) the free energy density is locally (but not globally) convex, corresponding to a metastable region where the system is protected against phase separation by a nucleation barrier. Note that in the high-density part of the *metastable* region the pressure can be negative (cf. footnote[8] in Chap. 4).

## 2.5.5. Mean-Field Renormalization of Distribution Functions[65]

The presence of additional higher-cumulant contributions (the correlation bonds) in the canonical perturbation series as compared to the grand-canonical one is in direct correspondence with the different role of the chemical potential in both perturbation series. In the canonical series, the auxiliary chemical potential $\tilde{\mu}$ is an effective parameter that is subject to renormalization, while in the grand-canonical one the chemical potential $\mu$ is a thermodynamic variable. In the canonical case, the "resummation" of higher-cumulant contributions leads to the consistent (i.e., such that the "correct" zero-temperature limit is preserved) renormalization of both the auxiliary chemical potential $\tilde{\mu}$ (via correlation bonds) and the single-particle spectrum $\{\varepsilon_i\}$ (via anomalous insertions), whereas in the grand-canonical case only the single-particle spectrum $\{\varepsilon_i\}$ gets renormalized (and the renormalization changes the zero-temperature limit in that case, see Fig. 2.15). This implies that the resummation has a larger effect on the grand-canonical case, and thus the only partially renormalized grand-canonical series [$A$, $A_{X_{[1]}}$, $A_{X_{[2]}}$, etc.] constitute less well-behaved perturbation expansions as compared to the canonical ones [$F$, $F_{X_{[1]}}$, $F_{X_{[2]}}$, etc.], irrespective of the presence of a liquid-gas phase transition. In the following, we examine these issues in more detail for the nuclear many-body system in terms of the resummation of first-order insertions ("mean-field renormalization"). In particular, we compare the results (from NN interactions only) from the grand-canonical and canonical perturbation series in the mean-field approximation where the contributions from skeletons beyond first order are neglected; this amounts to the restriction to terms with diagonal matrix elements $\bar{V}_{\text{NN}}^{ij,ij}$. In this approximation, the "bare" perturbation series for the grand-canonical potential density is given by

$$A(T,\mu) = \mathcal{A}(T,\mu) + \sum_{n=2}^{\infty} A_{n,\text{anomalous}}(T,\mu), \qquad (2.273)$$

where $A_{n,\text{anomalous}}$ is the contribution of order $n$ from anomalous diagrams with first-order parts. The "bare" canonical perturbation series for the free energy density is (in the mean-field approximation) given by

$$F(T,\tilde{\mu}) = \mathcal{F}(T,\tilde{\mu}) + \sum_{n=2}^{\infty} F_{n,\text{anomalous}}(T,\tilde{\mu}) + \sum_{n=2}^{\infty} F_{n,\text{corr.-bond}}(T,\tilde{\mu}), \qquad (2.274)$$

with $F_{n,\text{corr.-bond}}$ the sum of all correlation-bond contribution of order $n$ with first-order parts only. Since only diagonal matrix elements are considered, in each case the resummation of the higher-order terms leads to the "fully-renormalized" perturbation series, i.e., the self-consistent Hartree-Fock (SCHF) approximation. As we will see below, the SCHF approximation leads to a non-convex free energy density. This can be seen as an explicit proof that Eq. (2.273) is an divergent asymptotic expansion (since the free energy density from the "bare" grand-canonical series is necessarily convex).

---

[65] The (reverse) Legendre transform of the principal branch of $A(T, \mu, \Omega)$ produces the convex hull of $F(T, N, \Omega)$ [cf. Fig. 2.16], corresponding to the stable equilibrium configuration, $F_{\text{stable}}(T, N, \Omega)$. The Legendre transformation between $F_{\text{stable}}(T, N, \Omega)$ and $A_{\text{stable}}(T, \mu, \Omega)$ is then invertible. In the thermodynamic limit (cf. [83, 84, 202] for discussions of the finite-system case), the phase coexistence region corresponds to the point where $P^\infty_{\text{stable}}(T, \mu)|_T = -A^\infty_{\text{stable}}(T, \mu)|_T$ has a kink (cf. also Refs. [428, 265] and [216] where this feature is analyzed within the context of rigorous statistical mechanics; and Ref. [99] for a discussion of the issue of metastability in this context), corresponding to the pressure $P^\infty_{\text{binodal}}(T, \Omega)$ and the chemical potential $\mu_{\text{binodal}}(T, \Omega)$ of the coexisting phases; this leads to the double-tangent form (in the one-component case, cf. [301]) of $F_{\text{stable}}(T, N, \Omega)$ in the coexistence region, $F_{\text{binodal}}(T, N, \Omega) = A_{\text{binodal}}(T, \Omega) + N \mu_{\text{binodal}}(T, \Omega)$.

[63] When a self-consistent potential (cf. Sec. 2.5.2) is included the interactions are taken into account, but only in an average way; cf. also Refs. [48, 143] for discussions.

[64] In the thermodynamic limit, this corresponds to the free energy density $F(T, \rho)$ being a concave function of $\rho$ and the free energy per particle $\bar{F}(T, \bar{\Omega})$ a concave function of the volume per particle $\bar{\Omega} = \Omega/N$, i.e., $\frac{\partial F(T,\rho)^2}{\partial \rho^2} = \frac{\partial \mu(T,\rho)}{\partial \rho} = \frac{1}{\rho^2 \kappa_T} > 0$ and $\frac{\partial \bar{F}(T,\bar{\Omega})^2}{\partial \bar{\Omega}^2} = -\frac{\partial P(T,\bar{\Omega})}{\partial \bar{\Omega}} = \frac{1}{\bar{\Omega} \kappa_T} > 0$.

***Mean-Field Shift and Effective Mass.*** The resummation of first-order anomalous insertions leads to renormalization of the single-particle energies in terms of the first-order self-consistent self-energy $X_{1;k}(T,\mu)$. For (infinite) nuclear matter, $X_{1;k}(T,\mu)$ is to good accuracy a quadratic function of $k$, which leads to the well-known "effective-mass approximation" [143, 41]

$$\frac{k^2}{2M} + X_{1;k}(T,\mu) \simeq \frac{k^2}{2M^*(T,\mu)} + U(T,\mu). \qquad (2.275)$$

The quantity $M^*(T,\mu)$ is called the "(self-consistent) effective mass", and $U(T,\mu)$ is called the "(self-consistent) mean-field shift".

---

[65] In this section (and the remainder of the thesis) we work directly in the thermodynamic limit, so no dependence on $\Omega$ appears, the particle number $N$ is replaced by the particle density $\rho = N/\Omega$, and all quantities are replaced by their respective densities; note that we use the same notation for the "densities" as for the volume dependent quantities.

## 2. Many-Body Perturbation Theory

In the "effective-mass approximation", the renormalized distribution functions are given by

$$f^{-}_{X_1;k}(T,\mu) \simeq \frac{1}{1+\exp\left[\beta\left(\frac{k^2}{2M^*}+U-\mu\right)\right]}, \qquad \tilde{f}^{-}_{X_1;k}(T,\lambda) \simeq \frac{1}{1+\exp\left[\beta\frac{M}{M^*}\left(\frac{k^2}{2M}+\frac{\lambda^2}{2M}\right)\right]}, \qquad (2.276)$$

in the grand-canonical and the canonical case, respectively. In the canonical case, the additional renormalization of the auxiliary chemical potential has been brought into effect in terms of the parameter $\lambda(T,\rho)$ defined via $\rho(T,\lambda) = \sum_k \tilde{f}^{-}_{X_1;k}(T,\lambda)$; this implies that $\lambda \xrightarrow{T\to 0} k_F$. For comparison, the unrenormalized grand-canonical and canonical distribution functions are given by

$$f^{-}_{k}(T,\mu) = \frac{1}{1+\exp\left[\beta\left(\frac{k^2}{2M}-\mu\right)\right]}, \qquad \tilde{f}^{-}_{k}(T,\lambda_{\text{bare}}) = \frac{1}{1+\exp\left[\beta\left(\frac{k^2}{2M}+\frac{\lambda^2_{\text{bare}}}{2M}\right)\right]}, \qquad (2.277)$$

where $\lambda_{\text{bare}}(T,\rho)$ is defined via $\rho(T,\lambda_{\text{bare}}) = \sum_k \tilde{f}^{-}_{k}(T,\lambda_{\text{bare}})$, with $\lambda_{\text{bare}} \xrightarrow{T\to 0} k_F$. (Note that the change from $\tilde{f}^{-}_{k}$ to $\tilde{f}^{-}_{X_1;k}$ now corresponds to the renormalization of the particle mass $M$).

Now, the point is that for a system with significant mean-field effects (e.g., nuclear matter) the change from $\tilde{f}^{-}_{k}$ to $\tilde{f}^{-}_{X_1;k}$ is much smaller as the one from $f^{-}_{k}$ to $f^{-}_{X_1;k}$. This is a consequence of the presence of the mean-field shift $U(T,\mu)$ in the expression for $f^{-}_{X_1;k}$. In the canonical case the mean-field shift is effectively absorbed in the renormalization of the auxiliary chemical potential, and only a rescaling of the inverse temperature by a factor $M/M^*$ remains. From this, it can be inferred that the "bare" grand-canonical perturbation series [$\sim A$] constitutes a considerably less well-behaved perturbation series as the canonical one [$\sim F$].[66] Similar arguments apply for higher-order insertions and the partially renormalized grand-canonical series [$A_{X_1}$, $A_{X_{[2]}}$, etc.].

***Self-Consistent vs. Perturbative Results.*** To quantify the above assertions we now evaluate numerically the nuclear EoS obtained from the respective perturbation series, neglecting all skeleton contributions above first order (as well as 3N interactions), i.e., only (NN) mean-field effects are taken into account. The "fully-renormalized" free energy density is then given by the self-consistent Hartree-Fock expression, i.e., $F_{X_1}(T,\mu) = \mathcal{F}_{X_1}(T,\mu) - \sum_a f^{-}_{X_1;a} X_{1;a}$, with the particle density given by $\rho(T,\mu) = \sum_a f^{-}_{X_1;a}$.[67] The results for $U(T,\rho)$ and $M^*(T,\rho)/M$, obtained using the n3lo414 two-nucleon potential (cf. Table 1.1) are displayed in Fig. 2.17 for isospin-symmetric nuclear matter. Also shown are the results for $\tilde{\mu} = \lambda^2_{\text{bare}}/(2M)$ and $\mu = \lambda^2/(2M^*) - U$. One sees that compared to the prevalent scales (the ground-state energy per particle of at saturation density is $\bar{E}_{0,\text{sat}} \simeq -16\,\text{MeV}$) the mean-field shift $U(T,\rho)$ is sizeable, and gives the main contribution to the difference $\mu - \tilde{\mu}$, whose size is of the same order of magnitude. For comparison, we show also the "bare" results for $U(T,\rho)$ and $M^*(T,\rho)/M$, i.e., the results obtained at the first iteration step; one sees that the difference between the self-consistent and "bare" results is rather small.[68]

---

[66] See however the appendix A.1, in particular Table A.1, for a (somewhat contrived) counterexample.

[67] We compute these quantities as follows. We first calculate $X_1(T,\lambda)$ for fixed $\lambda$: we start with the perturbative self-energy $S_1(T,\lambda)$, use the results to extract $M^*(T,\lambda)$ and $U(T,\lambda)$, then evaluate $X_1(T,\lambda)$ with the distribution functions given by using $M^*(T,\lambda)$ in Eq. (2.276), then extract again the effective mass and the mean-field shift from the results, etc. This iterative procedure becomes stationary after about five to ten iterations. From the self-consistent results for $U(T,\lambda)$ and $M^*(T,\lambda)$ it is straightforward to compute $\rho(T,\lambda)$ and $F_{X_1}(T,\lambda)$. The self-consistent Hartree-Fock free energy $F_{X_1}(T,\lambda)$ is obtained by evaluating the expressions for $\mathcal{F}(T,\lambda)$ and $F_1(T,\lambda)$ with the renormalized distribution functions $\tilde{f}^{-}_{X_1;k}(T,\lambda)$ [cf. Eqs. (2.254) and (2.255)]. More details regarding the numerical evaluation of the various NN contributions are given in Sec. 3.1.

*2. Many-Body Perturbation Theory*

The mean-field results (with the n3lo414 two-nucleon potential) for the free energy density $F(T,\rho)$ at $T = 15$ MeV are shown in Fig. 2.18, i.e., the curves shown are the canonical results obtained at the Hartree-Fock level ["HF" in Fig. 2.18] and including the second-order higher-cumulant contributions ["HF+$\mathcal{K}$" in the plot], as well as the self-consistent Hartree-Fock ["SCHF"] results. The second-order higher-cumulant contributions, i.e., the anomalous contribution $F_{2,\text{anomalous}}(T,\rho)$ ["$\mathcal{K}_{\text{anom.}}$"] and the sum of the anomalous and the correlation-bond contribution $F_{2,\text{h}}(T,\rho) = F_{2,\text{anomalous}}(T,\rho) + F_{2,\text{corr.-bond.}}(T,\rho)$ ["$\mathcal{K}_{\text{anom.}} + \mathcal{K}_{\text{corr.}}$"], are shown explicitly (cf. Secs. 2.3.4 and 2.4.5 for the corresponding formulas). Moreover, we also show the first-order grand-canonical results ["HF (G.C.)"] and the first-order grand-canonical results with the second-order anomalous contribution included ["HF+$\mathcal{K}$ (G.C.)"].[69] For comparison, also the free Fermi gas results ("free") are displayed. In Fig. 2.18, one sees that the "HF", the "HF+$\mathcal{K}$", and the "SCHF" curves are very similar, with the "HF+$\mathcal{K}$" and the "SCHF" curves lying almost on top of each other. This shows that, at the mean-field level, the "bare" canonical perturbation series is well-converged already at second order also for large temperatures (at $T = 0$ there are no higher-cumulant insertions, so "HF" and "SCHF" are equivalent in that case, cf. also [143]). The grand-canonical curves on the other hand deviate substantially from the "SCHF" ones (which, within the mean-field approximation, correspond to the exact result).[70] The inclusion of the second-order (anomalous) contribution gives only a small improvement. As expected from the analysis of the renormalized distribution functions, the anomalous contribution "$\mathcal{K}_{\text{anom.}}$" is quite large (at high densities or large chemical potentials, respectively), but the sum of the second-order higher-cumulant contributions "$\mathcal{K}_{\text{anom.}} + \mathcal{K}_{\text{corr.}}$" is very small.

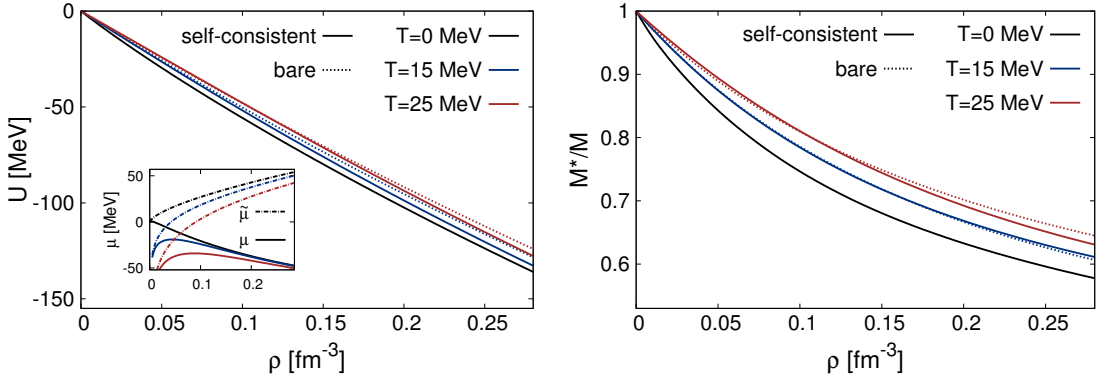

**Figure 2.17.:** Results for $U(T,\rho)$ and $M^*(T,\rho)/M$ (main plots) as well as $\tilde{\mu}(T,\rho)$ and $\mu(T,\rho)$ (inset), see text for details. In the main plots, the full lines correspond to the self-consistent Hartree-Fock results, and the dotted lines to the "bare" results.

---

[68] In the zero-temperature formalism the difference is (exactly) zero, i.e., HF=SCHF at $T = 0$, corresponding to the absence of anomalous contributions. A somewhat larger difference at finite $T$ however appears to have been observed [using a different NN potential and a slightly different method to compute $X_1(T,\lambda)$] in Ref. [346].

[69] The Legendre transformation from the grand-canonical potential to the "grand-canonical" free energy has been performed numerically using finite differences.

[70] Note that from the "SCHF" results the exact (within the mean-field approximation) chemical potential $\mu(T,\rho)$ can be computed from the thermodynamic relation $\mu(T,\rho) = \partial F(T,\rho)/\partial\rho$, which should agree [30] with the "microscopic" $\mu = \lambda^2/(2M^*) - U$. We have found that the results for $\mu(T,\rho)$ obtained in this way match the results obtained via $\mu = \lambda^2/(2M^*) - U$ to good accuracy; this can be seen to justify the use of the effective-mass approximation (at each iteration step) in the computation of the self-consistent Hartree-Fock self-energy.

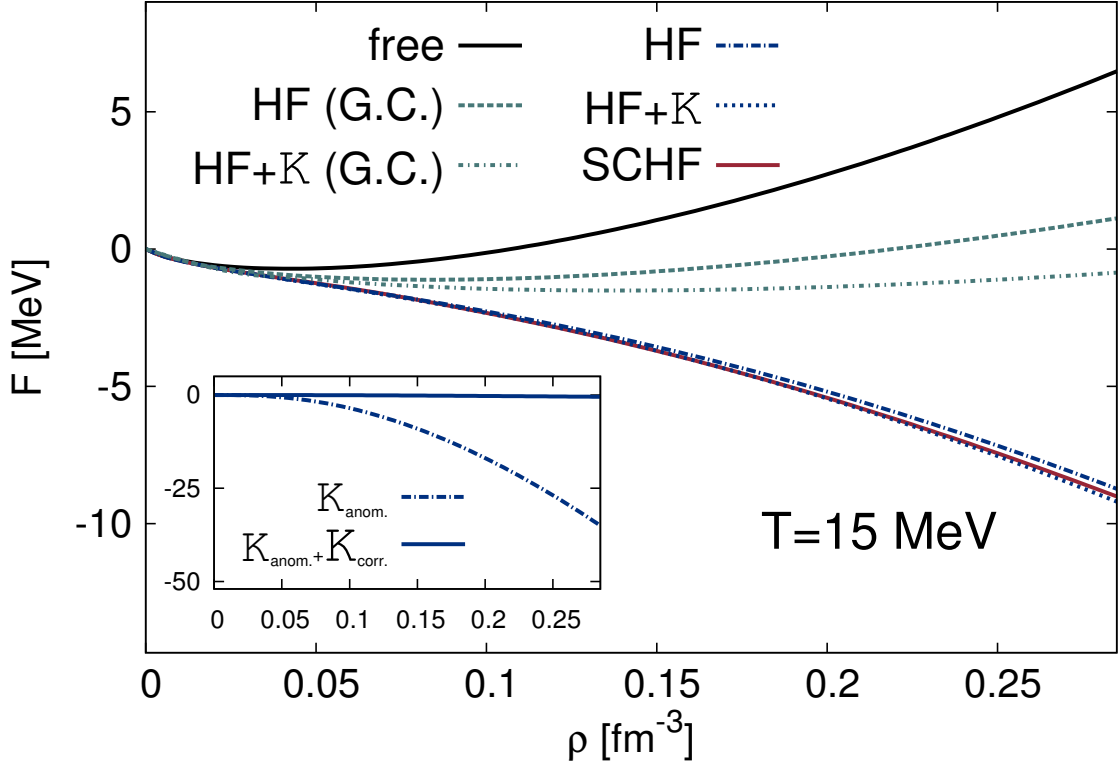

**Figure 2.18.:** Mean-field results[71] for the free energy density $F(T,\rho)$, see text for details. The inset shows the results for the second-order cumulant contributions (in the canonical framework). Only the "HF", "HF+$\mathcal{K}$" and "SCHF" curves have a nonconvex region. Note that the cumulant contributions "$\mathcal{K}_\text{anom.}$" and "$\mathcal{K}_\text{anom.} + \mathcal{K}_\text{corr.}$" (in the inset) are given in terms of the canonical perturbation series; the impact of "$\mathcal{K}_\text{anom.}$" in the grand-canonical case can be inferred from the difference of the "HF (G.C.)" and "HF+$\mathcal{K}$ (G.C.)" curves.

---

[71] We note that we have evaluated the "bare" grand-canonical perturbation series (for both SNM and PNM) also beyond the mean-field level, including the contributions from the chiral 3N interactions at N2LO [plots not shown]. The "mixed" (NN-3N) second-order anomalous contribution is then the dominant term for large values of $\mu$, respectively (cf. Fig. 3.6). This has the effect that the grand-canonical potential becomes positive at large chemical potential (roughly $\mu \gtrsim 40$ MeV), leading to a "maximal density" where the free energy density "curves back" and becomes multivalued (a completely unphysical feature).

# 3. Nuclear Many-Body Calculations

In the last chapter we have investigated the framework of many-body perturbation theory (MBPT) from a general point of view. The present chapter is now concerned with the application of MBPT in actual (numerical) nuclear many-body calculations; i.e., we want to compute the free energy per particle of infinite homogeneous nuclear matter $\bar{F}(T,\rho,\delta)$, where $\rho = \rho_n + \rho_p$ is the total nucleon density and $\delta = (\rho_n - \rho_p)/\rho$ is the isospin-asymmetry parameter ($\rho_{n/p}$ is the neutron/proton density).

As noted in the introduction, the dependence of $\bar{F}(T,\rho,\delta)$ on $\delta$ can as a first approximation be described in terms of the symmetry free energy $\bar{F}_{\text{sym}}(T,\rho) := \bar{F}(T,\rho,\delta=1) - \bar{F}(T,\rho,\delta=0)$. The first step in the problem of computing the (global) nuclear EoS is therefore to study the limiting cases $\delta = 0$ and $\delta = 1$, i.e., the EoS of isospin-symmetric nuclear matter (SNM) and pure neutron matter (PNM).

Since nuclear matter is expected to have no ferromagnetic instability [229, 228, 355], we consider only the spin-unpolarized case. Both SNM (given that isospin-symmetry breaking effects are neglected) and PNM are then essentially one-component systems. For a one-component system,[1] the ("bare") canonical perturbation series about the noninteracting Hamiltonian is given by[2]

$$\bar{F}(T,\tilde{\mu}) = \underbrace{\bar{\mathcal{F}}(T,\tilde{\mu})}_{\equiv \bar{F}_{\text{nonrel}}(T,\tilde{\mu})} + \lambda \bar{F}_1(T,\tilde{\mu}) + \lambda^2 \bar{F}_{2,\text{normal}}(T,\tilde{\mu}) + \lambda^2 \bar{F}_{2,\text{totanom}}(T,\tilde{\mu}) + O(\lambda^3), \quad (3.1)$$

where $\bar{\mathcal{F}}(T,\tilde{\mu}) \equiv \bar{F}_{\text{nonrel}}(T,\tilde{\mu})$ corresponds to the free energy per particle of a nonrelativistic free nucleon gas. As discussed in appendix A.1, (in the case of the "bare" series) it is useful to add an additional relativistic (kinematical) correction term $\bar{F}_{\text{corr}}(T,\tilde{\mu})$ to Eq. (3.1). This term is constructed via a perturbative expansion (to first order, in terms of the correlation-bond formalism) of the relativistic kinetic energy Hamiltonian $\mathcal{T}_{\text{rel}}$ about the nonrelativistic one $\mathcal{T}_{\text{nonrel}}$. The sum

$$\bar{F}_{\text{nonint}}(T,\tilde{\mu}) := \bar{F}_{\text{nonrel}}(T,\tilde{\mu}) + \bar{F}_{\text{corr}}(T,\tilde{\mu}) \quad (3.2)$$

then reproduces the free energy per particle of a relativistic free nucleon gas to high accuracy. In Eqs. (3.1) and (3.2), the auxiliary chemical potential $\tilde{\mu}$ is fixed by

$$\rho(T,\tilde{\mu}) = \int \frac{d^3k}{2\pi^3} \underbrace{\left[1 + \exp\left(\beta(k^2/(2M) - \tilde{\mu})\right)\right]^{-1}}_{\equiv n_k} = -g_\tau \alpha\, T^{3/2} \text{Li}_{3/2}(\tilde{x}). \quad (3.3)$$

where $g_\tau \in \{1,2\}$ is the isospin multiplicity, $\alpha = 2^{-1/2}(M/\pi)^{3/2}$ with $M \simeq 938.9\,\text{MeV}$ the average nucleon mass, $\tilde{x} = -\exp(\beta\tilde{\mu})$, and $\text{Li}_\nu(x) = \sum_{k=1}^{\infty} k^{-\nu} x^k$ is the polylogarithm of index $\nu$. Note that, for notational convenience, we have introduced a new notation for the (noninteracting) Fermi-Dirac distribution functions, i.e., "$n_k$" and "$\bar{n}_k$", where $\bar{n}_k = 1 - n_k$.

---

[1] For isospin-asymmetric matter, i.e., for a two-component system, there are two different auxiliary "chemical potentials" $\tilde{\mu}_n$ and $\tilde{\mu}_p$.
[2] We restrict the numerical computations to second-order MBPT in this thesis.

## 3. Nuclear Many-Body Calculations

In Eq. (3.1), the second-order "anomalous" (or higher-cumulant) contribution $\bar{F}_{2,\text{totanom}}(T,\tilde{\mu})$ is a purely thermal correction, i.e., $\bar{F}_{2,\text{totanom}} \xrightarrow{T\to 0} 0$. The canonical-ensemble approach is essential for a consistent treatment of such anomalous contributions. In particular, the "resummation" of these contribution to all orders leads to the self-consistent renormalization of the single-particle energies and the auxiliary chemical potential in the Fermi-Dirac distributions (cf. Sec. 2.5). In a grand-canonical approach, only the single-particle energies would be renormalized, which implies that these "resummations" have a much larger impact in that case.

Notably, in Sec. 2.5.5 we have found that in the mean-field approximation (with two-body interactions only) the effect of the resummation of anomalous contributions is negligible. This motivates the use of the unrenormalized canonical perturbation series for an initial study of the nuclear many-body problem, i.e, without fully addressing convergence issues and theoretical uncertainties.

The great advantage of the "bare" (unrenormalized) series is that no self-consistency requirements are involved. In particular, Eq. (3.3) can be easily inverted with respect to $\tilde{\mu}(T,\rho)$, enabling computations at fixed nucleon density $\rho$. In addition, the thermodynamic derivatives of the various terms in Eq. (3.1) can in principle all be calculated directly, i.e., by evaluating the explicit expressions obtained for the derivatives.[3] The effect of using self-consistently renormalized distribution functions should however be examined in more quantitative detail in future studies.

Motivated by the above discussion, in this chapter we investigate the application of chiral low-momentum two- and three-nucleon potentials in perturbative many-body calculations of the thermodynamic EoS of isospin-symmetric nuclear matter (SNM) and pure neutron matter (PNM), using unrenormalized canonical MBPT. In detail, the present chapter is organized as follows.

- In **Section 3.1** we discuss the partial-wave representation of the first- and second-order two-body contributions.

- In **Sec. 3.2** we discuss the approximative treatment of the (reducible) second-order contributions from three-body interactions in terms of an effective in-medium two-body potential.

- In **Sec. 3.3** we then examine the dependence of the results for the equation of state of SNM on the nuclear potential models, and benchmark the results against available empirical constraints.

- Selecting among the potentials those which describe best the empirical constraints, we study in more detail the thermodynamic EoS of SNM and PNM in **Sec. 3.4**.

- Finally, in **Sec. 3.5** we extract, from the SNM and PNM results, the symmetry free energy, entropy and internal energy, and study their density and temperature dependence.

---

[3] For the perturbative contributions an explicit evaluation of the derivatives is however not practical in terms of effort, since they can also be extracted numerically (using finite differences) when the numerical precision is high enough. It would be much more difficult to achieve the sufficient precision in a self-consistent scheme.

## 3.1. Partial-Wave Expansion

The numerical evaluation of the perturbative contributions from the NN potential can be facilitated by using partial-wave representation of the potential, i.e., the representation with respect to coupled states $|pJm_JLSTm_T\rangle$. The change from two-particle plane-wave states $|\vec{p}\,\sigma_1\tau_1\sigma_2\tau_2\rangle$ to partial-wave $|pJm_JLSTm_T\rangle$ states is done in the following steps:

$$|\vec{p}\,\sigma_1\tau_1\sigma_2\tau_2\rangle \xrightarrow{(i)} |\vec{p}\,Sm_STm_T\rangle \xrightarrow{(ii)} |pLm_LSm_STm_T\rangle \xrightarrow{(iii)} |pJm_JLSTm_T\rangle. \qquad (3.4)$$

The step (*i*) amounts to simple spin coupling and isospin coupling, step (*ii*) is associated with the expansion

$$|\vec{p}\rangle = \sum_{L,m_L} 4\pi i^L Y^*_{Lm_L}(\theta_p,\varphi_p)\,|pLm_L\rangle, \qquad (3.5)$$

and step (*iii*) corresponds to total angular momentum coupling. The spherical harmonics $Y_{L,m_L}(\theta,\varphi)$ are given by

$$Y_{L,m_L}(\theta,\varphi) := \sqrt{\frac{(2L+1)}{4\pi}\frac{(L-m_L)!}{(L+m_L)!}}\,P_L^{m_L}(\cos\theta)\,e^{im_L\varphi}, \qquad (3.6)$$

where $P_L^{m_L}(\cos\theta)$ are the associated Legendre polynomials. Carrying out these steps one arrives at

$$|\vec{p}\,\sigma_1\tau_1\sigma_2\tau_2\rangle = \sum_{J,L,S,T}\sum_{m_J,m_L,m_S,m_T} 4\pi i^L C^{Jm_J}_{Lm_LSm_S}C^{Sm_S}_{\sigma_1\sigma_2}C^{Tm_T}_{\tau_1\tau_2}|pJm_JLSTm_T\rangle, \qquad (3.7)$$

where $C^{Jm_J}_{Lm_LSm_S} \equiv \langle Jm_JLS|Lm_LSm_S\rangle$ are Clebsch-Gordan coefficients.[4] By the Wigner-Eckhart theorem and assuming charge independence, the partial-wave matrix elements of the (antisymmetrized) NN potential are given by[5]

$$\langle pJm_JLSTm_T|\bar{V}_{\text{NN}}|p'J'm_{J'}L'S'T'm_{T'}\rangle \equiv \delta_{J,J'}\delta_{m_J,m_{J'}}\delta_{S,S'}\delta_{T,T'}\delta_{m_T,m_{T'}}\,\langle p|\bar{V}_{\text{NN}}^{J,L,L',S,T}|p'\rangle, \qquad (3.8)$$

With respect to $|pJm_JLSTm_T\rangle$ states the antisymmetrizer is given by $1 - P_{12} = 1 - (-1)^{L+S+T}$. Thus, the nonvanishing partial-wave channels are suject to the condition

$$L + S + T \in \mathbb{O}\,(\text{odd numbers}). \qquad (3.9)$$

The "coupled channels" with components that are off-diagonal with respect to the angular momentum quantum number $L$ occur only in channels with $S = 1$ and $J = L + 1$, and are subject to

$$(L, L') \in \{(L, L), (L, L+2), (L+2, L), (L+2, L+2)\}. \qquad (3.10)$$

In the following, we give the partial-wave expressions for the perturbative many-body contributions at first and at second order, for the case of isospin-symmetric nuclear matter.[6]

---

[4] For simple spin and isospin coupling, the single-particle (iso)spins $s, t = 1/2$ are suppressed for notational convenience.

[5] For details regarding the partial-wave matrix elements of the different components of the NN potential (central, spin, spin-orbit, tensor, etc.) we refer to Refs. [139, 236].

[6] The corresponding expressions for pure-neutron matter (PNM) are given by Eqs. (3.16), (3.21) and (3.31) without the factor $(2T + 1)$, in Eq. (3.19) a factor $(2T + 1)/2$ is missig, in Eq. (3.30) a factor 2 is missing, and in Eq. (3.27) the partial-wave sum is restricted to $T = 1$ and $m_T = -1$ in the PNM case.

### 3. Nuclear Many-Body Calculations

***First-Order Contribution.*** The first-order (Hartree-Fock) contribution from the NN potential to the free energy density is given by

$$F_1^{\text{NN}}(T,\tilde{\mu}) = \frac{1}{2} \sum_{\sigma_1,\tau_1} \oint \frac{d3k_1}{(2\pi)^3} \sum_{\sigma_2,\tau_2} \oint \frac{d3k_2}{(2\pi)^3} \, n_{k_1} n_{k_2} \, \langle \vec{p}\sigma_1\tau_1\sigma_2\tau_2 | \bar{V}_{\text{NN}} | \vec{p}\,\sigma_1\tau_1\sigma_2\tau_2 \rangle . \tag{3.11}$$

Changing the integration variables to $\vec{p} = (\vec{k}_1 - \vec{k}_2)/2$ and $\vec{K} = (\vec{k}_1 + \vec{k}_2)/2$ and performing the partial-wave expansion we obtain[7]

$$F_1^{\text{NN}}(T,\tilde{\mu}) = \frac{1}{2} \int \frac{d3p}{(2\pi)^3} \int \frac{d3K}{(2\pi)^3} \, \mathcal{J}_{[\vec{k}_1,\vec{k}_2;\vec{p},\vec{K}]} \, n_{|\vec{K}-\vec{p}|} n_{|\vec{K}+\vec{p}|} \sum_{J,L,L',S,T} \langle p | \bar{V}_{\text{NN}}^{J,L,L',S,T} | p \rangle (4\pi)^2 i^{L-L'} \sum_{m_J,m_L,m_{L'},m_S,m_T}$$

$$\times \sum_{\sigma_1,\tau_1,\sigma_2,\tau_2} C^{\dagger\, Jm_J}_{Lm_L S m_S} C^{Jm_J}_{Lm_{L'} S m_S} C^{\dagger\, S m_S}_{\sigma_1 \sigma_2} C^{S m_S}_{\sigma_1 \sigma_2} C^{\dagger\, T m_T}_{\tau_1 \tau_2} C^{T m_T}_{\tau_1 \tau_2} Y_{Lm_L}(\theta_p,\varphi_p) Y^*_{L'm_{L'}}(\theta_p,\varphi_p),$$

$$\tag{3.12}$$

where the Jacobian is $\mathcal{J}_{[\vec{k}_1,\vec{k}_2;\vec{p},\vec{K}]} = 8$. Using the following identities (in that order)

- $\sum_{\sigma_1,\sigma_2} C^{\dagger\, S m_S}_{\sigma_1 \sigma_2} C^{S m_S}_{\sigma_1 \sigma_2} = 1$,

- $\sum_{\tau_1,\tau_2} C^{\dagger\, T m_T}_{\tau_1 \tau_2} C^{T m_T}_{\tau_1 \tau_2} = 1$,

- $\int_{-1}^{1} d\cos\theta_p \int_0^{2\pi} d\varphi_p Y_{Lm_L}(\theta_p,\varphi_p) Y^*_{L'm_{L'}}(\theta_p,\varphi_p) = \delta_{L,L'} \delta_{m_L m_{L'}}$,

- $\sum_{m_L,m_S} C^{\dagger\, Jm_J}_{Lm_L S m_S} C^{Jm_J}_{Lm_L S m_S} = 1$,

one arrives at

$$F_1^{\text{NN}}(T,\tilde{\mu}) = \frac{2}{\pi^3} \int_0^\infty dp\, p^2 \int_0^\infty dK\, K^2 \, \Xi(p,K) \sum_{J,L,S,T} (2J+1)(2T+1) \, \langle p | \bar{V}_{\text{NN}}^{J,L,L,S,T} | p \rangle, \tag{3.13}$$

where the function $\Xi(p,K)$ is given by

$$\Xi(p,K) = \int_{-1}^{1} d\cos\theta_K \, n_{|\vec{K}-\vec{p}|} n_{|\vec{K}+\vec{p}|} = \frac{\ln(1+e^{\eta+2x}) - \ln(e^{2x}+e^\eta)}{x(e^{2\eta}-1)}, \tag{3.14}$$

with $x = \beta \frac{Kp}{2M}$ and $\eta = \beta\left(\frac{K^2+p^2}{2M} - \tilde{\mu}\right)$.

In the zero-temperature case, the integration region for $\vec{K}$ is the volume of two identical spherical caps which are given by the intersection of a sphere of radius $\kappa_F$ by two parallel planes whose distance from the center of the sphere is set by $\vec{p}$ and $-\vec{p}$. We therefore have

$$\int_{|\vec{K}\pm\vec{p}|\leq\kappa_F} d^3K = 2\left(\frac{h\pi}{6}(3a^2+h^2)\right) = \frac{2\pi}{3}(\kappa_F - p)^2(2\kappa_F + p), \tag{3.15}$$

---

[7] Note that we use the same symbol "$T$" for the temperature and for the total isospin. Note also that here $\vec{K}$ denotes the average momentum and not the total momentum as in chapter 1.

where $a = \sqrt{\kappa_F^2 - p^2}$ is the radius of the base of the caps and $h = \kappa_F - p$ is the height of the caps. The expression for the first-order NN contribution to the ground-state energy per particle is then given by

$$\bar{E}^{\text{NN}}_{0;1}(k_F) = \frac{2}{\pi} \int_0^{k_F} dp\, p^2 \left(1 - \frac{3p}{2k_F} + \frac{p^3}{2k_F^3}\right) \sum_{J,L,S} (2J+1)(2T+1)\, \langle p | \bar{V}^{J,L,L,S,T}_{\text{NN}} | p \rangle. \qquad (3.16)$$

***First-Order Self-Energy.*** The first-order contribution to the (perturbative) self-energy from the NN potential is given by

$$S^{\text{NN}}_{1,k_1}(T, \tilde{\mu}) = \sum_{\sigma_2, \tau_2} \int \frac{d^3 k_2}{(2\pi)^3}\, n_{k_2}\, \langle \vec{k}_1 \vec{k}_2\, \sigma_1 \tau_1 \sigma_2 \tau_2 | \bar{V}_{\text{NN}} | \vec{k}_1 \vec{k}_1\, \sigma_2 \tau_2 \sigma_1 \tau_2 \rangle \qquad (3.17)$$

The partial-wave expanded form is

$$S^{\text{NN}}_{1,k_1}(T, \tilde{\mu}) = \int \frac{d^3 k_2}{(2\pi)^3}\, n_{k_2} \sum_{J,L,L',S,T} \langle p | \bar{V}^{J,L,L',S,T}_{\text{NN}} | p \rangle\, (4\pi)^2 i^{L-L'} \sum_{m_J, m_L, m_{L'}, m_S, m_T} \sum_{\sigma_2, \tau_2}$$
$$\times C^{\dagger\, Jm_J}_{Lm_L S m_S} C^{Jm_J}_{L'm_{L'} S m_S} C^{\dagger\, Sm_S}_{\sigma_1 \sigma_2} C^{Sm_S}_{\sigma_1 \sigma_2} C^{\dagger\, Tm_T}_{\tau_1 \tau_2} C^{Tm_T}_{\tau_1 \tau_2} Y_{Lm_L}(\theta_p, \varphi_p) Y^*_{L'm_{L'}}(\theta_p, \varphi_p). \qquad (3.18)$$

This expression can be simplified by noting that $C^{\dagger\, Sm_S}_{\sigma_1 \sigma_2} C^{Sm_S}_{\sigma_1 \sigma_2}$ takes the same values for each value of $\sigma_1$. We can thus average with respect to $\sigma_1$, leading to

- $\frac{1}{2} \sum_{\sigma_1, \sigma_2} C^{\dagger\, Sm_S}_{\sigma_1 \sigma_2} C^{Sm_S}_{\sigma_1 \sigma_2} = \frac{1}{2}$

The angle $\theta_p$ is completely determined by $k_2$ and $\theta_{k_2}$, therefore we can use $\varphi_{k_2} = \varphi_p$ and find

- $\int_0^{2\pi} d\varphi_{k_2} Y_{Lm_L}(\theta_p, \varphi_p) Y^*_{L'm_{L'}}(\theta_p, \varphi_p) = 2\pi \mathcal{Y}_{Lm_L}(\theta_p, \varphi_p) \mathcal{Y}^*_{L'm_L}(\theta_p, \varphi_p) \delta_{m_L, m_{L'}}$,

where $\mathcal{Y}_{L,m}(\theta)$ denotes the spherical harmonics without the azimuthal part $e^{im_L\varphi}$. Using the identities (in that order)

- $\sum_{m_J, m_S} C^{\dagger\, Jm_J}_{Lm_L S m_S} C^{Jm_J}_{L'm_L S m_S} = \frac{2J+1}{2L+1} \delta_{LL'}$,

- $\sum_{m_L} \mathcal{Y}_{Lm_L}(\theta_p, \varphi_p) \mathcal{Y}^*_{Lm_L}(\theta_p, \varphi_p) \frac{1}{2L+1} = \frac{1}{4\pi}$,

- $\sum_{\tau_2, m_T} C^{\dagger\, Tm_T}_{\tau_1 \tau_2} C^{Tm_T}_{\tau_1 \tau_2} = \frac{2T+1}{2}$,

one then finds

$$S^{\text{NN}}_{1,k_1}(T, \tilde{\mu}) = \frac{1}{4\pi} \int_0^\infty dk_2\, k_2^2\, n_{k_2} \int_{-1}^1 d\cos\theta_{k_2} \sum_{J,L,S} (2J+1)(2T+1) \left\langle p \left| \bar{V}^{J,L,L,S,T}_{\text{NN}} \right| p \right\rangle, \qquad (3.19)$$

where $p = |\vec{k}_1 - \vec{k}_2|/2$.

3. Nuclear Many-Body Calculations

*Second-Order Normal Contribution.* The expression for the second-order normal contribution from the NN potential is given by

$$F_{2,\text{normal}}^{\text{NN}}(T,\tilde{\mu}) = -\frac{1}{8} \left( \prod_{i=1}^{4} \sum_{\sigma_i,\tau_i} \int \frac{d^3k_i}{(2\pi)^3} \right) \langle 12|\bar{V}_{\text{NN}}|34\rangle \langle 34|\bar{V}_{\text{NN}}|12\rangle \frac{n_{k_1}n_{k_2}\bar{n}_{k_3}\bar{n}_{k_4} - \bar{n}_{k_3}\bar{n}_{k_4}n_{k_1}n_{k_2}}{\frac{k_3^2}{2M} + \frac{k_4^2}{2M} - \frac{k_1^2}{2M} - \frac{k_2^2}{2M}}$$
$$\times (2\pi)^3 \delta(\vec{k}_1 + \vec{k}_2 - \vec{k}_3 - \vec{k}_4), \tag{3.20}$$

where $|12\rangle \equiv |\vec{p}_1 \sigma_1 \tau_1 \sigma_2 \tau_2\rangle$ and $|34\rangle \equiv |\vec{p}_2 \sigma_3 \tau_3 \sigma_4 \tau_4\rangle$, with $\vec{p}_1 = (\vec{k}_1 - \vec{k}_2)/2$ and $\vec{p}_2 = (\vec{k}_3 - \vec{k}_4)/2$. The partial-wave representation of this expression is given by[8]

$$F_{2,\text{normal}}^{\text{NN}}(T,\tilde{\mu}) = -\frac{8}{\pi^2}M \int_0^\infty dp_1\, p_1^2 \int_{-1}^{1} d\cos\theta_1 \int_0^\infty dp_2\, p_2^2 \int_{-1}^{1} d\cos\theta_2 \int_0^\infty dK\, K^2 \frac{\mathcal{G}(p_1,p_2,K,\theta_1,\theta_2)}{p_2^2 - p_1^2}$$
$$\times \sum_{J,L_1,L_2,J',L_1',L_2,S} i^{L_2-L_1} i^{L_1'-L_2'} \langle p_1|\bar{V}_{\text{NN}}^{J,L_1,L_2,S,T}|p_2\rangle \langle p_2|\bar{V}_{\text{NN}}^{J',L_2',L_1',S,T}|p_1\rangle (2T+1) \sum_{m_J,m_S,m_S'} C(\theta_1,\theta_2). \tag{3.21}$$

where the function $C(\theta_1,\theta_2)$ is a product of spherical harmonics and Clebsch-Gordan coefficients:

$$C(\theta_1,\theta_2) = \mathcal{Y}_{L_1,(m_J-m_S)}(\theta_1)\mathcal{Y}_{L_2,(m_J-m_S')}(\theta_2)\mathcal{Y}_{L_2',(m_J-m_S')}(\theta_2)\mathcal{Y}_{L_1',(m_J-m_S)}(\theta_1)$$
$$\times C^{Jm_J}_{L_1(m_J-m_S)Sm_S} C^{Jm_J}_{L_2(m_J-m_S')Sm_S'} C^{J'm_J}_{L_2'(m_J-m_S')Sm_S'} C^{J'm_J}_{L_1'(m_J-m_S)Sm_S}, \tag{3.22}$$

where again $\mathcal{Y}_{L,m}(\theta)$ denotes the spherical harmonics without the azimuthal part $e^{im\varphi}$. The other function $\mathcal{G}(p_1,p_2,K,\theta_1,\theta_2)$ is given by

$$\mathcal{G}(p_1,p_2,K,\theta_1,\theta_2) = n_{|\vec{K}+\vec{p}_1|}n_{|\vec{K}-\vec{p}_1|}\bar{n}_{|\vec{K}+\vec{p}_2|}\bar{n}_{|\vec{K}-\vec{p}_2|} - \bar{n}_{|\vec{K}+\vec{p}_1|}\bar{n}_{|\vec{K}-\vec{p}_1|}n_{|\vec{K}+\vec{p}_2|}n_{|\vec{K}-\vec{p}_2|}, \tag{3.23}$$

where the angles $\theta_{1,2}$ are measured with respect to $\vec{K}$. Note that using the effective-mass approximation $\varepsilon_k + S^{\text{NN}}_{1;k} \simeq k^2/(2M^*) + U$, the effect of renormalizing the energy-denominator in terms of the first-order perturbative self-energy[9] amounts to multiplying Eq. (3.21) with a factor $M^*/M$.

*Second-Order Anomalous Contribution.* The second-order anomalous contribution from the NN potential is comprised of two terms, i.e.,

$$F_{2,\text{totanom}}^{\text{NN}}(T,\tilde{\mu}) = F_{2,\text{anomalous}}^{\text{NN}}(T,\tilde{\mu}) + F_{2,\text{corr.-bond.}}^{\text{NN}}(T,\tilde{\mu}), \tag{3.24}$$

where the first term corresponds to the second-order anomalous diagram (Fig. 2.4), or equivalently, to the equal-index $(\Gamma_1)^2$ contribution. It is given by

$$F_{2,\text{anomalous}}^{\text{NN}}(T,\tilde{\mu}) = -\frac{\beta}{2} \left( \prod_{i=1}^{3} \sum_{\sigma_i,\tau_i} \int \frac{d^3k_i}{(2\pi)^3} \right) n_{k_1}n_{k_2}\bar{n}_{k_2}n_{k_3} \langle 12|\bar{V}_{\text{NN}}|12\rangle \langle 23|\bar{V}_{\text{NN}}|23\rangle. \tag{3.25}$$

---

[8] Here, the particular form of the expression that results from the partial-wave expansion depends on the choice of the coordinate system. We choose a coordinate system where the average momentum $\vec{K} = (\vec{k}_1 + \vec{k}_2)/2 = (\vec{k}_3 + \vec{k}_4)/2$ is fixed in the $z$-direction.

[9] This corresponds to the resummation of normal "one-loop" insertions, which at zero-temperature is equivalent to the change from "bare" MBPT to Hartree-Fock perturbation theory, cf. Sec. 2.3 for details. Note also that in Sec. 2.5.5 we have found that at $T=15$ MeV the difference between the "bare" effective mass (perturbative self-energy) and the self-consistent one (self-consistent self-energy) is rather small, cf. Fig. 2.17.

3. Nuclear Many-Body Calculations

This can be written as

$$F^{\text{NN}}_{2,\text{anomalous}}(T,\tilde{\mu}) = -\frac{\beta}{2} \sum_{\sigma_2,\tau_2} \oint \frac{d3k_2}{(2\pi)^3} n_{k_2}\bar{n}_{k_2} \prod_{a\in\{1,3\}} \left( \sum_{\sigma_a\tau_a} \oint \frac{d3k_a}{(2\pi)^3} n_{k_a} \langle a2|\bar{V}_{\text{NN}}|a2\rangle \right). \quad (3.26)$$

The partial-wave representation of this expression is given by

$$F^{\text{NN}}_{2,\text{anomalous}}(T,\tilde{\mu}) = -\frac{16}{\pi^2}\beta \int_0^\infty dk\, k^2\, n_k\, \bar{n}_k$$

$$\times \left[ \int_0^\infty dp\, p^2 \sum_{J,L,L',S} i^{L-L'} \left\langle \tfrac{p}{2} \middle| \bar{V}^{J,L,L',S,T}_{\text{NN}} \middle| \tfrac{p}{2} \right\rangle \int_{-1}^1 d\cos\theta_p\, n_{|\vec{p}+\vec{k}|} \sum_{m_J,m_S,m_T} C'(\theta_p) \right]^2, \quad (3.27)$$

where $\theta_p \sphericalangle (\vec{p},\vec{k})$, and

$$C'(\theta_p) = \mathcal{Y}_{L,(m_J-m_S)}(\theta_p)\mathcal{Y}_{L',(m_J-m_S)}(\theta_p)\, C^{Jm_J}_{L(m_J-m_S),S m_S}\, C^{JM}_{L'(m_J-m_S),S m_S}\, C^{S m_S}_{1/2(m_S-1/2),S m_S}\, C^{T m_T}_{1/2(m_T-1/2),S m_S}. \quad (3.28)$$

The expression for the second-order correlation-bond contribution reads [cf. Eq. (2.196)][10]

$$F^{\text{NN}}_{2,\text{corr.-bond}}(T,\tilde{\mu}) = \left(\frac{\partial F^{\text{NN}}_1(T,\tilde{\mu})}{\partial \tilde{\mu}}\right)^2 \left[\frac{\partial^2 \mathcal{A}(T,\tilde{\mu})}{\partial \tilde{\mu}^2}\right]^{-1}, \quad (3.29)$$

where $\mathcal{A}(T,\tilde{\mu})$ is the expression for the noninteracting *nonrelativistic* grand-canonical potential density, cf. Eq. (2.29), evaluated with $\tilde{\mu}$ as the chemical potential. The term $\partial^2\mathcal{A}(T,\tilde{\mu})/\partial\tilde{\mu}^2$ is given by

$$\frac{\partial^2 \mathcal{A}(T,\tilde{\mu})}{\partial \tilde{\mu}^2} = -\frac{\partial \rho(T,\tilde{\mu})}{\partial \tilde{\mu}} = -\frac{2M}{\pi^2}\int_0^\infty dp\, n_p = 2\alpha\, T^{1/2} \text{Li}_{1/2}(\tilde{x}), \quad (3.30)$$

and from Eq. (3.13) one gets

$$\frac{\partial F^{\text{NN}}_1(T,\tilde{\mu})}{\partial \tilde{\mu}} = \frac{2}{\pi^3} \int_0^\infty dp\, p^2 \int_0^\infty dK\, K^2 \frac{\partial \Xi(p,K)}{\partial \tilde{\mu}} \sum_{J,L,S}(2J+1)(2T+1)\, \langle p|\bar{V}^{J,L,L,S,T}_{\text{NN}}|p\rangle. \quad (3.31)$$

We note that the $\tilde{\mu}$ derivative of $\Xi(p,K)$ can be given in closed form:

$$\frac{\partial \Xi(p,K)}{\partial \tilde{\mu}} = \frac{\beta}{x}\left(2\frac{\ln(1+e^{\eta+2x})-\ln(e^{2x}+e^\eta)}{(e^\eta-e^{-\eta})^2} + \frac{e^\eta(1-e^{4x})}{(e^{2\eta}-1)(e^\eta+e^{2x})(1+e^{\eta+2x})}\right), \quad (3.32)$$

where $x$ and $\eta$ are the same as in Eq. (3.14).

---

[10] In the Kohn-Luttinger inversion method, this contribution corresponds to the second-order "counterterm" [cf. Sec. 2.4.4].

## 3.2. Three-Body Contributions and Effective Two-Body Potential

Concerning the contributions from 3N interactions, in our nuclear many-body contributions we will calculate the first-order 3N contribution exactly. At second order, only the reducible diagrams are taken into account and evaluated in terms of an *approximative* effective in-medium two-body potential, called the "density dependent nucleon-nucleon" (DDNN) potential. This approach can be justified by the results of Refs. [232, 117], where it was found that in general the reducible contribution dominates.

*Effective Two-Body Potential.* As discussed in Sec. 2.3.7, the evaluation of reducible three-body diagrams (i.e., diagrams where all three-body vertices have at least one closed nucleon line) can be simplified by constructing an effective in-medium two-body potential, the "density dependent nucleon-nucleon" (DDNN) potential. The (antisymmetrized) DDNN potential $\bar{V}_{\text{DDNN}}$ is generated from the genuine three-nucleon potential $\bar{V}_{\text{3N}}$ by summing over the occupied (as specified by the corresponding distribution function) nucleon states associated with the closed line, cf. Eq. (2.137).

By construction, the matrix elements of the DDNN potential depend on the properties of the medium, i.e., on temperature $T$ as well as on the neutron and proton auxiliary chemical potentials $\tilde{\mu}_{\text{n/p}}$. In addition, since the nuclear medium defines a frame of reference, the DDNN potential depends also on the center-of-mass momentum $\vec{K} = (\vec{k}_1 + \vec{k}_2)/2$. This implies that the partial-wave expansion in terms of $|pJLST\rangle$ states is not applicable for $\bar{V}_{\text{DDNN}}$. To overcome this difficulty, in Ref. [206] the DDNN potential was constructed in the center-of-mass frame approximation where the dependence of $\bar{V}_{\text{DDNN}}$ on $\vec{K}$ is neglected. To test the quality of this approximation, in the following we calculate the first-order three-body contribution first exactly (using $\bar{V}_{\text{3N}}$) and then from $\bar{V}_{\text{DDNN}}$.[11]

Details regarding the construction of the DDNN potential can be found in [206, 71, 355]. The explicit expressions for the different components of the DDNN potential are given for general isospin-asymmetric nuclear matter (ANM) in the appendix A.3. We note that, in order to simplify the partial-wave expansion of the DDNN potential, the off-shell components of the DDNN potential are extrapolated from the on-shell ones following the prescription given in Ref. [206].

For each of the five sets of NN and 3N potentials of Table 1.1, the respective DDNN potential $\bar{V}_{\text{DDNN}}(\Lambda)$ is regularized using the same regulator used in the regularization of the corresponding NN potential $\bar{V}_{\text{NN}}(\Lambda)$, cf. Table 1.1 for the regulator properties.

*First-Order 3N Contribution (Exact).* The exact expression for the first-order three-body contribution to the free energy density of SNM is given by

$$F_1^{\text{3N}}(\Lambda_{\text{3N}}; T, \tilde{\mu}) = \left( \prod_{i=1}^{3} \sum_{\sigma_i, \tau_i} \int \frac{d^3k_i}{(2\pi)^3} \right) n_{k_1} n_{k_2} n_{k_3} \, f_{\text{3N}}(\vec{k}_1, \vec{k}_2, \vec{k}_3) \, \langle \mathbf{123} | \bar{V}_{\text{3N}} | \mathbf{123} \rangle. \tag{3.33}$$

The regulator $f_{\text{3N}}$ is usually taken to have the form (with $n = 2$) [191, 251]

$$f_{\text{3N}}(P, Q) = \exp\left[ -\frac{(P^2 + 3Q^2/4)^n}{(\Lambda_{\text{3N}})^{2n}} \right], \tag{3.34}$$

## 3. Nuclear Many-Body Calculations

with Jacobi momenta $\vec{P} = \frac{1}{2}(\vec{k}_1 - \vec{k}_2)$ and $\vec{Q} = \frac{2}{3}[\vec{k}_3 - \frac{1}{2}(\vec{k}_1 + \vec{k}_2)]$. For the first-order contribution from chiral N2LO 3N interactions, we have found the influence of this regulator to be negligible (see Fig. 3.3). Leaving out the regulator, the spin and isospin sums as well as the angular integrals in the expression for $F_1^{3N}(T, \tilde{\mu})$ can be carried out explicitly, leading to

$$F_1^{3N}(T, \tilde{\mu}) = \int_0^\infty dk_1 \frac{k_1}{2\pi^2} \int_0^\infty dk_2 \frac{k_2}{2\pi^2} \int_0^\infty dk_3 \frac{k_3}{2\pi^2} \left( \mathcal{K}_3^{(c_E)} + \mathcal{K}_3^{(c_D)} + \mathcal{K}_3^{(\text{Hartree})} + \mathcal{K}_3^{(\text{Fock})} \right) n_{k_1} n_{k_2} n_{k_3}, \quad (3.35)$$

where the kernels $\mathcal{K}^{(c_E)}$, $\mathcal{K}^{(c_D)}$, $\mathcal{K}^{(\text{Hartree})}$ and $\mathcal{K}^{(\text{Fock})}$ are given by [151, 55]

$$\mathcal{K}_3^{(c_E)} = -\frac{12 c_E}{f_\pi^4 \tilde{\Lambda}_\chi} k_1 k_2 k_3, \quad (3.36)$$

$$\mathcal{K}_3^{(c_D)} = \frac{3 g_A c_D}{f_\pi^4 \tilde{\Lambda}_\chi} k_3 \left( k_1 k_2 - \frac{m_\pi^2}{4} \ln \frac{m_\pi^2 + (k_1 + k_2)^2}{m_\pi^2 + (k_1 - k_2)^2} \right), \quad (3.37)$$

$$\mathcal{K}_3^{(\text{Hartree})} = \frac{3 g_A^2}{f_\pi^4} k_3 \Bigg[ 2 (c_3 - c_1) m_\pi^2 \ln \frac{m_\pi^2 + (k_1 + k_2)^2}{m_\pi^2 + (k_1 - k_2)^2} - 4 c_3 k_1 k_2$$
$$+ (c_3 - 2 c_1) m_\pi^4 \left( \frac{1}{m_\pi^2 + (k_1 + k_2)^2} - \frac{1}{m_\pi^2 + (k_1 - k_2)^2} \right) \Bigg], \quad (3.38)$$

$$\mathcal{K}_3^{(\text{Fock})} = \frac{g_A^2}{f_\pi^4 k_3} \bigg[ 3 c_1 m_\pi^2 H(k_1) H(k_2) + \left( \frac{c_3}{2} - c_4 \right) X(k_1) X(k_2) + (c_3 + c_4) Y(k_1) Y(k_2) \bigg]. \quad (3.39)$$

The functions $H(k_i)$, $X(k_i)$ and $Y(k_i)$ in the Fock-contribution are:

$$H(k_i) = k_i + \frac{k_3^2 - k_i^2 - m_\pi^2}{4 k_3} \ln \frac{m_\pi^2 + (k_i + k_3)^2}{m_\pi^2 + (k_i - k_3)^2}, \quad (3.40)$$

$$X(k_i) = 2 k_i k_3 - \frac{m_\pi^2}{2} \ln \frac{m_\pi^2 + (k_i + k_3)^2}{m_\pi^2 + (k_i - k_3)^2}, \quad (3.41)$$

$$Y(k_i) = \frac{k_i}{4 k_3} \left( 5 k_3^2 - 3 k_i^2 - 3 m_\pi^2 \right) + \frac{3 \left( k_i^2 - k_3^2 + m_\pi^2 \right)^2 + 4 m_\pi^2 k_3^2}{16 k_3^2} \ln \frac{m_\pi^2 + (k_i + k_3)^2}{m_\pi^2 + (k_i - k_3)^2}. \quad (3.42)$$

The many-body diagrams associated with the different kernels are depicted in Fig. 3.1. For pure neutron matter (PNM), the direct and exchange contributions proportional to $c_{E,D}$ cancel each other and the term proportional to $c_4$ vanishes due to its isospin structure, see Ref. [206] for details.

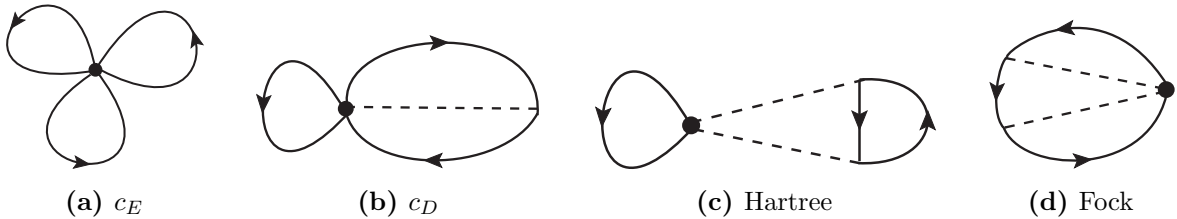

(a) $c_E$     (b) $c_D$     (c) Hartree     (d) Fock

**Figure 3.1.:** Contributions to $F_{1,3N}$ from chiral 3N interactions at N2LO in chiral effective field theory. Dashed lines represent pions.

---

[11] Recently, a new method was developed in Res. [188] where the $\vec{K}$ dependence of $\bar{V}_{\text{DDNN}}$ is averaged over. In Ref. [107] it was found that at high densities this method improves visibly upon the center-of-mass frame approximation.

### 3. Nuclear Many-Body Calculations

***Numerical Results* (3N).** The different potential sets involve (by construction) different values of $c_{E,D,1,3,4}$. It is therefore interesting to compare the components of the first-order three-body contribution proportional to different LECs. The results for the different components are shown in the left plot of Fig. 3.2 as functions of the nucleon density $\rho = \rho_n + \rho_p$ for $T = 0$ (using the n3lo500 values for $c_{E,D,1,3,4}$). One sees that the contributions from $c_3$ dominates (in SNM there is also a large contribution from $c_4$). Although the different components are different for PNM (where only $c_{1,3}$ contribute) and SNM, the (full) result for $\bar{F}_{3N}(T = 0, \rho)$ is almost the same for both cases, as shown in the inset in the left plot of Fig. 3.2. In the right plot of Fig. 3.2, the results for the five different potential sets are compared (for both SNM and PNM). One sees that for the other potentials sets the differences between the SNM and PNM results are larger: the difference is only slightly increased for n3lo414, but considerably larger for n3lo450 and in particular for VLK21 and VLK23 (the results for VLK21 and VLK23 are almost identical). The implications of the large size of the contributions from three-nucleon interactions for VLK21 and VLK23 (i.e., for the Nijmegen LECs) will be examined in detail in Sec. 3.3.

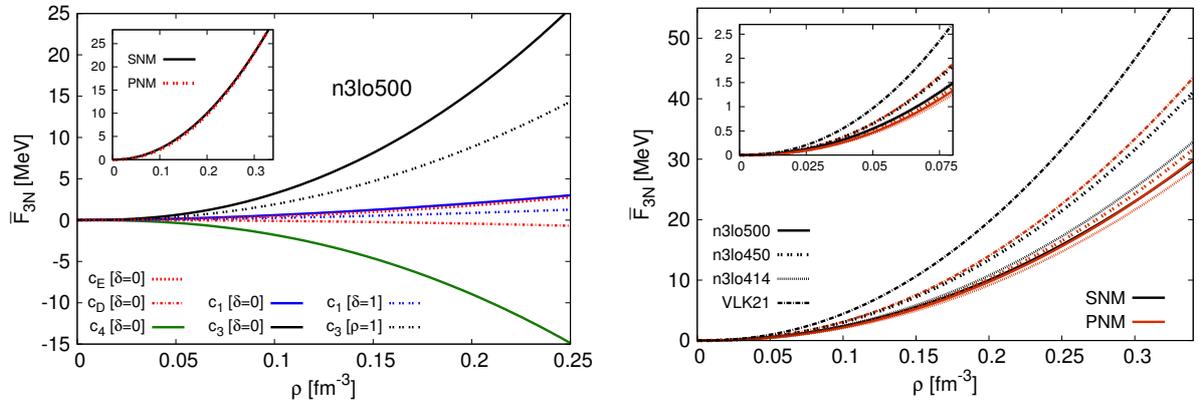

**Figure 3.2.:** Left plot: different components of the first-order contribution from the three-nucleon potential, $\bar{F}_{3N}$, evaluated for $T = 0$ and the n3lo500 LECs. The results for the components proportional to $c_1$ and $c_3$ are shown for both SNM and PNM. The inset shows the full results for SNM and PNM. Right plot: model dependence of $\bar{F}_{3N}$ at $T = 0$. The inset magnifies the behavior at low densities.

**DDNN *vs.* 3N *Results*.** To test the quality of the center-of-mass frame approximation for the DDNN potential, we compare in Figs. 3.3(a) and 3.3(b) the results (for SNM) for the first-order three-body contribution calculated using Eq. (3.35) with the results obtained using $\bar{V}_{\text{DDNN}}$ in the first-order NN contribution, Eq. (3.16). The quantity shown is the free energy per nucleon $\bar{F}(T, \rho) = \rho^{-1} F(T, \rho)$ as a function of the nucleon density $\rho$ for temperatures $T = (0, 25)$ MeV, obtained using the LECs corresponding to VLK21 and n3lo500, respectively. In each plot, the two insets magnify the behavior at low and at high densities. One sees that for sharp regulators the temperature dependence of the results obtained from Eq. (3.35) is similar to the one of the results obtained with the center-of-mass frame DDNN potential. For the relatively soft $n = 2$ regulator this is not the case in the high-density region, as is evident from Fig. 3.3(b) where the results for n3lo500 are shown. Nevertheless, the deviations are in all cases small, which suggests that it is justified to use $\bar{V}_{\text{DDNN}}$ for the computation of the second-order reducible diagrams.

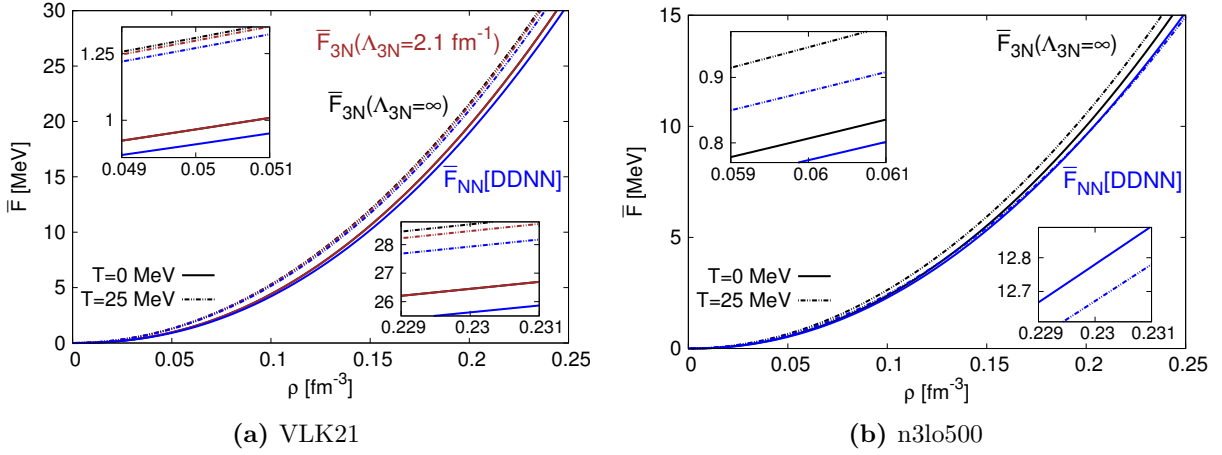

**Figure 3.3.:** First-order three-body contribution to the free energy per nucleon of SNM, calculated with genuine 3N interactions, $\bar{F}_{3N}$, and with the center-of-mass frame DDNN potential, $\bar{F}_{NN}[DDNN]$. In Fig. (a) we show also the results for the genuine 3N contribution with a Jacobi momentum regulator with $\Lambda_{3N} = 2.1\,\text{fm}^{-1}$ and $n = 10$.[12] At zero temperature the results for $\bar{F}_{3N}(\Lambda_{3N} = \infty)$ and $\bar{F}_{3N}(\Lambda_{3N} = 2.1\,\text{fm}^{-1})$ overlap (at this zoom level).

## 3.3. Model Dependence and Benchmarks

As a benchmark for both the many-body framework and the various sets of nuclear two- and three-nucleon potentials (cf. Table 1.1), we examine here the numerical results for the EoS of isospin-asymmetric nuclear matter (SNM). Special attention is paid to the influence of the contributions from three-nucleon interactions and the impact of the different second-order terms.[13]

### 3.3.1. Order-By-Order Results

In the following we examine the results for for the free energy per particle $\bar{F}(T, \rho)$ of SNM obtained at different orders in MBPT. The first- and second-order results for $\bar{F}(T, \rho)$, plotted as function of $\rho$ for $T = (0, 25)$ MeV, are shown in plots (a) and (b) of Fig. 3.4 for the case where only the two-nucleon potential is included. One sees that with NN interactions only, no saturation point, i.e., no local minimum of $\bar{E}_0(\rho) = \bar{F}(T = 0, \rho)$, is obtained, which is a well-known feature of low-momentum potentials [55]. (Contrary to "traditional", pre-$\chi$EFT ideas, cf. Refs. [42, 55]) this however must not be interpreted as a deficiency originating from the use of ("soft") low-momentum potentials, but as one that arises from the omission of 3N interactions. From the perspective of $\chi$EFT, the deficiency of the pure NN results is not surprising; without multi-nucleon interactions the description of the nuclear interaction is incomplete.

---

[12] To be precise, the regulator used to calculate the brown curves is given by $f(a, b) = \exp[-(a^2 + \tfrac{3}{4}b^2)^2/\Lambda_{3N}^4]$, where $a = \tfrac{1}{2}|k_1 - k_2|$ and $b = \tfrac{2}{3}|k_3 - \tfrac{1}{2}(k_1 + k_2)|$. As this regulator is more restrictive than the usual one where $a$ and $b$ are given by absolute values of (proper) Jacobi momenta, i.e., $a = |\vec{P}|$ and $b = |\vec{Q}|$, the effects of the latter are even smaller.

[13] The third-order ladder and ring contributions at zero temperature have been computed recently in Ref. [203] (for the ladders, see also Refs. [107, 251]). It was found that for low-momentum potentials they give only a comparatively small correction to the EoS. Since the focus of this thesis is an initial study of the full thermodynamic parameter space of the EoS, we use second-order MBPT and benchmark the results against available empirical constraints but do not consider the computationally more challenging third-order terms.

## 3. Nuclear Many-Body Calculations

Furthermore, in Fig. 3.4(a) one sees that at first order the n3lo500 two-nucleon potential gives a much smaller contribution to the free energy per particle as compared to the ("softer") n3lo414, n3lo450, VLK21 and VLK23 two-nucleon potentials, which points to the decreased perturbative quality of this NN potential. Even so, at second order (with the NN potential only) the model dependence is substantially reduced, and the results obtained with all five potential sets are in close agreement. The strong model-independence of the pure NN calculation at second order is likely coincidental; in fact, with third-order contributions included the deviations of the n3lo500 results become again more sizeable, cf. Ref. [90].

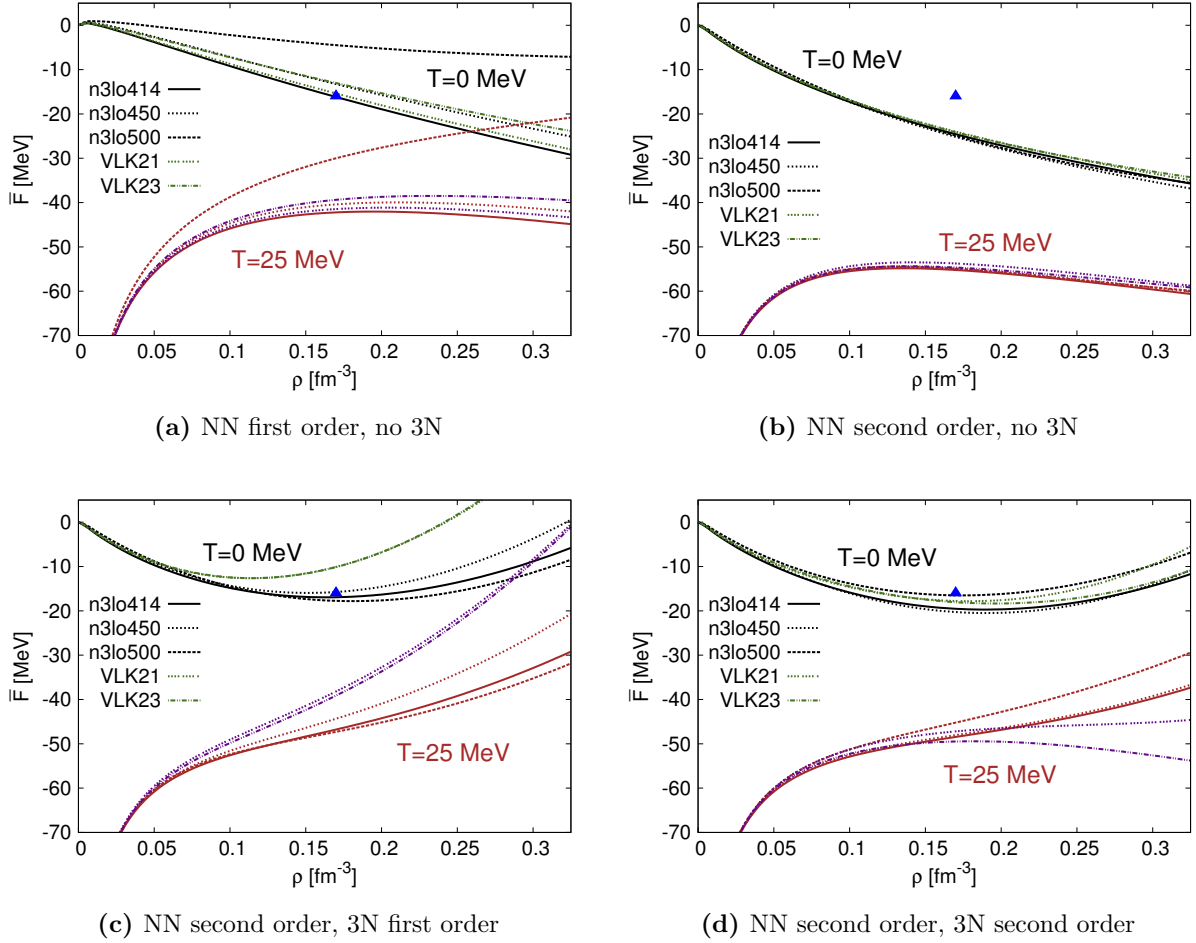

**Figure 3.4.:** Free energy per nucleon $\bar{F}(T,\rho)$ of SNM at different stages in MBPT, calculated using different sets of low-momentum NN and 3N potentials. The blue triangle marks the empirical saturation point $\bar{E}_{0,\text{sat}} \simeq -16\,\text{MeV}$, $\rho_{\text{sat}} \simeq 0.17\,\text{fm}^{-3}$ (cf. e.g., Refs. [42, 116]).

The results obtained from the different potential sets deviate again when (the chiral N2LO) 3N interactions are included at first order in MBPT, as can be seen in Fig. 3.4(c). For all potential sets, the inclusion of the first-order 3N contribution leads to a local minimum of $\bar{E}_0(\rho)$. The obtained saturation points are closer to the empirical one for the n3lo potential sets as compared to the VLK ones. The VLK results are still very similar; the deviations among the n3lo results are more sizeable, but considerably smaller than the differences between the VLK and n3lo results. The large deviations between the VLK and n3lo curves are entirely caused by the significantly larger size of the first-order 3N contribution for the VLK sets, which can be traced back to the larger value of $c_3$ in the Nijmegen LECs, cf. Figs. 3.2 and 3.3.

Finally, in Fig. 3.4(d) the second-order DDNN contributions are included. Here the results become again more model-independent, but only in the case of zero temperature. At $T = 25$ MeV the curves are now considerably flatter in the case of VLK21 and especially VLK23 as compared to the n3lo results. The reason for this behavior is the different size of the total second-order normal contribution from 3N interactions in each case (the numerical results for the different second-order contributions are examined in detail below). This contribution is much larger for the VLK potential sets; at zero temperature it balances the large first-order 3N contribution associated with the Nijmegen LECs, leading to results that are similar to those obtained with the n3lo LECs. However, because of the much more pronounced temperature dependence of the second-order DDNN contributions (as compared to the first-order three-body contribution), this balance turns into overcompensation at finite $T$, leading to the observed flattening of the $\bar{F}(T,\rho)$ curves with increasing temperature.[14] For both VLK21 and VLK23 this flattening leads to the crossing of pressure isotherms (for VLK23 the pressure $P = \rho^2 \partial \bar{F}/\partial \rho$ even becomes negative at high densities and high temperatures). Ultimately, the origin of this behavior lies in the large values of the Nijmegen LECs (in particular the large value of $c_3$).

### 3.3.2. Second-Order Contributions

To better understand the origin of the crossing of pressure isotherms found in the second-order results (for VLK21 and VLK23), we examine here in more detail the different second-order contributions for the various potential sets. The results for the different second-order normal contributions to the free energy per particle of SNM are depicted in Figs. 3.5, where $\bar{F}_{2,\text{normal}}[\text{NN}]$ denotes the contribution where both potentials are $\bar{V}_{\text{NN}}$, $\bar{F}_{2,\text{normal}}[\text{mixed}]$ is the contribution where one potential is given by $\bar{V}_{\text{NN}}$ and other one by $\bar{V}_{\text{DDNN}}(T,\rho)$, and $\bar{F}_{2,\text{normal}}[\text{DDNN}]$ denotes the case where both potentials are $\bar{V}_{\text{DDNN}}(T,\rho)$. Furthermore, we define $\bar{F}_{2,\text{normal}}[\text{total}] = \bar{F}_{2,\text{normal}}[\text{NN}] + \bar{F}_{2,\text{normal}}[\text{mixed}] + \bar{F}_{2,\text{normal}}[\text{DDNN}]$.

The results for the second-order anomalous terms are shown in Fig. 3.6 (for n3lo4500; the results for the other potential sets are similar); the contribution where both potentials are given by $\bar{V}_{\text{NN}}$ is denoted by $\bar{F}_{2,\text{anomalous}}[\text{NN}]$, the one with two $\bar{V}_{\text{DDNN}}(T,\rho)$ potentials is denoted by $\bar{F}_{2,\text{anomalous}}[\text{DDNN}]$, and the case where one interaction is given by $\bar{V}_{\text{NN}}$ and the other one by $\bar{V}_{\text{DDNN}}(T,\rho)$ is denoted by $\bar{F}_{2,\text{anomalous}}[\text{mixed}]$. As can be seen in Fig. 3.6, the size of these contributions is relatively large; in fact, in the high-density domain they give, together with the respective correlation-bond terms, the largest contributions in the perturbation series. However, the respective correlation-bond terms, the largest contributions in the perturbation series. However, the total higher-cumulant contributions, i.e., $\bar{F}_{\text{totanom}}[\ldots] = \bar{F}_{2,\text{anomalous}}[\ldots] + \bar{F}_{2,\text{corr.-bond}}[\ldots]$, are comparatively small in size and (as expected) decrease with temperature. We reiterate that the higher-cumulant terms correspond to the renormalization of the distribution functions; the large size of the (second-order) anomalous contributions and the respective correlation-bond terms as well as the substantial cancellation between these terms can be understood in terms of the effective-mass analysis of the mean-field contributions to the self-energy and the auxiliary chemical potential, cf. Sec. 2.5.5 for details.[15] In particular, the large size of the individual higher-cumulant terms reflects the inadequacy of the "conventional" grand-canonical version of MBPT.

---

[14] A similar (but more moderate, and without crossing pressure isotherms) flattening occurs also in the high-density domain of the results obtained from n3lo450 and n3lo414. It is entirely absent in the case of n3lo500, where the respective contribution is small (and has opposite sign).

*3. Nuclear Many-Body Calculations*

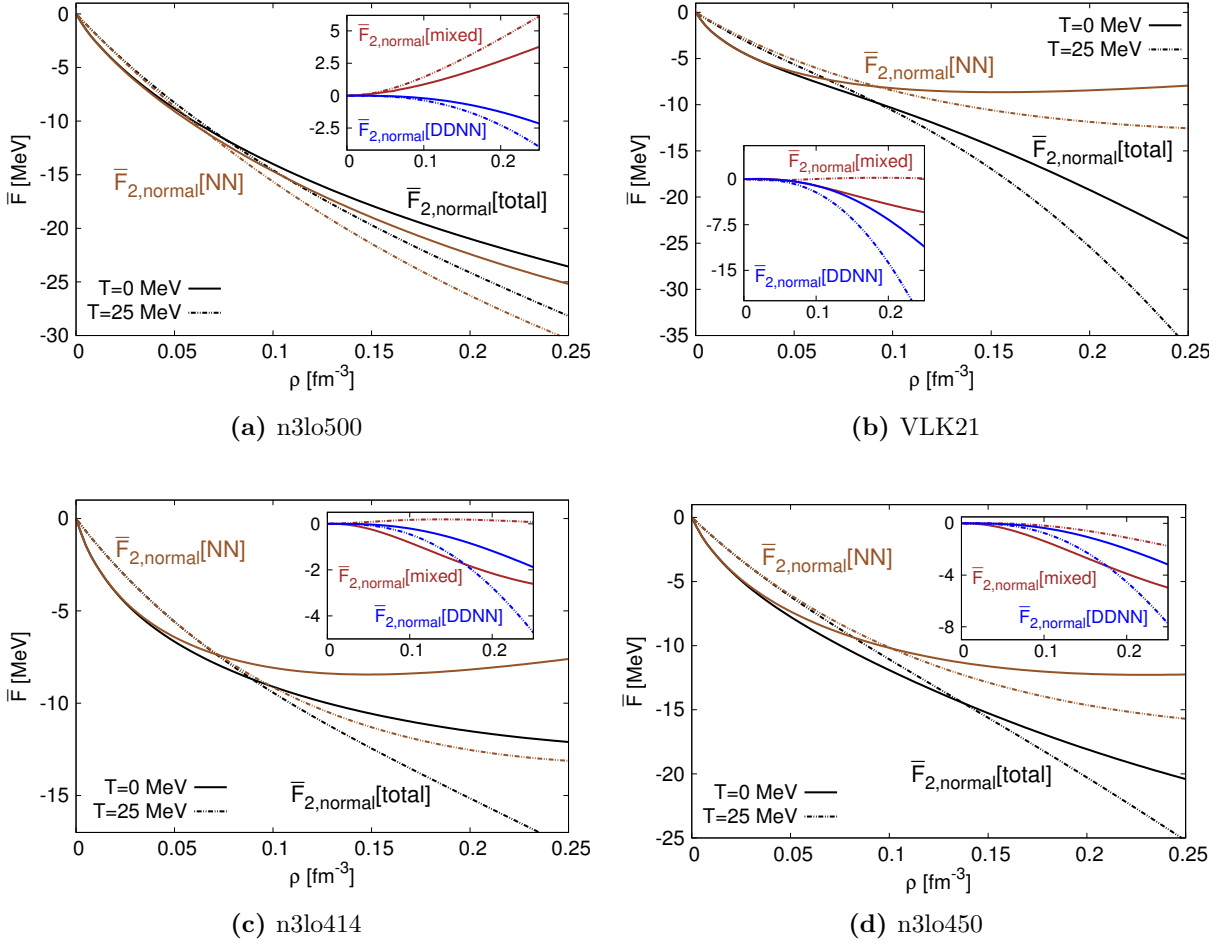

**Figure 3.5.:** Second-order normal contributions calculated for SNM from different potential sets. The insets show the contributions which arise from the temperature and density dependent DDNN potential.

In Fig. 3.5, one sees that the size of $\bar{F}_{2,\text{normal}}[\text{NN}]$ increases with the resolution scale; among the different NN potentials, n3lo500 gives rise to the largest contribution. For n3lo450 and VLK23 (not shown) as well as for n3lo414 and VLK21 the results for $\bar{F}_{2,\text{normal}}[\text{NN}]$ are very similar, and feature non-monotonic behavior with respect to the density and temperature dependence. In contrast, similar to the first-order 3N contributions, the pure DDNN contributions $\bar{F}_{2,\text{normal}}[\text{DDNN}]$ exhibit a continuous increase in magnitude with density as well as with temperature. The size of the $\bar{F}_{2,\text{normal}}[\text{DDNN}]$ contribution is noticeably larger for VLK21 (and for VLK23). The size of the sum of the total additional DDNN contributions, $\bar{F}_{2,\text{normal}}[\text{mixed}] + \bar{F}_{2,\text{normal}}[\text{DDNN}]$, is then also the largest in that case. As discussed above, for the VLK potentials the large second-order DDNN contributions counteracts the large first-order three-body contribution, which (due to the more pronounced temperature dependence of the second-order contributions) leads to the crossing of pressure isotherms. For the n3lo potentials $\bar{F}_{2,\text{normal}}[\text{DDNN}]$ is of smaller size; in the case of n3lo500 it is additionally suppressed by the mixed contribution $\bar{F}_{2,\text{normal}}[\text{mixed}]$, leading to an overall relatively small modification of the second-order normal contribution when the DDNN potential is included.

---

[15] We note that for PNM (results not shown) the size of the summed higher-cumulant contributions is even smaller ($\lesssim 0.1$ MeV in the same range of densities of temperatures as seen in Fig. 3.6), with the individual contributions (anomalous and correlation-bond) being about half as large as for SNM, corresponding to the smaller magnitude of mean-field effects in PNM.

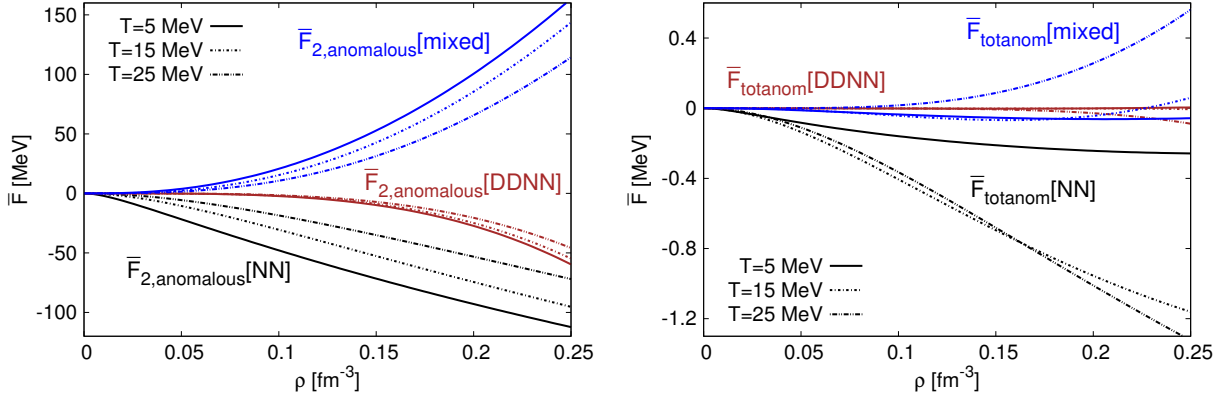

**Figure 3.6.:** (Left plot) Higher-cumulant contributions at second order arising from $\bar{V}_{\mathrm{NN}}$ and $\bar{V}_{\mathrm{DDNN}}(T,\rho)$. (Right plot) Summed second-order cumulant contributions (anomalous plus correlation bond). The results were obtained using n3lo450.

### 3.3.3. Effective-Mass Improved Results

Here, we examine the effect of including the effective-mass corrections $M^*/M$ in the evaluation of the second-order normal term. The effective-mass factors $M^*/M$ receive a contribution from both two-body and three-body interactions, i.e., their inclusion corresponds to the resummation of normal "one-loop" insertions from both two-body and three-body interactions ("two-loop insertions" in the latter case). We will find that in the case of n3lo414 and n3lo450 the inclusion of the $M^*/M$ factors leads to a better description of the saturation point. Note that *at zero temperature*[16] the $M^*/M$ results correspond to a second-order calculation in Hartree-Fock perturbation theory (HFPT), cf. the discussion of renormalized MBPT in Sec. 2.5. Since nuclear matter is subject to considerable mean-field effects, one can on general grounds consider the HFPT treatment to be more well-converged as compared to the bare second-order calculation.[17] In particular, this is supported by Ref. [352] where the zero-temperature results for n3lo414 (HFPT and bare second-order MBPT) have been compared to *configuration interaction Monte-Carlo* (CIMC) results (based on the same potential set), cf. Table IV in [352]; it was found that the $M^*/M$ improved results are indeed closer to the CIMC ones. In the case of the n3lo500 potential set, however, we will find that the $M^*/M$ factors in fact lead to a poorer description of the empirical saturation point. This behavior may be related to the decreased perturbative quality of the n3lo500 two-nucleon potential associated with the larger cutoff scale $\Lambda = 2.5$ fm$^{-3}$.

The SNM results for the temperature and density dependent effective-mass factors $M^*(T,\rho)/M$, calculated by substituting $\bar{V}_{\mathrm{NN}} + \frac{1}{2}\bar{V}_{\mathrm{DDNN}}$ for $\bar{V}_{\mathrm{NN}}$ in Eq. (3.19), are shown in Fig. 3.7. One sees that $M^*(T,\rho)/M$ decreases with density and increases with temperature and that $M^*/M \leq 1$. Hence, including the effective-mass factors leads to a reduction of the different second-order normal contributions.[18]

---

[16] HFPT corresponds to the renormalization of the single-particle spectrum in terms of the *self-consistent* first-order self-energy, which is equivalent to the inclusion of the (perturbative) $M^*/M$ correction only at $T = 0$.

[17] In particular, this suggests that the effect of the resummation of first-order higher-cumulant insertions in the second-order calculation (corresponding to second-order canonical HFPT) should be investigated in future studies. However, based on the first-order (SCHF) results of Sec. 2.5.5 as well as the moderate temperature dependence of the second-order normal contributions [cf. Eq. (2.276)], we expect no significant impact on the results.

[18] We note that higher-order contributions to the in-medium single-nucleon energies at or near zero temperature have been calculated from chiral nuclear potentials in [191, 207, 204, 73] (cf. also [41, 434]). The higher-order

*3. Nuclear Many-Body Calculations*

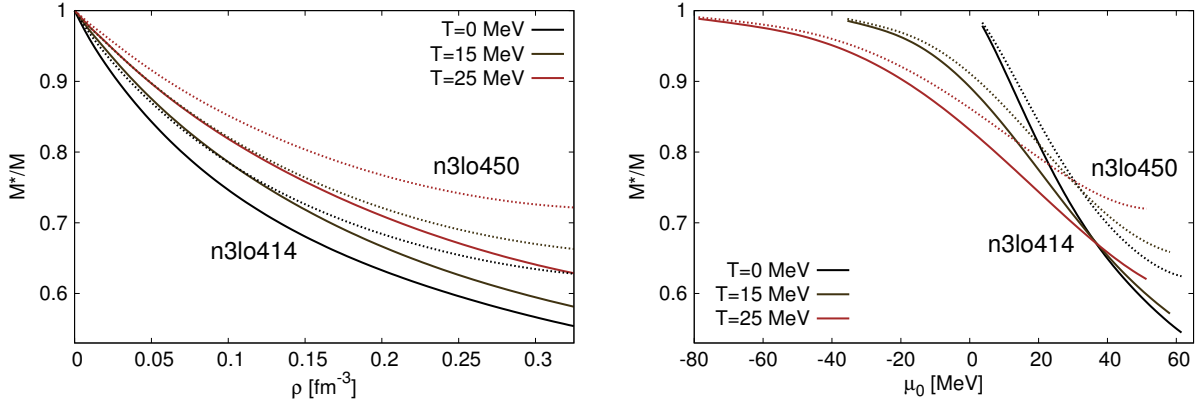

**Figure 3.7.:** Effective-mass ratio $M^*/M$ for SNM as a function of the nucleon density $\rho$ and the auxiliary chemical potential $\tilde{\mu}$ (denoted by $\mu_0$ in the plot), respectively, for different temperatures. Solid lines for n3lo414, dashed lines for n3lo450.

In Fig. 3.8 we then show the second-order results for $\bar{F}(T,\rho)$ with the effective-mass factors $M^*/M$ included. One sees that the "flatness problem" of the finite-temperature VLK results is no longer present, and at $T = 25\,\text{MeV}$ the VLK results and the ones obtained from n3lo450 and n3lo414 are in close agreement. At zero temperature nuclear matter is underbound with the VLK21 potential ($\bar{E}_{0,\text{sat}} = -12.73\,\text{MeV}$), the saturation density is somewhat small ($\rho_{\text{sat}} = 0.136\,\text{fm}^{-3}$), and the compression modulus is $K = 200\,\text{MeV}$. For VLK23 the saturation point is close to the empirical value, i.e., $\bar{E}_{0,\text{sat}} = -15.66\,\text{MeV}$ and $\rho_{\text{sat}} = 0.152\,\text{fm}^{-3}$, and the compression modulus $K = 260\,\text{MeV}$ is in agreement with empirical constraints (cf. Sec. 2.4). However, for both VLK21 and VLK23 the zero-temperature curves are now somewhat steep for densities just above saturation density, and the crossing of the pressure isotherms is therefore still present as can be seen in the second plot in Fig. 3.8.

In the case of n3lo450 and n3lo414, the pure second-order calculation resulted in nuclear matter that was slightly overbound at low temperatures. The $M^*/M$ correction reduces the large attractive contribution from the second-order normal term and improves the description of nuclear matter at zero temperature for the n3lo414 and n3lo450 potentials. By contrast, with n3lo500 the empirical saturation point is reproduced only in the bare second-order calculation without the $M^*/M$ correction (again, this agreement is likely coincidental).

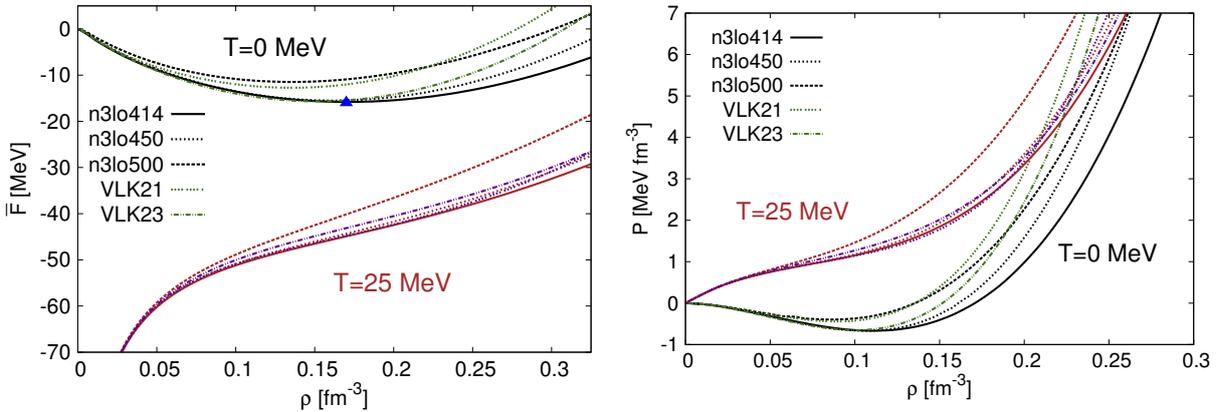

**Figure 3.8.:** Second-order results (SNM) with $M^*/M$ factors included for $\bar{F}(T,\rho)$ and $P(T,\rho)$. Only the pressure isotherms from VLK21 and the ones from VLK23 (green and purple curves) cross each other.

---

corrections increase the effective mass close to the bare mass in the vicinity of the Fermi surface.

### 3.3.4. Discussion

In this chapter, so far, we have computed the (single-phase constrained) thermodynamic EoS of isospin-symmetric nuclear matter (SNM) using NN and 3N chiral low-momentum potentials in second-order many-body perturbation theory (MBPT). As noted in Sec. 1.4, chiral nuclear potentials are not unique, and using various (sets of NN and 3N) potentials (in principle) enables systematic estmimates of theoretical uncertainties. However, one should be aware that artifacts can arise from various sources, e.g., from using potentials with unsuitable regulators or cutoff scales in MBPT (see also Refs. [117, 251] for studies of these effects), or from deficiencies in the fitting procedure carried out to fix the LECs [126].

In our calculations, we have seen that the differences in the results obtained from different potential sets are predominantly from the contributions associated with the three-body interactions, which depend sensitively on the choice of low-energy constants $c_E$, $c_D$ and $c_{1,3,4}$. The dominant three-body contributions are the ones which are proportional to $c_3$, and the crossing of pressure isotherm present in the VLK21 and VLK23 results can be linked mainly to the large value of this low-energy constant in the Nijmegen LECs.[19] This suggests that, concerning theoretical uncertainty estimates, a more systematic study of the uncertainties in the LECs may be useful. It should however be emphasized that we have considered only the leading order (N2LO) three-nucleon interactions. The N3LO multi-nucleon interactions have so far been fully included only in PNM calculations at $T = 0$ [386, 251, 106] (cf. also [231, 238]), but their implementation in SNM calculations remains a challenge at present. Both the inclusion of N3LO multi-nucleon interactions as well as the inclusion of higher-order contributions in MBPT may improve the results obtained with the VLK potentials. Moreover, the method used to determine the three-nucleon LECs (in the VKL case) does not fully (properly) account for "induced" (by the RG evolution) multi-nucleon interactions, cf. the discussion in Sec. 1.4. A more consistent RG treatment (e.g., employing SRG transformations [186, 223, 184, 154]) may also help to cure some of the deficiencies encountered with the Nijmegen LECs.

Our results may also indicate that in order to reduce the model dependence, it may be useful to use nuclear interactions based on $\chi$EFT with explicit $\Delta$ degrees of freedom. In the $\Delta$-less theory, contributions to the nuclear NN and 3N interactions from virtual $\Delta$ excitations are not included explicitly but only implicitly through the interaction vertices, i.e., parametrized by contact terms, leading to larger values of the LECs and (presumably) a larger strength of the 3N interaction in nuclear matter [210]. See also Ref. [274], where a reduction of the contribution to the nuclear EoS from 3N interactions was observed in BHF calculations with $\Delta$-improved chiral interactions.

Overall, the above considerations make clear that a quantitative analysis of theoretical uncertainties would require further investigations. In view of this, we restrict most of the subsequent computations to n3lo414 and n3lo450 because for these potential sets the many-body calculations can be considered to be free of artifacts, and for both sets the empirical saturation point is well-reproduced and thermodynamic consistency is satisfied.

---

[19] Strictly speaking, the crossing of pressure isotherms present in the VLK results is not an unphysical feature. The crossing implies a negative coefficient of thermal expansion $\alpha_\Omega = \Omega^{-1}(\partial\Omega/\partial T)_P$ at high densities; i.e., as follows from the relation $(\partial P/\partial T)_\rho = \alpha_\Omega/\kappa_T$ (where $\kappa_T$ is the isothermal compressibility, with $\kappa_T > 0$ outside the spinodal), there would be a large decrease in pressure when the temperature is increased at fixed density.

## 3.4. Thermodynamic Nuclear Equation of State

In the following we focus on the results obtained from n3lo414 and n3lo450 and examine in more detail the thermodynamic EoS of isospin-symmetric nuclear matter (SNM) and pure-neutron matter (PNM) obtained from these sets of chiral two- and three-nucleon potentials.[20] This is motivated by the results of Sec. 3.3, where we have found that for these potential sets the second-order calculation appears to be reasonable converged, and (with $M^*/M$ factors included) the empirical saturation point of isospin-symmetric nuclear matter (SNM) is reproduced to good accuracy. For comparison, we examine also the SNM results obtained from n3lo500 *without effective-mass contributions* only to study whether universal features at finite temperature can arise starting from a realistic zero-temperature EoS.

The results for the free energy per nucleon $\bar{F}$, the pressure $P$, the entropy per nucleon $\bar{S}$, and the internal energy per nucleon $\bar{E}$ of SNM ($\delta = 0$) and PNM ($\delta$) are shown in Figs. 3.9 and 3.10, respectively.[21] Note that, as a consequence of the $g_\tau$ factor in the density equation [Eq. (3.3)], at the same density higher energy scales are probed in PNM. Note also that the uncertainty bars obtained by varying the resolution scale (n3lo414 vs. n3lo450) are reduced in the case of PNM as compared to the SNM results, which is due to the decreased magnitude of nuclear interactions in PNM.

At $T = 0$ the entropy is zero and the system is in the ground state, i.e., $\bar{F}(T = 0) = \bar{E}(T = 0) = \bar{E}_0$, with $\bar{E}_0 \to 0$ for $\rho \to 0$. At finite temperature, however, $\bar{F}(T, \rho, \delta)$ diverges logarithmically ($\sim \ln \rho$) in the zero-density limit, which is entirely caused by the noninteracting contribution [cf. Eq. (A.12)]; the interaction contributions vanish for $\rho \to 0$.[22] At very low temperatures, the internal energy per particle of PNM increases monotonically with increasing density (as required by the absence of a liquid-gas instability in PNM), but otherwise there is a local minimum at finite density.

In Fig. 3.10, the green dashed lines correspond to the model-independent "virial" equation of state (VEoS) determined from neutron-neutron scattering phase shifts (cf. Sec. 3.4.2 for details). The VEoS lines end where the fugacity is $z = \exp(\mu/T) = 0.5$. PNM is thermodynamically stable, but for temperatures below a critical value $T_c$ the Eos of SNM has a region, the so-called *spinodal region*, where the free energy density $F(T, \rho)$ is a nonconvex function of $\rho$ and the system is unstable with respect to infinitesimal density fluctuations [4]. The boundary of the spinodal region (called the *spinodal*) is marked out explicitly in Fig. 3.9. The critical point is shown as a circle (full circle for n3lo414, open circle for n3lo450). Since the free energy density $F = \rho \bar{F}$ becomes convex for $\rho \to 0$ [cf. Eq. (A.13)], the zero-temperature endpoint of the low-density part of the spinodal is located at a finite value of $\rho$ (roughly, at $\rho \simeq 2 \cdot 10^{-4}$ fm$^{-3}$).

More details on the results shown in Figs. 3.9 and 3.10 are given below, i.e., in Secs. 3.4.1 and 3.4.2.

---

[20] The more involved case of isospin-asymmetric matter (ANM) is studied in chapters 3 and 4.
[21] The thermodynamic derivatives involved in the computation of the pressure and the entropy have been evaluated numerically using finite-difference methods (cf. Sec. 5.2.1).

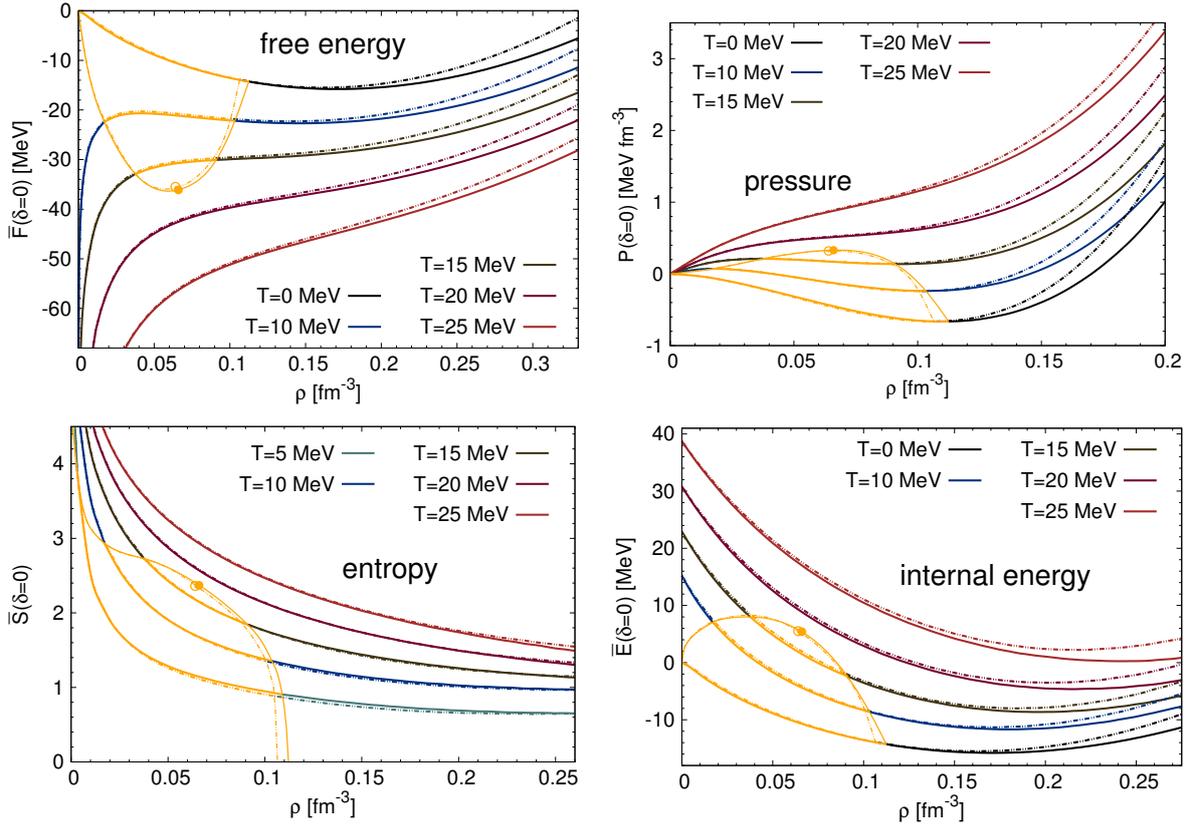

**Figure 3.9.:** Results for $\bar{F}(T,\rho)$, $P(T,\rho)$, $\bar{S}(T,\rho)$ and $\bar{E}(T,\rho)$ for SNM. Solid lines for n3lo414, dash-dot lines for n3lo450. The spinodal is marked out in yellow.

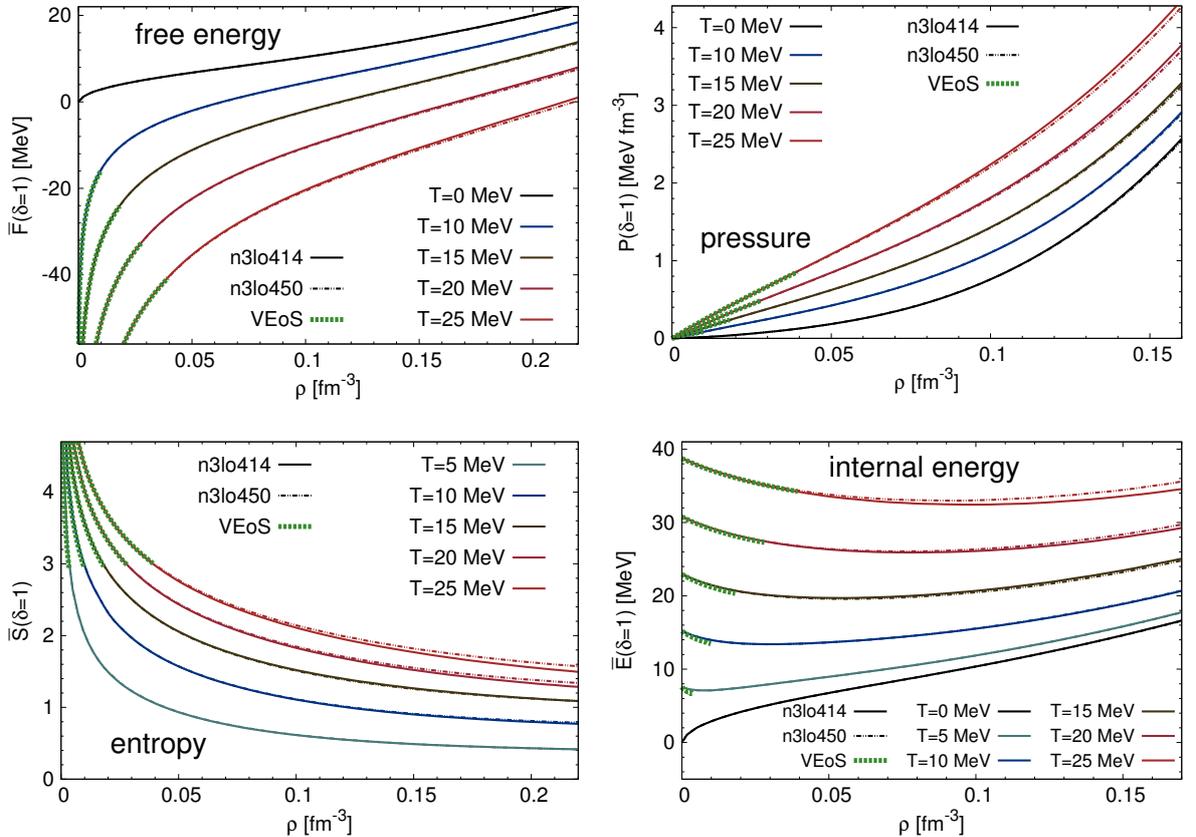

**Figure 3.10.:** Same as Fig. 3.9 but for PNM (including the VEoS results), see text for details.

### 3.4.1. Isospin-Symmetric Nuclear Matter

Here, we examine the liquid-gas phase transition of SNM, and compare the values for additional key quantities that characterize the obtained EoS, i.e., the compression modulus $K$ and the critical point coordinates $(T_c, \rho_c, P_c)$, with empirical estimates for these quantities.

*Liquid-Gas Equilibrium.* In the case of SNM, the nonconvexity of the (single-phase constrained) free energy density in the spinodal region entails that the chemical potential and pressure isotherms are nonmonotonic functions of $\rho$. This implies that (for temperatures $T < T_c$) there exist points $(\rho_\alpha(T), F_\alpha(T))$ and $(\rho_\beta(T), F_\beta(T))$ which have matching values of these quantities [denoted by $\mu_m(T)$ and $P_m(T)$] and therefore represent systems that can coexist in mutual thermodynamic equilibrium. The points $(\rho_\alpha(T), F_\alpha(T))$ and $(\rho_\beta(T), F_\beta(T))$ define a boundary (the coexistence boundary, or "binodal") that encloses the spinodal. In the interior of that boundary the equilibrium state corresponds to the coexistence of macroscopically large $\alpha$ and $\beta$ regions: the coexisting liquid and gas phases.

Between the spinodal and the binodal the (single-phase constrained) free energy density is locally convex (but not globally), corresponding to a metastable region where a finite disturbance (a "germ" or "nucleus" of the new phase) is required to induce phase separation (*nucleation*); in the interior of the spinodal, the system separates spontaneously (*spinodal decomposition*). For SNM (but not for $\delta > 0$, cf. Chap. 4) the binodal terminates at the critical point $(P_c, \rho_c, T_c)$ where the binodal touches the spinodal (cf. also Fig. 4.5), corresponding to critical behavior and a second-order transition point [323, 110]. The different parts of the region of thermodynamic instability are particularly exposed in the $P(T, \mu)$ curves. Here, the transition from metastability to the unstable region with nonconvex free energy density is marked out by sharp bends, and the single-phase constrained $P(T, \mu)$ curves are triple-valued for $T < T_c$ [double-valued at $T = 0$ where the gas phase is empty (vacuum), cf. also Sec. 4.2.2].

The values of $\mu_m(T)$ and $P_m(T)$ can be identified by constructing double tangents in the $\bar F(T, \nu)$ plots (where $\nu = 1/\rho$ is the volume per nucleon), i.e., for fixed temperature $T < T_c$ one finds values $\nu_a$ and $\nu_b$ (where $\nu_a > \nu_b$) for which[23]

$$\bar F(T, \nu_\alpha) - \bar F(T, \nu_\beta) = -P_m(T)(\nu_\alpha - \nu_\beta), \qquad \left.\frac{\partial \bar F(T, \nu)}{\partial \nu}\right|_{\nu_\alpha, \nu_\beta} = -P_m(T). \qquad (3.43)$$

The EoS corresponding to (stable) liquid-gas equilibrium is then given by substituting the single-phase constrained EoS with the double tangents, i.e., for $\rho \in [\rho_\alpha(T), \rho_\beta(T)]$ and $T < T_c$ the stable equilibrium configuration corresponds to $F(T, \rho) = \rho\, \mu_m(T) - P_m(T)$. The EoS resulting from this construction (the "Maxwell construction" [216]) is compared to the single-phase constrained results for $F(T, \rho, \delta = 0)$, $\mu(T, \rho, \delta = 0)$ and $P(\mu, T, \delta = 0)$ in Fig. 3.11. Since the Maxwell construction does not preserve the curvature of $\bar F(T, \rho)$ at the boundaries $\{\rho_\alpha(T), \rho_\beta(T)\}$ of the transition region, both $P(T, \rho)$ and $\mu(T, \rho)$ are not differentiable at these points. For $\rho \in [\rho_\alpha(T), \rho_\beta(T)]$ the chemical potential and the pressure are constant and their values given by $\mu_m(T)$ and $P_m(T)$, respectively. In the $P(T, \mu)$ diagrams the transition region thus corresponds to single points with coordinates $(P_m(T), \mu_m(T))$ [cf. also Fig. 2.16].

---

[22] The $\rho \to 0$ and the $T \to 0$ limits (of the free energy per particle) do not commute. This is plausible, since the $\rho \to 0$ at finite $T$ corresponds to an ideal classical gas, but the $T \to 0$ limit leads to the degenerate Fermi gas.

[23] Note that the Maxwell construction neglects the effect of the interface that divides the coexisting phases, corresponding to the bulk limit where the contribution from the dividing interface vanishes (cf. Sec. 4.1.1 for details). We emphasite also that the "usual" Maxwell construction is applicable only for SNM but not for ANM; in SNM neutrons and protons are thermodynamically indistinguishable (if charge-symmetry breaking effects

## 3. Nuclear Many-Body Calculations

The temperature-density phase diagrams obtained by applying the Maxwell construction to the n3lo414 and n3lo450 results (with the $M^*/M$ factors included) as well as the n3lo500 results (without $M^*/M$ factors) are shown in the fourth plot of Fig. 3.11. It should be noted that the labeling of the single-phase "liquid" and "gas" parts in the phase diagram corresponds to an only qualitative distinction in terms of a high- and a low-density fluid, respectively. Both liquids and gases have the same symmetries (spatial homogeneity and rotational isotropy), and hence are not distinguished in terms of a symmetry change. A strict distinction of the phases is possible only in the coexistence region, but any two single-phase points can (in principle) be connected by thermodynamic processes that do not cross the coexistence boundary. Concerning temperature-pressure phase diagrams one usually distinguishes "liquid", "gas", and "supercritical fluid" phases in terms of isobaric or isothermal processes that cross or do not cross (in the supercritical case) the coexistence line. For nuclear matter a more natural distinction would be one in terms self-bound states: starting from a given state in the high-density region (compressed nuclear matter), the freely (or, isentropically [301]) expanding system evolves to a (metastable) self-bound liquid state at finite density if the internal energy of the original state is not too high; states that are too energetic however evolve to "zero density".[24]

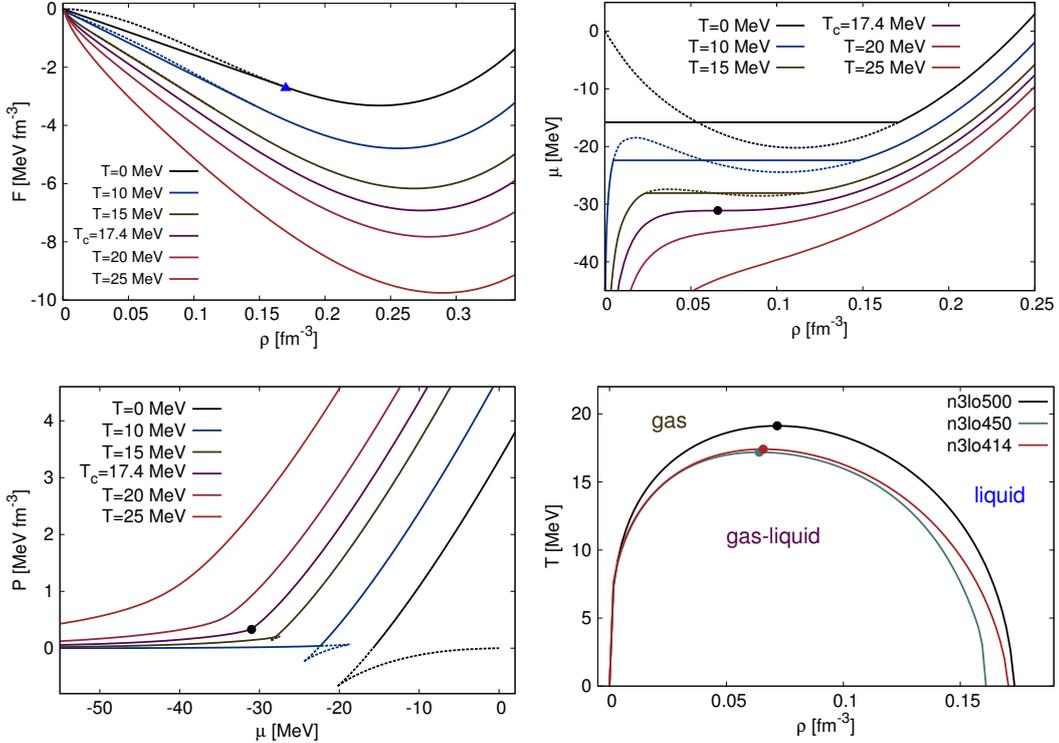

**Figure 3.11.:** Single-phase constrained (dashed & solid lines) vs. stable-equilibrium (solid lines) EoS of SNM (for n3lo414). The quantities shown are the free energy density $F(T, \rho)$, the (nucleon) chemical potential $\mu(T, \rho)$, and the pressure as function of chemical potential, $P(T, \mu)$. Also shown is the temperature-density phase diagram (for n3lo414, n3lo450 and n3lo500). The blue triangle and the (black) dot(s) mark the empirical saturation point and the obtained critical point(s), respectively.

---

are neglected), thus, thermodynamically SNM is a pure substance. ANM however is two-component system, and the "usual" Maxwell construction is not applicable to describe the "bulk" equilibrium configuration; see Chap. 4 for details.

[24] The properties of self-bound states are examined in Sec. 4.2.

*3. Nuclear Many-Body Calculations*

*"Observables"*. The obtained SNM equations of state can be characterized by a few key quantities, i.e., the saturation energy and density, the *compression modulus* (also known as the *incompressibility*, cf. [253]) $K = 9\rho^2 \partial^2 \bar{E}_0/\partial\rho^2|_{\rho=\rho_{sat}}$, and the coordinates of the critical point $(T_c, \rho_c, P_c)$. The values for these quantities obtained from the second-order calculation (including the $M^*/M$ factors) with n3lo414 and n3lo450 are shown in Table 3.1. For comparison, we also show the second-order n3lo500 results *without* $M^*/M$ factors. For each of these equations of state the saturation point coordinates ($\bar{E}_{0,sat}, \rho_{sat}$) are close to the (semi)empirical values obtained (e.g.,) from phenomenological Skyrme energy density functionals based on properties of finite nuclei [116]. The obtained values of $K$ are in overall agreement with the value $K \simeq 250 \pm 25$ MeV deduced from phenomenological mean-field models and from the analysis of nuclear giant monopole resonances (GMR) [407, 81, 243, 46, 430].[25] Estimates for the critical point have been made based on data from multifragmentation, nuclear fission, and compound nuclear decay experiments in (e.g.,) Refs. [241, 121, 308, 314]; our obtained critical point coordinates are in overall agreement with these estimates. In particular, the n3lo450 and n3lo414 results are very close to the values $T_c = 17.9 \pm 0.4$ MeV, $\rho_c = 0.06 \pm 0.02$ fm$^{-3}$, and $P_c = 0.31 \pm 0.07$ MeV fm$^{-3}$ inferred by Elliot *et al.* [121] from extrapolating multifragmentation and compound nuclear decay data to the bulk limit.

| | $\bar{E}_{0,sat}$ (MeV) | $\rho_{sat}$ (fm$^{-3}$) | $K$ (MeV) | $T_c$ (MeV) | $\rho_c$ (fm$^{-3}$) | $P_c$ (MeV fm$^{-3}$) |
|---|---|---|---|---|---|---|
| n3lo500 (no $M^*/M$) | -16.51 | 0.174 | 250 | 19.1 | 0.072 | 0.42 |
| n3lo450 ($M^*/M$) | -15.50 | 0.161 | 244 | 17.2 | 0.064 | 0.32 |
| n3lo414 ($M^*/M$) | -15.79 | 0.171 | 223 | 17.4 | 0.066 | 0.33 |

**Table 3.1.:** Saturation energy $\bar{E}_{0,sat}$ and density $\rho_{sat}$, compression modulus $K$, and the coordinates of the critical point $(T_c, \rho_c, P_c)$, corresponding to the EoS of SNM obtained from n3lo414, n3lo450 and n3lo500.[26] In the case of n3lo500 no $M^*$ corrections have been included.

### 3.4.2. Pure Neutron Matter

Here, we examine the PNM results shown in Fig. 3.10 in more detail. In particular, we discuss the comparison of the low-density PNM results with the results obtained from nonperturbative approaches to the many-body problem, i.e., the "virial" expansion and quantum Monte-Carlo methods.

*"Virial"*[27]*Expansion.* At very low energies, where higher partial waves are unimportant, the interaction between neutrons is characterized by the neutron-neutron $S$-wave scattering length $a_s$. In the regime where $a_s \simeq -18.5$ fm is large compared to the interparticle separation, $1 \ll |k_F a_s|$ (with $k_F$ the neutron Fermi momentum), a perturbative approach to neutron matter is not reliable (cf. e.g., Ref. [59]). The model-independent "virial" equation of state (VEoS) computed by

---

[25] Recently, however, a critical analysis of the extraction of $K$ from the GMR data was performed in [375], where it is claimed that $K$ should be somewhat larger, i.e., $250 \lesssim K/\text{MeV} \lesssim 315$.

[26] Note that the value of the so-called *critical compressibility factor* is $Z_c = P_c/(T_c \rho_c) \simeq 0.29$ for both n3lo414 and n3lo450. This is similar to the empirical values of $Z_c$ of various atomic or molecular fluids [97], but deviates from the value $Z_c = 0.375$ associated with equations of state of the van der Waals–Berthelot type [336].

[27] A better name would maybe be "fugacity expansion" (the usual virial expansion [292, 201] corresponds to organizing Eq. (3.44) into an expansion in powers of the density).

Horowitz and Schwenk in Ref. [213] from neutron-neutron scattering phase-shifts provides a benchmark for perturbative calculations of low-density neutron matter at nonzero temperature. The VEoS is based on the expansion of the grand-canonical expressions for the pressure and the density are expanded in powers of the fugacity $z = \exp(\mu/T)$, i.e.,

$$P(T, z, \delta = 1) = \frac{2T}{\lambda^3}\left[z + z^2 b_2(T) + O(z^3)\right], \qquad \rho(T, z, \delta = 1) = \frac{2}{\lambda^3}\left[z + 2z^2 b_2(T) + O(z^3)\right], \tag{3.44}$$

where $\mu$ is the neutron chemical potential, and $\lambda = \sqrt{2\pi/(MT)}$ is the neutron thermal wavelength. The second virial coefficient can be calculated as [213, 216]

$$b_2(T) = -\frac{\sqrt{2}}{8} + \frac{1}{\sqrt{2}\pi T}\int_0^\infty dE \, \exp[-E/(2T)]\,\delta_{\text{tot}}(E), \tag{3.45}$$

where $\delta_{\text{tot}}(E)$ is the sum of the isospin-triplet elastic scattering phase shifts at laboratory energy $E$. From the pressure and density as functions of the fugacity, the free energy per particle $\bar{F}$, the entropy per particle $\bar{S}$, and the internal energy per particle $\bar{E}$ follow again from standard thermodynamic relations (see [213] for details). The VEoS results for these quantities are shown as green dashed lines in Fig. 3.10.[28] One sees that in the case of $\bar{F}$, $P$ and $\bar{S}$ there are almost no visible deviations between the VEoS and the perturbative results. This seemingly perfect agreement is however misleading, because the discrepancies corresponding to the different treatment of the interactions in the "virial" and the perturbative approach are overpowered by the large size of the (nonrelativistic) free Fermi gas contribution. The deviations are more transparent in the case of the internal energy per particle due to cancellations of the free Fermi gas terms in the free energy and the entropy. The VEoS and perturbative results are closer at larger temperatures, since the EoS is less sensitive to the physics of large scattering lengths at higher momentum scales.

The differences between the VEoS and the perturbative results for the internal energy per particle are examined closer in Fig. 3.12 for $T = 10$ MeV. To depict the deviations more clearly we have subtracted the noninteracting contributions, i.e., the quantity shown is $\bar{E}_{\text{int}} = \bar{E} - \bar{E}_{\text{nonint}}$. The VEoS results include uncertainty bands obtained from estimating the neglected third virial coefficient as $|b_3(T)| \leq |b_2(T)|/2$. We also show the perturbative results at the Hartree-Fock (HF) level. One sees that compared to the HF results the second-order ($M^*$ improved) results are in much closer agreement with the VEoS results. The second-order calculation still slightly underpredicts the attraction present in the VEoS, in contrast to the pseudo-potential approach based on NN scattering phase shift data that was explored in Ref. [351]. We conclude that while the perturbative approach cannot fully capture the large-scattering-length physics of low-density PNM, the resulting errors are reasonably small when second-order contributions are included in MBPT.

***MBPT vs. QMC Results.*** In recent years, the zero-temperature EoS of PNM has been computed from chiral nuclear potentials within a variety of many-body frameworks [251, 424, 166, 106, 386, 192, 347, 181, 91, 13, 247, 191, 165]. In Fig. 3.13, we compare our results to ones obtained from perturbative calculations in Refs. [251, 386] (red band in Fig. 8 in [251], cf. also Fig. 1.10). In addition to the N2LO chiral three-neutron interactions, the calculations of Refs.

---

[28] Note that we have added the relativistic correction term to the VEoS lines. In particular, the zero-density limit of the internal energy per particle is given by $\bar{E}(T, \rho, \delta) \xrightarrow{\rho \to 0} \bar{E}_{\text{nonrel}}(T, \rho, \delta)|_{\rho \to 0} + \bar{E}_{\text{corr}}(T, \rho, \delta)|_{\rho \to 0} = 3T/2 + 15T^2/(8M)$.

[251, 386] also include the N3LO three- and four-neutron interactions. The uncertainty bands in their results were obtained by allowing large variations of the LECs parameterizing the multi-neutron interactions ($c_{1,3}$). One sees that the (almost overlapping) results from n3lo414 and n3lo450 lie within these bands. In Fig. 3.13 we also show results obtained from auxiliary-field quantum Monte Carlo simulations with chiral N3LO two-nucleon ("AFQMC [NN]") and N3LO two-nucleon plus N2LO three-nucleon potentials ("'AFQMC [NN+3N]") by Wlazłowski *et al.* [424]. The perturbative and the AFQMC results are very similar for densities $\rho \lesssim 0.006\,\text{fm}^{-3}$, where both are in close agreement with the fixed-node QMC calculations (based on the AV18 potential) of Gezerlis and Carlson [164]. However, at higher densities the EoS obtained from the AFQMC calculations (with 3N interactions included) is significantly more repulsive.

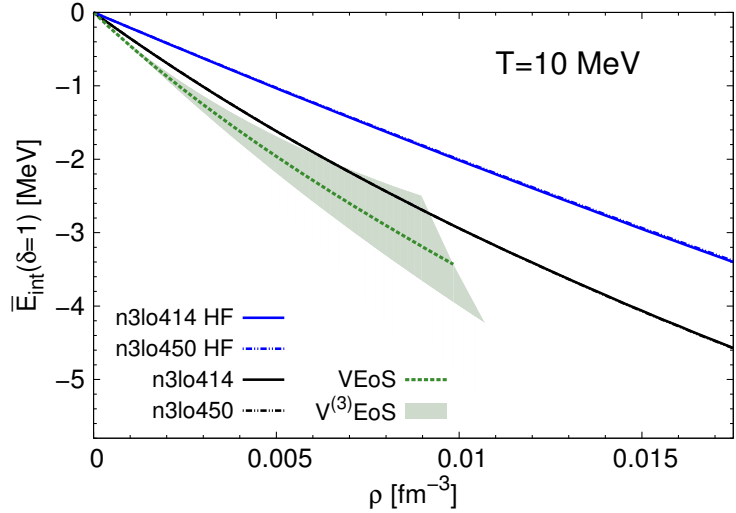

**Figure 3.12.:** Interaction contribution to the internal energy per particle of PNM at low densities. The different lines correspond to the results from n3lo414 and n3lo450 at first order (labeled "HF") and second order, as well as the VEoS at second order and with uncertainty bands obtained by estimating the third virial coefficient. The n3lo414 and n3lo450 results are almost identical.

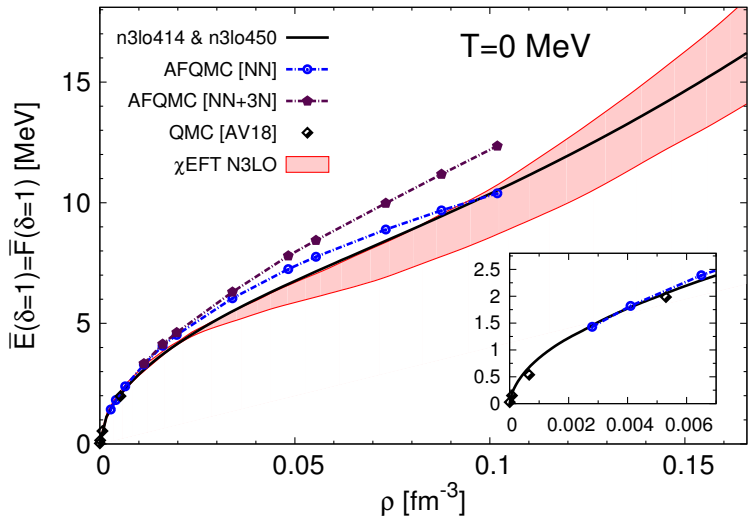

**Figure 3.13.:** Ground-state energy per particle of PNM, obtained from various many-body methods (see text for details). The inset magnifies the behavior at very low densities where quantum Monte Carlo simulations, labeled "QMC [AV18]", are expected to be most accurate.

## 3.5. Symmetry Free Energy, Entropy and Internal Energy

So far, we have considered only the limiting cases of isospin-symmetric nuclear matter (SNM) and pure neutron matter (PNM). For the description of general isospin-asymmetric nuclear matter (ANM), the isospin-asymmetry parameter $\delta = (\rho_n - \rho_p)/(\rho_n + \rho_p)$ is most useful. If charge-symmetry breaking effects are neglected, the various thermodynamic quantities are all even functions of $\delta$, and hence their Maclaurin expansion in terms of $\delta$ involves only even powers. It has been validated in numerous many-body calculations (cf. Refs. [58, 435, 434, 109, 405, 148]) that already the leading quadratic term in the expansion provides a good approximation to the exact $\delta$ dependence of the single-phase constrained nuclear EoS (at zero temperature). This implies that the EoS of ANM can as a first approximation be constructed by interpolating the SNM and PNM results, i.e., in the case of the free energy per particle it is

$$\bar{F}(T, \rho, \delta) \simeq \bar{F}(T, \rho, \delta = 0) + \bar{F}_{\text{sym}}(T, \rho)\,\delta^2, \tag{3.46}$$

and similar for (e.g.,) the entropy per particle and the internal energy per particle. In Eq. (3.46), $\bar{F}_{\text{sym}}(T, \rho) = \bar{F}(T, \rho, \delta = 1) - \bar{F}(T, \rho, \delta = 0)$ is called the *symmetry free energy*.[29] The *symmetry entropy* and the *symmetry internal energy* are related to the symmetry free energy via $\bar{S}_{\text{sym}} = -\partial \bar{F}_{\text{sym}}/\partial T$ and $\bar{E}_{\text{sym}} = \bar{F}_{\text{sym}} + T\bar{S}_{\text{sym}}$. Because of their important role in describing the overall isospin-asymmetry dependence, it is of interest to examine the density and temperature dependence of these quantities. The accuracy of Eq. (3.46) and the size of higher-order terms in the Maclaurin expansion in terms of $\delta$ will be examined in Chapter 5.

The n3lo414 and n3lo450 result (with $M^*/M$ factors) for $\bar{F}_{\text{sym}}$, $T\bar{S}_{\text{sym}}$, and $\bar{E}_{\text{sym}}$ in the left column of Fig. 3.14 as functions of $\rho$ at different temperatures are shown. In the insets we show the noninteracting contribution to these quantities, i.e.,

$$\bar{F}_{\text{nonint,sym}}(T, \rho) = \bar{F}_{\text{nonrel}}(T, \rho, 1) - \bar{F}_{\text{nonrel}}(T, \rho, 0) + \bar{F}_{\text{corr}}(T, \rho, 1) - \bar{F}_{\text{corr}}(T, \rho, 0), \tag{3.47}$$

in the case of the symmetry free energy. In the right column of Fig. 3.14 we show $\bar{F}_{\text{sym}}$, $T\bar{S}_{\text{sym}}$, and $\bar{E}_{\text{sym}}$ as functions of temperature at different densities. One sees that in the considered range of densities and temperatures, $\bar{F}_{\text{sym}}$ is a monotonic increasing function of density and temperature. The density and temperature dependence of $T\bar{S}_{\text{sym}}$ is more involved. At low densities $T\bar{S}_{\text{sym}}$ decreases monotonically with $T$, but for densities $\rho \gtrsim 0.2\,\text{fm}^{-3}$ a local minimum is found at around $T \simeq 10\,\text{MeV}$.[30] The results from n3lo414 and n3lo450 are very similar for densities well below saturation density, but at higher densities the dependence on the resolution scale becomes significant. In particular, the decrease in the slope of $\bar{F}_{\text{sym}}$ with increasing density is more pronounced in the n3lo450 results. The $T$ dependence of $T\bar{S}_{\text{sym}}$ approximately balances that of $\bar{F}_{\text{sym}}$, and as a result their sum, the symmetry internal energy $\bar{E}_{\text{sym}}$, increases with density but varies only very little with temperature. At densities near saturation density the deviations of $\bar{E}_{\text{sym}}(T, \rho \simeq \rho_{\text{sat}})$ from its value at zero temperature are below $0.5\,\text{MeV}$.[31]

---

[29] We emphasize that $\bar{F}(T, \rho, \delta = 0)$ corresponds to the *single-phase constrained* EoS of SNM, i.e., without the Maxwell construction applied. Eq. (3.46) is clearly valid only for the single-phase constrained system. The effect of the nuclear liquid-gas phase transition (as well as the presence of light nuclei at low densities, cf. also [212]) on the symmetry free energy (and the symmetry internal energy) has been examined in Ref. [399].

[30] We note that the temperature dependence of $\bar{F}_{\text{sym}}$ and $T\bar{S}_{\text{sym}}$ approaches linear behavior in the limit of vanishing density, $\bar{F}_{\text{sym}}(T, \rho \to 0) = -T\bar{S}_{\text{sym}}(T, \rho \to 0) = T \ln 2$. This follows from Eq. (5.22).

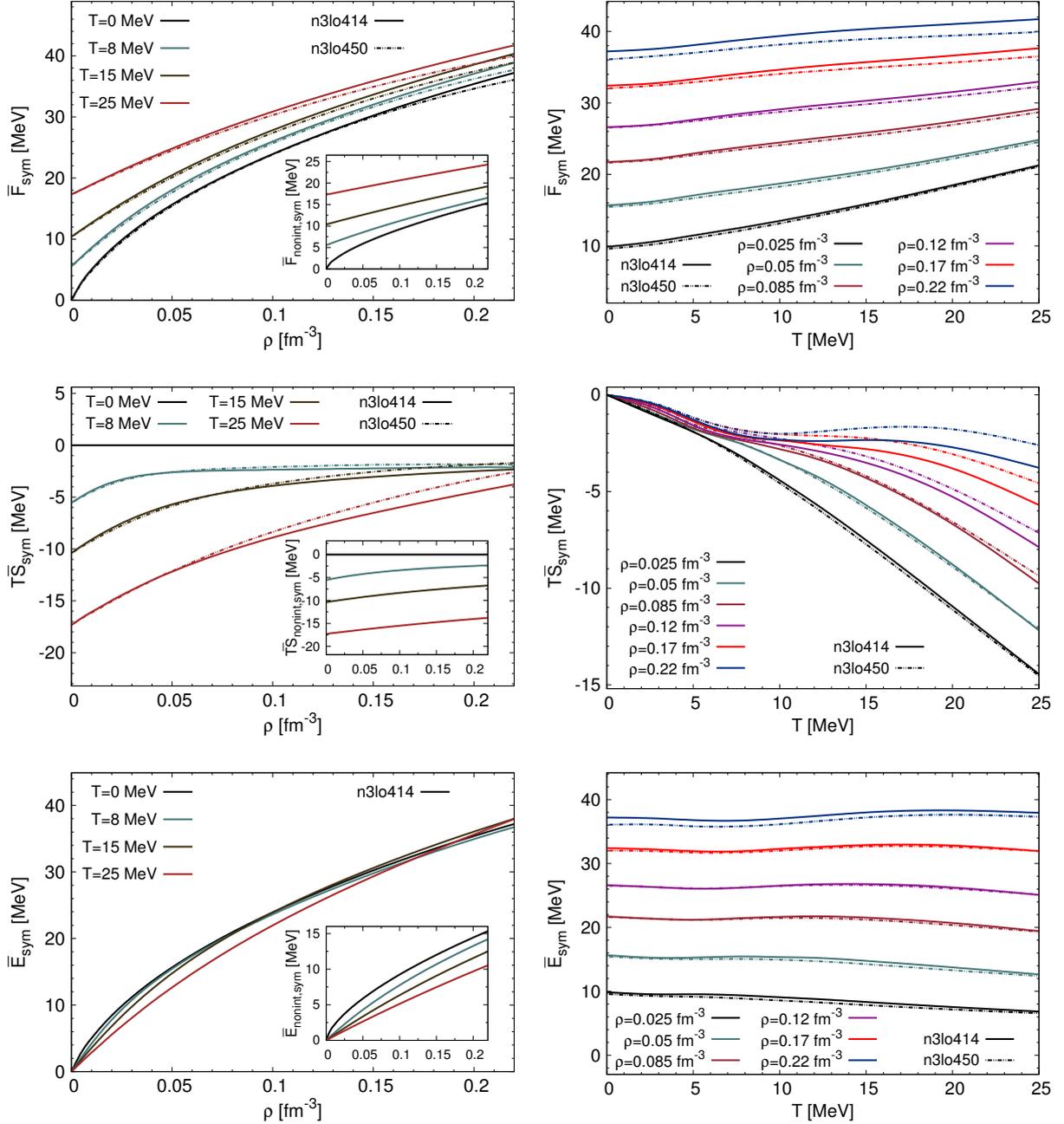

**Figure 3.14.:** Left column: results for $\bar{F}_{\text{sym}}(T,\rho)$, $T\bar{S}_{\text{sym}}(T,\rho)$ and $\bar{E}_{\text{sym}}(T,\rho)$, plotted as functions of density. The insets show the respective noninteracting contribution (free Fermi gas). Right column: $\bar{F}_{\text{sym}}(T,\rho)$, $T\bar{S}_{\text{sym}}(T,\rho)$, and $\bar{E}_{\text{sym}}(T,\rho)$ as functions of temperature at different densities. The lines are interpolated, with calculated data points at $T=(0,3,5,8,10,12,15,20,25)$ MeV.

---

[31] We note that, qualitatively, similar results (but differences can be seen) for the $T$ dependence of $\bar{F}_{\text{sym}}$ and $\bar{E}_{\text{sym}}$ were found in [399] (using a relativistic mean-field model), in contrast to the results for $\bar{F}_{\text{sym}}$ obtained within an in-medium $\chi$PT approach [144] (but note that a different definition of $\bar{F}_{\text{sym}}$ was used in Ref. [144]).

*3. Nuclear Many-Body Calculations*

In Fig. 3.15 we show the symmetry quantities with the noninteracting contributions subtracted, e.g., $\bar{F}_{\text{int,sym}}(T,\rho) = \bar{F}_{\text{sym}}(T,\rho) - \bar{F}_{\text{nonint,sym}}(T,\rho)$, as functions of $T$ at different densities.[32] In both cases the interaction contribution tends to counteract the $T$ dependence of the noninteracting contribution, as can be seen from the insets in Fig. 3.14. In the case of $\bar{F}_{\text{sym}}$ (and also $T\bar{S}_{\text{sym}}$) the noninteracting contribution dominates, but in the case of $\bar{E}_{\text{sym}}$ the size of the noninteracting contribution and the interaction contribution is more balanced, and the $T$ dependence of both contributions approximately cancels each other, leading to the observed approximate $T$ independence at densities near saturation density $\rho_{\text{sat}} \simeq 0.17 \text{ fm}^{-3}$.

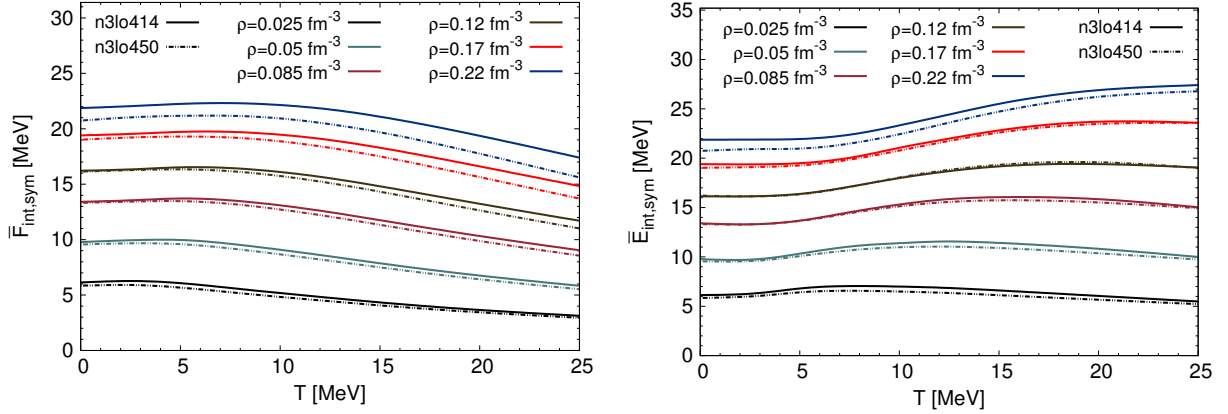

**Figure 3.15.:** Temperature dependence of $\bar{F}_{\text{sym,int}}$ and $\bar{E}_{\text{sym,int}}$ at different densities. The lines are interpolated, with calculated data points at $T = (0, 3, 5, 8, 10, 12, 15, 20, 25)$ MeV.

Finally, in Fig. 3.16 we compare our results for the symmetry (free) energy at zero temperature to the results obtained by Drischler *et al.* [109] from calculations of the EoS of neutron-rich matter using several renormalization group–evolved chiral nuclear two- and three-nucleon potentials. For comparison we also show the results of Akmal *et al.* [8] obtained from variational calculations using the AV18 two-nucleon [421] and the Urbana UIX three-nucleon potential [331].[33] While the results of Drischler *et al.* are compatible with our results, the calculations by Akmal *et al.* predict a symmetry energy that deviates visibly from the n3lo414 and n3lo450 results. In Fig. 3.16 we also show recent empirical constraints obtained from the analysis of isobaric analog states and neutron skins (IAS+NS) [94]. One sees that the n3lo414 and n3lo450 results lie in the IAS+NS bands in the entire constrained density region $0.04 \lesssim \rho/\text{fm}^{-3} \lesssim 0.16$.

For densities close to nuclear saturation density the symmetry (free) energy at zero temperature is usually [371] expanded around $J = \bar{F}_{\text{sym}}(T=0, \rho_{\text{sat}})$ in terms of $x = (\rho/\rho_{\text{sat}} - 1)/3$:

$$\bar{F}_{\text{sym}}(T=0, \rho) = J + Lx + \frac{1}{2} K_{\text{sym}} x^2 + O(x^3), \quad (3.48)$$

with $L = 3\rho_{\text{sat}} \partial \bar{F}_{\text{sym}}(T=0, \rho)/\partial\rho|_{\rho=\rho_{\text{sat}}}$ the "slope parameter", and $K_{\text{sym}} = 9\rho_{\text{sat}}^2 \partial^2 \bar{F}_{\text{sym}}(T=0, \rho)/\partial\rho^2|_{\rho=\rho_{\text{sat}}}$ the "symmetry incompressibility". The density where the ground-state energy

---

[32] We note that the temperature dependence of $\bar{F}_{\text{int,sym}}(T,\rho)$ comes almost entirely from $\bar{F}_{\text{int}}(T,\rho,\delta=0)$; the interaction contribution to the free energy per particle of pure neutron matter has only a very weak $T$ dependence (for n3lo414 and n3lo450 and for the considered range of densities and temperature).

[33] The results by Akmal *et al.* include relativistic boost corrections as well as an artificial correction term added to reproduce the empirical saturation point of SNM ("corrected" in Table VI. and "A18+$\delta v$+UIX*" in Table VII. of Ref. [8]).

## 3. Nuclear Many-Body Calculations

per particle of isospin-asymmetric nuclear matter has a local minimum is related to the parameters in the above expansion via $\rho_{\text{sat}}(\delta) \simeq \rho_{\text{sat}}[1 - 3L\delta^2/K]$ (cf. Ref. [81]).[34] The corresponding compression modulus $K(\delta)$ obeys the approximate relation

$$K(\delta) \simeq K + K_\tau \delta^2, \qquad K_\tau = K_{\text{sym}} - 6L, \tag{3.49}$$

where $K_\tau$ is usually called the "isobaric incompressiblity". The empirical values of $J = 29.0 - 32.7\,\text{MeV}$ and to a lesser degree also $L = 40.5 - 61.9\,\text{MeV}$ are relatively well constrained (values from [257], cf. also [370, 369, 263, 310, 311, 192, 386, 313]), whereas experimental extractions of $K_\tau$ suffer from large uncertainties. For instance, from measurements of neutron skin thicknesses [78] the value $K_\tau = -500^{+125}_{-100}\,\text{MeV}$ was extracted, which is compatible with the value $K_\tau = -550 \pm 100\,\text{MeV}$ inferred from the giant monopole resonance measured in Sn isotopes [270]. Theoretical studies using a selection of Skyrme interactions however led to an estimate of $K_\tau = -370 \pm 120\,\text{MeV}$ [81].[35] Our results for $J$, $L$ and $K_\tau$ are given in Table 3.2. They are in agreement with the mentioned constraints.

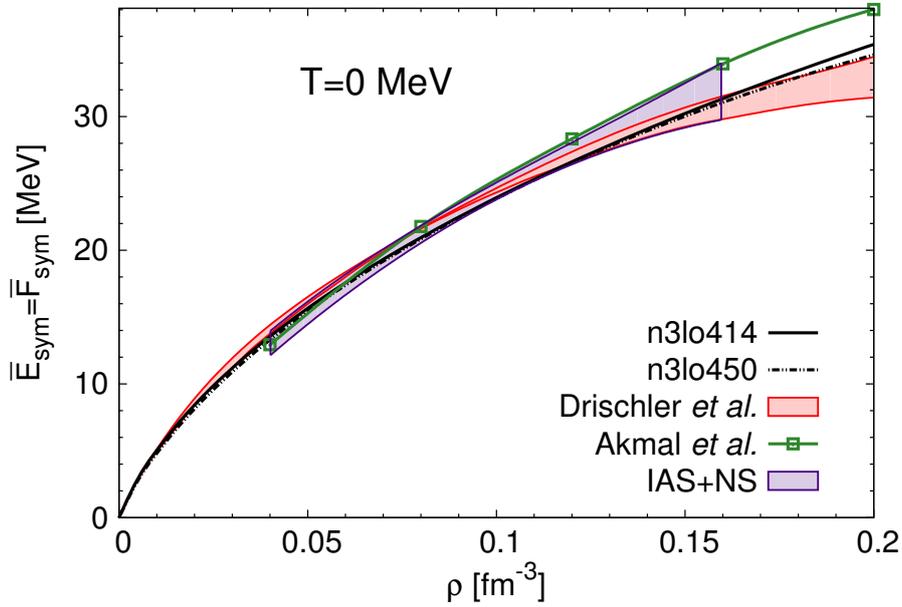

**Figure 3.16.:** Results for $\bar{F}_{\text{sym}}(T = 0, \rho)$, see text for details.

|  | $J$ (MeV) | $L$ (MeV) | $K_\tau$ (MeV) |
| --- | --- | --- | --- |
| n3lo414 | 32.51 | 53.8 | −424 |
| n3lo450 | 31.20 | 48.2 | −434 |

**Table 3.2.:** Symmetry energy at saturation density $J$, the slope parameter $L$, and the isobaric incompressibility $K_\tau$, corresponding to the n3lo414 and n3lo450 results.

---

[34] The densities $\rho_{\text{sat}}(\delta)$ correspond to stable self-bound states only for isospin asymmetries up to the neutron drip point, $\delta \leq \delta_{\text{ND}}$, cf. Secs. 4.2.2 and 4.2.3.
[35] We note that in each case a slightly different definition of $K_\tau$ is used, with the differences corresponding to higher-order terms in Eq. (3.49) and finite-size effects, cf. Ref. [81].

# 4. Nuclear Liquid-Gas Instability

In the last chapter we have obtained a "realistic" EoS of isospin-symmetric nuclear matter (SNM) and pure neutron matter (PNM) from the sets of chiral NN and 3N potentials "n3lo414" and "n3lo450". Compared to PNM, the main feature of the EoS of SNM is a region where the system is unstable with respect to the separation into two phases (liquid and gas). In this section we study how this region changes when the isospin asymmetry $\delta = (\rho_n - \rho_p)/\rho$ is increased, or equivalently, the proton fraction $Y = \rho_p/\rho = (1-\delta)/2$ is decreased.

If charge-symmetry breaking effects are neglected, SNM is effectively a pure substance with one species (nucleons).[1] In isospin-asymmetric nuclear matter (ANM), however, neutrons and protons are thermodynamically distinguishable, i.e., ANM is a mixture of two species. It is a generic feature of first-order liquid-gas phase transitions in mixtures that the composition of the coexisting phases deviates from the global composition. In the case of isospin-asymmetric nuclear matter this feature is often denoted as *isospin distillation* [426, 12, 111, 113]: in the transition region the system separates into two phases whose proton concentrations deviate from the global value of $Y$, with $0 \leq Y_\text{gas} < Y$ and $Y < Y_\text{liquid} < 0.5$ for the case $Y < 0.5$. These distillation effects are observed in intermediate-energy heavy-ion collision (cf. e.g., [426, 397, 273, 268]). As a consequence of the distillation effects in ANM, the "usual" Maxwell construction (cf. Sec. 3.4.1) is not applicable for $\delta \neq 0$, and the study of the liquid-gas phase transition of ANM is more involved compared to the simpler case of SNM. For this reason, we first review in **Section 4.1** the general concepts and principles involved in the thermodynamic analysis of a liquid-gas phase transition in mixtures. The discussion is based on Refs. [99, 296, 3, 34, 35, 33, 110].

In **Section 4.2** we then apply these concepts to examine the isospin-asymmetry dependence of the nuclear liquid-gas instability. We determine the trajectory of the critical temperature as a function of the isospin asymmetry, and study the properties of self-bound states. It should be emphasized that we are dealing with (bulk) nucleonic matter *without Coulomb interactions* [255]. The "switching off" of the Coulomb interaction is necessary to really have a first-order liquid-gas phase transition (coexistence of two bulk phases). For (finite) nuclear matter with Coulomb interactions, the liquid-gas phase transition is replaced by an instability with respect to the formation of finite-sized clusters, i.e., (possibly deformed) nuclei [23, 398, 399, 212, 328, 80], leading to the observed fragmentation events in intermediate-energy heavy-ion collisions. In the case of neutron stars (and supernova cores), also the presence of electrons (and muons) has to be taken into account. In particular, in the outer crust of neutron stars, the equilibrium configuration is given by a nuclear crystal lattice of iron ($^{56}$Fe) atoms. Further into the crust, the iron atoms are fully-ionized owing to the higher density ("pressure ionization"). In the inner regions of the crust, more neutron-rich nuclei appear as a result of electron captures ("neutronization"), and at densities $\rho \simeq \rho_{sat}/3$, neutrons drip out of the nuclei (cf. Sec. 4.2.2). At the bottom of the crust (i.e., at the boundary between inhomogeneous and homogeneous neutron-star matter), the competition between attractive nuclear and repulsive Coulomb interactions ("frustration") is expected to lead to the formation of spatially extended inhomogeneities with nontrivial structures (so-called "pasta" shapes) [358, 357, 287, 286, 339, 183, 272] and properties similar to (ordinary) liquid crystals [326, 80] (cf. also, e.g., [79]). This is similar to the structures formed in certain organic compounds (surfactants) as a result of the interplay of hydrophobic and hydrophilic components (cf. Ref. [80] pp.34-37, and [211]). "Pastas" are expected to appear also in supernova cores (and possibly, also in intermediate-energy heavy-in collisions). For a more thermodynamically oriented analysis of the effect of the (highly incompressible) charge-neutralizing background of electrons (and muons) on the properties of the nuclear liquid-gas instability, see also Ref. [112] (and [304, 114]). For general discussions of the thermodynamics and statistical mechanics of finite systems with long-range interactions, see e.g., Refs. [84, 321].

---

[1] More precisely, nuclear matter is a (peculiar type of) *azeotrope*, with the special feature that the proton fraction at the azeotropic point is (by charge symmetry) restricted to the value $Y_\text{azeotr.}(T,P) = 0.5$, i.e., the location of the azeotropic point is independent of the environmental parameters (e.g., temperature $T$ and pressure $P$), in contrast to usual (molecular) azeotropic substances [297]. Note also that the distillation effects in ANM have (using the Ehrenfest scheme) led the authors of Ref. [301] to the conclusion that the liquid-gas transition of ANM is a second-order phase transition, which is however invalid according to the modern classification in terms of latent heat [323] (cf. also Refs. [110, 199] for additional clarification).

Overall, the above considerations should bring into perspective that the nuclear liquid-gas phase transition (in the bulk sense) is an idealized process that does not apply to the real world without further qualification.

## 4.1. Thermodynamics of Phase Transitions

We consider an isolated macroscopic system with $K$ species (labeled $k = 1, \ldots, K$) that fills out a volume $\Omega$. If for a given configuration the entropy $S$ of the system decreases for all variations that are consistent with the isolation constraints (fixed energy $E$ and particle numbers $N_1, \ldots, N_K$, and fixed $\Omega$), then this configuration corresponds to an equilibrium state of the system, i.e.,

$$[\Delta S]_{E,\Omega,N_1,\ldots,N_K} < 0 \tag{4.1}$$

This fundamental principle [in a sense, a qualification of the (usual) second law of thermodynamics] is known as the **entropy maximum principle** [70] (cf. also Refs. [220, 221, 84] for a perspective based on information theory). If we consider a change described in terms of variables $x_1, x_2, \ldots, x_\nu$, the associated change of entropy can be expressed in terms of the Taylor expansion

$$\Delta S = \sum_{n=1}^{\infty} \frac{1}{n!} \delta^n S, \tag{4.2}$$

where the variation of order $n$ is given by

$$\delta^n S = \sum_{i_1=1}^{\nu} \cdots \sum_{i_n=1}^{\nu} \frac{\partial S}{\partial x_{i_1} \cdots \partial x_{i_n}} \delta x_{i_1} \cdots \delta x_{i_n}. \tag{4.3}$$

If the entropy is at a maximum in the original configuration, then it must be $\delta S = 0$, and $\delta^m S < 0$, where $m$ is the order of the lowest nonvanishing variation. The condition $\delta S = 0$ is called the **criterion of equilibrium**, and $\delta^m S < 0$ the **criterion of stability**. In general, for a system with a liquid-gas phase transition, the transition region is comprised of regions of metastable and unstable single-phase equilibrium, corresponding to different dynamical phase separation mechanisms: *nucleation* and *spinodal decomposition* (also called *spinodal fragmentation*) [4, 44, 3, 99, 239, 404, 83]. The surface delineating the metastable region (the *binodal* or coexistence boundary) where the stability criterion is satisfied locally (but not globally) is determined by the criterion of equilibrium (cf. [301] for a proof of this), and the criterion of stability determines the boundary of the unstable region (the *spinodal*).

## 4.1.1. Criteria for Phase Coexistence

For the case of the system containing two subsystems given by a liquid cluster $\beta$ immersed in a gas $\alpha$ (*drop system*), with dividing interface $\varsigma$, the condition of equilibrium is $\delta S = \delta S^{(\alpha)} + \delta S^{(\beta)} + \delta S^{(\varsigma)} = 0$. The relevant configuration changes correspond to heat or particle exchanges between the subsystems or changes in their relative size, as well as changes concerning the dividing interface. The expressions for $\delta S^{(\alpha)}$, $\delta S^{(\beta)}$ and $\delta S^{(\varsigma)}$ are then given by

$$\delta S^{(\alpha)} = \frac{1}{T^{(\alpha)}} \delta E^{(\alpha)} + \frac{P^{(\alpha)}}{T^{(\alpha)}} \delta \Omega^{(\alpha)} - \sum_{k=1}^{K} \frac{\mu_k^{(\alpha)}}{T^{(\alpha)}} \delta N_k^{(\alpha)}, \tag{4.4}$$

$$\delta S^{(\beta)} = \frac{1}{T^{(\beta)}} dE^{(\beta)} + \frac{P^{(\beta)}}{T^{(\beta)}} \delta \Omega^{(\beta)} - \sum_{k=1}^{K} \frac{\mu_k^{(\beta)}}{T^{(\beta)}} \delta N_k^{(\beta)}, \tag{4.5}$$

$$\delta S^{(\varsigma)} = \frac{1}{T^{(\varsigma)}} \delta E^{(\varsigma)} + \frac{P^{(\varsigma)}}{T^{(\varsigma)}} \delta \Omega^{(\varsigma)} - \frac{\sigma}{T^{(\varsigma)}} \delta A - \sum_{k=1}^{K} \frac{\mu_k^{(\varsigma)}}{T^{(\varsigma)}} \delta N_k^{(\varsigma)}, \tag{4.6}$$

where $A$ is the area and $\sigma$ the pressure (*surface tension*) associated with the dividing interface. The isolation constraint implies the following conditions:

$$\delta E = \delta E^{(\alpha)} + \delta E^{(\beta)} + \delta E^{(\varsigma)} = 0, \tag{4.7}$$

$$\delta \Omega = \delta \Omega^{(\alpha)} + \delta \Omega^{(\beta)} + \delta \Omega^{(\varsigma)} = 0, \tag{4.8}$$

$$\delta N_k = \delta N_k^{(\alpha)} + \delta N_k^{(\beta)} + \delta N_k^{(\varsigma)} = 0, \quad (k = 1, \ldots, K). \tag{4.9}$$

By the method of Lagrange multipliers one finds that the extremum of the entropy (subject to the isolation constraints) is given by the following conditions:

$$T^{(\alpha)} = T^{(\alpha)} = T^{(\varsigma)}, \tag{4.10}$$

$$\mu_k^{(\alpha)} = \mu_k^{(\alpha)} = \mu_k^{(\varsigma)}, \quad (k = 1, \ldots, K), \tag{4.11}$$

and

$$(P^{(\beta)} - P^{(\alpha)}) d\Omega^{(\beta)} + (P^{(\varsigma)} - P^{(\alpha)}) d\Omega^{(\varsigma)} - \sigma dA = 0, \tag{4.12}$$

which are generally referred to as the conditions for thermal, chemical, and mechanical equilibrium, respectively. Treating the interface as a thin layer one has $P^{(\varsigma)} \simeq P^{(\alpha)}$, and the condition for mechanical equilibrium simplifies:

$$P^{(\beta)} - P^{(\alpha)} = \sigma \frac{\partial A}{\partial \Omega^{(\beta)}}, \tag{4.13}$$

If the cluster $\beta$ is spherical in shape with radius $R$, then $\partial A/\partial \Omega^{(\beta)} = 2/R$, and we obtain

$$P^{(\beta)} - P^{(\alpha)} = \frac{2\sigma}{R}. \tag{4.14}$$

Eq. (4.14) is known as the *Laplace relation* [3]. If the subsystem $\beta$ is sufficiently large (i.e, $1/R \simeq 0$) the Laplace relation reduces to $P^{(\beta)} = P^{(\alpha)}$, which together with Eqs. (4.10) and (4.11) forms the famous **Gibbs conditions** for the mutual equilibrium of two bulk phases $\alpha$ and $\beta$.

## 4.1.2. Thermodynamic Stability Theory

To derive from the general stability criterion $\delta^m S < 0$ the stability criteria for arbitrary thermodynamic potentials, it is more convenient to use instead of the entropy maximum principle $[\Delta S]_{E,\Omega,N_1,\dots,N_K} < 0$ the equivalent **energy minimum principle** $[\Delta E]_{S,\Omega,N_1,\dots,N_K} > 0$ [99, 70]. Considering the stability of a simple (i.e., non-composite) system with respect to the separation into two phases $\alpha$ (liquid) and $\beta$ (gas), the expression for $\delta^2 E$ is given by

$$\delta^2 E = \delta^2(E^{(\alpha)} + E^{(\beta)}) = \delta^2 E^{(\alpha)} + \delta^2 E^{(\beta)}, \tag{4.15}$$

where the contribution from the interface is neglected. Using the shorthand notation

$$E^{(\alpha)}_{x_i x_j} \equiv \frac{\partial^2 E^{(\alpha)}}{\partial x_i^{(\alpha)} \partial x_j^{(\alpha)}}, \tag{4.16}$$

the expression for $\delta^2 E^{(\alpha)}$ reads

$$\delta^2 E^{(\alpha)} = E^{(\alpha)}_{SS}\left(\delta S^{(\alpha)}\right)^2 + 2 E^{(\alpha)}_{S\Omega}\delta S^{(\alpha)}\delta\Omega^{(\alpha)} + E^{(\alpha)}_{\Omega\Omega}\left(\delta\Omega^{(\alpha)}\right)^2 + \sum_{k=1}^{K}\sum_{l=1}^{K} E^{(\alpha)}_{N_k N_l}\delta N_k^{(\alpha)}\delta N_l^{(\alpha)}$$
$$+ \sum_{k=1}^{K} 2 E^{(\alpha)}_{N_k S}\delta N_k^{(\alpha)}\delta S^{(\alpha)} + \sum_{k=1}^{K} 2 E^{(\alpha)}_{N_k\Omega}\delta N_k^{(\alpha)}\delta\Omega^{(\alpha)}, \tag{4.17}$$

and similar for $\delta^2 E^{(\beta)}$. The condition that the total entropy $S$, volume $\Omega$, and particle numbers $N_1$ and $N_2$ remain unchanged gives rise to the conditions

$$\delta S^{(\alpha)} = -\delta S^{(\beta)}, \tag{4.18}$$
$$\delta\Omega^{(\alpha)} = -\delta\Omega^{(\beta)}, \tag{4.19}$$
$$\delta N_k^{(\alpha)} = -\delta N_k^{(\beta)}, \quad (k=1,\dots,K). \tag{4.20}$$

Furthermore, since the two subsystems (and hence their intensive properties) are originally identical, it is

$$\frac{E^{(\alpha)}_{x_i x_j}}{N^{(\alpha)}} = \frac{E^{(\beta)}_{x_i x_j}}{N^{(\beta)}}, \tag{4.21}$$

where $N^{(\alpha/\beta)} = N_1^{(\alpha/\beta)} + \dots + N_K^{(\alpha/\beta)}$. With these relations, the condition $\delta^2 E < 0$ is equivalent to $\delta^2 E^{(\alpha)} < 0$. Dropping the superscripts "$(\alpha)$", and introducing the notation $x_{1,\dots,K+2} \equiv S, \Omega, N_1, \dots, N_K$ and $y^{(0)}_{a,b} = E_{ab}$, we obtain as the stability criterion

$$\sum_{a=1}^{K+2}\sum_{b=1}^{K+2} y^{(0)}_{a,b}\delta x_a \delta x_b > 0. \tag{4.22}$$

By applying the sum-of-squares procedure, this becomes [34, 296]

$$\sum_{k=1}^{K+2} \frac{\mathfrak{a}_k}{\mathfrak{a}_{k-1}}\left(\delta Z_k\right)^2 > 0, \qquad \delta Z_k = \sum_{j=k}^{K+2} \frac{\mathfrak{g}_{kkj}}{\mathfrak{a}_k}\delta x_j, \tag{4.23}$$

where the coefficients $\mathfrak{a}_k$ and $\mathfrak{g}_{kkj}$ are given by $\mathfrak{a}_0 = 1$, and

$$\mathfrak{a}_k = \mathrm{Det} \begin{bmatrix} y_{1,1}^{(0)} & y_{1,2}^{(0)} & \cdots & y_{1,k}^{(0)} \\ y_{2,1}^{(0)} & y_{2,2}^{(0)} & \cdots & y_{2,k}^{(0)} \\ \vdots & \vdots & \ddots & \vdots \\ y_{k,1}^{(0)} & y_{k,2}^{(0)} & \cdots & y_{k,k}^{(0)} \end{bmatrix}, \qquad \mathfrak{g}_{kij} = \mathrm{Det} \begin{bmatrix} y_{1,1}^{(0)} & y_{1,2}^{(0)} & \cdots & y_{1,k-1}^{(0)} & y_{1,j}^{(0)} \\ y_{2,1}^{(0)} & y_{2,2}^{(0)} & \cdots & y_{2,k}^{(0)} & y_{2,j}^{(0)} \\ \vdots & \vdots & \ddots & \vdots & \vdots \\ y_{k-1,1}^{(0)} & y_{k-1,2}^{(0)} & \cdots & y_{k-1,k-1}^{(0)} & y_{k-1,j}^{(0)} \\ y_{i,1}^{(0)} & y_{i,2}^{(0)} & \cdots & y_{i,k-1}^{(0)} & y_{i,j}^{(0)} \end{bmatrix}, \quad (4.24)$$

where $i \geq k$ and $j \geq k$. Since the variations $\delta Z_k$ are all independent, and $(\delta Z_k)^2 \geq 0$, the stability condition is given by

$$\delta^2 E > 0 \quad \Leftrightarrow \quad \mathfrak{a}_{1,\ldots,K+2} > 0. \qquad (4.25)$$

Obviously, Eqs. (4.23) and (4.24) correspond to only one possible sum of squares, with the others given by permutations of the variables $S, \Omega, N_1, \ldots, N_K$ in the relation $x_{1,\ldots,K+2} = S, \Omega, N_1, \ldots, N_K$. The stability criteria given by Eq. (4.25) can be greatly simplified with the use of Legendre transformations. In fact, the ratio $\mathfrak{a}_k/\mathfrak{a}_{k-1}$ is given by (cf. Eqs. (31) and (35) of Ref. [33])

$$\frac{\mathfrak{a}_k}{\mathfrak{a}_{k-1}} = y_{kk}^{(k-1)}, \qquad (4.26)$$

where $y^{(k-1)}$ is the Legendre transform of order $k-1$ of $y^0$ into $(\xi_1, \ldots, \xi_{k-1}, x_k, \ldots, x_\nu)$ space. An important simplification is that the last term $\mathfrak{a}_\nu/\mathfrak{a}_{\nu-1} = y_{K+2,K+2}^{(K+1)}$ vanishes; for instance, for a pure substance ($K = 1$) one has (fixing the order of variables as $x_{1,2,3} = S, \Omega, N$):

$$y^{(0)}(x_1, x_2, x_3) = E(S, \Omega, N), \qquad (4.27)$$

$$y^{(1)}(\xi_1, x_2, x_3) = E(S, \Omega, N) + S\frac{\partial E(S, \Omega, N)}{\partial S} = F(T, \Omega, N), \qquad (4.28)$$

$$y^{(2)}(\xi_1, \xi_2, x_3) = F(T, \Omega, N) + \Omega\frac{\partial F(T, \Omega, N)}{\partial \Omega} = G(T, P, N). \qquad (4.29)$$

where $G = F + P\Omega$ is the Gibbs free energy. From $\partial G/\partial N = \mu$ and the Gibbs-Duhem relation $G = \mu N$ it follows that $y_{3,3}^{(2)} = 0$, and similar for a different ordering of variables. The relation $y_{K+2,K+2}^{(K+1)} = 0$ comes out automatically if the thermodynamic limit is performed, since then only $K + 1$ variables exist, i.e., $x_{1,\ldots,K+1} = s, \rho_1, \ldots, \rho_K$ (with $s = S/\Omega$ the entropy density) or $x_{1,\ldots,K+1} = \bar{S}, \bar{\Omega}_1, \ldots, \bar{\Omega}_K$ (with $\bar{S}$ the entropy per particle, and $\bar{\Omega}_k = \rho_k^{-1} = \Omega/N_k$ the volume per particle for each species). This implies that no analytic stability criterion exists for the thermodynamic potential associated with the Legendre transformation of order $K + 1$ (i.e., the grand-canonical potential). Therefore, an analytical calculation within the grand-canonical ensemble cannot produce an EoS with a spinodal instability, i.e., it cannot produce the single-phase constrained EoS of a system with a liquid-gas phase transition [cf. also Sec. 2.5.4].

The expression for $y_{kk}^{(k-1)}$ is related to the Legendre transform $y^{(k-2)}$ via [33, 34]

$$y_{k,k}^{(k-1)} = y_{k,k}^{(k-2)} - \frac{(y_{k,k-1}^{(k-2)})^2}{y_{k-1,k-1}^{(k-2)}}. \qquad (4.30)$$

From the stability criteria, Eq. (4.25), it follows that both $y_{k,k}^{(k-1)}$ and $y_{k-1,k-1}^{(k-2)}$ must be positive. Moreover, for a suitable ordering of variables, $y_{k,k}^{(k-2)} > 0$ is a stability criterion. Suppose $y_{k-1,k-1}^{(k-2)}$ decreases towards zero; then Eq. (4.25) shows that $y^{(k-1)}_{k,k}$ becomes negative before $y^{(k-2)}_{k-1,k-1}$. Hence, for a system with $K$ species the **necessary and sufficient stability criterion** is (as first derived by Beegle, Modell and Reid [34])

$$\boxed{y^{(K)}_{K+1,K+1} > 0} \qquad (4.31)$$

Since there are $(K + 2)!$ permutations of the thermodynamic variables $S, \Omega, N_1, \ldots, N_K$, Eq. (4.31) implies $(K + 2)!$ equivalent stability criteria for thermodynamic potentials associated with Legendre transformations of order $K$. Note also that Eq. (4.31) shows that the stability boundary is invariant under $K$ Legendre transformations.[2]

*Pure Substance.* For a pure substance ($K = 1$) the stability criterion is $y^{(1)}_{2,2} > 0$. This leads to the following stability critera for thermodynamic potentials

$$x_{1,2,3} = S, \Omega, N : \quad F_{\Omega\Omega} = -\left(\frac{\partial P}{\partial \Omega}\right)_{T,N} > 0, \qquad (4.32)$$

$$x_{1,2,3} = S, N, \Omega : \quad F_{NN} = -\left(\frac{\partial \mu}{\partial N}\right)_{T,\Omega} > 0, \qquad (4.33)$$

etc. Eqs. (4.32) and (4.33) imply that the system is stable if the free energy is a convex function of $N$ and $\Omega$.

*Binary Mixture.* For a two-component mixture ($K = 2$) one has $y^{(2)}_{3,3} > 0$, leading to

$$x_{1,2,3,4} = S, \Omega, N_1, N_2 : \quad G_{N_1 N_1} = \left(\frac{\partial \mu_1}{\partial N_1}\right)_{T,P,N_2} > 0, \qquad (4.34)$$

$$x_{1,2,3,4} = S, \Omega, N_2, N_1 : \quad G_{N_2 N_2} = \left(\frac{\partial \mu_2}{\partial N_2}\right)_{T,P,N_1} > 0, \qquad (4.35)$$

etc. The necessary and sufficient stability criterion for first-order Legendre transforms follows (using $k = 2$) from Eqs. (4.30) and (4.31), i.e, in the case of the free energy

$$F_{N_1 N_1} F_{N_2 N_2} - (F_{N_1 N_2})^2 > 0. \qquad (4.36)$$

In addition, the (necessary but not sufficient) criterion $y^{(1)}_{2,2} > 0$ implies that $F_{N_1 N_1} > 0$ and $F_{N_2 N_2} > 0$. Eq. (4.36) is equivalent to the condition that the determinant of the Hessian matrix $\mathcal{H}_{ij} = [\partial^2 F/(\partial N_i \partial N_j)]$ is positive, $\text{Det}[\mathcal{H}_{ij}] > 0$. The eigenvalues $\zeta_\pm$ of the Hessian matrix are given by

$$\zeta_\pm(T, \Omega, N_1, N_2) = \frac{1}{2} \text{Tr}[\mathcal{H}_{ij}] \pm \frac{1}{2} \sqrt{\left(\text{Tr}[\mathcal{H}_{ij}]\right)^2 - 4 \text{Det}[\mathcal{H}_{ij}]}. \qquad (4.37)$$

---

[2] But note that (in the single-phase constrained treatment) the properties of the critical point are in general not invariant under Legendre transformations. In particular, the isochoric heat capacity exhibits singular behavior at the critical point: $C_\Omega \sim (T - T_c)^{-\alpha}$, cf. Refs. [325, 419]. In terms of the single-phase constraint, this means that the "thermal stability" boundary where $C_\Omega^{-1} = T^{-1}(\partial^2 E/\partial S^2)_{\Omega, N_1, \ldots, N_K} = 0$ touches the spinodal at the critical point [but is otherwise in the interior of the spinodal, as is evident from Eq. (4.31)]. This feature cannot be obtained in an analytical calculation within the canonical ensemble, which is completely analogous to the impossiblity of obtaining an unstable region in an analytical calculation within the grand-canonical ensemble.

Requiring that the eigenvalues are positive give rise to the conditions

$$\zeta_- > 0 \quad \Leftrightarrow \quad \text{Det}[\mathcal{H}_{ij}] > 0, \tag{4.38}$$

$$\zeta_+ > 0 \quad \Leftrightarrow \quad \text{Tr}[\mathcal{H}_{ij}] < 0 \ \wedge \ \text{Det}[\mathcal{H}_{ij}] < -\frac{1}{2}\big(\text{Tr}[\mathcal{H}_{ij}]\big)^2. \tag{4.39}$$

The system is stable if both eigenvalues $\zeta_\pm$ are positive; for $\zeta_-$ this is evident from Eq. (4.36); if $\zeta_+$ is negative then $\text{Tr}[\mathcal{H}_{ij}] < 0$, which is impossible if the system is stable, since stability requires that both $\mathcal{H}_{11} = F_{N_1 N_1}$ and $\mathcal{H}_{22} = F_{N_2 N_2}$ are positive. Since by $y_{2,2}^{(1)} > 0$ it is also $F_{\Omega\Omega} > 0$, the stability of the system implies that the free energy is a strictly convex function of $\Omega$ (at fixed $N_1, N_2$) and of $N_1, N_2$ (at fixed $\Omega$), but (in contrast to the pure-substance case) only strict convexity with respect to $N_1, N_2$ constitutes a sufficient stability criterion. In particular, the "mechanical stability" criterion $\kappa_T^{-1} > 0$ [where $\kappa_T = \Omega F_{\Omega\Omega}$ is the isothermal compressibility] is a sufficient stability criterion only for a pure substance. Since $F_{\Omega\Omega} > 0$ implies $\kappa_T^{-1} > 0$, the inverse isothermal compressibility is positive in the stable region also for a mixture, but in that case $\kappa_T^{-1} > 0$ is not a sufficient stability criterion and the region where $\kappa_T^{-1} < 0$ is a subregion of the unstable region determined by $y_{3,3}^{(2)} \leq 0$.

## 4.2. Nuclear Liquid-Gas Phase Transition

Using the general theory discussed in the previous section we now examine the isospin-asymmetry dependence of the liquid-gas instability of bulk nuclear matter. The free energy per particle of (single-phase constrained) isospin-asymmetric nuclear matter (ANM) is calculated using the approximation

$$\bar{F}(T,\rho,\delta) \simeq \bar{F}_{\text{nonrel}}(T,\rho,\delta) + \bar{F}_{\text{corr}}(T,\rho,0) + \bar{F}_{\text{sym,corr}}(T,\rho)\,\delta^2 + \bar{F}_{\text{int}}(T,\rho,0) + \bar{F}_{\text{sym,int}}(T,\rho)\,\delta^2, \tag{4.40}$$

i.e., only the isospin-asymmetry dependence of the nonrelativistic free Fermi gas term $\bar{F}_{\text{nonrel}}$ is treated exactly, but the (leading) relativistic correction $\bar{F}_{\text{corr}}$ and the interaction contribution $\bar{F}_{\text{int}}$ are assumed to have a quadratic dependence on $\delta$ ("parabolic approximation").

Higher-order effects in the isospin-asymmetry dependence of $\bar{F}_{\text{int}}$ (and $\bar{F}_{\text{corr}}$) are examined in the next chapter; we will find that Eq. (4.40) can be expected to be reasonably accurate at the temperatures and densities relevant for the nuclear liquid-gas instability (the error of Eq. (4.40) increases with density and decreases with temperature, and receives its main contribution from 3N interactions, which are small at the relevant densities). In particular, the error of the parabolic approximation is smaller for $\bar{F}_{\text{int}}$ (and $\bar{F}_{\text{corr}}$) as compared to $\bar{F}_{\text{nonrel}}$, which motivates the exact treatment of the $\delta$ dependence of $\bar{F}_{\text{nonrel}}$.

The main motivation for Eq. (4.40) is however the fact that an approximative treatment of the isospin dependence via an expansion in terms of $\delta$ for $\bar{F}(T,\rho,\delta)$ misses the constraint that (for a given temperature) the spinodal terminates at a value $\delta < 1$; to enforce this constraint it is crucial to treat the $\delta$ dependence of $\bar{F}_{\text{nonrel}}(T,\rho,\delta)$ *exactly*. This issue as well as the results for $T_c(\delta)$ are discussed in Sec. 4.2.1. In Sec. 4.2.2 we then examine the properties of stable self-bound states at zero temperature. Self-bound states exist only for isospin-asymmetries and temperatures below certain values, and at nonzero temperature they are metastable (if surface effects are neglected); these issues are studied in Sec. 4.2.3.

### 4.2.1. Spinodal and Trajectory of Critical Temperature

To determine the spinodal of ANM we need to calculate the determinant, or equivalently the eigenvalues, of the Hessian matrix $\mathcal{H}_{ij}$ (cf. Sec. 4.1.2). The Hessian matrix is given by

$$\mathcal{H}_{ij}(T, \rho_1, \rho_2) = \left[\frac{\partial^2 F(T, \rho_1, \rho_2)}{\partial \rho_i \partial \rho_j}\right] = \left[\frac{\partial \mu_i(T, \rho_1, \rho_2)}{\partial \rho_j}\right], \quad i, j \in \{1, 2\}. \tag{4.41}$$

where $F(T, \rho_1, \rho_2)$ now denotes the free energy density of the system with component densities $\rho_{1,2}$. Th eigenvalues of the Hessian matrix are given by

$$\zeta_\pm(T, \rho_1, \rho_2) = \frac{1}{2}\left[\mathcal{H}_{11} + \mathcal{H}_{22} \pm ((\mathcal{H}_{11} - \mathcal{H}_{22})^2 + 4\mathcal{H}_{12}^2)^{1/2}\right]. \tag{4.42}$$

Since $\zeta_+ > 0$ is not a sufficient stability criterion (cf. Sec. 4.1.2), the relevant eigenvalue is $\zeta_-$. The signs of the eigenvalues are invariant under (linear) basis transformations. From the data $\bar{F}(T, \rho, \delta)$ the sign of $\zeta_-$ is readily evaluated using as independent density parameters $\rho_1 = \rho_n + \rho_p = \rho$ (nucleon density) and $\rho_2 = \rho_n - \rho_p = \rho \delta$ ("isospin-asymmetry density"). In this basis the Hessian matrix becomes diagonal at $\delta = 0$ with eigenvalues $\zeta_+ = (\partial^2 F/\partial \rho_2^2)_{T,\rho_1}|_{\rho_2=0} > 0$ and $\zeta_- = (\partial^2 F/\partial \rho_1^2)_{T,\rho_2}|_{\rho_2=0} = \rho^{-1}(\partial P/\partial \rho)_{T,\delta}|_{\delta=0}$ (this result depends on the neglect of charge-symmetry breaking effects). Hence, as required, in the case of SNM the region inside the spinodal corresponds to a negative isothermal compressibility $\kappa_T = \rho^{-1}(\partial \rho/\partial P)_{T,\delta}$, where $\kappa_T^{-1} > 0$ is a stability criterion for a pure substance.

**$\delta \to 1$ *Limit*.** The Hessian matrix can be written as $\mathcal{H}_{ij} = \mathcal{H}_{ij;\text{nonrel}} + \mathcal{H}_{ij;\text{corr}} + \mathcal{H}_{ij;\text{int}}$, where $\mathcal{H}_{ij;\text{nonrel}}$ denotes the contribution from the nonrelativistic free Fermi gas contributon to the free energy density, $F_{\text{nonrel}}$, and $\mathcal{H}_{ij;\text{int}}$ and $\mathcal{H}_{ij;\text{corr}}$ correspond to the interaction contribution $F_{\text{int}}$ and to the relativistic correction term $F_{\text{corr}}$, respectively. The nonrelativistic free Fermi gas contribution to the Hessian matrix components can be determined as follows. The neutron and proton contributions to $F_{\text{nonrel}}$ are functions of $T$ and $\tilde{\mu}_{\text{n/p}}$. Their total differentials at fixed $T$ are therefore given by

$$dF_{\text{nonrel}}^{\text{n/p}} = \left(\frac{\partial F_{\text{nonrel}}^{\text{n/p}}}{\partial \tilde{\mu}_{\text{n/p}}}\right)_T d\tilde{\mu}_{\text{n/p}}. \tag{4.43}$$

For notational convenience we define $\mathcal{A} := \rho \delta$. The relations $\rho_n = (\rho + \mathcal{A})/2$ and $\rho_p = (\rho - \mathcal{A})/2$ lead to the following expressions for the nucleon density as a function of $\mathcal{A}$ and either one of the auxiliary chemical potentials $\tilde{\mu}_{\text{n/p}}$:

$$\rho(\mathcal{A}, \tilde{\mu}_{\text{n/p}}) = \begin{cases} 2\rho_n(\tilde{\mu}_n) - \mathcal{A} \\ 2\rho_p(\tilde{\mu}_p) + \mathcal{A} \end{cases} = \begin{cases} -2\alpha T^{3/2} \text{Li}_{3/2}(\tilde{x}_n) - \mathcal{A} \\ -2\alpha T^{3/2} \text{Li}_{3/2}(\tilde{x}_p) + \mathcal{A} \end{cases}. \tag{4.44}$$

where $\tilde{x}_{\text{n/p}} = -\exp(\tilde{\mu}_{\text{n/p}}/T)$ and $\alpha = 2^{-1/2}(M/\pi)^{3/2}$. The total differential of $\rho(\mathcal{A}, \tilde{\mu}_{\text{n/p}})$ is given by

$$d\rho = \left(\frac{\partial \rho}{\partial \mathcal{A}}\right)_{\tilde{\mu}_{\text{n/p}}} d\mathcal{A} + \left(\frac{\partial \rho}{\partial \tilde{\mu}_{\text{n/p}}}\right)_{\mathcal{A}} d\tilde{\mu}_{\text{n/p}}. \tag{4.45}$$

This leads to the following expression for the total differentials of $\tilde{\mu}_{\text{n/p}}$ at fixed nucleon density:

$$[d\tilde{\mu}_{\text{n/p}}]_{d\rho=0} = -\frac{(\partial \rho/\partial \mathcal{A})_{\tilde{\mu}_{\text{n/p}}}}{(\partial \rho/\partial \tilde{\mu}_{\text{n/p}})_\mathcal{A}} d\mathcal{A} = \begin{cases} -\dfrac{1}{2\alpha T^{1/2} \text{Li}_{3/2}(\tilde{x}_n)} d\mathcal{A} \\ \dfrac{1}{2\alpha T^{1/2} \text{Li}_{3/2}(\tilde{x}_p)} d\mathcal{A} \end{cases}. \tag{4.46}$$

## 4. Nuclear Liquid-Gas Instability

The derivative of $F_{\text{nonrel}}^{\text{n/p}} = -\alpha T^{5/2}[\ln(-\tilde{x}_{\text{n/p}})\text{Li}_{3/2}(\tilde{x}_{\text{n/p}}) - \text{Li}_{5/2}(\tilde{x}_{\text{n/p}})]$ with respect to $\mathcal{A}$ at fixed $\rho$ is then given by

$$\left(\frac{\partial F_{\text{nonrel}}^{\text{n/p}}}{\partial \mathcal{A}}\right)_\rho = \left(\frac{\partial F_{\text{nonrel}}^{\text{n/p}}}{\partial \tilde{\mu}_{\text{n/p}}}\right)\left(\frac{\partial \tilde{\mu}_{\text{n/p}}}{\partial \mathcal{A}}\right)_\rho = \pm \frac{T \ln(-\tilde{x}_{\text{n/p}})}{2}. \tag{4.47}$$

The derivative of $F_{\text{nonrel}}^{\text{n/p}}$ with respect to $\rho$ at fixed $\mathcal{A}$ can be determined similarly, and similar for higher derivatives. One then arrives at the following expressions for the contribution from $F_{\text{nonrel}}$ to the Hessian matrix components:[3]

$$\mathcal{H}_{11;\text{nonrel}}(T,\rho,\delta) = -\frac{T^{-1/2}}{4\alpha}\left(\frac{1}{\text{Li}_{1/2}(\tilde{x}_n)} + \frac{1}{\text{Li}_{1/2}(\tilde{x}_p)}\right) = \mathcal{H}_{22;\text{nonrel}}(T,\rho,\delta), \tag{4.48}$$

$$\mathcal{H}_{12;\text{nonrel}}(T,\rho,\delta) = -\frac{T^{-1/2}}{4\alpha}\left(\frac{1}{\text{Li}_{1/2}(\tilde{x}_n)} - \frac{1}{\text{Li}_{1/2}(\tilde{x}_p)}\right). \tag{4.49}$$

In the limit $\delta \to 1$ the proton density vanishes, $\rho_p \to 0$, and the proton (auxiliary) chemical potential diverges (at finite $T$), $\tilde{\mu}_p \to -\infty$, thus $\text{Li}_{1/2}(\tilde{x}_p) \to 0$. Hence, the exact calculation of the nonrelativistic free Fermi gas contribution $F_{\text{nonrel}}(T,\rho,\delta)$ leads to divergent behavior of the Hessian components in the limit of vanishing proton concentration. The unstable region then vanishes at a value $\delta < 1$ for all values of $T$. This constraint is lost if the free Fermi gas contribution is approximated in terms of an expansion in powers of $\delta$. In fact, using the parabolic isospin-asymmetry approximation also for $F_{\text{nonrel}}$, the spinodal would start to cross the $\delta = 1$ line at around $T \simeq 5$ MeV (plot not shown), which is an unphysical feature.

**$Y^{5/3}$ Terms.** At zero temperature, the divergent behavior of the Hessian matrix elements in the limit $\delta \to 1$ arises due to the term $\sim (1-\delta)^{5/3} \sim (Y)^{5/3}$ in the proton part of $E_{0;\text{nonrel}}$, see Eq. (5.4). We mention here that a dependence on $Y$ of the form $\sim (Y)^{5/3}$ arises not only from the noninteracting contribution but also from the perturbative interaction contributions to the ground-state energy density [376] (cf. also Secs. 5.1.3 and 5.2.4). For instance, the np-channel Hartree-Fock contribution to the ground-state energy density from (central) NN interactions has the form

$$E_{0;1}^{\text{NN}} \sim \int \frac{d^3k_1 d^3k_2}{(2\pi)^6} v(|\vec{k}_1 - \vec{k}_2|)\,\Theta(k_F^n - k_1)\Theta(k_F^p - k_2), \tag{4.50}$$

where $k_F^{\text{n/p}}$ is the neutron/proton Fermi momentum, and $v(|\vec{k}_1 - \vec{k}_2|)$ a function incorporating the details of the interactions. The expansion of this expression with respect to the proton fraction $Y$ can be performed by setting $k_F^n = k_F(1-Y)^{1/3}$ and $k_F^p = k_F Y^{1/3}$ and reparametrizing the integrals, leading to

$$E_{0;1}^{\text{NN}} \sim \frac{k_F^6}{12\pi^2}\int_0^1 dx \left\{Y x^2 v(xk_F) + Y^{5/3}\frac{k_F}{10}\frac{\partial v(k_F)}{\partial k_F} - \frac{Y^2}{3}v(k_F) + O(Y^{7/3})\right\}, \tag{4.51}$$

i.e., the interaction contribution to the zero-temperature EoS involve also higher-order fractional powers of $Y$ [see also Eq. (5.69)]. In addition, at second order in MBPT also a logarithmic term $\sim Y^{7/3}\ln(Y)$ appears [cf. Eq. (5.69)].

## 4. Nuclear Liquid-Gas Instability

**$T_c(\delta)$ Trajectory.** From the parabolic approximation of the interaction contribution [cf. Eq. (4.40)], the matrix components $\mathcal{H}_{ij;\text{int}}$ are given by

$$\mathcal{H}_{11;\text{int}}(T,\rho,\delta) = \frac{\partial^2 F_{\text{int}}(T,\rho,0)}{\partial \rho^2} + \left(\frac{\partial^2 F_{\text{sym,int}}(T,\rho)}{\partial \rho^2} - \frac{4}{\rho}\frac{\partial F_{\text{sym,int}}(T,\rho)}{\partial \rho} + 3\frac{F_{\text{sym,int}}(T,\rho)}{\rho^2}\right)\delta^2, \tag{4.52}$$

$$\mathcal{H}_{12;\text{int}}(T,\rho,\delta) = 2\left(\frac{1}{\rho}\frac{\partial F_{\text{sym,int}}(T,\rho)}{\partial \rho} - 2\frac{F_{\text{sym,int}}(T,\rho)}{\rho^2}\right)\delta, \tag{4.53}$$

$$\mathcal{H}_{22;\text{int}}(T,\rho,\delta) = 2\frac{F_{\text{sym,int}}(T,\rho)}{\rho^2}, \tag{4.54}$$

and similar for $\mathcal{H}_{ij;\text{corr}}$. The derivatives with respect to $\rho$ have been calculated numerically using finite differences. The results obtained for the trajectory of the critical temperature $T_c(\delta)$ where $\zeta_-$ becomes positive definite for fixed $\delta$ are depicted in Fig. 4.1. The $T_c(\delta)$ trajectories are very similar for n3lo414 and n3lo450, and in both cases the critical lines end approximately at an isospin asymmetry $\delta_c^{\text{end}} \simeq 0.9994$, corresponding to a proton concentration $Y_c^{\text{end}} \simeq 3 \cdot 10^{-4}$. The value $\delta_c^{\text{end}} \simeq 0.9994$ exceeds the critical line endpoints obtained in Refs. [110, 141, 425] using different phenomenological mean-field models.

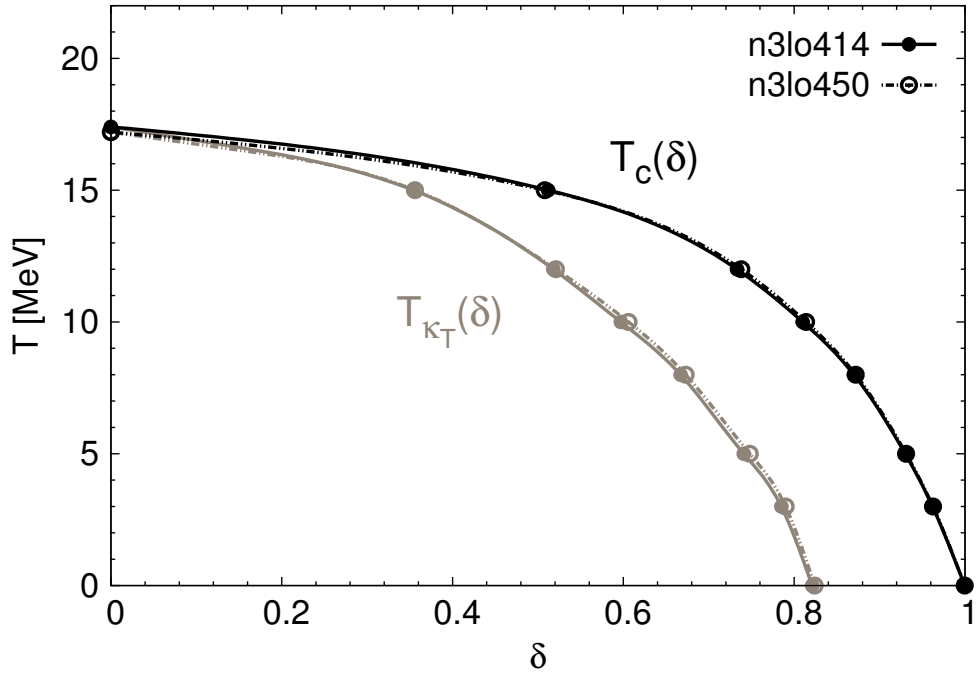

**Figure 4.1.:** Trajectories of the critical temperature $T_c(\delta)$ determined from the n3lo450 and n3lo414 results. The trajectories end at $\delta \simeq 0.9994$. Also shown are the trajectories of the temperature $T_{\kappa_T}(\delta)$ where the region with negative isothermal compressibility $\kappa_T$ vanishes at fixed $\delta$. The calculated data points are shown explicitly.

---

[3] The matrix elements $\mathcal{H}_{11}$ and $\mathcal{H}_{22}$ are identical only in the absence of interactions (and if the isospin-asymmetry dependence of the free Fermi gas term is treated exactly, i.e., without employing an expansion in terms of $\delta$).

*4. Nuclear Liquid-Gas Instability*

We note that in nuclear matter with $\delta \neq 0$ the coexistence region does *not* vanish at the critical temperature $T_c(\delta)$ (but the critical point is still a point on the binodal) but at a higher temperature $T_{\max}(\delta)$, the so-called maximum temperature [198, 110]. The existence of a neutron drip point (see Sec. 4.2.2) entails that at zero temperature the binodal extends to the pure substance over a finite region of densities or pressures. The trajectory of the maximum temperatures $T_{\max}(\delta)$ therefore reaches its zero-temperature endpoint at vanishing proton fraction, i.e., $\delta_{\max}^{\text{end}} = 1$ or $Y_{\max}^{\text{end}} = 0$.

For comparison, in Fig. 4.1 we show also the trajectories of the temperature $T_{\kappa_T}(\delta)$ where the region with negative $\kappa_T^{-1}$ vanishes at fixed $\delta$. The exact expression for the nonrelativistic free Fermi gas contribution to $\kappa_T^{-1}$ is given by

$$\kappa_T^{-1} = -\frac{\alpha T^{5/2}}{4}\left[\text{Li}_{3/2}(\tilde{x}_n) + \text{Li}_{3/2}(\tilde{x}_p)\right]^2 \left(\frac{(1+\delta)^2}{\text{Li}_{1/2}(\tilde{x}_n)} + \frac{(1-\delta)^2}{\text{Li}_{1/2}(\tilde{x}_p)}\right). \quad (4.55)$$

The contributions to $\kappa_T^{-1}$ from nuclear interactions as well as the from the relativistic correction term have been evaluated numerically using finite differences. For both n3lo414 and n3lo450 the $T_{\kappa_T}(\delta)$ trajectories end at approximately $\delta_{\kappa_T}^{\text{end}} \simeq 0.82$ or $Y_{\kappa_T}^{\text{end}} \simeq 0.09$, which exceeds the value $Y_{\kappa_T}^{\text{end}} \simeq 0.053$ extracted in Ref. [144] from calculations within the in-medium chiral-perturbation theory approach developed in Refs. [150, 237, 151] and the value $Y_{\kappa_T}^{\text{end}} \simeq 0.045$ obtained in Ref. [105] by evaluating a nucleon-meson model with the functional renormalization group (see Refs. [104, 105] for further details).

### 4.2.2. Stable Self-Bound Liquid

From the data $\bar{F}(T, \rho, \delta)$ the neutron and proton chemical potentials are obtained via

$$\mu_{\text{n/p}}(T, \rho, \delta) = \frac{\partial F(T, \rho, \delta)}{\partial \rho} \pm \frac{1 \mp \delta}{\rho}\frac{\partial F(T, \rho, \delta)}{\partial \delta}. \quad (4.56)$$

The results for $\mu_n(T, \rho, \delta)$ and $\mu_p(T, \rho, \delta)$ are displayed in Fig. 4.2 for temperatures $T = (0, 15)$ MeV. One sees that $\mu_n(T, \rho, \delta)$ increases and $\mu_p(T, \rho, \delta)$ decreases with $\delta$. At finite $T$ the chemical potentials diverge as $\rho \to 0$, but at zero temperature $\mu_{\text{n/p}} \to 0$ for $\rho \to 0$. The origin of this feature is again the asymptotically (for $\rho \to 0$) logarithmic density dependence of the auxiliary chemical potentials and the noninteracting contribution to the free energy per particle.

The binodal (coexistence boundary) is determined by the Gibbs conditions for the coexistence of two bulk phases $\alpha$ (liquid) and $\beta$ (gas) in mutual thermodynamic equilibrium (cf. Sec. 4.1):[4]

$$T^{(\alpha)} = T^{(\beta)}, \qquad P^{(\alpha)} = P^{(\beta)}, \qquad \mu_n^{(\alpha)} = \mu_n^{(\beta)}, \qquad \mu_p^{(\alpha)} = \mu_p^{(\beta)}. \quad (4.57)$$

The free energy density corresponding to stable two-phase equilibrium is then given by

$$F^{(\alpha)+(\beta)}(T, \rho^{(\alpha)}, \rho^{(\beta)}) = \lambda^{(\alpha)} F(T, \rho_n^{(\alpha)}, \rho_p^{(\alpha)}) + \lambda^{(\beta)} F(T, \rho_n^{(\beta)}, \rho_p^{(\beta)}), \quad (4.58)$$

where $\lambda^{(\alpha)} = 1 - \lambda^{(\beta)}$ is the volume fraction occupied by phase $\alpha$, and $\rho_i = \lambda^{(\alpha)}\rho_i^{(\alpha)} + \lambda^{(\beta)}\rho_i^{(\beta)}$, $i \in \{n, p\}$. The particle densities $\rho_i^{(\alpha)}$ in the two phases are such that $F^{(\alpha)+(\beta)}$ is minimized for given values of the "global" particle densities $\rho_i$ (corresponding to the single-phase constrained system). Whereas at finite $T$ liquid-gas equilibrium corresponds to finite values of $\rho$ and $Y = (1-\delta)/2$ in both phases, the vanishing of $\mu_{n/p}(T=0, \rho, \delta)$ at vanishing density entails that at $T=0$ the Gibbs conditions for the neutron and proton chemical potentials can in most cases not be satisfied, leading to a gas phase that is either empty (vacuum) or contains only neutrons (see Refs. [110, 262, 255] for more details on the binodal).

---
[4] The properties of the binodal of ANM have been examined in detail in Refs. [198, 110, 301, 141, 255, 24]; in particular, in Ref. [110] (cf. also [399] a method has been introduced that allows to reduce the two-dimensional "Gibbs construction" of the binodal to a one-dimensional Maxwell construction.

*4. Nuclear Liquid-Gas Instability*

The neutron drip point $\delta_{\mathrm{ND}}$ for *bulk* nuclear matter is given by the value of $\delta$ where the neutron chemical potential at vanishing temperature and pressure becomes positive. For isospin asymmetries $\delta \leq \delta_{\mathrm{ND}}$ an isolated (large) drop[5] of "cold" ($T=0$) liquid nuclear matter is stable (in equilibrium with the vacuum), defining a stable self-bound state. As seen from Fig. 4.2, neutron drip occurs at an isospin asymmetry $\delta_{\mathrm{ND}} \simeq 0.30$ or a proton concentration $Y_{p,\mathrm{ND}} = (1-\delta_{\mathrm{ND}})/2 \simeq 0.35$, which is similar to results obtained with Skyrme models [262, 255].

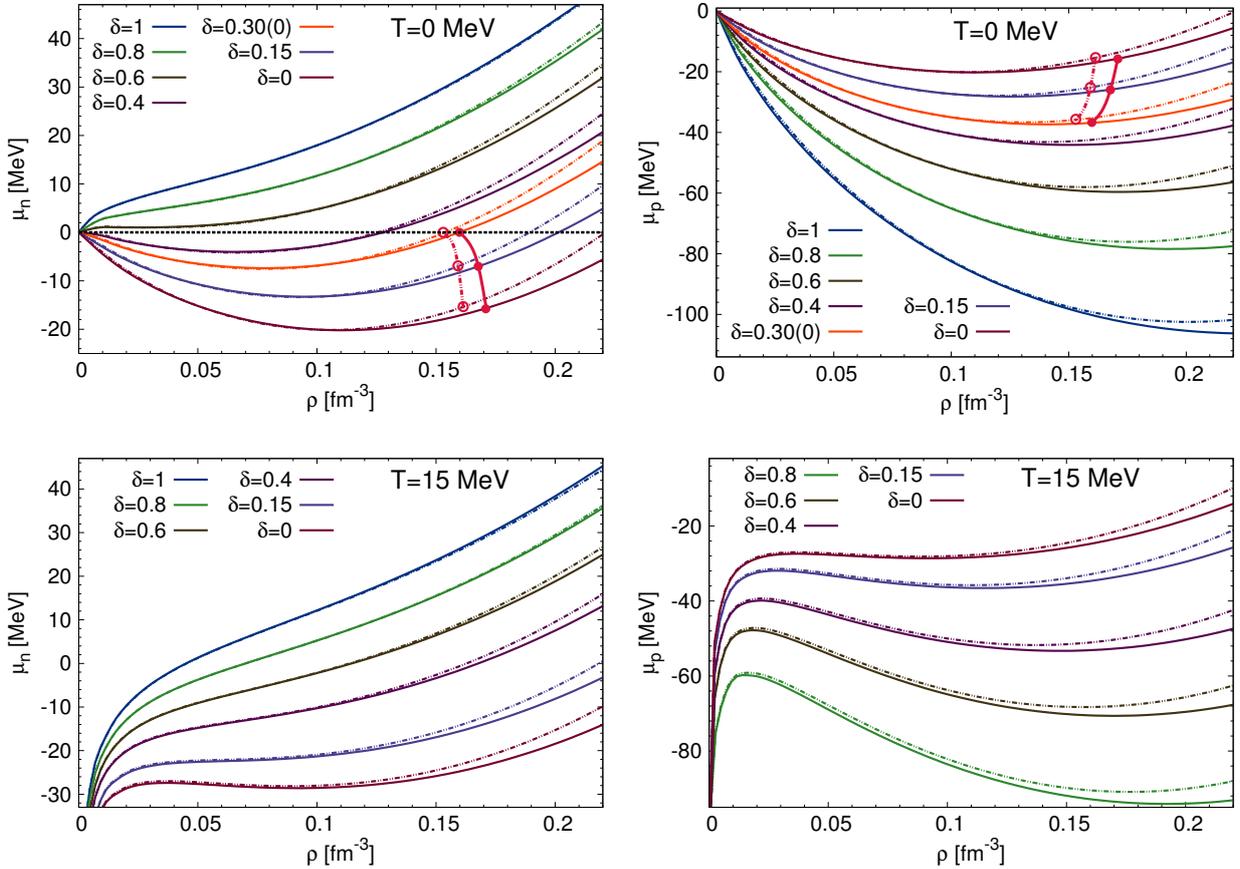

**Figure 4.2.:** Results for the (single-phase constrained) neutron and proton chemical potentials from n3lo414 (solid lines) and n3lo450 (dash-dot lines). Stable self-bound states are shown as thick dark red lines with circles (full circles for n3lo414, open circles for n3lo450); the lines end at the neutron drip point $\delta_{\mathrm{ND}} \simeq 0.30$.

---

[5] The drop is considered to be sufficiently large such that the surface tension can be neglected, cf. Eq. (4.14).

### 4.2.3. Metastable Self-Bound Liquid

The zero-temperature results obtained from Eq. (4.40) for the energy per particle $\bar{F}(T=0,\rho,\delta) = \bar{E}(T=0,\rho,\delta) = \bar{E}_0(\rho,\delta)$ and the pressure $P(T=0,\rho,\delta) = \rho^2 \partial \bar{E}_0(\rho,\delta)/\partial\rho$ are displayed in Fig. 4.3 as functions of the nucleon density $\rho = \rho_n + \rho_p$ for different values of $\delta$. The trajectory of the points where $\bar{E}_0(\rho,\delta)|_\delta$ has a local minimum (and thus, the pressure is zero) is shown explicitly. These points correspond to the properties of a (large) drop of "cold" liquid nuclear matter surrounded by vacuum. For $\delta \leq \delta_{\text{ND}} \simeq 0.30$ the "cold" drop is stable [the local energy minimum lies on the binodal], and for $\delta_{\text{ND}} < \delta < \delta_{\text{FP}} \simeq 0.66$ it is metastable [the local energy minimum lies between the binodal and the spinodal]. We refer to the point $\delta_{\text{FP}} \simeq 0.66$ where the trajectory of the local energy minima encounters the spinodal as the *fragmentation point* (FP). The energy per particle at neutron drip and at the fragmentation point is $\bar{E}_{0,\text{ND}} \simeq -13.0\,\text{MeV}$ and $\bar{E}_{0,\text{FP}} \simeq -3.1\,\text{MeV}$, respectively. For comparison, we follow the local energy minima also into the *unstable spinodal region*, i.e., we show the point where both derivatives of the single-phase constrained ground-state energy per particle vanish (saddle point, SP) at $\delta_{\text{SP}} \simeq 0.81$ as well as the local energy minimum with $\bar{E}_0 = 0$ at $\delta \simeq 0.76$.[6]

In Fig. 4.4, the results for $\bar{F}(T,\rho,\delta)$ and the $P(T,\rho,\delta)$ are shown for $T = (5,15)\,\text{MeV}$. At finite $T$, the trajectories of the local free energy minima lie entirely in the metastable region: an isolated drop of "hot" liquid nuclear matter has to be stabilized by a surrounding nucleon gas.[7] At $T = 5\,\text{MeV}$ the trajectory ends at $\delta_{\text{FP}} \simeq 0.61$, defining the fragmentation temperature for nuclear matter with proton concentration $Y_p \simeq 0.195$. The $T = 5\,\text{MeV}$ saddle point is located at $\delta_{\text{SP}} \simeq 0.68$ for n3lo414 and $\delta_{\text{SP}} \simeq 0.69$ for n3lo450. For $T = 15\,\text{MeV}$ no local free energy minimum exists: for $T \gtrsim 13.5\,\text{MeV}$ the single-phase constrained free energy per particle is a monotonic increasing function of density for all values of $\delta$ (cf. Fig. 4.6).

The relation between the spinodal, the binodal, and the trajectories of the local free energy minima is illustrated in Fig. 4.5.[8] The two plots in Fig. 4.5 represent isoplethal ($\delta = $ const) and isothermal cross sections of the respective surfaces (spinodal, binodal, surface of local free energy minima) in $(T,\rho,\delta)$ space (cf. also Refs. [301, 110]). In the isothermal plot we show also the surface with divergent isothermal compressibility $\kappa_T = \rho^{-1}(\partial\rho/\partial P)_{T,\delta}$, which corresponds to the violation of the stability criterion $\kappa_T^{-1} > 0$ for a pure substance. The saddle point (SP) where both derivatives of the $\bar{F}(T,\rho,\delta)$ with respect to $\rho$ vanish coincides with the fragmentation point (FP) where a liquid drop becomes unstable only for $\delta = 0$ where nuclear matter behaves like a pure substance. For $\delta \neq 0$ the more restrictive two-component stability criteria are needed ($\kappa_T^{-1} > 0$ is not a relevant stability criterion in that case), and the SP is located in the interior of the spinodal.

Finally, in Fig. 4.6 the trajectory of the fragmentation temperature $T_{\text{FP}}(\delta)$ is shown; for comparison we also show the trajectore of the saddle-point temperature $T_{\text{SP}}(\delta)$. The $T_{\text{FP}}(\delta)$ trajectory determines the range of temperatures and isospin asymmetries for which a self-bound liquid state exists for bulk nuclear matter.

---

[6] In other words, for $\delta \gtrsim 0.76$ the energy per particle is positive at all (finite) densities, and for $\delta \geq \delta_{\text{SP}}$ the pressure is a semipositive definite function of density.

[8] The sharp transition from stability to metastability (in terms of zero and nonzero $T$) is clearly an artifact associated with the thermodynamic limit and the neglect of surface effects. At this opportunity, we note also that the spinodal should be seen as a somewhat idealized concept (cf. Refs. [44, 99] for more detailed discussions).

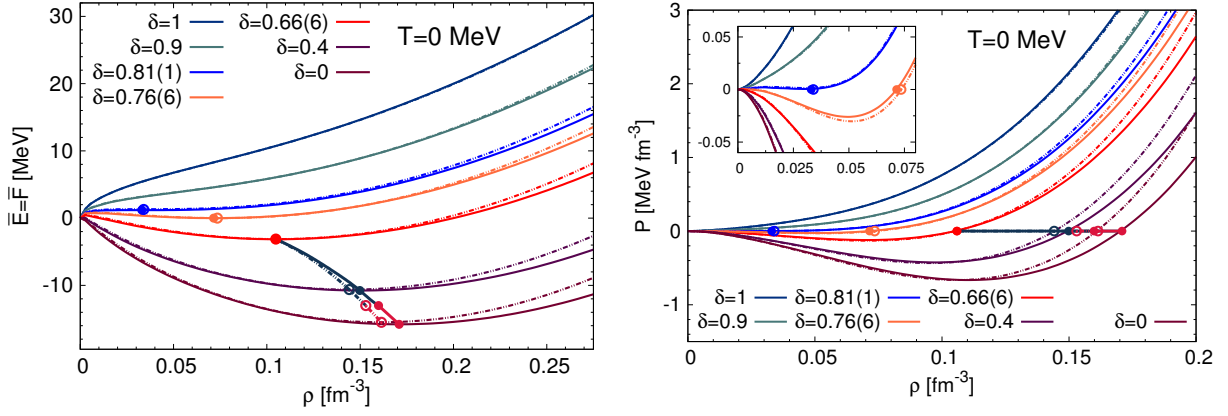

**Figure 4.3.:** Ground-state energy per particle $\bar{E}_0$ and pressure $P$ for (single-phase constrained) isospin-asymmetric nuclear matter, from n3lo414 (solid lines) and n3lo450 (dash-dot lines). The trajectories of self-bound states are shown as thick dark gray (dark red below neutron drip) lines with circles (n3lo414: full circles, n3lo450: open circles). The trajectories end at the fragmentation point $\delta_{\text{FP}} \simeq 0.66$. The inset magnifies the behavior of the pressure at low densities.

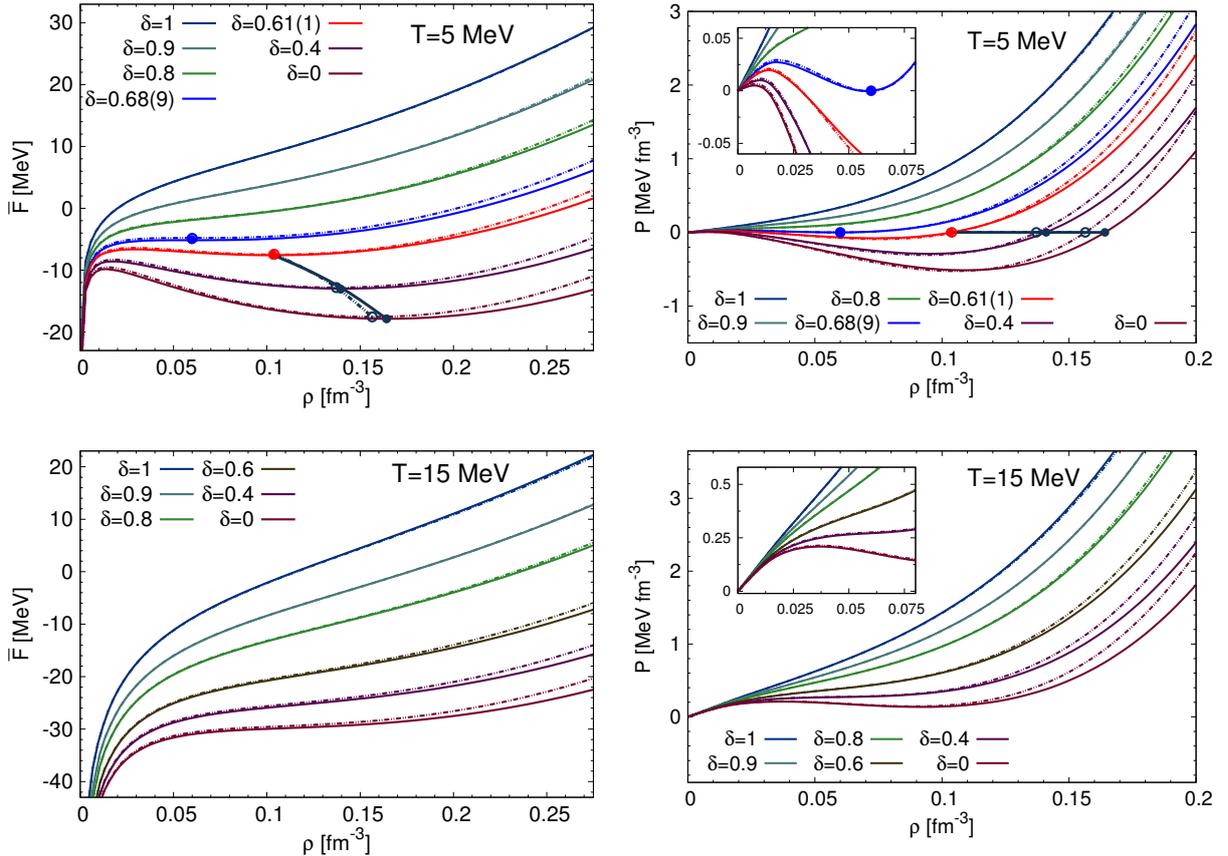

**Figure 4.4.:** (Single-phase constrained) free energy per particle $\bar{F}(T, \rho, \delta)$ and pressure $P(T, \rho, \delta)$ at temperatures $T = 5, 15\,\text{MeV}$; solid lines for n3lo414, dash-dot lines for n3lo450. The thick dark lines with circles depict the trajectories of the local free energy minima, up to the point where they encounter the spinodal. The insets magnify the behavior of the pressure at low densities.

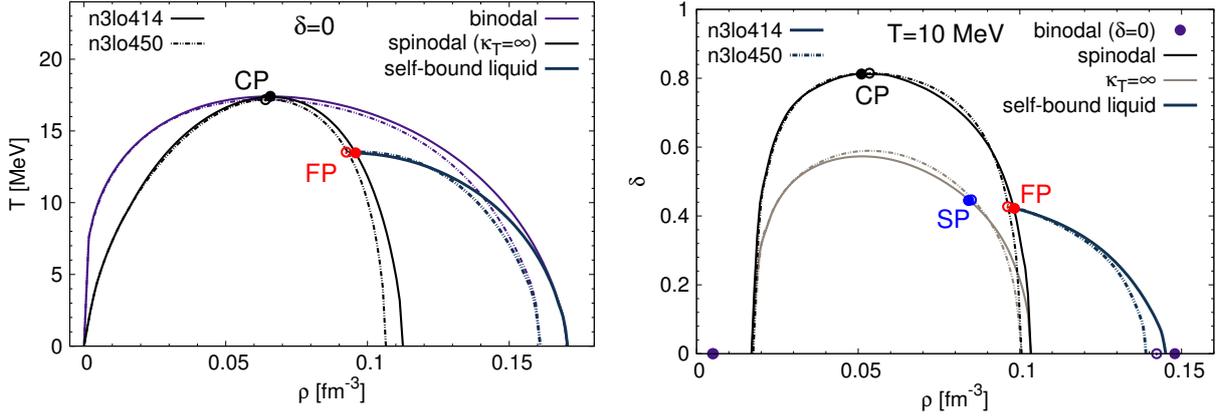

**Figure 4.5.:** Left plot: binodal, spinodal, and trajectory of local free energy minima ("self-bound liquid") in SNM ($\delta = 0$). Right plot: $T = 10\,\text{MeV}$ cross sections of the spinodal, the $\kappa_T = \infty$ boundary, and the surface of local free energy minima ("self-bound liquid"); only the $\delta = 0$ endpoints of the binodal are shown (we did not construct the binodal for $\delta \neq 0$). The critical points (CP), fragmentation points (FP) and saddle points (SP) are shown explicitly in both plots (in SNM the FP and the SP coincide).

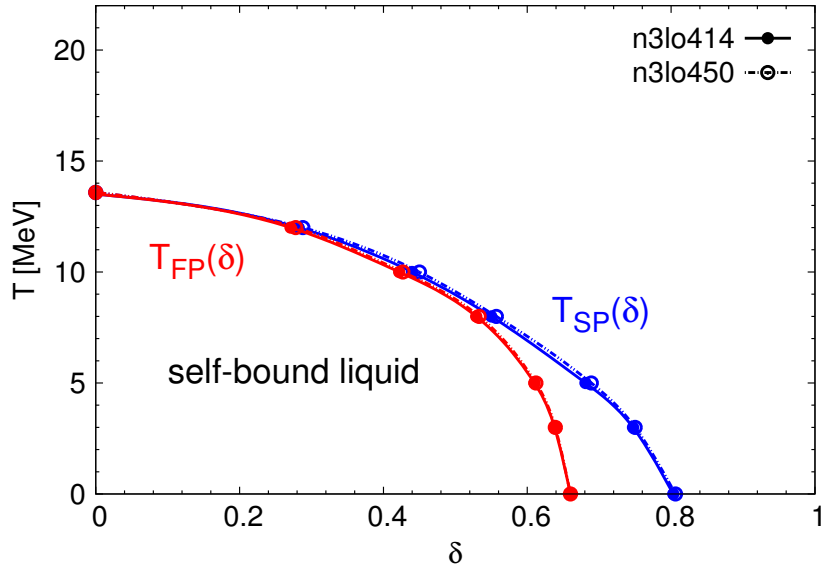

**Figure 4.6.:** Trajectory of the fragmentation temperature $T_{\text{FP}}(\delta)$ above which no (metastable) self-bound state of bulk nuclear matter can exist (lower red line). For comparison we also show the trajectory of the points where the (single-phase constrained) free energy per particle has a saddle point, $T_{\text{SP}}(\delta)$ (upper blue line). The calculated data points are shown explicitly.

---

[8] We note that the presence of a region with *negative pressure* ("tension" or "cavitation pressure", a form of superheat [99, 395]) in the metastable region (the region between the spinodal and the "self-bound liquid" line) is *not* unphysical. This feature is a generic property of liquids that are self-bound at low temperatures and a well-known feature of various molecular liquids [99, 364, 200, 432, 120, 365, 416]. It is consistent with general thermodynamic principles [276, 99]. States ("drops") in the negative-pressure region can (of course) not be in (mechanical) equilibrium with the vacuum, but this clearly does not imply a prompt collapse (or implosion); instead, the system (the "drop") is subject to (additional) cavitation and/or self-contraction effects that affect the nucleation dynamics (cf. e.g., [244, 120, 76, 77, 395]).

# 5. Isospin-Asymmetry Dependence of the Nuclear Equation of State

In this chapter, we examine in more detail the dependence of the (single-phase constrained) nuclear EoS on the isospin asymmetry $\delta = (\rho_n - \rho_p)/(\rho_n + \rho_p)$. Neglecting charge-symmetry breaking effects of the nuclear interactions [295, 131] as well as the neutron-proton mass difference $\Delta M/M \simeq 10^{-4}$, the nuclear EoS is invariant under the exchange of neutrons and protons, and therefore the Maclaurin expansion of the free energy per particle of (infinite homogeneous) nuclear matter with respect to the isospin asymmetry involves only even powers of $\delta$:

$$\bar{F}(T, \rho, \delta) \simeq \sum_{n=0}^{N} \bar{A}_{2n}(T, \rho)\, \delta^{2n} =: \bar{F}_{[2N]}(T, \rho, \delta), \tag{5.1}$$

where the different expansion (Maclaurin) coefficients $\bar{A}_{2n}(T, \rho)$ are given by

$$\bar{A}_{2n}(T, \rho) = \frac{1}{(2n)!} \frac{\partial^{2n} F(T, \rho, \delta)}{\partial \delta^{2n}}\bigg|_{\delta=0}. \tag{5.2}$$

This expansion is quoted in nearly every paper on nuclear physics or astrophysics where the EoS of neutron-rich nuclear matter is considered (e.g., [66, 320, 263, 107, 275, 368, 371, 261, 259, 109, 274, 275, 144, 190, 434]). The reason for the prevalence of Eq. (5.1) in the literature is obvious: it provides an explicit parametrization of the (otherwise not directly accessible) dependence of the EoS on the neutron-to-proton ratio, and hence allows for a straightforward extrapolation of a given EoS of isospin-symmetric nuclear matter to neutron-rich conditions.

In most cases, the expansion in $\delta$ was implemented in terms of the leading quadratic order only. Neglecting coefficients beyond $\bar{A}_2$ in Eq. (5.1), one obtains the relation

$$\bar{F}(T, \rho, \delta) \simeq \bar{F}(T, \rho, 0) + \bar{A}_2(T, \rho)\, \delta^2, \tag{5.3}$$

which, if globally enforced, yields $\bar{A}_2(T, \rho) \cong \bar{F}_{\text{sym}}(T, \rho)$, where $\bar{F}_{\text{sym}} := \bar{F}(\delta = 1) - \bar{F}(\delta = 1)$ is the symmetry free energy studied in Sec. 3.5. Eq. (5.3) is sometimes denoted as the "empirical parabolic law" [268]. The approximately quadratic isospin-asymmetry dependence of the EoS has (for the case of zero temperature) been validated in numerous many-body calculations with microscopic nuclear potentials (e.g., SCGF [148], MBPT [109], BHF [434, 435, 58, 405]). Nevertheless, it has been shown that higher-order terms in the isospin-asymmetry dependence can still have a significant influence on various properties of neutron stars such as the crust-core transition density or the threshold density for the direct URCA cooling process,[1] see Refs. [359, 368, 68]. This motivates a detailed investigation of the isospin-asymmetry dependence of the nuclear EoS at different temperatures and densities.

---

[1] The direct URCA process (i.e., the processes $n \rightarrow p + e^- + \bar{\nu}_e$ and $p + e^- \rightarrow n + \bar{\nu}_e$) can only proceed when the proton fraction is large enough (roughly, $Y = (1 - \delta)/2 \gtrsim 0.1$, cf. Ref. [368]). Because neutron stars cool much faster when the direct URCA process is allowed [427, 258], theoretical constraints on its threshold density are essential in describing the cooling of neutron stars.

*5. Isospin-Asymmetry Dependence of the Nuclear Equation of State*

Based on the "empirical parabolic law", the first expectation would be that the subleading terms in Eq. (5.1) are small. This expectation has been supported by *mean-field calculations* [359, 368, 68]: the quartic Maclaurin coefficient $\bar{A}_4$ was found to be of small size ($\lesssim 1$ MeV) at saturation density and $T = 0$.[2]

A recent work by N. Kaiser [234], however, has demonstrated that if perturbative contributions beyond the mean-field level (Hartree-Fock) are included, the higher-order coefficients $\bar{A}_{2n\geq 4}$ are, in fact, *singular* at zero temperature. In other terms, at higher orders in many-body perturbation theory $\bar{F}(T = 0, \rho, \delta)$ is not a smooth function of $\delta$, but of differentiability class $C^3$ only.[3] The origin of this feature lies in the energy denominators of the higher-order many-body contributions; i.e., at zero temperature, the integrands in the expressions for the higher-order terms diverge at the integration boundary where the energy denominators vanish.

Regarding finite temperatures, the result of Ref. [234] has an immediate consequence: because at finite $T$ there are no energy-denominator poles ("*cyclic formula*", cf. Sec. 2.2.2), the higher-order perturbative contributions to $\bar{F}(T \neq 0, \rho, \delta)$ are smooth ($C^\infty$) functions of $\delta$ (at $\delta = 0$), but at very low temperatures they cannot be analytic ($C^\omega$) since for $T \to 0$ their higher-order Maclaurin coefficients diverge.

For very low temperatures the Maclaurin expansion in terms of $\delta$ therefore represents an asymptotic expansion with zero radius of convergence.[4] This contrast the "natural" assumption that the Maclaurin expansion behaves convergent in the physical regime $\delta \in [-1, 1]$, and (as a consequence) that the inclusion of higher-order coefficients leads to systematic improvements also for very neutron-rich conditions.

The questions remain as to whether the radius of convergence becomes finite for higher temperatures and, in which region of the parameter space are isospin-asymmetry parametrizations $\bar{F}_{[2N]}(T, \rho, \delta)$ beyond the leading quadratic order useful. Furthermore, using a simple $S$-wave contact interaction in [234] a different expansion of the zero-temperature EoS that includes nonanalytic terms $\delta^{2n\geq 4} \ln |\delta|$ has been identified; however, the applicability of this expansion in the case of realistic (chiral) nuclear interactions has not been studied in [234].

Motivated by the situation described above, in this chapter we investigate in detail the isospin-asymmetry dependence of the EoS obtained from the sets of chiral nuclear potentials n3lo414 and n3lo450 in second-order MBPT. In particular, we examine the convergence behavior of the Maclaurin expansion [Eq. (5.1)] at finite temperature, as well as the applicability of a "logarithmic" expansion at zero temperature.

More specifically, in **Section 5.1** we study the isospin-asymmetry dependence of the noninteracting contribution to the EoS. The $\delta$ dependence of the interaction contributions is then investigated in **Sec. 5.2**. The quadratic, quartic and sextic Maclaurin coefficients $\bar{A}_{2,4,6}(T, \rho)$ for the interaction contributions are extracted numerically using higher-order finite-difference approximations. Since the various many-body contributions behave differently with respect to their $\delta$ dependence, we examine the Maclaurin coefficients for each contribution individually. Furthermore, we show that the singularity of the quartic coefficient at zero-temperature is approximately logarithmic also for the chiral nuclear potentials, and we extract the leading nonanalytic term $\bar{A}_{4,\log}(\rho)$ in the "logarithmic" expansion of the zero-temperature EoS. The convergence behavior of the Maclaurin expansion is summarized in **Sec. 5.3**. Finally, the various parametrizations obtained from the results for the Maclaurin coefficients and the leading coefficients in the "logarithmic" expansion, respectively, are compared to the full isospin-asymmetry dependence of the EoS in **Sec. 5.4**, where we also look at global isospin-asymmetry parametrizations constructed by fitting different polynomials as well as polynomials that include logarithmic terms to the exact results for the EoS.

---

[2] A small quartic coefficient was also reported in studies based on BHF(-type) calculations [72, 264]. Since in BHF calculations the effects of higher-order perturbative contributions are included (ladder resummation, cf. Sec. 5.2.4), the nonanalytic features present in higher-order MBPT should be visible there (in principle). The small quartic coefficients (based on BHF calculations) reported in Refs. [72, 264] may be due to the methods used to extract the quartic term (not stated in [72, 264]), i.e., polynomial fitting (cf. Sec. 5.4). See also Ref. [66], where a quartic polynomial has been fit to BHF results.

[3] As discussed in Sec. 5.1, additional features appear at the boundary of the physical region $\delta \in [-1, 1]$; i.e., at zero temperature the second derivative with respect to $\delta$ diverges at $\delta = 1$ [in terms of the proton fraction $Y = (1-\delta)/2$, this feature is associated with terms $\sim Y^{5/3}$], and at finite temperature the EoS is not differentiable at $\delta = 1$.

[4] A simple example [159] of a $C^\infty$ function $f(x)$ whose Maclaurin series has zero radius of convergence is $f(x) = \sum_{n=1}^\infty \exp(-n) \cos(n^2 x)$.

## 5.1. Isospin-Asymmetry Dependence of Free Nucleon Gas

Here, we examine the isospin-asymmetry dependence of the free Fermi gas term in the many-body perturbation series. The noninteracting contribution to the free energy per particle is given by $\bar{F}_{\text{nonint}}(T, \rho, \delta) = \bar{F}_{\text{nonrel}}(T, \rho, \delta) + \bar{F}_{\text{corr}}(T, \rho, \delta)$, where $\bar{F}_{\text{nonrel}}(T, \rho, \delta)$ corresponds to a nonrelativistic free nucleon gas, and $\bar{F}_{\text{corr}}(T, \rho, \delta)$ is the leading relativistic correction term [cf. the appendix A.1 for details]. We compute the quadratic, quartic, sextic and octic Maclaurin coefficients in the expansions of $\bar{F}_{\text{nonrel}}(T, \rho, \delta)$ and of $\bar{F}_{\text{corr}}(T, \rho, \delta)$ in powers of $\delta$. Furthermore, we examine the special features of the isospin-asymmetry dependence in the very neutron-rich region, in particular for $\delta \to 1$. The results show that an expansion in the third root of the proton fraction is applicable only at zero temperature.

### 5.1.1. Isospin-Asymmetry Derivatives

***Zero Temperature.*** The expressions for the noninteracting contributions to the ground-state energy density $E_{0;\text{nonint}}$ can be written explicitly as functions of $\rho$ and $\delta$, allowing the straightforward expansion in terms of $\delta$. The expressions for $E_{0;\text{nonrel}}$ and $E_{0;\text{corr}}$ are given by

$$E_{0;\text{nonrel}}(\rho, \delta) = \left(\frac{3\pi^2}{2}\right)^{5/3} \frac{\rho^{5/3}}{10\pi^2 M} \underbrace{\left((1+\delta)^{5/3} + (1-\delta)^{5/3}\right)}_{\Gamma_{\text{nonrel}}}, \tag{5.4}$$

$$E_{0;\text{corr}}(\rho, \delta) = -\left(\frac{3\pi^2}{2}\right)^{7/3} \frac{\rho^{7/3}}{56\pi^2 M^3} \underbrace{\left((1+\delta)^{7/3} + (1-\delta)^{7/3}\right)}_{\Gamma_{\text{corr}}}. \tag{5.5}$$

The general Taylor expansion of $\Gamma_{\text{nonrel}}(\delta)$ around a given point $\delta_0$ is given by $\Gamma_{\text{nonrel}}(\delta) = \sum_{n=0}^{\infty} \Gamma_{\text{nonrel},n}^{[\delta_0]} (\delta - \delta_0)^n$, and similar for $\Gamma_{\text{corr}}(\delta)$. The Maclaurin expansions ($\delta_0 = 0$) involve only even terms, with monotonously decreasing expansion coefficients

$$\Gamma_{\text{nonrel}}(\delta) = \sum_{n=0}^{\infty} \Gamma_{\text{nonrel},2n}^{[\delta_0=0]} \delta^{2n} = 2\sum_{n=0}^{\infty} \binom{5/3}{2n} \delta^{2n} = 2 + \frac{10}{9}\delta^2 + \frac{10}{9} \sum_{n=1}^{\infty} \delta^{2n} \prod_{k=0}^{2n-3} \frac{1+3k}{9+3k}, \tag{5.6}$$

$$\Gamma_{\text{corr}}(\delta) = \sum_{n=0}^{\infty} \Gamma_{\text{corr},2n}^{[\delta_0=0]} \delta^{2n} = 2\sum_{n=0}^{\infty} \binom{7/3}{2n} \delta^{2n} = 2 + \frac{28}{9}\delta^2 + \frac{28}{9} \sum_{n=2}^{\infty} \delta^{2n} \prod_{k=0}^{2n-3} \frac{3k-1}{9+3k}. \tag{5.7}$$

5. Isospin-Asymmetry Dependence of the Nuclear Equation of State

The radius of convergence of the Maclaurin expansions is $R_\delta^{[\delta_0=0]} = 1$,[5] which can be deduced form the ratios of sucessive Maclaurin coefficients. For $\Gamma_\text{nonrel}$ one has

$$\frac{\Gamma_{\text{nonrel},2n}^{[\delta_0=0]}}{\Gamma_{\text{nonrel},2(n+1)}^{[\delta_0=0]}} = \frac{18n^2 + 27n + 9}{18n^2 - 21n + 5} \xrightarrow{n\to\infty} 1, \tag{5.8}$$

and similar for $\Gamma_\text{corr}$. The general Taylor expansions around $\delta_0 \neq 0$ however have a smaller radius of convergence, and in general do not feature decreasing expansion coefficients, in particular for $\delta_0$ in the strongly neutron-rich region. For instance, for $\delta_0 = 0.9$ the leading coefficients in the expansion of $\Gamma_\text{nonrel}$ are given by $\Gamma_{\text{nonrel},(1,2,3,4,5,6)}^{[\delta_0=0.9]} \simeq (2.2, 1.6, 1.3, 4.4, 20.7, 114.9)$. This behavior is due to the proton contributions, i.e., the terms $\sim (1-\delta)^{5/3} = Y^{5/3}$ and $\sim (1-\delta)^{7/3} = Y^{7/3}$, which restrict the radius of convergence to $R_\delta^{[\delta_0=0.9]} = 0.1$.

***Finite Temperature.*** At finite temperature, the functional dependence on $\delta$ is given in terms of the neutron/proton (auxiliary) chemical potentials, so no straightforward expansion is possible in that case. Nevertheless, concise expressions for the $\delta$ derivatives (at fixed $T$ and $\rho$) can be derived from the expressions for $F_\text{nonrel}^\text{n/p}$ and $F_\text{corr}^\text{n/p}$ in terms of polylogarithms:

$$F_\text{nonrel}^\text{n/p}(T, \tilde\mu_\text{n/p}) = -\alpha T^{5/2}\Big( \ln(-\tilde x_\text{n/p}) \operatorname{Li}_{3/2}(\tilde x_\text{n/p}) - \operatorname{Li}_{5/2}(\tilde x_\text{n/p}) \Big), \tag{5.9}$$

$$F_\text{corr}^\text{n/p}(T, \tilde\mu_\text{n/p}) = \frac{15\alpha T^{7/2}}{8M} \operatorname{Li}_{7/2}(\tilde x_\text{n/p}), \tag{5.10}$$

where $\tilde x_\text{n/p} = -\exp(\tilde\mu_\text{n/p}/T)$ and $\alpha = 2^{-1/2}(M/\pi)^{3/2}$. We recall that for given values of $T$, $\rho$ and $\delta$ the (auxiliary) chemical potentials $\tilde\mu_\text{n/p}$ are uniquely determined by $\rho_\text{n/p} = -\alpha T^{3/2}\operatorname{Li}_{3/2}(\tilde x_\text{n/p})$. Using $\rho_\text{n/p} = \rho(1 \mp \delta)/2$, the total differential of $\rho(\delta, \tilde\mu_\text{n})$ and $\rho(\delta, \tilde\mu_\text{p})$, respectively, is given by

$$d\rho = \left(\frac{\partial \rho}{\partial \delta}\right)_{\tilde\mu_\text{n/p}} d\delta + \left(\frac{\partial \rho}{\partial \tilde\mu_\text{n/p}}\right)_\delta d\tilde\mu_\text{n/p}. \tag{5.11}$$

This leads to the following expressions for the total differentials of $\tilde\mu_\text{n}$ and $\tilde\mu_\text{p}$ at fixed nucleon density:

$$[d\tilde\mu_\text{n/p}]_{\rho=\text{const.}} = -\frac{(\partial\rho/\partial\delta)_{\tilde\mu_\text{n/p}}}{(\partial\rho/\partial\tilde\mu_\text{n/p})_\delta} d\delta = \begin{cases} \dfrac{T}{1+\delta}\dfrac{\operatorname{Li}_{3/2}(\tilde x_\text{n})}{\operatorname{Li}_{1/2}(\tilde x_\text{n})} d\delta \\[6pt] -\dfrac{T}{1+\delta}\dfrac{\operatorname{Li}_{3/2}(\tilde x_\text{p})}{\operatorname{Li}_{1/2}(\tilde x_\text{p})} d\delta \end{cases}. \tag{5.12}$$

The first derivative of $F_\text{nonrel}^\text{n/p}$ with respect to $\delta$ at fixed $\rho$ (and $T$) is then given by

$$\left(\frac{\partial F_\text{nonrel}^\text{n/p}}{\partial \delta}\right)_\rho = \left(\frac{\partial F_\text{nonrel}^\text{n/p}}{\partial \tilde\mu_\text{n/p}}\right)\left(\frac{\partial \tilde\mu_\text{n/p}}{\partial \delta}\right)_\rho = \begin{cases} -\dfrac{\alpha T^{5/2}}{1+\delta}\ln(-\tilde x_\text{n})\operatorname{Li}_{3/2}(\tilde x_\text{n}) \\[6pt] \dfrac{\alpha T^{5/2}}{1-\delta}\ln(-\tilde x_\text{p})\operatorname{Li}_{3/2}(\tilde x_\text{p}) \end{cases}. \tag{5.13}$$

Since the first derivative $F_{\text{nonrel}(1)}^\text{n/p} = (\partial F_\text{nonrel}^\text{n/p}/\partial\delta)_\rho$ has also an explicit $\delta$ dependence, the second derivative is given by

$$\left(\frac{\partial^2 F_\text{nonrel}^\text{n/p}}{\partial \delta^2}\right)_\rho = \left(\frac{\partial F_{\text{nonrel}(1)}^\text{n/p}}{\partial \delta}\right)_{\tilde\mu_\text{n/p}} + \left(\frac{\partial \bar F_{\text{nonrel}(1)}^\text{n/p}}{\partial \tilde\mu_\text{n/p}}\right)_\delta \left(\frac{\partial \tilde\mu_\text{n/p}}{\partial \delta}\right)_\rho, \tag{5.14}$$

---

[5] This feature is also evident from the branch points at $\delta = \pm 1$ of the explicit expressions given by Eqs. (5.4) and (5.5).

## 5. Isospin-Asymmetry Dependence of the Nuclear Equation of State

and similar for higher derivatives. One can write down the following general expression for the $n$-th derivative with respect to $\delta$ of $F_i^{n/p}$, $i \in \{\text{nonrel}, \text{corr}\}$, at fixed density (and temperature):

$$\left(\frac{\partial^n F_i^{n/p}}{\partial \delta^n}\right)_\rho = -\frac{(\pm 1)^n \alpha T^{5/2}(n-1)!}{(1 \pm \delta)^n} \mathscr{Y}_i^{(n)}(\tilde{x}_{n/p}), \tag{5.15}$$

where the functions $\mathscr{Y}_i^{(n)}$ are defined recursively as

$$\mathscr{Y}_i^{(n \geq 1)}(\tilde{x}_{n/p}) = \frac{\tilde{x}_{n/p}}{\max(n-1, 1)} \frac{\text{Li}_{3/2}(\tilde{x}_{n/p})}{\text{Li}_{1/2}(\tilde{x}_{n/p})} \frac{\partial}{\partial \tilde{x}_{n/p}} \mathscr{Y}_i^{(n-1)}(\tilde{x}_{n/p}) - (1 - \delta_{n,1}) \mathscr{Y}_i^{(n-1)}(\tilde{x}_{n/p}), \tag{5.16}$$

with $\delta_{k,l}$ the Kronecker delta. The expressions to start the recursion are

$$\mathscr{Y}_{\text{nonrel}}^{(0)}(\tilde{x}_{n/p}) = \ln(-\tilde{x}_{n/p}) \text{Li}_{3/2}(\tilde{x}_{n/p}) - \text{Li}_{5/2}(\tilde{x}_{n/p}), \qquad \mathscr{Y}_{\text{corr}}^{(0)}(\tilde{x}_{n/p}) = -\frac{15T}{8M}\text{Li}_{7/2}(\tilde{x}_{n/p}). \tag{5.17}$$

One then obtains for $\bar{A}_{\text{nonint},2n}$ the expression

$$\bar{A}_{\text{nonint},2n}(T, \tilde{x}) = \frac{T}{2n \, \text{Li}_{3/2}(\tilde{x})} \left(\mathscr{Y}_{\text{nonrel}}^{(2n)}(\tilde{x}) + \mathscr{Y}_{\text{corr}}^{(2n)}(\tilde{x})\right), \tag{5.18}$$

where $\tilde{x} = -\exp(\tilde{\mu}/T)$, with $\tilde{\mu}$ determined by $\rho = -2\alpha T^{3/2}\text{Li}_{3/2}(x)$.

*Asymptotics.* The nonrelativistic contribution to the free energy per particle is given by

$$\bar{F}_{\text{nonrel}} = \frac{F_{\text{nonrel}}^n + F_{\text{nonrel}}^p}{\rho_n + \rho_p} = T \frac{\ln(-\tilde{x}_n)\text{Li}_{3/2}(\tilde{x}_n) - \text{Li}_{5/2}(\tilde{x}_n) + \ln(-\tilde{x}_p)\text{Li}_{3/2}(\tilde{x}_p) - \text{Li}_{5/2}(\tilde{x}_p)}{\text{Li}_{3/2}(\tilde{x}_n) + \text{Li}_{3/2}(\tilde{x}_p)}. \tag{5.19}$$

This expression diverges in the limit of infinite temperature ($T \to \infty$) and in the limit of zero density ($\rho \to 0$). However, one can still extract the asymptotic behavior of the isospin asymmetry dependence. From $\rho_{n/p} = -\alpha T^{3/2}\text{Li}_{3/2}(\tilde{x}_{n/p})$ it follows that the two limits $T \to \infty$ (at fixed density) and $\rho \to 0$ (at fixed nonzero temperature) both correspond to $\tilde{\mu}_{n/p} \to -\infty$ and $\tilde{x}_{n/p} \to 0$. Using $\text{Li}_\nu(x) \xrightarrow{x \to 0} x$, the expression for $\bar{F}_{\text{nonrel}}$ in these limits is formally given by

$$[\bar{F}_{\text{nonrel}}]_{\rho \to 0} = [\bar{F}_{\text{nonrel}}]_{T \to \infty} = \frac{\tilde{\mu}_n \tilde{x}_n + \tilde{\mu}_p \tilde{x}_p}{\tilde{x}_n + \tilde{x}_p} - T. \tag{5.20}$$

The asymptotic behavior of the (auxiliary) chemical potentials is given by [cf. Eq. (A.12)]

$$[\tilde{\mu}_{n/p}]_{\rho \to 0} = [\tilde{\mu}_{n/p}]_{T \to \infty} = T \ln\left(\frac{\rho_{n/p}}{\alpha T^{3/2}}\right). \tag{5.21}$$

From this and using $\rho_{n/p} = \rho(1 \pm \delta)/2$ one arrives at the following expression for $[\bar{F}_{\text{nonrel}}]_{\rho \to 0}$ and $[\bar{F}_{\text{nonrel}}]_{T \to \infty}$ as functions of $T$, $\rho$ and $\delta$:

$$[\bar{F}_{\text{nonrel}}]_{\rho \to 0} = [\bar{F}_{\text{nonrel}}]_{T \to \infty} = T \ln\left(\frac{\rho}{2\alpha T^{3/2}}\right) + \frac{T}{2}\Big((1 + \delta)\ln(1 + \delta) + (1 - \delta)\ln(1 - \delta)\Big) - T. \tag{5.22}$$

The isospin-asymmetry dependent part of Eq. (5.22) contains the term $(1 - \delta)\ln(1 - \delta)$, whose first derivative diverges at $\delta = 1$. (Note that this term is associated with $F_{\text{nonrel}}^p$). Written in terms of the proton fraction $Y = (1 - \delta)/2$ it has the familiar form $Y \ln(Y)$, i.e., the isospin-asymmetry dependent part of Eq. (5.22) corresponds to the (usual) *entropy of mixing*. From Eq. (5.22), the asymptotic behavior of the Maclaurin coefficients at finite temperature is given by

$$[\bar{A}_{\text{nonint},2n}]_{\rho \to 0} = [\bar{A}_{\text{nonint},2n}]_{T \to \infty} = \frac{T}{2(2n!)} \frac{\partial^{2n}}{\partial \delta^{2n}} \Big((1 + \delta) \ln(1 + \delta) + (1 - \delta) \ln(1 - \delta)\Big)\Big|_{\delta=0}$$
$$= \frac{T}{2n(2n - 1)}, \tag{5.23}$$

which comes entirely from the logarithmic terms, $\sim \ln(-\tilde{x}_{\text{n/p}}) = \tilde{\mu}_{\text{n/p}}/T$, in the original expression for $\bar{F}_{\text{nonrel}}$. Note that $[\bar{A}_{\text{nonint},2n}]_{\rho \to 0}$ is finite although $[\bar{F}_{\text{nonrel}}]_{\rho \to 0}$ diverges logarithmically, $\sim \ln(\rho)$. In the $T \to \infty$ limit the Maclaurin coefficients diverge linearly with $T$, so their ratios are finite and have the same limiting values as obtained in the $\rho \to 0$ limit, i.e.,

$$\frac{\bar{A}_{\text{nonrel},2n}}{\bar{A}_{\text{nonrel},2(n+1)}}\Big|_{T \neq 0, \rho \to 0} = \frac{\bar{A}_{\text{nonrel},2n}}{\bar{A}_{\text{nonrel},2(n+1)}}\Big|_{T \to \infty} = \frac{2n^2 + 3n + 1}{2n^2 - n} \xrightarrow{n \to \infty} 1. \tag{5.24}$$

This shows that for $T \to \infty$ and $\rho \to 0$ the radius of convergence of the Maclaurin expansion of the noninteracting EoS is the same as the one for $T \to 0$, i.e., $R_\delta = 1$.

### 5.1.2. Behavior at Vanishing Proton Fraction

In Sec. 5.1.1 we have seen that at zero temperature the second derivative with respect to $\delta$ of the (nonrelativistic) free Fermi gas contribution to the free energy per particle diverges at $\delta = 1$. The divergent behavior is entirely caused by the proton contribution $\sim (1 - \delta)^{5/3}$, i.e.,

$$\frac{\partial^2}{\partial \delta^2}\Big[(1 + \delta)^{5/3} + (1 - \delta)^{5/3}\Big] = \frac{10}{9(1 + \delta)^{1/3}} + \frac{10}{9(1 - \delta)^{1/3}} \xrightarrow{\delta \to 1} \infty. \tag{5.25}$$

In the $\rho \to 0$ limit, however, already the first derivative of the (proton contribution to the) free energy per particle diverges at $\delta = 1$ for $T \neq 0$ (but is finite for $\delta \neq 1$):

$$\left(\frac{\partial \bar{F}_{\text{nonrel}}}{\partial \delta}\right)_{T \neq 0, \rho \to 0} = \frac{\partial}{\partial \delta} \frac{T}{2}\Big[(1 + \delta)\ln(1 + \delta) + (1 - \delta)\ln(1 - \delta)\Big] = \frac{T}{2}\Big[\ln(1 + \delta) - \ln(1 - \delta)\Big] \xrightarrow{\delta \to 1} \infty. \tag{5.26}$$

The question concerning the behavior for nonzero temperatures and densities arises. By Eq. (5.15), the first two $\delta$ derivatives of $F_{\text{nonrel}}$ are given by

$$F_{\text{nonrel}(1)} := \left(\frac{\partial F_{\text{nonrel}}}{\partial \delta}\right)_{T,\rho} = -\frac{\alpha T^{5/2}}{1 + \delta} \ln(-\tilde{x}_{\text{n}})\text{Li}_{3/2}(\tilde{x}_{\text{n}}) + \frac{\alpha T^{5/2}}{1 - \delta} \ln(-\tilde{x}_{\text{p}})\text{Li}_{3/2}(\tilde{x}_{\text{p}}), \tag{5.27}$$

$$F_{\text{nonrel}(2)} := \left(\frac{\partial F_{\text{nonrel}(1)}}{\partial \delta}\right)_{T,\rho} = -\frac{\alpha T^{5/2}}{(1 + \delta)^2} \frac{(\text{Li}_{3/2}(\tilde{x}_{\text{n}}))^2}{\text{Li}_{1/2}(\tilde{x}_{\text{n}})} - \frac{\alpha T^{5/2}}{(1 - \delta)^2} \frac{(\text{Li}_{3/2}(\tilde{x}_{\text{p}}))^2}{\text{Li}_{1/2}(\tilde{x}_{\text{p}})}, \tag{5.28}$$

where in each case the first term corresponds to the neutron part $F^{\text{n}}_{\text{nonrel}}$, and the second term to the contribution from protons $F^{\text{p}}_{\text{nonrel}}$. The derivatives of the neutron part are finite for $\delta \in [0, 1]$, but the derivatives of the proton part give an undefined expression "0/0" for $\delta \to 1$, which cannot be resolved by l'Hôpital's rule. One can however still evaluate Eq. (5.27) and (5.28) numerically for large proton fractions. The numerical results obtained for $\bar{F}_{\text{nonrel}(1)} = F_{\text{nonrel}(1)}/\rho$ at fixed temperatures $T = (0, 1, 10)$ MeV are plotted as functions of $[\ln(1 + \delta) - \ln(1 - \delta)]$ for different densities in Fig. (5.1). One sees that at finite temperatures, for very small proton fractions $F_{\text{nonrel}(1)}$ is to very high accuracy a linear function of $[\ln(1 + \delta) - \ln(1 - \delta)]$ for all values of $\rho$, indicating that the nondifferentiability of the free energy at $Y$ is maintained also for finite densities. For the second derivative one finds that $F_{\text{nonrel}(2)}$ diverges for $Y \to 0$, but $(1 - \delta) F_{\text{nonrel}(2)}$ approaches the value $T/2$ [cf. Eq. (5.26)] also for finite densities, cf. Fig. 5.3.

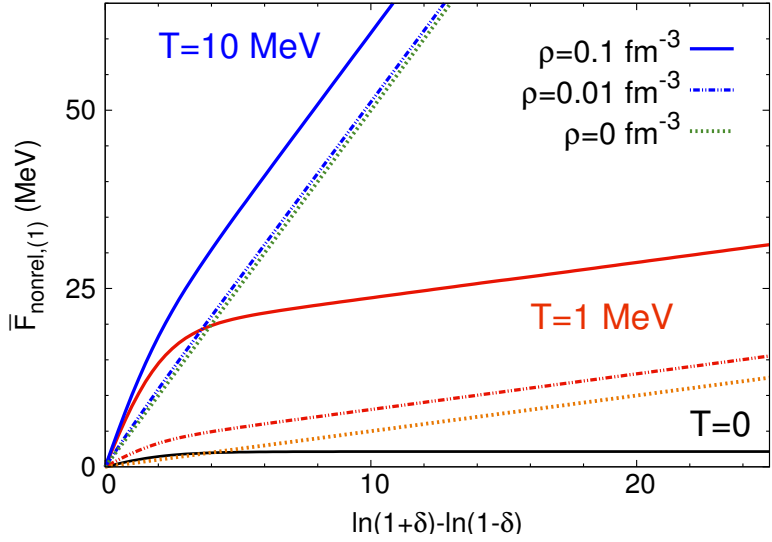

**Figure 5.1.:** $\bar{F}_{\text{nonrel}(1)}$ as a function of the derivative of the "entropy of mixing form".

In sum, one finds that $\bar{F}_{\text{nonrel}(1)}(T, \rho, \delta)$ is logarithmically divergent for $\delta \to 1$ at finite $T$ (but is finite at $T = 0$, which must be seen as a consequence of the entropy being zero at $T = 0$). The origin of this behavior is the same as for the logarithmic divergence of $\bar{F}_{\text{nonrel}}(T, \rho, \delta)$ as $\rho \to 0$ (at finite $T$), i.e., the divergence of the (auxiliary) chemical potentials for vanishing particle densities, $\mu_{\text{n/p}} \to -\infty$ for $\rho_{\text{n/p}} \to 0$.

As discussed above, the singularity of the isospin-asymmetry derivatives of the free energy per particle in the $\delta \to 1$ limit can be associated with the entropy of mixing. It is thus interesting to consider the isospin-asymmetry derivatives of the nonrelativistic Fermi gas contribution to the internal energy per particle $\bar{E} = \bar{F} + T\bar{S}$, with $\bar{S} = -\partial \bar{F}/\partial T$. From $E = -\frac{3\alpha T^{5/2}}{2}(\text{Li}_{5/2}(\tilde{x}_\text{n}) + \text{Li}_{5/2}(\tilde{x}_\text{n}))$ one obtains for the first two derivatives of the internal energy density the expressions

$$E_{\text{nonrel}(1)} := \left(\frac{\partial E_{\text{nonrel}}}{\partial \delta}\right)_{T,\rho} = -\frac{3\alpha T^{5/2}}{2(1+\delta)} \frac{(\text{Li}_{3/2}(\tilde{x}_\text{n}))^2}{\text{Li}_{1/2}(\tilde{x}_\text{n})} + \frac{3\alpha T^{5/2}}{2(1-\delta)} \frac{(\text{Li}_{3/2}(\tilde{x}_\text{p}))^2}{\text{Li}_{1/2}(\tilde{x}_\text{p})}, \quad (5.29)$$

$$E_{\text{nonrel}(2)} := \left(\frac{\partial E_{\text{nonrel}(1)}}{\partial \delta}\right)_{T,\rho} = -\frac{3\alpha T^{5/2}}{2} \sum_{\text{n,p}} \frac{1}{(1\pm\delta)^2}\left[\frac{(\text{Li}_{3/2}(\tilde{x}_{\text{n/p}}))^2}{\text{Li}_{1/2}(\tilde{x}_{\text{n/p}})} - \frac{(\text{Li}_{3/2}(\tilde{x}_{\text{n/p}}))^3 \text{Li}_{-1/2}(\tilde{x}_{\text{n/p}})}{(\text{Li}_{1/2}(\tilde{x}_{\text{n/p}}))^2}\right]. \quad (5.30)$$

For both of these expressions the proton part is undefined at $\delta = 1$. Evaluating Eq. (5.29) numerically, one finds that $\bar{E}_{\text{nonrel}(1)}$ remains finite at $\delta = 1$, cf. Fig. 5.2. The second derivative $\bar{E}_{\text{nonrel}(2)}$ however diverges at $\delta = 1$, as for $T = 0$. However, as seen in Fig. 5.3, the divergence is of the same form as the one for $\bar{F}_{\text{nonrel}(2)}$, i.e., the expression $(1 - \delta)\bar{E}_{\text{nonrel}(2)}$ is finite at $\delta = 1$ and approaches the value $3T/2$.

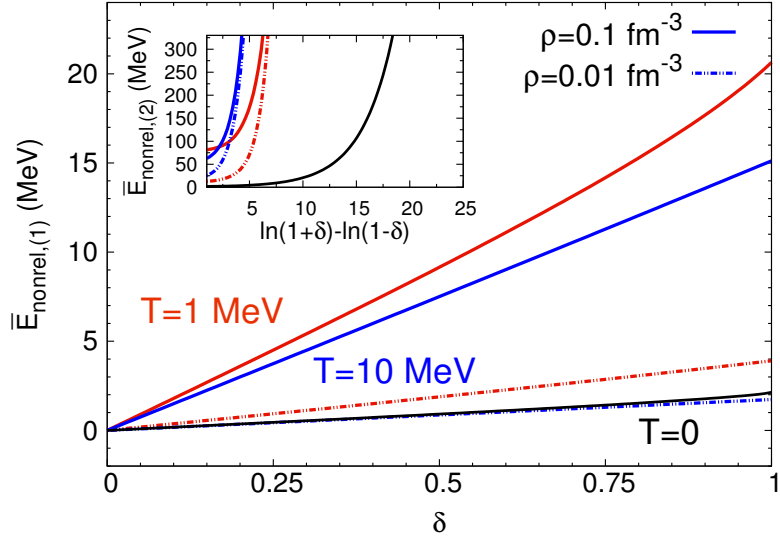

**Figure 5.2.:** Results for $\bar{E}_{\text{nonrel}(1)}$ (main plot) and $\bar{E}_{\text{nonrel}(2)}$ (inset). Note that $\bar{E}_{\text{nonrel}(1)}$ is a nonmonotonic function of $T$.

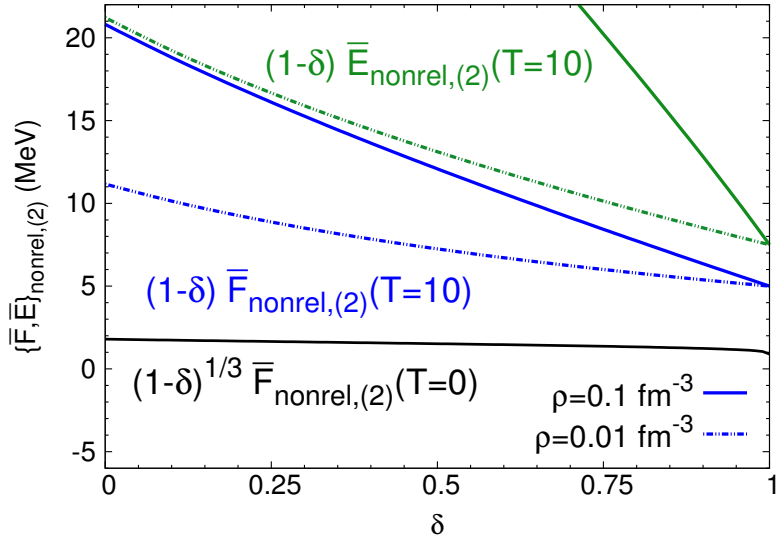

**Figure 5.3.:** Second isospin-asymmetry derivatives of the noninteracting contribution to the free energy per particle and the internal energy per particle, multiplied with a factor $(1-\delta)$ at finite $T$ and a factor $(1-\delta)^{1/3}$ at zero $T$, respectively.

In Sec. 4.2.1 we have found that these special features of the $\delta$ dependence in the very neutron-rich region are related to physical constraints on the boundary of the spinodal. Otherwise, the influence of these features, for example regarding properties of neutron stars,[6] is not clear at this stage and would have to be investigated in future studies. It is however obvious that these features cannot be captured by the usual Maclaurin expansion in terms of $\delta$. Moreover, while the results at zero temperature would suggest the applicability of a Maclaurin expansion in the third root of the proton fraction $y = Y^{1/3} = [(1-\delta)/2]^{1/3}$, the results discussed in this section show that such an expansion is not applicable for nonzero temperatures.

---

[6] Note also that in the expression for the proton chemical potential, the $\delta$ derivative of the free energy is multiplied with a factor $(1-\delta)$, cf. Eq. (4.56).

### 5.1.3. Evaluation of Expansion Coefficients

In the first part of this section we study the density and temperature dependence of the four leading coefficients in the Maclaurin expansion in $\delta^2$ of the noninteracting EoS. In the second part we introduce the expansion of the zero-temperature EoS in terms of the proton fraction $Y = (1 - \delta)/2$, and compare this expansion to the usual one in terms of the isospin asymmetry.

*Isospin-Asymmetry Expansion.* To identify the qualitative behavior of the $\delta^2$ expansion, it is useful to introduce the quantities $\xi = 1 - \frac{\bar{A}_2(T,\rho)}{\bar{F}_{\text{sym}}(T,\rho)}$ and $\zeta_{2N} = \frac{\sum_{n=2}^{N} \bar{A}_{2n}(T,\rho)}{\bar{F}_{\text{sym}}(T,\rho)}$. If the isospin-asymmetry expansion converges for $\delta \in [-1, 1]$ in a given region in the temperature-density plane, then $\zeta_{2N} \xrightarrow{N \to \infty} \xi$ in that region. From Eqs. (5.4) and (5.23), one finds that the nonrelativistic contribution $\bar{F}_{\text{nonrel}}(T, \rho, \delta)$ the quantities have the parameter-independent limiting values

$$\xi(T \to 0, \rho \neq 0) = 1 - \frac{10}{9(2^{5/3} - 2)} \simeq 0.054, \tag{5.31}$$

$$\xi(T \to \infty, \rho) = \xi(T \neq 0, \rho \to 0) = 1 - \frac{1}{\ln(4)} \simeq 0.279, \tag{5.32}$$

$$\zeta_{2N}(T \to 0, \rho \neq 0) = \frac{10}{9(2^{5/3} - 2)} \sum_{n=2}^{N} \prod_{k=0}^{2n-3} \frac{1 + 3k}{9 + 3k}, \tag{5.33}$$

$$\zeta_{2N}(T \to \infty, \rho) = \zeta_{2N}(T \neq 0, \rho \to 0)) = \frac{1}{\ln(2)} \sum_{n=3}^{2N} \frac{(-1)^{n+1}}{n}. \tag{5.34}$$

Note that in the $T \to \infty$ and $\rho \to 0$ cases, one finds explicitly that the expansion converges at the radius of convergence (alternating harmonic series). Comparing $\zeta_{2N}/\xi|_{\rho \neq 0, T \to 0} \in \{0.646, 0.814, 0.883, 0.918, 0.999\}$ and $\zeta_{2N}/\xi|_{T \to \infty} = \zeta_{2N}/\xi|_{T \neq 0, \rho \to 0} \in \{0.431, 0.604, 0.696, 0.754, 0.987\}$ for $N \in \{2, 3, 4, 5, 100\}$, one can deduce that the relative accuracy of the isospin-asymmetry expansion (truncated at a given order $N$) of $\bar{F}_{\text{nonrel}}$ decreases for increasing values of $T$ and decreasing values of $\rho$. To examine this behavior in more detail, we introduce the weight factors $\beta_{\text{nonint},2n} = \bar{A}_{\text{nonint},2n}/\bar{F}_{\text{nonint,sym}}$ as a means to specify the relative size of the different Maclaurin coefficients. The weight factors are related to the symmetry free energy via

$$\bar{F}_{\text{nonint,sym}}(T, \rho) = \sum_{n=1}^{\infty} \bar{A}_{\text{nonint},2n}(T, \rho) = \bar{F}_{\text{nonint,sym}}(T, \rho) \sum_{n=1}^{\infty} \beta_{\text{nonint},2n}(T, \rho). \tag{5.35}$$

The results for the first weight factor $\beta_{\text{nonint},2}$ as well as ones for the ratios $\beta_{\text{nonint},2n}/\beta_{\text{nonint},2(n+1)}$ [with $n = 1, 2, 3$] are displayed in Fig. 5.4. One sees that the ratios $\beta_{\text{nonint},2n}/\beta_{\text{nonint},2(n+1)}$ [or, equivalently, the ratios $\bar{A}_{\text{nonint},2n}/\bar{A}_{\text{nonint},2(n+1)}$] are at finite density monotonic decreasing functions of temperature and at finite temperature monotonic increasing functions of density, and the $T \to \infty$ limiting values are approached relatively quickly as the temperature is increased. The decrease of the convergence rate of the expansion of $\bar{F}_{\text{nonint}} = \bar{F}_{\text{nonrel}} + \bar{F}_{\text{corr}}$ with increasing temperature or decreasing density is entirely caused by the presence of the logarithmic terms in the nonrelativistic contribution $\bar{F}_{\text{nonrel}}$, i.e., by the entropy of mixing. The convergence rate of the expansion of the relativistic correction term alone increases with temperature, similar to the expansion of the noninteracting contribution to the internal energy per particle, $\bar{E}_{\text{nonint}} = \bar{E}_{\text{nonrel}} + \bar{E}_{\text{corr}}$. The weight factors associated with the expansion of $\bar{E}_{\text{nonint}}$ are denoted as $\alpha_{\text{nonint},2n}$; the results for $\alpha_{\text{nonint},2}$ and $\alpha_{\text{nonint},2}/\alpha_{\text{nonint},4}$ are shown in Fig. 5.5. One sees that indeed the convergence rate of the expansion of $\bar{E}_{\text{nonint}}$ increases strongly with increasing temperature.

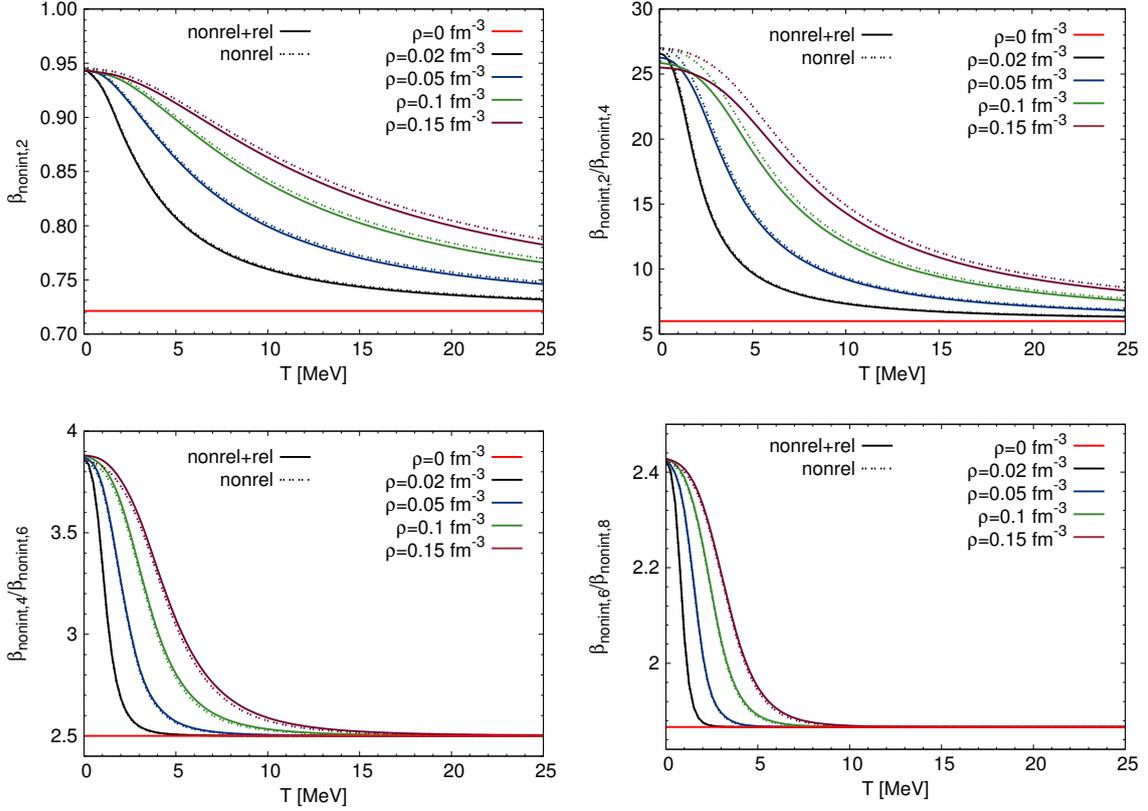

**Figure 5.4.:** Temperature dependence of the first weight factor $\beta_{\text{nonint},2}$ and the ratios $\beta_{\text{nonint},2n}/\beta_{\text{nonint},2(n+1)}$ for $n = 1, 2, 3$ at different densities. The full lines show the results with the relativistic correction term included ("nonrel+rel"), the dotted lines the nonrelativistic results ("nonrel"). Note that the deviations between the relativistically improved and the nonrelativistic results decrease with increasing values of $n$, indicating the opposite convergence behavior of the expansion in powers of $\delta^2$ of $F_{\text{nonrel}}$ and $F_{\text{corr}}$, respectively.

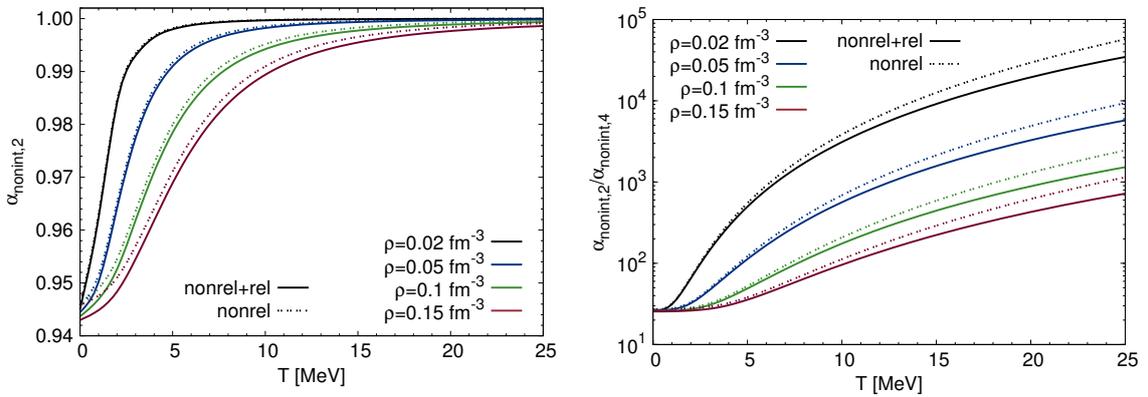

**Figure 5.5.:** Temperature dependence of the first weight factor $\alpha_{\text{nonint},2}$ and the ratio $\alpha_{\text{nonint},2}/\alpha_{\text{nonint},4}$, corresponding to the expansion in powers of $\delta^2$ of the noninteracting contributions to the internal energy per particle, $E_{\text{nonrel}}$ and $E_{\text{corr}}$.

*5. Isospin-Asymmetry Dependence of the Nuclear Equation of State*

**Proton-Fraction Expansion.** The matter in the interior of neutron stars has only a very small fraction of protons, roughly $Y_{\text{NS}} \simeq (0.05 - 0.1)$. For a given EoS, the value of $Y_{\text{NS}}$ is determined by the condition of beta equilibrium, and depends sensitively on the isospin-asymmetry dependence of the nuclear EoS in the very neutron-rich region. This suggests to consider instead of the "usual" isospin-asymmetry expansion [Eq. (5.1)] an expansion in terms of the proton fraction $Y = \rho_p/\rho = (1-\delta)/2$ about $Y = 0$. This "proton-fraction expansion" has been investigated at zero temperature in Ref. [376] using the in-medium $\chi$PT framework of Refs. [237, 151]. It was found that

- fractional powers of $Y$ emerge also from the interaction contributions [cf. also Sec. 4.2.1].[7]

- expect for the extremely neutron-rich region the EoS constructed from this expansion (including terms up to $\sim Y^2$) deviates considerably from the exact results.

In the following we show that the second feature is entirely due to the interaction contributions to the EoS, i.e., the proton-fraction expansion of the noninteracting contribution (at zero temperature) is well-converged at low orders in the entire physical region $Y \in [0, 1]$.

We start by introducing for a given contribution $\alpha$ to the EoS the general form of the expansions in terms of the isospin-asymmetry and in terms of the proton fraction about $\delta = 0$ and $Y = 0$, respectively, i.e.,

$$\bar{F}^{[\alpha]}(T, \rho, \delta) \sim \sum_{n=0}^{N_\delta} \bar{A}_n^{[\alpha]}(T, \rho)\, \delta^n, \qquad \bar{F}^{[\alpha]}(T, \rho, y) \sim \sum_{n=0}^{N_y} \bar{B}_n^{[\alpha]}(T, \rho)\, y^n, \qquad (5.36)$$

where we have formulated the second expansion in terms of in the third root of the proton fraction $y = Y^{1/3}$ to be able to deal with fractional powers of $Y$. To illustrate the differences regarding convergence of these expansions in the case of the interaction contribution, we examine in Fig. 5.6 the results obtained for the first-order contribution from 3N interactions $\bar{F}_1^{3N}(T, \rho, \delta)$ at zero temperature and $\rho = 0.15\,\text{fm}^{-3}$ (using n3lo414).[8] The quantities shown are the deviations $\Delta \bar{F}^{[\alpha]} := \bar{F}_{\text{approx}}^{[\alpha]} - \bar{F}^{[\alpha]}$, where $\bar{F}^{[\alpha]}$ corresponds to the exact results, and $\bar{F}_{\text{approx}}^{[\alpha]}$ to the results from the expansions (truncated at the order indicated in the leged). In the case of the $y$ expansion, the leading orders are given by $\sim (y^3, y^5, y^6)$, corresponding to $\sim (Y, Y^{5/3}, Y^2)$. One sees that the deviations become are very sizeable in the case of the $y$ expansion; including higher powers in the expansion leads to systematic improvements in the very neutron-rich region, but for $\delta \lesssim 0.9$ the $\delta$ expansion leads to better results. The convergence behavior of the respective expansions can be described in terms the functions $\mathcal{A}_{N_\delta}^{[\alpha]}(T, \rho, \delta)$ and $\mathcal{B}_{N_y}^{[\alpha]}(T, \rho, y)$ defined as

$$\mathcal{A}_{N_\delta}^{[\alpha]}(T, \rho, \delta) := 1 - \frac{\sum_{n=1}^{N_\delta} \bar{A}_n^{[\alpha]}(T, \rho)\, \delta^n}{\bar{F}^{[\alpha]}(T, \rho, \delta) - \bar{F}^{[\alpha]}(T, \rho, \delta = 0)}, \qquad (5.37)$$

$$\mathcal{B}_{N_y}^{[\alpha]}(T, \rho, y) := 1 - \frac{\sum_{n=1}^{N_y} \bar{B}_n^{[\alpha]}(T, \rho)\, y^n}{\bar{F}^{[\alpha]}(T, \rho, y) - \bar{F}^{[\alpha]}(T, \rho, y = 0)}. \qquad (5.38)$$

If the expansions converge, then $\mathcal{A}_{N_\delta}^{[\alpha]} \xrightarrow{N_\delta \to \infty} 0$ and $\mathcal{B}_{N_y}^{[\alpha]} \xrightarrow{N_y \to \infty} 0$, respectively. For $\bar{F}_1^{3N}(T, \rho, \delta)$ at $T = 0$ and $\rho = 0.15\,\text{fm}^{-3}$, the values of these functions at $\delta = 1$ and $Y = 0.5$, respectively, are

---

[7] Note that this feature cannot be captured by the "usual" isospin-asymmetry expansion.

[8] In the case of the first-order 3N contribution at zero temperature, a semi-analytical computation of the expansion coefficients is feasible [calculations by N. Kaiser]. For $\rho = 0.15\,\text{fm}^{-3}$, the values of the leading expansion coefficients are $\bar{A}_{2,4,6}^{[3N]} = (-0.26, -0.56, -0.08)$ MeV and $\bar{B}_{3,5,6}^{[3N]} = (6.88, -2.84, -16.19)$ MeV, respectively.

*5. Isospin-Asymmetry Dependence of the Nuclear Equation of State*

given by

$$\mathcal{A}^{[3N]}_{N_\delta=2,4,6}(T=0,\rho,\delta=1) = (\,0.71,\,-0.10,\,0.01\,), \tag{5.39}$$

$$\mathcal{B}^{[3N]}_{N_y=3,5,6}(T=0,\rho,y=0.5^{1/3}) = (\,-2.79,\,-1.80,\,2.66\,). \tag{5.40}$$

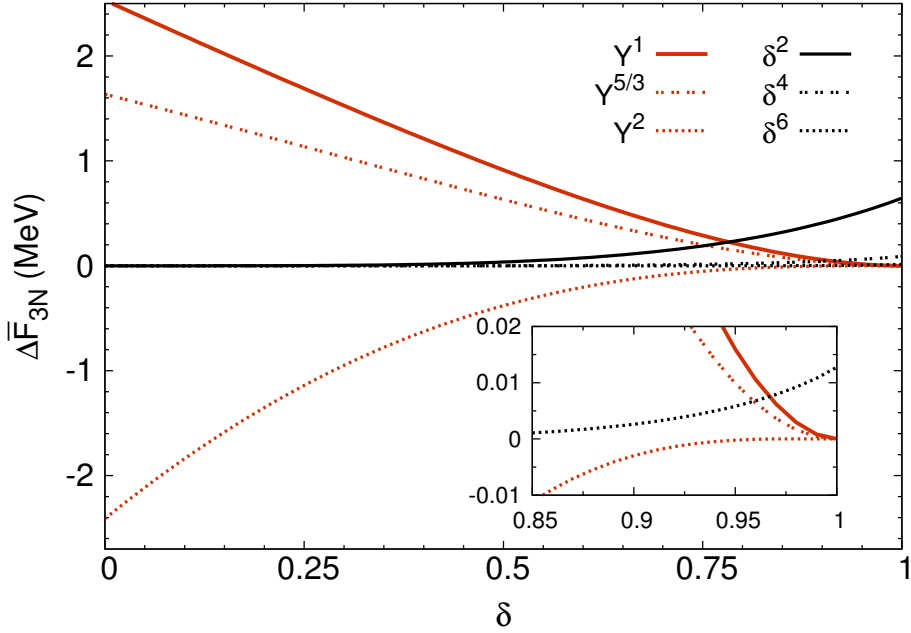

**Figure 5.6.:** Deviations $\Delta\bar{F} := \bar{F}_{\text{approx}} - \bar{F}$ of the approximations $\bar{F}_{\text{approx}}$ constructed from the truncated expansions[9] in $Y$ (red lines) and $\delta$ (black lines) from the exact results $\bar{F}$ for the case of the first-order contribution from 3N interactions $\bar{F}_1^{3N}(T,\rho,\delta)$ at zero temperature and $\rho = 0.15\,\text{fm}^{-3}$. The inset magnifies the behavior in the extremely neutron-rich region.

Using the functions $\mathcal{A}^{[\alpha]}_{N_\delta}$ and $\mathcal{B}^{[\alpha]}_{N_y}$, we now examine the two expansions given by Eq. (5.36) for the nonrelativistic free nucleon gas EoS $\bar{F}^{n+p}_{\text{nonrel}} = \bar{F}^n_{\text{nonrel}} + \bar{F}^p_{\text{nonrel}}$ at zero temperature. It is instructive to first study the neutron and proton contributions separately. Furthermore, we consider also the $\rho \to 0$ limit (at fixed $T \neq 0$), or equivalently, the $T \to \infty$ limit of $\bar{F}^{n/p}_{\text{nonrel}}$, which is formally given by [cf. Eq. (5.22)]

$$\bar{F}^{n/p}_{\text{nonrel}} \xrightarrow{T\to\infty} \left.\frac{\tilde{\mu}_{n/p}\tilde{x}_{n/p}}{\tilde{x}_n+\tilde{x}_p}\right|_{T\to\infty} = \underbrace{\frac{T}{2}(1\pm\delta)\ln(1\pm\delta)}_{\mathscr{X}_1^{n/p}} + \underbrace{\frac{T}{2}(1\pm\delta)\ln\left(\frac{\rho}{2\,\mathrm{e}\,\alpha\,T^{3/2}}\right)}_{\mathscr{X}_2^{n/p}}. \tag{5.41}$$

The second term $\mathscr{X}_2^{n/p}$ completely dominates the behavior of $\bar{F}^n_{\text{nonrel}}$ and $\bar{F}^p_{\text{nonrel}}$ as $\rho \to 0$ and $T \to \infty$, respectively. It gives only a contribution to the leading coefficient in the respective expansions, which leads to $\mathcal{A}^{n/p}_{N_\delta}(T \to \infty, \rho, \delta) = 0$ for $N_\delta \geq 1$ and $\mathcal{B}^{n/p}_{N_y}(T \to \infty, \rho, y) = 0$ for $N_y \geq 3$. The isospin-asymmetry dependent parts of $\mathscr{X}_2^{n/p}$ cancel each other, so the expansions of $\bar{F}^{n+p}_{\text{nonrel}}$ are completely determined by $\mathscr{X}_1^{n/p}$. For these reasons, in the following we examine for the separate neutron and proton contributions the expansions of the isolated first term $\mathscr{X}_1^{n/p}$.

---

[9] For example, the red dashed lines labeled "$Y^2$" correspond to the approximation $\bar{F}_{\text{approx}} = \bar{B}_0 + \bar{B}_3 y + \bar{B}_5 y^5 + \bar{B}_6 y^6 = \bar{B}_0 + \bar{B}_3 Y + \bar{B}_5 Y^{5/3} + \bar{B}_6 Y^2$.

*Neutron Contribution.* For the neutron contribution $\bar{F}^{\text{n}}_{\text{nonrel}}$, the asymptotic behavior of the nonvanishing coefficients (apart the one with $n = 0$) is given by

$$\bar{A}^{\text{n}}_n \xrightarrow{T \to 0} \left(\frac{3\pi^2}{2}\right)^{5/3} \frac{\rho^{2/3}}{10\pi^2 M} \prod_{k=1}^{n} \frac{8 - 3k}{3k}, \qquad \bar{A}^{\text{n}}_n \xrightarrow{T \to \infty} \begin{cases} \dfrac{T}{2}, & \text{for } n = 1 \\ \dfrac{(-1)^n T}{2n(n-1)}, & \text{for } n \geq 2 \end{cases}, \qquad (5.42)$$

$$\bar{B}^{\text{n}}_{3n} \xrightarrow{T \to 0} (3\pi^2)^{2/3} \frac{\rho^{2/3}}{10\pi^2 M} \prod_{k=1}^{n} \frac{3k - 8}{3k}, \qquad \bar{B}^{\text{n}}_{3n} \xrightarrow{T \to \infty} \begin{cases} -T - T\ln(2), & \text{for } n = 1 \\ \dfrac{T}{n(n-1)}, & \text{for } n \geq 2 \end{cases}, \qquad (5.43)$$

where in the case of the $T \to \infty$ results the term $\mathscr{X}^{\text{n}}_2$ is neglected. Note that radius of convergence of the proton-fraction expansion is $R_y = 1$, which is twice as large as the one of the isospin-asymmetry expansion $R_\delta = 1$. The asymptotic values of $\mathscr{A}^{\text{n}}_{N_\delta}(T, \rho, \delta = 1)$ are given by

$$\mathscr{A}^{\text{n}}_{N_\delta = 1,2,3,4,5,10}(T = 0, \rho, 1) = (\,0.233,\, -0.011,\, 0.007,\, -0.003,\, 0.002,\, -2 \times 10^{-4}\,), \quad (5.44)$$

$$\mathscr{A}^{\text{n}}_{N_\delta = 1,2,3,4,5,10}(T \to \infty, \rho, 1) = (\,0.279,\, -0.082,\, 0.038,\, -0.022,\, 0.014,\, -0.004\,), \quad (5.45)$$

and the ones of $\mathscr{B}^{\text{n}}_{N_y}(T, \rho, y = 0.5^{1/3})$ are given by[10]

$$\mathscr{B}^{\text{n}}_{N_y = 3,6,9,12,15,30}(T = 0, \rho, 0.5^{1/3}) = (\,-0.217,\, -0.014,\, -0.002,\, -6 \times 10^{-4},\, -2 \times 10^{-4},\, -10^{-6}\,), \quad (5.46)$$

$$\mathscr{B}^{\text{n}}_{N_y = 3,6,9,12,15,30}(T \to \infty, \rho, 0.5^{1/3}) = (\,-0.221,\, -0.041,\, -0.011,\, -0.003,\, -0.001,\, -10^{-5}\,). \quad (5.47)$$

One sees that in the case of the neutron contribution $F^{\text{n}}_{\text{nonrel}}$, the proton-fraction expansion converges faster as the one in the isospin-asymmetry. For both expansions the results indicate that the convergence rate decreases with temperature. Note also that while $\mathscr{B}^{\text{n}}_{N_y} < 0$, the sign of the functions $\mathscr{A}^{\text{n}}_{N_\delta}$ alternates with $N_\delta$.

*Proton Contribution.* In the case of the proton contribution $\bar{F}^{\text{p}}_{\text{nonrel}}$, the proton-fraction expansion is not applicable at finite temperature where the first derivative of $\bar{F}^{\text{p}}_{\text{nonrel}}$ with respect to $Y$ is singular, cf. Sec. 5.1.2. At zero temperature the "expansion" of $\bar{F}^{\text{p}}_{\text{nonrel}}$ in terms of $y$ has only one nonvanishing coefficient:

$$\bar{B}^{\text{p}}_5 \xrightarrow{T \to 0} \left(\frac{3\pi^2}{2}\right)^{5/3} \frac{\rho^{2/3}}{10\pi^2 M}, \qquad (5.48)$$

The isospin-asymmetry expansion of $\bar{F}^{\text{p}}_{\text{nonrel}}$ on the other hand is well-defined. The asymptotic values of the nonvanishing isospin-asymmetry coefficients are given by (the term $\mathscr{X}^{\text{p}}_2$ is neglected)

$$\bar{A}^{\text{p}}_n \xrightarrow{T \to 0} \left(\frac{3\pi^2}{2}\right)^{5/3} \frac{\rho^{2/3}}{10\pi^2 M} \prod_{k=1}^{n} \frac{3k - 8}{3k}, \qquad \bar{A}^{\text{p}}_n \xrightarrow{T \to \infty} \begin{cases} -\dfrac{T}{2}, & \text{for } n = 1 \\ \dfrac{T}{2n(n-1)}, & \text{for } n \geq 2 \end{cases}, \qquad (5.49)$$

---

[10] Note that to examine the two expansions on an equal footing we always compare the values of $\mathscr{A}^{[\alpha]}_{N_\delta}$ and $\mathscr{B}^{[\alpha]}_{N_y}$ for the same number of nonvanishing coefficients.

The radii of convergence are the same as for the neutron contribution. The zero-temperature values of the $\mathcal{A}_{N_\delta}^{\text{p}}$ function at $\delta = 1$ are[11]

$$\mathcal{A}_{N_\delta=1,2,3,4,5,10}^{\text{p}}(T=0,\rho,1) = (\ -0.667,\ -0.111,\ 0.049,\ -0.029,\ 0.019,\ 0.006\ ). \quad (5.50)$$

One sees that compared to the neutron contribution, the convergence rate of the $\delta$ expansion of the proton contribution at zero temperature is considerably decreased.

*Free Nucleon Gas.* Finally, we turn to the expansions of the complete nonrelativistic free nucleon gas EoS given by $\bar{F}_{\text{nonrel}}^{\text{n+p}} = \bar{F}_{\text{nonrel}}^{\text{n}} + \bar{F}_{\text{nonrel}}^{\text{p}}$. The zero-temperature values of $\mathcal{A}_{N_\delta}^{\text{n+p}}$ and $\mathcal{B}_{N_y}^{\text{n+p}}$ are given by[12]

$$\mathcal{A}_{N_\delta=2,4,6,8,10,20}^{\text{n+p}}(T=0,\rho,1) = (\ 0.054,\ 0.019,\ 0.010,\ 0.006,\ 0.004,\ 0.001\ ), \quad (5.51)$$

$$\mathcal{B}_{N_y=3,5,6,9,12,27}^{\text{n+p}}(T=0,\rho,0.5^{1/3}) = (\ -0.401,\ -0.025,\ -0.005,\ -0.001,\ -3\times10^{-4},\ -2\times10^{-6}), \quad (5.52)$$

One sees that while at first and second order ($N_\delta = 2, 4$ and $N_y = 3, 5$, respectively) the quadratic isospin-asymmetry expansion is more precise, the increase in precision for higher orders is much more pronounced in the case of the proton-fraction expansion.

**Summary.** To summarize, we have found for the noninteracting contribution to the free energy per particle of infinite homogeneous nuclear matter that

- the relative accuracy of the isospin-asymmetry expansion about $\delta = 0$ decreases substantially with increasing temperature.

- in the case of the neutron contribution, the expansion in terms of the proton fraction about $Y = 0$ has better converge properties as the one in the isospin asymmetry about $\delta = 0$. This expansion is however not applicable for the proton contribution at finite temperature due to the nonanalytic form of the $Y$ dependence [asymptotically given by $\sim Y \ln(Y)$] associated with the entropy of mixing.

- the nonanalytic form of the $Y$ dependence has a crucial influence on the isospin-asymmetry dependence of the nuclear liquid-gas instability [cf. Sec. 4.2.1].

These points make clear that the isospin-asymmetry dependence of the noninteracting term in the many-body perturbation series cannot be parametrized explicitly in a satisfactory way. Regarding the construction of a global nuclear EoS for astrophysical applications, this however does not constitute a serious obstacle since noninteracting contribution can easily be computed exactly. In the case of the interaction contributions, however, an exact treatment is computationally much more expensive. This motivates a detailed investigation of the isospin-asymmetry expansion about $\delta = 0$ of the interaction contributions (we have already seen in Fig. 5.6 that for the first-order 3N contribution at $T = 0$, this expansion is very well-converged at order $\delta^6$).

---

[11] Since the $\mathcal{X}_1^{\text{p}}$ part of Eq. (5.41) vanishes for $\delta \in \{0, 1\}$ the $T \to \infty$ limit of the $\mathcal{A}_{N_\delta}^{\text{p}}$ functions (without $\mathcal{X}_2^{\text{p}}$) is singular at $\delta = 1$.

[12] For completeness, the values of $\mathcal{A}_{N_\delta}^{\text{n+p}}$ in the $T \to \infty$ limit are $\mathcal{A}_{N_\delta=2,4,6,8,10,20}^{\text{n+p}}(T \to \infty, \rho, 1) = (\ 0.279,\ 0.158,\ 0.110,\ 0.085,\ 0.069,\ 0.035\ )$.

## 5.2. Isospin-Asymmetry Dependence of Interaction Contributions

Here, we examine the $\delta$ dependence of the interaction contributions to the free energy per particle $\bar{F}(T,\rho,\delta)$, computed in second-order MBPT. We use the sets of chiral low-momentum potentials n3lo414 and n3lo450, for which "realistic" results where obtained in Chap. 3 for the limiting cases of SNM and PNM. The interaction contributions to the quadratic, quartic and sextic Maclaurin coefficients $\bar{A}_{2,4,6}(T,\rho)$ are extracted using finite differences. We discuss the accuracy of the finite-difference method, and examine the results for each contribution individually. We will find that many-body contributions beyond the mean-field level give rise to coefficients $\bar{A}_{2n\geq 4}$ that diverge in the $T \to 0$ limit, and that this behavior is associated (for the most part) with the presence of a logarithmic term $\sim \delta^4 \ln|\delta|$ at zero temperature. Finally, a method that allows to extract (to good accuracy) the coefficient of the (leading) logarithmic term $\sim \delta^4 \ln|\delta|$ is introduced, benchmarked against the exact results for an $S$-wave contact interaction, and then applied to the results for the second-order EoS from n3lo414 and n3lo450.

### 5.2.1. Finite-Difference Methods

The interaction contributions to the Maclaurin coefficients $\bar{A}_{2n}(T,\rho)$ can in principle be computed from the explicit expressions obtained for the isospin-asymmetry derivatives (at fixed density and temperature) of the different many-body contributions. The length of these expressions however increases rapidly with the order of the derivative. To avoid the numerical evaluation of these lengthy expressions, we instead extract the isospin-asymmetry derivatives numerically using finite differences.

The general form of the $\mathcal{N}$-point central finite-difference approximation for $\bar{A}_{2n}(T,\rho)$ is (using a uniform grid with stepsize $\Delta\delta$ and grid length $N$):

$$\bar{A}_{2n}(T,\rho) \simeq \frac{1}{(2n)!(\Delta\delta)^{2n}} \sum_{k=0}^{N} (2-\delta_{k,0})\, \omega_{2n}^{N,k} \bar{F}(T,\rho,k\Delta\delta) =: \bar{A}_{2n}^{N,\Delta\delta}(T,\rho), \qquad (5.53)$$

where $\mathcal{N} = 2N + 1 \geq 2n + 1$. The finite-difference coefficients $\omega_{2n}^{N,k}$ are determined by the matching of Lagrange polynomials to the data and can be computed using the algorithm given in Ref. [147]. The formal order of accuracy of the finite-difference approximation $\bar{A}_{2n}^{N,\Delta\delta}(T,\rho)$ for $\bar{A}_{2n}(T,\rho)$ is $\Delta\delta^{2N-2n+2}$. Because $\bar{F}(T,\rho,\delta)$ can be computed only to a finite accuracy, $\Delta\delta$ cannot be chosen arbitrarily small without the results being affected by numerical noise. Varying $N$ and $\Delta\delta$ provides a means to test the validity of the results for $\bar{A}_{2n}(T,\rho)$. If the finite-difference approximation is valid, the result should not change under (moderate) variations of $N$ and $\Delta\delta$. Because the size of the higher-order derivatives as well as the numerical precision varies with the respective many-body contribution as well as the values of the external parameters $T$ and $\rho$, to avoid artifacts this variation needs to be carried out for every individual contribution and for every single EoS point. Carrying out this procedure and systematically increasing the precision of the numerical integration routine in the process, we were able to obtain accurate results (with respect to the second-order EoS, and with a further qualification concerning the second-order contribution, see footnote[14]) for the quadratic, quartic, and sextic (and to a lesser degree of precision also the octic) Maclaurin coefficients.[13]

Representative results obtained for the different many-body contributions to $\bar{A}_6^{N,\Delta\delta}(T,\rho)$ and $\bar{A}_8^{N,\Delta\delta}(T,\rho)$ are plotted in Fig. 5.7 for $T = (4, 5, 15)$ MeV and $\rho = (0.15, 0.30)\,\text{fm}^{-3}$. One sees that for the first-order contributions the finite-difference values are well converged for a large range of $\Delta\delta$ values. The numerical noise becomes visible only for very small values of $\Delta\delta$, and is more pronounced at larger temperatures. In the case of the first-order contributions the finite-difference values are well converged for a large region of $\Delta\delta$ values. At low temperatures the $\delta$ dependence of the first-order DDNN contribution approximately matches that of the first-order contribution with genuine 3N interactions, but this behavior deteriorates as the temperature is increased.

In the case of the second-order (normal) contributions, the results for $\bar{A}_{2n\geq 4}^{N,\Delta\delta}$ are similar the ones corresponding to the first-order contributions for high temperatures (i.e., $T = 15$ MeV in Fig. 5.7), but for lower temperatures (i.e., $T = 5$ MeV) a slight bending is observed as $\Delta\delta$ is increased. This bending gradually increases as the temperature is decreased and the density increased, cf. the results for $\bar{A}_6^{N,\Delta\delta}$ and $\bar{A}_8^{N,\Delta\delta}$ at $T = 4$ MeV and $\rho = 0.30$ fm$^{-3}$. The underlying reason for this behavior is the divergence of the Maclaurin expansion in the low-temperature and high-density regime (the divergent behavior of the expansion at low temperatures is examined in Secs. 5.2.3, 5.2.4 and 5.2.5). In that regime, the higher-order isospin-asymmetry derivatives $\partial^{2n}\bar{F}/\partial\delta^n$ become very large for small values of $\delta$. As a consequence, if the stepsize of the finite-difference approximation is chosen too large, the behavior near $\delta = 0$ is not resolved, which leads to the observed bending. However, as seen in Fig. 5.7, even in that case a plateau, i.e., region where the stepsize dependence of the finite-difference results approximately vanishes, can be found for small values of $\Delta\delta$. Note that in the region where the stepsize dependence nearly vanishes also the grid-length dependence is decreased.[14] Note also that we have restricted the "plateau analysis" to the np-channel contribution; in Secs. 5.2.3, 5.2.4 and 5.2.5 we will show that only this contribution is responsible for the nonanalytic behavior at low temperatures. Most of the subsequent discussion will therefore focus on the np-channel, and omit the small contribution to $\bar{A}_{2n\geq 4}$ from the nn- and pp-channels.

To check the values obtained for the second-order contribution to the higher-order Maclaurin coefficients in the low-temperature region, we have calculated $\bar{A}_{4,6,8}^{N,\Delta\delta}$ also by applying the finite-difference method iteratively, i.e., by evaluating finite differences of

$$\frac{\partial^n \bar{F}(T,\rho,\delta)}{\partial \delta^n} \simeq \frac{1}{(\Delta\delta)^n} \sum_{k=-N}^{N} \omega_n^{N,k} \bar{F}(T,\rho,\delta+k\Delta\delta). \tag{5.54}$$

The iterative method involves variable stepsizes and grid lengths at every iteration step and thus behaves differently concerning error systematics; for adequate values of $\Delta\delta$ and $N$ the iterative method gives matching results, cf. Fig. 5.7. Furthermore, we have extracted the Maclaurin coefficients for the (total) free energy per particle $\bar{F}(T,\rho,\delta)$ both by applying finite differences to the data $\bar{F}(T,\rho,\delta)$ and by summing the Maclaurin coefficients obtained for the individual many-body contributions: the results where found to agree very well (up to several digits for most cases), but overall the method of extracting Maclaurin coefficients individually was found to more precise (propagation of errors).

---

[13] We note that in the case of the first-order 3N contributions at zero temperature, $\bar{F}_{1,3N}(T=0,\rho,\delta)$, semi-analytical expressions for the Maclaurin coefficients can be derived [calculations by N. Kaiser]. The results for $\bar{A}_6(T=0)$ obtained in this way were found to match the results predicted by the finite-difference method up to four relevant digits.

[14] For the "highly divergent" regime [i.e., for low temperatures and high densities, cf. Fig. 5.16] we have determined the values of $\bar{A}_{2n\geq 4}$ by averaging the plateau values. The size of the plateau decreases with decreasing temperature, which is why we have restricted the analysis to $T \geq 2$ MeV (but for $T \geq 2$ we have always found a significant plateau). The validity of the extracted values for $\bar{A}_{2n\geq 4}$ is also evident from the improved description of the asymptotic behavior for $\delta \to 0$ with $\bar{F}_{4,6}(T,\rho,\delta)$, cf. Figs. 5.17 and 5.19 in Sec. 5.4.

## 5. Isospin-Asymmetry Dependence of the Nuclear Equation of State

Finally, we note that regarding the numerical extraction of isospin-asymmetry derivatives at $\delta \neq 0$, we have found that the finite differences become considerably less accurate (for fixed values of $\Delta\delta$ and $N$) for larger values of $\delta$. This feature can be expected to be related to the weaker convergence properties of the expansion in $\delta$ for increasing values of the expansion point $\delta_0$ [cf. Sec. 5.1.1], and the nonanalytic behavior of the isospin-asymmetry dependence in the neutron-rich region [cf. Secs. 4.2.1 and 5.1]. Moreover, concerning the extraction of the coefficients in the proton-fraction expansion [cf. Secs. 4.2.1 and 5.1] we have also found that the quality of the finite-difference method is further reduced. This issue has several aspects. One factor is the need to use forward differences[15] (which are less precise). In addition, when applied to derivatives with respect to the third root of the proton fraction the finite differences probe a narrower region of isospin asymmetries, with unevenly distributed data points. Another issue is of course the fact that the inferior convergence behavior (in the case of the interaction contributions) of the proton-fraction expansion and the large size of the higher-order coefficients in that expansion (see Ref. [376] and footnote[8]).

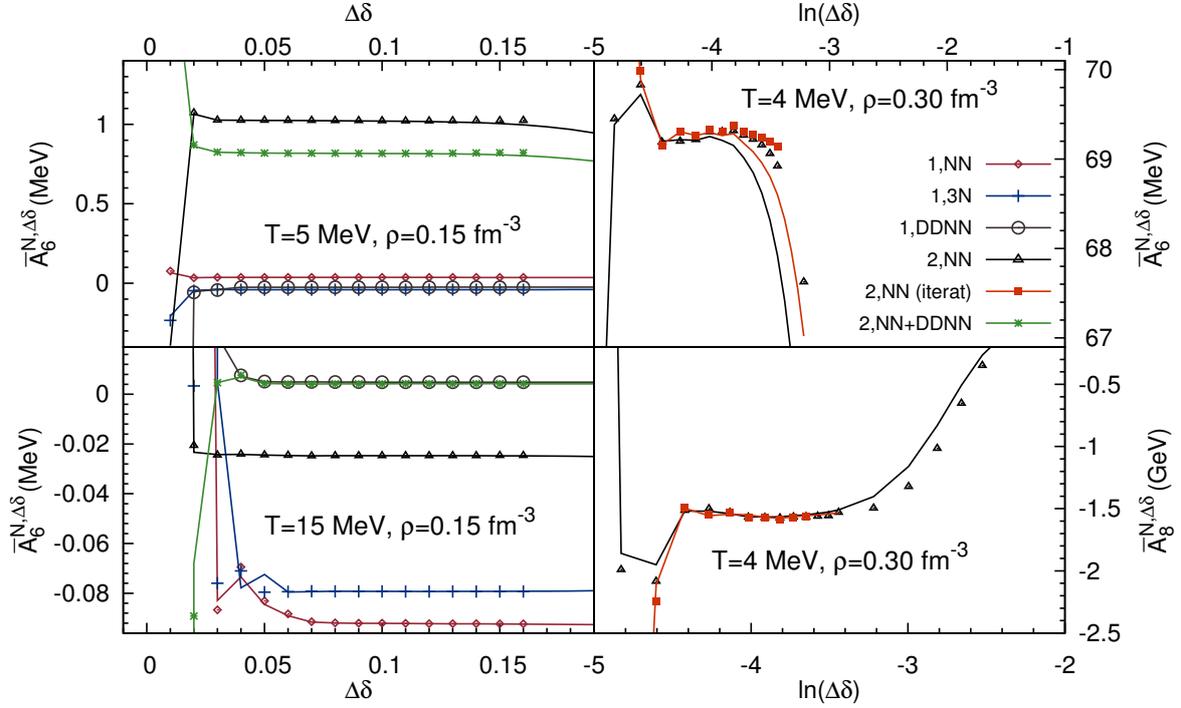

**Figure 5.7.:** Left column: finite-difference results for the first- and second-order normal contributions to $\bar{A}_6$ for $T = (5, 15)$ MeV and $\rho = 0.15\,\text{fm}^{-3}$ (calculated using n3lo414). The lines correspond to $N = 2 + n$, the points to $N = 3 + n$; in each the stepsize variation extends from $\Delta\delta = 0.01$ to $\Delta\delta = 1/N$. At $T = 15$ MeV the first-order NN and 3N contributions are given in units 10 keV. The approximate equality of the first-order DDNN and the second-order NN+DDNN results at $T = 15$ MeV is coincidental. Right column: $\bar{A}_6^{N,\Delta\delta}$ and $\bar{A}_8^{N,\Delta\delta}$ for the second-order NN contribution (np-channel only) at $T = 4$ MeV and $\rho = 0.30\,\text{fm}^{-3}$. Also shown are the results obtained from the iterative method (based on Eq. (5.54), with $n = 1$; see text for details). Note that $\bar{A}_8^{N,\Delta\delta}$ is given in units GeV.

---

[15] This applies to the case where the data is restricted to the physical regime $\delta \in [-1, 1]$. However, note that the EoS can formally be extended beyond the physical range by introducing complex proton auxiliary chemical potentials, which leads to a limiting condition of $\delta \leq 1 + \delta_{\text{"BEC"}}(T, \rho)$, with $\delta_{\text{"BEC"}}(T, \rho) = \sqrt{2}(M/\pi)^{3/2}T^{3/2}\rho^{-1}\text{Li}_{3/2}(1) \simeq 0.00248\,(T/\text{MeV})^{3/2}(\rho/\text{fm}^{-3})^{-1}$.

## 5.2.2. Hartree-Fock Results

In the following we examine the numerical results for $\bar{A}_{2,4,6}(T,\rho)$ associated with the first-order contributions from two- and three-nucleon interactions. We will find that at the Hartree-Fock level the Maclaurin coefficients are hierarchically ordered, $\bar{A}_2 > \bar{A}_4 > \bar{A}_6 \ (> \bar{A}_8)$, which indicates that the isospin-asymmetry expansion converges at this level.

**NN *Contribution.*** The partial-wave representation of the first-order NN contribution to the free energy density of isospin-asymmetric nuclear matter (ANM) is given by

$$F_1^{\text{NN}}(T,\tilde{\mu}_{\text{n}},\tilde{\mu}_{\text{p}}) = \frac{2}{\pi^3}\int_0^\infty dp\, p^2 \int_0^\infty dK\, K^2 \int_{-1}^1 d\cos\theta \sum_{J,L,S,m_T}(2J+1)\sum_{\tau_1\geq\tau_2,\tau_2}\delta_{m_T,\tau_1+\tau_2} n^{\tau_1}_{|\vec{K}+\vec{p}|}n^{\tau_2}_{|\vec{K}-\vec{p}|}\langle p|\bar{V}_{\text{NN}}^{J,L,L,S,T}|p\rangle, \tag{5.55}$$

The contributions from the neutron-neutron (nn), proton-proton (pp) and neutron-proton (np) channels are given by the sum over isospin indices.

The results for the quadratic, quartic and sextic Maclaurin coefficients of $F_{1,\text{NN}}(T,\rho,\delta)$ are displayed in the left column of Fig. 5.8. Also shown are the results for $F_{\text{sym}} - \bar{A}_2$. One sees that the quadratic coefficient $\bar{A}_2$ greatly outweighs the higher-order coefficients and matches the symmetry free energy $F_{\text{sym}}$ with high accuracy. Except for the quartic coefficient $\bar{A}_4$ the Maclaurin coefficients are monotonic increasing functions of density and decreasing functions of temperature. In the high-temperature regime the Maclaurin coefficients are hierarchically ordered, $\bar{A}_2 \gg \bar{A}_4 \gg \bar{A}_6 (\gg \bar{A}_8)$, but this behavior breaks down to some extent at low temperatures where $\bar{A}_4$ and $\bar{A}_6$ are of similar size. Note that the deviations between the n3lo414 and n3lo450 results are significantly reduced in the case of $\bar{A}_4$ and $\bar{A}_6$ as compared to $\bar{A}_2$.

**3N *Contribution.*** The first-order contribution arising from chiral N2LO three-nucleon interactions is given by (as discussed in Sec. 3.2, we can omit the usual Jacobi-momentum regulator):

$$F_1^{\text{3N}}(T,\tilde{\mu}_{\text{n}},\tilde{\mu}_{\text{p}}) = \rho^{-1}\int_0^\infty dk_1 \frac{k_1}{2\pi^2} \int_0^\infty dk_2 \frac{k_2}{2\pi^2} \int_0^\infty dk_3 \frac{k_3}{2\pi^2} \mathcal{X}(k_1,k_2,k_3), \tag{5.56}$$

where $\mathcal{X}(k_1,k_2,k_3) = \mathcal{X}^{(c_E)} + \mathcal{X}^{(c_D)} + \mathcal{X}^{(\text{Hartree})} + \mathcal{X}^{(\text{Fock})}$ and the different contributions $\mathcal{X}^{(c_E)}$, $\mathcal{X}^{(c_D)}$, $\mathcal{X}^{(\text{Hartree})}$ and $\mathcal{X}^{(\text{Fock})}$ correspond to the different components of the N2LO three-nucleon interaction, see Fig. 3.1 for the corresponding many-body diagrams. The explicit expressions for $\mathcal{X}^{(c_E)}$, $\mathcal{X}^{(c_D)}$, $\mathcal{X}^{(\text{Hartree})}$ and $\mathcal{X}^{(\text{Fock})}$ are given by

$$\mathcal{X}^{(c_E)} = \frac{1}{2}\mathcal{K}^{(c_E)} n^{\text{p}}_{k_1} n^{\text{n}}_{k_2}\left(n^{\text{p}}_{k_3} + n^{\text{n}}_{k_3}\right), \tag{5.57}$$

$$\mathcal{X}^{(c_D)} = \frac{1}{6}\mathcal{K}^{(c_E)}\left[\left(n^{\text{p}}_{k_1} + 2n^{\text{n}}_{k_1}\right)n^{\text{p}}_{k_2}n^{\text{n}}_{k_3} + \left(n^{\text{n}}_{k_1} + 2n^{\text{p}}_{k_1}\right)n^{\text{n}}_{k_2}n^{\text{p}}_{k_3}\right], \tag{5.58}$$

$$\mathcal{X}^{(\text{Hartree})} = \frac{1}{12}\mathcal{K}^{(\text{Hartree})}\left(n^{\text{p}}_{k_1}n^{\text{p}}_{k_2} + 4n^{\text{p}}_{k_1}n^{\text{n}}_{k_2} + n^{\text{n}}_{k_1}n^{\text{n}}_{p_2}\right)\left(n^{\text{p}}_{k_3} + n^{\text{n}}_{k_3}\right), \tag{5.59}$$

$$\mathcal{X}^{(\text{Fock})} = \frac{1}{6}\left(\mathcal{K}^{(\text{Fock},c_1)} + \mathcal{K}^{(\text{Fock},c_3)}\right)\left[\left(n^{\text{p}}_{k_1}n^{\text{p}}_{k_2} + 2n^{\text{n}}_{p_1}n^{\text{n}}_{k_2}\right)n^{\text{p}}_{p_3} + \left(2n^{\text{p}}_{k_1}n^{\text{p}}_{k_2} + n^{\text{n}}_{k_1}n^{\text{n}}_{k_2}\right)n^{\text{n}}_{k_3}\right]$$
$$+ \frac{1}{6}\mathcal{K}^{(\text{Fock},c_4)}\left[\left(n^{\text{p}}_{k_1} + 2n^{\text{n}}_{k_1}\right)n^{\text{p}}_{k_2}n^{\text{n}}_{k_3}\left(2n^{\text{p}}_{k_1} + n^{\text{n}}_{k_1}\right)n^{\text{n}}_{k_2}n^{\text{p}}_{k_3}\right], \tag{5.60}$$

where the kernels $\mathcal{K}^{(c_E)}$, $\mathcal{K}^{(c_D)}$ and $\mathcal{K}^{(\text{Hartree})}$ are given bin Eqs. (3.36)-(3.38), and $\mathcal{K}^{(\text{Fock},c_i)}$ is the part of the kernel $\mathcal{K}^{(\text{Fock})}$ given in Eq. (3.39) proportional to the low-energy constant $c_i$.

In the right column of Fig. 5.8 the results for the Maclaurin coefficients for $F_{1,3N}(T,\rho,\delta)$ are shown, as well as the corresponding results for $\bar{F}_{\text{sym}} - \bar{A}_2$. Overall, the temperature and density dependence of the first-order 3N Maclaurin coefficients is similar to the behavior of the first-order NN results (the $T$ and $\rho$ dependence of the "convergence rate" is opposite to the behavior of the noninteracting term). The results for the higher-order coefficients $\bar{A}_{2n\geq 4}$ are again of similar for n3lo414 and n3lo450, but (as in the NN case) larger deviations occur in the case of $\bar{A}_2$. Compared to the NN contributions, the 3N contribution to $\bar{A}_2$ is relatively small.[16] The quartic coefficient $\bar{A}_4$ is significantly larger in the 3N case, and gives the dominant contribution to $\bar{F}_{\text{sym}} - \bar{A}_2$, in particular at high temperatures where the sextic (and octic) coefficients are strongly suppressed (as in the NN case).

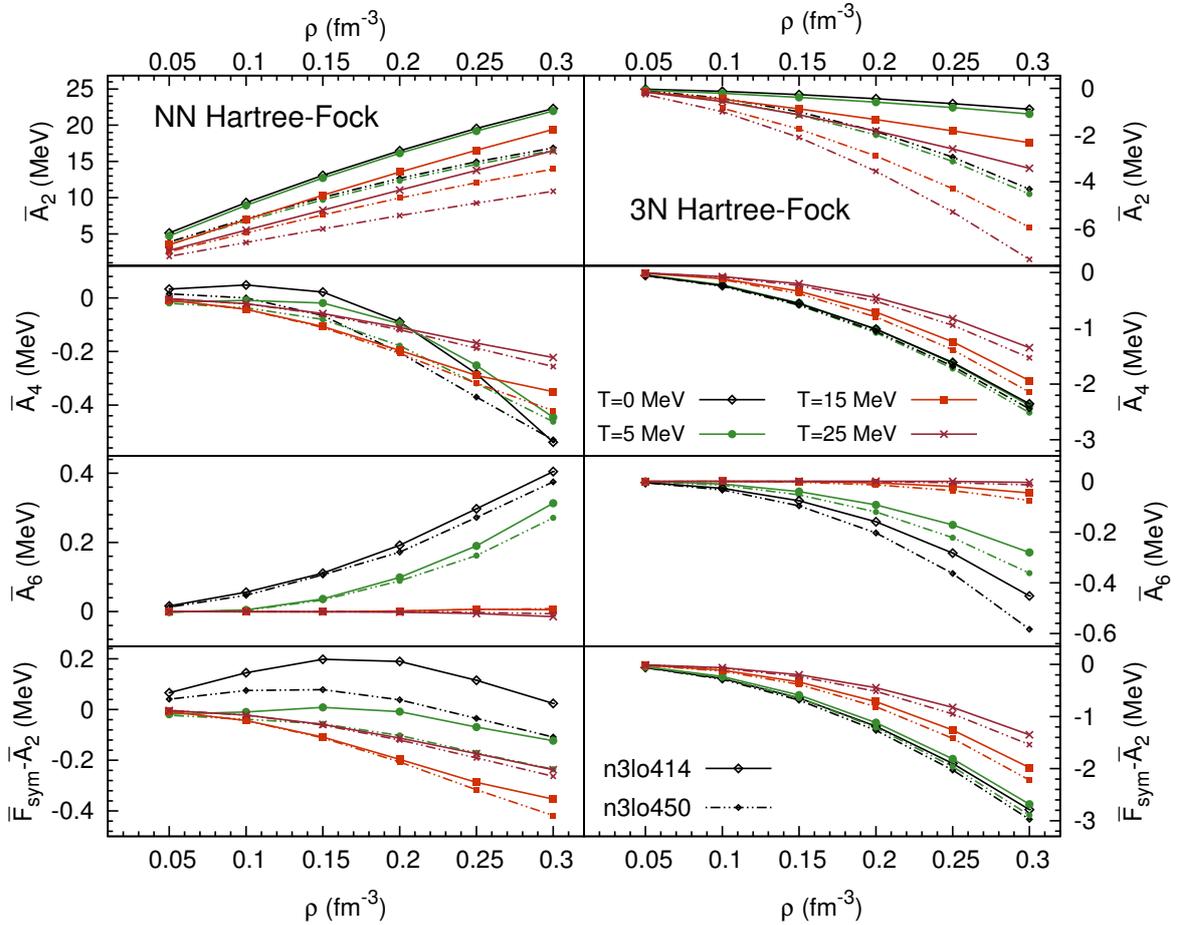

**Figure 5.8.:** Results for $\bar{A}_{2,4,6}$ for the first-order NN and 3N contributions (solid lines: n3lo414, dash-dot lines: n3lo450). Also shown are the results for $\bar{F}_{\text{sym}}(T,\rho) - \bar{A}_2(T,\rho)$.

---

[16] The relatively small size of the 3N contribution to $\bar{A}_2$ from the cancellation of contributions proportional to different low-energy constants (LECs). For different choices of the LECs (e.g., for VLK21 and VLK23) the overall size of the 3N contribution to $\bar{A}_2$ can be increased. We note that we have studied the first-order NN and 3N coefficients ($\bar{A}_{2,4,6}$) also for n3lo500, VLK21, and VLK23. For this larger set of potentials, a larger model dependence is visible, but the hierarchy as well as the overall density and temperature dependence of the 3N coefficients is maintained.

## 5.2.3. Second-Order Contribution[17]

The partial-wave representation of the second-order (normal) NN contribution to the free energy density of ANM is given by:

$$F_{2,\text{normal}}^{\text{NN}}(T, \tilde{\mu}_n, \tilde{\mu}_p) = -\frac{4}{\pi^2} \int_0^\infty dK\, K^2 \int_0^\infty dp_1\, p_1^2 \int_0^\infty dp_2\, p_2^2 \int_{-1}^1 d\cos\theta_1 \int_{-1}^1 d\cos\theta_2 \sum_{J,L_1,L_2} \sum_{J',L_1',L_2'} \sum_{S,m_T}$$

$$\times i^{L_2-L_1} i^{L_1'-L_2'} \sum_{\tau_1 \leq \tau_2} \delta_{m_T,\tau_1+\tau_2} \sum_{\tau_3 \leq \tau_4} \delta_{m_T,\tau_3+\tau_4} \sum_{m_J, m_S, m_{S'}} C(\theta_1, \theta_2)$$

$$\times \langle p_1 | \bar{V}_{\text{NN}}^{J,L_1,L_2,S,T} | p_2 \rangle \langle p_2 | \bar{V}_{\text{NN}}^{J',L_1',L_2',S,T} | p_1 \rangle \mathscr{G}(p_1, p_2, K, \theta_1, \theta_2), \quad (5.61)$$

where $\mathscr{G}(p_1, p_2, K, \theta_1, \theta_2)$ is given by

$$\mathscr{G} = \frac{n_{|\vec{K}+\vec{p}_1|}^{\tau_1} n_{|\vec{K}-\vec{p}_1|}^{\tau_2} (1 - n_{|\vec{K}+\vec{p}_2|}^{\tau_3})(1 - n_{|\vec{K}-\vec{p}_2|}^{\tau_4}) - (1 - n_{|\vec{K}+\vec{p}_1|}^{\tau_1})(1 - n_{|\vec{K}-\vec{p}_1|}^{\tau_2}) n_{|\vec{K}+\vec{p}_2|}^{\tau_3} n_{|\vec{K}-\vec{p}_2|}^{\tau_4}}{\varepsilon(|\vec{K}+\vec{p}_2|, \tau_3) + \varepsilon(|\vec{K}-\vec{p}_2|, \tau_4) - \varepsilon(|\vec{K}+\vec{p}_1|, \tau_1) - \varepsilon(|\vec{K}-\vec{p}_1|, \tau_2)}. \quad (5.62)$$

The function $C(\theta_1, \theta_2)$ collects Clebsch-Gordan coefficients and spherical harmonics, and is given by Eq. (3.22).

***Zero-Temperature Limit.*** At zero temperature the Fermi-Dirac distribution functions in Eq. (5.62) become step functions, which can be absorbed into the boundaries of the integrals. The first part of the numerator of $\mathscr{G}$ is associated with the following conditions on the angular integrals:

$$(1.i) \quad -\min(\alpha_1^{\tau_2}, 1) \leq \cos\theta_1 \leq \min(\alpha_1^{\tau_1}, 1),$$
$$(2.i) \quad -\min(-\alpha_2^{\tau_3}, 1) \leq \cos\theta_2 \leq \min(-\alpha_2^{\tau_4}, 1),$$

where $\alpha_i^\tau = [(k_F^\tau)^2 - K^2 - p_i^2]/(2K p_i)$, with $k_F^{n/p}$ the neutron/proton Fermi momentum. The integration region for $\theta_1$ vanishes unless the following two conditions are satisfied:

$$(1.ii) \quad -\alpha_1^{\tau_2} \leq \alpha_1^{\tau_1} \iff K^2 + p_1^2 \leq [(k_F^{\tau_1})^2 + (k_F^{\tau_2})^2]/2,$$
$$(1.iii) \quad \min(\alpha_1^{\tau_1}, \alpha_1^{\tau_2}) \geq -1 \iff \min(k_F^{\tau_1}, k_F^{\tau_2}) \geq |K - p_1|.$$

Similarly, the integration region for $\theta_2$ vanishes unless the following conditions are satisfied:

$$(2.ii) \quad \alpha_2^{\tau_3} \leq -\alpha_2^{\tau_4} \iff K^2 + p_2^2 \geq [(k_F^{\tau_3})^2 + (k_F^{\tau_4})^2]/2,$$
$$(2.iii) \quad \max(\alpha_2^{\tau_3}, \alpha_2^{\tau_4}) \leq 1 \iff \max(k_F^{\tau_3}, k_F^{\tau_4}) \leq K + p_2.$$

In addition, the following condition arises from the requirement that the intersection of the first two Fermi spheres is nonvanishing:

$$(1.iv) \quad K, p_1 \leq (k_F^{\tau_1} + k_F^{\tau_2})/2.$$

---

[17] Since the sum of the second-order anomalous terms is small (see Fig. 3.6), in the following we consider only the "normal" contribution at second order.

5. *Isospin-Asymmetry Dependence of the Nuclear Equation of State*

Fixing the order of the momentum integrals as in Eq. (5.61), the conditions 1.*ii,iii,iv* and 2.*ii,iii* become

$$(1.iv) \quad 0 \leq K \leq (k_F^{\tau_1} + k_F^{\tau_2})/2,$$
$$(1.ii + iii) \quad \max[0, K - \min(k_F^{\tau_1}, k_F^{\tau_2})] \leq p_1 \leq \min[K + \min(k_F^{\tau_1}, k_F^{\tau_2}), \kappa(\tau_1, \tau_2)],$$
$$(2.ii + iii) \quad \max[0, \kappa(\tau_3, \tau_4), \max(k_F^{\tau_3}, k_F^{\tau_4}) - K] \leq p_2 \leq \infty,$$

where $\kappa(\tau_1, \tau_2) = \kappa(\tau_3, \tau_4) = \sqrt{[(k_F^{\tau_1})^2 + (k_F^{\tau_2})^2]/2 - K^2}$. The second part of the numerator of $\mathcal{G}$ leads to the same condition on the integral boundaries, but with $(p_1, \theta_1)$ and $(p_2, \theta_2)$ interchanged. The energy denominator is antisymmetric under $(p_1, \theta_1) \leftrightarrow (p_2, \theta_2)$, therefore both parts of $\mathcal{G}$ yield identical contributions.

At finite temperature the integrand in Eq. (5.61) is smooth (cf. Sec. 2.2.2), but at zero temperature it diverges at the boundary of the integration region: the conditions $(1.ii+iii)$ and $(2.ii+iii)$ imly $p_1 \leq p_2$, and for $p_1 = p_2$ the energy denominator produces a pole. **This is the origin of the divergence of $\bar{A}_{2n \geq 4}(T, \rho)$ in the zero-temperature limit.**[18]

*Maclaurin Coefficients.* The np-channel results (using the NN potential only) for $\bar{A}_{2,4,6}(T, \rho)$ and $F_{\text{sym}}(T, \rho) - \bar{A}_2(T, \rho)$ are displayed in Fig. 5.9. Similar to the results for the first-order NN contribution, $F_{\text{sym}} - \bar{A}_2$ is small. Again the differences between the n3lo414 and n3lo450 results are significantly decreased for $\bar{A}_4$ and $\bar{A}_6$ as compared to $\bar{A}_2$. In the high-temperature and low-density region it is $\bar{A}_2 \gg \bar{A}_4 > \bar{A}_6 \, ( > \bar{A}_8)$, which indicates that the second-order contribution is an analytic function of the isospin asymmetry in this region. At high density and low temperature, however, this behavior breaks down, and terms beyond $\bar{A}_2$ diverge with alternating sign (as required by charge symmetry) in the zero-temperature limit.

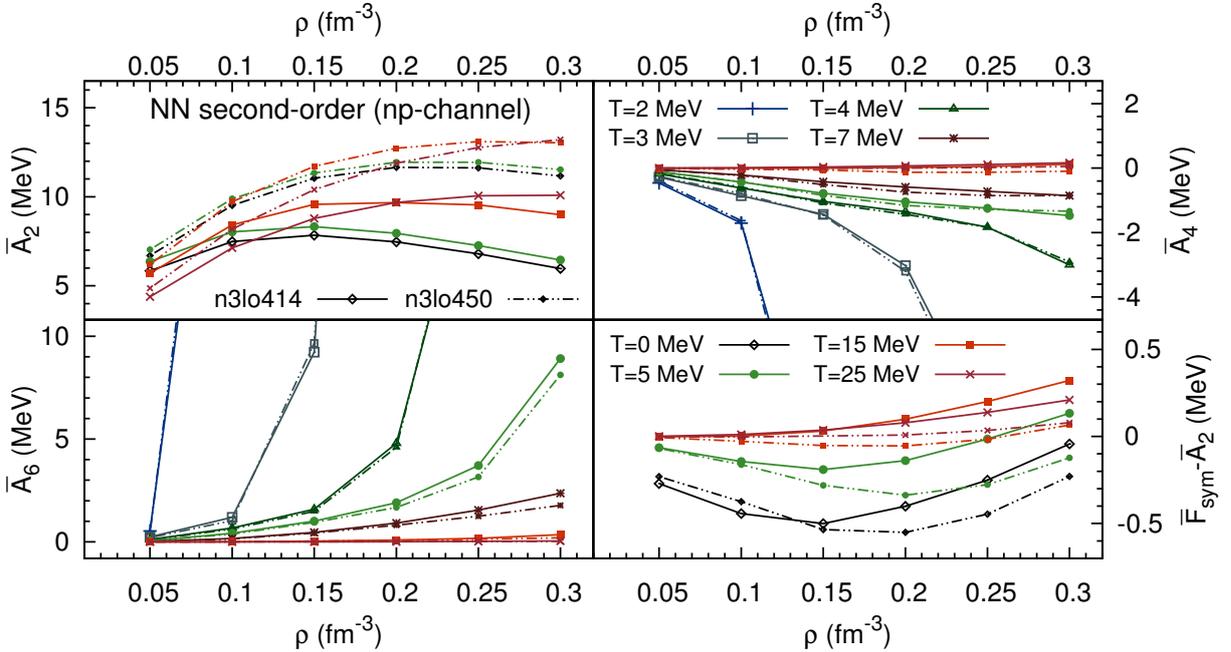

**Figure 5.9.:** Quadratic, quartic and sextic Maclaurin coefficients for the np-channel second-order contribution (with the NN potential only). Also shown are the results for $\bar{F}_{\text{sym}}(T, \rho) - \bar{A}_2(T, \rho)$.

---

[18] This feature is (somewhat) analogous to the singularity of the first derivative at $x = 0$ of the function $f(x) = \int_{x_0}^{x} dy \int_{y_0}^{y} dz \, \frac{1}{z} \sim x \ln|x|$.

## 5. Isospin-Asymmetry Dependence of the Nuclear Equation of State

In Table 5.1 we compare the results for the different second-order (normal) contributions to $\bar{A}_{2n}(T,\rho)$, i.e., the np-channel NN contribution, the total NN contribution, and the total contribution with the combined NN and DDNN potential. One sees that for the higher-order Maclaurin coefficients the difference between the total NN and the np-channel NN contribution almost vanishes, which indicates that the nn- and pp-channels are regular also for realistic nuclear interactions.[19] The deviations between the NN and the NN+DDNN results are more sizable, but the Maclaurin coefficients are still of similar order of magnitude. Because the numerical evaluation of the second-order contribution becomes more involved with the DDNN potential included we have restricted the detailed examination of the isospin-asymmetry expansion at second order to the (np-channel) NN contribution.

|  | $\bar{A}_0$ | $\bar{A}_2$ | $\bar{A}_4$ | $\bar{A}_6$ | $\bar{A}_8$ | $\bar{F}_{\text{sym}}$ |
|---|---|---|---|---|---|---|
| NN (np-channel) | -8.13 | 8.32 | -0.78 | 1.01 | -0.9 | 8.13 |
| NN (total) | -9.60 | 8.64 | -0.80 | 1.02 | -0.9 | 8.36 |
| NN+DDNN (total) | -10.80 | 10.48 | -1.24 | 0.82 | -0.8 | 9.63 |

**Table 5.1.:** Different second-order (normal) contributions to the Maclaurin coefficients and the symmetry free energy at $T = 5$ MeV and $\rho = 0.15$ fm$^{-3}$ (results for n3lo414), see text for details. The values of $\bar{A}_{2n}$ are given in units MeV.

For ANM, the partial-wave representation of the first-order perturbative self-energy is given by (cf. Sec. 3.1)

$$S_{1;k,\tau}(T,\tilde{\mu}_n,\tilde{\mu}_p) = \frac{1}{2\pi}\int_0^\infty dk'\, k'^2 n_{k'} \int_{-1}^1 d\cos\theta_{k'} \sum_{J,L,S,m_T} (2J+1)\, C^{\dagger\, Tm_T}_{\tau,(m_T-\tau)} C^{Tm_T}_{\tau,(m_T-\tau)} \qquad (5.63)$$

$$\times \left\langle \frac{|\vec{k}-\vec{k}'|}{2} \middle| \bar{V}^{J,L,L,S,T}_{\text{NN}} + \frac{1}{2}\bar{V}^{J,L,L,S,T}_{\text{DDNN}} \middle| \frac{|\vec{k}-\vec{k}'|}{2} \right\rangle.$$

In the calculations performed in the previous two chapters, we have included the first-order self-energy in the energy denominator of the second-order term using the effective-mass approximation; for ANM this reads

$$\frac{k^2}{2M} + S_{1;k,\tau}(T,\tilde{\mu}_n,\tilde{\mu}_p) \simeq \frac{k^2}{2M^*_\tau(T,\tilde{\mu}_n,\tilde{\mu}_p)} + U_\tau(T,\tilde{\mu}_n,\tilde{\mu}_p). \qquad (5.64)$$

The effective-mass corrections factor out of the energy denominator for the nn- and pp-channels, but not for the np-channel, leading to a more complicated integral structure in that case. It is clear that the overall temperature and density dependence of the Maclaurin coefficients is not affected (strongly) by including the first-order self-energy corrections. In the case of the nn- and pp-channels we have found explicitly that the magnitude of the finite differences are affected only little by including $M^*/M$ corrections. However, the accuracy of the results is significantly decreased when $M^*$ corrections are included, i.e., the effective-mass approximation does not provide sufficient accuracy when implemented in the numerical extraction of the Maclaurin coefficients. For this reason we have not included self-energy corrections in the detailed analysis of the $\delta$ dependence of the second-order contribution.

---

[19] Although at very low temperatures we were not able to obtain accurate results for the nn- and pp-channel $\bar{A}_{4,6,8}$ (this is due to the structure of $\mathscr{G}(p_1,p_2,K,\theta_1,\theta_2)$ and the associated impediment of the numerical precision at low temperatures), we have observed that the finite-difference results are small and do not show at all an approximately logarithmic stepsize dependence at zero temperature.

*5. Isospin-Asymmetry Dependence of the Nuclear Equation of State*

***Pairing & Thermodynamic Limit.*** Since its origin lies in the presence of an energy-denominator pole at the integration boundary, the property $|\bar{A}_{2n\geq 4}| \xrightarrow{T\to 0} \infty$ arises only in the thermodynamic limit $\{N,\Omega\} \to \infty$ where the reference spectrum becomes continuous. Furthermore, the coefficients $\bar{A}_{2n\geq 4}$ would remain finite also at $T=0$ if pairing correlations were taken into account: in the perturbation series about the BCS ground-state—obtained from the (usual) Bogoliubov-Valatin transformation of the single-particle basis [401, 57]— the sharp distribution functions (step functions) at $T=0$ are smeared througout a thickness $\Delta_k$ (the pairing gap) in energy (cf. Ref. [143] pp.326-336)).

In other terms, the higher-order "Maclaurin coefficients" (i.e., the finite differences, in the case where the system-size is not yet infinite) "diverge" in the combined limits $T \to 0$, $\{N,\Omega\} \to \infty$, and $\Delta_k \to 0$. This means that the Maclaurin expansion exists when $T = \epsilon_T$, $N = 1/\epsilon_N$, and $\Delta_k = \epsilon_\Delta$, where at least one $\epsilon_i$ is nonzero, but it constitutes a divergent asymptotic expansion if the $\epsilon_i$ are "sufficiently" small. For neutron-star matter, both $\epsilon_T$ and $\epsilon_N$ are certainly small, i.e., the thermodynamic and the zero-temperature limits are reasonable approximations. To get an idea regarding the "sufficient smallness" of $\epsilon_\Delta$, we consider the distribution function in the perturbation series about the BCS ground-state, i.e.,

$$n_k^{\text{BCS}} = \frac{1}{2}\Big[1 + \xi_k(\Delta_k^2 + \xi_k^2)^{-1/2}\Big], \qquad \bar{n}_k^{\text{BCS}} = \frac{1}{2}\Big[1 - \xi_k(\Delta_k^2 + \xi_k^2)^{-1/2}\Big]. \tag{5.65}$$

where $\xi_k = k^2/(2M) - k_F^2/(2M)$. The BCS distribution functions $n_k^{\text{BCS}}(\Delta_k, k_F)$ are compared for different pairing gaps $\Delta_k$ to the Fermi-Dirac distributions $n_k(T, \tilde{\mu})$ for different temperatures $T$ in Fig. 5.10, for $k_F = 1.0\,\text{fm}^{-3}$ and $\tilde{\mu} = k_F^2/(2M)|_{k_F=1.0\,\text{fm}^{-3}}$, respectively. One sees that for typical values $\Delta_k \simeq (1-5)$ MeV (cf. e.g., Refs. [291, 108, 103, 61, 98, 250]) the deviation of the BCS distribution functions from the step function $\Theta(k_F - k)$ is comparable to or less than that of the Fermi-Dirac distributions for $T \lesssim 3-5$ MeV where the dependence on $\delta$ is still distinctly nonanalytic.

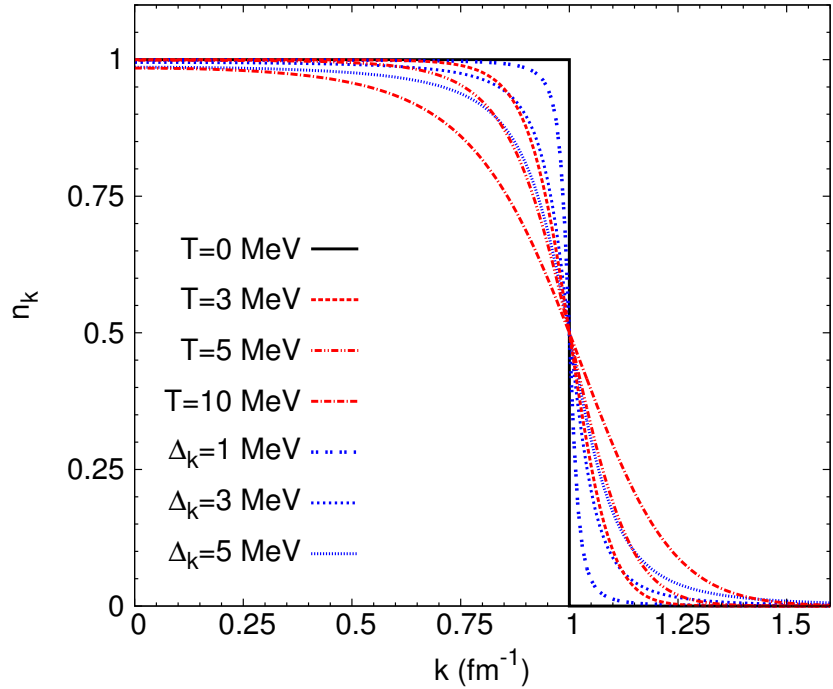

**Figure 5.10.:** BCS distribution functions (blue lines) for different pairing gaps $\Delta_k$ vs. Fermi-Dirac distributions (black and red lines) for different temperatures $T$.

### 5.2.4. Exact Results for *S*-Wave Contact Interaction[20]

To better understand the divergent behavior in the higher-order Maclaurin coefficients we have encountered in the previous section for the chiral nuclear potentials, we consider now the simpler case of an *S*-wave contact interaction $V_{\text{contact}} = \pi M^{-1}(a_s + 3a_t + (a_t - a_s)\vec{\sigma}_1 \cdot \vec{\sigma}_2)$.

***Second-Order Contribution.*** For the *S*-wave contact interaction the (dimensionally regularized [235]) second-order (normal) contribution can be written as

$$F_{2,\text{normal}}(T,\tilde{\mu}_n,\tilde{\mu}_p) = \frac{1}{2\pi^4 M}\Big(a_s^2\,\Gamma^{\text{nn}}(T,\tilde{\mu}_n) + a_s^2\,\Gamma^{\text{pp}}(T,\tilde{\mu}_p) + (3a_t^2 + a_s^2)\,\Gamma^{\text{np}}(T,\tilde{\mu}_n,\tilde{\mu}_p)\Big), \quad (5.66)$$

where the functions $\Gamma^{\text{nn/pp/np}}$ are defined as

$$\Gamma^{\text{nn/pp/np}} = \int_0^\infty dk_3 \int_{-1}^1 dy \int_0^\infty dq \fint_0^\infty dk_1 \int_{-1}^1 dx \frac{k_3^2 k_1^2 q}{k_3 y - k_1 x} \times \begin{cases} 2n_{k_1}^{\text{n}} n_{k_2}^{\text{n}} n_{k_3}^{\text{n}} \\ 2n_{k_1}^{\text{p}} n_{k_2}^{\text{p}} n_{k_3}^{\text{p}} \\ n_{k_1}^{\text{n}} n_{k_2}^{\text{p}} n_{k_3}^{\text{n}} + n_{k_1}^{\text{p}} n_{k_2}^{\text{n}} n_{k_3}^{\text{p}} \end{cases}, \quad (5.67)$$

with $k_2 = (k_3^2 - 2k_3 qy + q^2)^{1/2}$. The dashed integral denotes the principal value. At zero temperature the integrals in Eq. (5.67) can be resolved in closed form, leading to[21]

$$\Gamma^{\text{nn/pp}}(T=0,\rho,\delta) = \frac{4k_F^7}{105}(11 - 2\ln(2))(1 \pm \delta)^{7/3}, \quad (5.68)$$

$$\begin{aligned}\Gamma^{\text{np}}(T=0,\rho,\delta) = &\frac{k_F^7}{420}\sum_\pm (1 \pm \delta)^{7/3}\bigg[-\frac{8(1 \mp \delta)^{5/3}}{(1 \pm \delta)^{5/3}} + \frac{66(1 \mp \delta)}{(1 \pm \delta)} + \frac{30(1 \mp \delta)^{1/3}}{(1 \pm \delta)^{1/3}} \\ &+ \Big(-\frac{35(1 \mp \delta)^{4/3}}{(1 \pm \delta)^{4/3}} + \frac{42(1 \mp \delta)^{2/3}}{(1 \pm \delta)^{2/3}} - 15\Big)\ln|K_1| + \frac{8(1 \mp \delta)^{7/3}}{(1 \pm \delta)^{7/3}}\ln|K_2^\mp|\bigg],\end{aligned} \quad (5.69)$$

with $k_F = (3\pi^2\rho/2)^{1/3}$ the nucleon Fermi momentum. The np-channel contribution involves terms proportional to $\ln|K_1|$ and $\ln|K_2^\mp|$, with $K_1 = [(1+\delta)^{1/3} + (1-\delta)^{1/3}]/[(1+\delta)^{1/3} - (1-\delta)^{1/3}]$ and $K_2^\mp = (1 \mp \delta)^{2/3}/[(1+\delta)^{2/3} - (1-\delta)^{2/3}]$. Both $K_1$ and $K_2^\mp$ exhibit a Laurent series with principal part $\sim \delta^{-1}$. Expanding the logarithms $\ln|K_1|$ and $\ln|K_2^\mp|$ around the principal part and the remaining $\delta$ dependent terms in Eq. (5.69) around $\delta = 0$ one obtains a series of the form[22]

$$\bar{F}(T=0,\rho,\delta) = \bar{A}_0(T=0,\rho) + \bar{A}_2(T=0,\rho)\delta^2 + \sum_{n=2}^\infty \bar{A}_{2n,\text{reg}}(\rho)\delta^{2n} + \sum_{n=2}^\infty \bar{A}_{2n,\log}(\rho)\delta^{2n}\ln|\delta|. \quad (5.70)$$

From Eq. (5.70) it follows for the isospin-asymmetry derivatives of $\bar{F}(T=0,\rho,\delta)$:

$$\begin{aligned}\frac{1}{(2n)!}\frac{\partial^{2n}\bar{F}(T,\rho,\delta)}{\partial\delta^{2n}}\bigg|_{n\geq 2,T=0,\delta\to 0} &= \text{const.} + \bigg[A_{2n,\log}\ln|\delta| - \sum_{k=2}^{n-1}\frac{(2k)!(2n-2k-1)!}{(2n)!}\frac{A_{2k,\log}}{\delta^{2(n-k)}}\bigg]_{\delta\to 0} \\ &= -\infty \times \text{sgn}(A_{4,\log}),\end{aligned} \quad (5.71)$$

---

[20] Note that here we use the sign convention for scattering lengths as in Refs. [233, 234].

[21] The calculations leading to Eqs. (5.68) and (5.69) were first performed by N. Kaiser [234].

[22] Note that the expansion of the np-channel contribution in terms of the third root of the proton fraction involves, in addition to terms proportional to $Y^{\nu/3}$ with $\nu \in \mathbb{N}/\{1,2,4\}$, a logarithmic term $Y^{7/3}\ln(Y)$, but (in contrast to the $\delta$ expansion) no higher-order logarithmic terms.

i.e., the degree of divergence increases with $n$, and all higher-order derivatives diverge with equal sign. This behavior is impossible for the $T \to 0$ limit of the higher-order Maclaurin coefficients, i.e., the $T \to 0$ and the $\delta \to 0$ limits of the isospin-asymmetry derivatives $\partial^{2(2n+1)} \bar{F}_2/\partial \delta^{2(2n+1)}$ cannot commute for $n \geq 1$:

$$\left.\frac{\partial^{2(2n+1)} \bar{F}_{2,\text{normal}}}{\partial \delta^{2(2n+1)}}\right|_{n\geq 1, T=0, \delta\to 0} \neq \left.\frac{\partial^{2(2n+1)} \bar{F}_{2,\text{normal}}}{\partial \delta^{2(2n+1)}}\right|_{n\geq 1, \delta=0, T\to 0}. \tag{5.72}$$

This is explained as follows. As can be inferred from Eq. (5.71), at zero temperature the higher-order isospin-asymmetry derivatives $\partial^{2n} \bar{F}_2/\partial \delta^{2n}$, with $n \geq 2$, all have positive (isospin-asymmetry) slope and negative (isospin-asymmetry) curvature for $\delta = 0 + \epsilon$. This behavior is impossible at finite temperature where $\bar{F}_2 \in C^\infty$, and hence $\partial^{2n+1} \bar{F}_2/\partial \delta^{2n+1} = 0$ at $\delta = 0$ (by charge symmetry). If $\partial^4 \bar{F}_2/\partial \delta^4$ has positive slope for $T \neq 0$ and $\delta \to 0$ then it must be convex at $\delta = 0$, thus $\partial^6 \bar{F}_2/\partial \delta^6$ can only diverge with positive sign and curvature for $\delta = 0$ and $T \to 0$, etc.

***Ladder Resummation.*** As mentioned earlier, the origin of the logarithmic terms in the isospin-asymmetry dependence of the second-order (normal) contribution at zero temperature lies in the energy denominators. This suggests that logarithmic terms should arise from all higher-order diagrams that feature energy denominators (i.e., higher-order skeletons and insertions with higher-order skeleton subdiagrams). In fact, in Ref. [203] it was found, using again an $S$-wave contact interaction, that also the third-order skeletons give a contribution to the quartic Maclaurin coefficient that is singular at $T = 0$. One may now ask whether logarithmic terms arise also in self-consistent schemes (i.e., BHF and SCGF, cf. Sec. 2.5) that effectively resum to all orders various classes of perturbative many-body contributions. To investigate this question, we consider the result of Ref. [233] (cf. also Ref. [230]) where the following expression for the all-order-sum of the ladder diagrams (including the second-order normal diagram) was derived:

$$\bar{E}_{0,\text{resum}}(k_F^n, k_F^p) = -\frac{24}{\pi M[(k_F^n)^3 + (k_F^p)^3]}\left(\Gamma_{\text{resum}}^{\text{nn}}(a_s) + \Gamma_{\text{resum}}^{\text{pp}}(a_s) + \Gamma_{\text{resum}}^{\text{np}}(a_s) + 3\Gamma_{\text{resum}}^{\text{np}}(a_t)\right), \tag{5.73}$$

where

$$\Gamma_{\text{resum}}^{\text{nn/pp}}(a_s) = (k_F^{\text{n/p}})^5 \int_0^1 ds\, s^2 \int_0^{\sqrt{1-s^2}} d\kappa\, \kappa \arctan \frac{I(s,\kappa)}{(a_s k_F^{\text{n/p}})^{-1} + \pi^{-1} R(s,\kappa)}, \tag{5.74}$$

$$\Gamma_{\text{resum}}^{\text{np}}(a_{s/t}) = \int_0^{(k_F^n+k_F^p)/2} dP\, P^2 \int_{q_{\min}}^{q_{\max}} dq\, q \arctan \frac{\Phi(P,q,k_F^n,k_F^p)}{(a_{s/t})^{-1} + (2\pi)^{-1}[k_F^n R(P/k_F^n, q/k_F^n) + k_F^p R(P/k_F^p, q/k_F^p)]}. \tag{5.75}$$

The functions $I(s,\kappa)$, $R(s,\kappa)$ and $\Phi(P,q,k_F^n,k_F^p)$ are given by

$$I(s,\kappa) = \kappa\,\Theta(1 - s - \kappa) + \frac{1 - s^2 - \kappa^2}{2s}\Theta(s + \kappa - 1), \tag{5.76}$$

$$R(s,\kappa) = 2 + \frac{1 - (s-\kappa)^2}{2s} \ln\frac{1 + s + \kappa}{|1 - s - \kappa|} + \frac{1 - s^2 - \kappa^2}{2s} \ln\frac{1 + s - \kappa}{1 - s + \kappa}, \tag{5.77}$$

$$\Phi(P,q,k_F^n,k_F^p) = \begin{cases} q, & \text{for } P+q < k_F^p \\ \dfrac{(k_F^p)^2 - (P-q)^2}{4P}, & \text{for } k_F^p < P+q < k_F^n \wedge |P-q| < k_F^p \\ \dfrac{(k_F^n)^2 + (k_F^p)^2 - 2(P^2-q^2)}{4P}, & \text{for } k_F^n < P+q \wedge P^2+q^2 < \dfrac{(k_F^n)^2+(k_F^p)^2}{2} \end{cases}.$$
(5.78)

The integration boundaries in the expression for $\Gamma_{\text{resum}}^{\text{np}}(a)$ are given by

$$q_{\min} = \max\left(0, P - k_F^p\right), \qquad q_{\max} = \min\left(k_F^p + P, \sqrt{[(k_F^n)^2 + (k_F^p)^2]/2 - P^2}\right). \tag{5.79}$$

Setting $\rho = 0.20\,\text{fm}^{-3}$, $a_s = 19.0\,\text{fm}$, $a_t = -5.4\,\text{fm}$ and $M = 938\,\text{MeV}$, we examine the isospin-asymmetry dependence of $\bar{E}_{0,\text{resum}}$ in the same way as for the second-order results based on chiral interactions, see Sec. 5.2.5 for details. The quartic finite-difference results for the spin-triplet ($\sim a_t$) np-channel as well as for the sum of the nn- and pp-channel contributions are displayed in Fig. 5.11. Also shown are the spin-triplet np-channel results for the quadratic Maclaurin coefficient (which is regular). The numerical noise is caused by the fact that the numerical integration converges rather slowly due to the poles in the argument of the arctan functions. However, one still sees clearly that the np-channel results for $\bar{A}_4^{N,\Delta\delta}$ have an approximately logarithmic dependence on $\Delta\delta$, whereas no logarithmic dependence appears in the nn- and pp-channel results. The extracted values for the quartic coefficients in the "logarithmic" series are $\bar{A}_{4,\log} \simeq -7$ MeV and $\bar{A}_{4,\text{reg}} \simeq 1$ MeV. Substracting the extracted logarithmic term from the np-channel results leads to quartic finite differences that are approximately stepsize and grid-length indeppendent for stepsizes $\Delta\delta \gtrsim 0.15$. Note that the ratio $\bar{A}_{4,\text{reg}}/\bar{A}_{4,\log} \simeq -0.14$ is considerably reduced as compared to the second-order results where $\bar{A}_{4,\text{reg}}/\bar{A}_{4,\log} = [3 - 60\ln(3) + 4\ln(2)]/60 \simeq -1.002$. We note that qualitatively, the finite-difference results for the spin-singlet np-channel (with $a_s \simeq 19.0$ fm) are very similar to the ones for the spin-triplet channel (with $a_t \simeq -5.4$ fm).

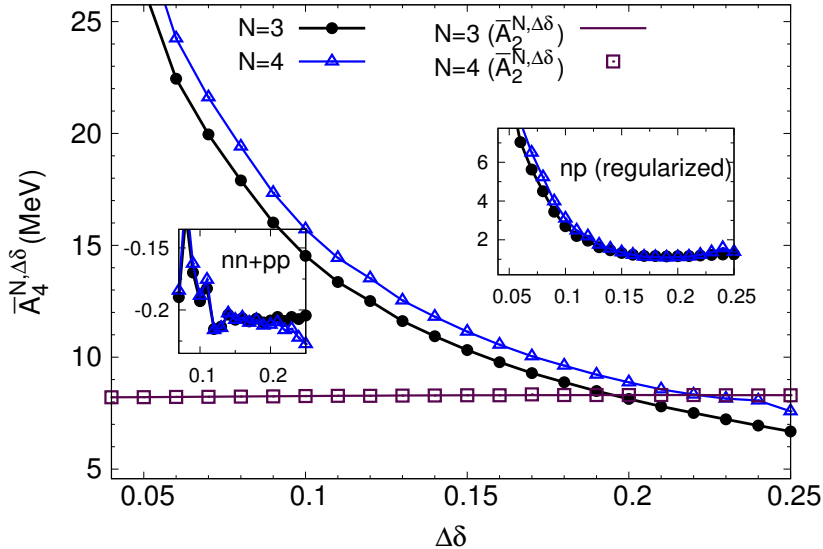

**Figure 5.11.:** Main plot: quadratic and quartic finite-difference results for the spin-triplet np-channel of $\bar{E}_{\text{resum}}$. Top inset: quartic finite-difference results (np-channel) with the logarithmic term subtracted. Bottom inset: quartic finite-difference results for the combined nn- and pp-channel contributions. The results were obtained setting $\rho = 0.20\,\text{fm}^{-3}$, $a_s = 19.0\,\text{fm}$, $a_t = -5.4\,\text{fm}$ and $M = 938\,\text{MeV}$.

## 5. Isospin-Asymmetry Dependence of the Nuclear Equation of State

The dependence on $\delta$ of the spin-triplet np-channel contribution to $\bar{E}_{0,\text{resum}}$ is then studied in Fig. 5.12. In the extremely neutron-rich region, $\bar{E}_{0,\text{resum}}(\delta)$ has a maximum,[23] a feature that is absent in the first- and second-order results (and in the nn- and pp-channels of $\bar{E}_{0,\text{resum}}$). In addition, there may even be a kink at $\delta \simeq 0.992$. Also shown in Fig. 5.12 are the results obtained from the parabolic approximation $\bar{E}_{0,\text{parabolic}}(\delta) = \bar{E}_0(\delta = 0) + \bar{E}_{0,\text{sym}}\delta^2$ as well as the approximations $\bar{E}_{0,\text{quadratic}}(\delta) = \bar{E}_0(\delta = 0) + \bar{A}_2\delta^2$ and $\bar{E}_{0,\text{quartic+log}}(\delta) = \bar{E}_0(\delta = 0) + \bar{A}_2\delta^2 + \bar{A}_{4,\text{reg}}\delta^4 + \bar{A}_{4,\log}\delta^4 \ln|\delta|$. The $\bar{E}_{0,\text{quartic+log}}(\delta)$ curve is very close to the exact results for $\delta \lesssim 0.5$, but in the very neutron-rich region large deviations occur. The deviations of the parabolic ($\sim \bar{E}_{0,\text{parabolic}}$) and quadratic ($\sim \bar{E}_{0,\text{quadratic}}$) approximations from the exact results ($\sim \bar{E}_{0,\text{resum}}$) are considerably larger as compared to the first- and second-order results, which is reflected in the relatively large size of $\xi = 1 - \bar{A}_2/\bar{E}_{0,\text{sym}}$, i.e., $\xi_{\text{resum,np}} \simeq 1 - 8.3/15.6 \simeq 0.47$; for comparison, for the second-order contribution ($\sim \Gamma^{\text{np}}$) it is $\xi \simeq 0.13$, and for the sum of the nn- and pp-channels of $\bar{E}_{0,\text{resum}}$ it is $\xi_{\text{resum,nn+pp}} \simeq 1 - 5.8/6.1 \simeq 0.05$ (for $a_s \simeq 19.0$ fm). If one assumes that Eq. (5.70) is valid also for $\bar{E}_{0,\text{resum}}$, then $\sum_{n\geq 3}^{\infty} \bar{A}_{2n,\text{reg}} = \bar{E}_{0,\text{sym}} - \bar{A}_2 - \bar{A}_{4,\text{reg}} \simeq 6$ MeV, i.e., there is a considerable contribution to to $\bar{E}_{0,\text{sym}}$ from $\bar{A}_{2n\geq 6,\text{reg}}$ terms.

To summarize, the results discussed in this section suggest that nonanalytic terms of the form $\delta^{2n\geq 4} \ln|\delta|$ are a generic feature of MBPT (in the thermodynamic limit, and if pairing effects are neglected, cf. the discussion at the end of Sec. 5.2.3), and nonanalyticities in the isospin-asymmetry dependence should arise also in self-consistent schemes such as BHF or SCGF.

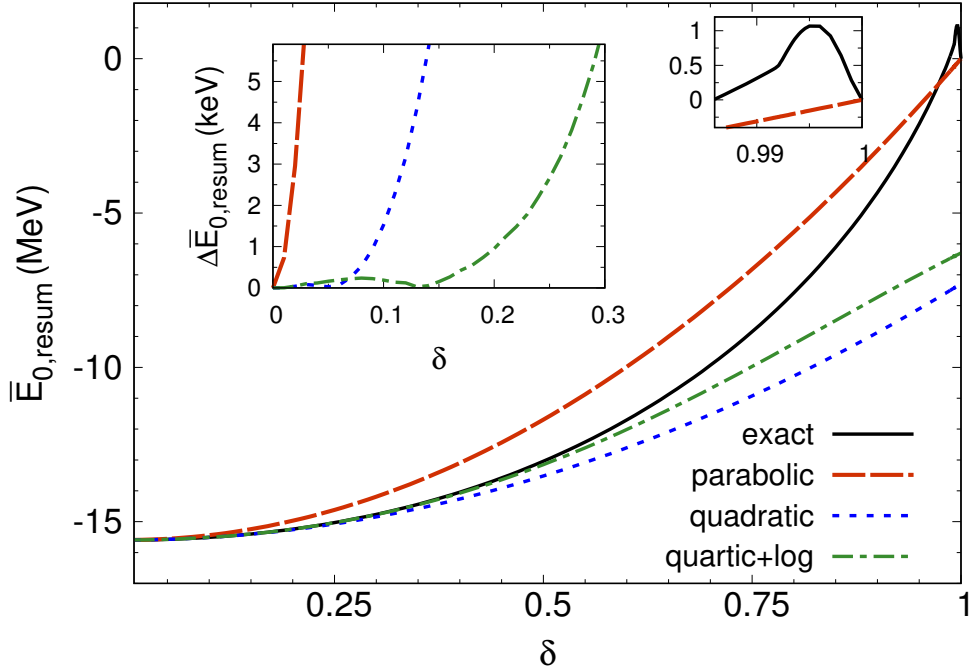

**Figure 5.12.:** Interaction contribution to the ground-state energy per particle from the resummation of ladder diagrams with a spin-triplet $S$-wave contact interaction, see text for details. The left inset shows the absolute value of the deviation $|\Delta\bar{E}_{0,\text{resum}}| := |\bar{E}_{0,\text{resum}} - \bar{E}_{0,\text{approx}}|$ of the various approximations from the exact results for small neutron excesses ($\delta \in [0, 0.3]$). The small right inset magnifies the behavior in the extremely neutron-rich region. The results were obtained setting $\rho = 0.20\,\text{fm}^{-3}$, $a_t = -5.4$ fm and $M = 938$ MeV.

---

[23] This feature depends on the scattering lengths, e.g., it is absent in the unitary limit $a \to \infty$.

## 5.2.5. Extraction of Leading Logarithmic Term

In the last section, we have found that for an $S$-wave contact interaction the singularity of the higher-order Maclaurin coefficients at zero temperature may in general be associated with logarithmic terms $\sim \delta^{2n} \ln(\delta)$. It remains to show that the singularity of $\bar{A}_{2n\geq 4}(T=0)$ is logarithmic also in the case of realistic chiral nuclear interactions.

Eq. (5.70) results in a stepsize dependence of the following form in the quartic and sextic finite-difference formulas:

$$\bar{A}_4^{N,\Delta\delta} = \bar{A}_{4,\text{reg}} + C_1^4(N)\bar{A}_{4,\text{log}} + \bar{A}_{4,\text{log}} \ln(\Delta\delta) + C_2^4(N)\bar{A}_{6,\text{log}}\Delta\delta^2 + O(\Delta\delta^4), \tag{5.80}$$

$$\bar{A}_6^{N,\Delta\delta} = \bar{A}_{6,\text{reg}} + C_1^6(N)\bar{A}_{4,\text{log}}\Delta\delta^{-2} + \bar{A}_{6,\text{log}} \ln(\Delta\delta) + C_2^6(N)\bar{A}_{6,\text{log}} + O(\Delta\delta^2), \tag{5.81}$$

where the numbers $C_i^{2n}(N)$ are determined by the respective finite-difference coefficients $\omega_{2n}^{N,k}$. From Eqs. (5.80) and (5.81) the leading logarithmic term is given by

$$\Xi_4(N_1, N_2, \Delta\delta) := \frac{\bar{A}_4^{N_1,\Delta\delta} - \bar{A}_4^{N_2,\Delta\delta}}{C_1^4(N_1) - C_1^4(N_2)} \simeq \bar{A}_{4,\text{log}}, \tag{5.82}$$

$$\Xi_6(N_1, N_2, \Delta\delta) := \frac{\bar{A}_6^{N_1,\Delta\delta} - \bar{A}_6^{N_2,\Delta\delta}}{C_1^6(N_1) - C_1^6(N_2)} \Delta\delta^2 \simeq \bar{A}_{4,\text{log}}, \tag{5.83}$$

where the leading correction is proportional to $\bar{A}_{6,\text{log}}\Delta\delta^2$. For the $S$-wave contact interaction, where $\bar{A}_{4,\text{log}}/\bar{A}_{6,\text{log}} \simeq 2.60$, Eqs. (5.82) and (5.83) reproduce the exact value of $A_{4,\text{log}}$ to high accuracy. The n3lo414 results for $\bar{A}_{4,6}^{N,\Delta\delta}$ are shown for grid lengths $N = 3, 4, 5$ in Figs. 5.13 and 5.14. For sufficiently large stepsizes[24] the results for $\bar{A}_4^{N,\Delta\delta}$ and $\bar{A}_6^{N,\Delta\delta}$ approximately exhibit the logarithmic and inverse quadratic stepsize dependence, respectively, expected from Eqs. (5.80) and (5.81). As expected from the analytic results for the $S$-wave contact interaction, this feature is absent for the nn- and pp-channel contributions. The results for $\Xi_4(N_1, N_2, \Delta\delta)$ and $\Xi_6(N_1, N_2, \Delta\delta)$ are approximately constant (for sufficiently large stepsizes) and give similar values of $\bar{A}_{4,\text{log}}$. The extracted values of $\bar{A}_{4,\text{log}}(\rho)$ (averages of the values obtained from $\Xi_4$ and $\Xi_6$ as well as different stepsizes $\Delta\delta$ and grid lengths $N_1$ and $N_2$) are given in Tables 5.2 and 5.3. One sees that the magnitude of both $\bar{A}_{4,\text{log}}(\rho)$ and $\bar{A}_{4,\text{nonlog}}(\rho)$ increases monotonically with density (we have checked this behavior for numerous additional values of $\rho$).

Using the approximate results obtained for $\bar{A}_{4,\text{log}}(\rho)$, we compute for both the chiral interactions and the spin-triplet ($\sim a_t$) $S$-wave contact interaction the quartic and sextic finite differences corresponding to

$$\bar{F}_{2,\text{normal}}^{\text{NN,regularized}}(T=0, \rho, \delta) := \bar{F}_{2,\text{normal}}^{\text{NN}}(T=0, \rho, \delta) - \bar{A}_{4,\text{log}}(\rho)\, \delta^4 \, \ln|\delta|. \tag{5.84}$$

The results are plotted in the upper insets of Figs. 5.13 and 5.14. One sees that for the "regularized" second-order term the (approximately logarithmic and inverse quadratic, respectively) stepsize dependence is removed, and for sufficiently large stepsizes $\bar{A}_4^{N,\Delta\delta} \simeq A_{4,\text{reg}}$ is approximately constant. Adding the Hartree-Fock results for $\bar{A}_4(T=0, \rho)$, the extracted values for $A_{4,\text{reg}}(\rho)$ are given in Table 5.2 (for comparison, the Hartree-Fock results are also given separately).

---

[24] Due to the divergence of the integrand at the integration boundary, in the case of the second-order (normal) contribution the numerical noise is significantly amplified as compared to the case of higher temperatures (note that we have restricted the extraction of the finite-temperature Maclaurin coefficients to $T \geq 2$ MeV).

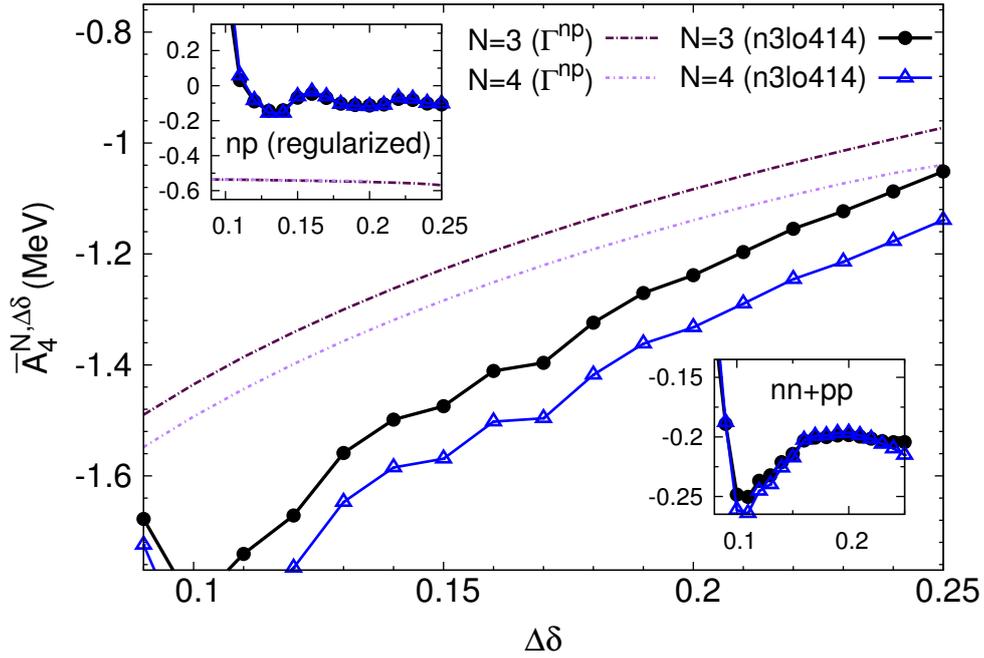

**Figure 5.13.:** Main plot: quartic finite-difference results for the np-channel second-order NN contribution at zero temperature (n3lo414, $\rho = 0.20 \, \text{fm}^{-3}$) and for the (suitably scaled) expression given by Eq. (5.69). Top inset: analogous results for the np-channel contribution with the logarithmic term subtracted. Bottom inset: (combined) results for the nn- and pp-channel contributions.

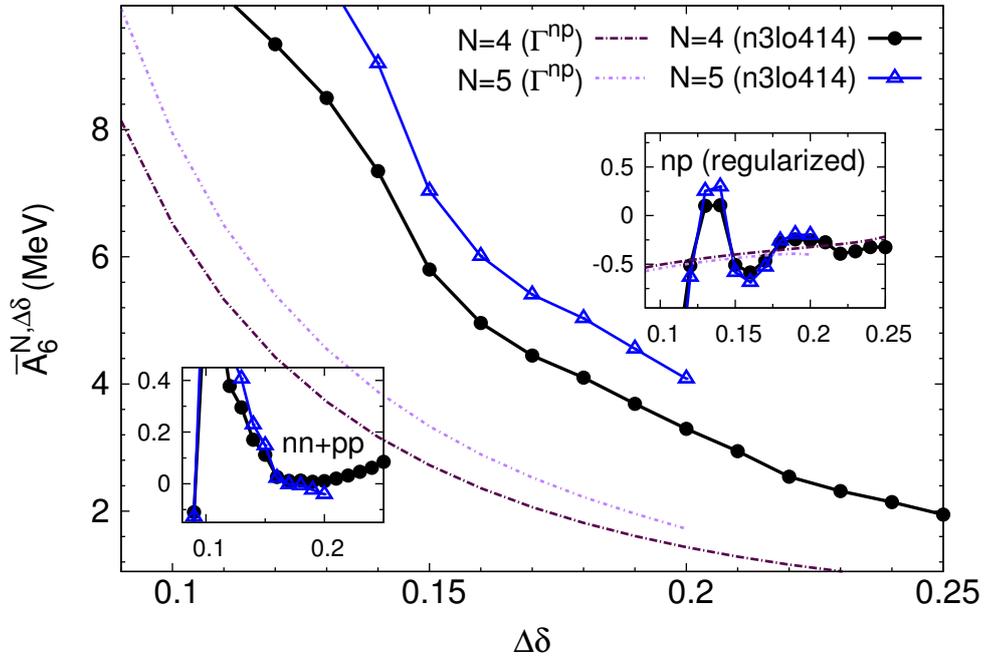

**Figure 5.14.:** Same as Fig. 5.14 but for the sextic finite differences.

As an additional check, we have extracted $\bar{A}_{4,\text{reg}}$ and $\bar{A}_{4,\log}$ also by fitting the coefficients $a_{\text{reg}}$ and $a_{\log}$ of a function $f_N(\Delta\delta) = a_{\text{reg}} + a_{\log}[C_1^4(N) + \ln(\Delta\delta)]$ to the unregularized results for $\bar{A}_4^{N,\Delta\delta}$, cf. Eq. (5.80); the results for $\bar{A}_{4,\text{reg}}(\rho)$ and $\bar{A}_{4,\log}(\rho)$ obtained in that way were found to match those given in Table 5.2. Note that the relative size of the second-order contribution to $\bar{A}_{4,\text{reg}}$ is smaller for the chiral interactions as compared to the spin-triplet $S$-wave contact interaction, where $\bar{A}_{4,\text{reg}}/\bar{A}_{4,\log} = [3 - 60\ln(3) + 4\ln(2)]/60 \simeq -1.002$.

| $\rho$ (fm$^{-3}$) | 0.05 | 0.10 | 0.15 | 0.20 | 0.25 | 0.30 |
|---|---|---|---|---|---|---|
| $\bar{A}_2$ (MeV) | 16.48 | 24.41 | 32.03 | 36.94 | 41.38 | 44.83 |
| $\bar{A}_{4,\text{reg}}^{\text{HF}}$ (MeV) | 0.19 | 0.15 | -0.09 | -0.56 | -1.27 | -2.18 |
| $\bar{A}_{4,\text{reg}}$ (MeV) | -0.0 | -0.2 | -0.3 | -0.7 | -1.0 | -1.4 |
| $\bar{A}_{4,\log}$ (MeV) | 0.4 | 0.8 | 1.3 | 1.5 | 2.0 | 2.4 |

**Table 5.2.:** Extracted values of $\bar{A}_{4,\log}(\rho)$ and $\bar{A}_{4,\text{reg}}(\rho)$ for n3lo414. The numbers in front (brackets) correspond to n3lo414 (n3lo450). The statistical errors (with respect to stepsize and grid length variations) of the results are of order $\pm 0.1$ MeV. For comparison we also show the results for $\bar{A}_2(T=0,\rho)$. Note that $\bar{A}_{4,\text{reg}}(\rho)$ includes the Hartree-Fock level results for the quartic coefficient $\bar{A}_{4,\text{reg}}^{\text{HF}}(\rho)$, which are shown separately in the second column.

| $\rho$ (fm$^{-3}$) | 0.05 | 0.10 | 0.15 | 0.20 | 0.25 | 0.30 |
|---|---|---|---|---|---|---|
| $\bar{A}_2$ (MeV) | 16.03 | 24.92 | 31.41 | 36.08 | 39.30 | 41.28 |
| $\bar{A}_{4,\text{reg}}^{\text{HF}}$ (MeV) | 0.17 | 0.09 | -0.20 | -0.72 | -1.42 | -2.28 |
| $\bar{A}_{4,\text{reg}}$ (MeV) | -0.0 | -0.2 | -0.6 | -1.1 | -1.4 | -1.8 |
| $\bar{A}_{4,\log}$ (MeV) | 0.4 | 0.8 | 1.1 | 1.3 | 1.8 | 1.9 |

**Table 5.3.:** Same as Table 5.2 but for n3lo450.

We have examined the isospin-asymmetry dependence of the second-order term also for model interactions of the one-boson exchange kind (results not shown) where a semi-analytical treatment is possible (and therefore the numerical precision can be further increased), cf. [208] for details on the model interactions. It was found that the expansion in $\delta$ is again well-behaved for the nn- and pp-channels, but in the case of the np-channel the quartic and sextic finite differences exhibit (to good accuracy) a logarithmic and inverse quadratic stepsize dependence at $T=0$, respectively). Overall, we can conclude that Eq. (5.70) should be valid for generic interactions.[25]

---

[25] See also Ref. [234], where it was found that the second-order contribution to the quartic Maclaurin coefficient from one-pion exchange is singular.

## 5.3. Threshold for Convergence of the Maclaurin Expansion

Having studied the isospin-asymmetry dependence of the various many-body contribution individually, we now discuss the results for the Maclaurin coefficients $\bar{A}_{2,4,6}(T,\rho)$ associated with the full second-order results (only the dominant contributions at second-order are included, cf. Sec. 5.2.3). The results for $\bar{A}_{2,4,6}(T,\rho)$ obtained from the two sets of chiral two- and three-nucleon potentials n3lo414 and n3lo450 are displayed in Fig. 5.15. Also shown is the difference between the quadratic coefficients $\bar{A}_2(T,\rho)$ and the symmetry free energy $\bar{F}_{\text{sym}}(T,\rho) = \bar{F}(T,\rho,1) - \bar{F}(T,\rho,0)$. Overall, the results from n3lo414 and n3lo450 are very similar; the largest differences occur for the quadratic coefficient $\bar{A}_2$ at high densities. Compared to the results for $\bar{F}_{\text{sym}}(T,\rho)$ shown in Sec. 3.5, the temperature dependence of $\bar{A}_2(T,\rho)$ is decreased in the case of n3lo414 and inverted for n3lo450 (at densities above saturation density).

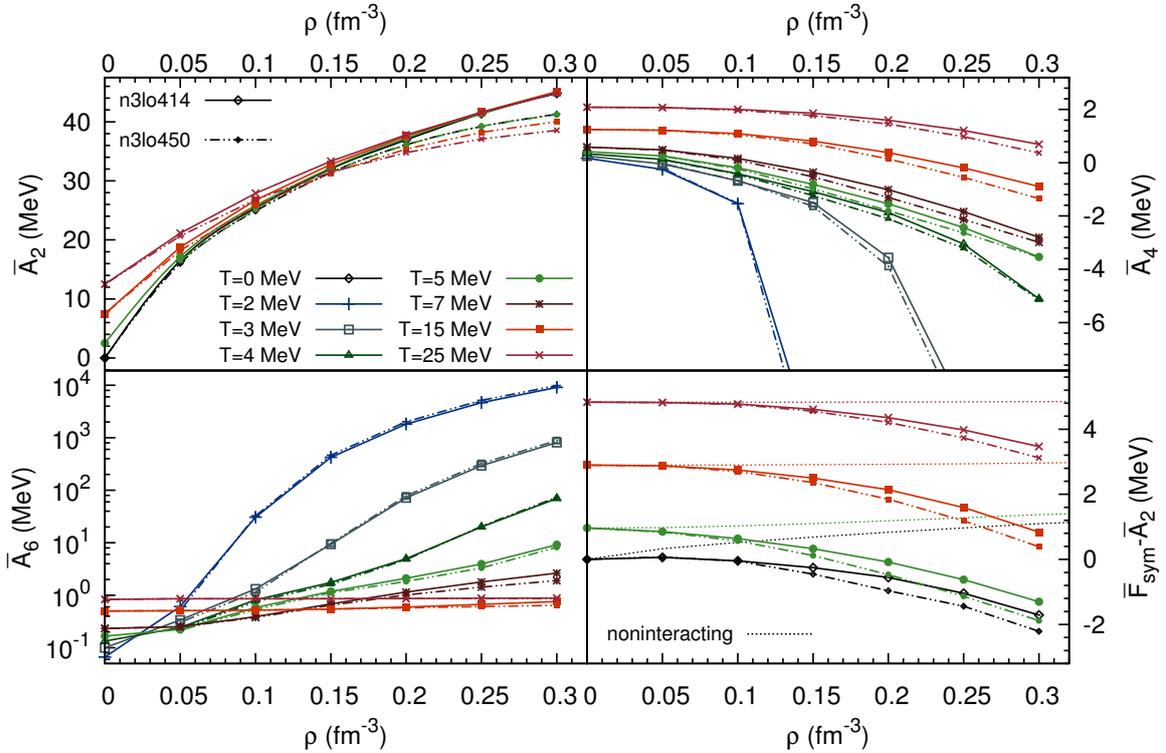

**Figure 5.15.:** Quadratic, quartic and sextic Maclaurin coefficients for the free energy per particle $\bar{F}(T,\rho,\delta)$ of infinite homogeneous nuclear matter; (solid lines: n3lo414, dash-dot lines: n3lo450). Also shown is the difference between quadratic coefficient and the symmetry free energy (the dotted lines correspond to the results for a noninteracting nucleon gas).

The deviations between $\bar{F}_{\text{sym}}$ and $\bar{A}_2$ increase with temperature, but whereas for a noninteracting nucleon gas $\bar{F}_{\text{sym}} - \bar{A}_2$ (slightly) increases with density, for the interacting system $\bar{F}_{\text{sym}} - \bar{A}_2$ decreases with density. The quantity $F_{\text{sym}} - \bar{A}_2$ is significantly increased for the high temperatures relevant for astrophysical simulations of core-collapse supernovæ, and receives its main contributions from the noninteracting term and the first-order 3N contribution, with the respective contributions carrying opposite signs (at high temperatures or low densities the noninteracting contribution dominates). The increase of $F_{\text{sym}} - \bar{A}_2$ with increasing temperature

*5. Isospin-Asymmetry Dependence of the Nuclear Equation of State*

is in line with the previous results that the convergence rate of the expansion of the noninteracting contribution decreases significantly with temperature, cf. Sec. 5.1.3. The accuracy of the quadratic/parabolic approximation of the various interaction contributions however is increased at higher temperatures, see also Figs. 5.8 and 5.9.

In the high-temperature and low-density region the Maclaurin coefficients obey $\bar{A}_2 > \bar{A}_4 > \bar{A}_6 (> \bar{A}_8)$, and accordingly $|\xi - \zeta_6| < |\xi - \zeta_4|$, where (again) $\xi := 1 - \frac{\bar{A}_2(T,\rho)}{\bar{F}_{\text{sym}}(T,\rho)}$ and $\zeta_{2N} := \frac{\sum_{n=2}^{N} \bar{A}_{2n}(T,\rho)}{\bar{F}_{\text{sym}}(T,\rho)}$. This behavior breaks down when the temperature is decreased and the density is increased, leading to $\bar{A}_2 \ll \bar{A}_4 \ll \bar{A}_6 (\ll \bar{A}_8)$ at high densities and low temperatures. In the sense that the expansion coefficients are hierarchically ordered at high temperatures and low densities the Maclaurin expansion with respect to the isospin-asymmetry can be rated as a convergent series in that regime, and as a divergent asymptotic series in the low-temperature and high-density region. To that effect, one can loosely identify a threshold line that separates the two regions. This line (roughly corresponding to $|\xi - \zeta_4| = |\xi - \zeta_6|$) is sketched in Fig. 5.16. Note that since the divergent behavior is more pronounced for $\bar{A}_{2(n+1)}$ than for $\bar{A}_{2n}$ (cf. also Fig. 5.7), one can expect that this threshold line rises when the isospin-asymmetry expansion is probed at increasing orders.[26]

The strongly divergent behavior of the higher-order Maclaurin coefficients below the threshold line arises solely (in second-order MBPT) from the second-order contribution (cf. Sec. 5.2.3), whose Maclaurin coefficients diverge with alternating sign (for even and odd values of $n$) and increasing order of divergence for increasing values of $n$ in the zero-temperature limit. In contrast, at the Hartree-Fock level the coefficients are hierarchically ordered $\bar{A}_2 > \bar{A}_4 > \bar{A}_6 (> \bar{A}_8)$ also at high densities and low temperatures, and the Mauclaurin expansion is overall well-converged at that level.[27] With the second-order term included the free energy per particle is a nonanalytic (smooth for $T \neq 0$) function of the isospin asymmetry $\delta$ in the low-temperature regime, and the Maclaurin expansion in terms of $\delta$ constitutes a divergent asymptotic expansion in that regime. The divergent behavior in the low-temperature regime and the convergent behavior for high temperatures is evident in Figs. 5.17, 5.18, 5.19 amd 5.20, which are discussed in Sec. 5.4.

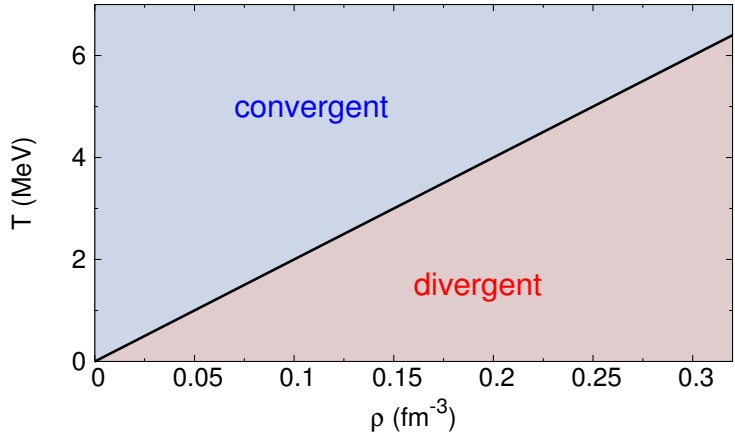

**Figure 5.16.:** Threshold line for the convergence of the Maclaurin expansion of the free energy per particle in terms of $\delta$, see text for details.

---

[26] For instance, the line that (roughly) corresponds to $|\xi - \zeta_4| = |\xi - \zeta_6|$ approximately crosses the point $(T/\text{MeV}, \rho/\text{fm}^{-3}) \simeq (5, 0.25)$, whereas the one for $|\xi - \zeta_6| = |\xi - \zeta_8|$ crosses the point $(T/\text{MeV}, \rho/\text{fm}^{-3}) \simeq (5, 0.20)$.

[27] This behavior is in agreement with the results for the quartic Maclaurin coefficient obtained in mean-field theory calculations with phenomenological NN interactions [68, 359, 81], i.e., $[\bar{A}_4(T = 0, \rho_{\text{sat}})]_{\text{mean-field,NN}} \lesssim 1$ MeV.

5. Isospin-Asymmetry Dependence of the Nuclear Equation of State

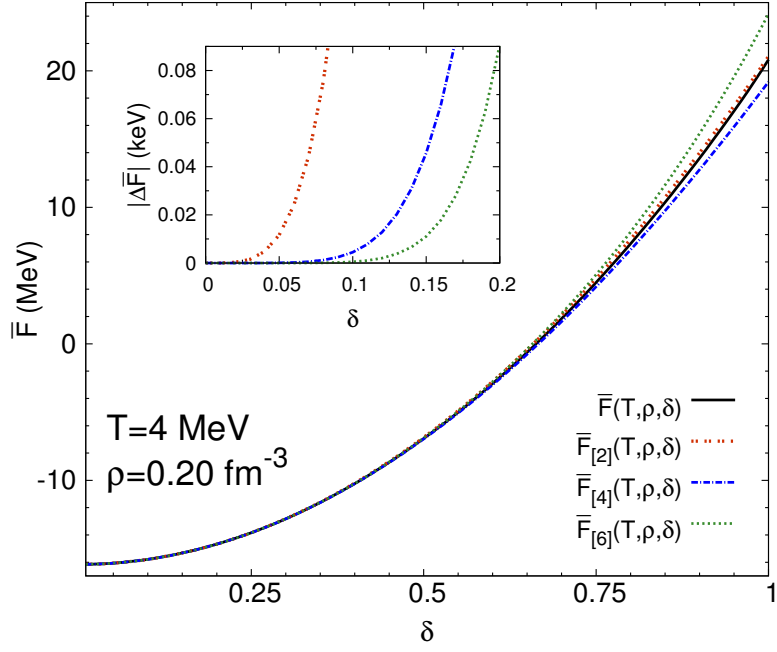

**Figure 5.17.:** Full isospin-asymmetry dependent free energy per particle $F(T, \rho, \delta)$ versus the quadratic, quartic and sextic polynomial isospin-asymmetry approximations at $T = 4\,\text{MeV}$ and $\rho = 0.20\,\text{fm}^{-3}$; results for n3lo414. The inset shows the absolute value of the deviation $|\Delta \bar{F}| := |\bar{F} - \bar{F}_{[2N]}|$ of the various approximations from the exact results for small neutron excesses ($\delta \in [0, 0.2]$). The values of the Maclaurin coefficients are $\bar{A}_{2,4,6}/\textbf{MeV} \simeq (37.21, -1.87, 4.98)$.

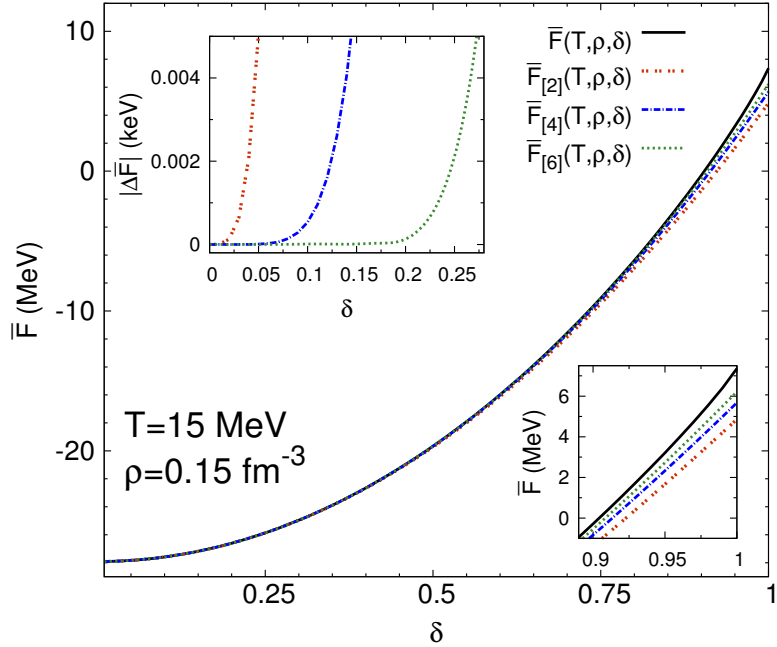

**Figure 5.18.:** Same as Fig. 5.21 but for $T = 15\,\text{MeV}$ and $\rho = 0.15\,\text{fm}^{-3}$ (results for n3lo414), with an additional inset for the very neutron-rich region (note the behavior of $F(T, \rho, \delta)$ compared to the approximations). Note that the error plot (upper inset, $|\Delta \bar{F}|$) involves units keV (as in the other Figures). The values of the Maclaurin coefficients are $\bar{A}_{2,4,6}/\textbf{MeV} \simeq (35.28, 0.81, 0.55)$.

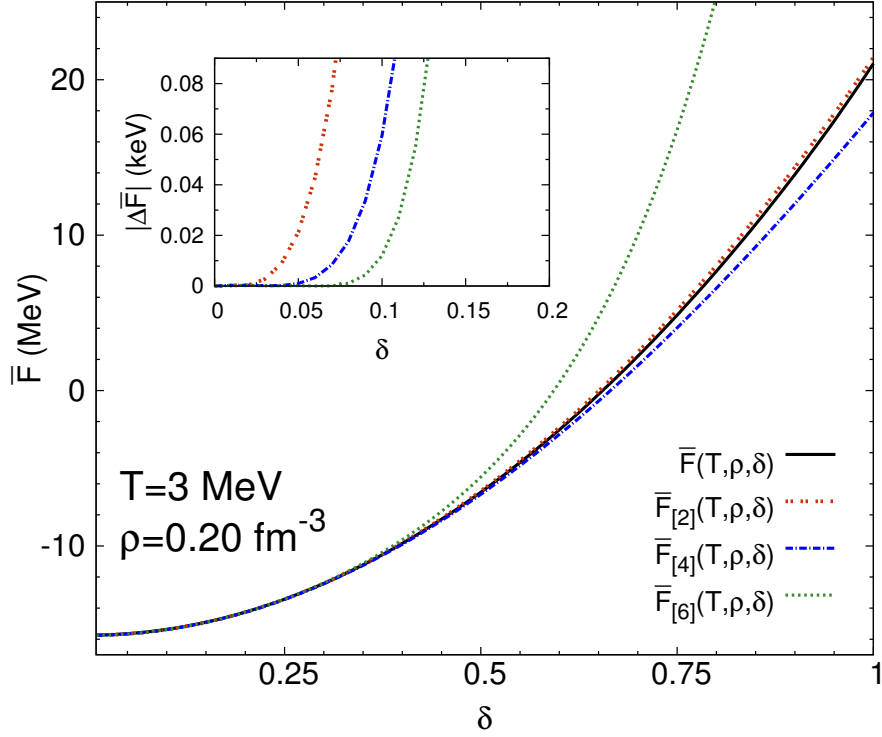

**Figure 5.19.:** Same as Fig. 5.17 but for $T = 3$ MeV and $\rho = 0.20$ fm$^{-3}$; results for n3lo414. The values of the Maclaurin coefficients are $\bar{A}_{2,4,6}/\text{MeV} \simeq (37.18, -3.56, 71.6)$.

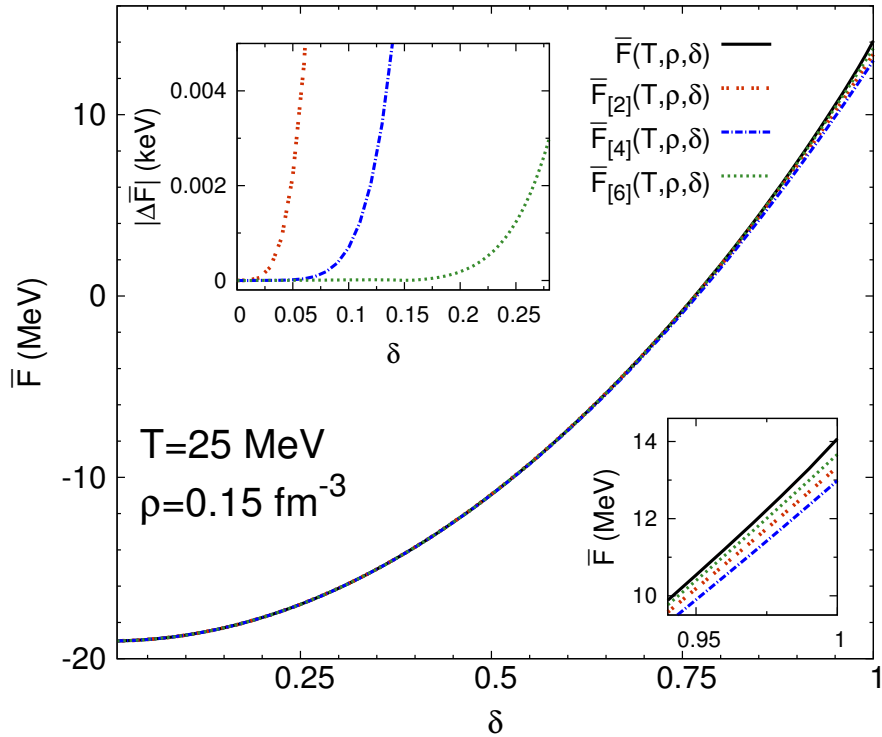

**Figure 5.20.:** Same as Fig. 5.18 but for $T = 25$ MeV and $\rho = 0.15$ fm$^{-3}$; results for n3lo414. The values of the Maclaurin coefficients are $\bar{A}_{2,4,6}/\text{MeV} \simeq (33.10, -0.36, 0.68)$.

## 5.4. Isospin-Asymmetry Parametrizations

Here, we examine the various higher-order isospin-asymmetry parametrizations constructed from the Maclaurin exansion (for $T \neq 0$) and the "logarithmic" expansion (for $T = 0$), respectively. In addition, we examine how the nonanalytic features of the isospin-asymmetry dependence influence the results obtained from global fits to the computed EoS.

*Maclaurin Expansion.* The behavior of the higher-order isospin-asymmetry approximations $\bar{F}_{[4],[6]}(T, \rho, \delta)$ depends on the convergence behavior of the Maclaurin expansion of the higher-order interaction contribution, as depicted by the threshold line of Fig. 5.16 separating the "divergent" and "convergent" regime. In contrast, the overall accuracy of the quadratic approximation $\bar{F}_{[2]}(T, \rho, \delta)$ depends only on the size of $\bar{F}_{\text{sym}} - \bar{A}_2$, which has no connection to threshold line of Fig. 5.16. In Figs. 5.17 to 5.20, we compare the quadratic, quartic and sextic approximations $\bar{F}_{[2],[4],[6]}(T, \rho, \delta)$ to the exact results $\bar{F}(T, \rho, \delta)$ for different temperatures and densities; i.e.,

- Fig. 5.17: $T = 4\,\text{MeV}$ and $\rho = 0.20\,\text{fm}^{-3}$, ("divergent regime"),
- Fig. 5.18: $T = 15\,\text{MeV}$ and $\rho = 0.15\,\text{fm}^{-3}$, ("convergent regime"),
- Fig. 5.19: $T = 3\,\text{MeV}$ and $\rho = 0.20\,\text{fm}^{-3}$ ("divergent regime"),
- Fig. 5.20: $T = 25\,\text{MeV}$ and $\rho = 0.15\,\text{fm}^{-3}$, ("convergent regime"),

where the first two cases correspond to the "divergent regime", and the second two cases to the "convergent regime" of Fig. 5.16. The different behavior in the "divergent regime" and the "convergent regime" are clearly visible. In the convergent regime, the Maclaurin coefficients are (overall) hierarchically ordered,[28] and the quartic and sextic coefficients lead to systematic improvements also in the very neutron-rich region. In the divergent regime, the inclusion of the quartic and sextic coefficients leads to an improved approximation for very small isospin asymmetries, but this behavior breaks down for larger values of $\delta$ where the sextic and to a lesser extent also the quartic approximations deviate significantly from the exact result. In other terms, the "critical" value $\delta_{\text{threshold}}$ of $\delta$ for which the accuracy of the higher-order approximations becomes inferior to the leading quadratic one decreases below the threshold line of Fig. 5.16, with $\delta_{\text{threshold}} \xrightarrow{T \to 0} 0$, and "exceeds" $\delta = 1$ for sufficiently large temperatures or low densities.

*"Logarithmic" Expansion.* In the zero-temperature case the following approximation can be constructed from the values for $\bar{A}_{4,\text{log}}$ and $\bar{A}_{4,\text{reg}}$ extracted in Sec. 5.2.5:

$$\bar{F}_{[4,\text{log}]}(T = 0, \rho, \delta) := \bar{A}_0(0, \rho) + \bar{A}_2(0, \rho)\,\delta^2 + \bar{A}_{4,\text{reg}}(\rho)\,\delta^4 + \bar{A}_{4,\text{log}}(\rho)\,\delta^4 \ln |\delta|. \tag{5.85}$$

To identify the effect of the logarithmic term we consider also the quartic approximation of the zero-temperature EoS without the logarithmic term, i.e.,

$$\bar{F}_{[4,\text{nonlog}]}(T = 0, \rho, \delta) := \bar{A}_0(0, \rho) + \bar{A}_2(0, \rho)\,\delta^2 + \bar{A}_{4,\text{reg}}(\rho)\,\delta^4. \tag{5.86}$$

---

[28] In the convergent regime, the small deviations concerning the hierarchy are due to balancing effects from the contributions associated with the different many-body contributions.

The deviations $\Delta \bar{F} := \bar{F} - \bar{F}_{\text{approx}}$ corresponding to the approximations given by $\bar{F}_{[2]}$, $\bar{F}_{[4,\log]}$ and $\bar{F}_{[4,\text{nonlog}]}(T=0,\rho,\delta)$ are plotted in Figs. 5.22 and 5.23 for different densities.[29] Additional details are shown in Fig. 5.21 for $\rho = 0.25\,\text{fm}^{-3}$. One sees that including both the quartic and the logarithmic term considerably improves the description of the isospin-asymmetry dependence at zero temperature, as compared to the quadratic approximation. Both the $\bar{F}_{[4,\log]}(T=0,\rho,\delta)$ and the $\bar{F}_{[4,\text{nonlog}]}(T=0,\rho,\delta)$ approximation improve upon $\bar{F}_{[2]}(T=0,\rho,\delta)$ in all cases. Except for the n4lo414 results at densities around $\rho \simeq 0.2\,\text{fm}^{-3}$, the $\bar{F}_{[4,\log]}(T=0,\rho,\delta)$ approximation further improves upon $\bar{F}_{[4,\log]}(T=0,\rho,\delta)$, and the error $\Delta\bar{F} := \bar{F} - \bar{F}_{[4,\log]}$ does not exceed 10 keV except for isospin-asymmetries $\delta \gtrsim 0.8$ and densities above saturation density $\rho_{\text{sat}} \simeq 0.17\,\text{fm}^{-3}$. The offbeat behavior of the n3lo414 results behavior for $\rho \simeq 0.2\,\text{fm}^{-3}$ can be attributed to balancing effects between different many-body contributions (since it is not present in Fig. 5.24).

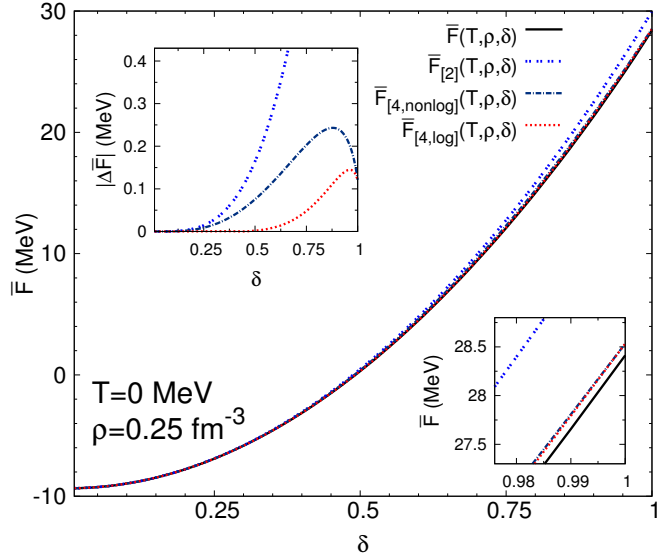

**Figure 5.21.:** Similar to Fig. 5.18 but for $T=0$ and $\rho = 0.25\,\text{fm}^{-3}$, and with Eqs. (5.85) and (5.86) as the higher-order isospin-asymmetry approximations (results for n3lo450). The lower inset shows the absolute value of the deviation $|\Delta\bar{F}| := |\bar{F} - \bar{F}_{\text{approx}}|$ of the various approximations from the exact results (for $\delta \in [0,1]$). The lower inset depicts the behavior in the very neutron-rich region.

For the higher-order approximations, the errors shown in Figs. 5.22 amd 5.23 may however include balancing effects between contributions from different many-body terms. To identify such effects we show in Figs. 5.24 and 5.25 the deviations of the various approximations for the second-order term alone, i.e., the quantities shown are $\Delta\bar{F}_2 := \bar{F}_2 - \bar{F}_{2;\text{approx}}$. The deviations for results are very similar for n3lo414 and n3lo450. For not too large values of $\delta$, the logarithmic approximation $\bar{F}_{[4,\log]}(T=0,\rho,\delta)$ always gives the best approximation, and the errors are below 10 keV for $\delta \gtrsim 0.8$, which can be seen as an additional validation of our approximate results for $\bar{A}_{4,\text{reg}}$ and $\bar{A}_{4,\log}$ (and confirms that the subleading logarithmic terms can be expected to be small). At very high densities, however, the deviations in the very neutron-rich region are considerably larger for $\bar{F}_{[4,\log]}(T=0,\rho,\delta)$ and $\bar{F}_{[4,\text{nonlog}]}(T=0,\rho,\delta)$ as compared to $\bar{F}_{[2]}(T=0,\rho,\delta)$. Note that this feature is not present in Figs. 5.22 and 5.23, so there is indeed a balancing of errors. In particular, the errors associated with $\bar{F}_{[4,\log]}$ and $\bar{F}_{[4,\text{nonlog}]}$ are smaller for the total free energy as compared to the second-order term alone, i.e., $|\Delta\bar{F}| < |\Delta\bar{F}_2|$ for the higher-order approximations.

---

[29] It should be emphasized that the higher-order approximations at zero temperature based on Eq. (5.70) must be strictly distinguished from those for the finite-temperature EoS based on the Maclaurin expansion Eq. (5.1).

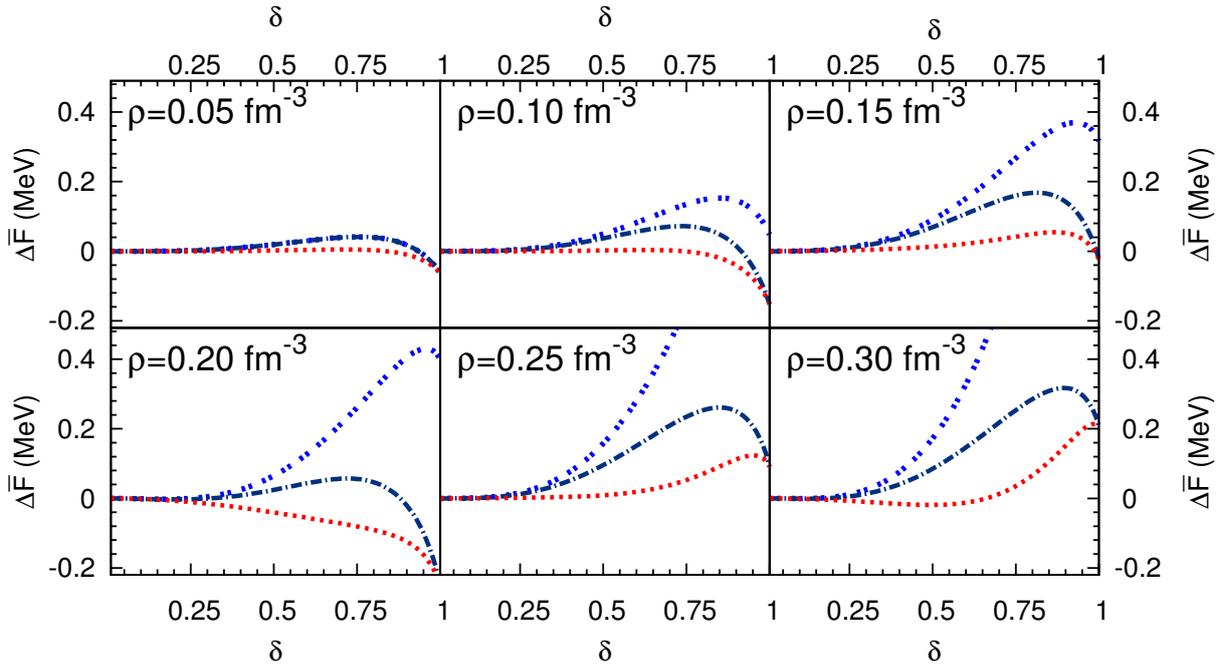

**Figure 5.22.:** Deviation $\Delta\bar{F} := \bar{F} - \bar{F}_{\text{approx}}$ of the various approximations (quadratic, quartic with/without logarithmic term) from the exact results at $T=0$ for different densities (results for n3lo414). Color coding as in Fig. 5.21.

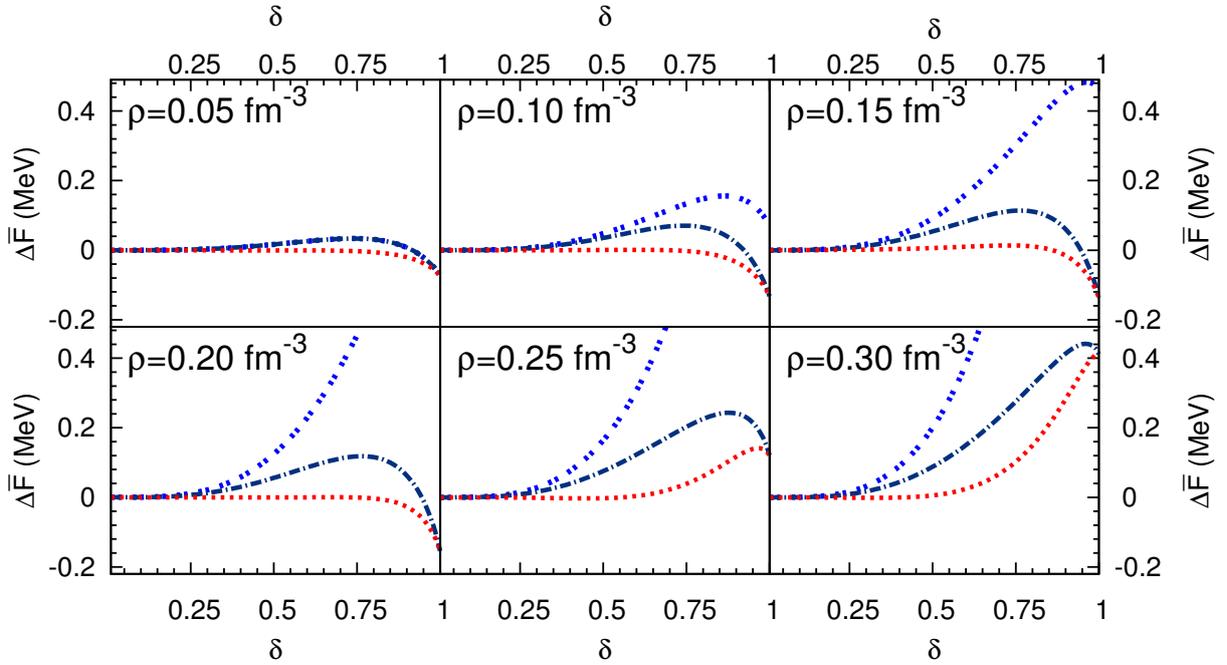

**Figure 5.23.:** Same as Fig. 5.22 but for n3lo450.

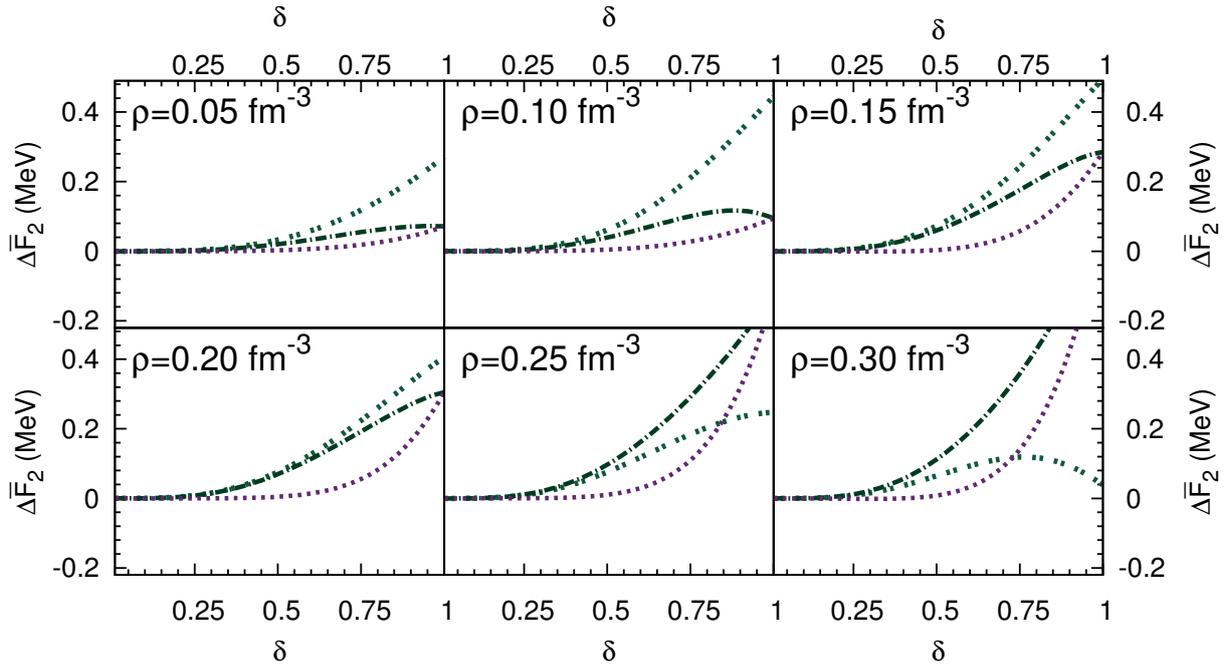

**Figure 5.24.:** Same as Fig. 5.22 but for the (isolated) second-order contribution $\bar{F}_2(T = 0, \rho, \delta)$ (results for n3lo414). For better distinction against Fig. 5.22 the coloring has been slightly changed, but the line shapes for the different approximations are the same as in Fig. 5.22.

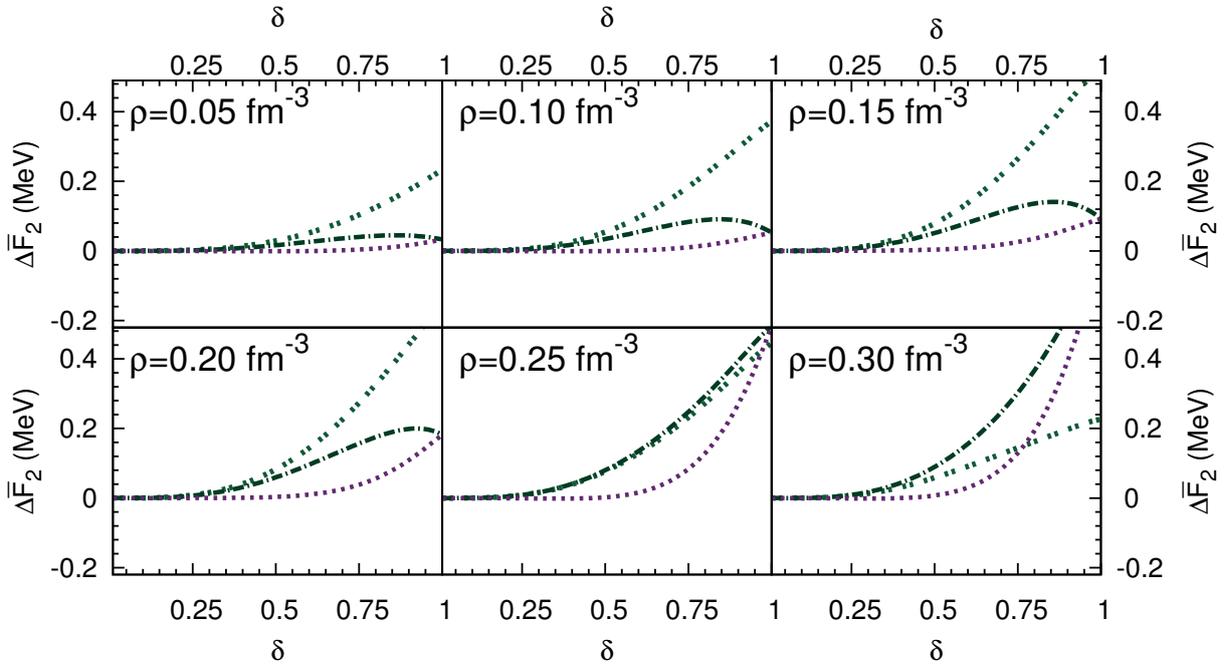

**Figure 5.25.:** Same as Fig. 5.24 but for n3lo450.

## 5. Isospin-Asymmetry Dependence of the Nuclear Equation of State

***Polynomial Fits.*** Regarding the construction of approximative parametrizations of the isospin-asymmetry dependence of the EoS via fits, the following points are obvious:

- A global fit leads to a better *global* approximation of the results (compared to a truncated expansion about $\delta = 0$), but the asymptotic behavior at $\delta = 0$ is described less accurately.

- The results obtained by fitting a polynomial to a nonanalytic function depend strongly on the fitting procedure.

- The fine-tuning of fit parameters (associated with the fit) increases with the number of fit parameters unless the fit function approaches the functional form of the data.

To elaborate these points in more detail we consider a polynomial

$$\bar{F}_{\text{fit}}^{[M]}(T, \rho, \delta) := \sum_{n=0}^{M} \bar{B}_{2n}(T, \rho)\, \delta^{2n}, \tag{5.87}$$

and determine for different polynomials lengths $M$ the coefficients $\bar{B}_{2n}$ by means of a global unweighted (i.e., the data is distributed uniformly in the regime $\delta \in [0, 1]$, and all data points carry equal weights) least-squares fit to the results for the (np-channel) second-order term, i.e., to the data $\{\bar{F}_{2,\text{normal}}(T, \rho, k\Delta\delta),\ k \in [0, N_{\text{data}}]\}$, where the stepsize is $\Delta\delta = 0.01$ and the grid length is $N_{\text{data}} = 100$.[30] The results for the fit coefficients $\bar{B}_{2n}$ are shown in Tables 5.4 and 5.5 for representative points in the convergent and the divergent regime of the EoS, respectively. In the convergent regime the fit coefficients are hierarchic, and satisfy $\bar{B}_{2n} \simeq \bar{A}_{2n}$ to good accuracy. In the divergent regime, however, only the leading two coefficients satisfy $\bar{B}_{0,2} \simeq \bar{A}_{0,2}$. The values of $\bar{B}_{2n \geq 6}$ on the other hand depend strongly on $M$, and their values become very large for large $M$. Notably, the quartic fit coefficient $\bar{B}_4$ is relatively small also for the divergent regime. To identify the origin of this behavior, we have fitted the coefficients $\bar{B}_{2n}$ also to functions of the form

$$f(\delta) := a_1 \delta^2 + a_2 \delta^4 \ln|\delta|, \tag{5.88}$$

which correspond to the truncated "logarithmic" expansion, cf. Eq. (5.70). For $a_1 > a_2$, $f(\delta)$ is an approximately quadratic function. Crucially, however, the fit results obtained for $\bar{B}_4$ are almost independent of $a_1$ if $a_2$ is fixed, and $\bar{B}_4$ is always consideraly smaller as compared to the higher-order fit coefficients $\bar{B}_{2n \geq 6}$ (for sufficiently large values of $M$ where they appear), and the results are similar to the results obtained from fits to the second-order term (in the divergent regime). Including additional logarithmic terms in $f(\delta)$ does not change this behavior. This makes evident that the nonanalytic behavior is for the most part resolved in the fit coefficients $\bar{B}_{2n \geq 6}$.

The parametrizations corresponding to the global polynomial fits are compared to the exact results for $\bar{F}_{2,\text{normal}}$ in Figs. 5.26 and 5.27. In the convergent regime [Fig. 5.26] the polynomial fits describe the data very well already for low values of $M$, and increasing $M$ leads to systematic improvements. The errors of the fits are overall considerably larger in the divergent regime [Fig. 5.27]. For very large values of $M$ the fits appear to converge to a curve that is very close to the exact results, expect for the very-neutron rich region. This feature can be expected to be related to the nonanalytic behavior of the isospin-asymmetry dependence at $\delta = 1$. In a global fit, the higher-order fit coefficients resolve predominantly the "stronger" nonanalyticity at $\delta = 0$ which affects the EoS for a wider range of isospin asymmetries.

*5. Isospin-Asymmetry Dependence of the Nuclear Equation of State*

|               | $M=2$   | $M=3$   | $M=4$            | $M=5$              | $M=12$  | $M=13$  | $M=14$  |
|---------------|---------|---------|------------------|--------------------|---------|---------|---------|
| $\bar{B}_0$/MeV  | -8.818  | -8.818  | -8.818           | -8.818             | -8.818  | -8.818  | -8.818  |
| $\bar{B}_2$/MeV  | 8.780   | 8.782   | 8.782            | 8.782              | 8.782   | 8.782   | 8.782   |
| $\bar{B}_4$/MeV  | 0.038   | 0.032   | 0.032            | 0.032              | 0.032   | 0.032   | 0.032   |
| $\bar{B}_6$/keV  | —       | 4.45    | 4.40             | 4.40               | 4.53    | 4.27    | 4.55    |
| $\bar{B}_8$/keV  | —       | —       | $3\times10^{-2}$ | $2\times10^{-2}$   | 1.2     | 1.8     | 2.1     |
| $\bar{B}_{10}$/eV | —      | —       | —                | $-3\times10^{-3}$  | 7.9     | 13.0    | 18.3    |
| $\bar{B}_0$/MeV  | -9.602  | -9.603  | -9.603           | -9.603             | -9.603  | -9.603  | -9.603  |
| $\bar{B}_2$/MeV  | 9.554   | 9.571   | 9.571            | 9.571              | 9.571   | 9.571   | 9.571   |
| $\bar{B}_4$/MeV  | 0.046   | -0.006  | -0.007           | -0.007             | -0.007  | -0.007  | -0.007  |
| $\bar{B}_6$/MeV  | —       | 0.038   | 0.038            | 0.040              | 0.040   | 0.040   | 0.040   |
| $\bar{B}_8$/keV  | —       | —       | -0.4             | -2.0               | -3.1    | 0.4     | -3.8    |
| $\bar{B}_{10}$/keV | —     | —       | —                | 0.7                | 7.3     | -16.6   | -17.4   |

**Table 5.4.:** Polynomial fit coefficients $\bar{B}_{2,4,6,8,10}$ for the second-order contribution for temperatures $T = 25$ MeV (upper rows) and $T = 15$ MeV (lower rows), each for $\rho = 0.15\,\text{fm}^{-3}$. For the higher-order fits ($M \geq 12$), the coefficients $\bar{B}_{2n\geq 12}$ are all small in magnitude, $\lesssim 0.1$ MeV ($\lesssim 0.1 - 0.5$ MeV for $M = 14$) in both cases. For comparison, the values of the Maclaurin coefficients (obtained from the finite-difference analysis) are $\bar{A}_{0,2,4,6,8}/\text{MeV}|_{T=25\,\text{MeV},\rho=0.15\,\text{fm}^{-3}} \simeq (-8.818, 8.782, 0.0319, 0.004, 0.0000)$ and $\bar{A}_{0,2,4,6,8}/\text{MeV}|_{T=15\,\text{MeV},\rho=0.15\,\text{fm}^{-3}} \simeq (-9.603, 9.571, 0.007, 0.04, -0.002)$.

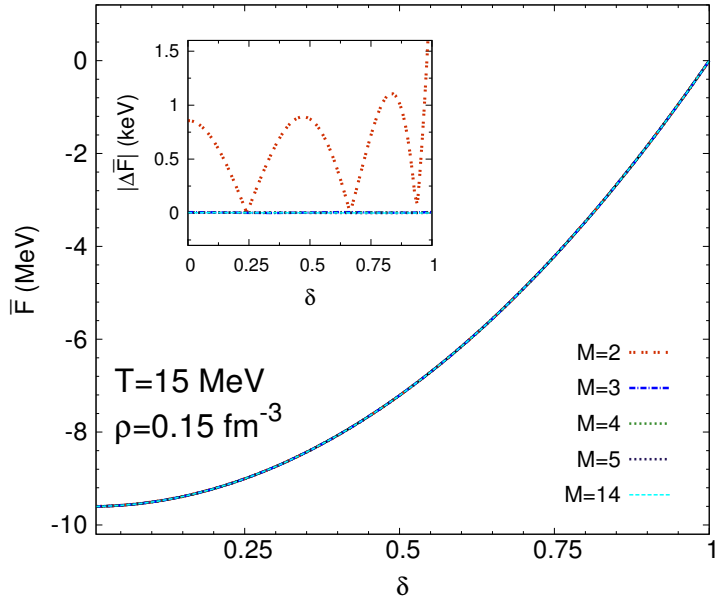

**Figure 5.26.:** Polynomial fits to the second-order term for $T = 15$ and $\rho = 0.15\,\text{fm}^{-3}$ (results for n4lo414). The inset shows the absolute value of the deviation $|\Delta \bar{F}| := |\bar{F} - \bar{F}_{\text{fit}}^{[M]}|$ of the fits from the exact results (for $\delta \in [0, 1]$). The behavior for $T = 25$ MeV and $\rho = 0.15\,\text{fm}^{-3}$ (plot not shown) is very similar.

---

[30] Note that we consider very large polynomial lengths $M$ only for illustrative purposes. By construction, increasing the number of fit parameters increases the overall precision of the fit (in the sum-of-squares sense), but for the divergent regime the high degree of fine-tuning involved inhibits the use of high-precision fits for global EoS parametrizations where the $T$ and $\rho$ dependent fit coefficients would have to be interpolated.

| | $M=2$ | $M=3$ | $M=4$ | $M=5$ | $M=12$ | $M=13$ | $M=14$ |
|---|---|---|---|---|---|---|---|
| $\bar{B}_0$/MeV | -6.339 | -6.352 | -6.354 | -6.355 | -6.355 | -6.355 | -6.355 |
| $\bar{B}_2$/MeV | 5.853 | 6.111 | 6.176 | 6.247 | 6.265 | 6.266 | 6.266 |
| $\bar{B}_4$/MeV | 0.45 | -0.31 | -0.67 | -1.28 | -1.56 | -1.61 | -1.60 |
| $\bar{B}_6$/MeV | — | 0.55 | 1.16 | 2.97 | 5.75 | 6.90 | 6.79 |
| $\bar{B}_8$/MeV | — | — | -0.32 | -2.50 | -23.7 | -37.1 | -35.5 |
| $\bar{B}_{10}$/MeV | — | — | — | 0.91 | 109.0 | 202.9 | 189.7 |
| $\bar{B}_0$/MeV | 7.323 | 7.332 | 7.333 | 7.334 | 7.334 | 7.334 | 7.334 |
| $\bar{B}_2$/MeV | 7.510 | 7.694 | 7.738 | 7.776 | 7.830 | 7.831 | 7.833 |
| $\bar{B}_4$/MeV | -0.21 | -0.76 | -0.99 | -1.32 | -2.43 | -2.56 | -2.65 |
| $\bar{B}_6$/MeV | — | 0.40 | 0.80 | 1.77 | 13.21 | 16.07 | 18.49 |
| $\bar{B}_8$/MeV | — | — | -0.3 | -1.5 | -90.4 | -131.0 | -176.1 |
| $\bar{B}_{10}$/MeV | — | — | — | 0.5 | 335.9 | 568.7 | 838.9 |

**Table 5.5.:** Polynomial fit coefficients $B_{2,4,6,8,10}$ for the second-order contribution for $T=4$ MeV and $\rho = 0.30\,\text{fm}^{-3}$ (upper rows) as well as $T=0$ MeV and $\rho = 0.15\,\text{fm}^{-3}$ (lower rows). For the higher-order fits ($M \geq 12$), the coefficients $\bar{B}_{2n\geq 12}$ are all in the GeV scale (but not hierarchical) for both cases. The values of the corresponding Maclaurin coefficients and coefficients in the "logarithmic expansion", respectively, are $\bar{A}_{0,2,4,6}/\text{MeV}|_{T=4\,\text{MeV},\rho=0.30\,\text{fm}^{-3}} \simeq (-6.355, 6.278, 3.00, 69.2)$, $\bar{A}_8/\text{GeV}|_{T=4\,\text{MeV},\rho=0.30\,\text{fm}^{-3}} \simeq 1.5$ as well as $\bar{A}_{0,2}/\text{MeV}|_{T=0\,\text{MeV},\rho=0.15\,\text{fm}^{-3}} \simeq (-7,334, 7.83)$, $\bar{A}_{4,\text{reg}}/\text{MeV}|_{T=0\,\text{MeV},\rho=0.15\,\text{fm}^{-3}} \simeq -0.2$, and $\bar{A}_{4,\text{log}}/\text{MeV}|_{T=0\,\text{MeV},\rho=0.15\,\text{fm}^{-3}} \simeq 1.3$.

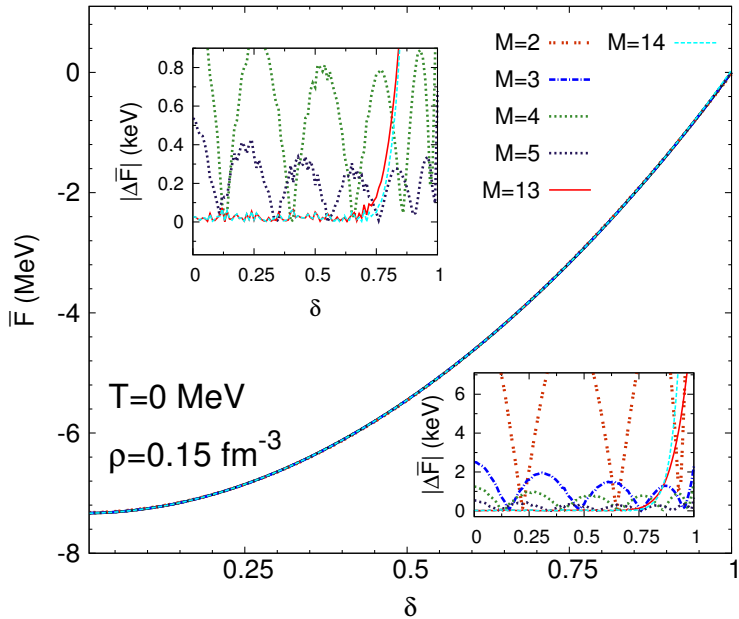

**Figure 5.27.:** Same as Fig. 5.26 but for $T=0$ and $\rho = 0.15\,\text{fm}^{-3}$. The two insets show the deviations $|\Delta \bar{F}| := |\bar{F} - \bar{F}_{\text{fit}}^{[M]}|$ at two different zoom levels. For clarity, in the upper inset only the results for $M \in \{4, 5, 13, 14\}$ are shown. The behavior for $T=4$ MeV and $\rho = 0.30\,\text{fm}^{-3}$ (plot not shown) is very similar.

## 5. Isospin-Asymmetry Dependence of the Nuclear Equation of State

***"Logarithmic" Fits.*** It is now interesting to consider the case where logarithmic terms are included in the fit function, i.e.,

$$\bar{F}_{\text{fitlog}}^{[M_{\text{log}}]}(T, \rho, \delta) := \bar{B}_0 + \bar{B}_2 \delta^2 + \sum_{n=2}^{M_{\text{log}}} \bar{B}_{2n,\text{nonlog}} \delta^{2n} + \sum_{n=2}^{M_{\text{log}}} \bar{B}_{2n,\text{log}} \delta^{2n}. \quad (5.89)$$

In particular, we want to perform an intial study concerning the question as to whether the inclusion logarithmic terms can be useful regarding the construction of an *approximative* parametrization of the isospin-asymmetry dependence at low but nonzero temperatures. For this purpose, we compare the previous results based on a polynomial fit function to the ones obtained by fitting of the above "logarithmic" polynomial [Eq. (5.89)] to the low-temperature results for the second-order term.

The values obtained for the "logarithmic" fit coefficients (from a global least-squares fit) for different values of $M_{\text{log}}$ are shown in Tables 5.6 and 5.7 for $T = 0$ and $\rho = 0.15 \, \text{fm}^{-3}$ as well as $T = 4$ and $\rho = 0.30 \, \text{fm}^{-3}$. One sees that the higher-order fit coefficients $\bar{B}_{2n,\text{nonlog}}$ and $\bar{B}_{2n,\text{log}}$ are substantially smaller as compared to the higher-order ones $\bar{B}_{2n \geq 4}$ in the polynomial fit [cf. Table 5.5]. They are larger at nonzero temperature, which points to the higher degree of fine-tuning involved in that case; however, they are still significantly smaller as compared to the corresponding ones for the polynomial fit without logarithmic terms.

In Figs. 5.28 and 5.29 we then compare the exact results for the second-order term to the ones corresponding to the "logarithmic" fits. One sees that compared to the polynomial fits, the "logarithmic" fits lead to a overall better description of the data for zero temperature and (to a lesser extent) also for low but nonzero temperatures.

Altogether, while strictly speaking a logarithmic dependence on $\delta$ is present only at $T = 0$, the above findings indicate that including logarithmic terms may be useful to improve the description of the isospin-asymmetry dependence of the EoS also for low but nonzero temperatures. In particular, this may be useful for the construction of approximative global EoS parametrizations since (except for the case of zero temperature) in that regime no well-behaved expansion exists.

|  | $\bar{B}_0$ | $\bar{B}_2$ | $\bar{B}_{4,\text{nonlog}}$ | $\bar{B}_{4,\text{log}}$ | $\bar{B}_{6,\text{nonlog}}$ | $\bar{B}_{6,\text{log}}$ | $\bar{B}_{8,\text{nonlog}}$ | $\bar{B}_{8,\text{log}}$ |
|---|---|---|---|---|---|---|---|---|
| $M_{\text{log}} = 2$ | -7.334 | 7.820 | -0.488 | 0.918 | — | — | — | — |
| $M_{\text{log}} = 3$ | -7.335 | 7.858 | 0.383 | 1.802 | -0.907 | 1.059 | — | — |
| $M_{\text{log}} = 4$ | -7.334 | 7.837 | -0.699 | 0.987 | -0.705 | -1.661 | 0.902 | -0.509 |

**Table 5.6.:** "Logarithmic" fit coefficients for the second-order contribution for $T = 0$ MeV and $\rho = 0.15 \, \text{fm}^{-3}$ (results for n4lo414), see text for details. The fit coefficients are all given in units MeV. For comparison, the values extracted from the finite-difference analysis for corresponding coefficients in the "logarithmic expansion" are $\bar{A}_{0,2}/\textbf{MeV} \simeq -7, 334, 7.83$, $\bar{A}_{4,\text{reg}}/\textbf{MeV} \simeq -0.2$, and $\bar{A}_{4,\text{log}}/\textbf{MeV} \simeq 1.3$.

|  | $\bar{B}_0$ | $\bar{B}_2$ | $\bar{B}_{4,\text{nonlog}}$ | $\bar{B}_{4,\text{log}}$ | $\bar{B}_{6,\text{nonlog}}$ | $\bar{B}_{6,\text{log}}$ | $\bar{B}_{8,\text{nonlog}}$ | $\bar{B}_{8,\text{log}}$ |
|---|---|---|---|---|---|---|---|---|
| $M_{\text{log}} = 2$ | -6.355 | 6.289 | 0.063 | 1.291 | — | — | — | — |
| $M_{\text{log}} = 3$ | -6.356 | 6.372 | 1.635 | 2.979 | -1.654 | 1.863 | — | — |
| $M_{\text{log}} = 4$ | -6.355 | 6.283 | -0.084 | 0.965 | 5.138 | 4.578 | -4.982 | 5.561 |

**Table 5.7.:** "Logarithmic" fit coefficients (in units MeV) for the second-order contribution for $T = 4$ MeV and $\rho = 0.30 \, \text{fm}^{-3}$ (results for n4lo414), see text for details.

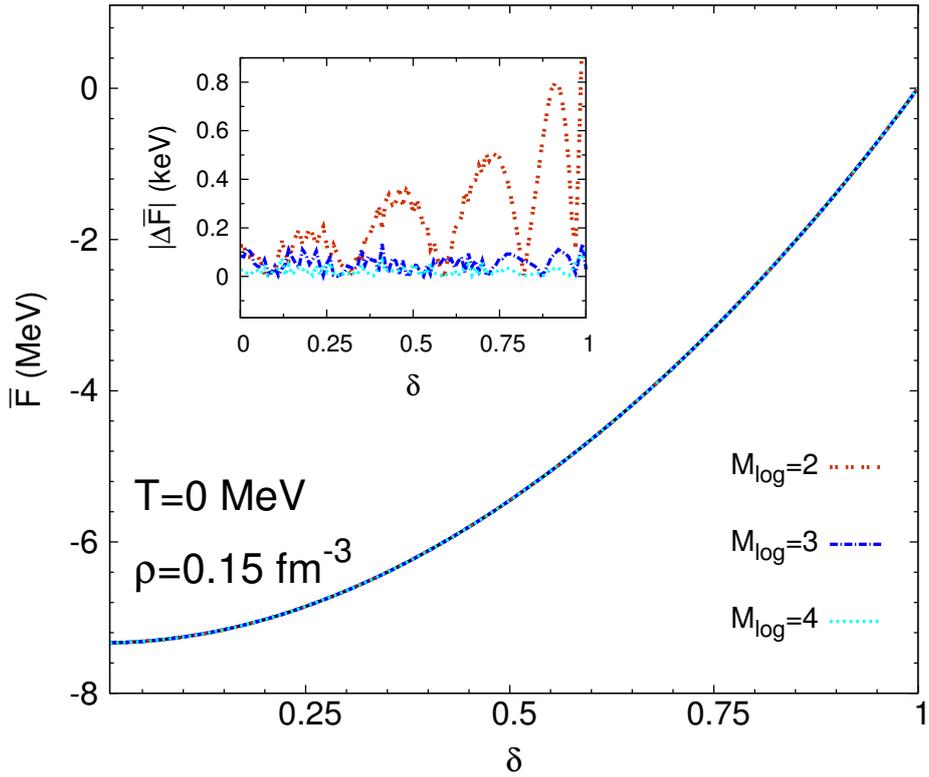

**Figure 5.28.:** "Logarithmic" fits to the second-order term for $T = 0$ and $\rho = 0.15\,\mathrm{fm}^{-3}$ (results for n4lo414). The inset shows the absolute value of the deviation $|\Delta \bar{F}| := |\bar{F} - \bar{F}_{\mathrm{fitlog}}^{[M_{\mathrm{log}}]}|$ of the fits from the exact results (for $\delta \in [0, 1]$).

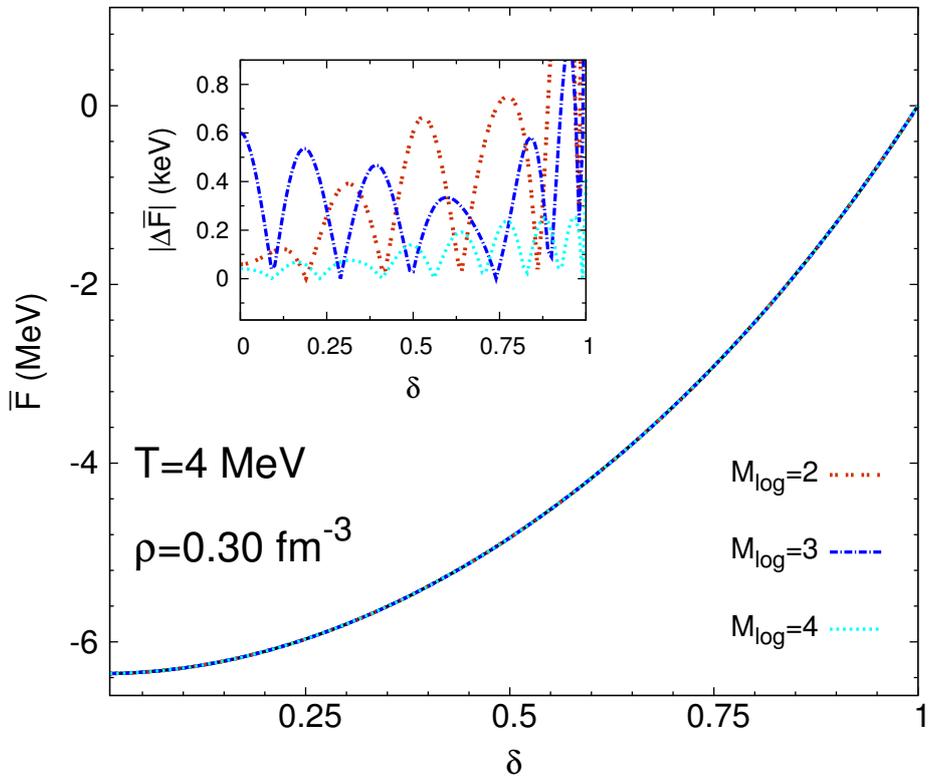

**Figure 5.29.:** Same as Fig. 5.28 but for $T = 4$ and $\rho = 0.30\,\mathrm{fm}^{-3}$.

# Conclusion

In this thesis, we have investigated the thermodynamics of nuclear matter using many-body perturbation theory (MBPT). The consistent generalization of zero-temperature MBPT to finite temperatures has been investigated, and we have found that the (usual) grand-canonical approach is invalid on general accounts [cf. Secs. 2.5.4 and 2.5.5]. Building on existing work [49, 65, 215, 18, 278, 246, 277, 150, 394], we have shown that a consistent thermodynamic MBPT can be constructed, starting from the canonical ensemble, by evaluating truncated correlation functions in terms of a Legendre transformation (correlation-bond formalism). The structure of MBPT at higher orders and the relations between the canonical approach and the "standard" grand-canonical formalism were investigated, in particular, we have studied the cancellation of the additional anomalous contributions present in finite-temperature MBPT (but not in the zero-temperature formalism) in terms of the self-consistent renormalization of the single-particle basis.

Using this formalism (canonical finite-temperature MBPT), we have then computed the free energy per particle $\bar{F}(T, \rho, \delta)$ of infinite homogeneous nuclear matter using several sets of low-momentum (N3LO) two- and (N2LO) three-nucleon potentials constructed within the framework of chiral effective field theory ($\chi$EFT). The pertinent low-energy constants (LECs) that parametrize the potentials were fit—in Refs. [90, 91, 125, 312, 343, 342]—to nucleon-nucleon phase shifts and properties of light nuclei. The results for isospin-symmetric nuclear matter have been benchmarked against available empirical constrains, and we have found that the potentials sets ("n3lo414" and "n3lo450") which match constraints from zero-temperature bulk properties best give also consistent results at finite temperature. However, for some potential sets ("VLK21" and "VLK23") we have found that even if a reasonable zero-temperature EoS is obtained in second-order MBPT, thermodynamic inconsistencies can arise. The origin of this feature has been traced back to the different values of the LECs which parametrize the three-nucleon potential. In the case of the "VLK" potentials, however, the requirement of a consistent evolution to low-momentum scales of multi-nucleon interactions was not fully (properly) accounted for. These results highlight the importance of a consistent treatment of two- and multi-nucleon interactions, and suggest that one should try to better constrain the values of the LECs in the future.

Using the "n3lo414" and "n3lo450" potentials, we have then computed the EoS of pure neutron matter, and we have found that the results at low fugacities are in good agreement with the model-independent virial EoS determined from neutron-neutron phase shifts. In addition, we have examined the symmetry free energy $\bar{F}_{\text{sym}}(T, \rho) = \bar{F}(T, \rho, \delta = 1) - \bar{F}(T, \rho, \delta = 0)$, a quantity that gives an approximate measure for the dependence of the nuclear EoS on the isospin asymmetry $\delta = (\rho_n - \rho_p)/(\rho_n + \rho_p)$. The results for $\bar{F}_{\text{sym}}(T = 0, \rho)$ have been found to be consistent with empirical constraints from neutron skin thicknesses and isobaric analog states. Moreover, we have studied the liquid-gas phase transition of isospin-asymmetric nuclear matter, focussing on the new aspects of the phase transition that arise in the case of nonvanishing isospin asymmetry ("isospin distillation"). In particular, we have found that the exact isospin-asymmetry dependence of the noninteracting contribution to the EoS sets essential constraints on the properties of the liquid-gas instability in the very neutron-rich region. In the case of finite temperature, this is related to the entropy of mixing, i.e., a dependence on $\delta$ of the form $\sim [(1+\delta)\ln(1+\delta) + (1-\delta)\ln(1-\delta)]$, and at zero temperature to terms $\sim (1-\delta)^{(2\nu+1)/3}|_{\nu \geq 2}$.

Finally, we have investigated the Maclaurin expansion of the free energy per particle in terms of $\delta$. We have extracted the quadratic, quartic, and sextic expansion (Maclaurin) coefficients $\bar{A}_{2,4,6}(T,\rho)$ from using higher-order finite-difference approximations, and we have shown that the accuracy of the leading-order quadratic approximation is decreased both at high temperatures and at high densities, which is mainly due to the noninteracting term (free nucleon gas) and the (first-order) contribution from three-nucleon interactions. Regarding the higher-order terms in the expansion, we have found that the coefficients beyond the quadratic one diverge in the zero-temperature limit. This feature has its origin in the energy denominators of the higher-order many-body contributions (and is therefore absent at the Hartree-Fock level). At $T = 0$, the energy denominators cause a divergence of integral kernels at the boundary of the respective integration regions, which leads to a singular quartic isospin-asymmetry derivative at $\delta = 0$. In other terms, in MBPT the zero-temperature EoS is not a smooth function of $\delta$ but only of differentiability class $C^3$.[31]

At finite temperature the integral kernels have no singularities (if treated properly), so the EoS is a smooth ($C^\infty$) function of the isospin asymmetry for $T \neq 0$, but since $|\bar{A}_{2n\geq 4}| \xrightarrow{T\to 0} \infty$, the dependence on $\delta$ is nonanalytic at low temperatures and the Maclaurin expansion in $\delta$ constitutes a divergent asymptotic expansion (in that regime). However, we have found that the Maclaurin coefficients become hierarchically ordered for higher temperatures, which indicates that the expansion converges in the high-temperature region. In sum, we have found

$$\bar{F}(T,\rho,\delta) \sim \sum_{n=0}^{\infty} \bar{A}_{2n}(T,\rho)\,\delta^{2n} = \begin{cases} \text{singular}, & \text{if } (T,\rho) \in \{(T,\rho)_{\text{divergent}}\} \\ \bar{F}(T,\rho,\delta), & \text{if } (T,\rho) \in \{(T,\rho)_{\text{convergent}}\} \end{cases},$$

where the divergent regime $\{(T,\rho)_{\text{divergent}}\}$ corresponds to low temperatures and high densities, as specified by the threshold line of Fig. 5.16. At zero temperature, the nonanalytic behavior of the isospin-asymmetry dependence is associated with terms of the form $\sim \delta^{2n\geq 4} \ln|\delta|$. We have extracted the coefficients of the leading logarithmic term $\sim \delta^4 \ln|\delta|$ and the "regular" quartic term $\sim \delta^4$ from the analysis of linear combinations of higher-order finite differences, and we have found that including these terms leads to an overall improved parametrization of the isospin-asymmetry dependence of the zero-temperature EoS.

The question arises of course as to whether the nonanalytic behavior (at low temperatures) of the isospin-asymmetry dependence is a "genuine" feature of the nuclear EoS, or only a feature of MBPT, i.e., one that arises from probing higher-order derivatives of a perturbation series. It is, however, not clear whether this question can be resolved since exact solutions are available in many-body theory only for very special models [29, 290, 431] and genuinely nonperturbative approaches such as QMC (presumably) at present lack the numerical precision for the extraction of higher-order isospin-asymmetry derivatives. Nevertheless, our results suggest that the relevance of nonanalyticities (and in particular, logarithmic terms) in the isospin-asymmetry for (e.g.,) propertes of neutron stars should be investigated further in future research.[32]

---

[31] The property $|\bar{A}_{2n\geq 4}| \xrightarrow{T\to 0} \infty$ relies on the thermodynamic limit as well as the neglect of pairing correlations. For a finite system the higher-order "Maclaurin coefficients" (i.e., the finite differences, since the spectrum is discrete) are finite at zero temperature, but diverge in the thermodynamic limit. Hence, for sufficiently large particle numbers the dependence of the finite-system EoS on $\delta = (N_n - N_p)/(N_n + N_p)$ is still "nonanalytic", analogous to the behavior of the infinite system at low temperatures. Similar considerations apply for the case where pairing correlations are included in MBPT by expanding about the BCS ground state [cf. Sec. 5.2.3].

Altogether, the work presented in this thesis provides the basis for future research headed towards the construction of a $\chi$EFT-based EoS for applications in nuclear astrophysics. For applications in simulations of core-collapse supernovæ and binary neutron-star mergers, a fine-meshed numerical EoS table that covers a wide range of densities, temperatures, and isospin-asymmetries would be required. This presents a challenge to the framework used in this thesis. In particular, the implementation of the DDNN potential and the effective nucleon mass ($M^*$) in such extensive computations requires special attention. In the case of the DDNN potential, an expansion of its matrix elements in terms of $\delta$ may be useful. For isospin-asymmetric matter, the (isospin-asymmetry dependent) $M^*$ terms do not factor out of the energy denominators, leading to changes in the integral structure which significantly increase the CPU time needed to numerically evaluate the higher-order contributions. Testing different approximations that amend this feature may be useful in this respect. In addition, as a complementary approach to a brute-force computation of a fine-meshed numerical EoS table, one could try to construct explicit approximative parametrizations of the (interaction contributions to the) EoS via fits to the computed data.

The applicability of the perturbative framework employed in this thesis is restricted to densities below about $\rho \sim 2\, \rho_{\text{sat}}$. A major issue regarding astrophysical applications is therefore also the extrapolation of the chiral EoS to the high densities probed in core-collapse supernovæ and binary neutron-star mergers. Different approaches should be considered for that purpose. One possibile approach would be to use phenomenological mean-field models (that can reach to higher densities and temperatures) and constrain their parameters by matching to the $\chi$EFT-based results [352]. In view of the results presented in this thesis, isospin-asymmetry dependent terms beyond the usual quadratic ones should be taken into account in the matching.

Finally, to conclude the present thesis, we assess that, while far from being a closed chapter, the nuclear many-body problem represents an interesting and fruitful subject in (applied) theoretical physics. Several decades after the discovery of the atomic nucleus, the theoretical basis for its description has been established in terms of QCD, and nearly a century after its historical emergence, the subject of theoretical nuclear physics has been cast into new form in terms of $\chi$EFT. The application of these theories in the study of dense nuclear matter still represents a challenge, but promising new developments have been made in recent times, incentive results have been obtained, and one can look forward to further progress on behalf of present and future research activities.

---

[32] Note, however, that for extremely neutron-rich systems the influence of logarithmic terms is rather small [cf. Fig. 5.21]; for instance, the proton fraction in neutron stars is more sensitive to a term $\sim \delta^4$ as compared to a term $\sim \delta^4 \ln|\delta|$.

# A. Appendix

## A.1. Free Nucleon Gas

The free nucleon gas can of course be treated fully relativistically, but since the nuclear interactions are formulated nonrelativistically a more consistent approach is to start with the nonrelativistic Fermi gas, and include relativistic (kinematical) effects by appropriate correction terms.[1] At zero temperature the construction of relativistic corrections is straightforward, but a consistent generalization of these corrections for finite temperatures is nontrivial. The situation becomes much clearer when the nonrelativistic expansion is formulated as a many-body perturbation theory (MBPT) problem: based on the expansion $\varepsilon_{\text{rel}}(k) = \varepsilon_{\text{nonrel}}(k) + \sum_{\nu=1}^{\infty} \varepsilon_{\text{corr}(\nu)}(k) = \frac{k^2}{2M} - \frac{k^4}{8M^3} + \frac{k^6}{16M^5} - \frac{5k^8}{128M^7} + \ldots$ we write the relativistic kinetic energy Hamiltonian as $\mathcal{T}_{\text{rel}} = \mathcal{T}_{\text{nonrel}} + \mathcal{T}_{\text{corr}}$, where the reference Hamiltonian $\mathcal{T}_{\text{nonrel}} = \sum_k \varepsilon_{\text{nonrel}}(k) a_k^\dagger a_k$ represents the nonrelativistic system, and $\mathcal{T}_{\text{corr}}$ is given by a sum of hierachically ordered "perturbation Hamiltonians" $\mathcal{T}_{\text{corr}(\nu)}$:

$$\mathcal{T}_{\text{corr}} = \sum_{\nu=1}^{\infty} \lambda^\alpha \mathcal{T}_{\text{corr}(\nu)} = \sum_{\nu=1}^{\infty} \lambda^\alpha \sum_k \varepsilon_{\text{corr}(\nu)}(k) a_k^\dagger a_k. \tag{A.1}$$

For clarity, we restrict the discussion to the one-species case where (in the thermodynamic limit) it is $\sum_k = \frac{g_\sigma}{2\pi^2} \int_{k=0}^{\infty} dk\, k^2$, with spin multiplicity $g_\sigma = 2$. From the one-species results for the free energy density $F(T, \rho_{\text{n/p}})$, the free energy density of the two-component system with total density $\rho = \rho_n + \rho_p$ is given by $F(T, \rho_n, \rho_p) = F(T, \rho_n) + F(T, \rho_p)$.

### A.1.1. Relativistic and Nonrelativistic Results

Here we briefly summarize the relativistic and nonrelativistic free Fermi gas results. If not otherwise indicated the single-particle energies in $n_k$ are the nonrelativistic ones $\varepsilon_{\text{nonrel}}(k) = k^2/(2M)$.

***Zero Temperature.*** For a given energy-momentum relation the Fermi momentum $k_F$ is related to the Fermi energy $\varepsilon_F$ via $\varepsilon(k_F) = \varepsilon_F$. The expression for the particle density $\rho(k_F)$ as a function of the Fermi momentum $k_F$ is independent of the energy-momentum relation $\varepsilon(k)$, i.e.,

$$\rho(k_F) = \sum_k \Theta(\varepsilon_F - \varepsilon(k)) = \frac{1}{\pi^2} \int_0^{k_F} dk\, k^2 = \frac{k_F^3}{3\pi^2}. \tag{A.2}$$

For a given energy-momentum relation $\varepsilon(k)$ the ground-state energy density is given by

$$E_0(k_F) = \sum_k \varepsilon(k)\Theta(\varepsilon_F - \varepsilon(k)) = \frac{1}{\pi^2} \int_0^{\infty} dk\, k^2 \varepsilon_k \Theta(\varepsilon_F - \varepsilon(k)) = \frac{1}{\pi^2} \int_0^{k_F} dk\, k^2 \varepsilon(k), \tag{A.3}$$

---

[1] In addition, since it necessarily (cf. Sec. A.1.2) leads to nonrelativistic single-particle energies [$\varepsilon_{\text{nonrel}}(k) = k^2/(2M)$] in the expressions for the various many-body contributions, the "nonrelativistic+corrections" approach is much more convenient regarding the inclusion of self-energy effects in MBPT (effective-mass approximation, cf. Sec. 3.1).

## A. Appendix

and the chemical potential is given by $\mu(k_F) = \partial E_0(k_F)/\partial\rho = \varepsilon(k_F)$. So far everything is general. Inserting $\varepsilon_{\text{nonrel}}(k) = k^2/(2M)$ and $\varepsilon_{\text{rel}}(k) = \sqrt{k^2 + M^2} - M$, respectively, and resolving the integral in Eq. (A.3), gives $E_{0;\text{nonrel}}(k_F) = k_F^5/(10\pi^2 M)$ and

$$E_{0;\text{rel}}(k_F) = \frac{1}{24\pi^2}\left[-8Mk_F^3 + 3k_F\left(M^2 + 2k_F^2\right)\sqrt{k_F^2 + M^2} - 3M^4 \ln\left(\frac{k_F + \sqrt{k_F^2 + M^2}}{M}\right)\right]. \quad (A.4)$$

*Thermodynamics (General).* The general expression for the grand-canonical potential density is given by

$$A(T,\mu) = -\frac{T}{\pi^2}\int_0^\infty dk\, k^2 \ln\left[1 + e^{-\beta(\varepsilon(k)-\mu)}\right]. \quad (A.5)$$

The grand-canonical expression for the particle density is then given by

$$\rho(T,\mu) = -\frac{\partial A(T,\mu)}{\partial \mu} = \frac{1}{\pi^2}\int_0^\infty dk\, k^2 \frac{1}{1 + \exp\left[\beta(\varepsilon(k)-\mu)\right]}, \quad (A.6)$$

and the free energy density is given by $F(T,\mu) = \frac{\mu}{\pi^2}\int_0^\infty dk\, k^2 \frac{1}{1+\exp[\beta(\varepsilon_k-\mu)]} - \frac{T}{\pi^2}\int_0^\infty dk\, k^2 \ln\left[1 + e^{-\beta(\varepsilon(k)-\mu)}\right]$. Using l'Hôpital's rule to evaluate the zero-temperature limit of $A(T,\mu)$ we obtain

$$F(T,\mu) \xrightarrow{T\to 0} \frac{\mu}{\pi^2}\int_0^\infty dk\, k^2 \Theta(\mu - \varepsilon(k)) + \frac{1}{\pi^2}\int_0^\infty dk\, k^2(\varepsilon(k) - \mu)\,\Theta(\mu - \varepsilon(k)) = E_0(k_F), \quad (A.7)$$

with $\varepsilon(k_F) = \mu$. The expression for $A(T,\mu)$ can be given in an alternative form by making use of the identity $\int_0^\infty dk\, k^2 \ln[X(k)] = -\int_0^\infty dk\, k^3 \frac{\partial X(k)/\partial k}{3X(k)}$, which, when applied to Eq. (A.5), leads to

$$A(T,\mu) = -\frac{1}{3\pi^2}\int_0^\infty dk\, k^3 \frac{\partial \varepsilon(k)/\partial k}{1 + \exp\left[\beta(\varepsilon(k)-\mu)\right]}. \quad (A.8)$$

In the relativistic case no simplification of these formulas is possible.

*Thermodynamics (Nonrelativistic).* In the nonrelativistic case, $\varepsilon_{\text{nonrel}}(k) = k^2/(2M)$, Eq. (A.8) is given by

$$A(T,\mu) = -\frac{1}{3\pi^2}\int_0^\infty dk\, \frac{k^4}{M} n_k = \alpha T^{5/2}\text{Li}_{5/2}(x), \quad (A.9)$$

where $x = -\exp(\mu/T)$, $\alpha = 2^{-1/2}(M/\pi)^{3/2}$, and $\text{Li}_\nu(x) = \sum_{k=1}^\infty k^{-\nu} x^k$ is the polylogarithm of index $\nu$, which has the integral representation [5]

$$\text{Li}_\nu(x) = -\frac{1}{\Gamma(\nu)\, 2^{\nu-1}\, T^\nu M^{\nu-3}}\int_0^\infty dk\, k^{2\nu-1}\, n_k \xrightarrow{T\to 0} -\frac{1}{\Gamma(\nu)\, 2^{2\nu-1}\, T^\nu M^{\nu-3}}(k_F)^{2\nu}, \quad (A.10)$$

## A. Appendix

with $\Gamma(\nu)$ the (Euler) gamma function, which for half-integer indices is given by $\Gamma(1/2) = \sqrt{\pi}$, $\Gamma(1/2 - n) = \sqrt{\pi}(2n-1)!!/2^n$, and $\Gamma(1/2-n) = \sqrt{\pi}(-2)^n/(2n-1)!!$, where $n \in \mathbb{N}$. The nonrelativistic expression for the density is given by

$$\rho(T,\mu) = \frac{1}{3\pi^2}\int_0^\infty dk\, k^2 \frac{\partial n_k}{\partial \mu_{n/p}} = \frac{1}{\pi^2}\int_0^\infty dk\, k^2 n_k = -\alpha\, T^{3/2} \text{Li}_{3/2}(x), \qquad \text{(A.11)}$$

where we have used $\partial n_k/\partial \mu = -(M/k)\,\partial n_k/\partial k$ and partial integration. The polylogarithmic expression can also be obtained directly from $\partial \text{Li}_\nu(x)/\partial x = x^{-1}\text{Li}_{\nu-1}(x)$. For densities $\rho < -\alpha T^{3/2}\text{Li}_{3/2}(-1) \simeq 3.6 \times 10^{-4}\,(T/\text{MeV})^{3/2}\,\text{fm}^{-3}$ the chemical potential $\mu$ is negative, and its asymptotic behavior in the limit of vanishing density is given by

$$\mu(T,\rho) \xrightarrow{\rho \to 0} T \ln\left(\alpha^{-1} T^{-3/2} \rho\right), \qquad \text{(A.12)}$$

which, using $\text{Li}_\nu(x) \xrightarrow{x \to 0} x$, follows from inverting Eq. (A.11) in the limit $\mu \to -\infty$. From Eqs. (A.9) and (A.11) the polylogarithmic expression for the free energy density is given by

$$F(T,\mu) = -\alpha T^{5/2}\Big(\ln(-x)\,\text{Li}_{3/2}(x) - \text{Li}_{5/2}(x)\Big) \xrightarrow{T \to 0} \frac{k_F^5}{10\pi^2 M}. \qquad \text{(A.13)}$$

with $\mu \xrightarrow{T \to 0} k_F^2/(2M)$. Note that Eq. (A.12) implies that the free energy per particle $\bar{F}(T \neq 0, \rho)$ diverges logarithmically, $\sim \ln(\rho)$, for $\rho \to 0$. The internal energy density is given by $E = F - T(\partial F/\partial T)_\rho = -\frac{3\alpha T^{5/2}}{2}\text{Li}_{5/2}(x) \xrightarrow{T \to 0} E_0(k_F)$; the zero-density limit of the internal energy per particle is $\bar{E} \xrightarrow{\rho \to 0} 3T/2$, which corresponds to the EoS of a nonrelativistic classical ideal gas.

### A.1.2. Relativistic Corrections[2]

We now apply the different forms of many-body perturbation theory (zero-temperature, grand-canonical, canonical) to the set of Hamiltonians $\mathcal{T}_{\text{nonrel}}$ and $\mathcal{T}_{\text{corr}}$. Note that *only the "bare" versions of MBPT are applicable for this particular problem*. The *partially* renormalized versions involve the renormalized single-particle energies $\varepsilon_{\text{nonrel}(n)} = \varepsilon_{\text{nonrel}} + \sum_{\nu=1}^{n}\varepsilon_{\text{corr}(\nu)}$ in the Fermi-Dirac distributions. This is problematic, concerning the behavior for large values of $k$ (the state-sums are unrestricted, and the correction terms $\varepsilon_{\text{corr}(\nu)}$ are well-behaved only for sufficiently small momenta) of the nonrelativistic expansion $\varepsilon_{\text{rel}}(k) \simeq \varepsilon_{\text{nonrel}(n)}(k)$, which is well-behaved (for large $k$) only for $n = 0$; in particular, $\varepsilon_{\text{nonrel}(n)}(k) \xrightarrow{k \to \infty} -\infty$ for odd $n$, leading to divergent results.

***Zero-Temperature Formalism.*** The zero-temperature formalism [applied to the perturbation given by Eq. (A.1)] is equivalent to the expansion of Eq. (A.4) in powers of $k_F$, i.e.,

$$E_{0;\text{rel}}(k_F) = \underbrace{\frac{1}{10\pi^2 M}(k_F)^5}_{E_{0;\text{nonrel}}} - \underbrace{\frac{1}{56\pi^2 M^3}(k_F)^7}_{E_{0;\text{corr}(1)}} + \underbrace{\frac{1}{144\pi^2 M^5}(k_F)^9}_{E_{0;\text{corr}(2)}} - \underbrace{\frac{5}{1408\pi^2 M^7}(k_F)^{11}}_{E_{0;\text{corr}(3)}} + \dots. \qquad \text{(A.14)}$$

---

[2] We note here that the first-order canonical relativistic correction term $F_{\text{corr}(1)}$ was first constructed (in an *ad hoc* fashion) in Ref. [150]. Note also that the description of the implementation of $F_{\text{corr}(1)}$ [cf. Eq. (A.20)] in the Kohn-Luttinger method is somewhat messed up in [Phys. Rev. C, 89 (2014), p. 064009] (in particular, Eqs. (4) and (5) of that reference are not consistent with each other).

## A. Appendix

The nonrelativistic approximation of order $n$ is then given by $E_{0;\text{nonrel}(n)}(k_F) := E_{0;\text{nonrel}}(k_F) + \sum_{\nu=1}^{n} E_{0;\text{corr}(\nu)}(k_F)$.

***Grand-Canonical MBPT.*** The grand-canonical perturbation series corresponds to the nonrelativistic expansion of the momentum distribution $n_k^{\text{rel}} = 1/[1 + \exp(\beta((k^2+M^2)^{1/2} - M - \mu))]$ and the term $\partial \varepsilon_{\text{rel}}(k)/\partial k$ in Eq. (A.8). The nonrelativistic expansion of the relativistic momentum distribution is given by

$$n_k^{\text{rel}} = n_k + \frac{\partial n_k}{\partial \varepsilon_{\text{nonrel}}} \sum_{\nu=1}^{\infty} \varepsilon_{\text{corr}(\nu)} + \frac{1}{2!} \frac{\partial^2 n_k}{\partial^2 \varepsilon_{\text{nonrel}}} \left( \sum_{\nu=1}^{\infty} \varepsilon_{\text{corr}(\nu)} \right)^2 + \ldots. \quad (A.15)$$

Inserting this expansion and the expansion of $\partial \varepsilon_{\text{rel}}(k)/\partial k$ into Eq. (A.8) leads to the nonrelativistic expansion of the noninteracting grand-canonical potential $A_{\text{nonrel}(n)}(T,\mu) = \mathcal{A}(T,\mu) + \sum_{\nu=1}^{n} A_{\text{corr}(\nu)}(T,\mu)$. The first two correction terms are given by[3]

$$A_{\text{corr}(1,2)}(T,\mu) \in \left\{ \frac{15\alpha T^{7/2}}{8M} \text{Li}_{7/2}(x), \frac{105\alpha T^{9/2}}{128 M^2} \text{Li}_{9/2}(x) \right\} \quad (A.16)$$

The corresponding corrections to the expression for the particle density are obtained either via $\rho_{\text{corr}(n)}^{\text{G.C.}} = -\partial A_{\text{corr}(n)}/\partial \mu$, or directly by expanding the Fermi-Dirac distributions in Eq. (A.6), i.e.,

$$\rho_{\text{corr}(1,2)}^{\text{G.C.}}(T,\mu) \in \left\{ -\frac{15\alpha T^{5/2}}{8M} \text{Li}_{5/2}(x), -\frac{105\alpha T^{7/2}}{128 M^2} \text{Li}_{7/2}(x) \right\} \xrightarrow{T \to 0} \left\{ \frac{k_F^5}{8\pi^2 M^2}, \frac{k_F^7}{128 \pi^2 M^4} \right\}, \quad (A.17)$$

with $\mu \xrightarrow{T \to 0} \varepsilon_{\text{nonrel}}(k_F) = k_F^2/(2M)$. The corrections to the free energy density are given by

$$F_{\text{corr}(1)}^{\text{G.C.}}(T,\mu) = -\frac{15\alpha T^{7/2}}{8M} \left( \ln(-x) \text{Li}_{5/2}(x) - \text{Li}_{7/2}(x) \right) \xrightarrow{T \to 0} \frac{5}{112 \pi^2 M^3} (k_F)^7, \quad (A.18)$$

$$F_{\text{corr}(2)}^{\text{G.C.}}(T,\mu) = -\frac{105\alpha T^{9/2}}{128 M^2} \left( \ln(-x) \text{Li}_{7/2}(x) - \text{Li}_{9/2}(x,) \right) \xrightarrow{T \to 0} \frac{7}{2304 \pi^2 M^5} (k_F)^9. \quad (A.19)$$

Note that $F_{\text{corr}(n)}^{\text{G.C.}}(T = 0, \mu) \neq E_{0;\text{corr}(n)}(k_F)$, but this should *not* be interpreted offhand as a deficiency of the grand-canonical approach towards the construction of relativistic corrections (see Table A.1 and the discussion below).

***Canonical MBPT.*** The canonical perturbation series for the set of Hamiltonians $\mathcal{T}_{\text{nonrel}}$ and $\mathcal{T}_{\text{corr}}$ is given by $F(T,\tilde{\mu}) = \mathcal{F}(T,\tilde{\mu}) + \sum_{n=1}^{\infty} \lambda^n F_{\text{corr}(n)}(T,\tilde{\mu})$, where the auxiliary chemical potential is in one-to-one correspondence with the density via $\rho(T,\tilde{\mu}) = -\alpha T^{3/2} \text{Li}_{3/2}(\tilde{x})$, with $\tilde{x} = -\exp(\beta\tilde{\mu})$. The leading two terms are given by $\mathcal{F}(T,\tilde{\mu}) = F_{\text{nonrel}}(T,\tilde{\mu})$ and $F_{\text{corr}(1)}(T,\tilde{\mu}) = A_{\text{corr}(1)}(T,\tilde{\mu})$, i.e.,

$$\mathcal{F}(T,\tilde{\mu}) = -\alpha T^{5/2} \left( \ln(-\tilde{x}) \text{Li}_{3/2}(\tilde{x}) - \text{Li}_{5/2}(\tilde{x}) \right), \qquad F_{\text{corr}(1)}(T,\tilde{\mu}) = \frac{15\alpha T^{7/2}}{8M} \text{Li}_{7/2}(\tilde{x}), \quad (A.20)$$

where $\mathcal{F}(T,\tilde{\mu}) \xrightarrow{T \to 0} E_{0;\text{nonrel}}(k_F)$ and $F_{\text{corr}(1)}(T,\tilde{\mu}) \xrightarrow{T \to 0} E_{0;\text{corr}(1)}(k_F)$, with $\tilde{\mu} \xrightarrow{T \to 0} \varepsilon_{\text{nonrel}}(k_F) = k_F^2/(2M)$. The corrections beyond first order are given by $F_{\text{corr}(\nu)}(T,\tilde{\mu}) = A_{\text{corr}(\nu)}(T,\tilde{\mu}) + F_{\nu,\text{h.-c.}}(T,\tilde{\mu})$, with $F_{\nu,\text{h.-c.}}(T,\tilde{\mu})$ the higher-cumulant contribution of order $\nu$. The second correction is given by

$$F_{\text{corr}(2)}(T,\tilde{\mu}) = \frac{105\alpha T^{9/2}}{128 M^2} \text{Li}_{9/2}(\tilde{x}) - \frac{225\alpha T^{9/2}}{128 M^2} \frac{\text{Li}_{5/2}(\tilde{x}) \text{Li}_{5/2}(\tilde{x})}{\text{Li}_{1/2}(\tilde{x})} \xrightarrow{T \to 0} E_{0;\text{corr}(2)}(k_F). \quad (A.21)$$

---

[3] For example, $A_{\text{corr}(2)}(T,\mu) = \frac{1}{3\pi^2} \int_0^{\infty} dk\, k^3 \left( n_k \frac{\partial \varepsilon_{\text{corr}(2)}}{\partial k} + \frac{\partial n_k}{\partial \varepsilon_{\text{nonrel}}} \varepsilon_{\text{corr}(1)} \frac{\partial \varepsilon_{\text{corr}(1)}}{\partial k} + \left( \frac{\partial n_k}{\partial \varepsilon_{\text{nonrel}}} \varepsilon_{\text{corr}(2)} + \frac{1}{2!} \frac{\partial^2 n_k}{\partial \varepsilon_{\text{nonrel}}^2} \varepsilon_{\text{corr}(1)}^2 \right) \frac{\partial \varepsilon_{\text{nonrel}}}{\partial k} \right) = -\frac{1}{128\pi^2} \int_0^{\infty} dk \frac{k^8}{M^5} n_k = \frac{105\alpha T^{9/2}}{128 M^2} \text{Li}_{9/2}(x)$.

## A. Appendix

For the corresponding corrections to the internal energy density one obtains the expressions

$$E_{\text{corr}(1)}(T,\tilde{\mu}) = -\frac{75\alpha T^{7/2}}{16M}\text{Li}_{7/2}(\tilde{x}) + \frac{45\alpha T^{7/2}}{16M}\frac{\text{Li}_{5/2}(\tilde{x})\text{Li}_{3/2}(\tilde{x})}{\text{Li}_{1/2}(\tilde{x})}, \quad (A.22)$$

$$E_{\text{corr}(2)}(T,\tilde{\mu}) = \frac{525\alpha T^{9/2}}{256M^2}\text{Li}_{9/2}(\tilde{x}) - \frac{315\alpha T^{9/2}}{256M^2}\frac{\text{Li}_{7/2}(\tilde{x})\text{Li}_{3/2}(\tilde{x})}{\text{Li}_{1/2}(\tilde{x})} - \frac{1125\alpha T^{9/2}}{256M^2}\frac{\text{Li}_{5/2}(\tilde{x})\text{Li}_{5/2}(\tilde{x})}{\text{Li}_{1/2}(\tilde{x})}$$
$$+ \frac{675\alpha T^{9/2}}{256M^2}\frac{\text{Li}_{3/2}(\tilde{x})}{\text{Li}_{1/2}(\tilde{x})}\left(\frac{2\text{Li}_{5/2}(\tilde{x})\text{Li}_{3/2}(\tilde{x})}{\text{Li}_{1/2}(\tilde{x})} - \frac{\text{Li}_{5/2}(\tilde{x})\text{Li}_{5/2}(\tilde{x})\text{Li}_{-1/2}(\tilde{x})}{\text{Li}_{1/2}(\tilde{x})\text{Li}_{1/2}(\tilde{x})}\right). \quad (A.23)$$

The zero-density limits of the corrections to the internal energy per particle are given by

$$\bar{E}_{\text{corr}(1,2)}(T,\rho) \xrightarrow{\rho\to 0} \left\{\frac{15T^2}{8M}, \frac{-15T^3}{16M^2}\right\}. \quad (A.24)$$

This should be compared with the expansion in powers of $T$ of the internal energy per particle of a relativistic classical ideal gas, i.e.,

$$\bar{E}_{\text{classical}}(T) = 3T + M\frac{K_1(M/T)}{K_2(M/T)} - M = \frac{3T}{2} + \frac{15T^2}{8M} - \frac{15T^3}{8M^2} + \frac{135T^4}{128M^3} + \ldots \quad (A.25)$$

where $K_{1,2}(M/T)$ are modified Bessel functions of the second kind. One sees that the zero-density limit of the first correction reproduces the first-order term in the classical expansion, but the second-order term is off by a factor $1/2$ (the origin of this feature is not clear).

*Numerical Analysis.* The numerical results obtained for the noninteracting contribution to the chemical potential $\mu(T,\rho,\delta=0)$ and the free energy per particle $\bar{F}(T,\rho,\delta=0)$ are shown for $T=25\,\text{MeV}$ and $\rho=(0.04,0.16,0.32)\,\text{fm}^{-3}$ in Table A.1. For comparison, we also show the results for unreasonably large densities $\rho=(2.0,4.0)\,\text{fm}^{-3}$. One sees that even at very large densities, all of the various higher-order nonrelativistic approximations systematically improve upon the nonrelativistic approximation without additional correction terms. Somewhat surprisingly (considering the results of Sec. 2.5), compared to the canonical correction terms the grand-canonical ones ("G.C.") lead to results that are closer to the relativistic results (in particular, the error of the second-order grand-canonical approximation is negligible even at the highest densities). This however is not in conflict with the analysis of Sec. 2.5, because of the additional expansion of the "perturbation Hamiltonian" $\mathcal{T}_{\text{corr}}$. In fact, using $\mathcal{T}_{\text{corr}} = \lambda\sum_k(\varepsilon_{\text{rel}}(k) - \varepsilon_{\text{nonrel}}(k))a_k^\dagger a_k$ as the perturbation Hamiltonian one finds that the canonical perturbation series is indeed closer to the exact (relativistic) results as compared the grand-canonical one [but in much weaker form compared to the calculations of Sec. 2.5.5 (note that the analysis in terms of the mean-field shift does not apply here)]. For the scales of interest, the first-order canonical correction term leads to results that are reasonably close to the relativistic ones. Because a canonical correction term is much easier to incorporate in the correlation-bond formalism, in this thesis we have computed the noninteracting contribution in MBPT as

$$\boxed{F_{\text{nonint}}(T,\tilde{\mu}) := F_{\text{nonrel}}(T,\tilde{\mu}) + F_{\text{corr}}(T,\tilde{\mu})} \quad (A.26)$$

where $F_{\text{nonrel}} := \mathcal{F}$ and $F_{\text{corr}} := F_{\text{corr}(1)}$ are given by Eq. (A.20). The results obtained from this approximation as well as the relativistic and the nonrelativistic ones are plotted as functions of density in Fig. A.1 for different temperatures. Notably, the deviations between the relativistic results and the ones obtained from Eq. (A.26) increase only very little with temperature.

| $\rho$ (fm$^{-3}$) | 0.04 | 0.16 | 0.32 | 2.0 | 4.0 |
|---|---|---|---|---|---|
| $\mu_{\text{rel}}$ (MeV) | -25.72 | 16.95 | 44.48 | 176.9 | 271.8 |
| $\mu_{\text{nonrel}}$ (MeV) | -24.27 | 19.11 | 47.72 | 195.8 | 313.4 |
| $\mu_{\text{nonrel}(1)}^{\text{G.C.}}$ (MeV) | -25.70 | 16.97 | 44.53 | 177.3 | 273.0 |
| $\mu_{\text{nonrel}(2)}^{\text{G.C.}}$ (MeV) | -25.72 | 16.95 | 44.48 | 176.9 | 271.8 |
| $\mu_{\text{nonrel}(1)}$ (MeV) | -25.75 | 16.87 | 44.29 | 173.2 | 258.9 |
| $\mu_{\text{nonrel}(2)}$ (MeV) | -25.72 | 16.95 | 44.50 | 177.9 | 277.1 |
| $\bar{F}_{\text{rel}}$ (MeV) | -52.09 | -13.38 | 9.18 | 102.1 | 164.5 |
| $\bar{F}_{\text{nonrel}}$ (MeV) | -50.75 | -11.70 | 11.37 | 111.4 | 184.1 |
| $\bar{F}_{\text{nonrel}(1)}^{\text{G.C.}}$ (MeV) | -52.07 | -13.36 | 9.22 | 102.3 | 165.1 |
| $\bar{F}_{\text{nonrel}(2)}^{\text{G.C.}}$ (MeV) | -52.10 | -13.38 | 9.18 | 102.2 | 164.5 |
| $\bar{F}_{\text{nonrel}(1)}$ (MeV) | -52.11 | -13.42 | 9.10 | 100.8 | 159.0 |
| $\bar{F}_{\text{nonrel}(2)}$ (MeV) | -52.09 | -13.38 | 9.19 | 102.4 | 166.0 |

**Table A.1.:** Noninteracting contribution to the chemical potential $\mu(T,\rho)$ and the free energy per particle $\bar{F}(T,\rho)$, respectively, of isospin-symmetric nuclear matter at $T = 25$ MeV for different densities. The relativistic results are compared to the ones obtained from the various nonrelativistic approximations, see text for details.

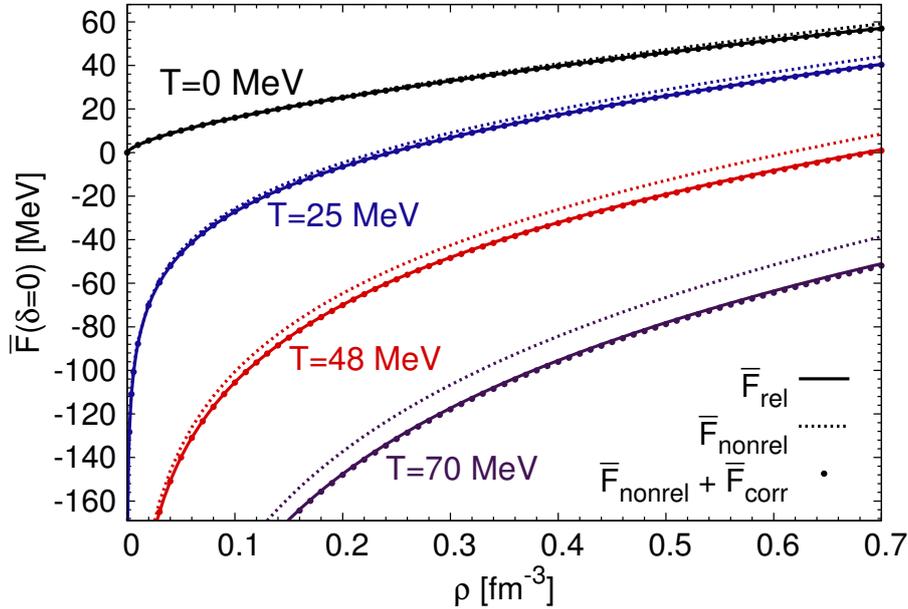

**Figure A.1.:** Noninteracting contribution to $\bar{F}(T,\rho,\delta=0)$ for $T=(0,25,48,70)$ MeV. The solid lines show the relativistic results ($\bar{F}_{\text{rel}}$), the dashed lines the nonrelativistic ones ($\bar{F}_{\text{nonrel}}$), and the dots the nonrelativistic results with the first-order canonical correction term included ($\bar{F}_{\text{nonrel}}+\bar{F}_{\text{corr}}$).

## A.2. Realization of Chiral Symmetry

An elementary aspect concerning the construction of the chiral effective Lagrangian $\mathscr{L}_{\chi\text{EFT}}$ is the transformation behavior of pions and nucleons with respect to the chiral symmetry group $G_\chi = \text{SU}(2)_L \times \text{SU}(2)_R$. In the following we briefly describe these transformation properties and construct the leading terms of $\mathscr{L}_{\chi\text{EFT}}$, following mostly Refs. [356, 88, 69, 163, 283], cf., e.g., also [413, 115] for additional details.

### A.2.1. Pions

A QFT with a "spontaneously broken" symmetry group $G$ has the property that the quantum fields do not transform according a linear representation of $G$; instead, the symmetry is realized nonlinearly. In the case of the chiral symmetry group $G_\chi$, the fact that chiral transformations of the three pion fields $\pi^+, \pi^-, \pi^0$ cannot be realized by a linear representation of $G_\chi$ can be seen as follows. The Lie algebra of $G_\chi = \text{SU}(2)_L \times \text{SU}(2)_R$ is isomorphic to that of SO(4), whose smallest faithful representation is four-dimensional. The triplet of pion fields however amounts to only three coordinates. Instead of a linear representation, one must therefore consider a general *realization* of $G_\chi$ of the form

$$\boldsymbol{\pi} \xrightarrow{g \in G_\chi} \phi_g(\boldsymbol{\pi}), \qquad \phi_g(0) = 0 \iff g \in \text{SU}(2)_V. \tag{A.27}$$

As shown in [88] the condition that only the (diagonal) subgroup $\text{SU}(2)_V$ leaves the *origin* 0 invariant guarantees that the group realization becomes linear when restricted to $\text{SU}(2)_V$ (called the *stability group* in this context). Under the full $\text{SU}(2)_L \times \text{SU}(2)_R$ the pion fields transform nonlinearly.

*General Discussion.* We consider a physical system described by an effective Lagrangian that is invariant under a symmetry group $G$, while the vacuum of the system is invariant only under the subgroup $H$. The Nambu-Goldstone bosons associated with this spontaneous symmetry breaking—we call them pions—are described by a multi-component vector $\boldsymbol{\pi}$ whose components are coordinates on a manifold $M_\pi$. A transformation of the points on $M_\pi$ under the symmetry $G$ is defined by a mapping $\phi$ which uniquely associates with each pair $(g, \boldsymbol{\pi}) \in G \times M_\pi$ an element $\phi_g(\boldsymbol{\pi}) \in M_\pi$ with the properties

- $\phi_e(\boldsymbol{\pi}) = \boldsymbol{\pi} \quad \forall \boldsymbol{\pi} \in M_\pi, \quad e \equiv$ identity element of $G$.

- $\phi_{g_1} \circ \phi_{g_2}(\boldsymbol{\pi}) = \phi_{g_1 g_2}(\boldsymbol{\pi}) \quad \forall g_1, g_2 \in G, \quad \forall \boldsymbol{\pi} \in M_\pi$.

This mapping defines an *operation* of the group $G$ on $M_\pi$, which will however, because of the missing linearity condition $\phi_g(\lambda \boldsymbol{\pi}) = \lambda \phi_g(\boldsymbol{\pi})$, not constitute a representation of $G$, but only a *realization* of $G$ on $M_\pi$. As mentioned above, we require in addition that the origin be invariant under transformations restricted to the subgroup $H$:

$$\phi_h(0) = 0 \quad \forall h \in H. \tag{A.28}$$

Following [356], we now show that there exists an isomorphism $\xi$ between $M_\pi$ and the set of all *left cosets* $\{gH \mid g \in G\}$ (also known as the quotient space $G/H$), i.e.,

$$\xi : M_\pi \to G/H, \quad \boldsymbol{\pi} \mapsto gH, \tag{A.29}$$

## A. Appendix

where for one element $g$ of $G$, the set $gH = \{gh \mid h \in H\}$ defines the left coset of $g$, one element of $G/H$. From Eq. (A.28) it follows that

$$\phi_{gh}(0) = \phi_g \circ \phi_h(0) = \phi_g(\phi_h(0)) = \phi_g(0) \qquad \forall \; g \in G, \; h \in H. \tag{A.30}$$

Moreover, the mapping $\phi$ is injective with respect to the cosets, which can be seen by considering two elements $g$ and $g'$ of $G$, with $g' \notin gH$. Assuming that $\phi_g(0) = \phi_{g'}(0)$ leads to

$$0 = \phi_e(0) = \phi_{g^{-1}g}(0) = \phi_{g^{-1}}(\phi_g(0)) = \phi_{g^{-1}}(\phi_{g'}(0)) = \phi_{g^{-1}g'}(0). \tag{A.31}$$

By Eq. (A.28) this implies that $g^{-1}g' \in H$, or equivalently $g' \in gH$, which contradicts the assumption. This shows that the mapping

$$G/H \to M_\pi, \quad \tilde{g}h \mapsto \phi_{\tilde{g}h}(0) \qquad \forall \; \tilde{g} \in G, \; h \in H \tag{A.32}$$

is bijective, and we can for each $\boldsymbol{\pi} \in M_\pi$ identify a corresponding coset $\tilde{g}H$ by the relation $\phi_{\tilde{g}h}(0) = \boldsymbol{\pi}$. The isomorphism between pion fields and left cosets is thus given by:

$$\xi(\boldsymbol{\pi}) = \tilde{g}H \quad \Leftrightarrow \quad \phi_{\tilde{g}h}(0) = \boldsymbol{\pi}. \tag{A.33}$$

The transformation behaviour of the pion fields is now uniquely determined (up to an appropriate choice of parametrization) by the transformation of its coset representation, i.e.,

$$\begin{array}{ccc} M_\pi \xrightarrow{\phi_g} M_\pi & & \boldsymbol{\pi} \xmapsto{\phi_g} \boldsymbol{\pi}' \\ \xi \downarrow \quad \downarrow \xi & & \xi \downarrow \quad \downarrow \xi \\ G/H \xrightarrow{g} G/H & & \tilde{g}H \xmapsto{g} g\tilde{g}H \end{array} \quad . \tag{A.34}$$

*Application to Chiral Symmetry.* In the case of the chiral symmetry group $G_\chi$ we have

$$G_\chi = \mathrm{SU}(2)_L \times \mathrm{SU}(2)_R = \{(L,R) \mid L \in \mathrm{SU}(2), \; R \in \mathrm{SU}(2)\}, \tag{A.35}$$
$$H = \mathrm{SU2}_V = \{(V,V) \mid V \in \mathrm{SU}(2)\}, \tag{A.36}$$

where $(X,Y)$ specifies the transformation behaviour of the left-handed, $q_L \to X q_L$, and the right-handed quark fields, $q_R \to Y q_R$. The left coset of an element $\tilde{g} \in G_\chi$, $\tilde{g}H = \{(\tilde{L}V, \tilde{R}V) \mid V \in H\}$, can be uniquely characterized through the SU2-matrix $U = \tilde{R}\tilde{L}^\dagger$:

$$(\tilde{L}V, \tilde{R}V) = (\tilde{L}V, \tilde{R}\tilde{L}^\dagger \tilde{L}V) = (1, \tilde{R}\tilde{L}^\dagger)\underbrace{(\tilde{L}V, \tilde{L}V)}_{\in H} \quad \Rightarrow \quad \tilde{g}H = (1, U)H. \tag{A.37}$$

The transformation of $U$ under $g = (L, R)$ is given by multiplication in the left coset

$$g\tilde{g}H = (L, RU)H = (1, RUL^\dagger)(L, L)H = (1, RUL^\dagger)H. \tag{A.38}$$

Hence, $U \xrightarrow{g} U' = RUL^\dagger$, and it suffices to specify an isomorphic mapping of the pion fields to the space of unitary matrices $U$ with determinant one. An often used parametrization is[4]

$$U = \exp\left(\frac{i\boldsymbol{\tau} \cdot \boldsymbol{\pi}}{f_\pi}\right), \tag{A.39}$$

---

[4] The general form of the parametrization of $U$ by pion field is given by $U = 1 + \frac{i}{f_\pi}(\boldsymbol{\tau} \cdot \boldsymbol{\pi}) - \frac{1}{2f_\pi^2}\pi^2 - \frac{i\alpha}{f_\pi^3}(\boldsymbol{\tau} \cdot \boldsymbol{\pi})^3 + \frac{8\alpha - 1}{8f_\pi^4}\pi^4 + \ldots$, with $\alpha$ an arbitrary coefficient [283]. The "exponential" parametrization given by Eq. (A.39) is equivalent to $\alpha = 1/6$. Another popular choice is the parametrization $U = (\sqrt{1 - \pi^2} + i\boldsymbol{\tau} \cdot \boldsymbol{\pi})/f_\pi$, which corresponds to $\alpha = 0$.

## A. Appendix

with $\boldsymbol{\tau} = (\tau^1, \tau^2, \tau^3)$ are Pauli matrices, $f_\pi$ is pion decay constant, and

$$\boldsymbol{\pi} = \begin{pmatrix} \frac{1}{\sqrt{2}}(\pi^+ + i\pi^-) \\ \frac{i}{\sqrt{2}}(\pi^+ - i\pi^-) \\ \pi^0 \end{pmatrix} \tag{A.40}$$

the vector which collects the pion fields. The freedom of choice regarding the parametrization of $U$ by pion fields has an effect only on (unobservable) off-shell amplitudes, but physical on-shell amplitudes are invariant under a change of parametrization [130, 283]. For a given parametrization the group operation $\phi$ is given by

$$\begin{array}{ccc} M_\pi \xrightarrow{\phi_g} M_\pi & & \boldsymbol{\pi} \xmapsto{\phi_g} \boldsymbol{\pi}' \\ \downarrow \xi \quad \downarrow \xi & , & \downarrow \xi \quad \downarrow \xi \\ M_U \xrightarrow{g} M_U & & U \xmapsto{g} RUL^\dagger \end{array} \tag{A.41}$$

where $M_U$ is the space of $U$-matrices (which is not a vector space), i.e.,

$$M_U = \{U : \mathbb{M} \to \mathrm{SU}(2) \,\big|\, x \mapsto U(x)\}, \tag{A.42}$$

where $\mathbb{M}$ denotes Minkowski space. For transformations under the subgroup $\mathrm{SU}(2)_V$ the pions should transform linearly. Using the parametrization given by Eq. (1.21), this can be checked by expanding $U$ in a power series:

$$VUV^\dagger = V \sum_{\alpha=0}^\infty \left(\frac{i\boldsymbol{\tau} \cdot \boldsymbol{\pi}}{f_\pi}\right)^\alpha V^\dagger = \sum_{\alpha=0}^\infty \left(\frac{iV\boldsymbol{\tau} \cdot \boldsymbol{\pi} V^\dagger}{f_\pi}\right)^\alpha. \tag{A.43}$$

Hence it is $\boldsymbol{\tau} \cdot \boldsymbol{\pi} \to V(\boldsymbol{\tau} \cdot \boldsymbol{\pi})V^\dagger$, which is a linear representation of $\mathrm{SU}(2)_V$. That pions transform nonlinearly under $g \in G_\chi$ with $g \notin \mathrm{SU}(2)_V$ can for example be seen by considering the transformation $g_A \in G_\chi$ with $g_A \notin \mathrm{SU}(2)_V$ defined in terms of left- and right-handed quark fields $\psi_{L/R} = \frac{1}{2}(1 \mp \gamma_5)\psi$ as

$$\begin{pmatrix} \psi_L \\ \psi_R \end{pmatrix} \xrightarrow{g_A} A \begin{pmatrix} \psi_L \\ \psi_R \end{pmatrix}, \tag{A.44}$$

where $A = \exp\left(-i\Theta \frac{\boldsymbol{\tau}}{2}\gamma_5\right)$. Using the Weyl representation of Dirac spinors, this can be decomposed as

$$\psi_L \xrightarrow{g_A} \exp\left(-i\Theta \frac{\boldsymbol{\tau}}{2}\right)\psi_L \equiv A_L \psi_L, \qquad \psi_R \xrightarrow{g_A} \exp\left(i\Theta \frac{\boldsymbol{\tau}}{2}\right)\psi_R \equiv A_R \psi_R. \tag{A.45}$$

Hence, it is $U \to A_L U A_R^\dagger$ under that particular transformation, and by expanding $U$ as in Eq. (A.43) we do not obtain a linear transformation of pion fields, but instead a nonlinear one. The pionic part of the relativistic chiral Lagrangian $\mathcal{L}_{\pi\pi}$ is the organized in powers of $\partial_\mu U$, with the leading-order (chiral dimension $d = 2$) terms given by [157, 115]

$$\mathcal{L}_{\pi\pi}^{(d=2)} = \frac{f_\pi^4}{4} \mathrm{Tr}\left[\partial_\mu U \partial^\mu U^\dagger + m_\pi^2(U + U^\dagger)\right], \tag{A.46}$$

where the first term constitutes the most general $G_\chi$ invariant Lagrangian with minimal number of derivatives, and the second term (the mass term) breaks $G_\chi$ explicitly. Expanding $U$ in a power series and leaving out constant terms yields the purely pionic part of $\mathcal{L}_{\chi\mathrm{EFT}}^{A=0}$ [first line in Eq. (1.21)].

# A. Appendix

## A.2.2. Nucleons

There are many possibilities to realize chiral symmetry for the nucleon field (cf. [163] pp. 93-96). They all lead to the same physics. One particularly convenient choice is given by

$$\Psi \to K(L, R, U)\Psi = \sqrt{LU^\dagger R^\dagger} R \sqrt{U} \Psi, \tag{A.47}$$

where $\Psi = (p, n)$ is the relativistic nucleon field, with isospin components given by the proton Dirac field $p$ and the neutron Dirac field $n$. Because $K$ depends on the pion-field matrix $U(\boldsymbol{\pi}(x))$ the above equation defines a local transformation law. To see more explicitly how $\Psi$ transforms under $SU(2)_L \times SU(2)_R$ we consider an infinitesimal transformation $K = \exp(i\boldsymbol{\gamma}\boldsymbol{\tau})$, and use the parametrizations $L = \exp(i(\boldsymbol{\alpha} - \boldsymbol{\beta})\boldsymbol{\tau})$ and $R = \exp(i(\boldsymbol{\alpha} + \boldsymbol{\beta})\boldsymbol{\tau})$, where $\boldsymbol{\alpha}$ and $\boldsymbol{\beta}$ are infinitesimal. Using the Baker-Campbell-Hausdorff formula $e^x e^y = e^z$, where $z$ is given by

$$z = x + y + \frac{1}{2}[x, y] - \frac{1}{12}[[x, y], x - y] - \frac{1}{24}\left[[[x, y], x], y\right] + \ldots, \tag{A.48}$$

we find the following relation:

$$\boldsymbol{\gamma} = \boldsymbol{\alpha} - \frac{\boldsymbol{\beta} \times \boldsymbol{\pi}}{2 f_\pi} + \frac{((\boldsymbol{\beta} \times \boldsymbol{\pi}) \times \boldsymbol{\pi}) \times \boldsymbol{\pi}}{6 f_\pi^3} + O(\pi^4). \tag{A.49}$$

This shows that for a general transformation under $SU(2)_L \times SU(2)_R$ the transformed nucleon field is a nonlinear function of pion fields. In contrast, for a transformation under $SU(2)_V$ with $\boldsymbol{\beta} = 0$ we obtain $K = V$ (up to order $\pi^3$, for infinitesimal transformations). This can be shown to be true also in the general case by observing that $\zeta_1 := \sqrt{VU^\dagger V^\dagger} \in SU(2)$ and $\zeta_2 := V\sqrt{U^\dagger}V^\dagger \in SU(2)$ both lead to the same expression, $\zeta_{1/2}^2 := VU^\dagger V^\dagger \in SU(2)$, when squared. The square root of an $SU(2)$-matrix is unique up to a sign, therefore it is $\zeta_1 = \pm\zeta_2$. Using the parametrization $V = \exp(i\boldsymbol{\alpha}\boldsymbol{\tau})$ one can show by explicit calculation that it is indeed $\zeta_1 = \zeta_2$. We therefore have

$$K(V, V, U) = V\sqrt{U^\dagger}V^\dagger V \sqrt{U} = V. \tag{A.50}$$

Hence, $\Psi$ transforms linearly as an isospin doublet under $SU(2)_V$. Flavor symmetry between up- and down-quarks has become isospin symmetry between protons and neutrons. Since nucleons transform locally under $SU(2)_L \times SU(2)_R$, one needs a chirally covariant derivative:

$$\mathcal{D}_\mu = \partial_\mu + \Gamma_\mu, \qquad \Gamma_\mu = \frac{i}{4 f_\pi^2} \boldsymbol{\tau} \cdot (\boldsymbol{\pi} \times \partial_\mu \boldsymbol{\pi}) + O(\pi^4), \tag{A.51}$$

where $\Gamma_\mu$ is the so-called *chiral connection*. The leading terms in the relativistic pion-nucleon Lagrangian have chiral dimension $d = 1$ and are given by [158]:

$$\mathcal{L}_{\pi N}^{(d=1)} = \bar{\Psi}\left(i\gamma^\mu \mathcal{D}_\mu - M + \frac{g_A}{2}\gamma^\mu \gamma_5 u_\mu\right)\Psi, \tag{A.52}$$

where $M$ is the nucleon mass (in the chiral limit), $\Psi$ is the Dirac field representing the nucleon, and $g_A$ is the axial-vector strength, which can be measured in neutron $\beta$-decay, $n \to p e^- \bar{\nu}_e$, $g_A \simeq 1.26$ [40]. In addition to the chiral derivative and the mass term, the above Lagrangian also includes a coupling term which involves the axial vector quantity $u_\mu$, which is given by[5]

$$u_\mu = -\frac{1}{f_\pi}\boldsymbol{\tau} \cdot \partial_\mu \boldsymbol{\pi} + \frac{4\alpha - 1}{2 f_\pi^3}(\boldsymbol{\tau} \cdot \boldsymbol{\pi})(\boldsymbol{\pi} \cdot \partial_\mu \boldsymbol{\pi}) + \frac{\alpha}{f_\pi^3}\boldsymbol{\pi}^2(\boldsymbol{\tau} \cdot \partial_\mu \boldsymbol{\pi}) + O(\pi^4). \tag{A.53}$$

## A. Appendix

The explicit form of Eq. (A.52) is then

$$\mathcal{L}_{\pi N}^{(d=1)} = \bar{\Psi}\left(i\gamma^\mu \partial_\mu - M - \frac{1}{4f_\pi^2}\gamma^\mu \boldsymbol{\tau} \cdot (\boldsymbol{\pi} \times \partial_\mu \boldsymbol{\pi}) - \frac{g_A}{2f_\pi}\gamma^\mu\gamma_5 \boldsymbol{\tau} \cdot \partial_\mu \boldsymbol{\pi} + \ldots\right)\Psi. \quad (A.54)$$

The term proportional to $g_A/2f_\pi$ is the pseudo-vector coupling of one pion to the nucleon, and the quadratic term proportional to $1/4f_\pi^2$ is known as the Weinberg-Tomozawa coupling. At chiral dimension $d = 2$ the pion-nucleon Lagrangian is given by

$$\mathcal{L}_{\pi N}^{(d=2)} = \sum_{i=1}^{4} c_i \, \bar{\Psi} \mathcal{O}_i^{(2)} \Psi, \quad (A.55)$$

where $c_i$ are the first low-energy constants. The various operators $\mathcal{O}_i^{(2)}$ are such that they represent all terms at chiral dimension two that are consistent with chiral symmetry and Lorentz invariance. For the explicit form of these operators we refer to Ref. [283]. The relativistic treatment of nucleons leads to certain problems [356], which can be avoided by treating nucleons as heavy static sources [222, 39]. We begin by reparametrizing the nucleon four-momentum:

$$p^\mu = Mv^\mu + l^\mu, \quad (A.56)$$

where $v^\mu$ is the relativistic four-velocity satisfying $v^\mu v_\mu = 1$, and $l^\mu$ a small residual momentum, i.e. $v^\mu l_\mu \ll M$. We define the projection operators

$$P_v^\pm = \frac{1 \pm \gamma_\mu v^\mu}{2}, \qquad P_v^+ + P_v^- = 1, \quad (A.57)$$

where $P^+$ projects on the large component of the nucleon field, which can be seen by considering its Fourier expansion

$$\Psi(x) = \sum_{\sigma,\tau} \int d^3p \, u(x; \vec{p}, \sigma, \tau) \, a^\dagger(\vec{p}, \sigma, \tau), \qquad l = 1, 2, 3, 4. \quad (A.58)$$

Here, $\sigma$ is the spin projection quantum number, $\tau$ the isospin projection quantum number, and $a^\dagger(\vec{p}, \sigma, \tau)$ the operator which creates a nucleon with momentum $\vec{p}$ and quantum numbers $\sigma$ and $\tau$. The coefficient $u(x; \vec{p}, \sigma, \tau)$ is given by the plane-wave solution of the free Dirac equation

$$(i\gamma^\mu \partial_\mu - M)\, u(x) = 0. \quad (A.59)$$

In the Dirac representation it is given by (we omit the isospin part)

$$u(x) = \sqrt{\frac{p_0 + M}{2M}} \begin{pmatrix} \chi_s \\ \frac{\vec{\sigma} \cdot \vec{p}}{p_0 + M}\chi_s \end{pmatrix} e^{-ip^\mu x_\mu}, \quad (A.60)$$

where $p_0 = \sqrt{\vec{p}^2 + M}$, $\vec{\sigma} = (\sigma_1, \sigma_2, \sigma_3)$ are the Pauli spin matrices, and $\chi_s$ is the Pauli spinor for spin-1/2. The operator $P^+$ projects on the large upper component of $u(x)$, and $P^-$ on the lower component ($\sim \vec{\sigma} \cdot \vec{p}$), which is small in the static limit. Defining the *velocity dependent fields*

$$N = e^{iMv^\mu x_\mu} P_v^+ \Psi, \qquad h = e^{iMv^\mu x_\mu} P_v^- \Psi, \quad (A.61)$$

---
[5] Here, $\alpha$ is an arbitrary coefficient that corresponds to different parametrizations of $U$ by pions field, cf. footnote[4].

## A. Appendix

we can write the nucleon field as

$$\Psi = e^{-iM v^\mu x_\mu}(N + h). \tag{A.62}$$

The Euler-Lagrange equations corresponding to $\mathscr{L}_{\pi N}^{(d=1)}$ allow to express the small component $h$ in terms of $N$. These corrections enter at second order and are suppressed by a factor $M^{-1}$. Hence, the heavy-baryon projected first-order pion-nucleon Lagrangian is given by Eq. (A.54) with the substitutions

$$\bar{\Psi} \to e^{iM v^\mu x_\mu} N^\dagger, \qquad \Psi \to e^{-iM v^\mu x_\mu} N. \tag{A.63}$$

Assuming that $v^\mu = (1, 0, 0, 0)$ this leads to

$$\mathscr{L}_{\pi N}^{(d=1)} = N^\dagger \left( i\partial_0 - \frac{1}{4f_\pi^2} \boldsymbol{\tau} \cdot (\boldsymbol{\pi} \times \partial_0 \boldsymbol{\pi}) - \frac{g_A}{2f_\pi} \boldsymbol{\tau} \cdot (\vec{\sigma} \cdot \vec{\nabla}) \boldsymbol{\pi} \right. \\
\left. - \frac{g_A(4\alpha - 1)}{4f_\pi^3} (\boldsymbol{\tau} \cdot \boldsymbol{\pi})[\boldsymbol{\pi} \cdot (\vec{\sigma} \cdot \vec{\nabla})\boldsymbol{\pi}] + \frac{g_A \alpha}{2f_\pi^3} \boldsymbol{\pi}^2 [\boldsymbol{\tau} \cdot (\vec{\sigma} \cdot \vec{\nabla})\boldsymbol{\pi}] + \ldots \right) N, \tag{A.64}$$

where the ellipsis stands for terms involving four or more pion fields. In addition to pion-nucleon interactions, nucleon contact terms are needed in the chiral Lagrangian to compensate the divergences arising from loop-diagrams, and to parametrize the unresolved short-distance dynamics. Invariance under parity transformations restricts the nucleon contact interactions to involve only even powers of derivatives:

$$\mathscr{L}_{NN} = \mathscr{L}_{NN}^{(d=0)} + \mathscr{L}_{NN}^{(d=2)} + \mathscr{L}_{NN}^{(d=4)} + \ldots \tag{A.65}$$

The lowest-order two-nucleon contact Lagrangian is given by (with low-energy constants $C_S$ and $C_T$)

$$\mathscr{L}_{NN}^{(d=0)} = -\frac{1}{2} C_S (N^\dagger N)(N^\dagger N) - \frac{1}{2} C_T (N^\dagger \vec{\sigma} N) \cdot (N^\dagger \vec{\sigma} N), \tag{A.66}$$

which generates the leading-order (LO) two-nucleon contact interaction [cf. Eq. (1.33)]. At the next order $d = 2$ there are also three-nucleon contact terms, and contact terms with additional pions:

$$\mathscr{L}_{NN}^{(d=2)} = -\frac{D}{4f_\pi} (N^\dagger N) \left( N^\dagger \left[ \boldsymbol{\tau} \cdot (\vec{\sigma} \cdot \vec{\nabla}) \boldsymbol{\pi} \right] N \right) - \frac{1}{2} E (N^\dagger N)(N^\dagger \boldsymbol{\tau} N) \cdot (N^\dagger \boldsymbol{\tau} N) + \ldots, \tag{A.67}$$

where the ellipsis represents further two-nucleon contact interactions. The part proportional to the low-energy constant $D$ gives rise to a vertex involving two in- and outgoing nucleon lines as well as one pion line, while the $E$ part represents a pure three-nucleon contact interaction.

# A. Appendix

## A.3. Effective In-Medium Two-Body Potential

Here, we given the explicit expressions for the effective in-medium two-body (DDNN) potential constructed from the next-to-next-to-leading order (N2LO) chiral three-nucleon interactions (for more details, we refer to Refs. [206, 355, 205, 71]). The expressions are valid for the center-of-mass frame approximation [206, 205, 71] where the dependence on the total nucleon momentum is neglected. Within the canonical finite-temperature many-body perturbation series, the effective two-body potential depends on the temperature $T$ as well as the neutron and proton auxiliary chemical potentials $\tilde{\mu}_n$ and $\tilde{\mu}_p$. It can be decomposed as $V_{\text{DDNN}}(T, \tilde{\mu}_n, \tilde{\mu}_p) = \sum_{i=1}^{6} V_{\text{DDNN}}^{(i)}$, where the different contributions correspond to the respective diagrams shown in Fig. A.2.

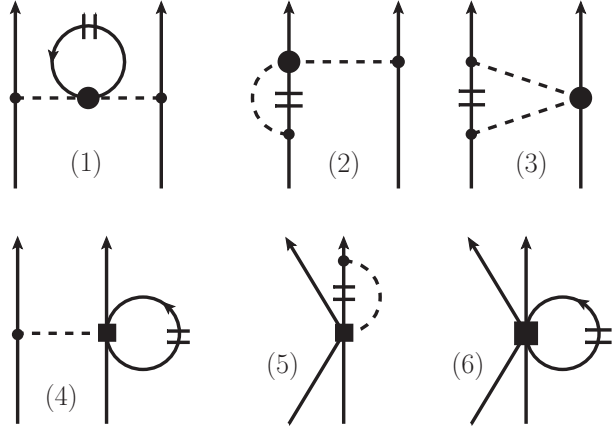

**Figure A.2.:** (From Ref. [355]) Diagrams for the effective DDNN potential corresponding to $i = 1, \ldots, 6$ in $V_{\text{DDNN}}^{(i)}$. The double dashes (the "medium insertion") indicate that the momenta of the associated nucleon lines are distributed according the Fermi-Dirac distribution.

The first three contribution to the DDNN potential come from the two-pion exchange part of the N2LO chiral 3N interaction. The contribution from the Pauli-blocked self-energy diagram is

$$V_{\text{DDNN}}^{(1)} = \frac{g_A^2 M}{8 f_\pi^4} \boldsymbol{\tau_1} \cdot \boldsymbol{\tau_2} \frac{\vec{\sigma}_1 \cdot \vec{q}\, \vec{\sigma}_2 \cdot \vec{q}}{(m_\pi^2 + q^2)^2} (2 c_1 m_\pi^2 + c_3 q^2)(\rho_n + \rho_p), \tag{A.68}$$

and the one from the Pauli blocked vertex correction is

$$\begin{aligned}
V_{\text{DDNN}}^{(2)} =& \frac{g_A^2 M}{64 \pi^2 f_\pi^4} \boldsymbol{\tau_1} \cdot \boldsymbol{\tau_2} \frac{\vec{\sigma}_1 \cdot \vec{q}\, \vec{\sigma}_2 \cdot \vec{q}}{(m_\pi^2 + q^2)^2} \Big\{ -4 c_1 m_\pi^2 \big[\Gamma_0^+(p) + \Gamma_1^+(p)\big] - (c_3 + c_4) \\
& \times \big[q^2 \Gamma_0^+(p) + 2 q^2 \Gamma_1^+(p) + q^2 \Gamma_3^+(p) + 4 \Gamma_2^+(p)\big] + 4 c_4 \big[2 \pi^2 (\rho_n + \rho_p) - m_\pi^2 \Gamma_0^+(p)\big]\Big\} \\
& + \frac{g_A^2 M}{128 \pi^2 f_\pi^4} (\tau_1^3 + \tau_2^3) \frac{\vec{\sigma}_1 \cdot \vec{q}\, \vec{\sigma}_2 \cdot \vec{q}}{(m_\pi^2 + q^2)^2} \Big\{ -4 c_1 m_\pi^2 \big[\Gamma_0^-(p) + \Gamma_1^-(p)\big] - (c_3 - c_4) \\
& \times \big[q^2 \Gamma_0^-(p) + 2 q^2 \Gamma_1^-(p) + q^2 \Gamma_3^-(p) + 4 \Gamma_2^-(p)\big] + 4 c_4 \big[2 \pi^2 (\rho_n - \rho_p) + m_\pi^2 \Gamma_0^-(p)\big]\Big\} \\
& + \frac{g_A^2 M}{64 \pi^2 f_\pi^4} (\tau_1^3 - \tau_2^3) i (\vec{\sigma}_1 - \vec{\sigma}_2) \cdot (\vec{p} \times \vec{q}) \frac{1}{m_\pi^2 + 4 p^2 - q^2} \\
& \times \Big\{ 4 c_1 m_\pi^2 \big[\Gamma_0^-(p) + \Gamma_1^-(p)\big] + c_3 (4 p^2 - q^2)\big[\Gamma_0^-(p) + 2 \Gamma_1^-(p) + \Gamma_3^-(p)\big]\Big\}. \tag{A.69}
\end{aligned}$$

## A. Appendix

The last term $\sim (\tau_1^3 - \tau_2^3)$ in $V_{\text{DDNN}}^{(2)}$ leads to spin singlet-triplet mixing [355] and has been neglected in the calculations performed in this thesis. The Pauli blocked two-pion exchange contribution reads

$$\begin{aligned}
V_{\text{DDNN}}^{(3)} = &\frac{g_A^2 M}{128\pi^2 f_\pi^4} \Big\{ -12c_1 m_\pi^2 \big[2\Gamma_0^+(p) - (2m_\pi^2 + q^2)G_0^+(p,q)\big] \\
&- 3c_3 \big[8\pi^2(\rho_n + \rho_p) - 4(2m_\pi^2 + q^2)\Gamma_0^+(p) - 2q^2\Gamma_1^+(p) + (2m_\pi^2 + q^2)^2\Gamma_0^+(p)\big] \\
&+ 4c_4 \,\boldsymbol{\tau_1}\cdot\boldsymbol{\tau_2}\,(\vec{\sigma}_1\cdot\vec{\sigma}_2 q^2 - \vec{\sigma}_1\cdot\vec{q}\,\vec{\sigma}_2\cdot\vec{q})G_2^+(p,q) - (3c_3 + c_4\boldsymbol{\tau_1}\cdot\boldsymbol{\tau_2})\,i(\vec{\sigma}_1 + \vec{\sigma}_2)\cdot(\vec{q}\times\vec{p}) \\
&\times \big[2\Gamma_0^+(p) + 2\Gamma_1^+(p) - (2m_\pi^2 + q^2)[G_0^+(p,q) + 2G_1^+(p,q)]\big] \\
&- 12c_1 m_\pi^2\, i(\vec{\sigma}_1 + \vec{\sigma}_2)\cdot(\vec{q}\times\vec{p})\big[G_0^+(p,q) + 2G_1^+(p,q)\big] \\
&+ 4c_4\,\boldsymbol{\tau_1}\cdot\boldsymbol{\tau_2}\,\vec{\sigma}_1\cdot(\vec{q}\times\vec{p})\,\vec{\sigma}_2\cdot(\vec{q}\times\vec{p})\big[G_0^+(p,q) + 4G_1^+(p,q) + 4G_3^+(p,q)\big]\Big\} \\
&+ \frac{g_A^2 M}{128\pi^2 f_\pi^4}\,(\tau_1^3 + \tau_2^3)\,\Big\{4c_1 m_\pi^2\big[2\Gamma_0^-(p) - (2m_\pi^2 + q^2)G_0^-(p,q)\big] \\
&+ c_3\big[8\pi^2(\rho_n - \rho_p) - 4(2m_\pi^2 + q^2)\Gamma_0^-(p) - 2q^2\Gamma_1^-(p) + (2m_\pi^2 + q^2)^2\Gamma_0^-(p)\big] \\
&- 4c_4\,(\vec{\sigma}_1\cdot\vec{\sigma}_2 q^2 - \vec{\sigma}_1\cdot\vec{q}\,\vec{\sigma}_2\cdot\vec{q})G_2^-(p,q) + (c_3 + c_4)\,i(\vec{\sigma}_1 + \vec{\sigma}_2)\cdot(\vec{q}\times\vec{p}) \\
&\times \big[2\Gamma_0^-(p) + 2\Gamma_1^-(p) - (2m_\pi^2 + q^2)[G_0^-(p,q) + 2G_1^-(p,q)]\big] \\
&+ 4c_1 m_\pi^2\,i(\vec{\sigma}_1 + \vec{\sigma}_2)\cdot(\vec{q}\times\vec{p})\big[G_0^-(p,q) + 2G_1^-(p,q)\big] \\
&- 4c_4\,\vec{\sigma}_1\cdot(\vec{q}\times\vec{p})\,\vec{\sigma}_2\cdot(\vec{q}\times\vec{p})\big[G_0^-(p,q) + 4G_1^-(p,q) + 4G_3^-(p,q)\big]\Big\}. \quad (A.70)
\end{aligned}$$

Next, the one-pion exchange contributions proportional to the low-energy constant $D$ are given by

$$V_{\text{DDNN}}^{(4)} = D\frac{g_A^2 M}{64}\Big\{-2\boldsymbol{\tau_1}\cdot\boldsymbol{\tau_2}\,(\rho_n + \rho_p) + (\tau_1^3 + \tau_2^3)(\rho_n - \rho_p)\Big\}\frac{\vec{\sigma}_1\cdot\vec{q}\,\vec{\sigma}_2\cdot\vec{q}}{m_\pi^2 + q^2}, \quad (A.71)$$

$$\begin{aligned}
V_{\text{DDNN}}^{(5)} = &D\frac{g_A^2 M}{128\pi^2}\Big\{\boldsymbol{\tau_1}\cdot\boldsymbol{\tau_2}\Big[2\vec{\sigma}_1\cdot\vec{\sigma}_2\Gamma_2^+(p) + \big[\vec{\sigma}_1\cdot\vec{\sigma}_2\,(2p^2 - q^2/2) + \vec{\sigma}_1\cdot\vec{q}\,\vec{\sigma}_2\cdot\vec{q}\,(1 - 2p^2/q^2) \\
&- \frac{2}{q^2}\vec{\sigma}_1\cdot(\vec{q}\times\vec{p})\,\vec{\sigma}_2\cdot(\vec{q}\times\vec{p})\big]\big[\Gamma_0^+(p) + 2\Gamma_1^+(p) + \Gamma_2^+(p)\big]\Big] + 12\pi^2(\rho_n + \rho_p) - 6m_\pi^2\Gamma_0^+(p)\Big\} \\
&+ D\frac{g_A^2 M}{256\pi^2}(\tau_1^3 + \tau_2^3)\Big\{2\vec{\sigma}_1\cdot\vec{\sigma}_2\Gamma_2^-(p) + \big[\vec{\sigma}_1\cdot\vec{\sigma}_2\,(2p^2 - q^2/2) + \vec{\sigma}_1\cdot\vec{q}\,\vec{\sigma}_2\cdot\vec{q}\,(1 - 2p^2/q^2) \\
&- \frac{2}{q^2}\vec{\sigma}_1\cdot(\vec{q}\times\vec{p})\,\vec{\sigma}_2\cdot(\vec{q}\times\vec{p})\big]\big[\Gamma_0^-(p) + 2\Gamma_1^-(p) + \Gamma_2^-(p)\big] + 4\pi^2(\rho_n - \rho_p) + 2m_\pi^2\Gamma_0^-(p)\Big\}. \\
&\quad (A.72)
\end{aligned}$$

Finally, the contribution from the contact 3N interaction proportional to the low-energy constant $E$ is

$$V_{\text{DDNN}}^{(6)} = E\frac{3M}{16}\Big\{-2(\rho_n + \rho_p) + (\tau_1^3 + \tau_2^3)(\rho_n - \rho_p)\Big\}. \quad (A.73)$$

In Eqs. (A.69), (A.70) and (A.72), the functions $\Gamma_j^\pm$ correspond to integrals over a pion propagator. They are given by

$$\Gamma_0^\pm(p) = \frac{1}{2p}\int_0^\infty dk\,k\,\ln\left(\frac{m_\pi^2 + (p+k)^2}{m_\pi^2 + (p-k)^2}\right)\big[n_k^p \pm n_k^n\big], \quad (A.74)$$

## A. Appendix

$$\Gamma_1^{\pm}(p) = \frac{1}{4p^3} \int_0^{\infty} dk\, k \left\{ 4pk - (m_\pi^2 + p^2 + k^2) \ln\left(\frac{m_\pi^2 + (p+k)^2}{m_\pi^2 + (p-k)^2}\right) \right\} \left[n_k^{\text{p}} \pm n_k^{\text{n}}\right], \quad (A.75)$$

$$\Gamma_2^{\pm}(p) = \frac{1}{16p^3} \int_0^{\infty} dk\, k \left\{ 4pk(m_\pi^2 + p^2 + k^2) - \left[m_\pi^2 + (p+k)^2\right]\left[m_\pi^2 + (p-k)^2\right] \right.$$
$$\left. \times \ln\left(\frac{m_\pi^2 + (p+k)^2}{m_\pi^2 + (p-k)^2}\right) \right\} \left[n_k^{\text{p}} \pm n_k^{\text{n}}\right], \quad (A.76)$$

$$\Gamma_3^{\pm}(p) = \frac{1}{16p^5} \int_0^{\infty} dk\, k \left\{ -12pk(m_\pi^2 + p^2 + k^2) + \left[3(m_\pi^2 + p^2 + k^2)^2 - 4p^2 k^2\right] \right.$$
$$\left. \times \ln\left(\frac{m_\pi^2 + (p+k)^2}{m_\pi^2 + (p-k)^2}\right) \right\} \left[n_k^{\text{p}} \pm n_k^{\text{n}}\right], \quad (A.77)$$

The functions $G_j^{\pm}(p)$ result from integrals over two pion propagators. They are given by Eqs. (19)-(22) of Ref. [206] together with

$$G_{0,*,**}^{\pm}(p) = \frac{1}{2p} \int_0^{\infty} dk\, \frac{\{k, k^3, k^5\}}{\sqrt{A(p) + q^2 k^2}} \ln\left(\frac{qk + \sqrt{A(p) + q^2 k^2}}{\sqrt{A(p)}}\right) \left[n_k^{\text{p}} \pm n_k^{\text{n}}\right], \quad (A.78)$$

where $A(p) = (m_\pi^2 + (p+k)^2)(m_\pi^2 + (p-k)^2)$.

Numerical simulations of core-collapse supernovæ and binary neutron-star mergers probe the nuclear EoS over a wide range of densities, temperature, and isospin asymmetries. Within the framework employed in this thesis, the computation of such a global nuclear EoS requires the computation of the matrix elements of the antisymmetrized effective two-body potential $\bar{V}_{\text{DDNN}}(T, \rho, \delta)$ for each value of $T$, $\rho$ and $\delta$. To facilitate this procedure, it may be useful to employ the isospin-asymmetry expansion of the nn-, pp- and np-channel components of $\bar{V}_{\text{DDNN}}(T, \rho, \delta)$, i.e.,

$$\bar{V}_{\text{DDNN}}^{\text{nn/pp}}(T, \rho, \delta) \simeq \sum_{n=0}^{N} \nu_n^{\text{nn/pp}}(T, \rho)\, \delta^n, \quad (A.79)$$

$$\bar{V}_{\text{DDNN}}^{\text{np}}(T, \rho, \delta) \simeq \sum_{n=0}^{N} \nu_{2n}^{\text{np}}(T, \rho)\, \delta^{2n}, \quad (A.80)$$

where the different expansion coefficients are given by

$$\nu_n^{\text{nn/pp/np}}(T, \rho) = \frac{1}{n!} \left.\frac{\partial^n \bar{V}_{\text{DDNN}}^{\text{nn/pp/np}}(T, \rho, \delta)}{\partial \delta^n}\right|_{\delta=0}. \quad (A.81)$$

Note that the expansion of the nn- and pp-channel components involves also odd term in $\delta$. In Figs. A.3 and A.4, selected nn- and np-channel partial-wave amplitudes of the antisymmetrized effective two-body potential are shown for $\delta = 0, 0.5, 0.9$. Also shown are the results for the expansion coefficients, $\nu_{1,2,3}^{\text{nn}}(T, \rho)$ and $\nu_{2,4,6}^{\text{np}}(T, \rho)$, respectively. For $\delta = 0.9$, we compare the full isospin-asymmetry dependent results with the ones obtained from Eqs. (A.79) and (A.80), truncated at order $N = 3$. One sees that the expansion of the effective two-body potential behaves similar to the one of the first-order many-body contribution from two-nucleon interactions [$\bar{F}_{1,\text{NN}}$], i.e., the coefficients $\nu_n^{\text{nn/pp}}(T, \rho)$ and $\nu_{2n}^{\text{np}}(T, \rho)$ are hierarchically ordered, and the expansion is well-converged at low orders.

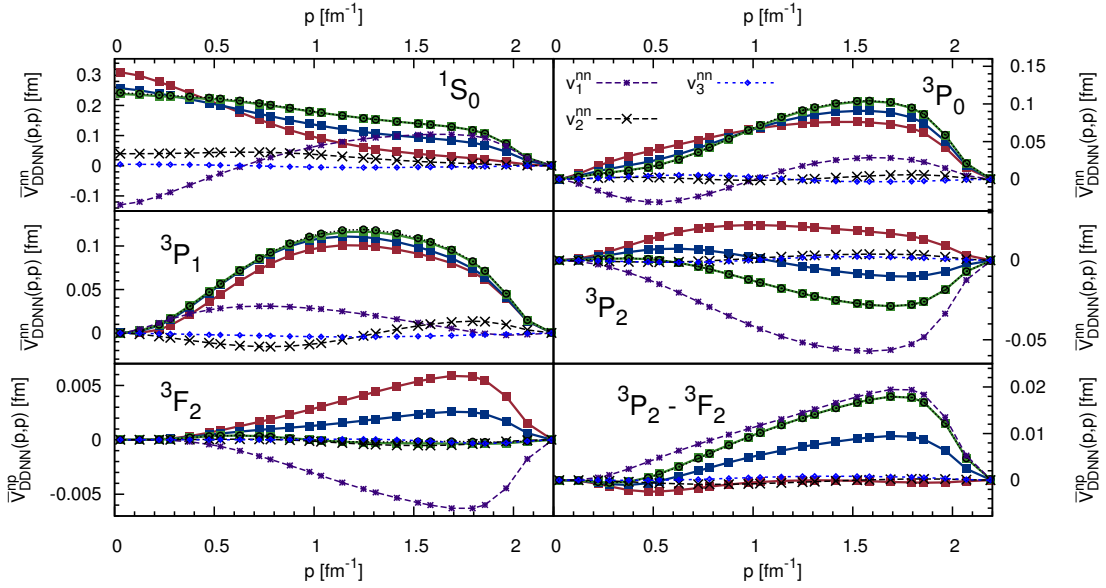

**Figure A.3.:** Same as Fig. A.4 but for the nn-channel, see text for details.

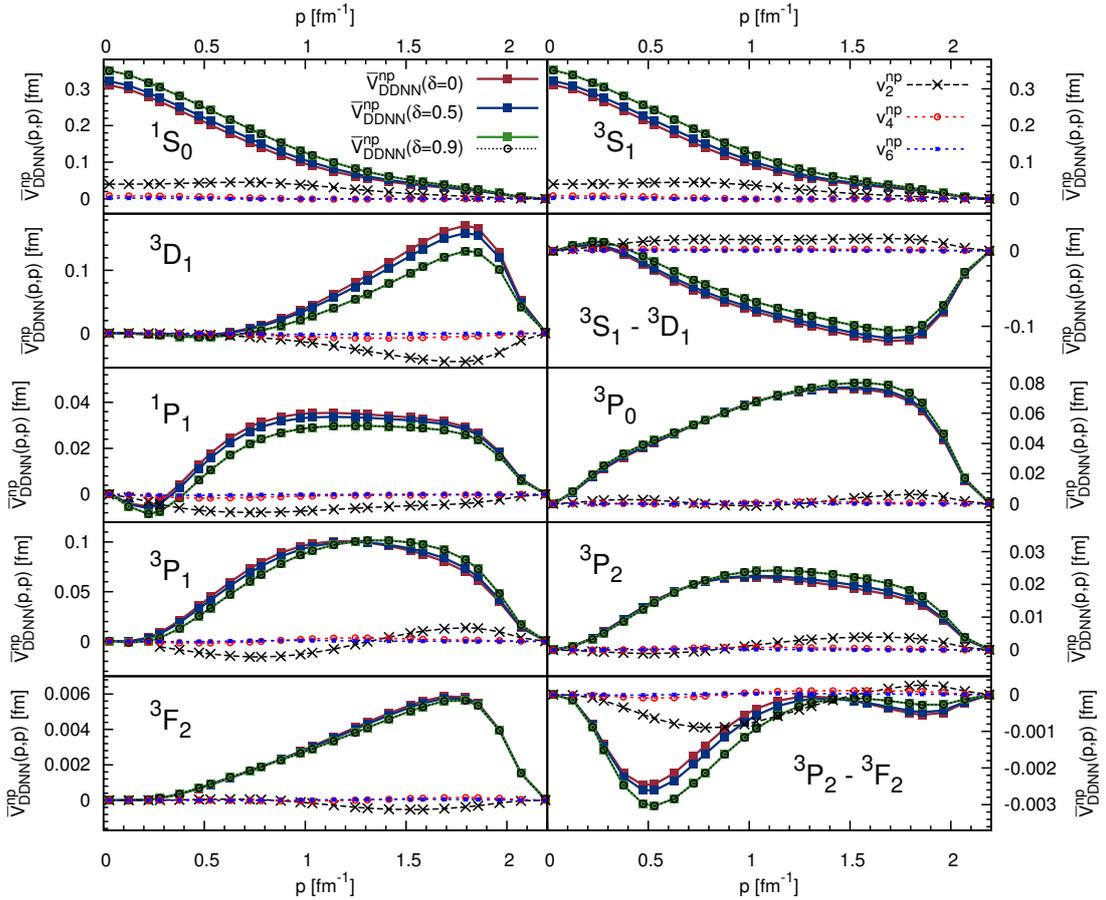

**Figure A.4.:** Selected partial-wave matrix elements of the np-channel n3lo414 DDNN interaction at $T = 5\,\text{MeV}$ and $\rho = 0.15\,\text{fm}^{-3}$, see text for details. In the case of $\delta = 0.9$, the green solid lines show the exact matrix elements, and the black dotted lines the ones obtained from the Maclaurin expansion in $\delta$ (truncated at order $N = 3$).

# B. Bibliography

## B. Bibliography

# Acknowledgements

I am grateful to my thesis supervisor Norbert Kaiser for having given me the opportunity to do the research that led to this thesis. I thank him for his invaluable guidance and support, his confidence in my research, for being always available for questions and discussions, and for providing his expertise, in chiral effective field theory in particular and theoretical physics (as well as mathematics) in general, whenever the situation called for his advice. I also thank him for trusting me with assisting his various lectures.

I thank my co-supervisor Jeremy W. Holt for being a reliable and supportive contact and a fantastic coauthor. In particular, I thank him for sharing his expertise in nuclear physics in many discussions, for motivating various projects and research directions that have contributed strongly in shaping the present thesis, and for his invaluable help regarding several technical and computational issues that have challenged its success. I thank him also for inviting me to the ECT* workshop on neutron stars and to the ICNT workshop.

I thank Wolfram Weise for constructive and encouraging comments and stimulating discussions at numerous seminars, for valueable input and words of wisdom for my first paper, for numerous invitations to ECT*, for his appreciation (apparently) of Pink Floyd, and for inviting me to his fantastic birthday symposium and to the ECT* Doctoral Training Program (DTP).

I thank all the ECT* DTP lecturers and the organizers and all the students that participated; in particular, I thank Guiseppe Colucci for Guinness-fueled conversations and a few very valuable advices; Miguel Gullón, Heiko Möller and Sanjin Benić for joint mountaineering expeditions; Daniele Viganò and Miguel Gullón for initiating a commemorable (and possibly, illegal) three-player soccer match (at an ungodly hour and a doubtful venue); Kota Masuda for having been a pleasant roommate. I apologize to Kota Masuda for my scattiness leading to unintentional fare-beating on his side during his visit to Munich (—if you were caught, please tell me, I'll repay!).

I thank Arianna Carbone and Christian Drischler for many discussions, and Christian Drischler for offering caffeinated refreshment during my first visit to Darmstadt. I thank Achim Schwenk for inviting me to present my research in the Darmstadt theory seminar and everyone in his group for a pleasant stay, in particular Kai Hebeler and Joel Lynn for sharing their office with me.

I thank Ingo Tews and Christian Drischler for exploring the East Lansing Party Zone with me.

I thank Sanjay Reddy and Chuck Horowitz for organizing the INT summer school, the other lecturers and all other people involved in the organization of that summer school.

Many thanks to my PhD colleagues Paul Springer, Stefan Petschauer and Susanne Strohmeier for providing an enjoyable work environment and for accomodating me into their office after I had to vacate my previous office in consequence of higher powers.

I thank my late-stage office mates Bolin Huang and Qibo Chen for numerous intense table-tennis matches that helped to spice up the work hours.

I thank Stefan Petschauer for destroying me not too much (that was taken care of by the metal lattice fence) in an imbalanced tennis match, and both Stefan Petschauer and Paul Springer for many physics and non-physics discussions and activities. Both must also be thanked for co-operating in various lecture assistantships (many thanks also to Andreas Trautner for this support in the AQFT tutorial).

Many thanks to Susanne Tillich for providing refreshment at numerous seminars and for her tiredless support concerning the administrative issues related to my various research travels.

Finally, I thank all my ex-colleagues for various discussions and occasional joint consumptions; in particular, I thank Matthias Drews for having set up numerous licentious festivities, and Robert Lang for providing entertainment with his peculiar (and slightly decadent) humor.

Last but not least, I thank my friends and family, in particular my mother for having supported my studies over the years.